\documentclass{aa}  

\usepackage{graphicx}
\usepackage{txfonts,textcomp}

\usepackage{natbib}
\usepackage{hyperref}
\usepackage{pdflscape}
\usepackage{ltcaption}
\usepackage{capt-of}
\hypersetup{
    colorlinks=True,
    linkcolor=blue,
    citecolor=blue,
    urlcolor=blue
    }

\newcommand{\nh}{NH$_3$}
\newcommand{\kms}{km s$^{-1}$}
\newcommand{\ucold}{10$^{15}$ cm$^{-2}$}
\newcommand{\tk}{$T_{\rm K}$}
\newcommand{\tr}{$T_{\rm rot}$}

\newcommand{\oneone}{(1,1)}
\newcommand{\twotwo}{(2,2)}
\newcommand{\cyx}{Cygnus X}
\newcommand{\simi}{$\sim$}
\newcommand{\uden}{cm$^{-3}$}
\newcommand{\um}{$\mu$m}
\newcommand{\water}{H$_2$O}
\newcommand{\htwo}{H\begin{small}II\end{small}}
\newcommand{\lwb}{$\Delta v_{\rm blended}$}
\newcommand{\lwi}{$\Delta v_{\rm int}$}
\newcommand{\vlsr}{$v_{\rm lsr}$}
\newcommand{\td}{$T_{\rm dust}$}
\newcommand{\tg}{$T_{\rm gas}$}

\newcommand{\ra}{$\to$}

\newcommand{\vnt}{$\sigma_{v,{\rm NT}}$}
\newcommand{\vng}{$\sigma_{v,{\rm ng}}$}
\newcommand{\msun}{M$_\odot$}
\newcommand{\mng}{$\mathcal{M}_{\rm ng}$}
\newcommand{\mnt}{$\mathcal{M}_{\rm NT}$}

\DeclareUnicodeCharacter{2212}{-}
\DeclareUnicodeCharacter{FF0C}{+}

\graphicspath{{./}{figures/}}

\def\gridline#1{\vskip6pt\hbox to\hsize{#1}\vskip6pt}

\begin{document} 

   \title{Surveys of clumps, cores, and condensations in Cygnus X}

   \subtitle{Temperature and nonthermal velocity dispersion revealed by VLA NH3 observations\thanks{Table of the physical parameters of NH$_3$ fragments is only available in electronic form at the CDS via anonymous ftp to cdsarc.u-strasbg.fr (130.79.128.5) or via http://cdsweb.u-strasbg.fr/cgi-bin/qcat?J/A+A/* after publishing.}}

   \author{X. Zhang\inst{1,2,3}\orcid{0000-0002-8078-1841} \and K. Qiu\inst{1,3}\orcid{0000-0002-5093-5088} \and Q. Zhang\inst{4}\orcid{0000-0003-2384-6589} \and Y. Cao\inst{1,3}\orcid{0000-0002-6368-7570} \and Y. Cheng\inst{5}\orcid{0000-0002-8691-4588} \and J. Liu\inst{5}\orcid{0000-0002-4774-2998} \and Y. Wang\inst{1,3}\orcid{0000-0001-6630-0944} \and X. Lu\inst{6}\orcid{0000-0003-2619-9305} \and X. Pan\inst{1,3,4}\orcid{0000-0003-1337-9059}}
   \institute{School of Astronomy and Space Science, Nanjing University, 163 Xianlin Avenue, Nanjing 210023, PR China\\
              \email{kpqiu@nju.edu.cn} \and 
    Max-Planck-Institut für Radioastronomie, Auf dem Hügel 69, 53121 Bonn, Germany \and
   Key Laboratory of Modern Astronomy and Astrophysics (Nanjing University), Ministry of Education, Nanjing 210023, PR China \and
   Center for Astrophysics $|$ Harvard \& Smithsonian, 160 Concord Avenue, Cambridge, MA 02138, USA \and
   National Astronomical Observatory of Japan, 2-21-1 Osawa, Mitaka, Tokyo, 181-8588, Japan \and
   Shanghai Astronomical Observatory, Chinese Academy of Sciences, 80 Nandan Road, Shanghai 200030, PR China}

   \titlerunning{\nh\ observations of Cygnus X}

   \date{Received January 21, 2023; accepted February 15, 2024}

  \abstract
   {The physical properties, evolution, and fragmentation of massive dense cores (MDCs, \simi\ 0.1 pc) are fundamental pieces in our understanding of high-mass star formation.}
   {We aim to characterize the temperature, velocity dispersion, and fragmentation of the MDCs in the \cyx\ giant molecular cloud and to investigate the stability and dynamics of these cores.}
   {We present the Karl G. Jansky Very Large Array (VLA) observations of the \nh\ ($J,K$) = \oneone\ and \twotwo\ inversion lines towards 35 MDCs in \cyx, from which we calculated the temperature and velocity dispersion. We extracted 202 fragments (\simi\ 0.02 pc) from the \nh\ \oneone\ moment-0 maps with the {\small GAUSSCLUMPS} algorithm. We analyzed the stability of the MDCs and their \nh\ fragments through evaluating the corresponding kinetic, gravitational potential, and magnetic energies and the virial parameters.}
   {The MDCs in Cygnus X have a typical mean kinetic temperature \tk\ of \simi\ 20 K. Our virial analysis shows that many MDCs are in subvirialized states, indicating that the kinetic energy is insufficient to support these MDCs against their gravity. The calculated nonthermal velocity dispersions of most MDCs are at transonic to mildly supersonic levels, and the bulk motions make only a minor contribution to the velocity dispersion. Regarding the \nh\ fragments, with \tk\ \simi\ 19 K, their nonthermal velocity dispersions are mostly trans-sonic to subsonic. Unless there is a strong magnetic field, most \nh\ fragments are probably not in virialized states. We also find that most of the \nh\ fragments are dynamically quiescent, while only a few are active due to star formation activity.}
   {}


   \maketitle

\section{Introduction}\label{sec:1}

High-mass stars ($>$ 8 \msun) play a vital role in shaping their host galaxies via their intense feedback effects in the form of powerful outflows, strong stellar winds, and violent UV radiation \citep{2018ARA&A..56...41M}. However, the formation of high-mass stars is still an open question. Unlike the star formation model built in low-mass stars \citep{1993prpl.conf....3S}, accretion onto high-mass protostars is difficult to model because of the radiation pressure problem \citep{1987ApJ...319..850W} and the violent wind \citep{2015arXiv151103457K}. In observations, it is difficult to catch the high-mass stellar embryos as their birthplaces are usually far away and deeply embedded \citep{2007ARA&A..45..481Z}. The small populations and short timescales of critical evolutionary phases mean that the sample of high-mass stellar embryos is even more insufficient \citep{2007ARA&A..45..481Z}.

Several candidate theories of high-mass star formation have been proposed, such as competitive accretion in protostellar clusters \citep{1997MNRAS.285..201B,2001MNRAS.323..785B}, mergers of protostars \citep{1998MNRAS.298...93B}, turbulent core accretion \citep{2003ApJ...585..850M}, and inertial inflow \citep{2020ApJ...900...82P}. To verify these theories, it is necessary to observationally investigate the initial conditions of high-mass star formation. Observations with a range of tracers in a variety of wavelengths and resolutions have been carried out towards massive dense cores (MDCs), the possible birth places of high-mass stars \citep[e.g.,][]{2002ApJ...566..945B,2007AAP...476.1243M,2014ApJ...790...84L,2016PASA...33...30R,2019ApJS..241....1C,2019A&A...627A..85Z}. By characterizing the physical conditions of high-mass star formation with observations, we can test and provide practical constraints to the corresponding theories. 

The CENSUS project (Surveys of Clumps, CorEs, and CoNdenSations in CygnUS X, PI: Keping Qiu) is dedicated to a systematic study of the 0.01--10 pc hierarchical molecular structures and the high-mass star formation process in the molecular complex \cyx. In order to investigate the temperature structures and dynamic conditions of MDCs in our project, we make use of the \nh\ \oneone\ and \twotwo\ inversion emission lines, which trace gas with critical density of \simi\ 10$^4$ cm$^{-3}$ and excitation temperature of \simi\ 10--30 K \citep{1983ARA&A..21..239H,2015PASP..127..266M}. These lines are widely used for studying the initial conditions of star formation. Previous \nh\ surveys find a mean temperature of $\lesssim$ 15 K in less-evolved structures and \simi\ 20 K in protostellar cores with significant star-forming activity \citep[e.g.,][]{2006A&A...450..569P,2006A&A...450..607W,2013A&A...552A..40C,2019ApJ...874..147K}. The velocity dispersion of these lines is found to be dominated by nonthermal components in most cases \citep[e.g.,][]{2014ApJ...790...84L,2019MNRAS.483.3146B}. Furthermore, analyses of the dynamical stability of star-forming structures at various scales and evolutionary stages can be done with these lines via the virial analysis \citep[e.g.,][]{1999ApJS..125..161J,2020A&A...638A.105Z}. Therefore, a high-resolution unbiased \nh\ survey towards the MDCs in the \cyx\ cloud complex offers a remarkable observational sample that can be used to unravel the initial conditions of high-mass star formation.

The Cygnus X cloud, located at 1.4 kpc from the Sun \citep{2012A&A...539A..79R}, is one of the richest and most active molecular and \htwo\ complexes in the Galaxy. It is massive ($\sim 4 \times 10^6 M_\odot$) and large (extends over $\sim$ 100 pc in diameter) \citep{1992ApJS...81..267L}, and consists of substantial \htwo\ regions \citep{1991A&A...241..551W} and OB associations \citep{2001A&A...371..675U}. All these features make \cyx\ an ideal object for studying high-mass star formation. Several surveys and numerous case studies towards \cyx\ have been carried out over recent decades \citep[e.g.,][]{2007AAP...476.1243M,2012A&A...541A..79G,2019ApJ...883..156T}. The CENSUS project has published several works on Cygnus X: \citet{2019ApJS..241....1C} generated a complete sample of MDCs in Cygnus X and performed an analysis of their far-infrared emission; \citet{2022ApJ...927..185W} provide radio continuum views of the MDCs; \citet{2023arXiv231217455P} provide a survey of the possible circumstellar disks in Cygnux X; \citet{2023arXiv231204880Y} analyzed the high-velocity outflow using Submillimeter Array (SMA) observations of SiO (5-4). In the CENSUS project, we adopt the terminology that clumps, cores, and condensations correspond to cloud structures of a physical full width at half maximum (FWHM) size of \simi\ 1, \simi\ 0.1, and \simi\ 0.01 pc, respectively, which is also widely used in other studies \citep[e.g.,][]{2018ApJ...853....5P,2019A&A...627A..85Z}.

As a part of CENSUS, we report an unbiased \nh\ survey towards 37 fields in Cygnus X observed with the Karl G. Jansky Very Large Array (VLA) with a spatial resolution of \simi\ 0.02 pc. This paper is organized as follows. Section \ref{sec:2} outlines details of the sample and observational setups. In Sect. \ref{sec:3}, we describe the methods we used to analyze the \nh\ data. The derived physical parameters and their distributions and correlations are presented in Sect. \ref{sec:4}. The \nh\ morphology, a virial analysis, the dynamical states of MDCs, and a comparison between dust and gas temperatures are discussed in Sect. \ref{sec:5}. Finally, the study is summarized  in Sect. \ref{sec:6}, where we outline our conclusions.

\section{Sample and observations} \label{sec:2}

\subsection{Sample selection}

The interferometer observations of the CENSUS project \citep[PI: Keping Qiu, also see ][]{2019ApJS..241....1C} were designed to target most MDCs in Cygnus X \citep{2022ApJ...927..185W}. We first selected 42 sources with estimated masses of greater than 30 \msun\ from \citet{2007AAP...476.1243M}, though one source (S26) was later found to be outside of Cygnus X \citep{2012A&A...539A..79R} and another (N58) should have a much lower mass (see Table \ref{tab:obpro} for more details). In this work, we exclude 10 sources in the DR 21 region, which were observed by the VLA in a mosaic setup (Project ID: 14A-241; PI: Qizhou Zhang) and will be presented in a separate work. The other 32 sources, as well as an additional source (NW12, see Table \ref{tab:obpro} for more details), were observed with the VLA in 29 fields. In the present work, we also include VLA observations of another eight fields covering bright JCMT SCUBA-2 sources \citep{2019ApJS..241....1C} located in the OB2 association region\footnote{these sources were not included by \citet{2007AAP...476.1243M} as the OB2 region has a lower extinction while \citet{2007AAP...476.1243M} extracted sources from high-extinction regions.}. Here, VLA observations of a total of 37 fields are therefore presented (Table \ref{tab:obpro}). As part of the CENSUS project, \citet{2021ApJ...918L...4C} constructed a column density map of the entire Cygnus X complex based on SED fitting of multiband Herschel observations and extracted more than 8000 cores with the \emph{getsources} algorithm. In the following analysis, we adopt the parameters of dense cores from \citet{2021ApJ...918L...4C}.

\subsection{Very Large Array}\label{subsec:2.1}

The \nh\ data were obtained from the NRAO\footnote{The National Radio Astronomy Observatory is a facility of the National Science Foundation operated under cooperative agreement by Associated Universities, Inc.} VLA program 17A-107 (PI: Keping Qiu), which observed with 27 antennas in the D-array configuration. The program 17A-107 is an unbiased and high-resolution survey dedicated to observing the \nh\ (J,K)=(1,1), (2,2), (3,3), (4,4), and (5,5) rotation-inversion transitions, CCS (1,2--2,1) and CCS (6,5--5,5) emissions, and \water\ maser lines for MDCs in \cyx. The baselines of the array range from 0.035 km to 1.03 km, corresponding to the largest recoverable angular scale of $\sim$ 66$^{\prime \prime}$ (0.45 pc for \cyx) and an angular resolution of $\sim$ 3$^{\prime \prime}$.1 (0.02 pc for \cyx), respectively. The correlator setup provides eight independent spectral windows with a spectral channel spacing of 7.8 kHz and a bandwidth of 8 MHz, corresponding to a velocity resolution of 0.1 km s$^{-1}$ and a velocity coverage of 100 km s$^{-1}$ for the \nh\ \oneone\ and \twotwo\ lines. 

The VLA data were calibrated using the Common Astronomy Software Applications (CASA) package \citep{2007ASPC..376..127M}. The flux calibrator and bandpass calibrator were both set as the quasar 1331+305 (3C286). Gain calibration was performed with observations of J2007+4029. The calibrated data were imaged using the CASA task \texttt{clean}, gridded with a pixel size of 1$^{\prime \prime}$. The rms of these fields is $\lesssim$ 0.013 Jy beam$^{-1}$ in a 0.1 km s$^{−1}$ channel width. The coordinates of the field centers and noise levels are also listed in Table \ref{tab:obpro}. The \nh\ \oneone\ and \twotwo\ emission lines are the main interests of this paper. The rest frequencies of the \oneone\ and \twotwo\ emission lines are 23.6945 and 23.72263 GHz, while upper energy levels are 23.4 and 64.9 K, respectively. Observations of the other spectral lines in 17A-107 are beyond the scope of this paper, and are listed in Appendix \ref{app:A}. 

\subsection{MIR archive data from the \emph{Spitzer} Space Telescope}\label{subsec:2.2}

We use the \emph{Spitzer} IRAC 3.6, 4.5, and 8.0 \um\ data from the \emph{Spitzer} Legacy Survey of the \cyx\ Complex\footnote{\url{https://www.cfa.harvard.edu/cygnusX/}} \citep{2010AAS...21541401K} in this paper, which are downloaded from the NASA/IPAC Infrared Science Archive\footnote{\url{http://irsa.ipac.caltech.edu/frontpage/}}. All these data cover the entire \cyx\ region, with a pixel size of 2$^{\prime \prime}$ and 1$\sigma$ rms noise levels of 25.1, 27.1, and 47.9 MJy$\cdot$sr$^{-1}$, respectively.

\section{Methods} \label{sec:3}

We checked detections in all data cubes before fitting the \nh\ lines. In order to determine whether there is detection in a spectrum, the \nh\ \oneone\ data cubes are smoothed to a velocity resolution of 0.7 km s$^{-1}$, or seven spectral channels in our observations, to improve signal-to-noise ratios (S/Ns). The criterion of significant detection within a beam is set as S/N $\geq$ 5 in the smoothed \nh\ \oneone\ data. The corresponding original spectra were fitted in the following processes if their smoothed spectra exceeded the threshold. The detections and noise levels in all the fields are reported in Table \ref{tab:detec}. The detections in fields are marked as marginal if the emissions are only detected in one or two beams. We adopt the distance of 1.4 kpc \citep{2012A&A...539A..79R} for all the fields except for Field 21 (also known as AFGL2591), of which the distance is 3.3 kpc \citep{2012A&A...539A..79R}. 

\definecolor{Yellow}{rgb}{1,1,0}
\begin{table*}
\caption{Observational fields and related MDCs \label{tab:obpro}}
\centering
\begin{tabular}{c|ll|ccc}
\hline\hline
Field & MDCs (Cao21)\tablefootmark{a} & MDCs (Motte07)\tablefootmark{b} & R.A. & Decl. & Beam\tablefootmark{c} \\
 &  &  & (J2000) & (J2000) & ($^{\prime \prime}\times ^{\prime \prime}$; $^{\circ}$) \\
\hline
1 & 220 & N3 & 20:35:34.6 & 42:20:08.8 & 3.3$\times$2.6; -29.0 \\
2 & 341 & N63 & 20:40:05.4 & 41:32:13.1 & 3.1$\times$2.5; -24.0 \\
3 & 274 584 & N6 & 20:36:07.3 & 41:39:58.0 & 4.5$\times$3.1; -69.0 \\
4 & 725 1919 & N10 & 20:35:52.2 & 41:36:23.0 & 4.5$\times$3.1; -69.0 \\
5 & 248 774\tablefootmark{d} 1669 & N12 & 20:36:57.4 & 42:11:27.0 & 3.5$\times$3.4; 61.0 \\
6 & 714 & N14 & 20:37:00.9 & 41:34:57.0 & 4.3$\times$3.0; -69.0 \\
7 & 1267 3188 6479 & N24 & 20:38:04.6 & 42:39:54.0 & 3.5$\times$3.3; 61.0 \\
8 & 698 1179 & N56 & 20:39:16.9 & 42:16:07.0 & 3.6$\times$3.3; 58.0 \\
9 & 532 & N58\tablefootmark{e} & 20:39:25.9 & 41:20:01.0 & 3.5$\times$3.4; -11.0 \\
10 & 801 & N65 & 20:40:28.4 & 41:57:11.0 & 4.8$\times$3.2; -72.0 \\
11 & 684 & N68 & 20:40:33.5 & 41:59:03.0 & 4.6$\times$3.1; -71.0 \\
12 & 4315 4797 & N69 & 20:40:33.7 & 41:50:59.0 & 4.5$\times$3.1; -69.0 \\
13 & 327 619 742\tablefootmark{d} & NW1 NW2 & 20:19:39.0 & 40:56:45.0 & 4.4$\times$3.0; -63.0 \\
14 & 640 675 & NW5 & 20:20:30.5 & 41:21:40.0 & 3.5$\times$3.0; -70.0 \\
15 & 839\tablefootmark{e} & NW12\tablefootmark{f} & 20:24:14.3 & 42:11:43.0 & 3.5$\times$3.0; -70.0 \\
16 & 310 & NW14 & 20:24:31.7 & 42:04:23.0 & 3.6$\times$3.0; -70.0 \\
17 & 507 753 & S7 S8 & 20:20:38.6 & 39:38:00.0 & 4.1$\times$3.0; -80.0 \\
18 & 798 & S10 & 20:20:44.4 & 39:35:20.0 & 4.0$\times$2.9; -75.0 \\
19 & 874 & S15 & 20:27:14.0 & 37:22:58.0 & 3.5$\times$3.1; -61.0 \\
20 & 1201 1537 & S18 S20 & 20:27:26.7 & 37:22:45.0 & 3.5$\times$3.0; -61.0 \\
21 & 277\tablefootmark{g} & S26\tablefootmark{g} & 20:29:24.6 & 40:11:19.1 & 3.2$\times$2.6; 54.0 \\
22 & 723 & S29 & 20:29:58.8 & 40:15:58.0 & 3.3$\times$3.0; 87.0 \\
23 & 509 & S30 & 20:31:12.6 & 40:03:16.0 & 3.3$\times$2.9; -87.0 \\
24 & 351 & S32 & 20:31:20.3 & 38:57:16.0 & 3.3$\times$2.9; -86.0 \\
25 & 1225 & S34 & 20:31:57.8 & 40:18:30.0 & 4.7$\times$3.3; -75.0 \\
26 & 1454 2210 & S37 & 20:32:28.6 & 40:19:41.5 & 3.5$\times$3.0; -70.0 \\
27 & 892 & S41 & 20:32:33.4 & 40:16:43.0 & 3.5$\times$3.0; -70.0 \\
28 & 540 & S43 & 20:32:40.8 & 38:46:31.0 & 3.3$\times$2.9; -89.0 \\
29 & 1460 & -- & 20:35:00.0 & 41:34:57.0 & 5.0$\times$2.9; -69.0 \\
30 & 608 & -- & 20:34:00.0 & 41:22:25.0 & 4.9$\times$2.9; -68.0 \\
31 & 302 520 & -- & 20:35:09.5 & 41:13:30.0 & 4.3$\times$3.0; -71.0 \\
32 & 340 & -- & 20:32:22.5 & 41:07:56.0 & 4.0$\times$3.0; -69.0 \\
33 & 2976 & -- & 20:32:17.0 & 41:09:10.0 & 4.4$\times$3.0; 86.0 \\
34 & 214 247 & -- & 20:30:28.5 & 41:15:55.0 & 3.6$\times$3.0; -69.0 \\
35 & 370 893 & -- & 20:28:09.4 & 40:52:50.0 & 4.0$\times$3.0; -61.0 \\
36 & -- & -- & 20:28:05.5 & 40:51:17.0 & 3.9$\times$3.0; -59.0 \\
37 & 1112 1231 1244 & N30 N32 & 20:38:36.4 & 42:37:34.8 & 3.3$\times$2.6; -29.0 \\
\hline
\end{tabular}
\tablefoot{\tablefoottext{a}{MDCs from \citet{2021ApJ...918L...4C} covered by the field.}
\tablefoottext{b}{MDCs in \citet{2007AAP...476.1243M} covered by the field, extracted from 1.2 mm continuum data from IRAM 30-m telescope (with effective angular resolution of \simi\ 11$^{\prime\prime}$ and spatial resolution of \simi\ 0.09 pc).}
\tablefoottext{c}{The beams are shown as ``major axes $\times$ minor axes; position angles''.}
\tablefoottext{d}{Dense core 774 and 742 have mass of 25.0 and 19.4 \msun, slightly lower than 30 \msun. However, both of them have remarkable \nh\ detections in our observations. We therefore took them into the following analysis and do not distinguish them from other MDCs.}
\tablefoottext{e}{The mass of N58 could be highly overestimated in \citet{2007AAP...476.1243M} due to the significant contribution of free-free emission in this core. N58 is therefore excluded from the sample in \citet{2022ApJ...927..185W}. However, MDC 532 is included in this work as the SED-fitted mass of MDC 532 is 68.6 \msun, which is still larger than 30 \msun.}
\tablefoottext{f}{NW12 is an interesting isolated core in its parental clump. It is covered by VLA 14A-107 though its mass is lower than 30 \msun. The mass of MDC 839 extracted by \citet{2021ApJ...918L...4C} is also lower than 30 \msun.} 
\tablefoottext{g}{S26, also known as AFGL 2591, is excluded from our sample due to its different distance from Cygnux X region, of namely 3.3 kpc \citep{2012A&A...539A..79R}.}
}
\end{table*}

\begin{table*}
\caption{Detections in each field \label{tab:detec}}
\centering
\begin{tabular}{c|ccccccc|cccc|cccc|c}
\hline\hline
Field & & \multicolumn{5}{c}{\nh\ detection\tablefootmark{a}} & & & \multicolumn{2}{c}{r.m.s. level} & & & \multicolumn{2}{c}{Smoothed S/N (peak)\tablefootmark{b}} & & H$_2$O masers\tablefootmark{c} \\
\cline{3-7} \cline{10-11} \cline{14-15}
 & & (1,1) & (2,2) & (3,3) & (4,4) & (5,5) & & & (mJy Beam$^{-1}$) & (K) & & & (1,1) & (2,2) & & \\
 \hline
1 & & $\surd$ & $\surd$ & $\times$ & $\times$ & $\times$ & & & 11.5 & 3.0 & & & 11.7 & 11.0 & & -- \\
2 & & $\surd$ & $\surd$ & $\times$ & $\times$ & $\times$ & & & 9.7 & 2.8 & & & 7.9 & 6.0 & & -- \\
3 & & $\surd$ & $\surd$ & -- & -- & -- & & & 7.5 & 1.2 & & & 25.2 & 16.8 & & 1 \\
4 & & $\surd$ & $\surd$ & $\surd$ & -- & -- & & & 8.0 & 1.3 & & & 19.9 & 15.9 & & -- \\
5 & & $\surd$ & $\surd$ & $\surd$ & $\surd$ & * & & & 6.5 & 1.2 & & & 32.2 & 20.3 & & 1 \\
6 & & $\surd$ & $\surd$ & $\surd$ & * & -- & & & 7.8 & 1.3 & & & 12.1 & 7.2 & & 1 \\
7 & & $\surd$ & $\surd$ & -- & -- & -- & & & 6.5 & 1.2 & & & 18.7 & 10.2 & & 2 \\
8 & & $\surd$ & $\surd$ & $\surd$ & * & -- & & & 6.1 & 1.1 & & & 27.4 & 17.2 & & 1 \\
9 & & $\surd$ & $\surd$ & -- & -- & -- & & & 9.2 & 1.7 & & & 9.8 & 7.7 & & -- \\
10 & & $\surd$ & $\surd$ & $\surd$ & -- & -- & & & 9.5 & 1.4 & & & 16.5 & 11.1 & & -- \\
11 & & $\surd$ & $\surd$ & $\surd$ & -- & -- & & & 9.2 & 1.4 & & & 19.5 & 9.2 & & 1 \\
12 & & $\surd$ & $\surd$ & -- & -- & -- & & & 9.1 & 1.4 & & & 17.4 & 7.6 & & 1 \\
13 & & $\surd$ & $\surd$ & $\surd$ & -- & -- & & & 8.5 & 1.4 & & & 26.4 & 19.4 & & -- \\
14 & & $\surd$ & $\surd$ & -- & -- & -- & & & 8.4 & 1.7 & & & 14.7 & 5.7 & & 1 \\
15 & & $\surd$ & -- & -- & -- & -- & & & 8.4 & 1.7 & & & 7.4 & -- & & -- \\
16 & & $\surd$ & $\surd$ & * & -- & -- & & & 8.2 & 1.6 & & & 15.0 & 6.9 & & 1 \\
17 & & $\surd$ & $\surd$ & $\surd$ & -- & -- & & & 6.8 & 1.2 & & & 33.7 & 20.3 & & 2 \\
18 & & $\surd$ & $\surd$ & -- & -- & -- & & & 6.6 & 1.3 & & & 20.9 & 10.8 & & -- \\
19 & & $\surd$ & $\surd$ & -- & -- & -- & & & 8.2 & 1.6 & & & 20.9 & 14.7 & & -- \\
20 & & $\surd$ & $\surd$ & -- & -- & -- & & & 8.8 & 1.9 & & & 17.3 & 13.7 & & 2 \\
21 & & $\surd$ & $\surd$ & $\surd$ & $\surd$ & $\surd$ & & & 7.6 & 1.9 & & & 14.9 & 9.3 & & 2 \\
22 & & $\surd$ & $\surd$ & -- & -- & -- & & & 6.8 & 1.5 & & & 15.0 & 7.1 & & -- \\
23 & & $\surd$ & $\surd$ & $\surd$ & -- & -- & & & 7.0 & 1.6 & & & 22.2 & 13.2 & & 3 \\
24 & & $\surd$ & $\surd$ & $\surd$ & -- & -- & & & 7.4 & 1.7 & & & 16.7 & 8.8 & & -- \\
25 & & $\surd$ & $\surd$ & -- & -- & -- & & & 8.4 & 1.2 & & & 18.6 & 8.6 & & 3 \\
26 & & $\surd$ & $\surd$ & -- & -- & -- & & & 7.8 & 1.6 & & & 17.4 & 10.9 & & -- \\
27 & & $\surd$ & $\surd$ & $\surd$ & -- & -- & & & 7.5 & 1.5 & & & 22.2 & 20.6 & & -- \\
28 & & $\surd$ & $\surd$ & -- & -- & -- & & & 7.4 & 1.7 & & & 23.0 & 13.9 & & -- \\
29 & & $\surd$ & $\surd$ & -- & -- & -- & & & 11.6 & 1.7 & & & 11.6 & 7.6 & & -- \\
30 & & -- & -- & -- & -- & -- & & & 11.8 & 1.8 & & & -- & -- & & -- \\
31 & & $\surd$ & $\surd$ & -- & -- & -- & & & 8.5 & 1.4 & & & 11.8 & 8.3 & & -- \\
32 & & $\surd$ & $\surd$ & -- & -- & -- & & & 7.0 & 1.2 & & & 22.8 & 12.4 & & 1 \\
33 & & $\surd$ & * & -- & -- & -- & & & 8.9 & 1.5 & & & 13.6 & 7.7 & & -- \\
34 & & $\surd$ & $\surd$ & $\surd$ & -- & -- & & & 6.5 & 1.3 & & & 26.8 & 19.5 & & 1 \\
35 & & -- & -- & -- & -- & -- & & & 10.8 & 2.0 & & & -- & -- & & -- \\
36 & & -- & -- & -- & -- & -- & & & 11.2 & 2.1 & & & -- & -- & & -- \\
37 & & $\surd$ & $\surd$ & $\times$ & $\times$ & $\times$ & & & 12.0 & 3.1 & & & 11.3 & 10.4 & & -- \\
\hline
\end{tabular}
\tablefoot{\tablefoottext{a}{$\surd$ for remarkable detection, * for marginal detection, and -- for nondetection. $\times$ shows the lack of related cubes as we only obtain \nh\ \oneone\ and \twotwo\ cubes for Fields 1, 2, and 37.}
\tablefoottext{b}{Nondetection cases are shown as --}
\tablefoottext{c}{The number of \water\ maser spots.}}
\end{table*}

\subsection{Ammonia line fitting} \label{subsec:3.1}

Following the theoretical framework in \citet{2015PASP..127..266M}, we built spectral models of the \nh\ \oneone\ and \twotwo\ emission with the centroid velocity, line width, excitation temperature, \nh\ column density in the upper transition state, and beam-filling factor as free parameters and with the following assumptions: (1) the same beam-filling factor and excitation temperature for all hyperfine transitions; (2) local thermodynamic equilibrium (LTE). The \nh\ \oneone\ and \twotwo\ data cubes were fitted simultaneously using the model to derive the aforementioned physical parameters. 

We can get the relationship between the brightness temperature and excitation temperature as
\begin{equation} \label{equ:rt}
T_{\rm B}=f(T_{\rm ex}-T_{\rm bg})(1-{\rm e}^{-\tau_{\nu}})
,\end{equation}
where $f$ is the beam-filling factor and T$_{bg}$ is the background temperature. We take the temperature of cosmic microwave background (CMB) as T$_{bg}$, which is 2.73K. The optical depth $\tau_{\nu}$ can be written as
\begin{equation}
\tau_\nu=\frac{hc^3N_{u}A_{ul}}{8\pi\nu^2k_{\rm B}T_{\rm ex}}\Phi_\nu,
\end{equation}
where $N_u$ is the column density of molecules in the upper transition state. Here, we assume a Gaussian profile,
\begin{equation}
\Phi_{\nu}=\frac{1}{\sqrt[]{2{\rm \pi}}\Delta\nu}{\rm e}^\frac{(\nu-\nu_0)^2}{2(\Delta\nu)^2},
\end{equation}
where $\nu_0$ is the central frequency of the emission. The Einstein coefficient $A_{ul}$ is
\begin{equation}
A_{ul}\equiv\frac{64{\rm \pi}^4\nu^3\mu_{lu}^2}{3hc^3}\cdot\frac{K^2}{J(J+1)}\cdot R_s
,\end{equation}
where $\mu_{lu}$ and $R_s$ are the permanent dipole moment of \nh\ and the relative strengths of the hyperfine structures, respectively. In our model, $R_s$ is set as 1/2, 5/36, and 1/9 for the main, inner satellite, and outer satellite components of \nh\ \oneone\ emission, respectively. As for the main and inner satellite components of \nh\ \twotwo\ emission, $R_s$ is set as 0.796 and 0.052, respectively. We can derive the total \nh\ column density $N_{\rm tot}$ from the relationship between $N_{\rm u}$ and $N_{\rm tot}$ (Boltzmann distribution),
\begin{equation}
\frac{N_{\rm tot}}{N_{\rm u}}=\frac{Q_{\rm rot}}{g_u}{\rm e}^\frac{E_u}{{\rm k}T_{\rm ex}},
\end{equation}
where the rotation partition function $Q_{\rm rot}\approx 168.7\sqrt[]{T_{\rm ex}^3/B^2C}$ (B = 298.117 GHz, C = 186.726 GHz), and $g_u$ is the total degeneracy for the upper energy level, which can be found in \citet{2015PASP..127..266M}.

Using these formulas, we built a blended spectral model rather than the all-hyperfine model after comparing their fitted results. This model consists of five Gaussian components for \nh\ \oneone\ emission and three Gaussian components for \nh\ \twotwo\ emission. We do not include the two outer satellite components in our model as most NH$_3$ (2,2) lines do not show remarkable satellite components in our data. The excitation temperature and the column density are derived only when the S/Ns of the \twotwo\ moment-0 maps are greater than 2 in order to avoid large uncertainties.

\begin{figure}
    \centering 
    \includegraphics[width=8.5cm]{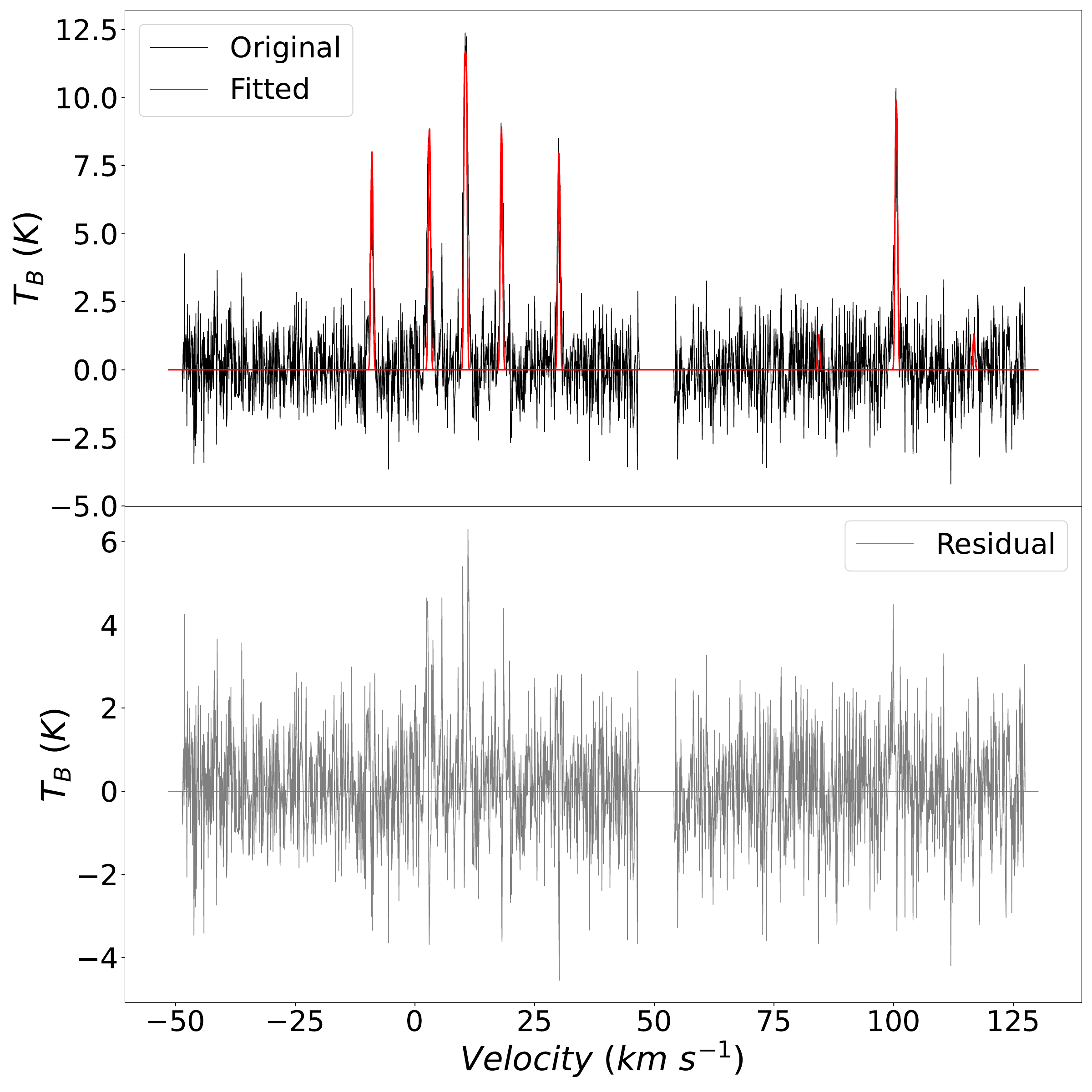} 
    \caption{Example five-component-fitting results on \nh\ \oneone\ and \twotwo\ emission (pixel coordinate (142,131) in Field 3). Upper panel: Original spectrum and the fitted result are colored in black and red, respectively. Lower panel: Residual spectrum of this fitting result.}
    \label{fig:smd}
\end{figure}

We then fitted the \nh\ \oneone\ and \twotwo\ simultaneously with five Gaussian components for \oneone\ emission and three Gaussian components for \twotwo\ emission. An example result is shown in Fig. \ref{fig:smd}.

\subsection{Temperature threshold and kinetic temperature}

Regarding the \nh\ inversion emission lines, the excitation temperature refers to the rotation temperature of \nh, \tr. We set the upper limit and lower limit of \tr\ in the fitting processes to 50 K and 8 K, respectively. The lower limit refers to the typical temperature of interstellar molecular clouds without star-forming activity, that is, $\sim$ 10K, and the upper bound corresponds to the fact that the \nh\ \oneone\ and \twotwo\ lines cannot be a good temperature probe when the \tr\ of the gas exceeds $\sim$ 30 K. 

To obtain the kinetic temperature, \tk, from \tr, we employ the empirical approximation provided by \citet{2011ApJ...742...95O},
\begin{equation}
T_{\rm K}=6.06e^{0.061(T_{\rm rot}/{\rm K})}.
\end{equation}
The temperature mentioned hereafter refers to the kinetic temperature.

\citet{2004A&A...416..191T} and \citet{2005ApJ...620..823S} also derived \tr--\tk\ relations. Comparing these results, we find that the empirical approximation from \citet{2011ApJ...742...95O} applies for a wider range of \tr\ (\tr\ $\lesssim$ 500 K), while the ranges are 5 K < \tr\ < 20 K and \tk\ $\ll$ 41 K for the relations in \citet{2004A&A...416..191T} and \citet{2005ApJ...620..823S}, respectively. In our results, a remarkable number of detections are hotter than 20 K, which is why we choose the relation from \citet{2011ApJ...742...95O}.

\subsection{Intrinsic line width and velocity dispersion}\label{subsec:ILC}

The line width derived from our fitting is blended by several hyperfine lines, because we do not conduct the full-hyperfine fitting. To derive the intrinsic line width that truly reflects the gas motion, we firstly subtract the channel width (0.1 km s$^{-1}$) from the FWHM line widths derived from the line fitting  quadratically. We then apply the following empirical relation of \citet{1998ApJ...504..207B} to eliminate the effect of hyperfine-line blending:
\begin{equation}
\Delta v_{\rm int}=\Sigma _{n=0}^3C_n(\tau _0)\Delta v_{\rm blended}^n,
\end{equation}
where $C_n(\tau _0)$ is given as
\begin{equation}
C_n(\tau _0)=K_{n0}+K_{n1}{\rm e}^{-K_{n2}\tau _0}.
\end{equation}
Here, $\tau_0$ is the total optical depth of \nh\ \oneone\ emission. We take the optical depth of the \nh\ \oneone\ main line, $\tau_{11,{\rm m}}$, as the approximation of $\tau_0$. The conversion coefficient, $K_{nm}$, is listed in Table \ref{tab:coef}. As a polyfit result, this relation has its limitations for large FWHM line widths, which will make \lwi\ larger than \lwb. In this case, we use \lwb\ as an approximation of \lwi; that is, \lwi\ $\approx$ \lwb. Furthermore, we also tested the full-hyperfine fitting in Fields 3 and 5 and compared the fitted line width with that derived from our fitting results. In Field 3, the mean and median differences between the line widths from these two fitting methods are 0.034 and 0.027 km s$^{-1}$ (5.4\% and 6.7\% for relative deviations, respectively). Regarding Field 5, these values are 0.062 (2.4\%) and 0.008 (1.5\%) km s$^{-1}$, respectively. The empirical relation is therefore reliable and the differences are totally negligible.

\begin{table}
\caption{Values of $K_{nm}$ \label{tab:coef}}
\centering
$$ 
\begin{tabular}{c|cccc}
\hline\hline
  & \multicolumn{4}{c}{n} \\
\cline{2-5}
m & 0 & 1 & 2 & 3   \\
\hline
0 & -0.3308 & 0.8572 & -0.1062 & 0.0152 \\
1 & -0.6072 & 1.8001 & -0.9679 & 0.2179 \\
2 & 0.1808  & 0.1932 & 0.2479  & 0.2950 \\
\hline
\end{tabular}
 $$ 
\end{table}

The intrinsic line width can be further decomposed into \lwi $^2$ = 8$ln2$ $\times$ ($\sigma_{v_{\rm t,NH_3}}^2$ + $\sigma_{v,{\rm NT}}^2$), where $\sigma_{v_{\rm t,NH_3}}$ is the thermal velocity dispersion of \nh\ and \vnt\ is the nonthermal component. We can calculate the thermal velocity dispersion via $\sigma_{v_{\rm t,m}} = 0.288\ {\rm km\ s^{-1}} \cdot (m/m_{\rm H})^{−1/2} \cdot (T_{\rm K}/10 K)^{1/2}$. Thus, we can derive the nonthermal velocity dispersion $\sigma_{v,{\rm NT}}$. Taking the molecular weight per mean free particle of typical interstellar molecular gas as 2.33 $m_{\rm H}$ \citep{2008A&A...487..993K}, we can calculate the thermal velocity dispersion of the whole molecular gas, $\sigma_{v_{\rm t,ISM}}$. We then derive the total velocity dispersion for the whole molecular gas from 
\begin{equation}
    \sigma_v=\sqrt{\sigma_{v,{\rm NT}}^2+\sigma_{v_{\rm t,ISM}}^2}.
\end{equation}
We refer to $\sigma_v$ when we discuss velocity dispersion in the following context.

\subsection{Uncertainties of the results}\label{subsec:3.4}

The uncertainties of the derived physical parameters could originate from both the observational process and the data reduction, including the missing flux, the calibration errors, the systematic errors of the calculating algorithms we use, and so on.
The uncertainty of kinetic temperatures comes from two processes: the calculation of \tr\ and the transformation from \tr\ to \tk. Here, \tr\ is mainly determined by the ratio of the main components of the \nh\ \oneone\ and \twotwo\ inversion lines. In this process, the uncertainty of \tr\ is \simi\ 10.3/S/N K, where S/N is the signal-to-noise ratio of the \nh\ \twotwo\ inversion line \citep{2003ApJ...587..262L}. The S/Ns of our results range from \simi\ 2 to $\gtrsim$ 15, corresponding to the rms errors from \simi\ 5.2 to $\lesssim$ 0.69 K. For the transformation process, we expect an uncertainty of \simi\ 4 K according to \citet{2011ApJ...742...95O}.

Regarding the velocity dispersion, there are three physical parameters involved in its calculation: the blended line width, the optical depth, and the kinetic temperature. The optical depth has little effect on the calculation of intrinsic line width \citep{1998ApJ...504..207B}. According to \citet{2014ApJ...790...84L}, the uncertainty of the blended line width can be written as 
\begin{equation}
\frac{\delta\Delta v}{\Delta v}=\frac{1}{2}\frac{\delta F}{F}
,\end{equation}
where \emph{F} is the flux intensity. $\delta F/F$ is estimated to be \simi\ 30\% in our data, indicating an uncertainty on the blended line width of \simi\ 15\%. The empirical relation between \lwi\ and \lwb\ would increase this uncertainty to about 30\%. Coupled with the uncertainty of kinetic temperatures, the uncertainty of the velocity dispersion could be $\gtrsim$ 30\%.

\section{Results} \label{sec:4}

\subsection{Moment-0 paps and fragment extraction} \label{sec:4.1}

\begin{figure}
\gridline{\fig{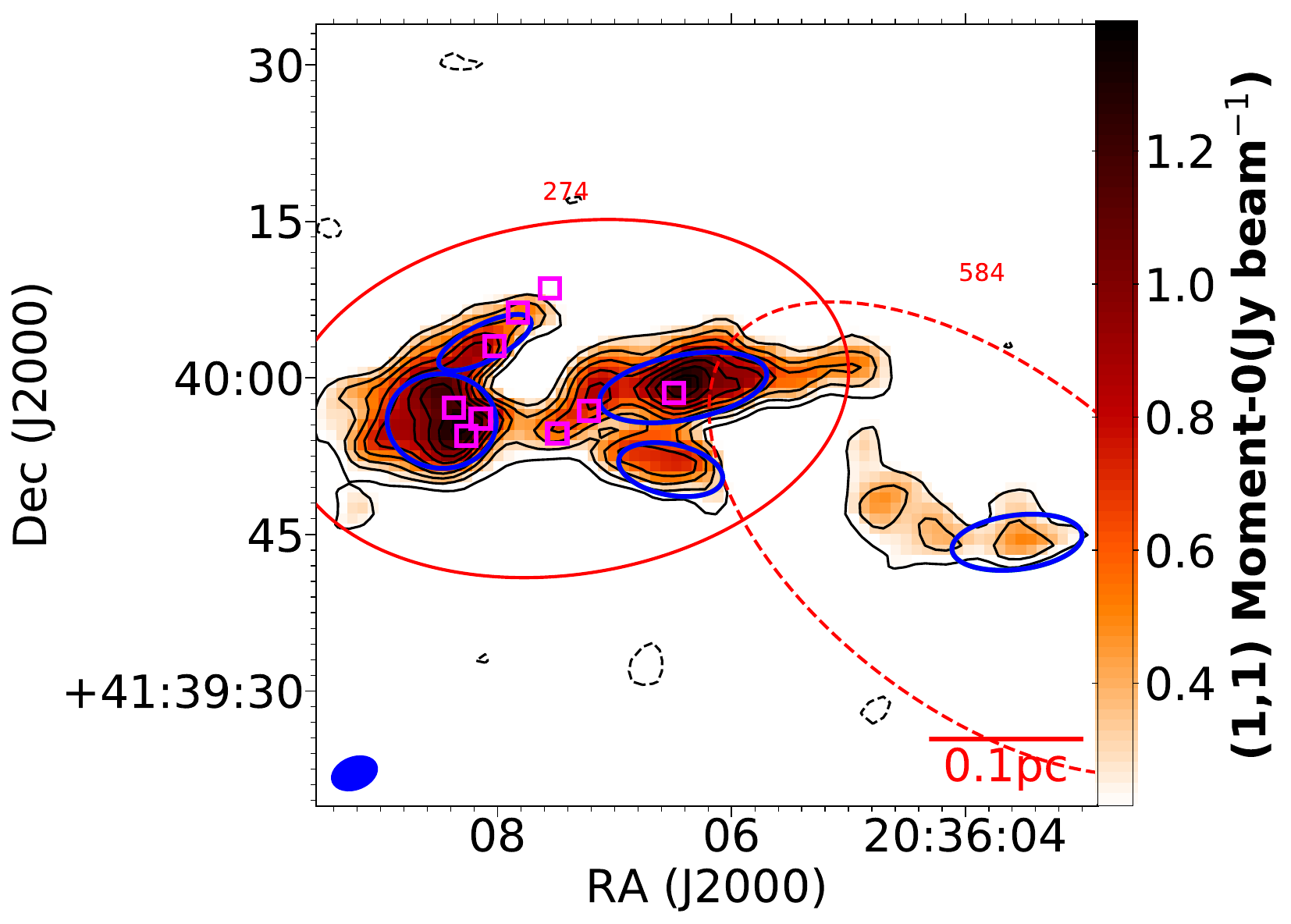}{0.44\textwidth}{(a)}}
\gridline{\fig{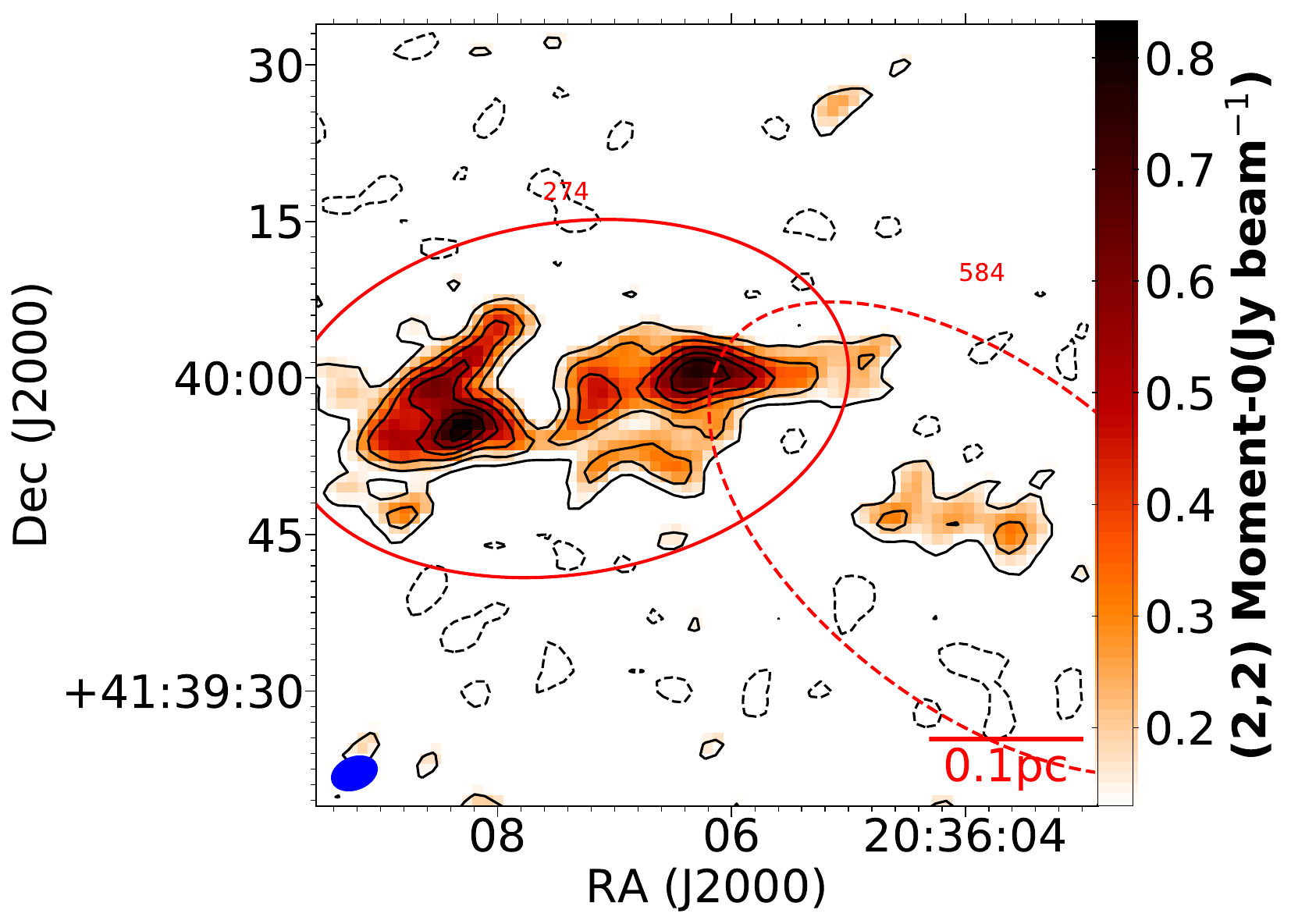}{0.44\textwidth}{(b)}}
\caption{(a) Moment-0 map of the \nh\ (1,1) line for Field 3. The contour is plotted from 3$\sigma$ level and in steps of 2$\sigma$. The red ellipses show the MDCs identified from the H$_2$ column density map (resolution \simi\ 20$^{\prime \prime}$) by \citet{2021ApJ...918L...4C}, and the blue ellipses illustrate the fragments extracted using the {\small GAUSSCLUMPS} algorithm. The magenta boxes show the condensations identified from the SMA 1.3 mm continuum. (b) Moment-0 map of the \nh\ (2,2) line for Field 3. The contour is plotted from 2$\sigma$ level and in steps of 2$\sigma$ \label{fig:m0}}
\end{figure}

In order to show the distributions of the spectral lines, we produce the moment-0 maps of all fields by integrating their intensities in the velocity ranges of the \nh\ \oneone\ main components. The \nh\ \oneone\ and \twotwo\ moment-0 maps of Field 3 are shown as an example in Fig. \ref{fig:m0}. The \nh\ \oneone\ and \twotwo\ moment-0 maps are plotted from 3$\sigma$ and 2$\sigma$ levels, respectively. Moment-0 maps of other fields and the other emission lines are presented in Appendix \ref{app:A}. 

As mentioned in Sect. \ref{sec:1}, the \nh\ \oneone\ moment-0 maps could roughly illustrate the distribution of cold dense gas in our MDCs. We extracted 202 fragments from the \nh\ \oneone\ moment-0 maps using the Starlink \begin{small}GAUSSCLUMPS\end{small} algorithm \citep{2007ASPC..376..425B,2014ASPC..485..391C}. The field 21 was excluded from the extraction as it is not at the distance of Cygnus X. The results of the identified \nh\ fragments in Field 3 are also plotted in Fig. \ref{fig:m0} with blue ellipses. In addition, the sizes of \nh\ fragments are defined as the deconvolved radius, following
\begin{equation}
R=\sqrt[]{ab-\frac{1}{4{\rm ln}2}beam_{\rm maj}beam_{\rm min}},
\end{equation}
where $a$ and $b$ are the semi-major and semi-minor axes of \nh\ fragments. 

\begin{figure}
    \centering
    \includegraphics[width=8.5cm]{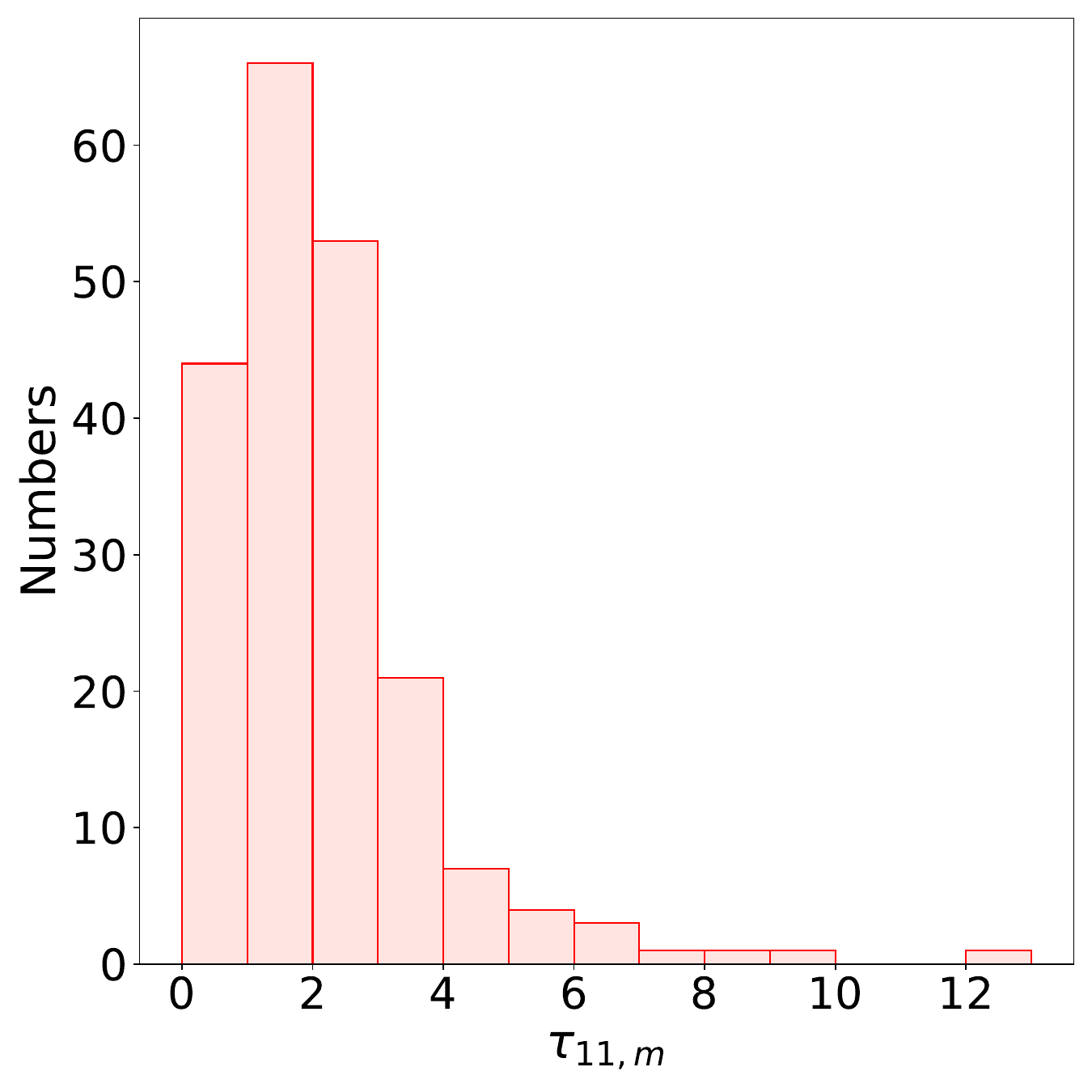}
    \caption{Histogram of the average optical depth of the main component of the \nh\ \oneone\ line for identified \nh\ fragments.}
    \label{fig:tau}
\end{figure}

We also show the positions of the condensations identified in the SMA 1.3 mm continuum data from \citet{2021ApJ...918L...4C}. There are 148 condensations overlapping with the fields in this paper. The majority of them (103 out of 148) are distributed in the identified \nh\ fragments, indicating that the \nh\ fragments can probe the density structures well in most cases. As shown in Fig. \ref{fig:tau}, the\ \oneone\ main components of most \nh\ fragments have moderate optical depths, and the mean and median values are 2.2 and 1.9. Under such a moderate optical depth, the \nh\ \oneone\ lines might show different morphologies from the density profiles of the dust continuum emission, which could account for the condensations uncorrelated to the \nh\ fragments. In conclusion, the majority of the \nh\ fragments can be considered as density structures.

\subsection{Maps of the derived physical parameters} \label{sec:4.2}

\begin{figure*} 
\gridline{\fig{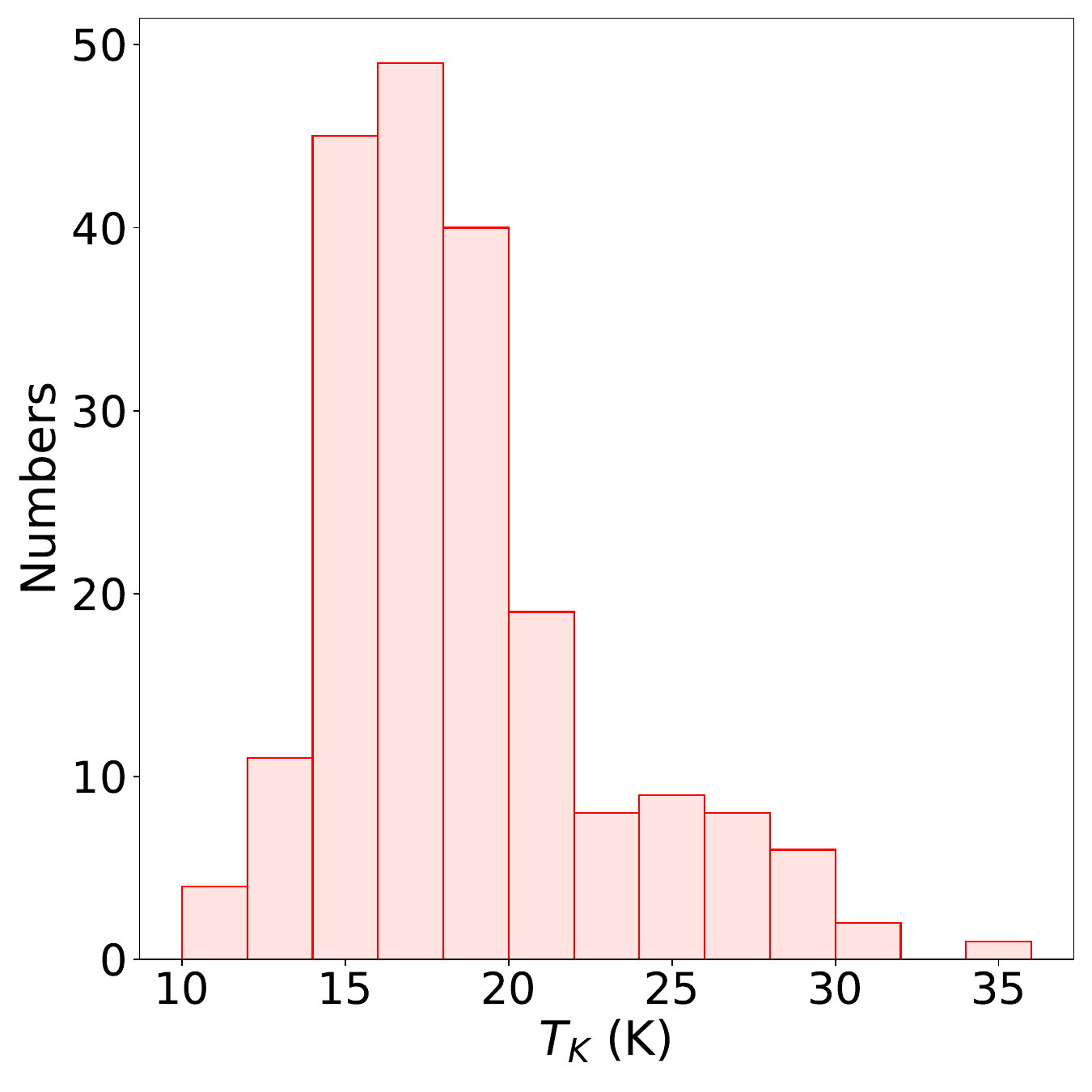}{0.33\textwidth}{(a)}
          \fig{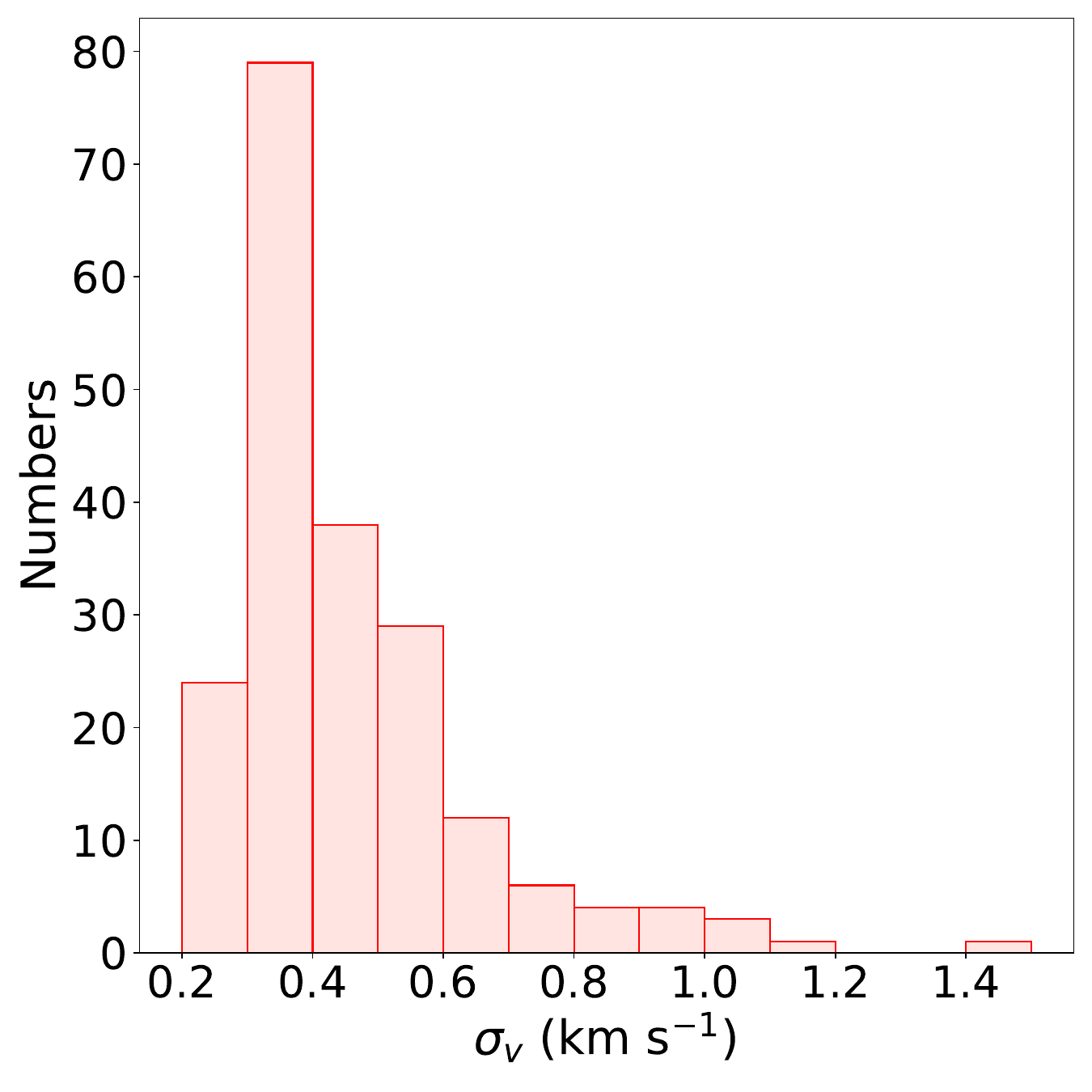}{0.33\textwidth}{(b)}
          \fig{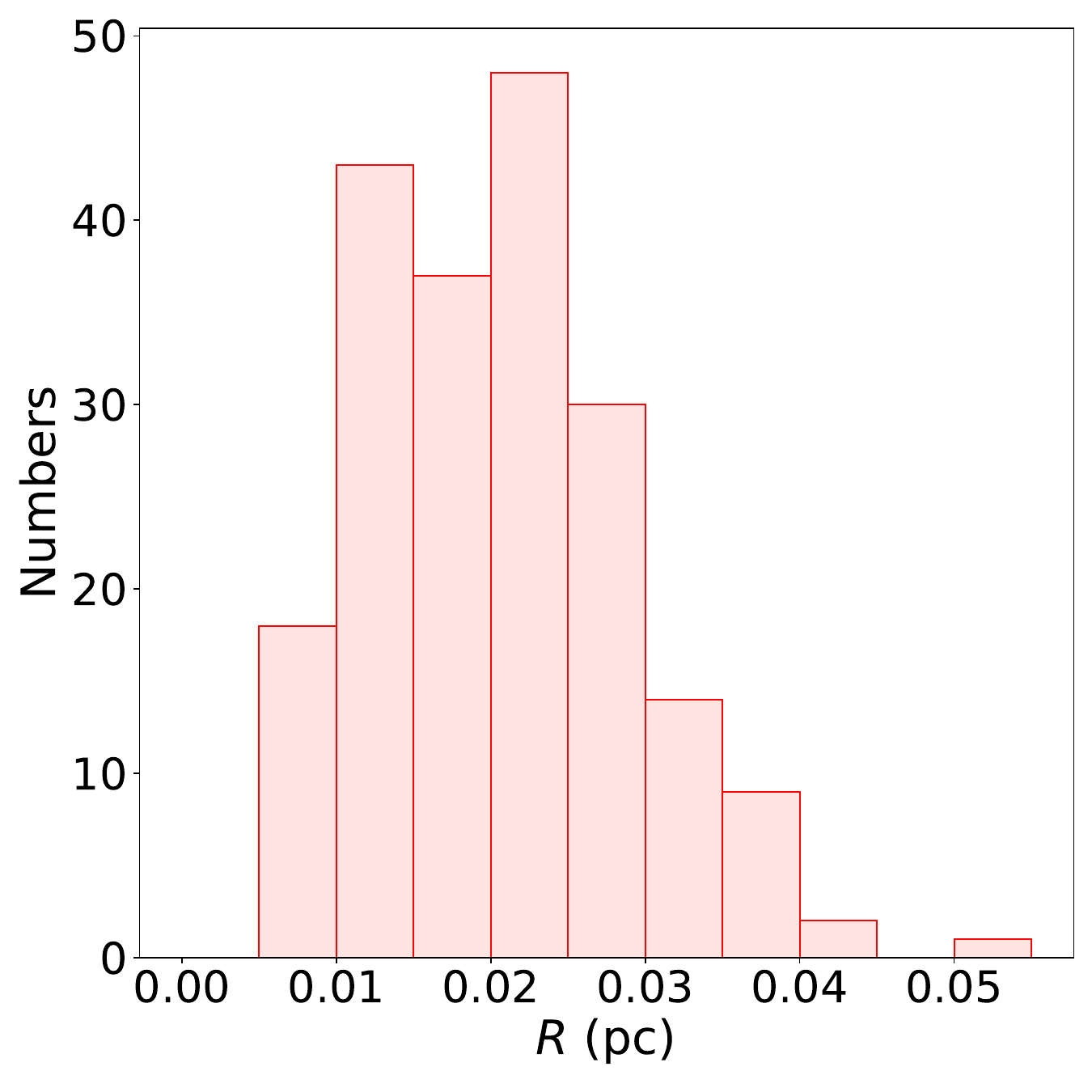}{0.33\textwidth}{(c)}
          }
\gridline{\fig{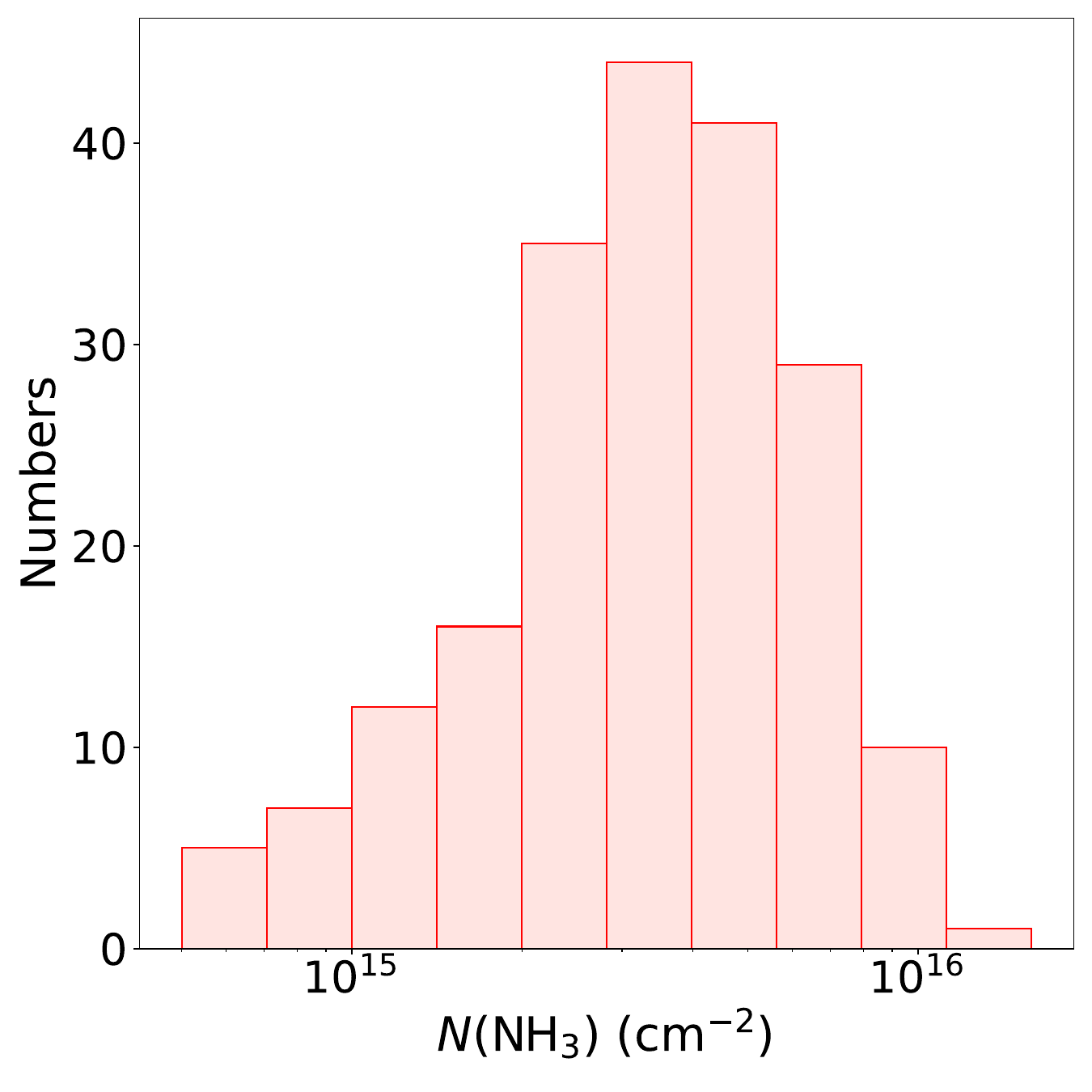}{0.33\textwidth}{(d)}
          \fig{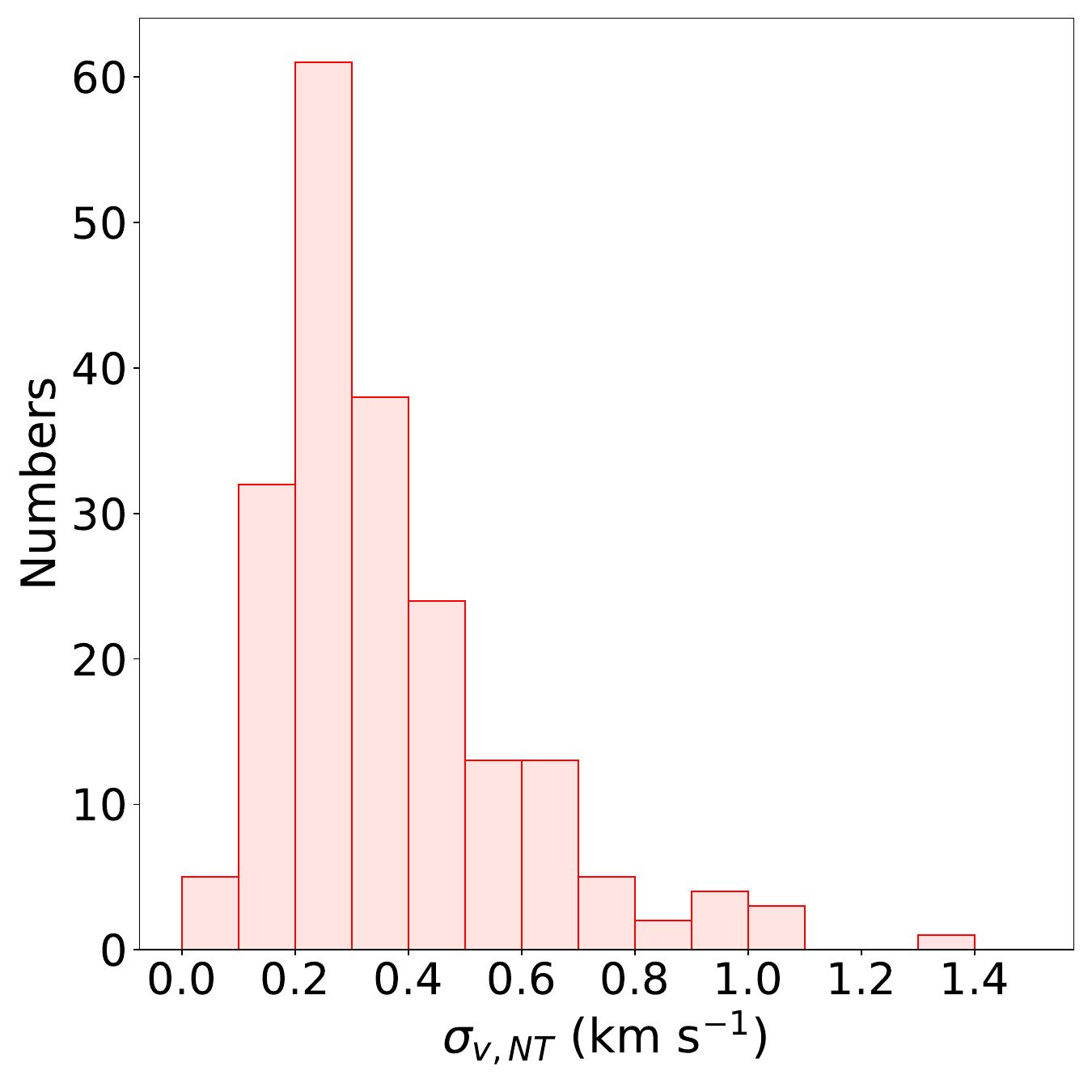}{0.33\textwidth}{(e)}
          \fig{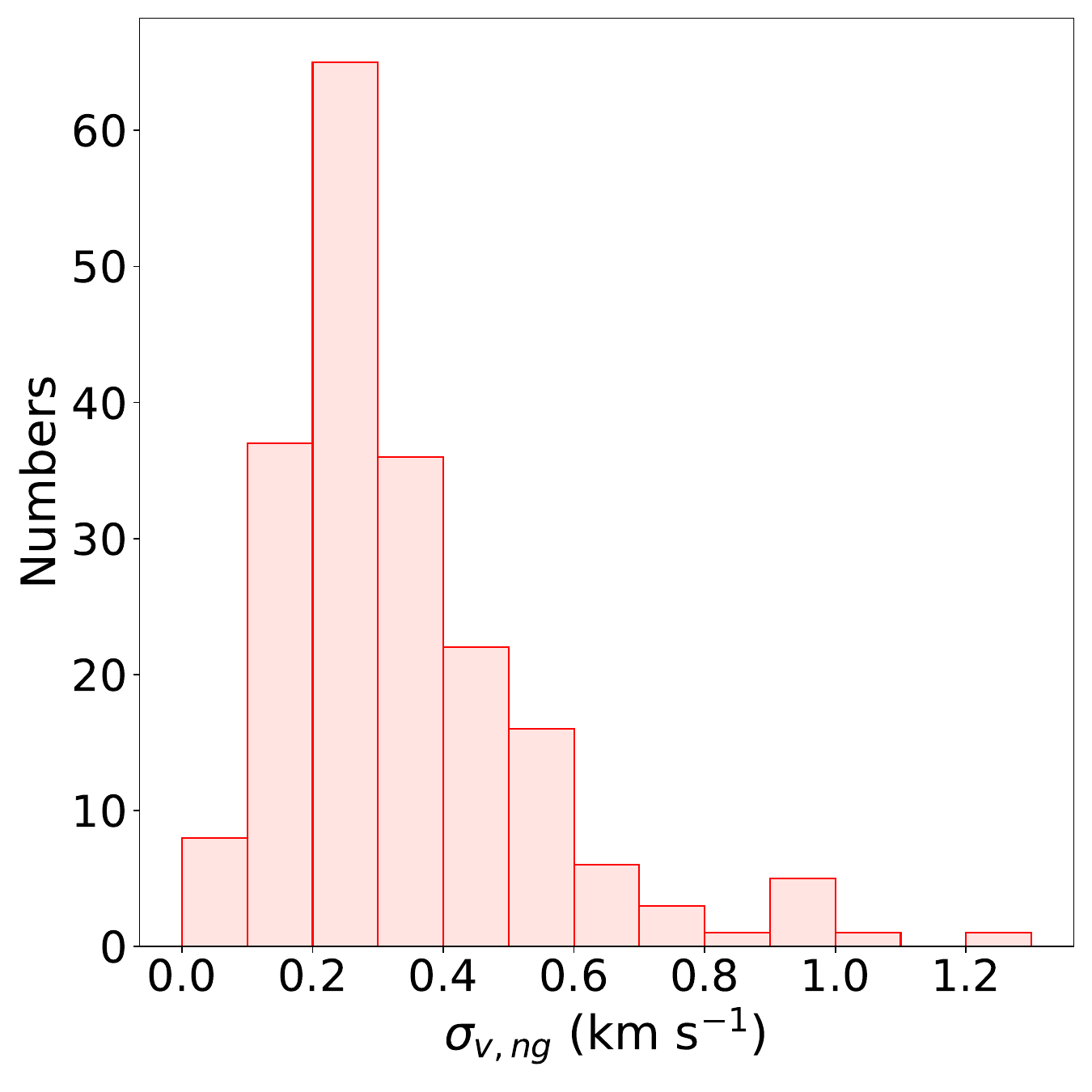}{0.33\textwidth}{(f)}
          }
\caption{Distributions of the physical parameters of \nh\ fragments. (a) Histogram of kinetic temperatures \tk. (b) Histogram of total velocity dispersions $\sigma_v$ (including bulk motions and using the thermal component of the entirety of the molecular gas). (c) Histogram of deconvolved radii of \nh\ fragments $R_{\rm dec}$. (d) Histogram of \nh\ column density $N$(\nh). (e) Histogram of nonthermal velocity dispersion with bulk motions $\sigma_{v,{\rm NT}}$. (f) Histogram of nonthermal velocity dispersion without bulk motions $\sigma_{v,{\rm ng}}$. \label{fig:dis}}
\end{figure*}

The line-fitting process of the \nh\ \oneone\ and \twotwo\ emission lines provides the spatial distributions of the kinetic temperature, centroid velocity, velocity dispersion, optical depth of the \nh\ \oneone\ main component, \nh\ column density, and the beam-filling factor in all the fields, which allow us to investigate the temperature and dynamic conditions of the MDCs in Cygnus X and their fragments. We present maps of these derived parameters in Appendix \ref{app:B}, and the maps of Field 3 in Fig. \ref{fig:props} as an example. Ranges of the derived physical parameters are listed by each field in Table \ref{tab:depro}. Using these maps, we obtain the average kinetic temperature, \nh\ column density, velocity dispersion, and its nonthermal component for all the \nh\ fragments. The contribution from the velocity gradients within the \nh\ fragments, which is usually considered as the bulk motion, is taken into account when calculating the velocity dispersion, $\sigma_v$, and its nonthermal component, \vnt. We also fit the velocity gradients \citep[following the method in e.g.,][]{1993ApJ...406..528G,2018A&A...616A.111W} and deduct them from the nonthermal velocity dispersions, which are provided as \vng. The detailed calculation can be found in Appendix \ref{app:E}. A brief statistic analysis of these physical parameters is reported in Table \ref{tab:avepro} and the histograms of their distributions are shown in Fig. \ref{fig:dis}. The detailed physical parameters of the \nh\ fragments are available in our online materials at the CDS.

\begin{figure*}
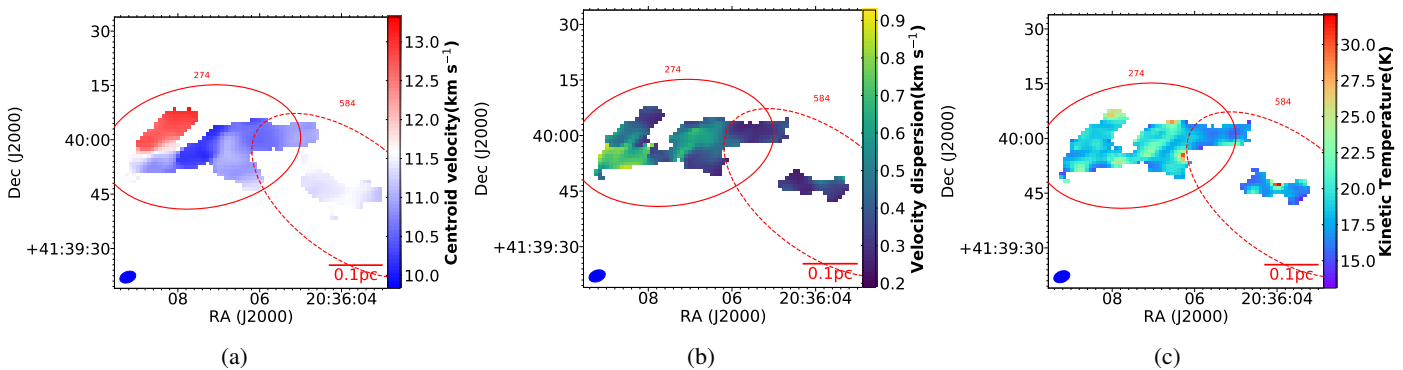

\gridline{\fig{N5_Peakv.pdf}{0.33\textwidth}{(a)}
          \fig{N5_vdis.pdf}{0.33\textwidth}{(b)}
          \fig{N5_Tk.pdf}{0.33\textwidth}{(c)}
          }
\caption{Physical parameter maps of Field 3. (a) Centroid velocity map. (b) Velocity dispersion map. (c) Kinetic temperature map. The ellipses are the same as those in Fig. \ref{fig:m0} \label{fig:props}}
\end{figure*}

\begin{table*}
\caption{Ranges of derived physical properties in each field\label{tab:depro}}
\centering
\begin{tabular}{c|cccccc}
\hline\hline
Field & $v_{peak}$ & $\sigma_v$ & $T_K$ & $\tau_{11,{\rm m}}$ & $N({\rm NH_3})$ & \vnt\tablefootmark{a} \\
 & (km s$^{-1}$) & (km s$^{-1}$) & (K) &  & (10$^{15}$ cm$^{-2}$) & (km s$^{-1}$)\\
\hline
1 & 13.61--15.15 & 0.26--1.05 & 14.72--62.22 & 0.16--4.81 & 0.95--17.52 & 0.06--0.98\\ 
2 & -5.54--{-}3.87 & 0.26--1.01 & 13.84--29.30 & 0.22--6.44 & 1.05--9.81 & 0.12--0.95\\ 
3 & 10.11--13.19 & 0.25--0.87 & 14.00--35.04 & 0.17--5.37 & 0.47--14.50 & 0.01--0.83\\ 
4 & -4.00--{-}1.02 & 0.23--1.48 & 13.11--38.01 & 0.11--7.16 & 0.38--12.57 & 0.05--1.44\\ 
5 & 13.97--16.96 & 0.26--1.85 & 11.63--66.97 & 0.05--7.03 & 0.33--53.44 & 0.11--1.80\\ 
6 & -3.71--{-}0.33 & 0.66--1.82 & 18.28--32.81 & 0.03--66.72 & 0.63--18.62 & 0.59--1.80\\ 
7 & -3.12--{-}1.70 & 0.20--0.94 & 10.94--24.27 & 0.11--13.58 & 0.34--9.83 & 0.01--0.91\\ 
8 & 17.34--20.36 & 0.25--1.36 & 12.53--30.12 & 0.10--6.01 & 0.40--15.49 & 0.10--1.33\\ 
9 & -3.03--{-}2.42 & 0.23--0.59 & 13.97--31.05 & 0.15--4.06 & 0.40--7.22 & 0.03--0.54\\ 
10 & -7.48--{-}5.46 & 0.22--0.84 & 12.58--36.26 & 0.13--4.53 & 0.34--6.29 & 0.02--0.78\\ 
11 & -6.97--{-}5.16 & 0.22--1.53 & 13.38--39.98 & 0.13--4.77 & 0.30--10.81 & 0.02--1.49\\ 
12 & -6.06--{-}3.70 & 0.23--0.56 & 11.96--28.93 & 0.10--5.46 & 0.25--8.51 & 0.03--0.50\\ 
13 & -0.60--1.88 & 0.23--0.74 & 13.22--45.89 & 0.10--9.63 & 0.35--10.20 & 0.02--0.64\\ 
14 & 6.75--8.59 & 0.23--0.72 & 13.30--31.18 & 0.14--4.45 & 0.36--5.50 & 0.01--0.68\\ 
16 & 5.52--6.43 & 0.26--1.39 & 13.46--22.84 & 0.15--4.61 & 0.55--11.62 & 0.11--1.36\\ 
17 & 0.31--3.11 & 0.21--1.39 & 11.85--87.87 & 0.03--30.02 & 0.35--29.15 & 0.01--1.29\\ 
18 & -1.28--0.59 & 0.21--0.80 & 12.51--30.34 & 0.12--4.33 & 0.28--10.43 & 0.02--0.77\\ 
19 & -2.52--{-}0.15 & 0.21--0.61 & 11.54--27.32 & 0.18--33.36 & 0.30--21.23 & 0.01--0.56\\ 
20 & -3.22--{-}0.81 & 0.27--0.83 & 15.99--53.48 & 0.14--5.11 & 0.44--11.40 & 0.02--0.79\\ 
21\tablefootmark{b} & -7.03--{-}4.59 & 0.41--1.37 & 16.21--127.96\tablefootmark{c} & 0.14--2.70 & 0.74--79.52 & 0.31--1.30\\ 
22 & 4.70--7.14 & 0.23--1.00 & 11.56--18.29 & 0.24--5.72 & 0.57--9.71 & 0.02--0.98\\ 
23 & 3.88--8.07 & 0.24--1.18 & 11.12--41.60 & 0.14--18.47 & 0.42--39.98 & 0.01--1.14\\ 
24 & -1.43--1.12 & 0.25--0.82 & 14.48--28.72 & 0.10--3.01 & 0.44--8.91 & 0.10--0.78\\ 
25 & -0.19--2.90 & 0.20--0.59 & 9.87--19.20 & 0.21--83.00 & 0.75--69.41 & 0.01--0.56\\ 
26 & -0.57--1.99 & 0.22--0.52 & 9.87--27.98 & 0.17--7.70 & 0.38--7.72 & 0.01--0.45\\ 
27 & 4.68--4.99 & 0.30--0.50 & 21.34--53.26 & 0.12--2.12 & 0.45--6.85 & 0.05--0.41\\ 
28 & -1.97--0.65 & 0.22--0.70 & 12.60--34.68 & 0.19--10.97 & 0.47--16.27 & 0.03--0.64\\ 
29 & -3.55--0.15 & 0.24--0.84 & 9.99--25.74 & 0.27--5.64 & 0.50--12.22 & 0.12--0.80\\ 
31 & 7.96--9.13 & 0.22--0.58 & 12.91--40.77 & 0.12--4.65 & 0.31--6.13 & 0.03--0.47\\ 
32 & -12.30--{-}10.53 & 0.23--0.75 & 11.59--21.83 & 0.14--8.34 & 0.51--8.90 & 0.05--0.71\\ 
33 & -10.52--{-}8.97 & 0.24--0.47 & 14.24--24.56 & 0.18--4.36 & 0.52--4.89 & 0.08--0.38\\ 
34 & -5.39--{-}2.19 & 0.26--1.30 & 14.70--57.77 & 0.08--3.04 & 0.31--11.74 & 0.07--1.27\\ 
37\tablefootmark{d} & 6.54--10.29 & 0.24--1.90 & 11.47--127.96\tablefootmark{c} & 0.04--36.71 & 0.95--8.00 & 0.09--1.79\\
\hline
\end{tabular}
\tablefoot{\tablefoottext{a}{The nonthermal velocity dispersion}
\tablefoottext{b}{Also known as AFGL 2591}
\tablefoottext{c}{Derived from the upper limit of \tr, 50 K.}
\tablefoottext{d}{Field 37 is fitted with two velocity components}}
\end{table*}

Statistically, the \nh\ fragments have a mean deconvolved radius of 0.02 pc thanks to the high spatial resolution of our observations. Recently, \citet{2019A&A...627A..85Z} obtained the same value for the typical radius of condensations from their PdBI 1.3 mm continuum observations. In the other two papers of CENSUS \citep{2023arXiv231217455P,2023arXiv231204880Y}, the properties of condensations in the \cyx\ regions are explored in more detail using submillimeter continuum and spectral observations from SMA.

\begin{figure} 
    \centering 
    \includegraphics[width=8.5cm]{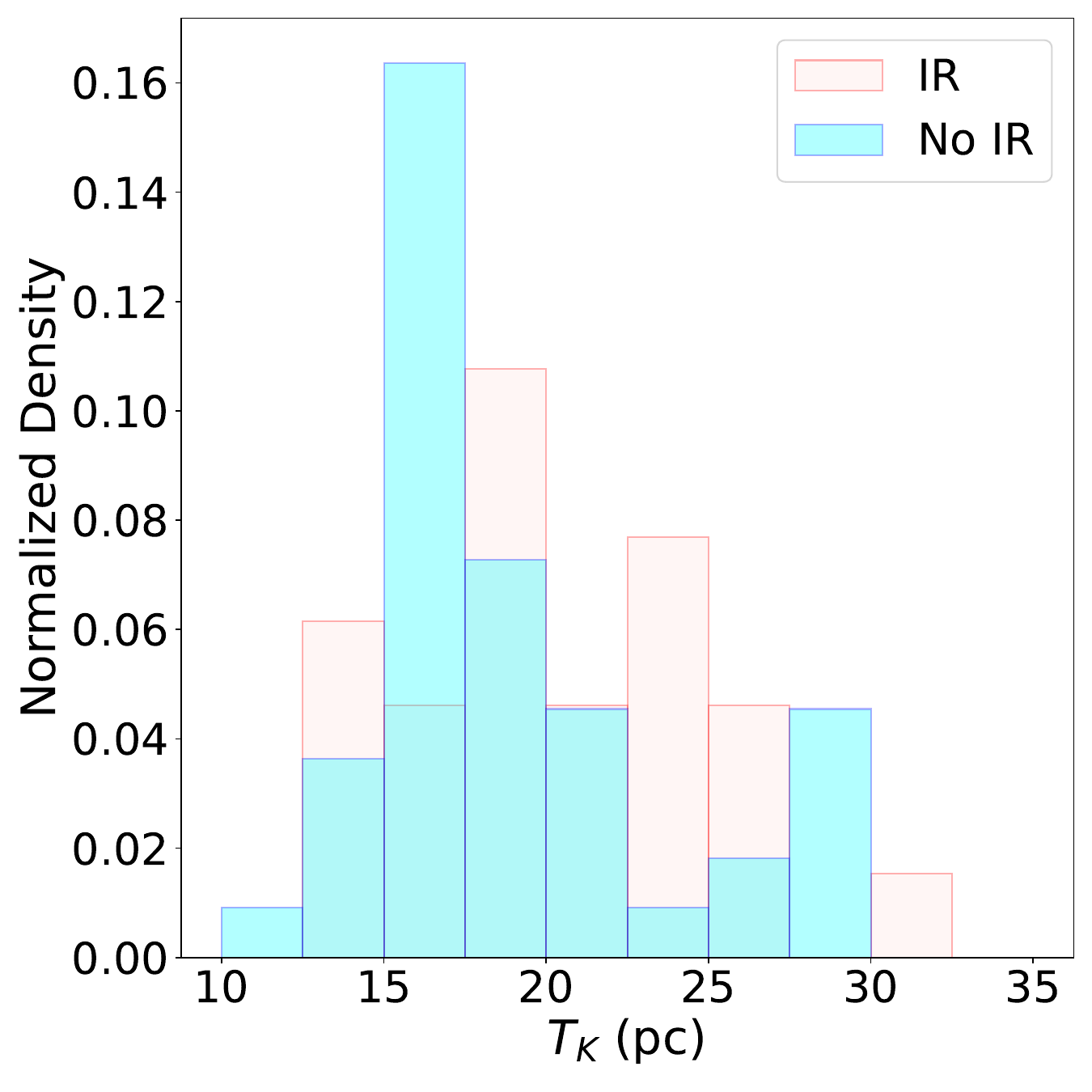} 
    \caption{Histogram of average kinetic temperatures of the \nh\ fragments coinciding with the condensations of 1.3mm continuum emission. Those matched with IR sources are colored in red, while the rest are colored in cyan.}
    \label{fig:Tbn}
\end{figure}

The temperatures of \nh\ fragments range from \simi\ 11.0 K to 34.1 K, with mean and median values of 18.6 K and 17.5 K, respectively. The mean temperature of the condensations in \citet{2019A&A...627A..85Z}, which are selected from several massive proto-cluster clumps, is 28.3 K, which is higher than the value found in the present study. \citet{2014ApJ...790...84L} derived a mean temperature of \simi\ 18.3 K for \nh\ dense cores (with a typical size of \simi\ 0.08 pc) in 62 high-mass star-forming regions, which is consistent with our results. Temperature is one of the commonly used indicators of the evolutionary stage of molecular structures. For example, \citet{2013MNRAS.432.3288S} calculated the mean temperatures of the quiescent starless cores and the protostellar cores (with typical sizes of \simi\ 0.05 pc) as 18.8 K and 28.8 K, respectively. To verify this relation in our data, we matched the \nh\ fragments with the condensations from \citet{2021ApJ...918L...4C} and the IR sources from the Cygnus X Archive catalog \citep[the \emph{Spitzer} Cygnus X Legacy Survey,][]{2010AAS...21541401K}. The temperature distribution of the \nh\ fragments coincident with the continuum condensations (70 out of 202) is shown in Fig. \ref{fig:Tbn}, indicating that these fragments are more likely to be density structures. These condensation-related fragments are divided into two groups: those associated and not associated with the IR sources (44 and 26 in number). The mean temperature of the former group is 20.4$\pm$4.5 K, slightly higher than that of the latter group of 19.1$\pm$4.4 K. Also, their median temperature  values are 19.4 K and 17.3 K. In general, those with IR sources are considered to be more evolved than those without IR sources. We performed a two-sample Kolmogorov-Smirnov test on the temperature distribution of these two groups and the p-value is calculated as 0.18. The p-value is higher than the threshold 0.05, indicating the two groups are not distinguishable. As shown in Fig. \ref{fig:Tbn}, there are also considerable overlaps in temperatures between the two groups, suggesting that there is no significant difference in internal heating for the sources in the two groups. Instead, we find that the morphology of the \nh\ fragments is more correlated with the evolutionary stage; we discuss this topic in Sect. \ref{sec:5.1}.

The mean values of $\sigma_v$, \vnt\, and \vng\ are 0.46$\pm$0.20 km s$^{-1}$, 0.37$\pm$0.23 km s$^{-1}$, and 0.34$\pm$0.21 km s$^{-1}$, respectively. The typical velocity dispersion of the condensations in \citet{2019A&A...627A..85Z} is \simi\ 0.9 km s$^{-1}$. Considering that their temperature is also higher than that of our NH3 fragments, the larger velocity dispersion could also result from them being a more evolved sample. In contrast, our results are comparable to those of quiescent starless cores (0.44 km s$^{-1}$ for nonthermal component) in \citet{2013MNRAS.432.3288S}. The cores in \citet{2014ApJ...790...84L} have a smaller typical velocity dispersion of \simi\ 0.46 km s$^{-1}$. Detailed discussions on the dynamic states of the MDCs are given in Sect. \ref{sec:5.3}. The mean \nh\ column density of the \nh\ fragments is 3.9 $\times$ 10$^{15}$ cm$^{-2}$, of the same order of magnitude as the values found by \citet{2013MNRAS.432.3288S} (\simi\ 10$^{15}$ cm$^{-2}$) and \citet{2014ApJ...790...84L} (\simi\ 2.3 $\times\ 10^{15}$ cm$^{-2}$).

\begin{table*}
    \caption{Statistics of physical parameters of \nh\ fragments \label{tab:avepro}}
    \centering
    \begin{tabular}{c|cccccc}
        \hline\hline
         &
        $R$ & $\sigma_v$ & \vnt & \vng & $T_{\rm K}$ & $N({\rm NH_3})$ \\
         & pc & km s$^{-1}$ & km s$^{-1}$ & km s$^{-1}$ & K & 10$^{15}$ cm$^{-2}$\\
        \hline
        Mean & 0.021 & 0.46 & 0.37 & 0.34 & 18.6 & 3.9 \\
        Medium & 0.021 & 0.40 & 0.30 & 0.29 & 17.5 & 3.6 \\
        \hline
    \end{tabular}
\end{table*}

\subsection{Correlation analysis} \label{sec:4.3}

\begin{figure*}
\gridline{\fig{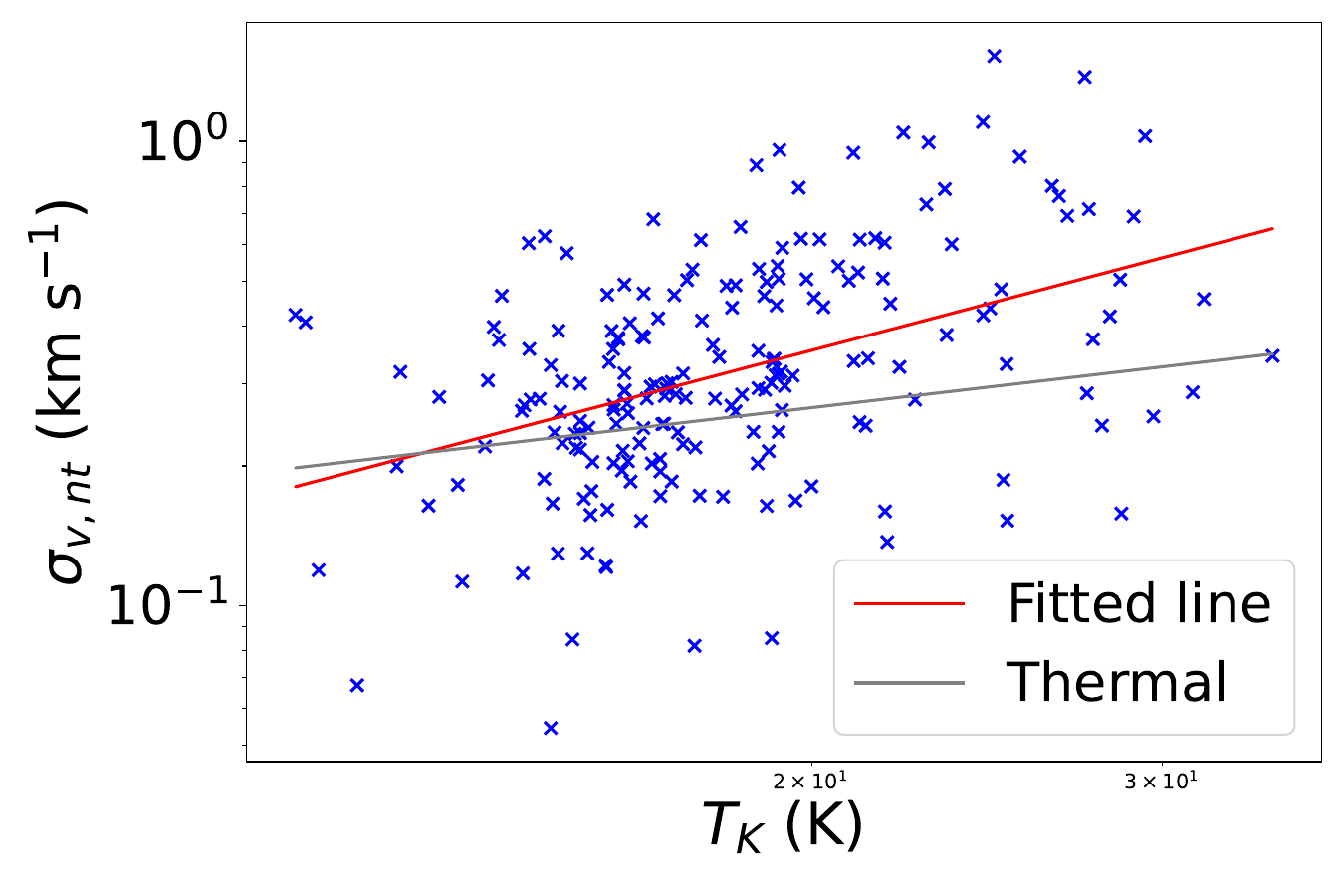}{0.45\textwidth}{(a)}
          \fig{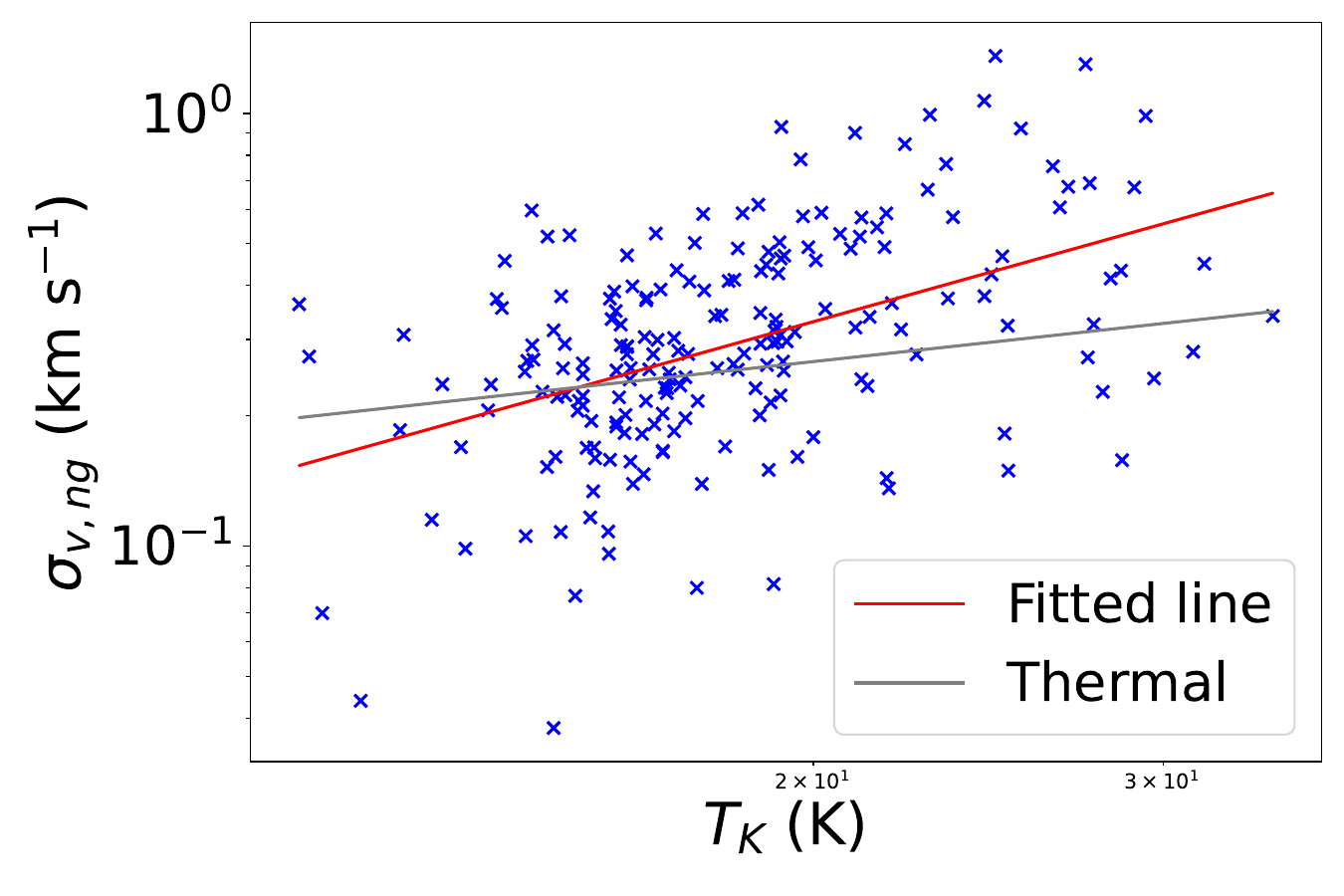}{0.45\textwidth}{(b)}
          }
\gridline{\fig{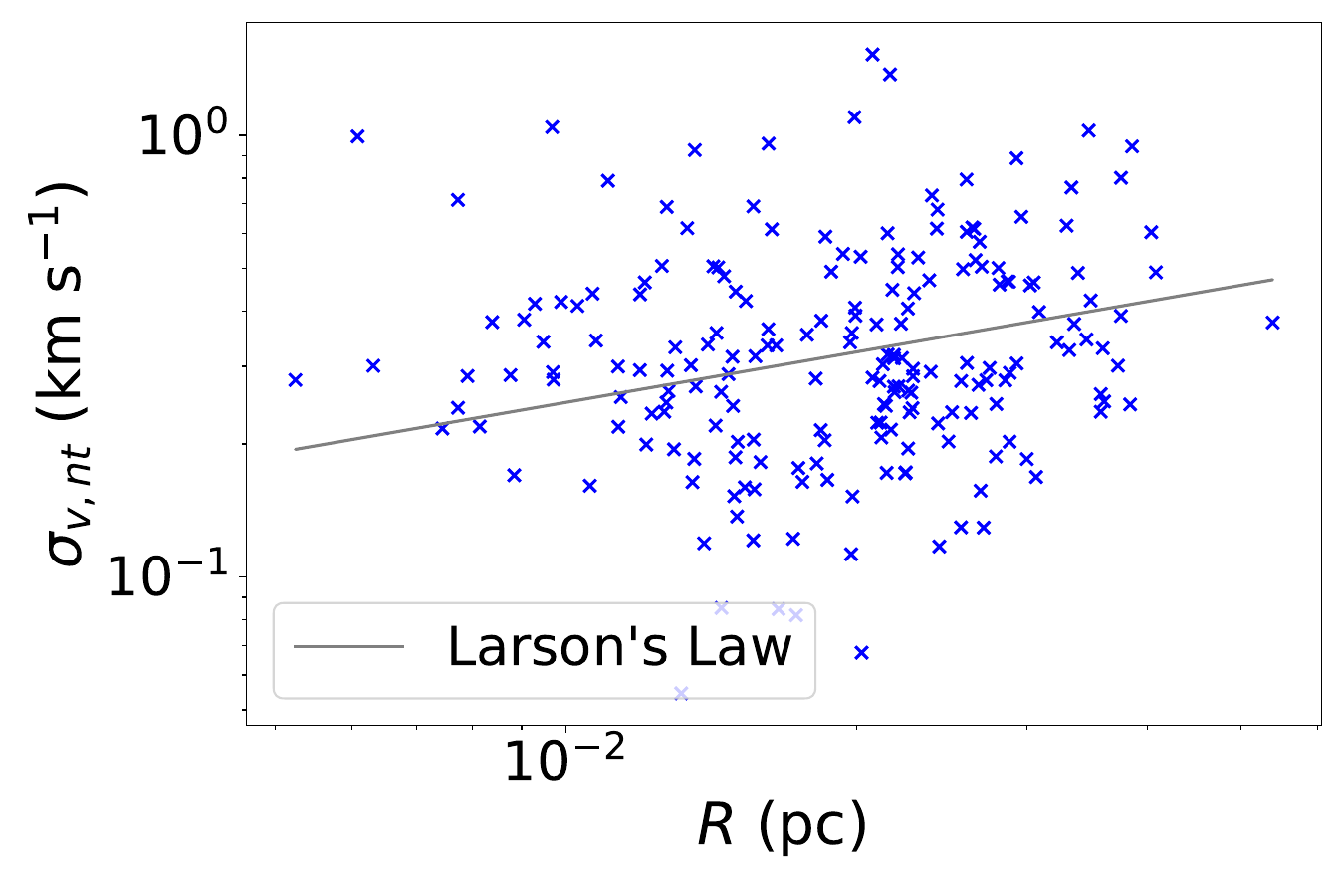}{0.45\textwidth}{(c)}
          \fig{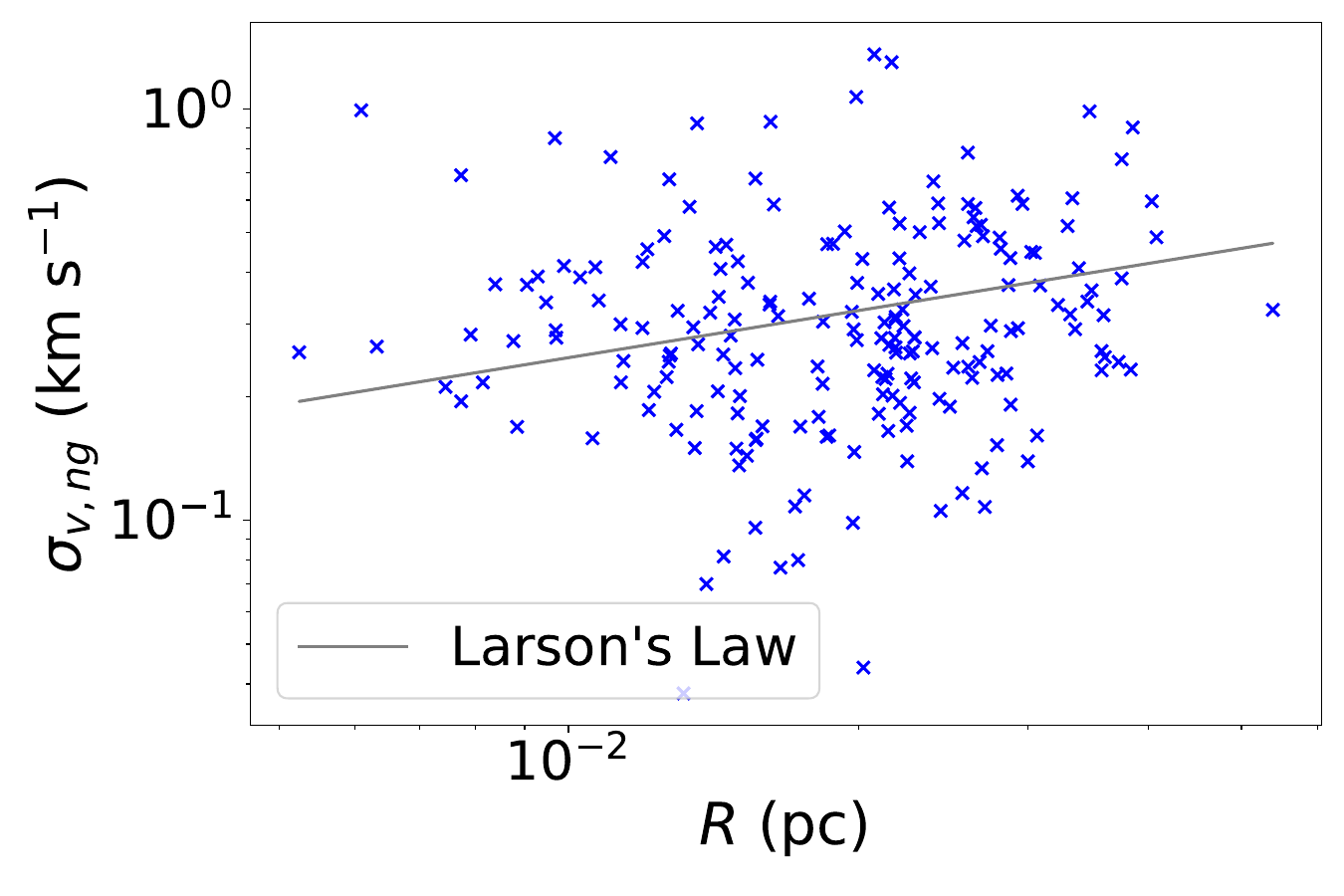}{0.45\textwidth}{(d)}
          }
\caption{Correlations between physical parameters of \nh\ fragments. Panels (a) and (b) plot nonthermal velocity dispersions including or excluding the bulk motions versus kinetic temperature. The red and gray lines show the fitted result of the linear regression and the thermal velocity dispersion. There are positive correlations in both of them. Panels (c) and (d) plot nonthermal velocity dispersions including or excluding the bulk motions versus deconvolved radius. The gray line shows Larson's law. \label{fig:cor}}
\end{figure*}

In this section, we present  analyses of the correlations between physical parameters of the \nh\ fragments. We find one possible correlation between the nonthermal velocity dispersion and the kinetic temperature in our results, which are presented in Fig. \ref{fig:cor}. 

Figures \ref{fig:cor} (a) and (b) show \vnt\ and \vng\ versus kinetic temperature. Though the distributions show large scatters, both of them show weak positive correlations: \vnt\ = 0.012$T_{\rm K}^{1.13\pm 0.17}$ with a correlation coefficient of $r$ = 0.43, and \vng\ = 0.007$T_{\rm K}^{1.28\pm 0.17}$ with $r$ = 0.47. Similar correlations have been reported in previous \nh\ and H$_2$CO observations, such as \tk\ $\varpropto$ $\Delta v^{0.53\pm 0.14}$ in \citet{2006A&A...450..607W}, \tr\ $\varpropto$ $\Delta v^{0.65}$ in \citet{2013MNRAS.432.3288S}, $\Delta v$ $\varpropto$ \tr$^{1.16\pm 0.29}$ in \citet{2014ApJ...790...84L}, and \tk\ $\varpropto$ $\sigma_{v, {\rm NT}}^{1.01\pm 0.25}$ in \citet{2018A&A...609A..16T}. This relation is interpreted 
by \citet{2006A&A...450..607W} and \citet{2018A&A...609A..16T} as the turbulent heating effect, which converts the turbulent energy into the internal energy of gas \citep{1985A&A...142..381G}. In particular, \citet{2018A&A...609A..16T} obtain a positive correlation between temperature and nonthermal velocity dispersion with the APEX H$_2$CO observations. Using the equations of turbulent temperature from \citet{2013A&A...550A.135A},  \citet{2018A&A...609A..16T} suggest that the turbulent heating could play an important role in widely existing massive star formation regions on scales of \simi\ 0.06–2 pc. However, such an effect usually dominates in the central molecular zone \citep{2016A&A...586A..50G,2016A&A...595A..94I} or other regions with strong nonthermal motions, which is not the case in our sample \citep[e.g., \simi\ 0.6-4 \kms\ in][much larger than our result of \simi\ 0.4 \kms]{2018A&A...609A..16T}. Using
the same method as \citet{2013A&A...550A.135A}, we find a turbulent temperature of only \simi\ 8 K, which is comparable to the background temperature (\simi\ 10 K) of cold molecular gas. Therefore, the turbulent heating does not seem to be remarkable in our results. \citet{2014ApJ...790...84L} linked temperature with velocity dispersion using the black-body radiation law ($L \sim T^4$), the mass--luminosity relation ($L \sim M^\alpha$), and the virial equilibrium approximation ($M \sim \sigma^2$), which gives an index varying from 0.67 to 2. Nevertheless, the assumptions of black-body radiation and virial equilibrium are questionable in molecular cores, and the mass--luminosity relation is very uncertain. In our results, the relation is considered as a natural consequence of the star-forming activity. An \nh\ fragment with stronger star-forming activity tends to be hotter and to have stronger infall and shearing motions that contribute to its nonthermal velocity dispersion. Conversely, an \nh\ fragment with weaker star-forming activity would be cooler and would have a smaller nonthermal velocity dispersion. This scenario is also consistent with the fact that our derived temperatures and velocity dispersions are higher at the positions of water masers and radio sources from \citet{2022ApJ...927..185W}, where strong star-forming activities reside.

We also plot \vnt\ and \vng\ versus the radius of fragments in panels (c) and (d). Such a relation is known as one of Larson's laws \citep{1981MNRAS.194..809L}, and is found in a number of works \citep[e.g.,][]{1999ApJS..125..161J,2006A&A...450..607W,2014ApJ...790...84L}. However, this relation is not found in our results. The main reason for the absence of these correlations could be the narrow range of the radius in our sample, which is within an order of magnitude. On the other hand, our mean value of \vng\, 0.34 km s$^{-1}$ coincides with the prediction of Larson's laws of 0.32 km s$^{-1}$ at the scale of 0.02 pc. 

\section{Discussion} \label{sec:5}

\subsection{Morphology} \label{sec:5.1}

\subsubsection{Comparison between ammonia observations and H$_2$ column density\label{sec:5.1.1}}

The \nh\ \oneone\ emissions in our data show diverse morphologies in the moment-0 maps, such as filaments (Field 28), core-like structures (Field 6), hub-like structures (Field 13), and dispersed features (Field 17). The morphology of the \nh\ \oneone\ emission can be affected by protostellar feedback during core evolution. During the star formation process, increasing feedback from protostars can distort their envelopes \citep{2018ARA&A..56...41M} and heat the surrounding gas. The \nh\ \oneone\ line gets much weaker and becomes undetectable when the gas temperature rises beyond 100 K \citep{1983ARA&A..21..239H}. Therefore, the morphology of the \nh\ in MDCs could evolve from a concentrated distribution at the peak of the H$_2$ density profile (hereafter, the M1 stage) to a dispersed shape around the protostar (hereafter, M2), and finally become undetectable (hereafter, M3). In order to construct an evolutionary sequence of the MDCs in Cygnus X, we compared the morphology of the \nh\ \oneone\ emission to that of the $\rm H_2$ column density \citep[from the column density map, hereafter N-map, derived from spectral energy distribution (SED) fitting in][]{2019ApJS..241....1C} of the MDCs, and classified the MDCs with infrared emissions. We present our findings in the following subsection.

\begin{figure}
    \centering
    \gridline{\fig{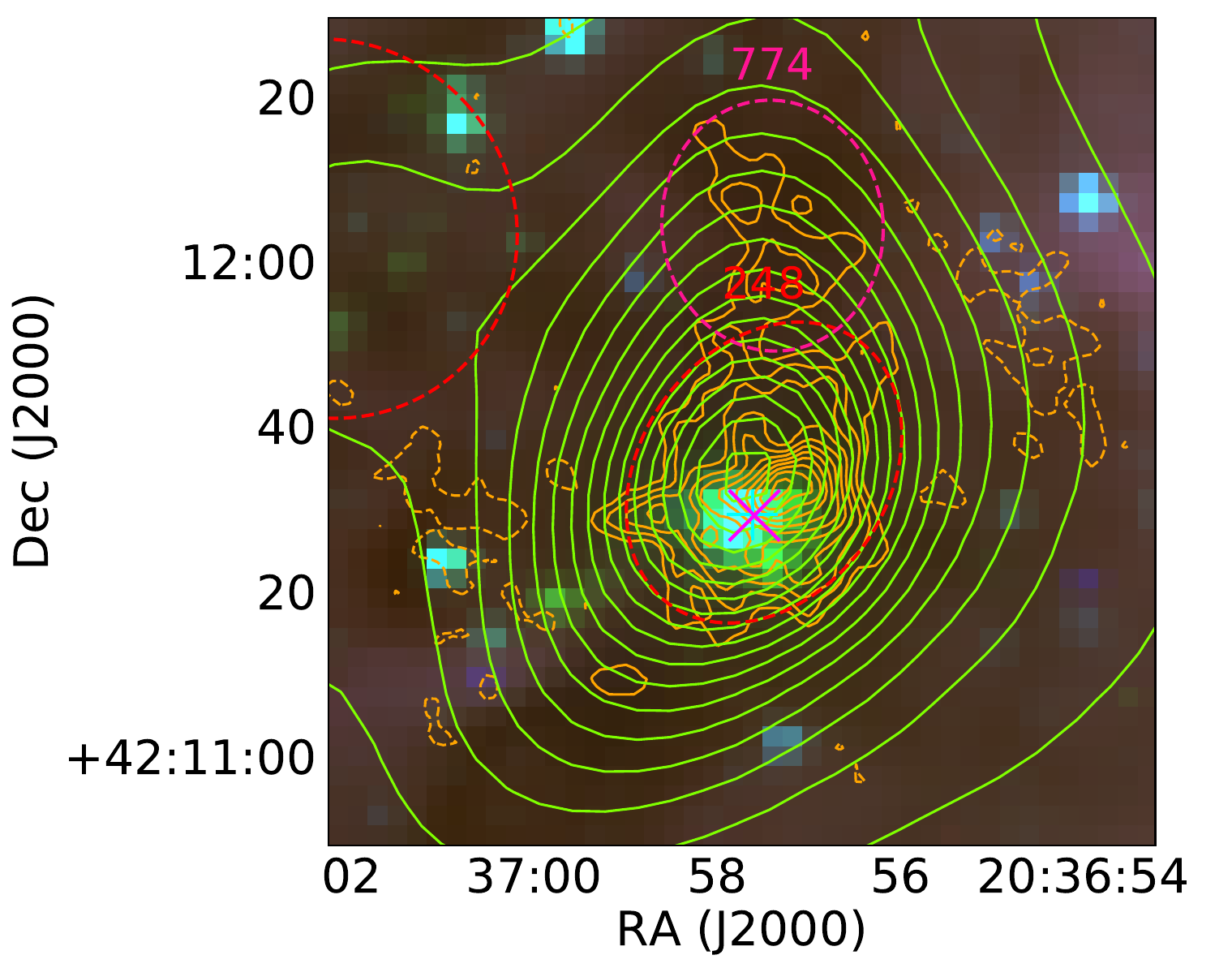}{0.45\textwidth}{MDC 774}}
    \gridline{\fig{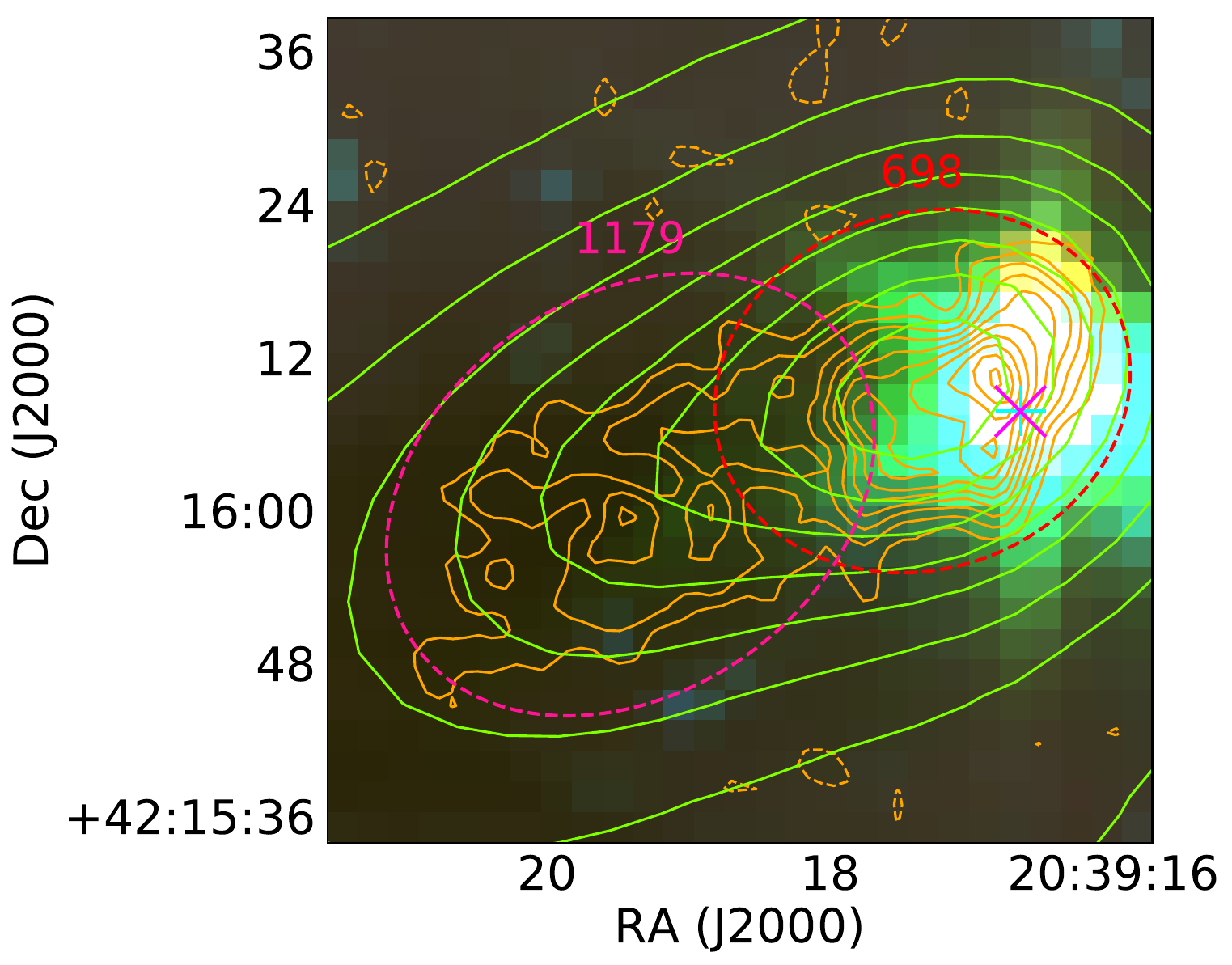}{0.45\textwidth}{MDC 1179}
    }
    \gridline{\fig{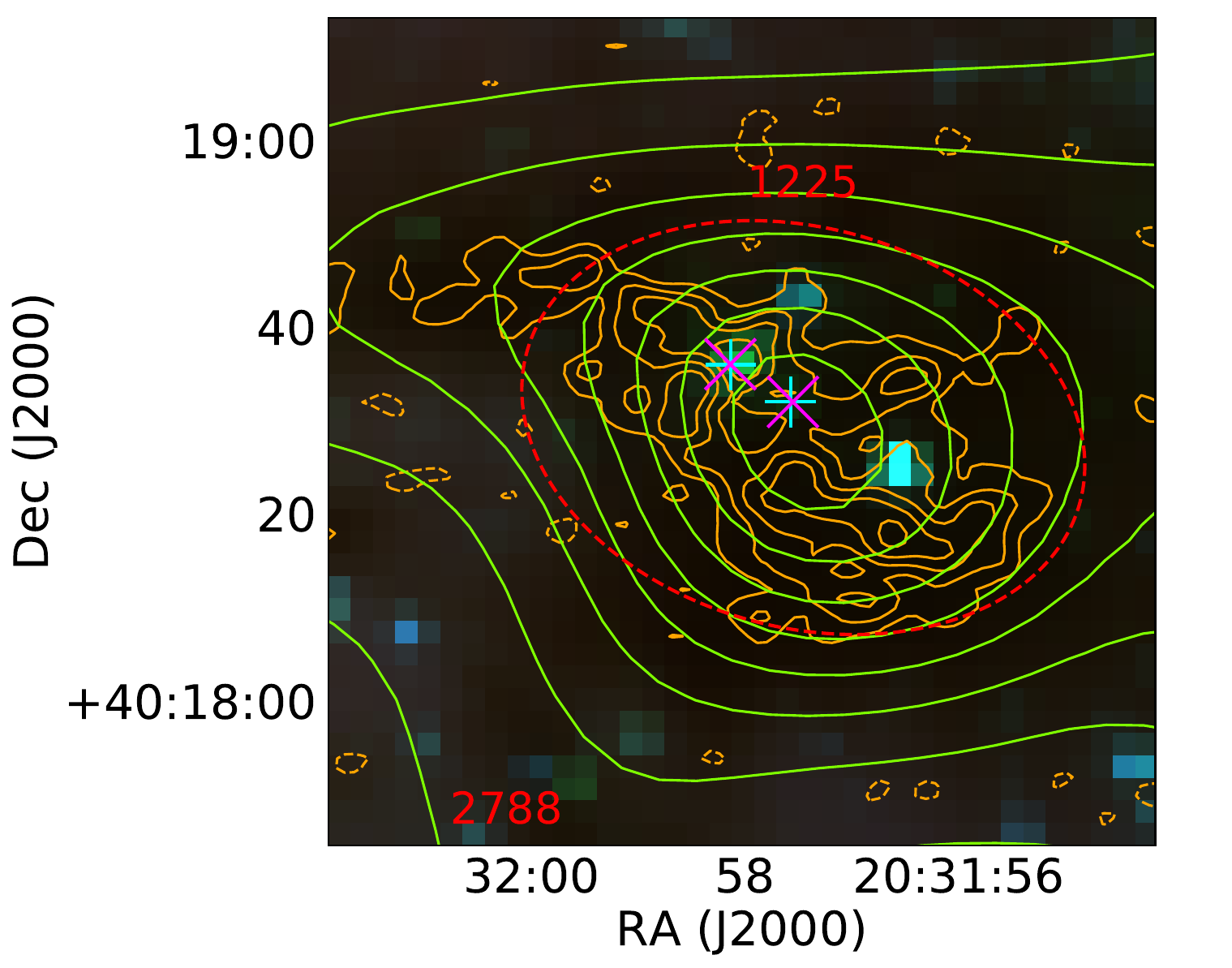}{0.45\textwidth}{MDC 1225}
    }
    \caption{ MDCs at the M1 stage (i.e., the \nh\ \oneone\ emission is roughly concentrated around the column density peaks). The background is pseudo three-color images of each field, with \emph{Spitzer} 8, 4.5, and 3.6\um\ emission colored in red, green, and blue, respectively. The contours of H$_2$ column density and \nh\ (1,1) integrated emission are over-plotted with green and orange, respectively. Cyan and magenta crosses mark the positions of any \water\ masers or radio continuum sources, respectively. The starless, IR-quiet, and IR-bright MDCs are shown as pink dashed, red dashed, and red solid ellipses, respectively. More figures are showed in Appendix \ref{app:C}.}
    \label{fig:m1} 
\end{figure}

As shown in Fig. \ref{fig:m1} and Appendix \ref{app:C}, the \nh\ \oneone\ emission is strongly correlated with dense regions of the N-map in most fields, indicating that the \nh\ \oneone\ line emission generally traces the dense structures in molecular gas. Nevertheless, the difference in resolution (\simi\ 3$^{\prime \prime}$.1 for \nh\ data and \simi\ 20$^{\prime \prime}$ for the N-map) leads to a more complicated morphology of \nh\ \oneone\ emission than that of the N-map for our MDCs. We also find that many of the peaks of the \nh\ \oneone\ emission show slight offsets from the peaks in the N-map, which could result from the depletion of \nh\ around protostars \citep{2011ApJ...740...45T,2017ApJ...849...68S} and from the difference in resolution between the N-map and the \nh\ cubes.

\begin{figure*}
    \centering
    \gridline{\fig{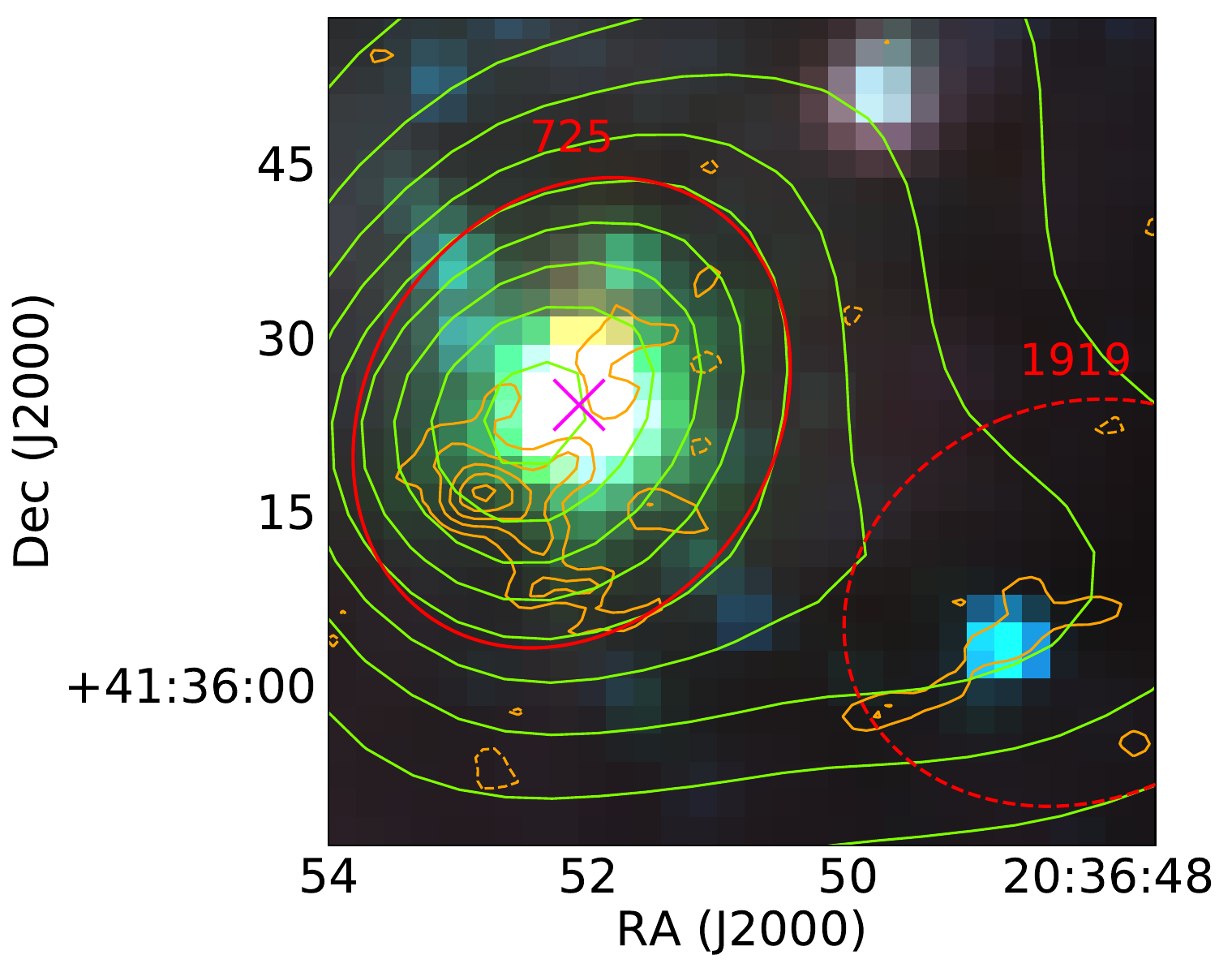}{0.4\textwidth}{MDC 725}
    \fig{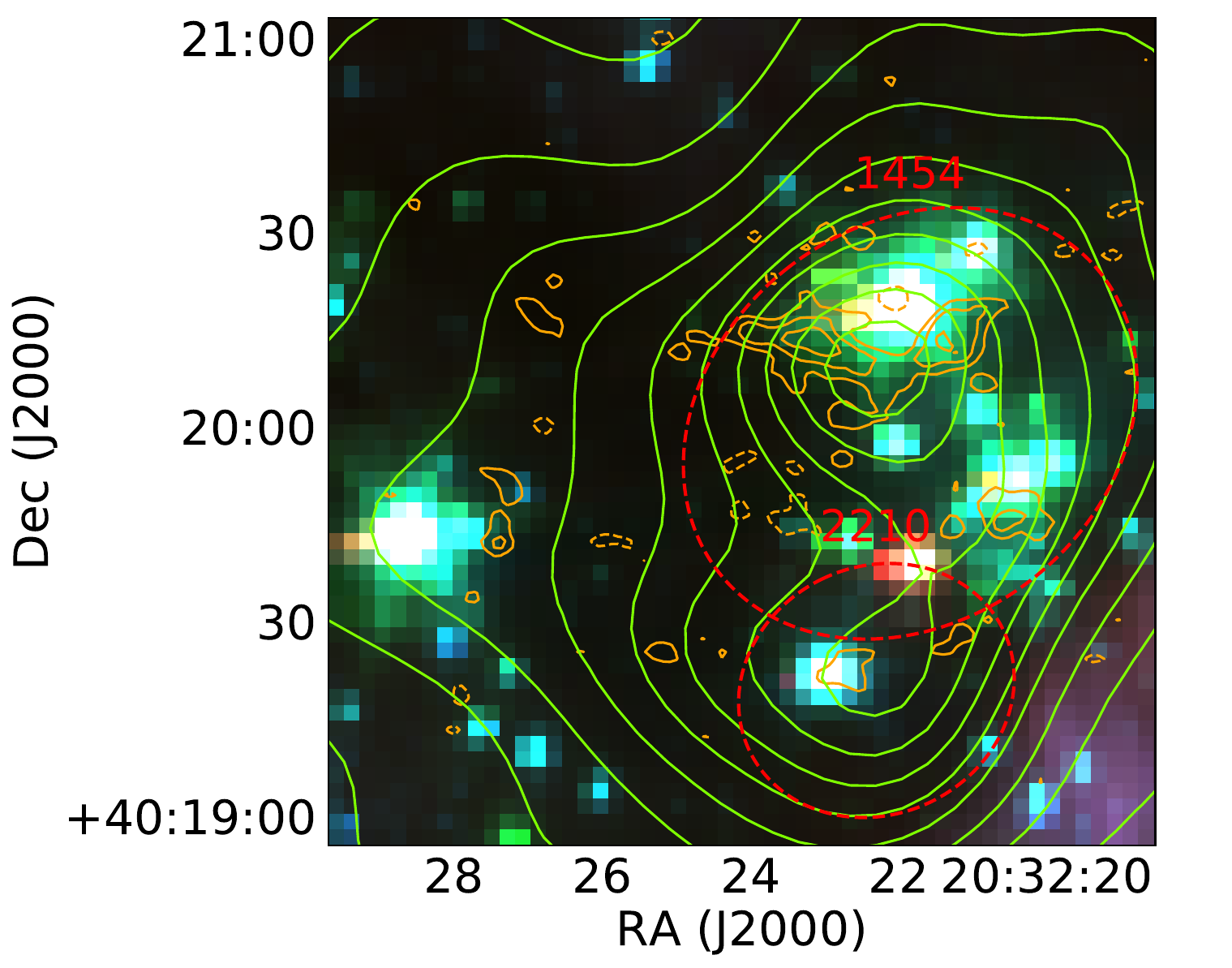}{0.4\textwidth}{MDC 1454}
    }
    \gridline{\fig{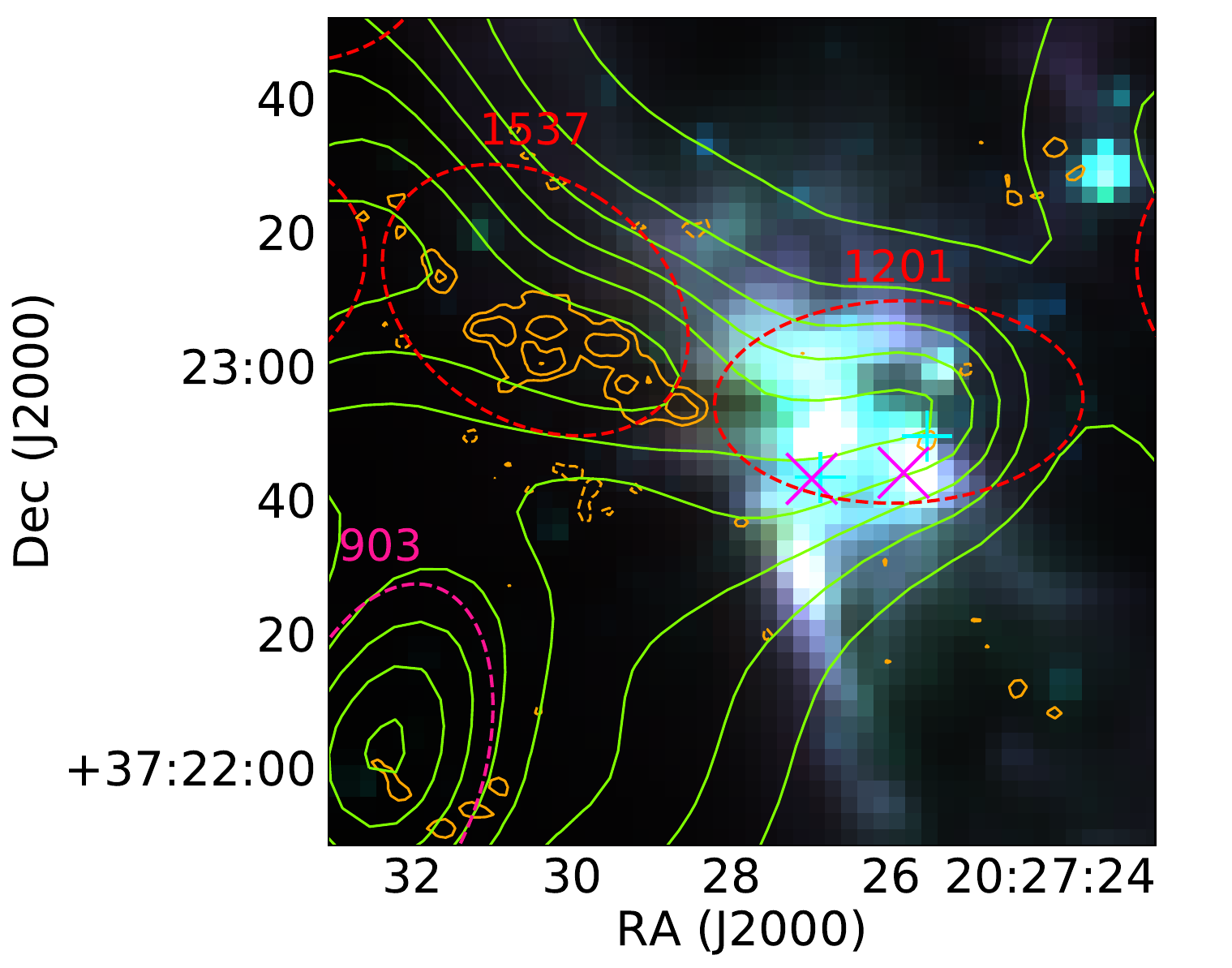}{0.4\textwidth}{MDC 1201}
    \fig{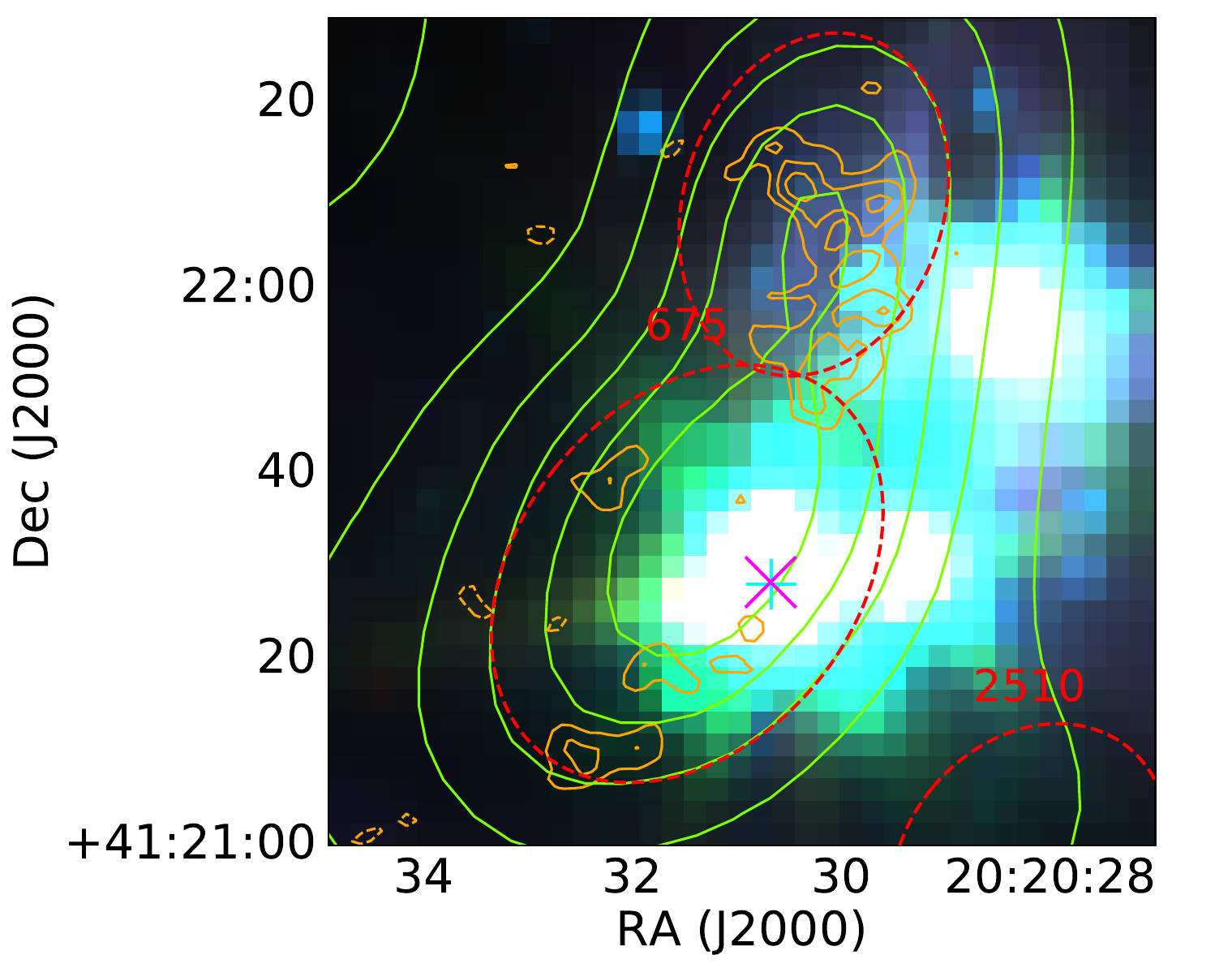}{0.4\textwidth}{MDC 675}
    }
    \caption{MDCs at the M2 stage (MDC 725 and 1454) and M3 stage (MDC 1201 and 675). Marks are the same as those in Fig. \ref{fig:m1}. For MDC 1201, there is nearly no \nh\ \oneone\ emission in it. As for MDC 675, \nh\ \oneone\ emission is only detected in a few isolated beams around the density peak. We therefore consider both of them to be MDCs at the M3 stage.}
    \label{fig:m23}
\end{figure*}

In light of their morphologies, we find that most (27 out of 35) MDCs are at the M1 stage. This result is consistent with our expectations, because the majority of our MDCs are not very evolved. As for the MDCs at the M2 stage (8 out of 35), we find evident star-forming activity (radio continuum emission, water masers, or bright sources on Spitzer images) in all of them, which corroborates the mentioned mechanism for distorted \nh\ \oneone\ morphology. The MDCs at the M3 stage are not included in Table \ref{tab:ir} because of their nondetection with the \nh\ \oneone\ line. Figures \ref{fig:m1} and \ref{fig:m23} present some example MDCs in each stage.

\subsubsection{A possible evolutionary sequence in our sample} \label{sec:5.1.2}

The evolution of MDCs is a fundamental piece in our understanding of high-mass star formation. There have been many studies focusing on this topic,  both observational and theoretical \citep[e.g.,][]{2009ApJS..181..360C,2014ApJ...787..113B,2019MNRAS.486.4508M}. Based on theoretical frameworks and observations, a simple four-stage evolutionary scenario of MDCs is proposed as CDMC\ra HDMC\ra DAMS\ra FIMS by \citet{2007ARA&A..45..481Z}, where CDMC, HDMC, DAMS and FIMS stand for cold dense massive core, hot dense massive core, disk-accreting main sequence star, and final main sequence star, respectively. Recently, \citet{2018ARA&A..56...41M} updated the star-formation pictures in different scales: (1) at the scale of 1--10 pc, the gas is fed into clumps by the inflow on ridges and hubs; (2) at the scale of \simi\ 0.1 pc, the beginning of star-forming activity will make starless MDCs grow into protostellar MDCs; (3) the stellar embryos in high-mass protostars grow from low-mass ones into high-mass ones, which transform the high-mass protostars from IR-quiet into IR-bright at \simi\ 0.02 pc scales; and (4) the \htwo\ region phase at 0.01--10 pc scales is the end of this scenario. 

Although the evolution of high-mass protostars cannot be properly determined based on their infrared SEDs ---as is possible for the low-mass ones--- due to their complicated environments \citep{1993ApJ...406..122A,2010A&A...522A..40M}, the infrared properties can still serve as an indicator of the evolution of embedded high-mass protostars. In order to obtain a better estimation of evolutionary stage, we adopt the infrared classification of \citet{2022ApJ...927..185W}\footnote{The IR-bright and quiet MDCs are classified according to their \emph{Spitzer} 24 \um\ fluxes or MSX 21 \um\ fluxes when 24 \um\ data are saturated. The criteria are 23 Jy for 24 \um\ flux and 17 Jy for 21 \um\ flux. The starless candidates are selected from IR-quiet MDCs, which are not associated with any water maser, radio source, or IR sources in 2MASS, \emph{Spitzer}, \emph{Herschel}, and MSX catalogs.}. The MDCs are divided into three groups: 8 IR-bright MDCs, 25 IR-quiet MDCs, and 2 starless MDCs. The IR-bright MDCs correspond to MDCs with possible embedded high-mass protostar(s), while the IR-quiet ones only harbor low-mass protostars. Regarding the MDCs without any star-forming activity according to the criterion of \citet{2022ApJ...927..185W}, we take them as starless MDC candidates. To illustrate the evolutionary stages of MDCs, we combine the data at hand, which include \nh\ emission data, the N-map, water masers, radio continuum sources, and mid-infrared pseudo three-color maps (8, 4.5, and 3.6 \um\ emission colored in red, green, and blue, respectively), into a composite image for each field (see Figures \ref{fig:m1} and \ref{fig:m23}). 

Figure \ref{fig:m1} shows three MDCs at the M1 stage. The MDC 774 and 1179 are the only two starless MDCs in our sample. In the IR-quiet MDC 1225, we find two radio sources and two water masers, indicating star-forming activities and the existence of protostar(s). As the protostar or protostars grow, stronger feedback distorts the morphology of \nh, which could be the situation shown in Fig. \ref{fig:m23}. The MDC 725 and 1454 are considered to be at stage M2. The MDC 725 is an IR-bright MDC with a radio source; that is, the high-mass stellar embryo or embryos have formed in it, while the MDC 1454 is an IR-quiet MDC with \nh\ \oneone\ emission distorted near the density peak. The MDC 1201 and 675 illustrate the latest stages in our sample, where most of the \nh\ is destroyed and \nh\ \oneone\ emission is almost undetected within them. The MDC 1201 and 675 are IR-bright and IR-quiet, respectively. In conclusion, the morphology of the \nh\ \oneone\ emission could be a possible indicator for the evolutionary stage of MDCs.

\subsection{Virial analysis} \label{sec:5.2}

The virial parameter, which is the ratio between the kinetic energy and the gravitational energy, is an important indicator of the dynamical conditions of molecular structures, especially for those in early evolutionary stages. Typically, the turbulent core accretion model requires a virialized MDC as the birth place of the high-mass star \citep{2002Natur.416...59M}. Previous studies in which a virial analysis was carried out using different gas tracers at different size scales found that the distribution of virial parameters varies greatly with the sizes of molecular structures \citep[e.g.,][]{1981MNRAS.194..809L,2001ApJ...551..852H,2019ApJ...875...24C}. The virial parameters of molecular clouds (\simi\ 10 pc) are mostly $\gg$ 2, which is the critical value of an isothermal sphere in hydrostatic equilibrium  embedded in a pressurized medium \citep{1956MNRAS.116..351B,1955ZA.....37..217E,1981MNRAS.194..809L}, while virial (1 $\leq$ $\alpha_{vir}$ $\leq$ 2) and subvirial structures ($\alpha_{vir}$ $<$ 1) are found in the fragments (\simi\ 0.01--1 pc) within molecular clouds \citep[e.g.,][]{2011A&A...530A.118P,2013ApJ...768L...5L}. In this section, we perform a virial analysis of MDCs and try to understand their stability.

\subsubsection{Calculation of virial parameters} \label{5.2.1}

As defined in \citet{1992ApJ...395..140B}, the virial parameter, $\alpha_{\rm vir}$, is in the form of
\begin{equation} \label{equ:vir}
    \alpha_{\rm vir} \equiv \frac{5\sigma_v^2R}{{\rm G}M},
\end{equation}
where $\sigma_v$ is the velocity dispersion and $R$ is the radius of the structure. In this work, the total velocity dispersion of the whole molecular gas, $\sigma_v$, is obtained from the \nh\ results, and the mass of MDCs is provided by \citet{2021ApJ...918L...4C}. We therefore calculate the virial parameters for MDCs that have both \nh\ \oneone\ and \twotwo\ detections. It is worth noting that we include the contributions from both turbulence and kinematics in the calculation of $\sigma_v$. Detailed calculation processes are listed in Appendix \ref{app:E}.

\begin{figure}
    \centering
    \includegraphics[width=8.5cm]{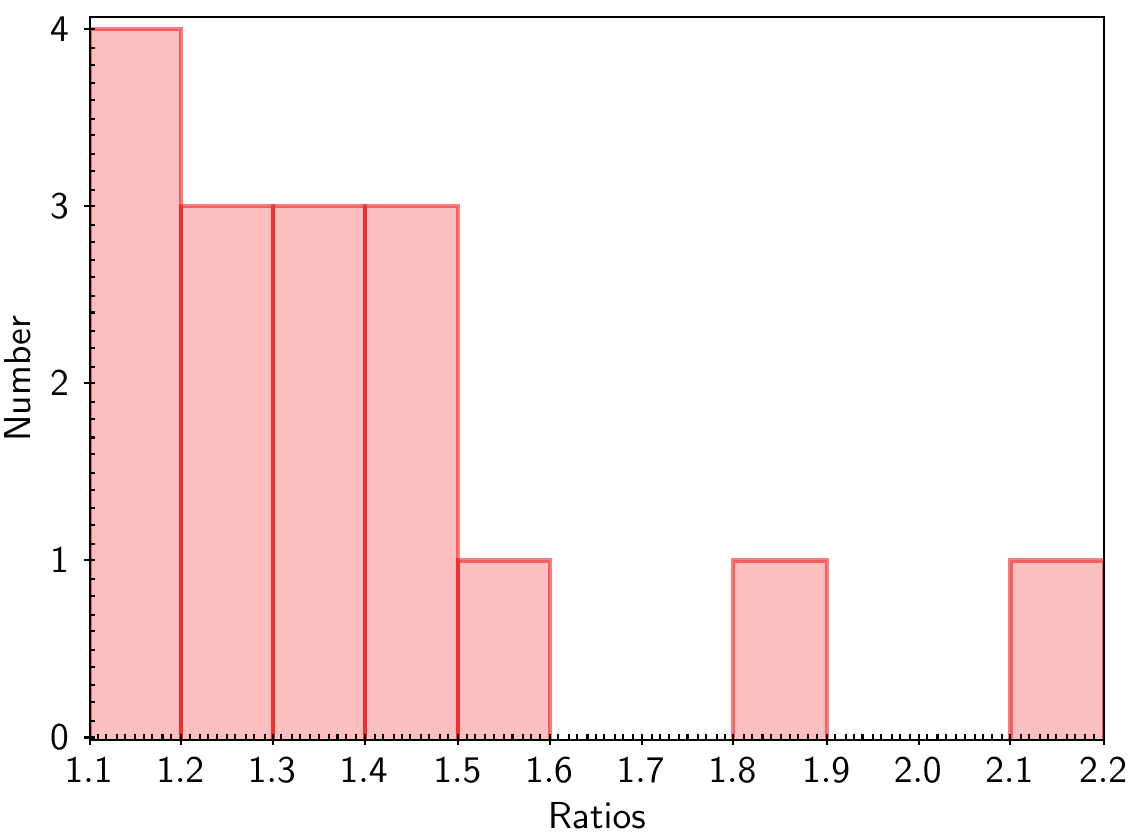}
    \caption{Distribution of the ratios between $R_{sel}/M_{sel}$ and $R_{total}/M_{total}$}
    \label{fig:rat}
\end{figure}

In many MDCs, the detected NH3 emission only covers part of the spatial extent of MDCs (such as the MDC 725 in Fig. \ref{fig:m23}). For these cases, we take the sizes of their detected parts, $R_{\rm sel}$, and calculate the corresponding mass, $M_{\rm sel}$, by multiplying their areas by the column densities taken from the N-map. Using $R_{\rm sel}$ and $M_{\rm sel}$, the virial parameters are calculated for the detected parts of the MDCs, which are called ``partial" virial parameters in our analysis. The partial virial parameters are considered to be an upper limit value for their host MDCs. Such overestimation comes from two aspects, $R/M$ and $\sigma_v$. First, we compare $R_{\rm sel}/M_{\rm sel}$ of the selected parts and $R_{\rm total}/M_{\rm total}$ of their host MDCs. Figure \ref{fig:rat} shows that all the ratios between $R_{\rm sel}/M_{\rm sel}$ and $R_{\rm total}/M_{\rm total}$ are larger than 1, suggesting $R/M$ is overestimated in the partial virial parameters. Second, as shown in Appendices \ref{app:A} and \ref{app:B}, the velocity dispersion is usually enhanced where the integral \nh\ (1,1) emission is strong in most sources. The undetected parts should therefore have lower velocity dispersion than the detected parts, indicating the velocity dispersion is also overestimated in the partial virial parameters.

\subsubsection{Distribution of virial parameters}

Through the above calculations, we determined the virial parameters for 35 MDCs, including 14 full ones and 21 partial ones, as listed in Table \ref{tab:ir}. The mean and median values of the virial parameter are 1.0 and 0.9, respectively, with a standard deviation of 1.0. The majority of MDCs (23 out of 35) have a virial parameter of smaller than 1. Moreover, most full virial parameters (11 out of 14) are smaller than 1, with only one exceeding 2. Furthermore, the partial virial parameters, mostly distributed around 1, are considered as upper limits in our analysis, indicating that the majority of MDCs could be subvirialized. This result is consistent with previous findings that many MDCs are moderately to strongly subvirial \citep[e.g.,][]{2011A&A...530A.118P,2012A&A...544A.146W,2013ApJ...768L...5L,2014ApJ...790...84L,2017ApJ...850....3K,2019ApJ...874..147K,2020MNRAS.491.4310T,2015ApJ...804..141Z,2016ApJ...833..209O,2020ApJ...896..110L}. Recently, \citet{2021ApJ...922...87S} proposed three possible sources of bias that could result in a false subvirial result when calculating virial parameters: neglected bulk motions, an overestimated density profile correction factor, and mass contribution from the fore- and/or background. We tried to exclude these sources of bias in our calculation: (1) The bulk motions are considered in the total kinetic energy of MDCs; (2) the density profile correction factor is taken as 1 as defined in \citet{2021ApJ...922...87S}, which could result in an overestimated virial parameter rather than an underestimated one; and (3) the mass of our MDCs is derived from the SED fitting results of \citet{2019ApJS..241....1C}, where the fore- and background are removed in the fitting processes. Our subvirial results should therefore not be affected by these sources of bias and we consider them to be trustworthy.

\begin{figure}
    \centering
    \includegraphics[width=8.5cm]{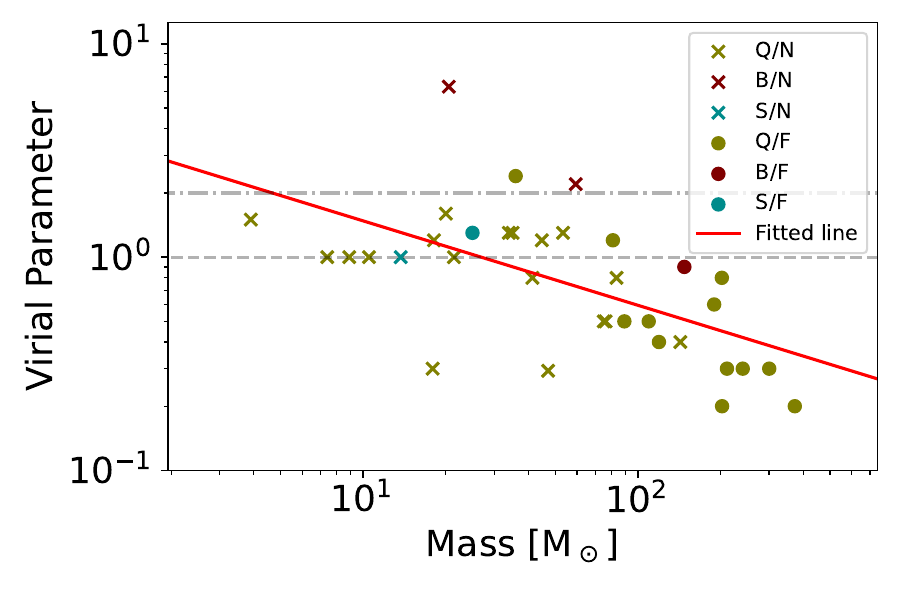}
    \caption{Virial parameters vs. mass of the 35 MDCs in Cygnus X. In the legends, ``B'', ``Q'', and ``S'' refer to different infrared classes, IR-bright, IR-quiet, and starless, respectively. ``F'' and ``N'' are the same as those in Table \ref{tab:ir}. The red line shows the least-linear-square fitting result. The gray dashed lines show where virial parameters are equal to 1 and 2.}
    \label{fig:av}
\end{figure}

As a result, our MDCs appear to be mostly subvirialized, which is not expected based on the turbulence core accretion model \citep{2002Natur.416...59M}. This suggests that MDCs could undergo dynamical collapse in nonequilibrium in the early stages of star formation unless there is additional support. Regarding high-mass star formation, the subvirial prestellar cores, according to the simulation work by \citet{2019ApJ...887..108R}, are more likely to form massive stars than virialized ones due to their higher accretion rates and lower levels of fragmentation. Additionally, an anti-correlation between the virial parameter and mass has been found in many works using only a single gas tracer \citep[e.g.,][]{2008ApJ...672..410L,2009ApJ...696..298F}. We also plot virial parameter versus mass for our sample of MDCs in Fig. \ref{fig:av}. However, the Pearson's correlation coefficient of the fitted line is only 0.18, which indicates no correlation. This is mainly due to the fact that the mass range of the MDCs in our sample is very small compared with other works \citep[e.g.,][]{2010ApJ...723..492R,2012A&A...544A.146W}, namely \simi\ 10$^1$--10$^2$ \msun. This trend also disappears in the combination of works with different gas tracers and different samples \citep{2013ApJ...779..185K}. \citet{2018A&A...619L...7T} interpret this trend as an observational bias due to the limitation of gas tracers.

\begin{figure}
    \centering
    \includegraphics[width=\hsize]{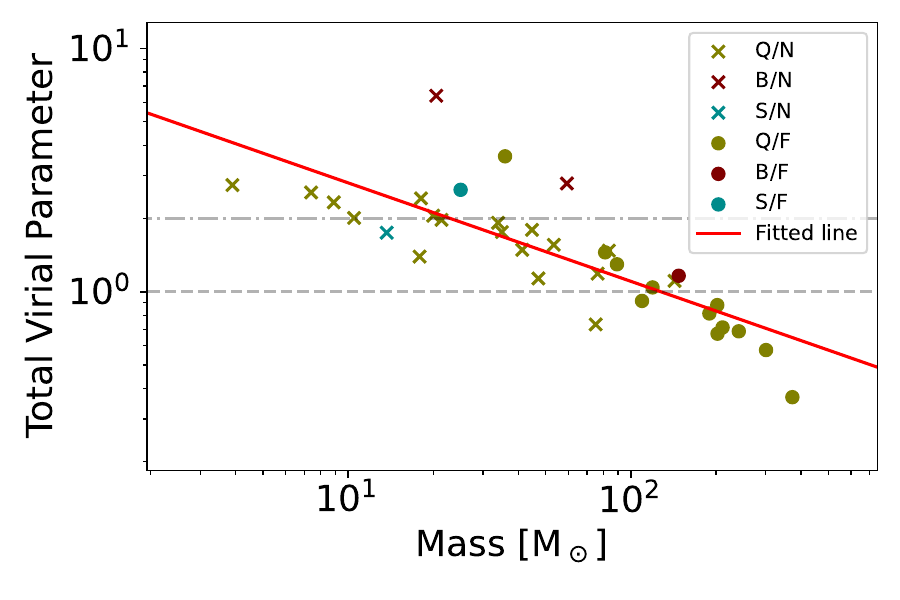}
    \caption{Total virial parameter v.s. mass. The legends are the same as those in Fig. \ref{fig:av}.\label{fig:ab}}
\end{figure}

As mentioned above, the subvirial MDCs would collapse if there were no additional support. Previous studies suggested that the magnetic field could provide significant support in MDCs \citep[e.g.,][]{2014A&A...567A.116F,2019FrASS...6....3H,2020ApJ...895..142L,2022ApJ...925...30L}. Following the methods in \citet{2020ApJ...895..142L}, the total virial mass ---which includes the contribution of magnetic field--- can be calculated as
\begin{equation}
M_{{\rm k}+B}=\sqrt{M_B^2+(\frac{M_{\rm k}}{2})^2}+\frac{M_{\rm k}}{2},
\end{equation}
where $M_{\rm k}$ = $\alpha_{\rm vir} M$ is the kinetic virial mass, and $M_B$ here is
\begin{equation}
M_B=\frac{\pi R^2B}{\sqrt{\frac{9}{10}\mu_0\pi G}}.
\end{equation}
We then calculated the total virial parameter, $\alpha_{{\rm k}+B}=M_{{\rm k}+B}/M$, by assuming a typical magnetic field strength of \simi\ 0.5 mG \citep{2014MNRAS.438..663H}. The results are shown in Fig. \ref{fig:ab} for comparison with the original results in Fig. \ref{fig:av}.

\begin{landscape}
\begin{table}
\caption{Properties of MDCs with \nh\ Detection \label{tab:ir}}
\centering
\begin{tabular}{c|cccccccccccccc}
\hline\hline
MDC & IR-class  & Morph\tablefootmark{a} & \nh\tablefootmark{b}  & $R_{\rm sel}$\tablefootmark{c} & $M_{\rm sel}$\tablefootmark{d} & $\sigma_v$ & \vnt & \vng & $T_{\rm dust}$\tablefootmark{e} & $T_{\rm gas}$ & $\alpha_{\rm vir}$ & $\mathcal{M}_{\rm NT}$ & $\mathcal{M}_{\rm ng}$ & $N_{\rm H_2}$\tablefootmark{f}\\
 &  &  &  & (pc) & (M$_{\odot}$) & (km s$^{-1}$) & (km s$^{-1}$) & (km s$^{-1}$) & (K) & (K) & & & & ($10^{22}$ cm$^{-2}$) \\
\hline
220 & IR-quiet & M1 & N & 0.082 & 75.1 & 0.61 & 0.52 & 0.47 & 17.8 & 27.3 & 0.5 & 1.7 & 1.5 & 15.8 \\
274 & IR-bright & M2 & N & 0.122 & 59.4 & 0.97 & 0.93 & 0.76 & 21.3 & 19.0 & 2.2 & 3.6 & 2.9 & 5.7 \\
584 & IR-quiet & M1 & N & 0.055 & 7.4 & 0.34 & 0.23 & 0.23 & 20.3 & 17.4 & 1.0 & 0.9 & 0.9 & 3.5 \\
725 & IR-bright & M2 & F & 0.133 & 147.6 & 0.91 & 0.87 & 0.74 & 24.0 & 21.3 & 0.9 & 3.2 & 2.7 & 11.8 \\
248 & IR-quiet & M1 & F & 0.117 & 202.1 & 1.06 & 1.01 & 0.99 & 18.2 & 26.2 & 0.8 & 3.3 & 3.3 & 21.0 \\
774 & Starless & M1 & F & 0.098 & 25.0 & 0.53 & 0.46 & 0.45 & 16.4 & 18.9 & 1.3 & 1.8 & 1.7 & 3.7 \\
714 & IR-bright & M1 & N & 0.059 & 20.5 & 1.37 & 1.35 & 1.23 & 22.7 & 24.8 & 6.3 & 4.5 & 4.1 & 8.3 \\
1267 & IR-quiet & M1 & F & 0.155 & 202.5 & 0.44 & 0.37 & 0.31 & 15.3 & 14.7 & 0.2 & 1.6 & 1.4 & 12.0 \\
3188 & IR-quiet & M1 & N & 0.067 & 17.9 & 0.27 & 0.39 & 0.31 & 16.3 & 14.1 & 0.3 & 1.7 & 1.4 & 5.7 \\
698 & IR-quiet & M1 & F & 0.103 & 81.1 & 0.89 & 0.85 & 0.75 & 17.0 & 20.2 & 1.2 & 3.2 & 2.8 & 10.9 \\
1179 & IR-quiet & M1 & F & 0.123 & 35.9 & 0.78 & 0.73 & 0.55 & 16.0 & 17.5 & 2.4 & 2.9 & 2.2 & 3.4 \\
532 & IR-quiet & M2 & N & 0.055 & 10.5 & 0.41 & 0.30 & 0.28 & 23.2 & 21.4 & 1.0 & 1.1 & 1.0 & 4.9 \\
801 & IR-quiet & M2 & N & 0.117 & 76.3 & 0.55 & 0.48 & 0.45 & 17.8 & 18.4 & 0.5 & 1.9 & 1.7 & 7.9 \\
684 & IR-quiet & M1 & F & 0.117 & 109.5 & 0.64 & 0.59 & 0.54 & 17.6 & 19.3 & 0.5 & 2.3 & 2.1 & 11.4 \\
327 & IR-quiet & M1 & N & 0.097 & 47.1 & 0.35 & 0.26 & 0.25 & 20.9 & 16.4 & 0.3 & 1.1 & 1.0 & 7.1 \\
619 & Starless & M1 & N & 0.057 & 13.7 & 0.45 & 0.36 & 0.31 & 22.2 & 20.3 & 1.0 & 1.3 & 1.1 & 6.0 \\
742 & IR-quiet & M2 & N & 0.077 & 21.4 & 0.50 & 0.42 & 0.35 & 27.0 & 24.9 & 1.0 & 1.4 & 1.2 & 5.1 \\
310 & IR-quiet & M1 & N & 0.063 & 20.0 & 0.66 & 0.60 & 0.60 & 20.5 & 18.0 & 1.6 & 2.4 & 2.4 & 7.1 \\
507 & IR-quiet & M1 & F & 0.138 & 372.5 & 0.68 & 0.63 & 0.61 & 21.4 & 21.2 & 0.2 & 2.3 & 2.2 & 27.8 \\
753 & IR-quiet & M1 & F & 0.128 & 189.5 & 0.87 & 0.83 & 0.65 & 17.8 & 23.0 & 0.6 & 2.9 & 2.3 & 16.4 \\
798 & IR-quiet & M1 & F & 0.137 & 89.3 & 0.52 & 0.46 & 0.43 & 19.2 & 17.6 & 0.5 & 1.8 & 1.7 & 6.8 \\
874 & IR-quiet & M1 & F & 0.152 & 211.0 & 0.61 & 0.55 & 0.40 & 17.9 & 15.9 & 0.3 & 2.3 & 1.7 & 13.0 \\
1537 & IR-quiet & M1 & N & 0.099 & 44.7 & 0.67 & 0.62 & 0.58 & 25.3 & 23.7 & 1.2 & 2.1 & 2.0 & 6.5 \\
723 & IR-quiet & M1 & N & 0.087 & 33.9 & 0.66 & 0.62 & 0.57 & 16.0 & 14.5 & 1.3 & 2.7 & 2.5 & 6.4 \\
509 & IR-quiet & M2 & F & 0.155 & 300.6 & 0.72 & 0.67 & 0.64 & 18.2 & 18.9 & 0.3 & 2.6 & 2.4 & 17.8 \\
351 & IR-quiet & M1 & N & 0.081 & 35.0 & 0.69 & 0.64 & 0.49 & 18.3 & 18.2 & 1.3 & 2.5 & 1.9 & 7.6 \\
1225 & IR-quiet & M1 & N & 0.131 & 83.6 & 0.66 & 0.62 & 0.43 & 14.3 & 12.3 & 0.8 & 3.0 & 2.1 & 6.9 \\
1454 & IR-quiet & M2 & N & 0.158 & 142.7 & 0.59 & 0.52 & 0.37 & 16.2 & 17.4 & 0.4 & 2.1 & 1.5 & 8.1 \\
892 & IR-quiet & M1 & N & 0.038 & 3.9 & 0.37 & 0.18 & 0.18 & 31.6 & 29.6 & 1.5 & 0.6 & 0.5 & 3.9 \\
540 & IR-quiet & M1 & F & 0.159 & 240.8 & 0.63 & 0.58 & 0.47 & 15.5 & 16.3 & 0.3 & 2.4 & 1.9 & 13.5 \\
302 & IR-quiet & M2 & N & 0.08 & 18.1 & 0.48 & 0.39 & 0.35 & 22.3 & 20.3 & 1.2 & 1.5 & 1.3 & 4.0 \\
520 & IR-quiet & M1 & N & 0.057 & 8.9 & 0.36 & 0.24 & 0.20 & 23.4 & 18.3 & 1.0 & 0.9 & 0.8 & 3.9 \\
340 & IR-quiet & M1 & F & 0.139 & 119.3 & 0.57 & 0.52 & 0.47 & 16.1 & 15.0 & 0.4 & 2.3 & 2.0 & 8.8 \\
214 & IR-quiet & M1 & N & 0.084 & 53.4 & 0.84 & 0.78 & 0.77 & 21.2 & 27.2 & 1.3 & 2.5 & 2.5 & 10.7 \\
247 & IR-quiet & M1 & N & 0.092 & 41.3 & 0.56 & 0.46 & 0.41 & 21.3 & 20.1 & 0.8 & 1.7 & 1.5 & 6.9 \\
\hline
\end{tabular}
\tablefoot{
\tablefoottext{a}{The morphology of \nh. MDCs at the M3 stage are not listed because of their nondetection.}
\tablefoottext{b}{Distribution of \nh\ within MDCs. ``F'' for \nh\ emission covering almost the whole spatial extent of MDCs and ``N'' for \nh\ emission only covering part of the spatial extent of MDCs.}
\tablefoottext{c}{The effective radii of selected regions.}
\tablefoottext{d}{The mass of selected regions.}
\tablefoottext{e}{Refer to \citet{2021ApJ...918L...4C}}
\tablefoottext{f}{Measured from the N-map in \citet{2019ApJS..241....1C}}
}
\end{table}
\end{landscape}

The mean and median values of the total virial parameter are 2.2 and 2.1, with a standard deviation of 1.2. The majority of total virial parameters are approximately 2 \citep[as shown in][the critical virial parameter is \simi\ 2 for Bonnor-Ebert spheres]{2013ApJ...779..185K}, indicating that most MDCs are expected to be in stability with a typical magnetic field strength of 0.5 mG. In addition, if there is a strong magnetic field strength up to 1 mG, the total virial parameter would increase by \simi\ 1-2 in value. In addition, \citet{2022ApJ...925...30L} find an anti-correlation between the magnetic virial parameter ($\alpha_B=M_B/M$) and the column density of H$_2$. Considering the typical column density of our MDCs, of namely \simi\ 10$^{23}$ \uden, the magnetic virial parameters are \simi\ 0.5 in \citet{2022ApJ...925...30L}, indicating that the magnetic field could provide remarkable support against the gravity. In our results, the nonthermal velocity dispersion of MDCs is lower than that predicted by \citet{2002Natur.416...59M}, \simi\ 1.65($m$/30\msun)$^{1/4}\Sigma^{1/4}$ km s$^{-1}$. Subsequent studies using this model \citep[e.g.,]{2005ApJ...630..250K} also proposed significant support from the magnetic field against the gravity at the scale of MDCs. The Cygnus X North and South regions are also covered by the lower resolution \nh\ survey (KEYSTONE) presented in \citet{2019ApJ...884....4K}. The virial parameter of the \nh\ cores presented by these authors ($\gtrsim$ 0.1 pc, referring to the \nh\ ``leaves'' in that paper) show higher values than our results. A significant part of the \nh\ cores are at supervirial states in \citet{2019ApJ...884....4K}. Therefore, the parent structures of our \nh\ fragments have larger virial parameters, indicating that the accretion of ``cores'' may not be dominated by the gravity. The gas might be fed to the ``core'' via the filament or the initial inflow suggested by \citet{2018ARA&A..56...41M} and \citet{2020ApJ...900...82P}.

\subsection{Dynamical states in star-forming regions}\label{sec:5.3}

\begin{figure*}
\gridline{\fig{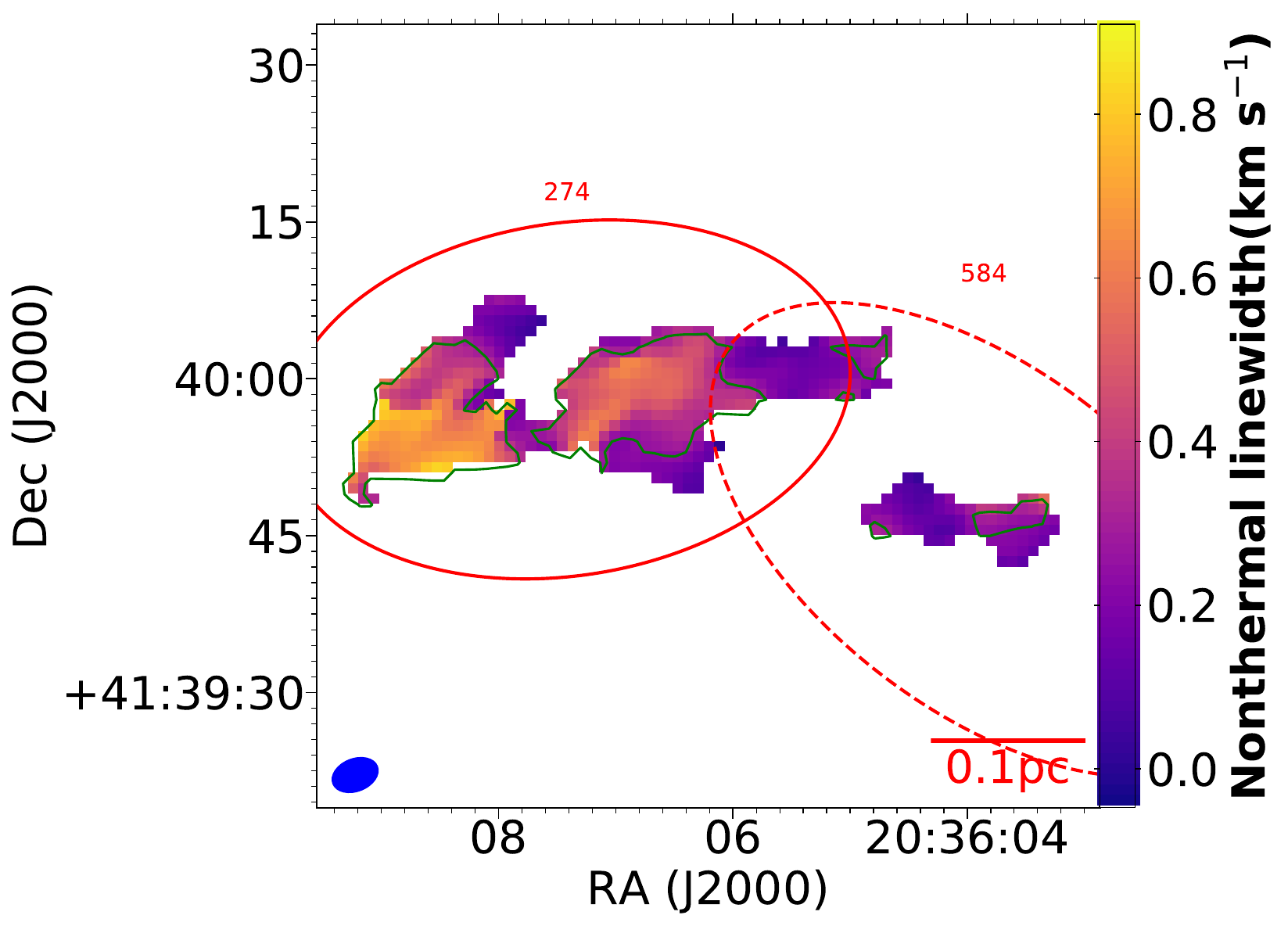}{0.33\textwidth}{Field 3}
          \fig{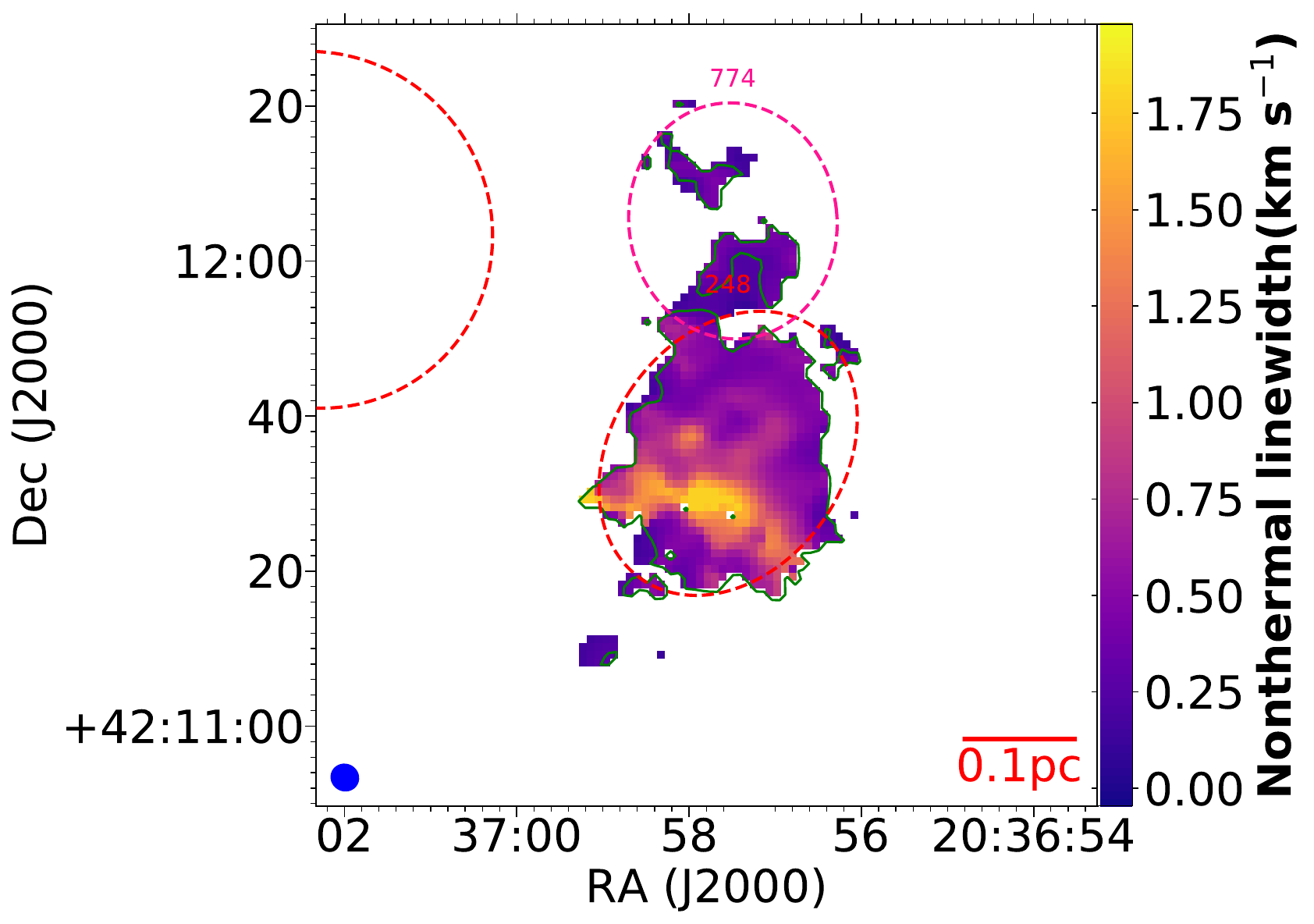}{0.33\textwidth}{Field 5}
          \fig{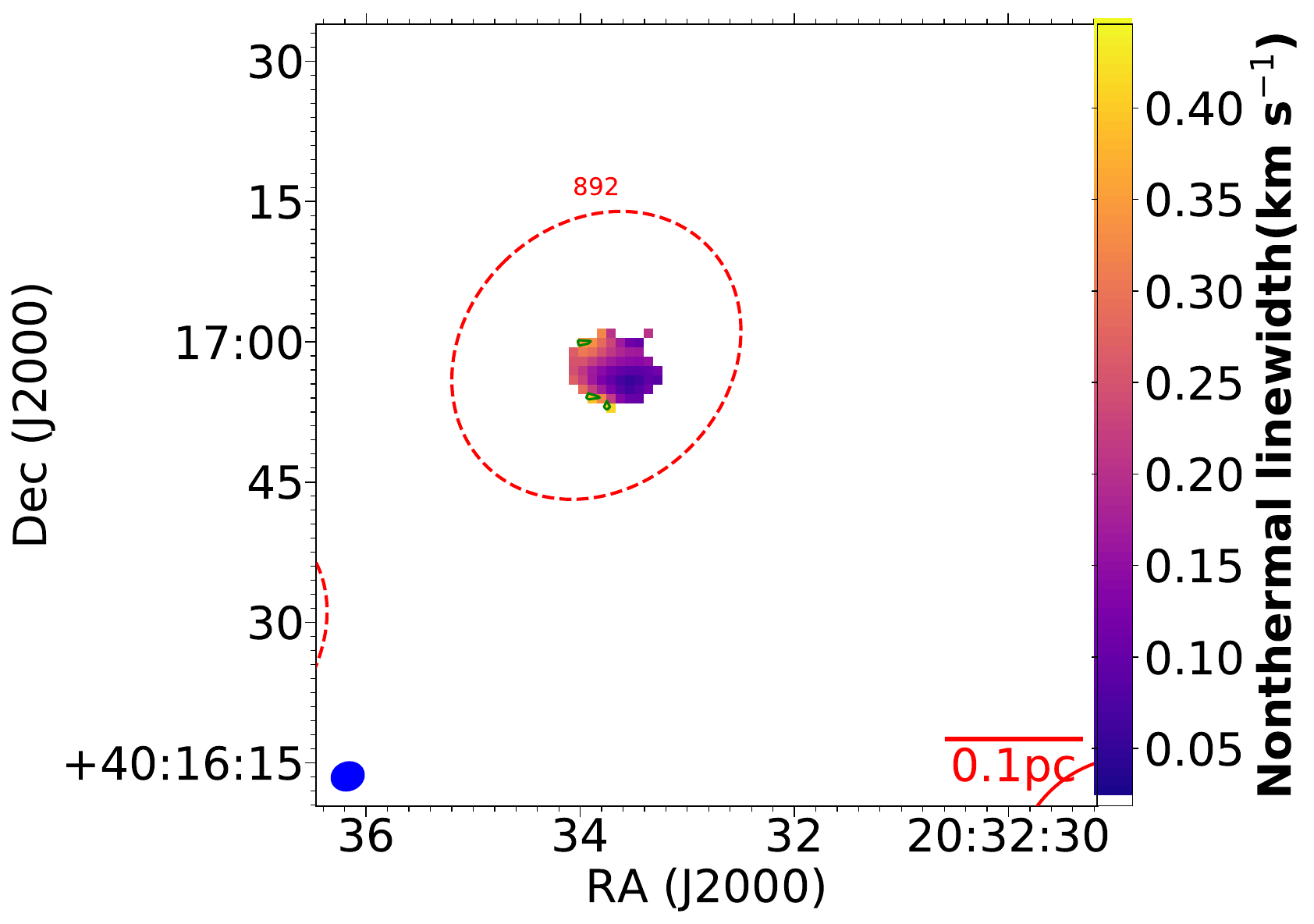}{0.33\textwidth}{Field 26}
          }
\caption{Nonthermal velocity dispersion maps of Fields 3, 5, and 26. The green contours show the regions with Mach numbers of greater than 1. The ellipses are the same as those in Fig. \ref{fig:m0}}
\label{fig:NT}
\end{figure*}

We produced nonthermal velocity dispersion maps of each field and mark the regions where the nonthermal velocity dispersion is higher than the sound speed with green contours (Fig. \ref{fig:NT}). We find the dynamical states vary among different MDCs. As shown in Fig. \ref{fig:NT}, the MDC 248 is dominated by the trans-sonic to supersonic nonthermal velocity dispersion ($\mathcal{M}_{\rm NT}\ \geq$ 1), while the nonthermal velocity dispersion in  MDC 892 is almost entirely subsonic. The majority of MDCs consist of both supersonic and subsonic regions, such as MDCs 274, 584, and 774. Maps of other fields are reported in Appendix \ref{app:D}.

\begin{figure*} 
\centering
\gridline{\fig{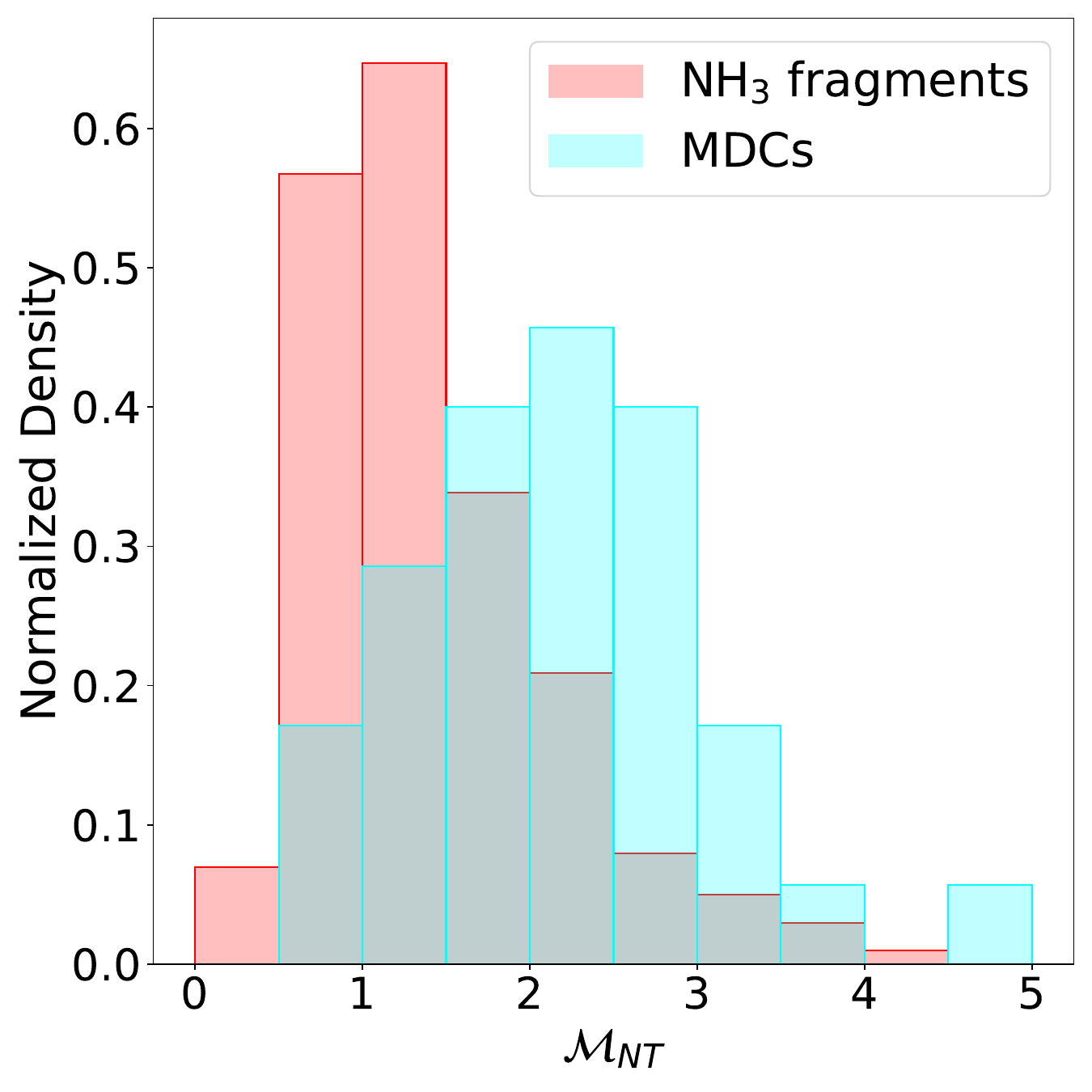}{0.4\textwidth}{(a)}
\fig{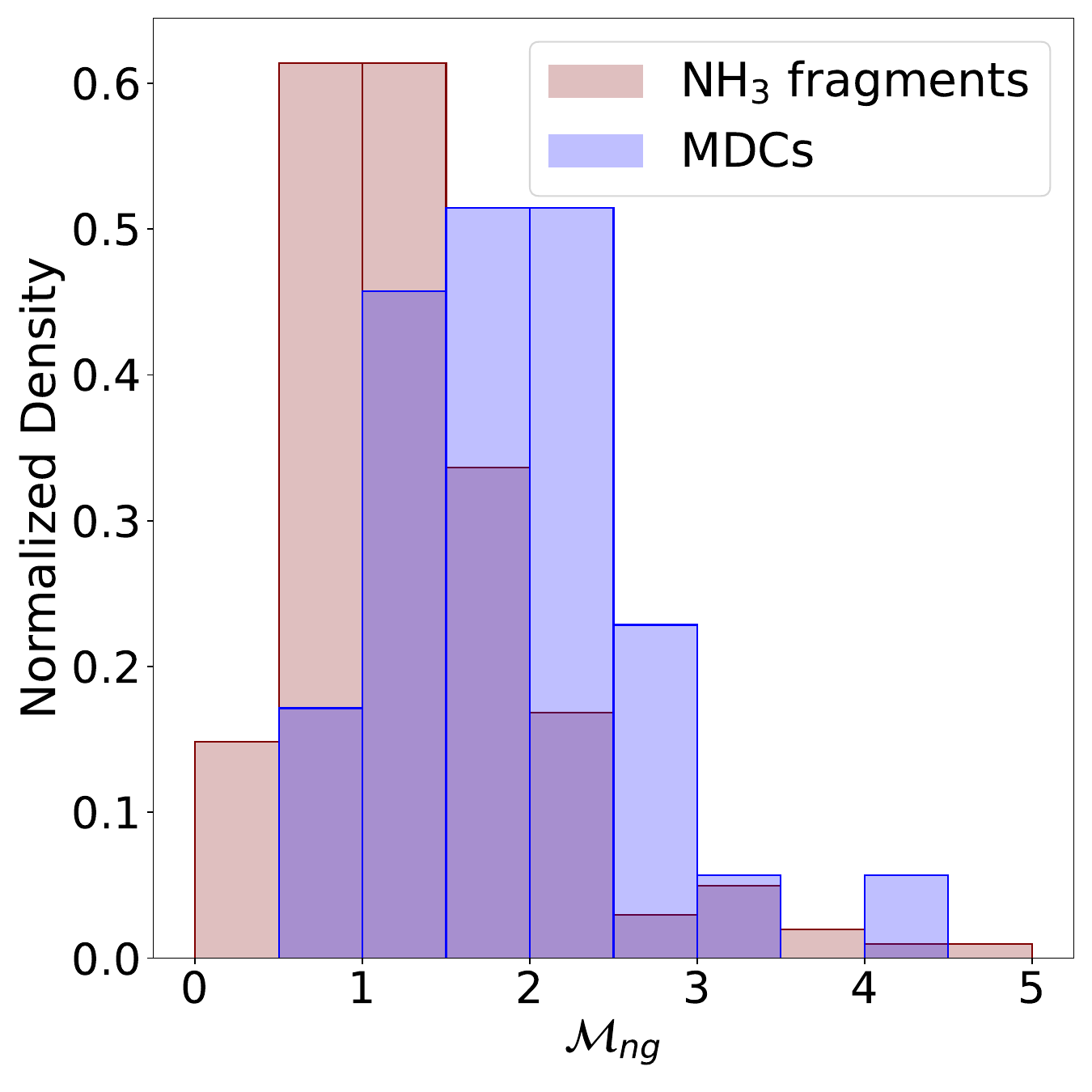}{0.4\textwidth}{(b)}}
\caption{Normalized density distribution of $\mathcal{M}_{\rm NT}$ and $\mathcal{M}_{\rm ng}$ in \nh\ fragments and MDCs. The contribution of velocity gradient is included in (a) and excluded in (b)\label{fig:mach}}
\end{figure*}

In Sect. \ref{sec:4.2}, we provide the \vnt\ and \vng\ of \nh\ fragments, where the resolved bulk motions are included and excluded, respectively. We also calculate \vnt\ and \vng\ for MDCs. Then, using \vnt\ and \vng, we obtain the Mach numbers, $\mathcal{M}_{\rm NT}$ and $\mathcal{M}_{\rm ng}$, for both \nh\ fragments and MDCs (Fig. \ref{fig:mach}). For MDCs, $\mathcal{M}_{\rm NT}$ has mean and median values of 2.2 and 2.3, with a standard deviation of 0.8, while the mean and median values of $\mathcal{M}_{\rm ng}$ are 1.9 and 1.9, with a standard deviation of 0.7. As for \nh\ fragments, the majority of both $\mathcal{M}_{\rm NT}$ and $\mathcal{M}_{\rm ng}$ are smaller than 2 (trans-sonic to subsonic). The mean and median values of $\mathcal{M}_{\rm NT}$ are 1.4 and 1.2, with a standard deviation of 0.8, and the mean and median values of $\mathcal{M}_{\rm ng}$ are 1.3 and 1.1, with a standard deviation of 0.7. These distributions of \nh\ fragments are similar to previous results \citep[e.g.,][]{2018A&A...611L...3S,2021RAA....21...24Y,2020ApJ...896..110L} The differences between \mnt\ and \mng\ of MDCs are more remarkable than those of \nh\ fragments, suggesting that the bulk motions have a larger contribution to the nonthermal velocity dispersions of MDCs than those of \nh\ fragments. In contrast, the $\mathcal{M}_{\rm ng}$ and $\mathcal{M}_{\rm NT}$ of \nh\ fragments are close. Moreover, both panels show a decreasing trend in Mach number from MDCs to \nh\ fragments.

Recently, high-spatial-resolution observations revealed similar transonic and subsonic nonthermal velocity dispersion in both low-mass and high-mass star-forming regions, such as L1517 \citep{2011A&A...533A..34H}, NGC 1333 \citep{2017A&A...606A.123H}, the Orion integral filament \citep{2018A&A...610A..77H}, OMC 2/3 \citep{2021RAA....21...24Y}, and NGC 6334S \citep{2020ApJ...896..110L}. Comparing N$_2$H$^+$ $J$=1-0 observations from the Nobeyama 45m telescope and ALMA, \citet{2021RAA....21...24Y} propose the nonthermal motions would transit from supersonic motions to subsonic motions at a scale of \simi\ 0.05 pc. In our results, the transition occurs at a scale of \nh\ fragments, \simi\ 0.04 pc, which is consistent with the result of \citet{2021RAA....21...24Y}. These latter authors interpret this transition as a resolution-dependent bias, where more bulk motions are involved in larger beams and give a greater nonthermal velocity dispersion. The large-scale dispersion measured from Nobeyama observations, $\sigma_{\rm ls}$, is decomposed in \citet{2021RAA....21...24Y} into small-scale dispersion measured from ALMA observations, $\sigma_{\rm ss}$, bulk motion probed by dispersion between the peak velocity of each ALMA beam, $\sigma_{\rm bm}$, and residual dispersion, $\sigma_{\rm rd}$, with the relation $\sigma_{\rm ls}^2$=$\sigma_{\rm ss}^2$+$\sigma_{\rm bm}^2$+$\sigma_{\rm rd}^2$. Comparing $\mathcal{M}_{\rm ng}$ of MDCs with that of \nh\ fragments as shown in the panel (b) of Fig. \ref{fig:mach}, where the bulk motions are deducted, the decreasing trend of $\mathcal{M}_{\rm ng}$ from MDCs to \nh\ fragments still remains. The differences between the $\mathcal{M}_{\rm ng}$ of MDCs and that of \nh\ fragments could refer to $\sigma_{\rm rd}$. A possible explanation  for $\sigma_{\rm rd}$ is turbulent vortex, as put forward by \citet{2021RAA....21...24Y}. In most cases, the decrease in nonthermal velocity dispersion from large scale to small scale is also thought to be the dissipation of turbulence, as suggested by the relation between velocity dispersion and size at lower spatial resolutions (\simi\ 0.1 -- 100 pc) in \citet{1981MNRAS.194..809L}. Considering that the typical size of \nh\ fragments (\simi\ 0.04 pc) is larger than the typical Kolmogorov scale in molecular clouds (\simi\ 10$^{-5}$ -- 10$^{−4}$ pc) \citep{2007ARA&A..45..565M,2011ApJ...737...13K,2018ApJ...864..116Q}, the turbulent dissipation could be a possible cause for the decreasing trend of $\mathcal{M}_{\rm ng}$.  

Furthermore, the lower Mach numbers imply that magnetic fields can play a more important role if the fragments are in a virialized state. However, as shown in \citet{2022ApJ...925...30L}, $\alpha_B$ would decrease with increasing column density. Considering that $\alpha_B$ is about 0.5 in our MDCs, the gravity could be more dominant than the magnetic field in these fragments (usually the peaks on the column density map). As estimated by \citet{2018A&A...611L...3S}, magnetic fields with strengths over \simi\ 1 mG are needed to achieve virialization in these fragments. However, such a strong magnetic fields have only been detected in a small number of star-forming structures \citep[e.g.,][]{2007MNRAS.382..699C,2007ApJ...668..331F,2021ApJ...912..159P}. Therefore, the fragments of MDCs are probably not in virialized states.

\subsection{Gas and dust temperature} \label{sec:5.4}

In general, there are two ways to trace the temperature of molecular clouds: from the dust continuum emission or from spectral line emission of a certain molecule. As the temperatures of MDCs are also provided in \citet{2021ApJ...918L...4C} (\td\ column in Table \ref{tab:ir}), which were calculated from the SED fitting of the dust continuum data, we compared these temperatures with the temperatures derived from the \nh\ data; we present our findings below.

\begin{figure}
    \centering
    \includegraphics[width=9cm]{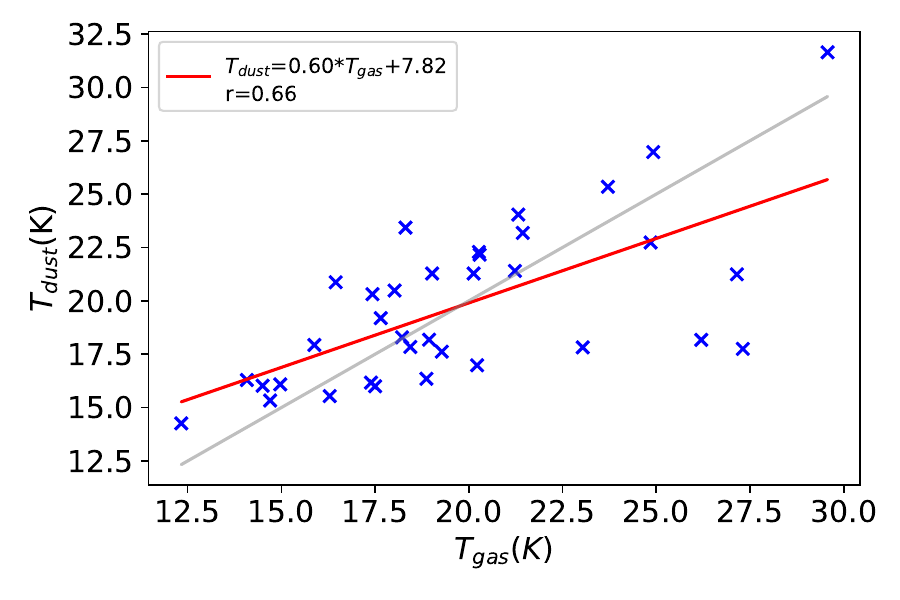}
    \caption{Correlation between \tg\ and \td. The red line shows the linear regression fitting results, and the gray line shows where \td\ = \tg.}
    \label{fig:TT}
\end{figure}

In order to compare the \nh\ temperature and the dust temperature in MDCs, the \nh\ data cubes are convolved to the resolution of temperature map in \citet{2019ApJS..241....1C} and the \nh\ temperatures of MDCs are rederived (\tg\ column in Table \ref{tab:ir}). The correlation between them is plotted in Fig. \ref{fig:TT}. The gray line illustrates where \tg\ = \td\ and the red line presents the linear regression fitting result,
\begin{equation}
    T_{\rm dust} = 0.60 T_{\rm gas} + 7.82,
\end{equation}
with the correlation coefficient r = 0.66. According to this relation, the gas temperature is mostly higher than the dust temperature when \tg\ $\gtrsim$ 20 K. In the scatter plot, the turning point is also around 20 K. We find that \td\ is almost equal to \tg\ when \tg\ $\lesssim$ 20 K and deviates from \tg\ when \tg\ $\gtrsim$ 20 K. 

The gas and dust are expected to be well coupled ---that is, they can be characterized by the same temperature--- within a molecular core with a density of $\gtrsim$ 10$^{4.5}$ \uden\ \citep{2001ApJ...557..736G,2004ApJ...614..252Y}. This coupling has been confirmed by many works \citep[e.g.,][]{2010ApJ...717.1157D,2019MNRAS.483.5355M}. However, although all the MDCs in our sample have densities higher than 10$^{5}$ \uden\ , \tg\ and \td\ are not thermally well coupled. Similarly, in high-mass star-forming regions, \tg\ is usually measured to be higher than \td, which is the case for example for the quiescent clumps in \citet{2014ApJ...786..116B}, the S 140 in \citet{2015A&A...580A..68K}, and the W40 in \citet{2020A&A...643A.178T}. 

The origin of the excess gas temperature is a complex open question. In the photon-dominated region S 140, \citet{2015A&A...580A..68K} found an excess of \simi\ 5-15 K in their deeply embedded regions, where \td\ and \tg\ are expected to be well coupled in the KOSMA-$\tau$ PDR model \citep{2013A&A...549A..85R}. \citet{2015A&A...580A..68K} interpret these differences as the result of inhomogeneous clumpy structures, which could lead to a deeper penetration of the UV radiation and heating of the gas in the embedded cores. In addition, differences between \tg\ and \td\ can also arise from different layers traced by the gas and the dust \citep{2014ApJ...786..116B}. Such differences arise from both different excitation conditions and different angular resolutions. The \nh\ \oneone\ and \twotwo\ lines are usually excited in the dense gas, while the dust continuum emission arises from a range of densities. In the present work, the resolutions of the \nh\ data and the temperature map from \citet{2019ApJS..241....1C} are quite different. Only a few MDCs have sufficient temperature pixels as shown in Fig. \ref{fig:props} and Appendix \ref{app:B}. The spatially extended \nh\ emission, which usually has lower temperatures, is missed due to the filtering effect of VLA. In our results, \tg\ deviates from \td\ at \tg\ $\gtrsim$ 20 K (Fig. \ref{fig:TT}). The MDCs with \tg\ $\gtrsim$ 20 K should have internal heating from the star-forming activity, but this heating is diluted by the large beam in the dust temperature data. Therefore, on average, we obtain a higher \tg\ than \td\ in our results.

\section{Summary} \label{sec:6}

The initial conditions of the birth places of high-mass stars are of utmost importance in the study of high-mass star formation. We present VLA \nh\ \oneone\ and \twotwo\ observations of MDCs in \cyx\ at a distance of 1.4 kpc. As a part of the CENSUS project, we provide the temperature and dynamical information derived from the simultaneous line fitting of the \nh\ \oneone\ and \twotwo\ inversion emission lines.

We used the MDCs catalog from \citet{2021ApJ...918L...4C} in our analysis. We compared the morphology between the H$_2$ column density map from \citet{2019ApJS..241....1C} and \nh\ \oneone\ emission within each MDC and provide a possible indicator for estimating their evolutionary stages. We also performed a virial analysis for each MDC to evaluate the stability of the MDCs. The dynamical states of the MDCs are discussed based on their nonthermal velocity dispersions. We also compare the gas temperature with the dust temperature of MDCs. In addition, we identified 202 fragments from the \nh\ \oneone\ moment-0 maps using the \begin{small}GAUSSCLUMPS\end{small} algorithm. We derived the physical parameters of these \nh\ fragments and conducted a correlation analysis among them. We also analyzed the dynamical states of the \nh\ fragments and compared them with those of the MDCs. The main results of this paper can be summarized as follows:

\begin{enumerate}
    \item The \nh\ fragments have a mean deconvolved radius of 0.02 pc, a mean velocity dispersion of 0.46 km s$^{-1}$, a mean nonthermal velocity dispersion of 0.37 km s$^{-1}$, a mean kinetic temperature of 18.6 K, and a mean \nh\ column density of 3.9 $\times$ \ucold. Comparison with the temperature of structures in previous works indicates that the \nh\ fragments are mostly early evolved structures.
    \item We find a correlation between nonthermal velocity dispersion and kinetic temperature. This correlation is likely the result of different levels of star-forming activity. A molecular structure with more active star-forming activity is usually hotter and with larger velocity dispersion.
    \item We perform a morphological analysis of the \nh\ \oneone\ emission and column density map of MDCs. We propose a possible evolutionary sequence of MDCs based on the \nh\ \oneone\ emission: in the early, relatively little-evolved MDCs, the \nh\ emission is concentrated around density peaks;  the \nh\ emission is subsequently dispersed by the star-forming activity, after which, in the evolved MDC, the \nh\ emission becomes undetected.
    \item We perform a virial analysis of all MDCs. The majority of the MDCs of our sample are subvirialized, indicating that the kinetic energy cannot support the MDCs alone in our sample. Considering the support from magnetic fields, MDCs could attain their stability states if there were a typical magnetic field strength of \simi\ 0.5 mG.
    \item We investigate the dynamic states of MDCs and the \nh\ fragments with their nonthermal Mach numbers. The Mach numbers show a decreasing trend from MDCs to \nh\ fragments, which could be an indication of the dissipation of turbulence. When the resolved bulk motions are excluded from the nonthermal velocity dispersion, we find the \nh\ fragments are mostly subsonic to trans-sonic, while the majority of MDCs are transonic.
    \item The gas temperature derived from \nh\ \oneone\ and \twotwo\ lines is compared with the dust temperature calculated by SED fitting from \citet{2019ApJS..241....1C}. The two are well coupled when the gas temperature is $\lesssim$ 20 K, while the gas temperature is mostly larger than the dust temperature when the gas temperature is $\gtrsim$ 20 K.
\end{enumerate}

\begin{acknowledgements}
This work is supported by National Key R\&D Program of China No. 2022YFA1603103, No. 2017YFA0402604, the National Natural Science Foundation of China (NSFC) grant U1731237, and the science research grant from the China Manned Space Project with No. CMS-CSST-2021-B06. This research made use of $APLpy$, an open-source plotting package for Python. \citep{2012ascl.soft08017R} This research made use of $Astropy$, a community-developed core Python package for Astronomy. \citep{2013A&A...558A..33A,2018AJ....156..123A}  This study used observations from the Spitzer Space Telescope Cygnus X Legacy Survey (PID 40184, P.I. Joseph L. Hora). 
\end{acknowledgements}

\bibliographystyle{aa}
\bibliography{references}

\clearpage
\onecolumn

\begin{appendix}
\section{Overview of observations of detected spectral lines}\label{app:A}

\begin{longtable*}{@{}ccc@{}}
\endhead
\caption*{\textbf{Fig. A.1.} continued.}
\endfoot
\caption*{\textbf{Fig. A.1.} Moment-0 maps with remarkable detection of \nh\ (J,K) = (1,1), (2,2), (3,3), (4,4), and (5,5) inversion lines among our observations. The (1,1) maps are plotted from a 3 rms level and in steps of 2$\sigma$, while others are plotted from the 2 rms level and in steps of 2$\sigma$. The IR-bright, IR-quiet, and starless MDCs are shown as red solid, red dashed, and pink dashed ellipses, respectively. The smaller ellipses present the fragments extracted using the \begin{small}GAUSSCLUMPS\end{small} algorithm. The blue ones are included in the parameter analysis while the green ones are not. \label{fig:A.1}}
\endlastfoot
\includegraphics[width=.3\textwidth]{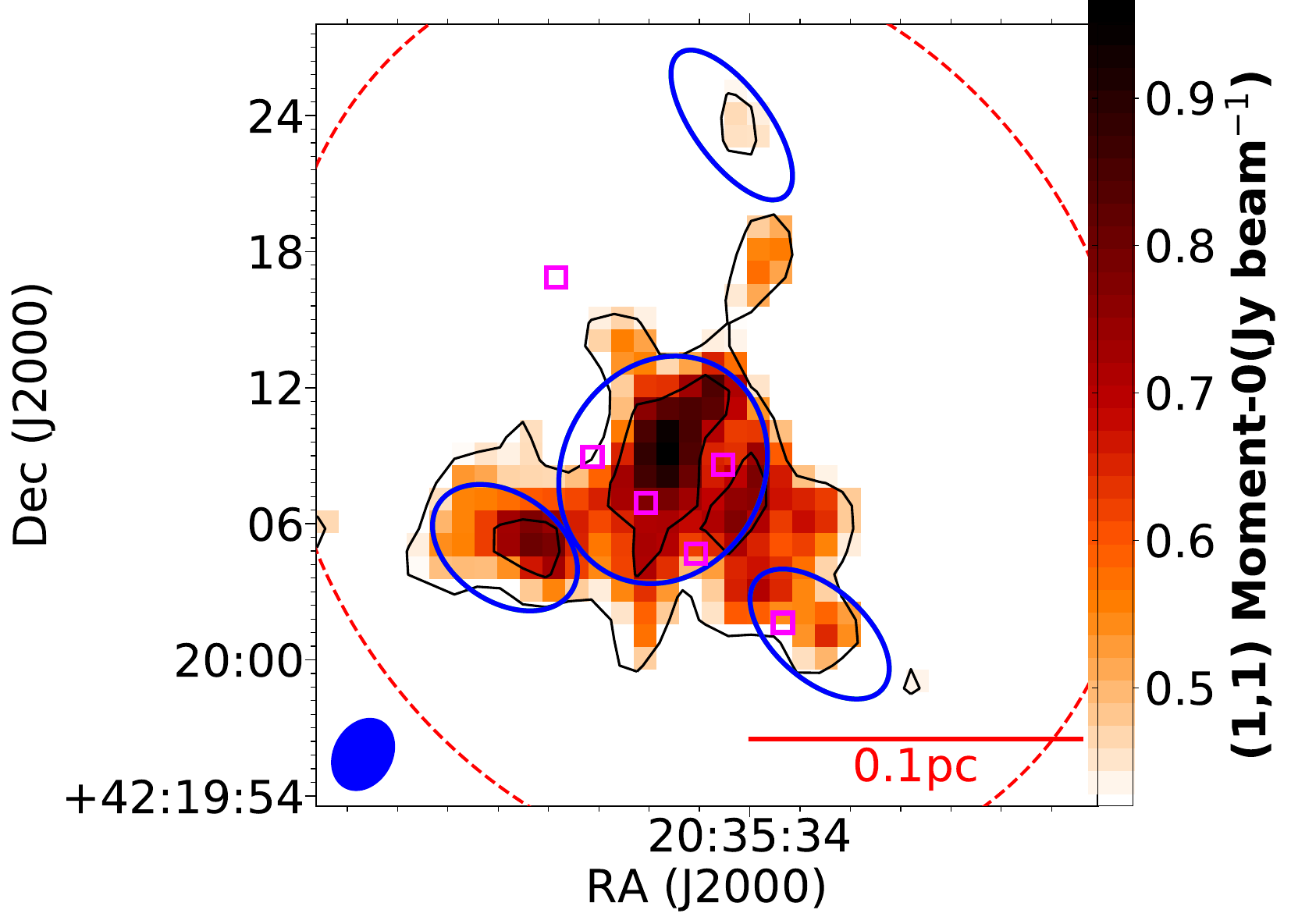} &
\includegraphics[width=.3\textwidth]{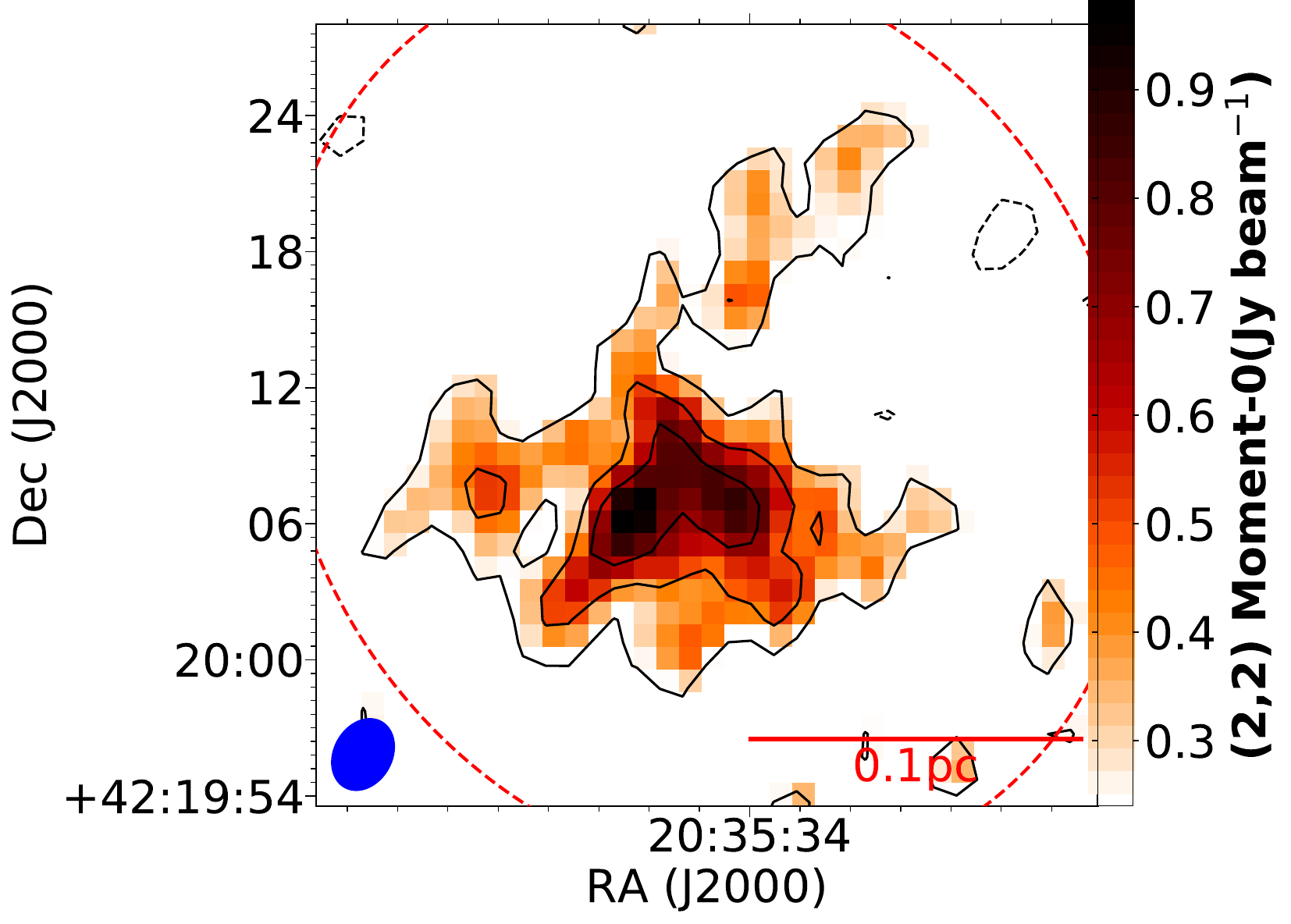} &
 \\
 & Field 1 & \\
\includegraphics[width=.3\textwidth]{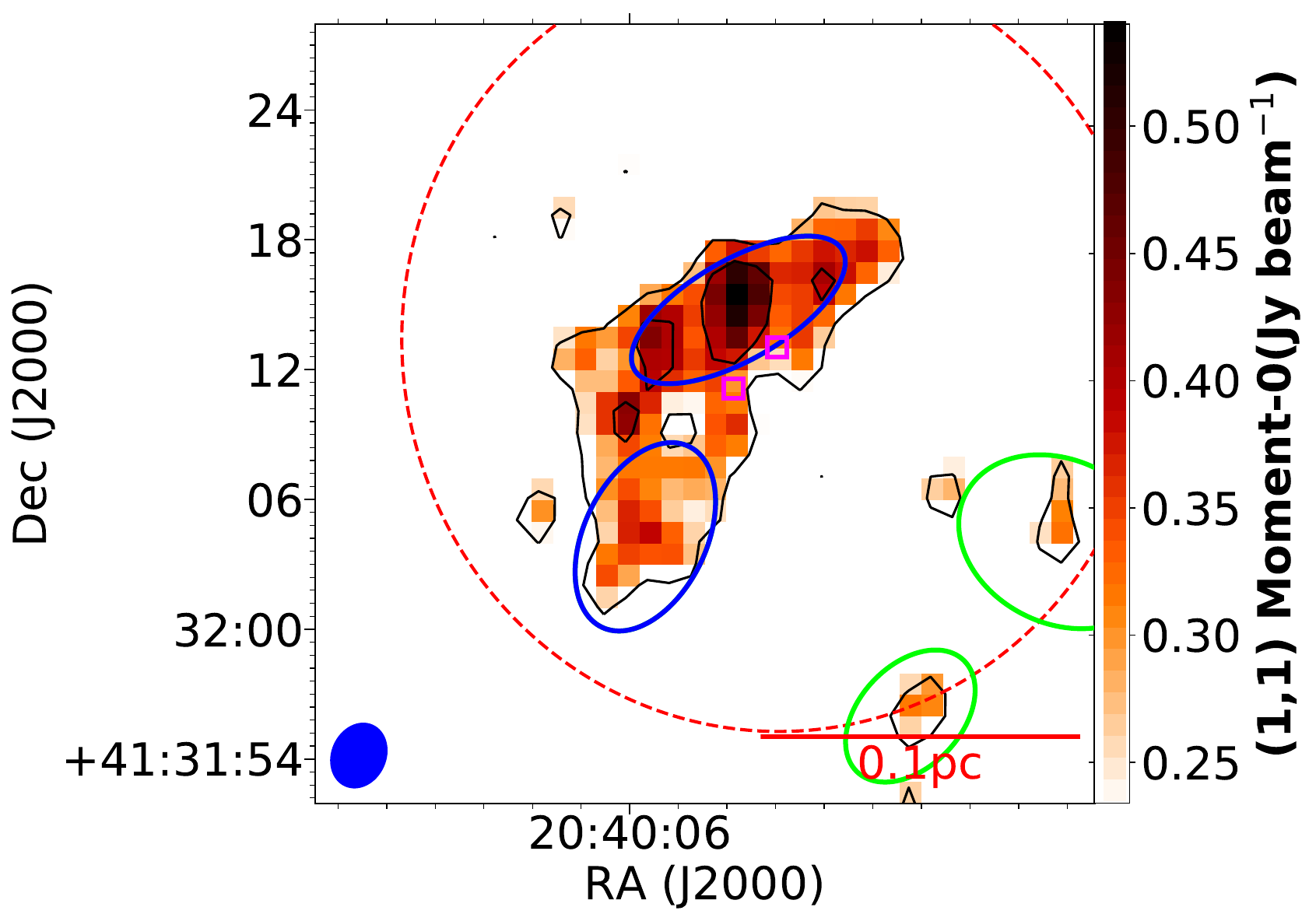} &
\includegraphics[width=.3\textwidth]{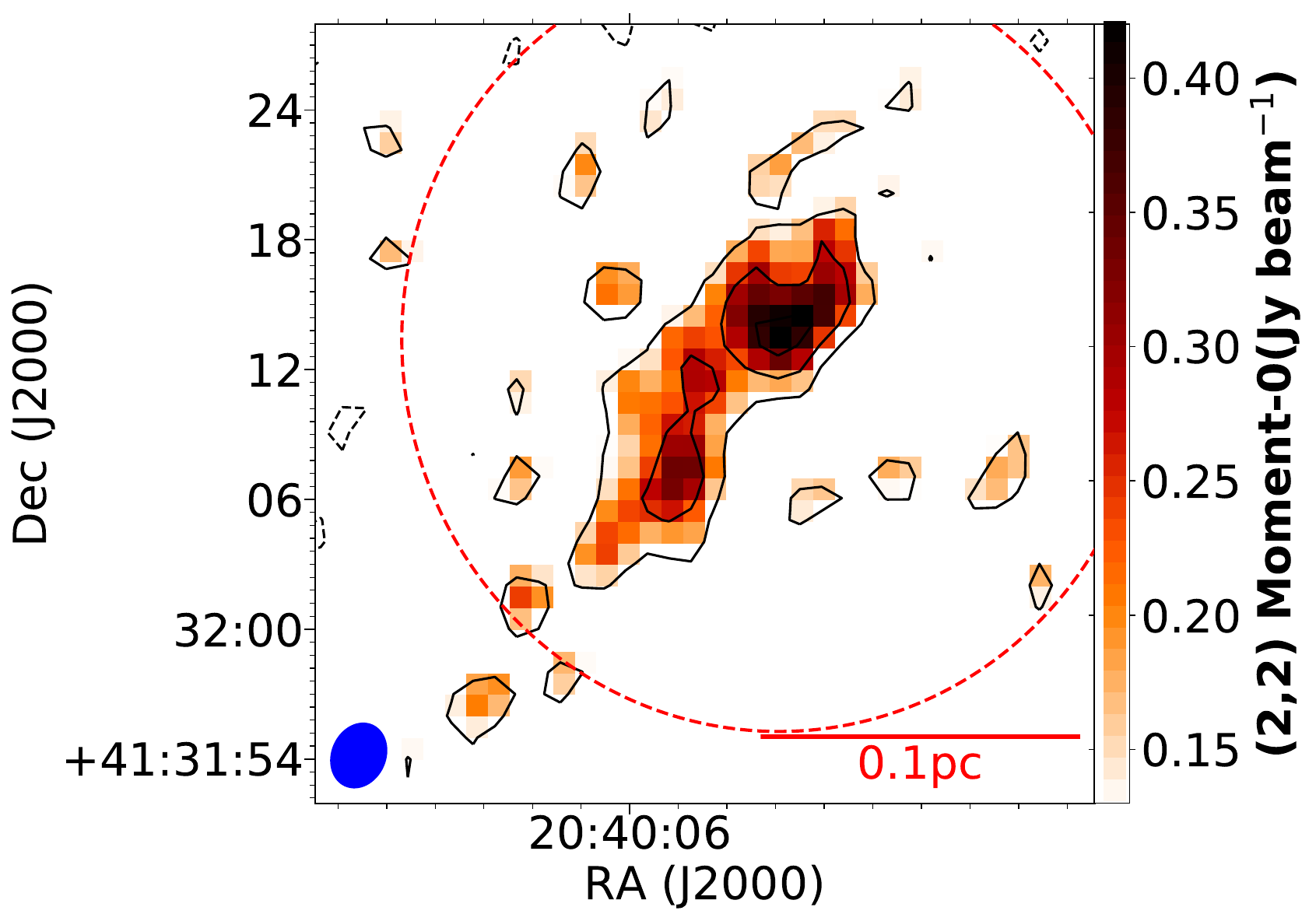} &
 \\
 & Field 2 & \\
\includegraphics[width=.3\textwidth]{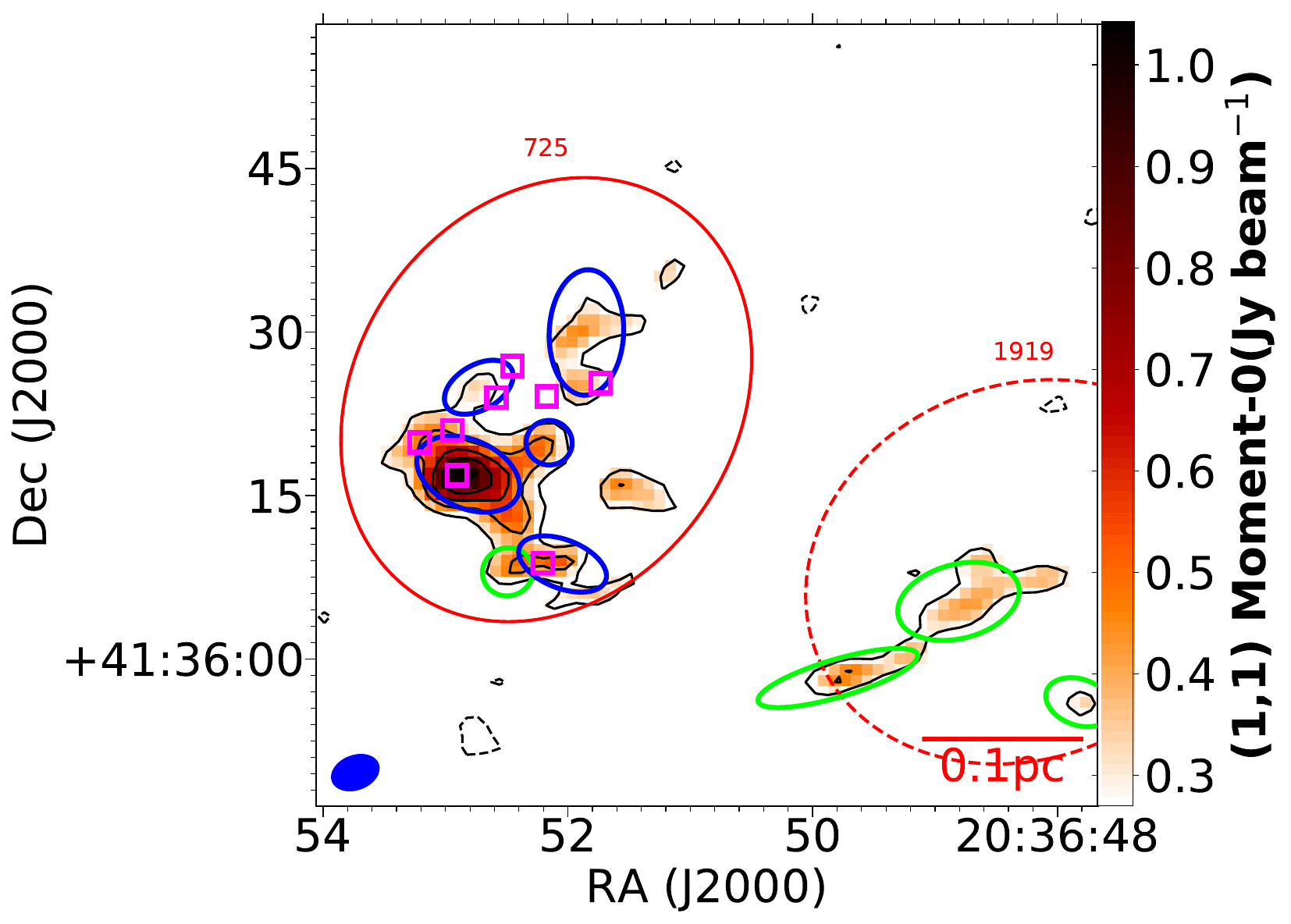} &
\includegraphics[width=.3\textwidth]{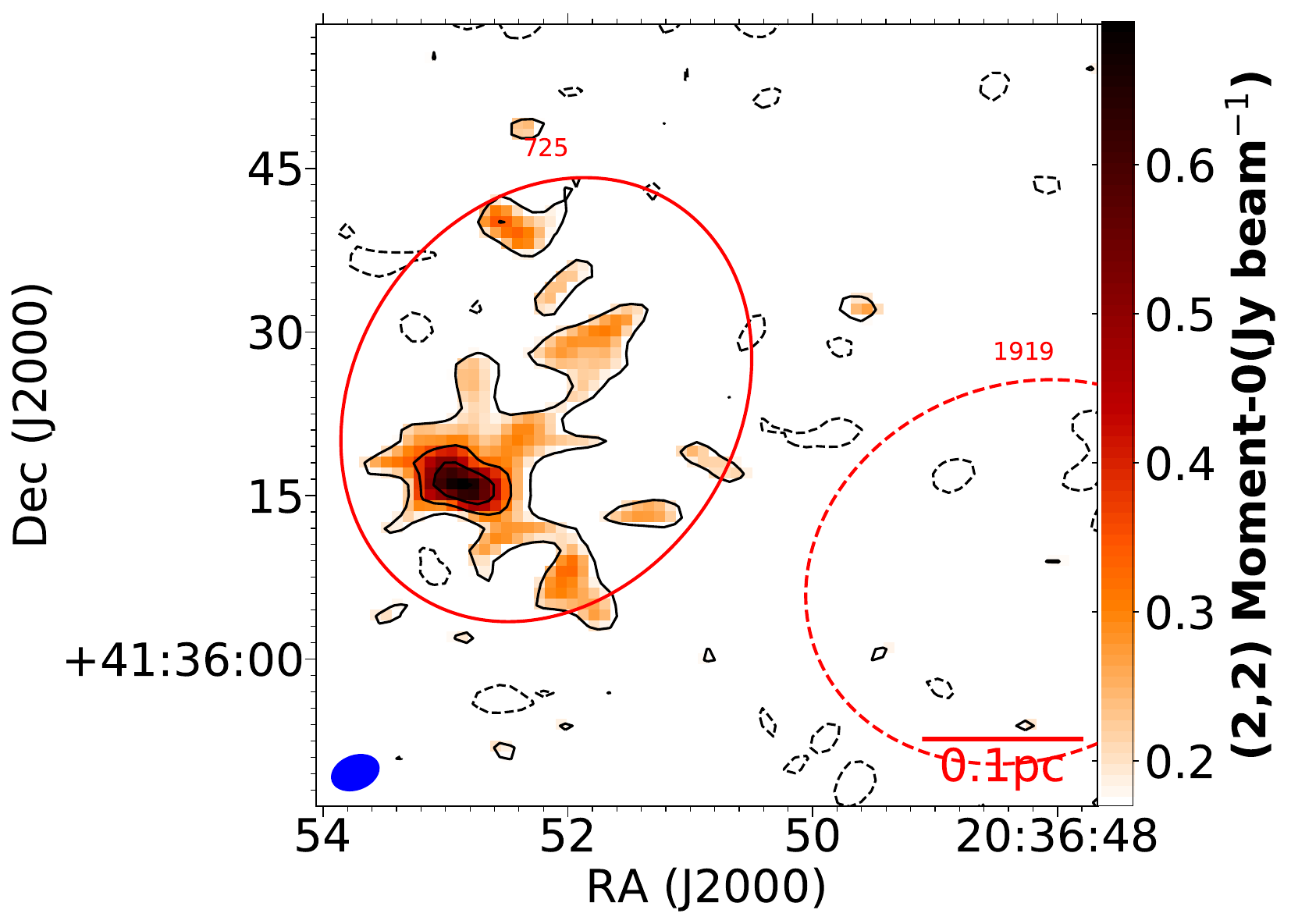} &
\includegraphics[width=.3\textwidth]{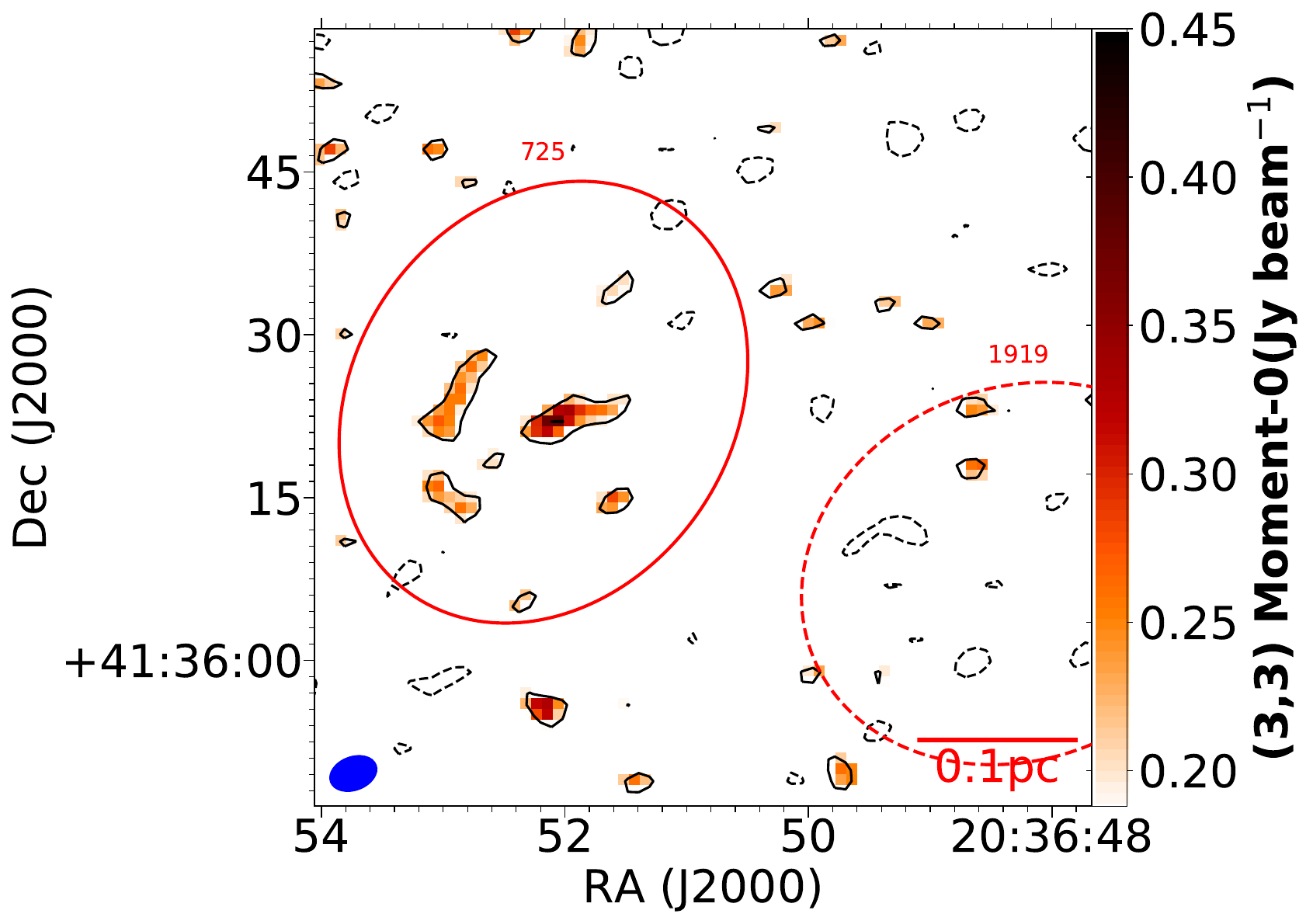}\\
 & Field 4 & \\
\includegraphics[width=.3\textwidth]{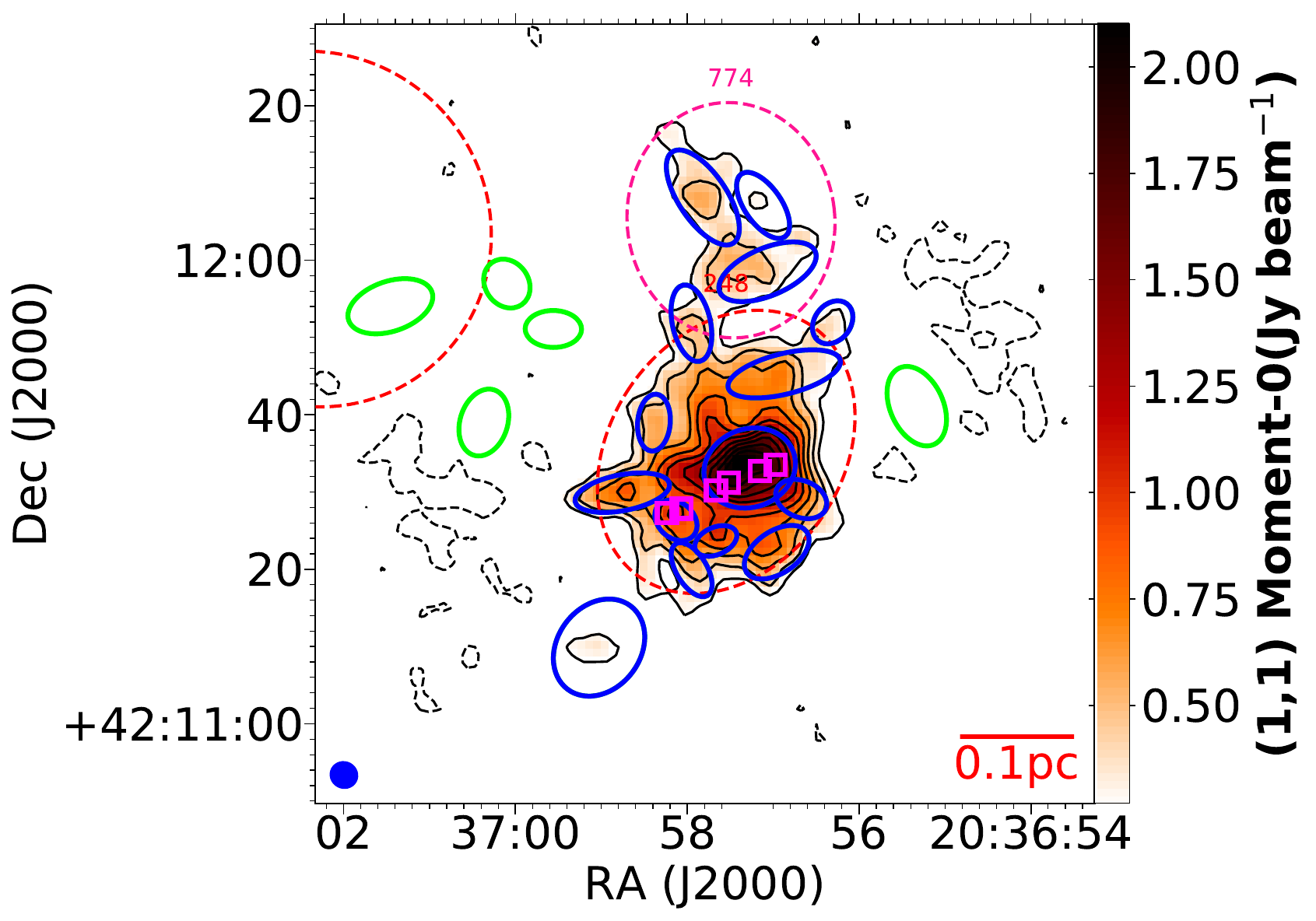} &
\includegraphics[width=.3\textwidth]{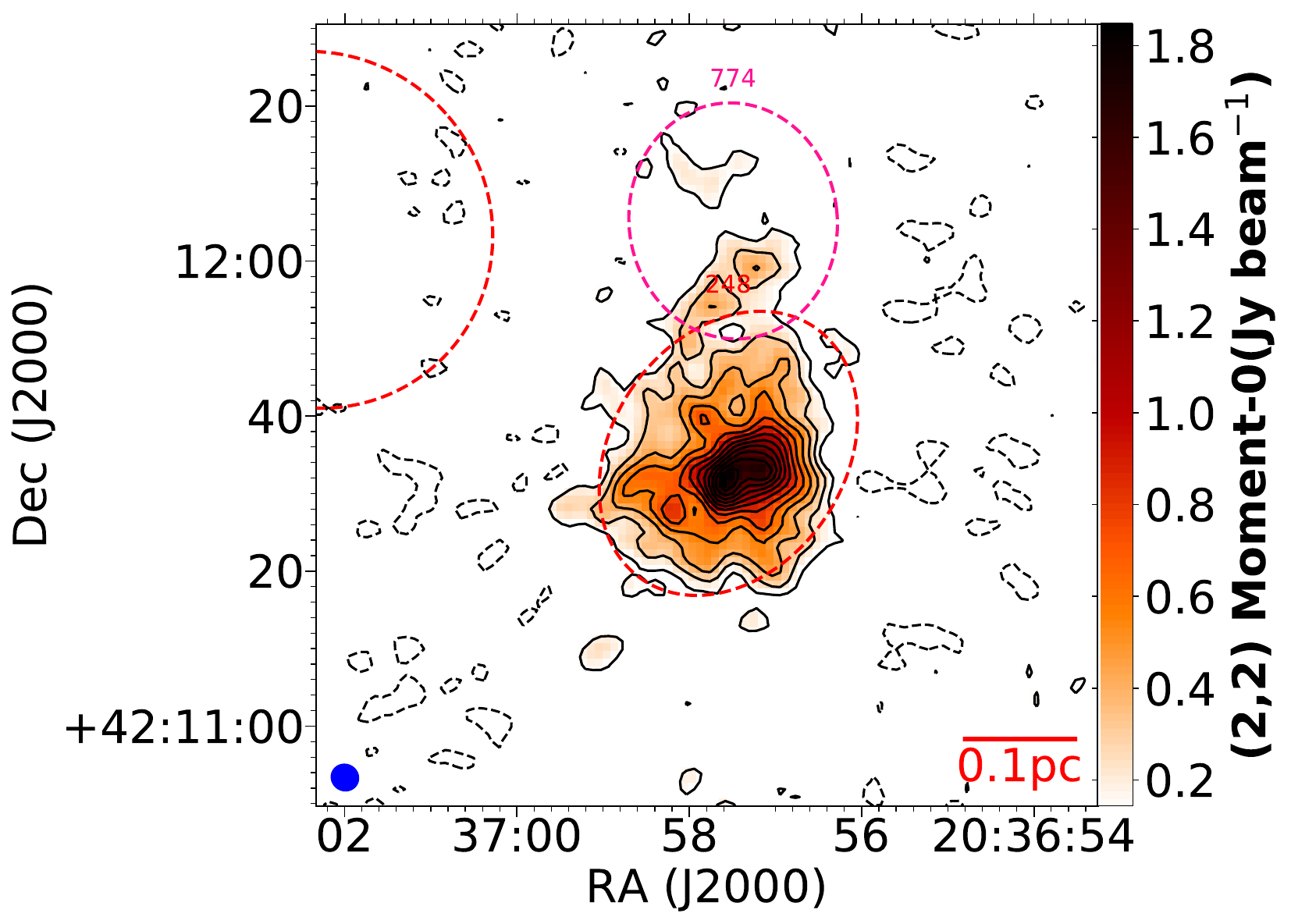} &
\includegraphics[width=.3\textwidth]{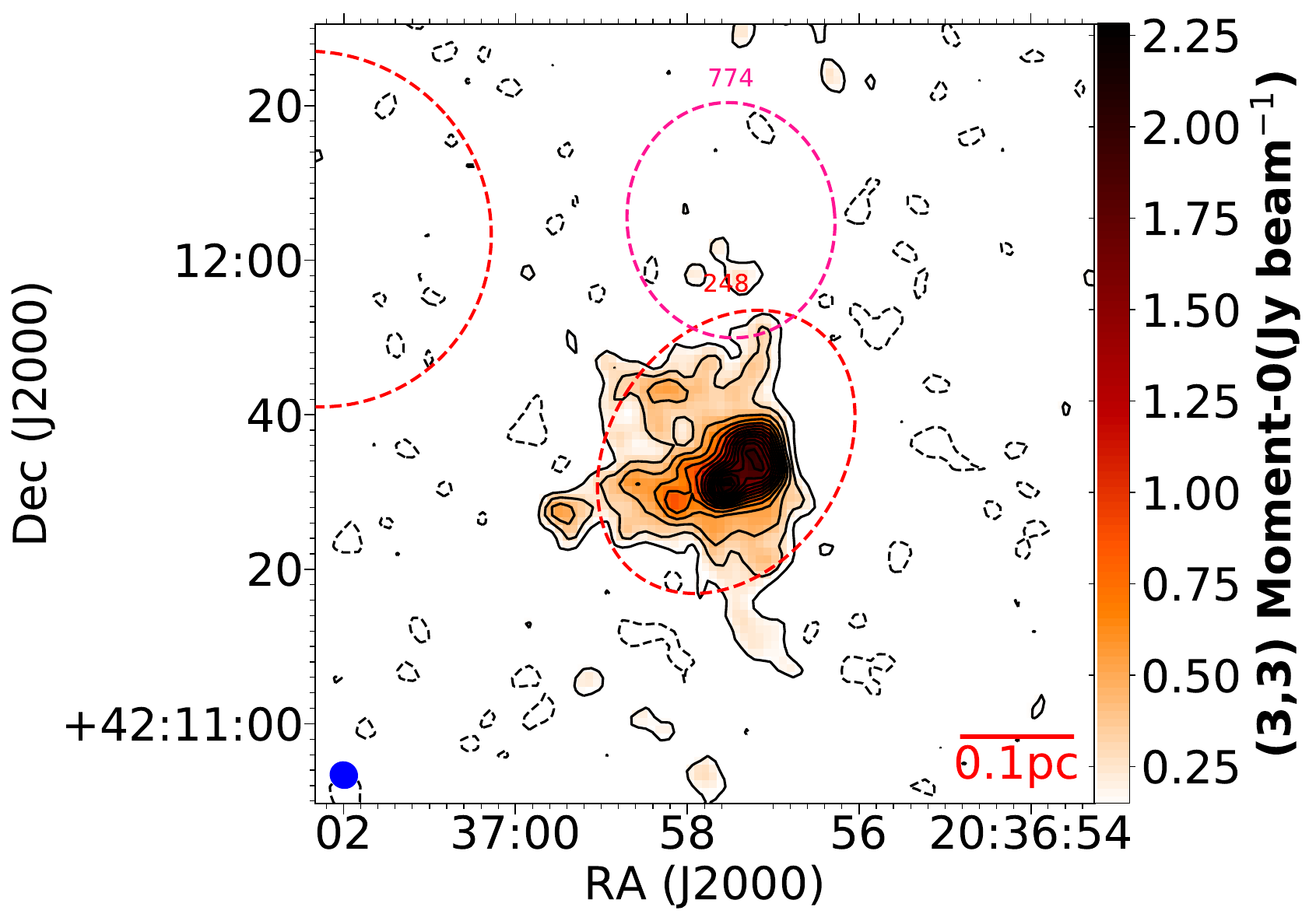} \\
\includegraphics[width=.3\textwidth]{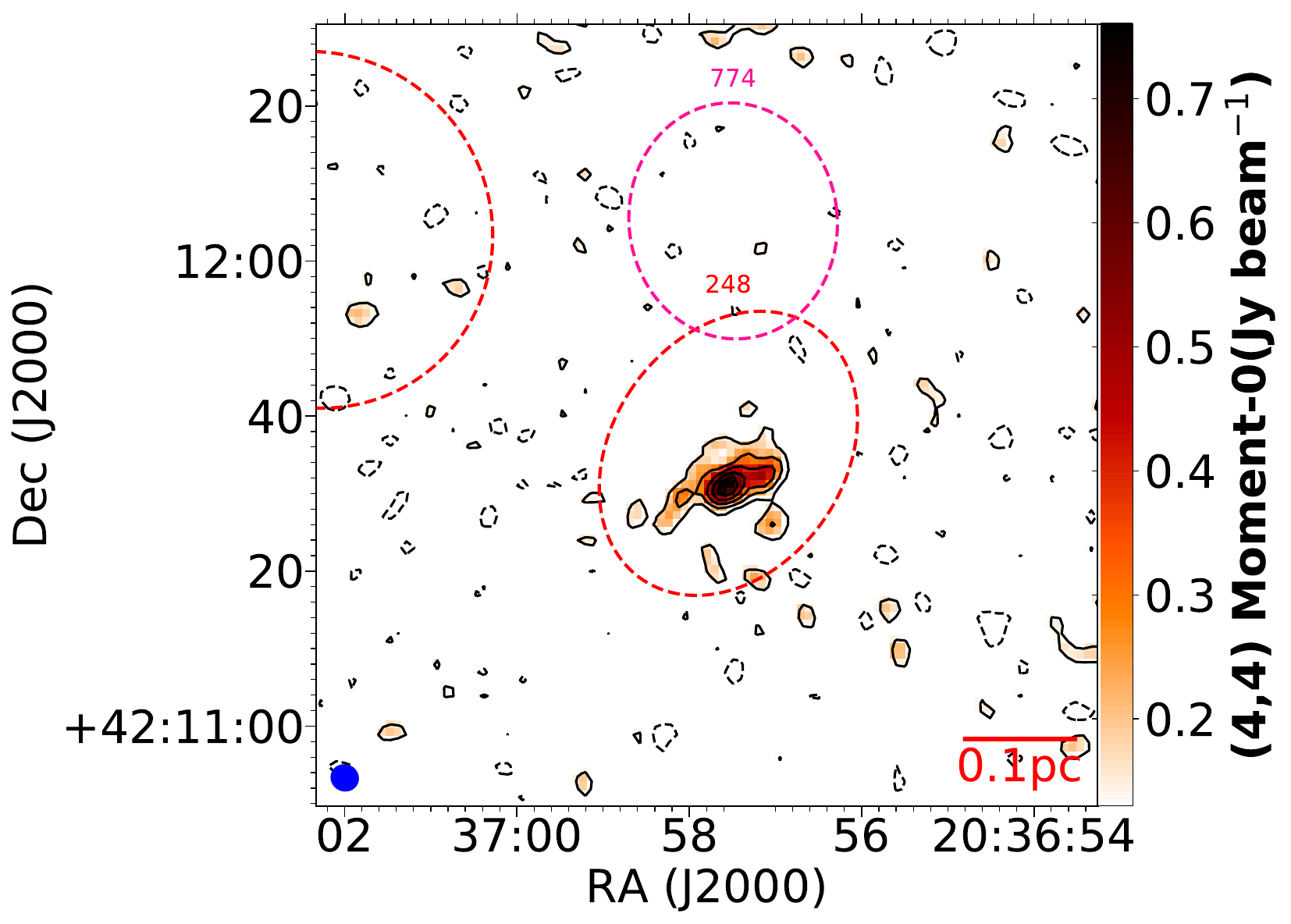} &
\includegraphics[width=.3\textwidth]{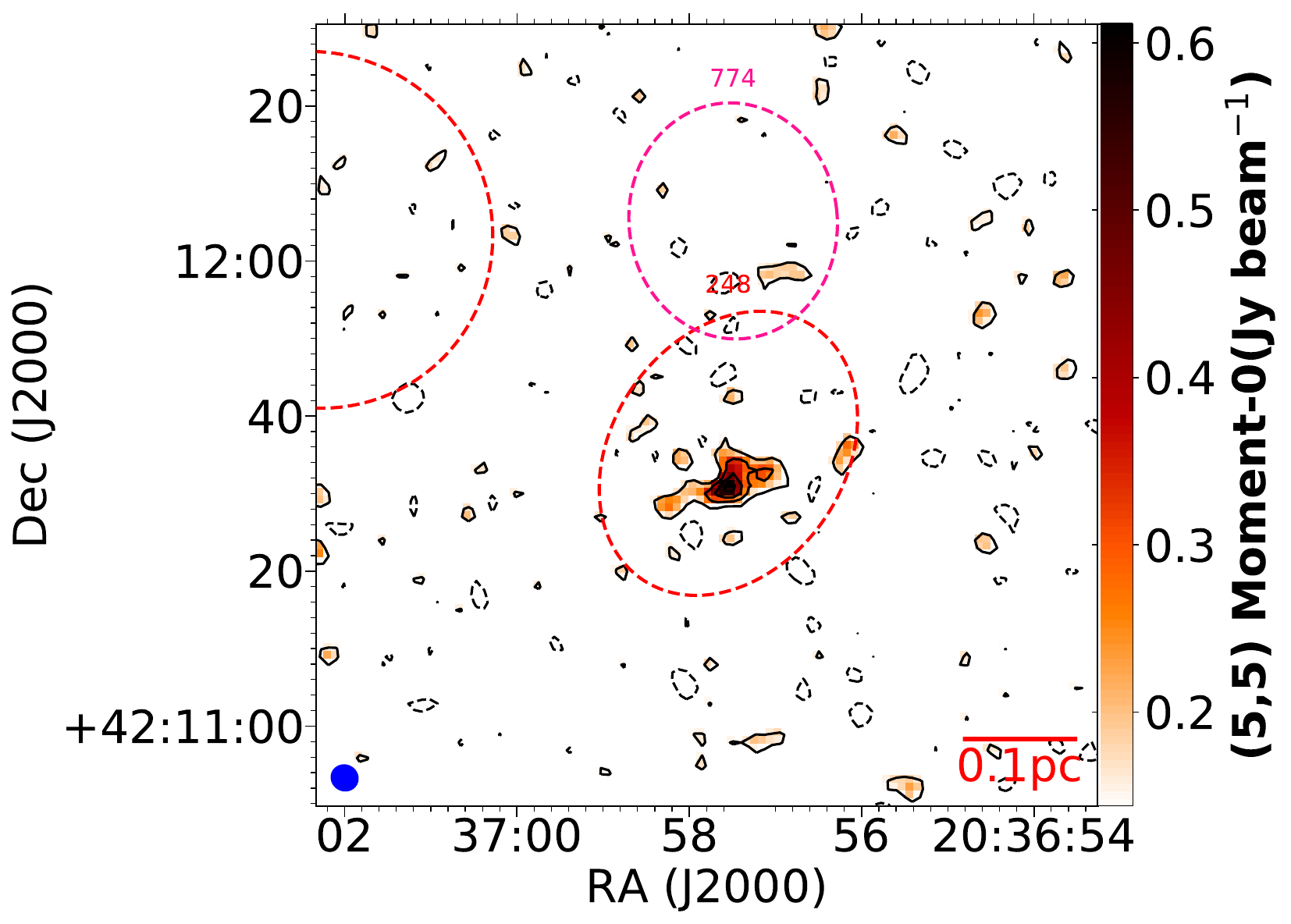} &
 \\
 & Field 5 & \\
\includegraphics[width=.3\textwidth]{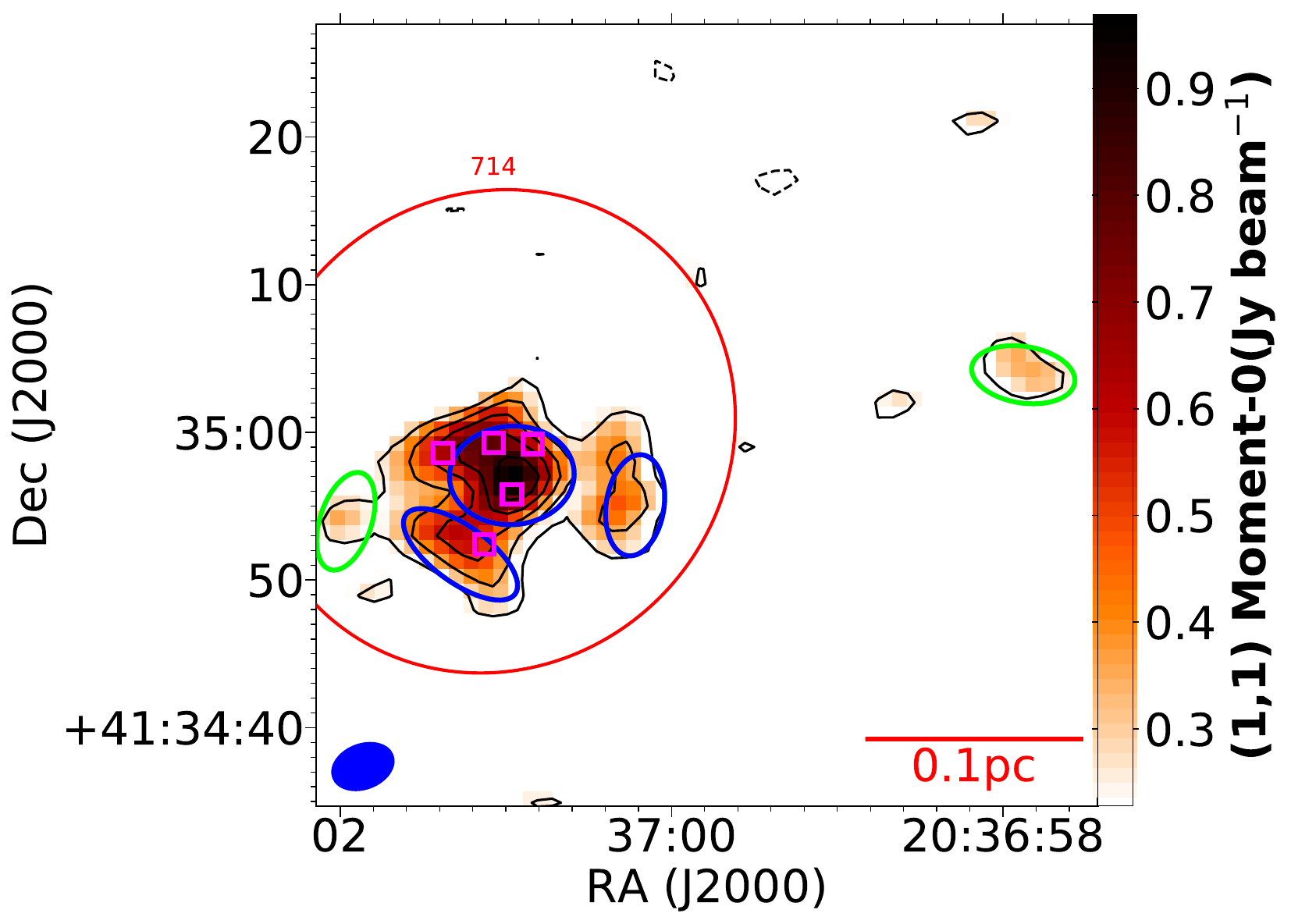} &
\includegraphics[width=.3\textwidth]{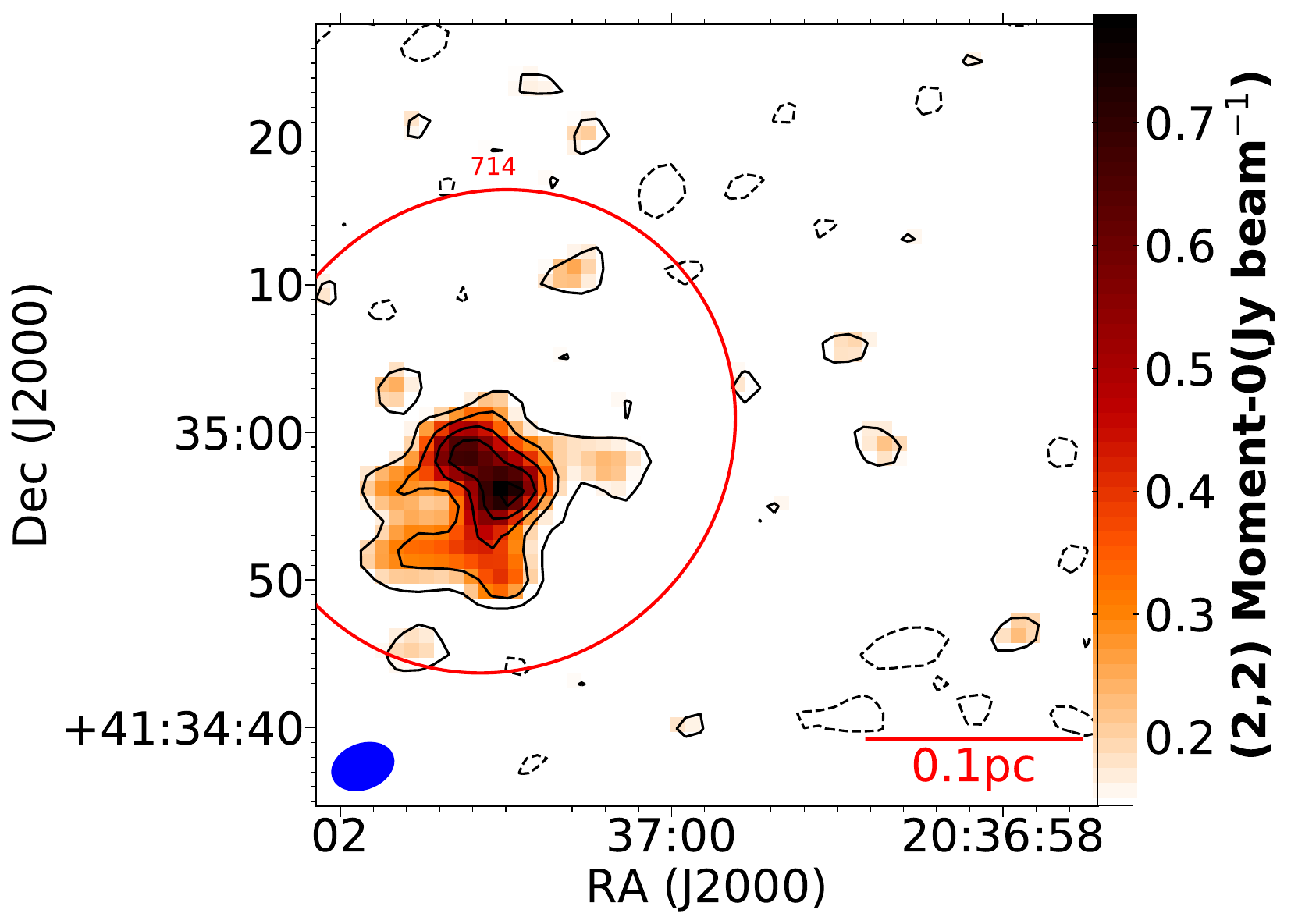} &
\includegraphics[width=.3\textwidth]{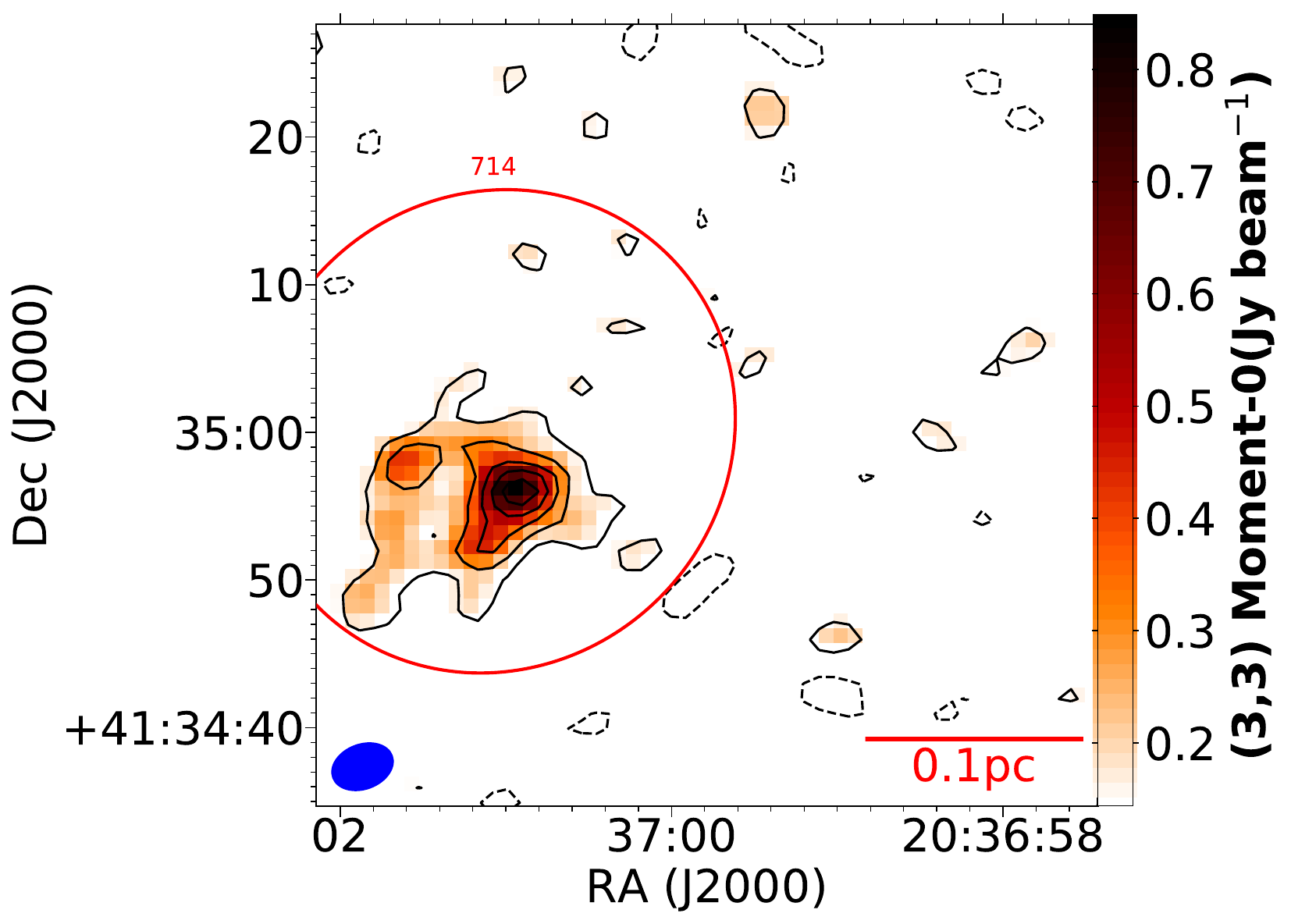} \\
\includegraphics[width=.3\textwidth]{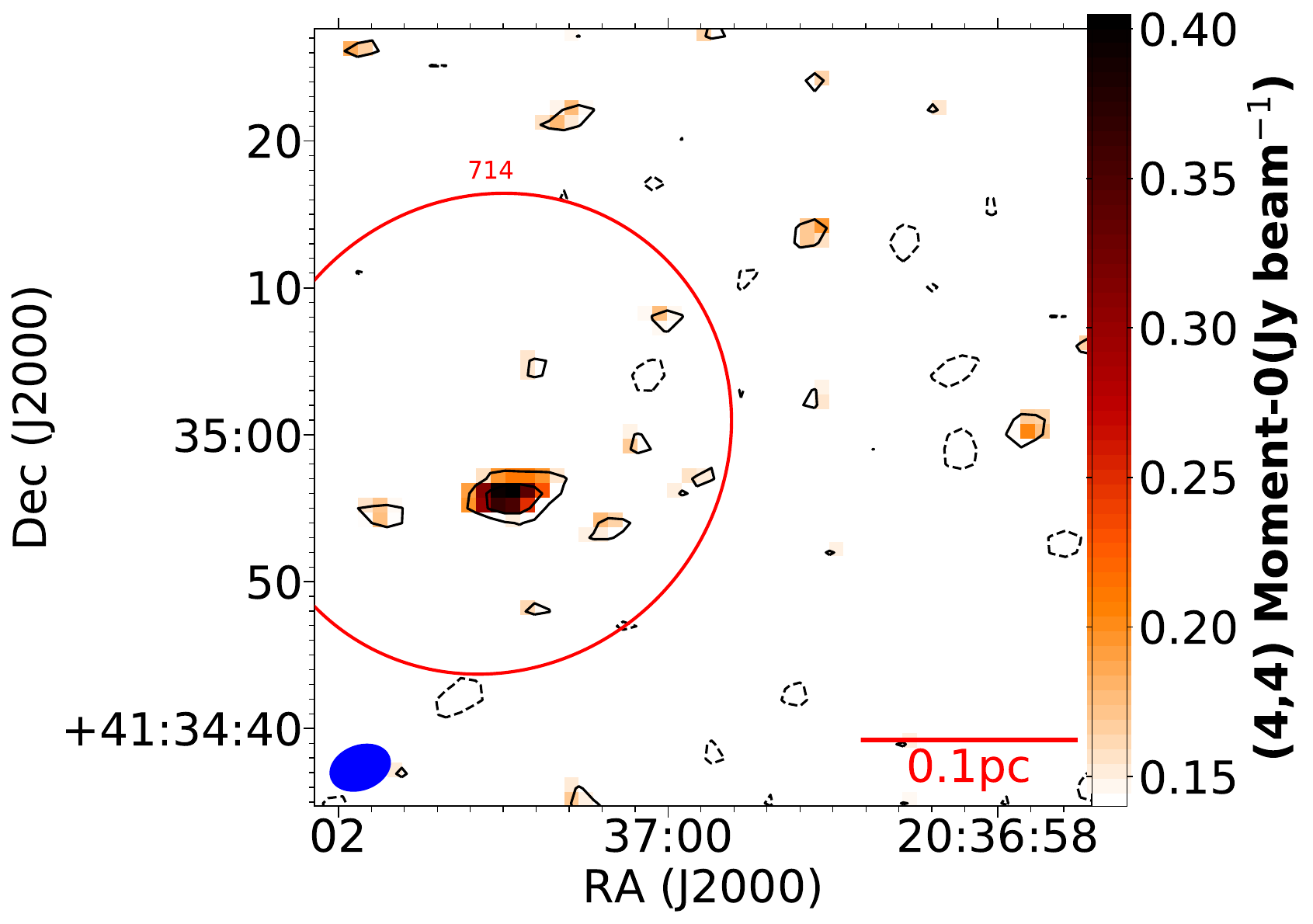} &
 &
 \\
 & Field 6 & \\
\includegraphics[width=.3\textwidth]{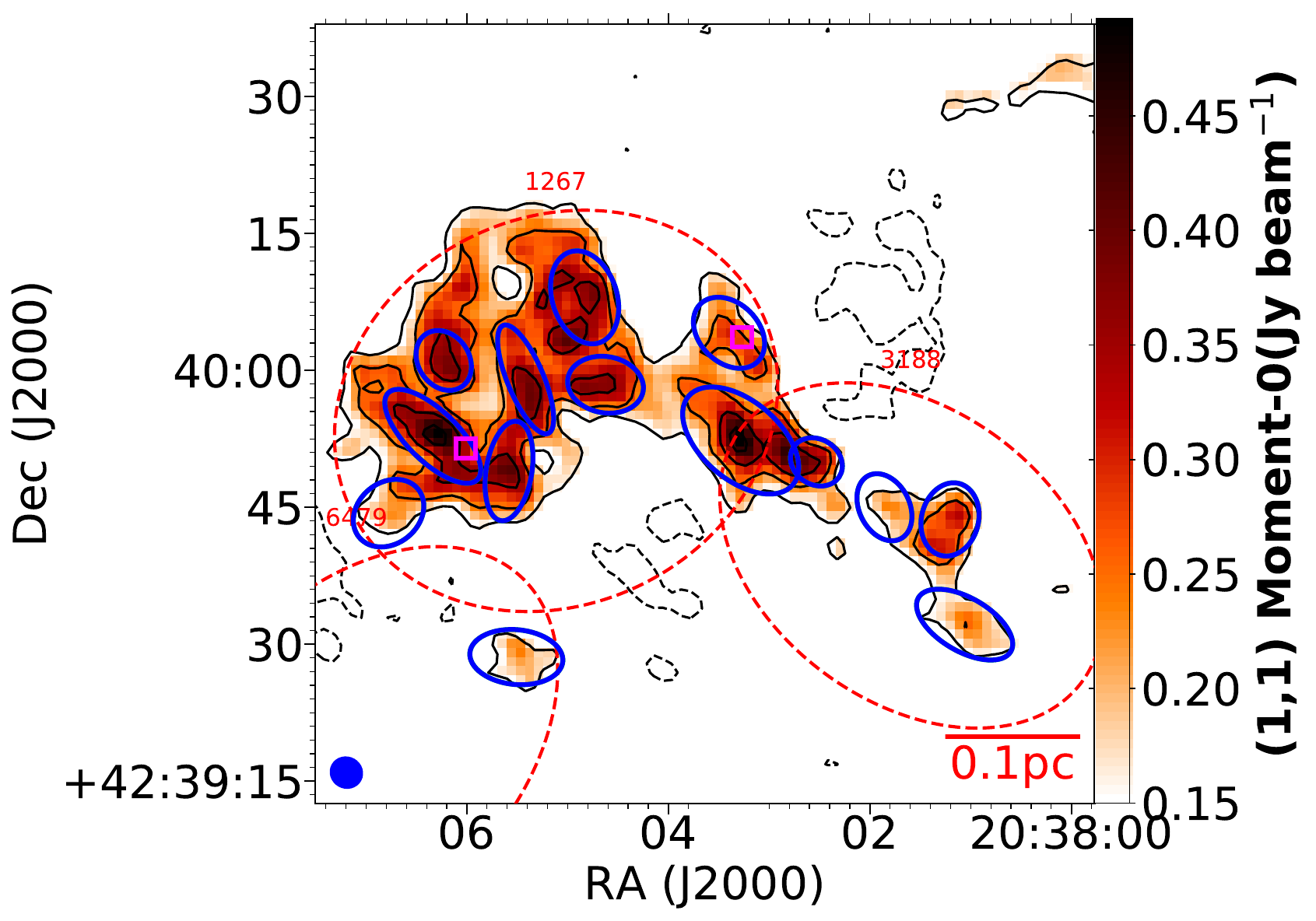} &
\includegraphics[width=.3\textwidth]{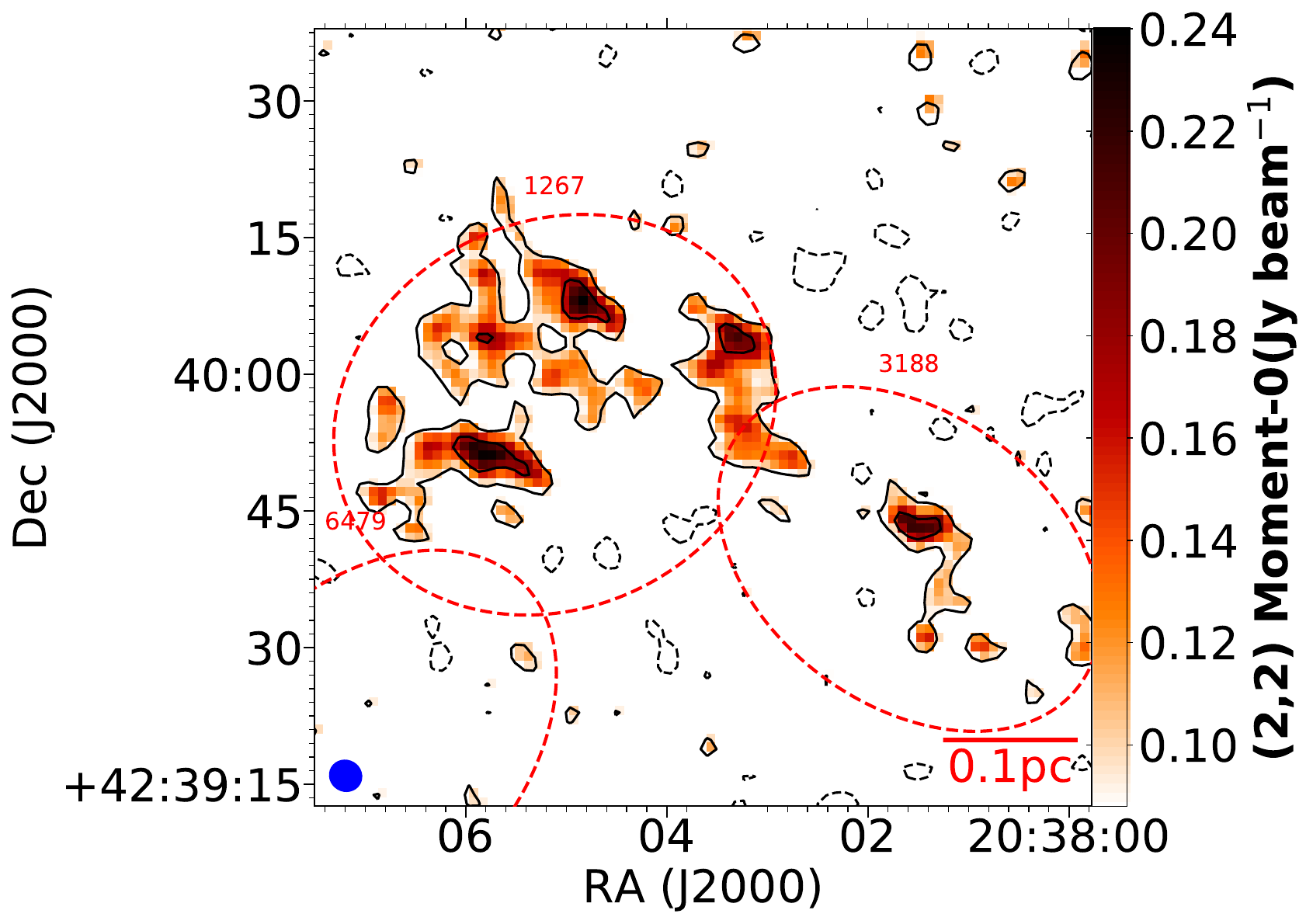} &
 \\
 & Field 7 & \\
\includegraphics[width=.3\textwidth]{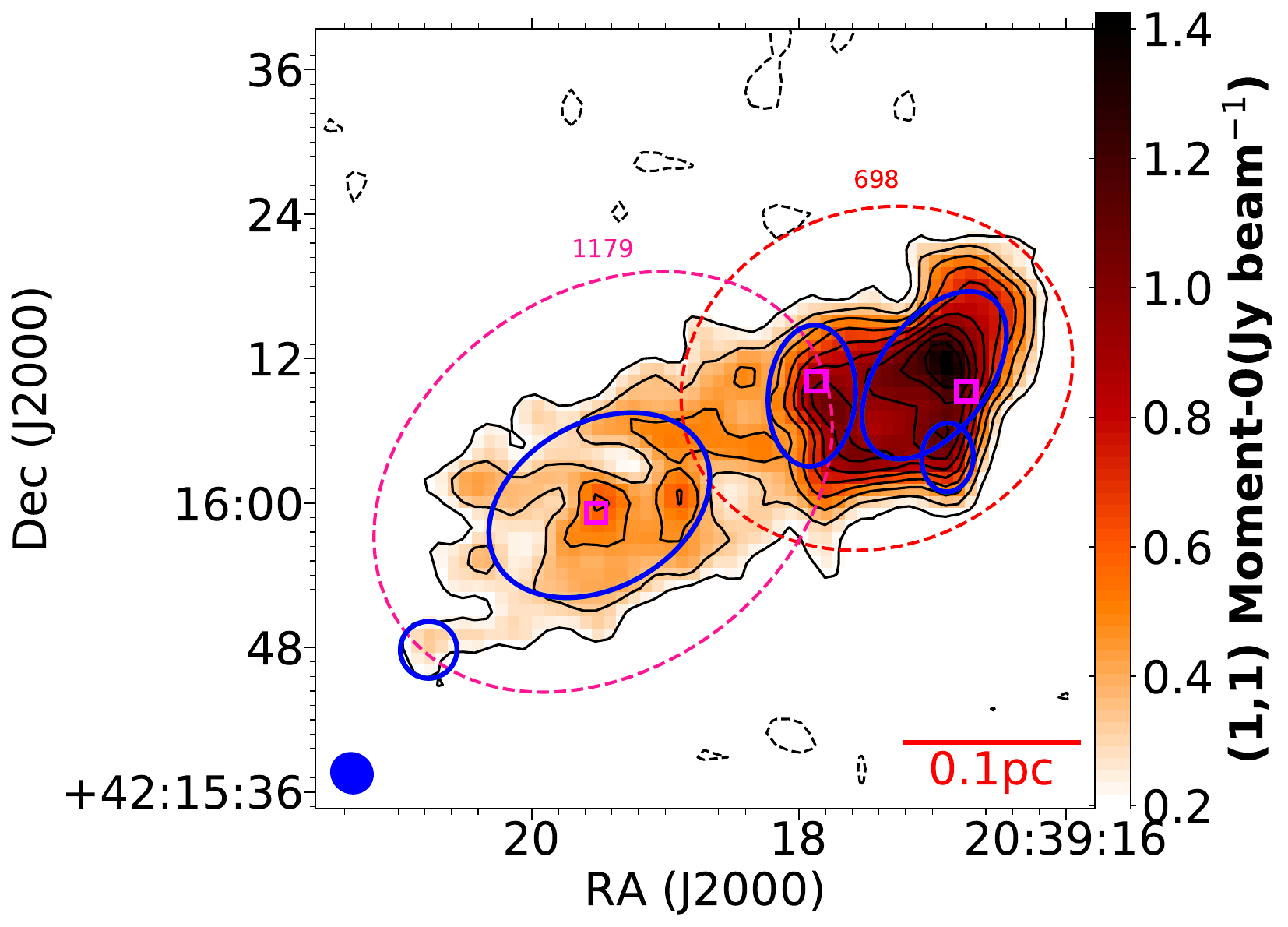} &
\includegraphics[width=.3\textwidth]{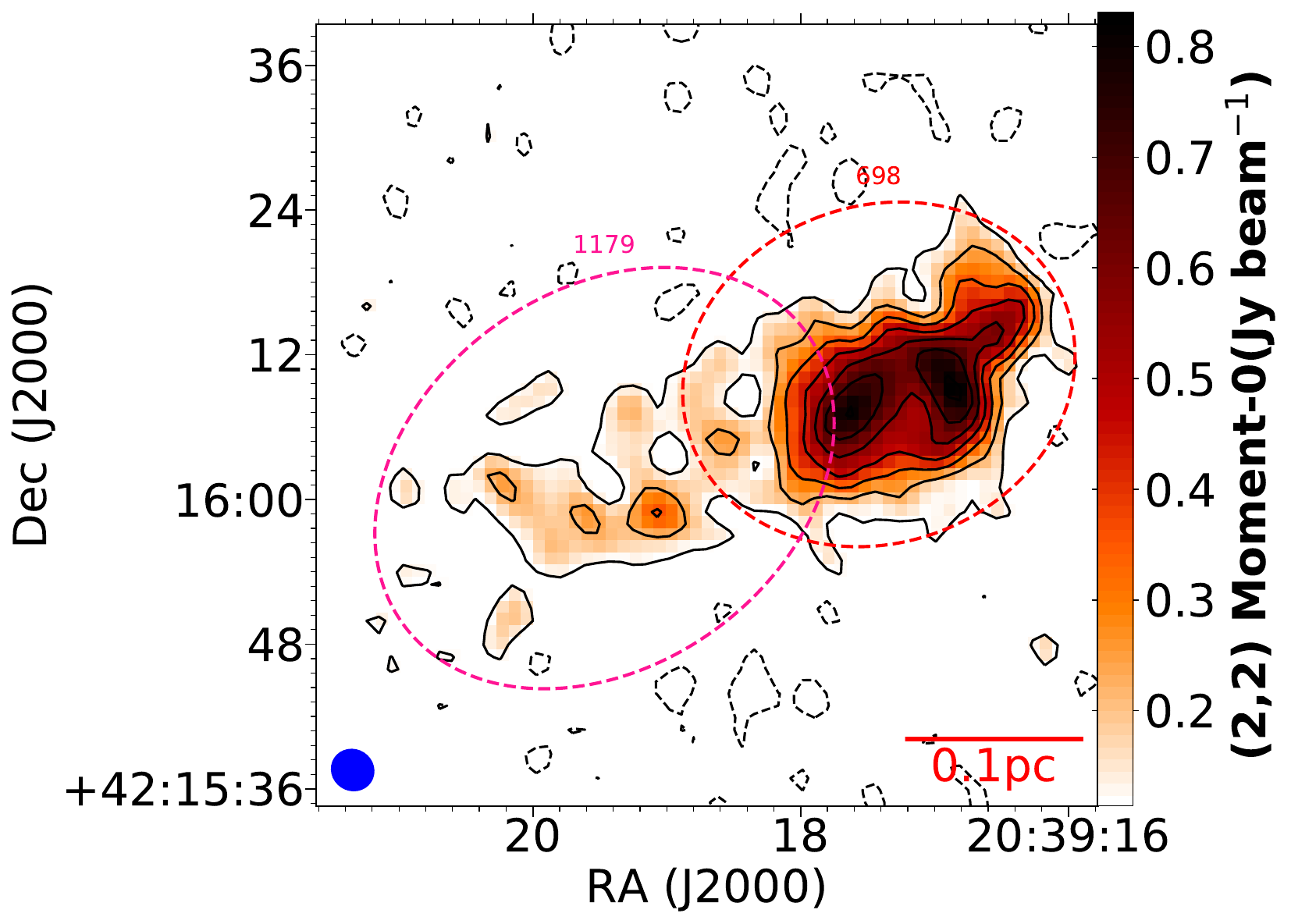} &
\includegraphics[width=.3\textwidth]{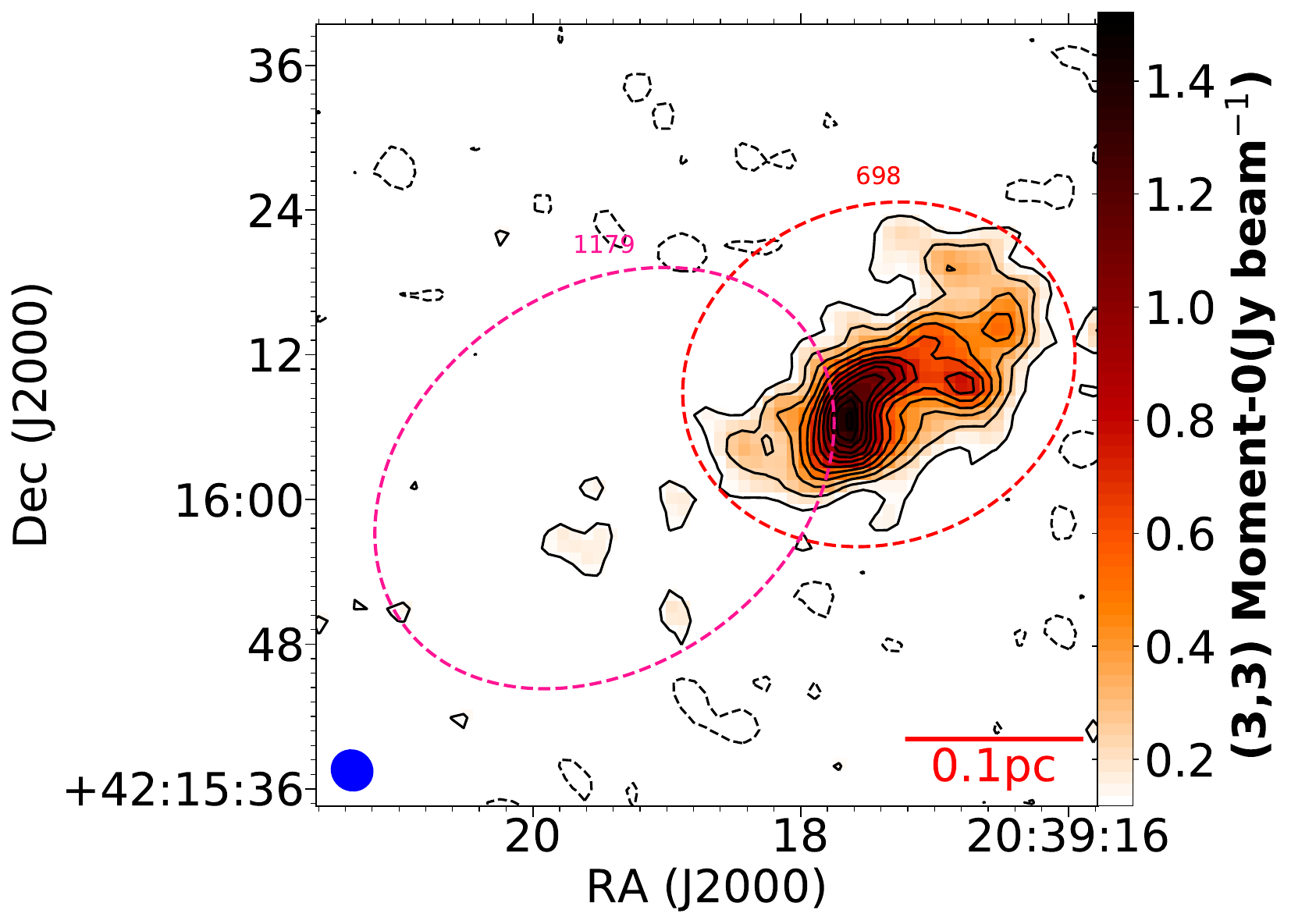} \\
\includegraphics[width=.3\textwidth]{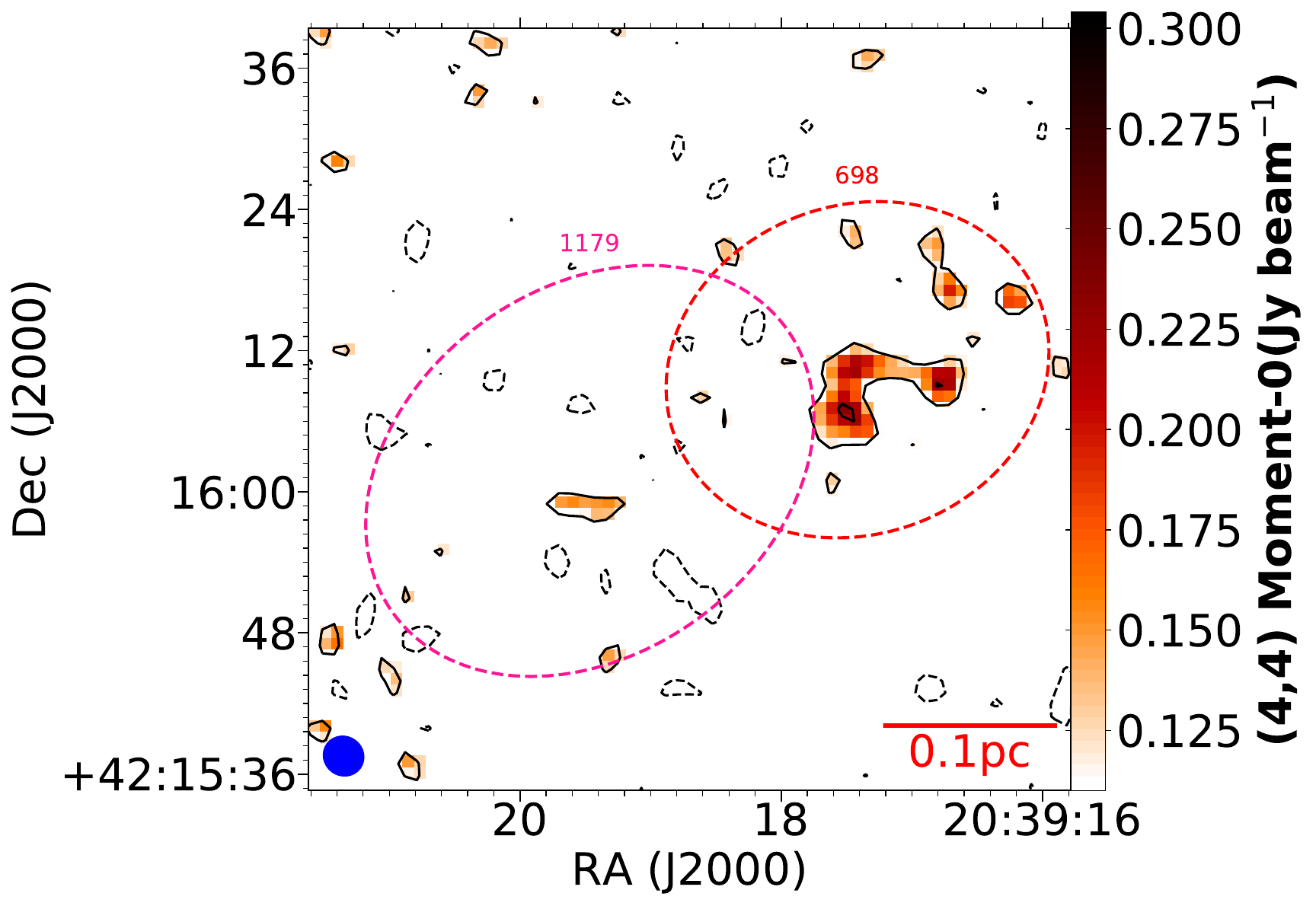} &
 &
 \\
 & Field 8 & \\
\includegraphics[width=.3\textwidth]{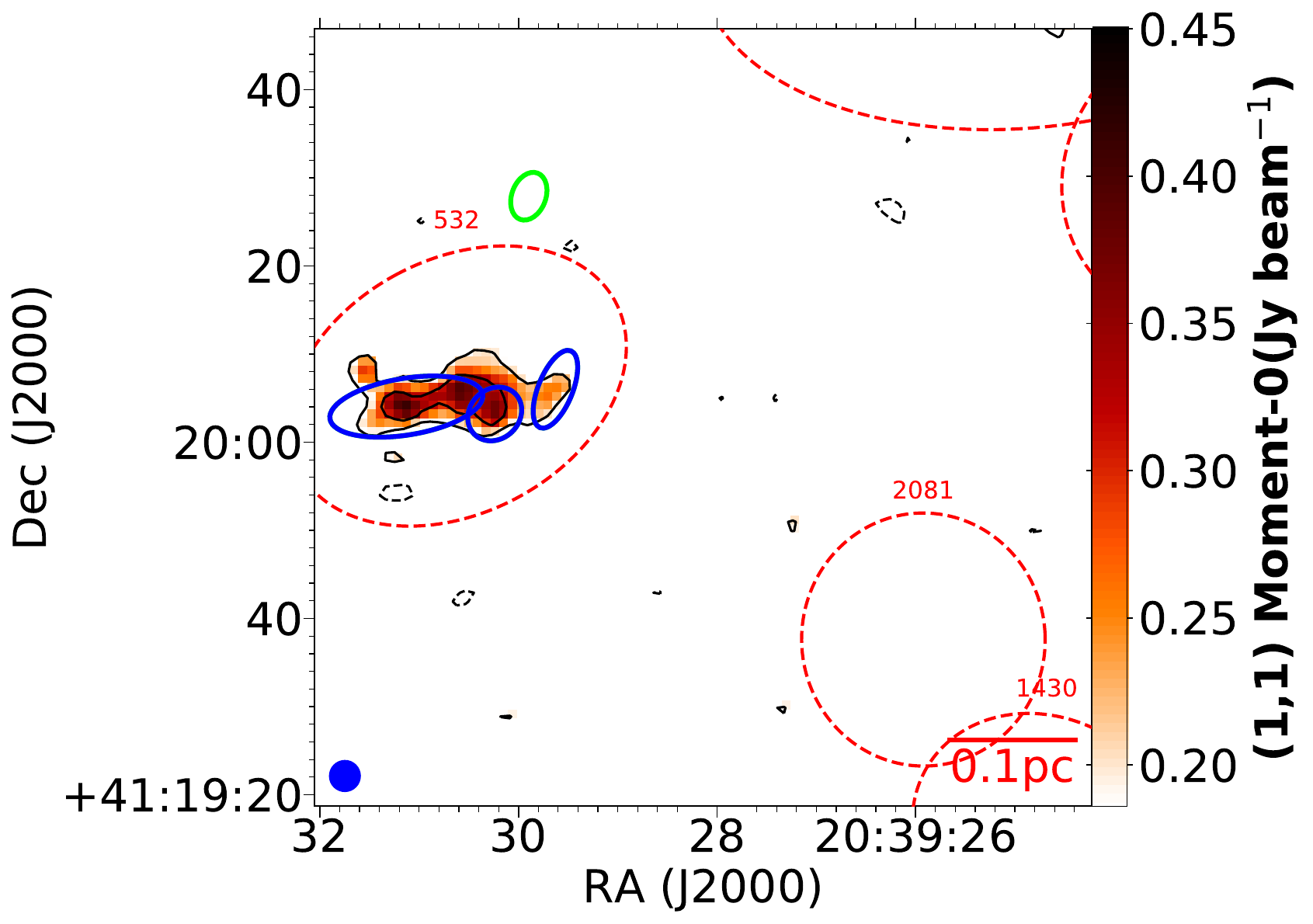} &
\includegraphics[width=.3\textwidth]{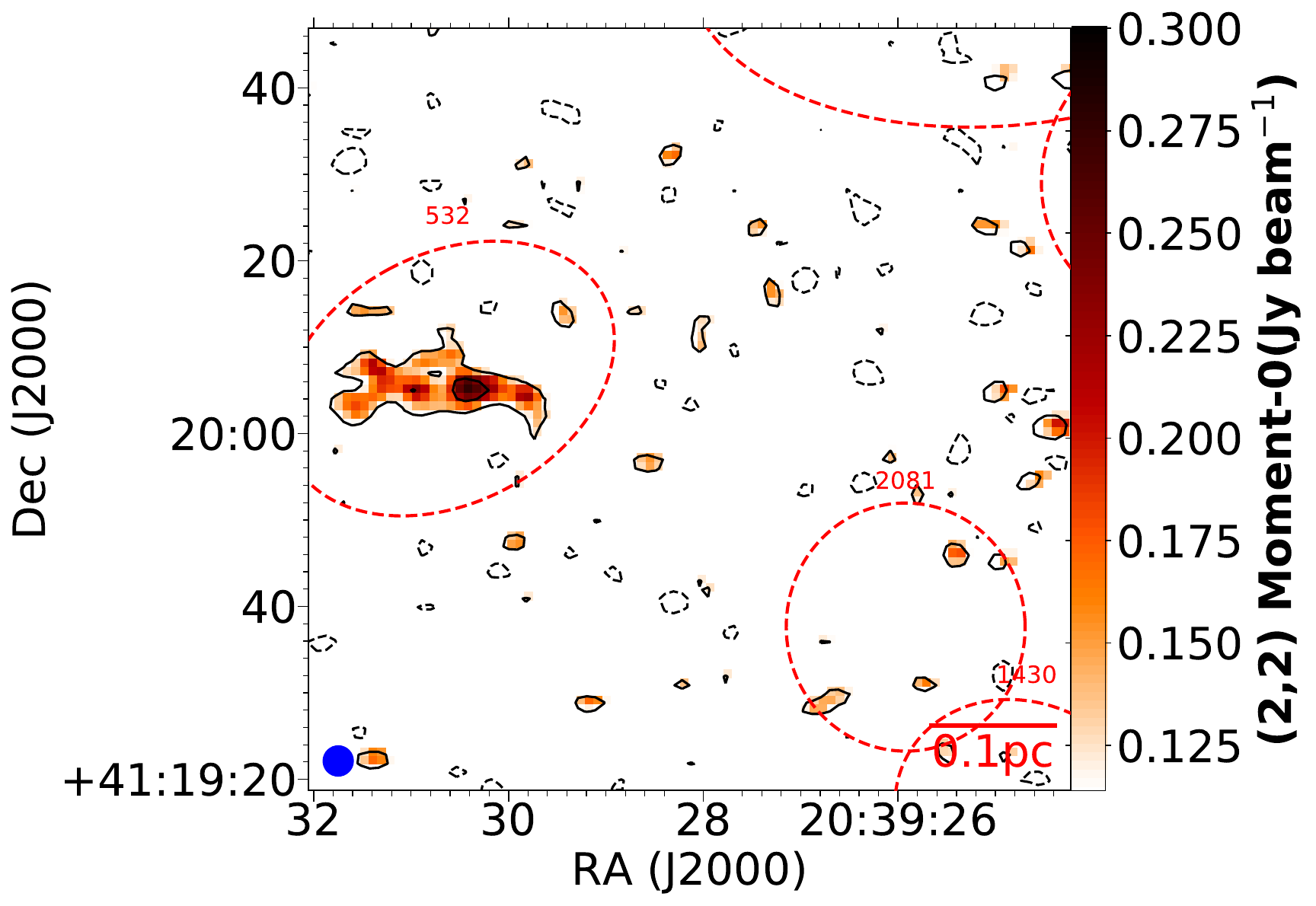} &
 \\
 & Field 9 & \\
\includegraphics[width=.3\textwidth]{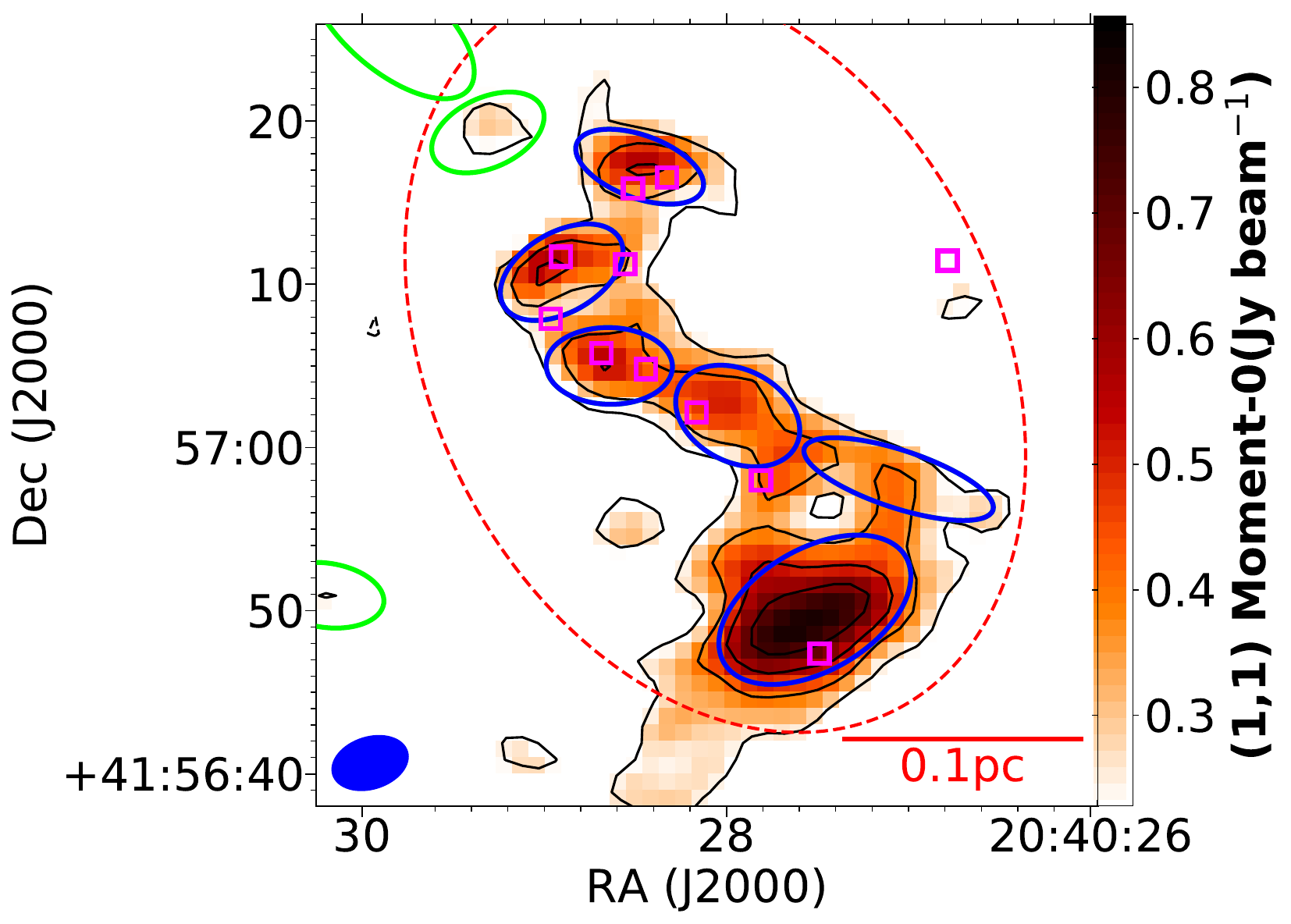} &
\includegraphics[width=.3\textwidth]{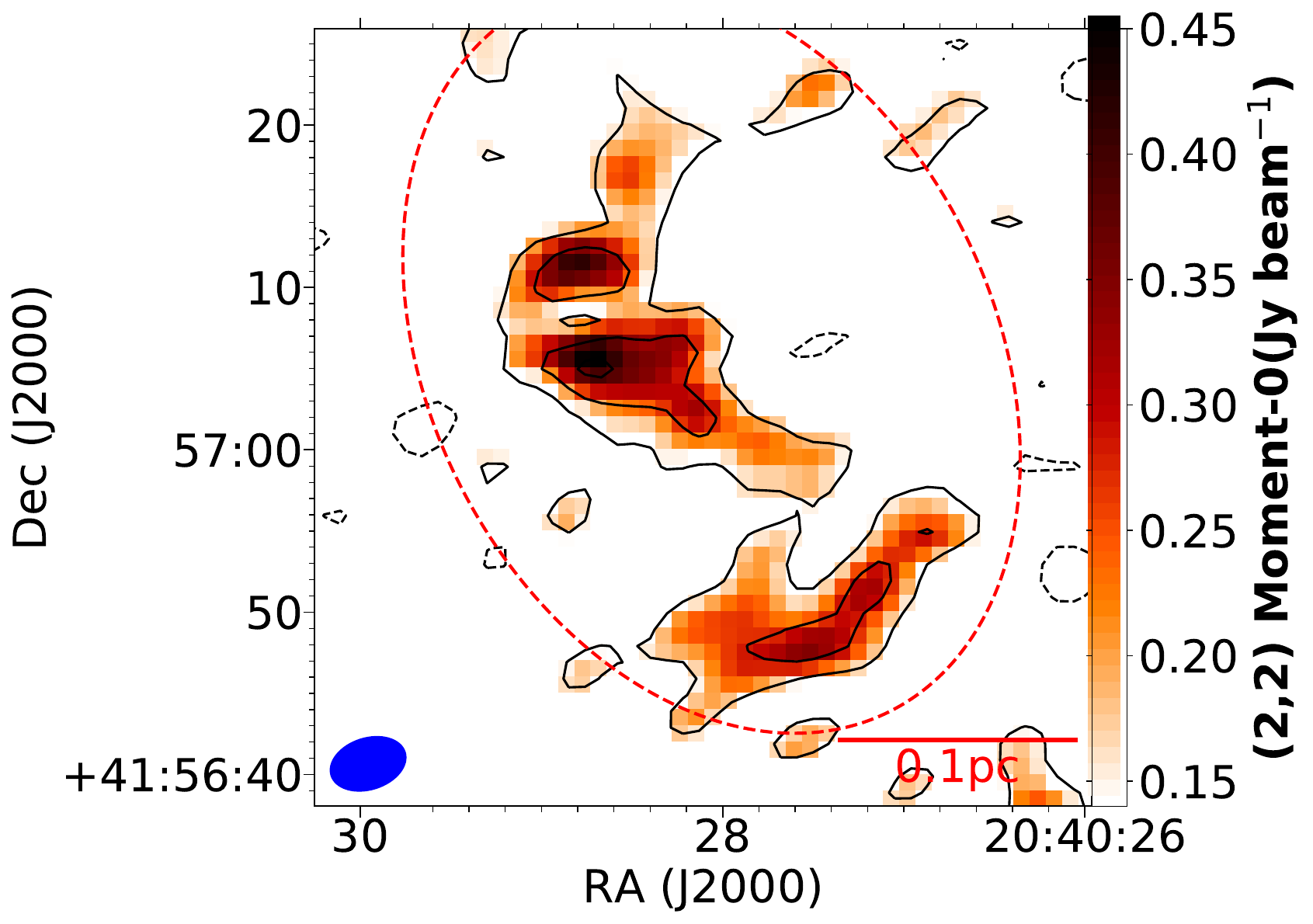} &
\includegraphics[width=.3\textwidth]{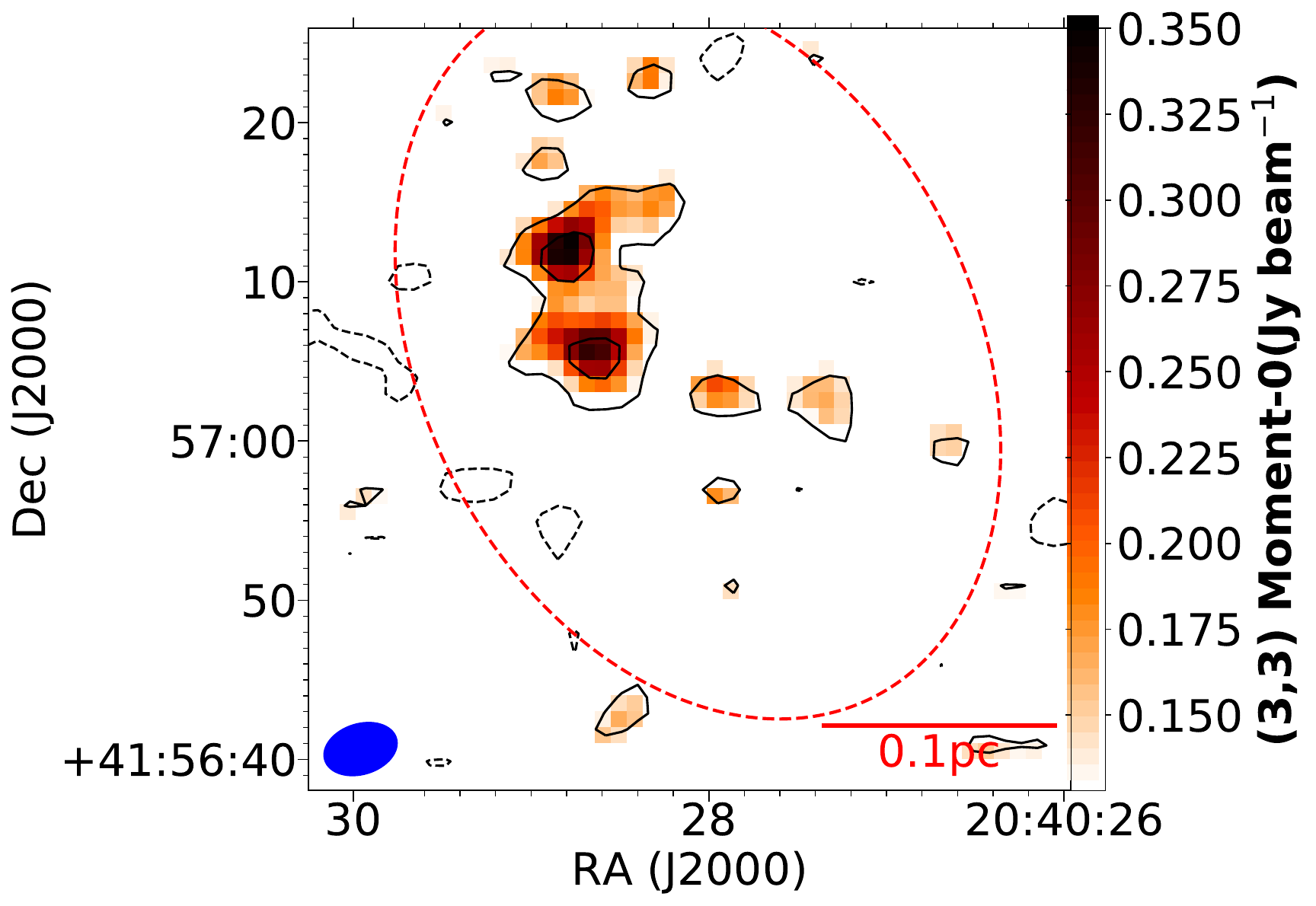} \\
 & Field 10 & \\
\includegraphics[width=.3\textwidth]{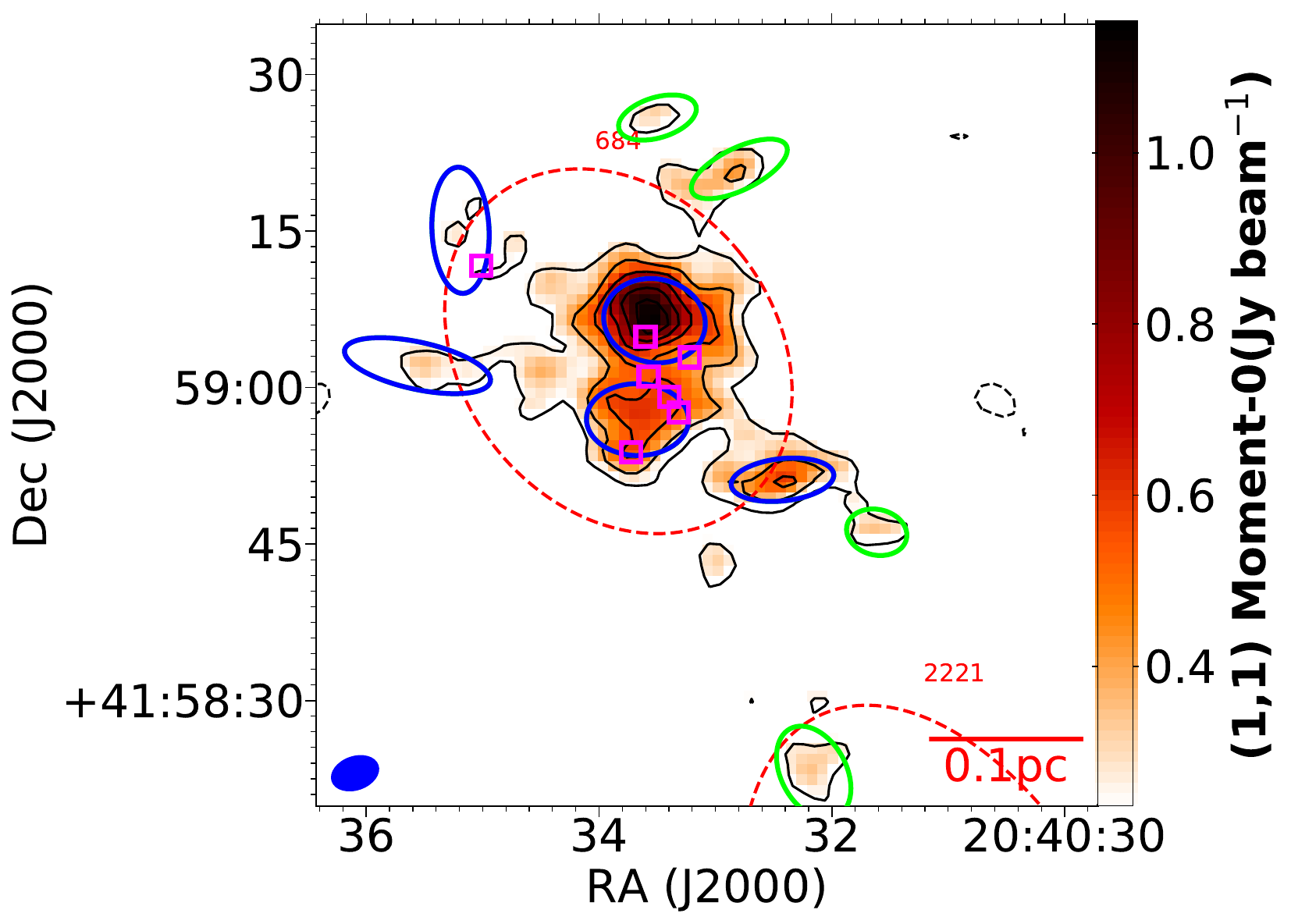} &
\includegraphics[width=.3\textwidth]{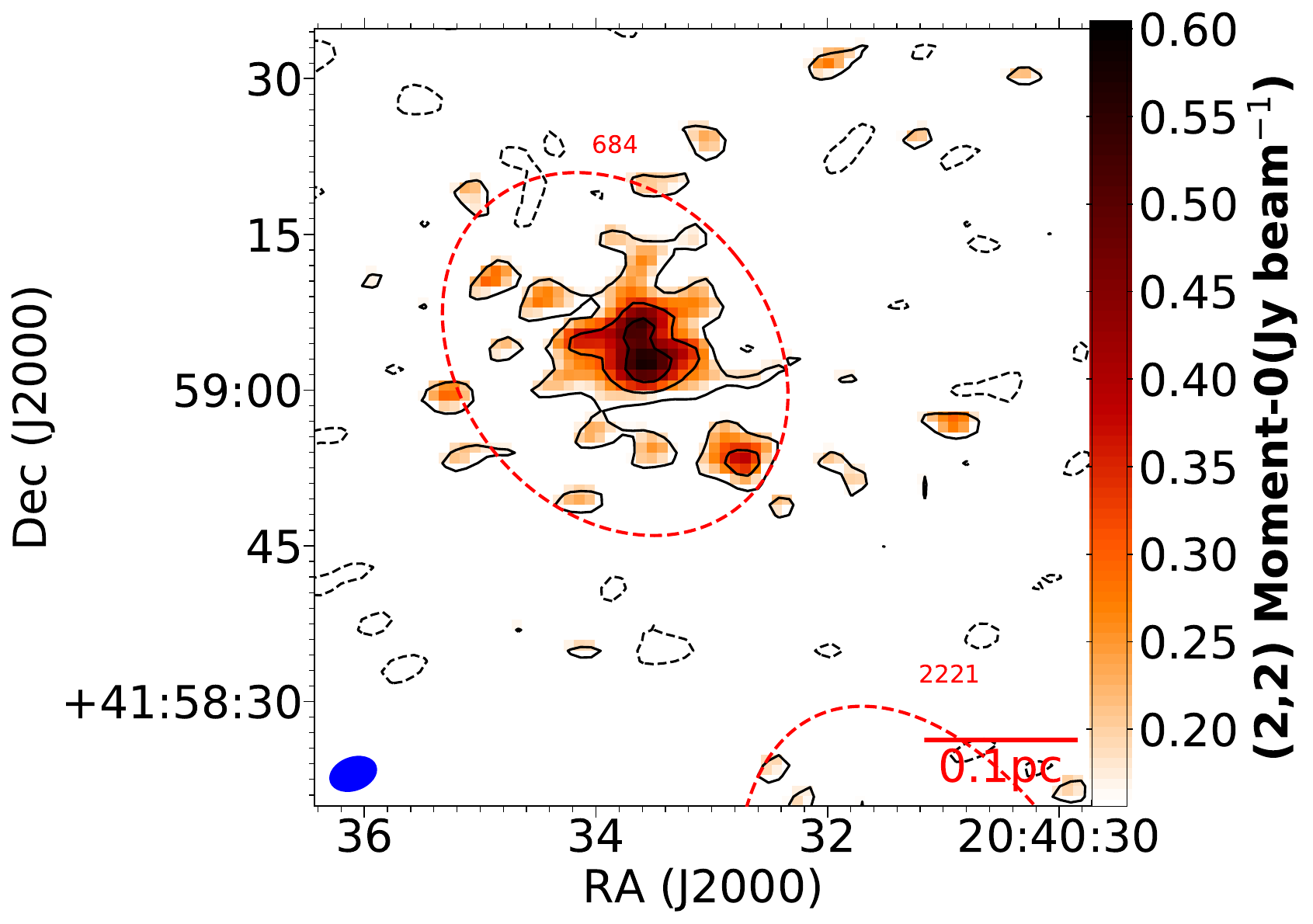} &
\includegraphics[width=.3\textwidth]{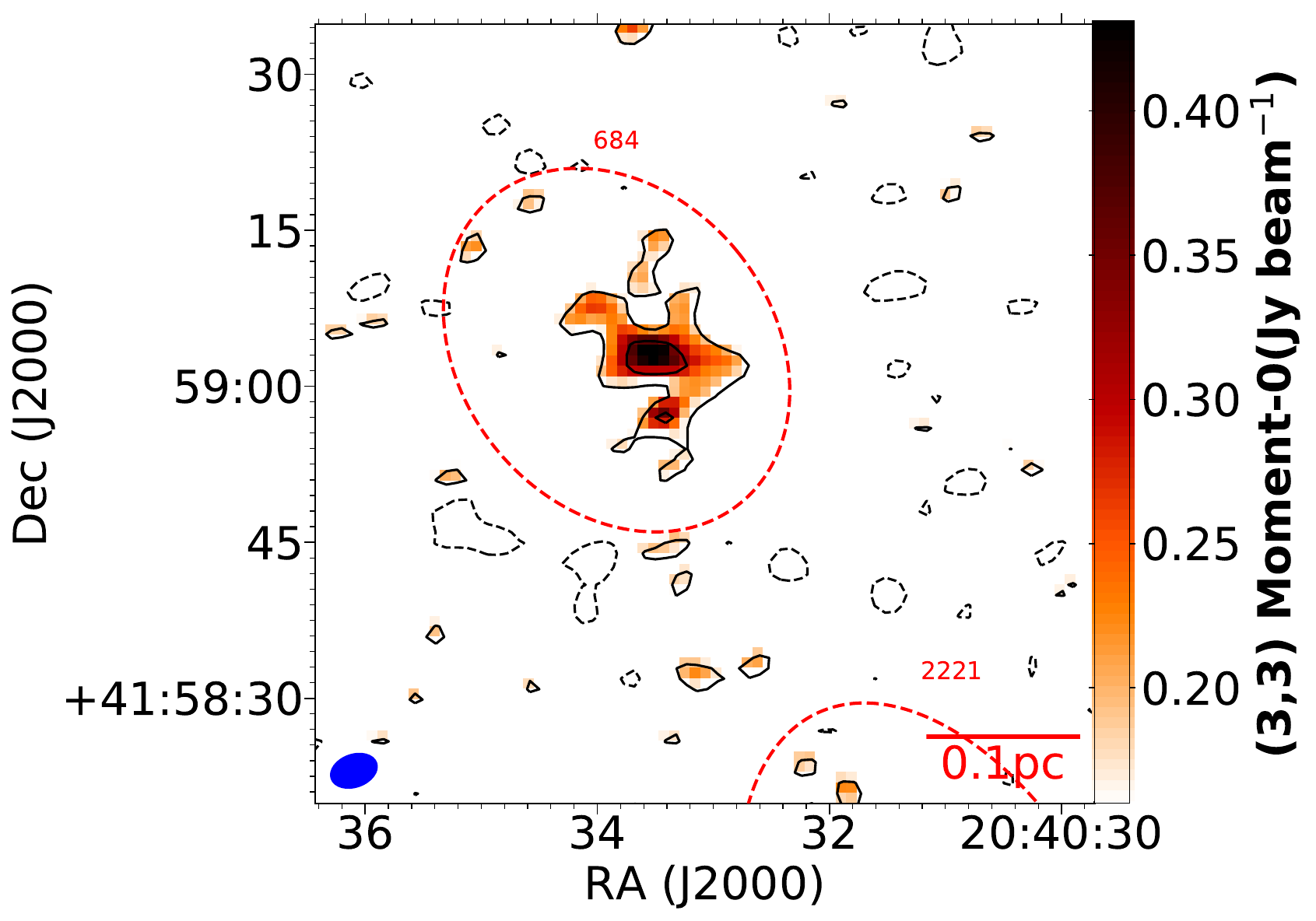} \\
 & Field 11 & \\
\includegraphics[width=.3\textwidth]{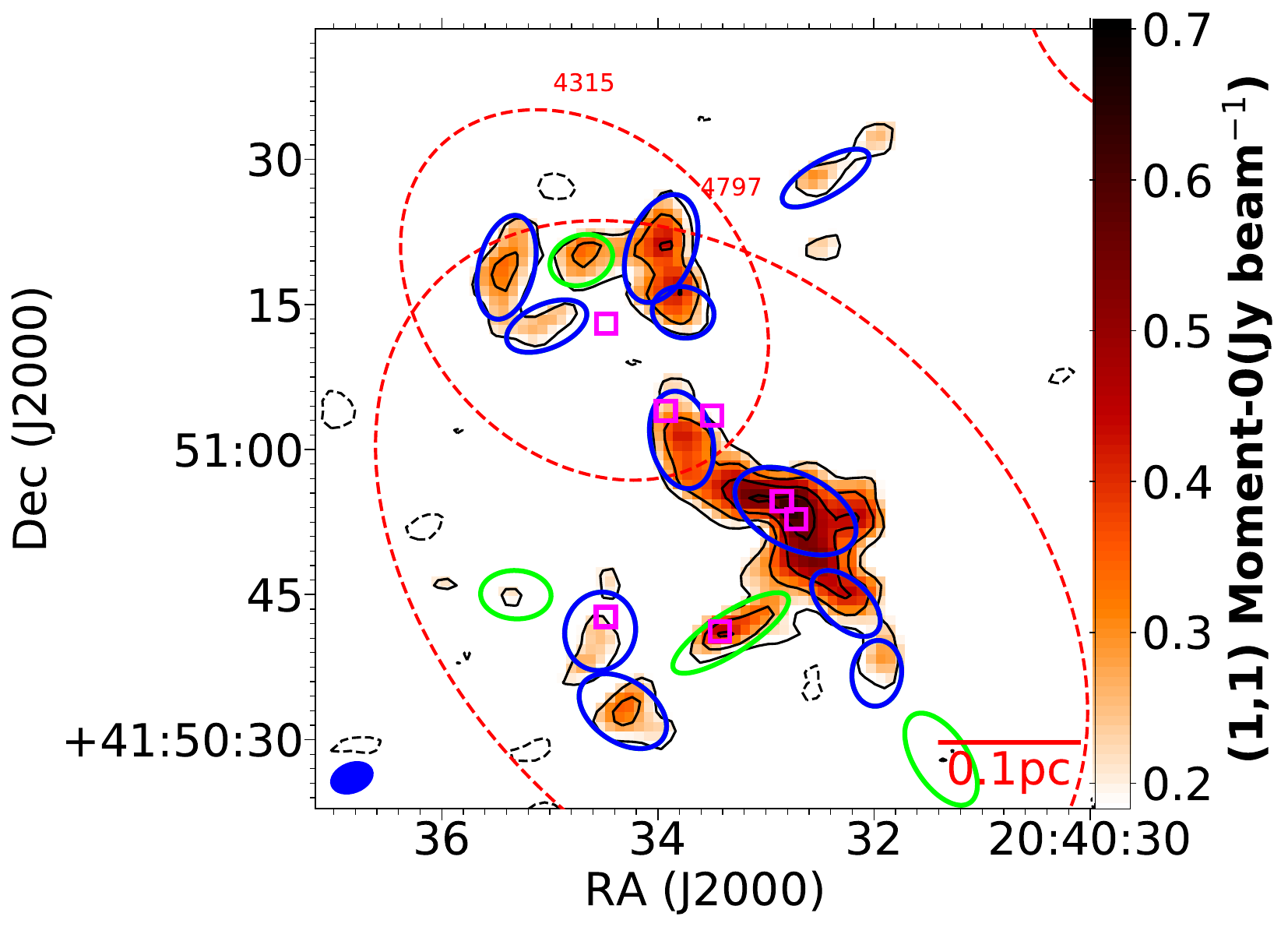} &
\includegraphics[width=.3\textwidth]{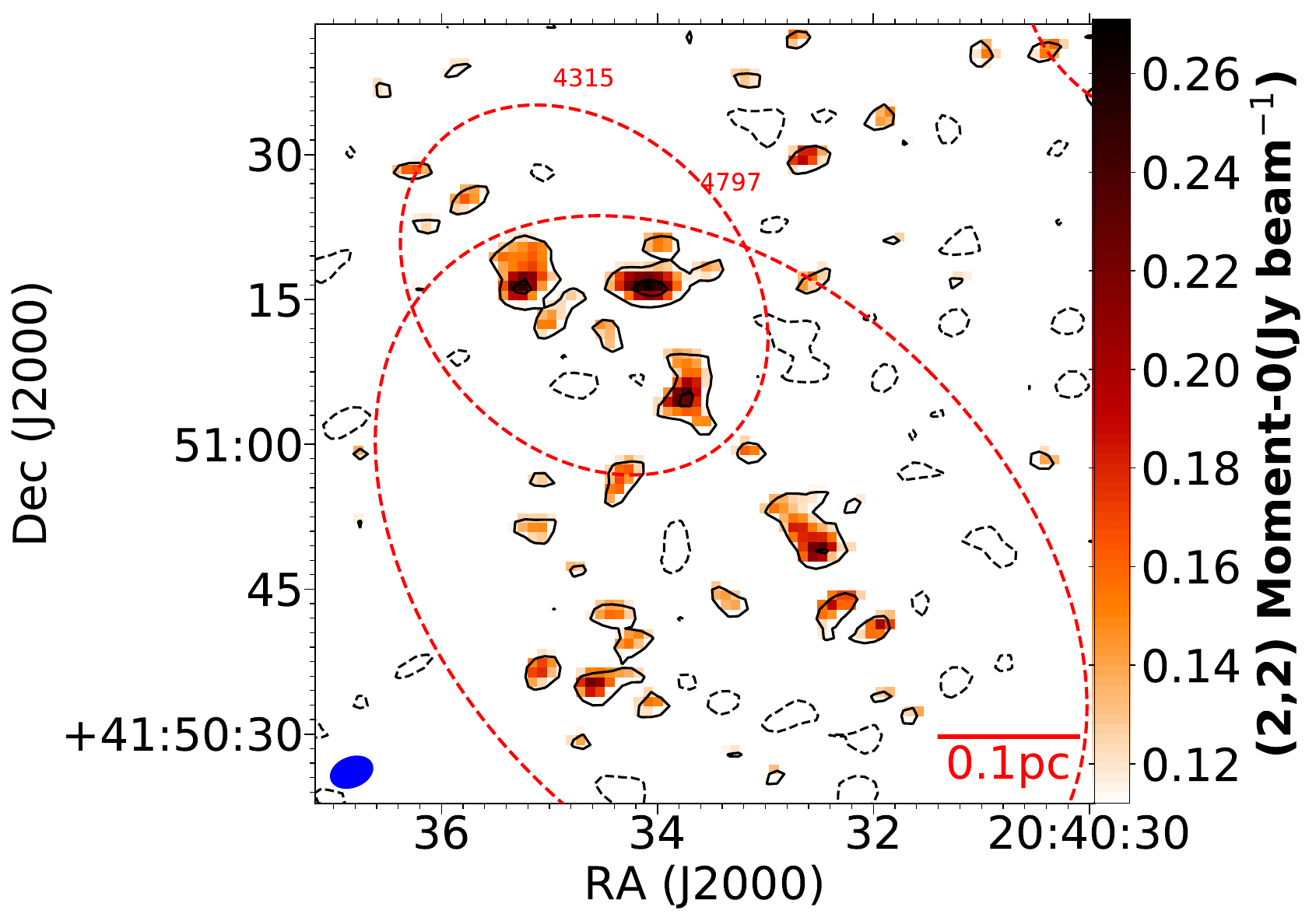} &
 \\
 & Field 12 & \\
\includegraphics[width=.3\textwidth]{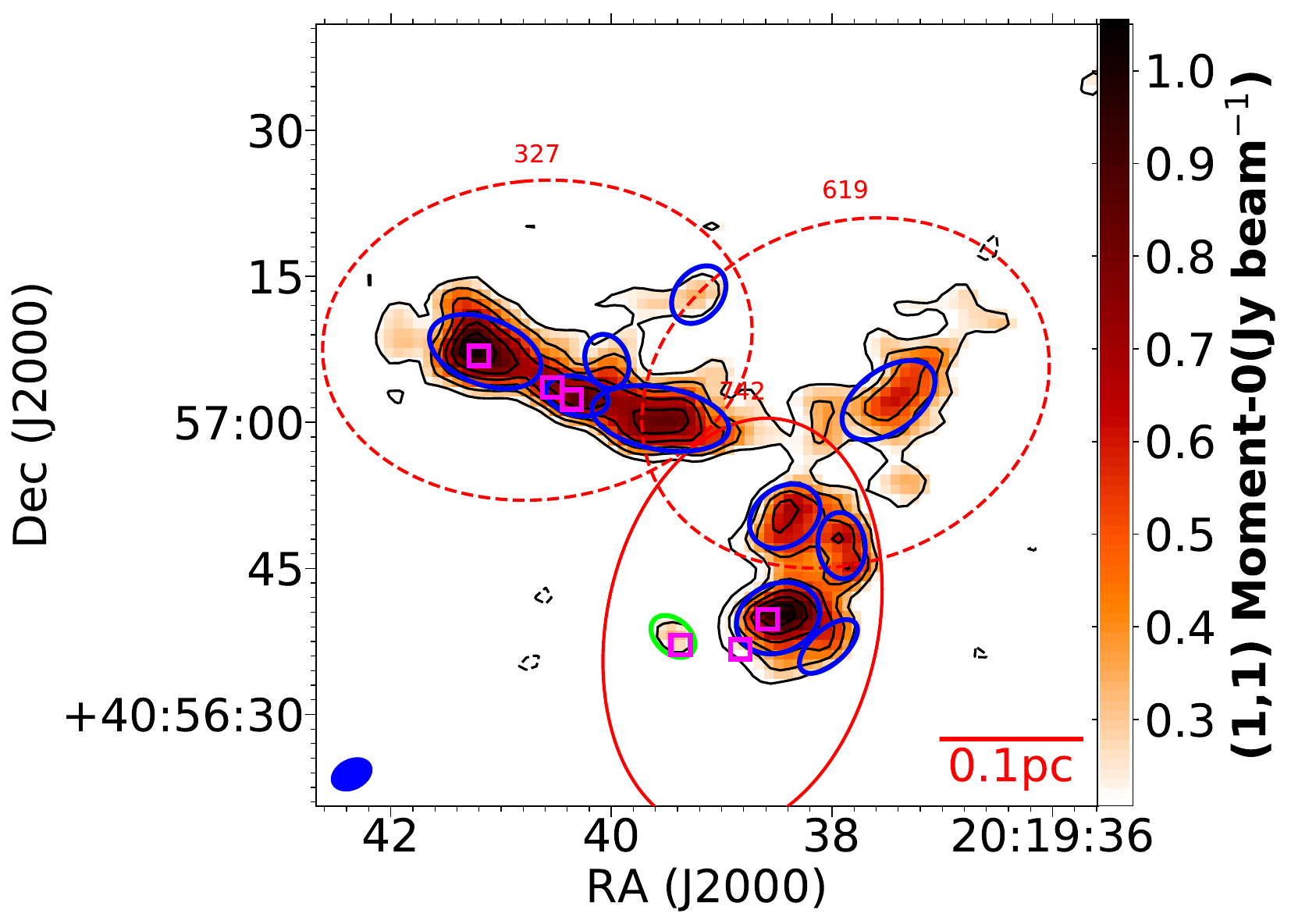} &
\includegraphics[width=.3\textwidth]{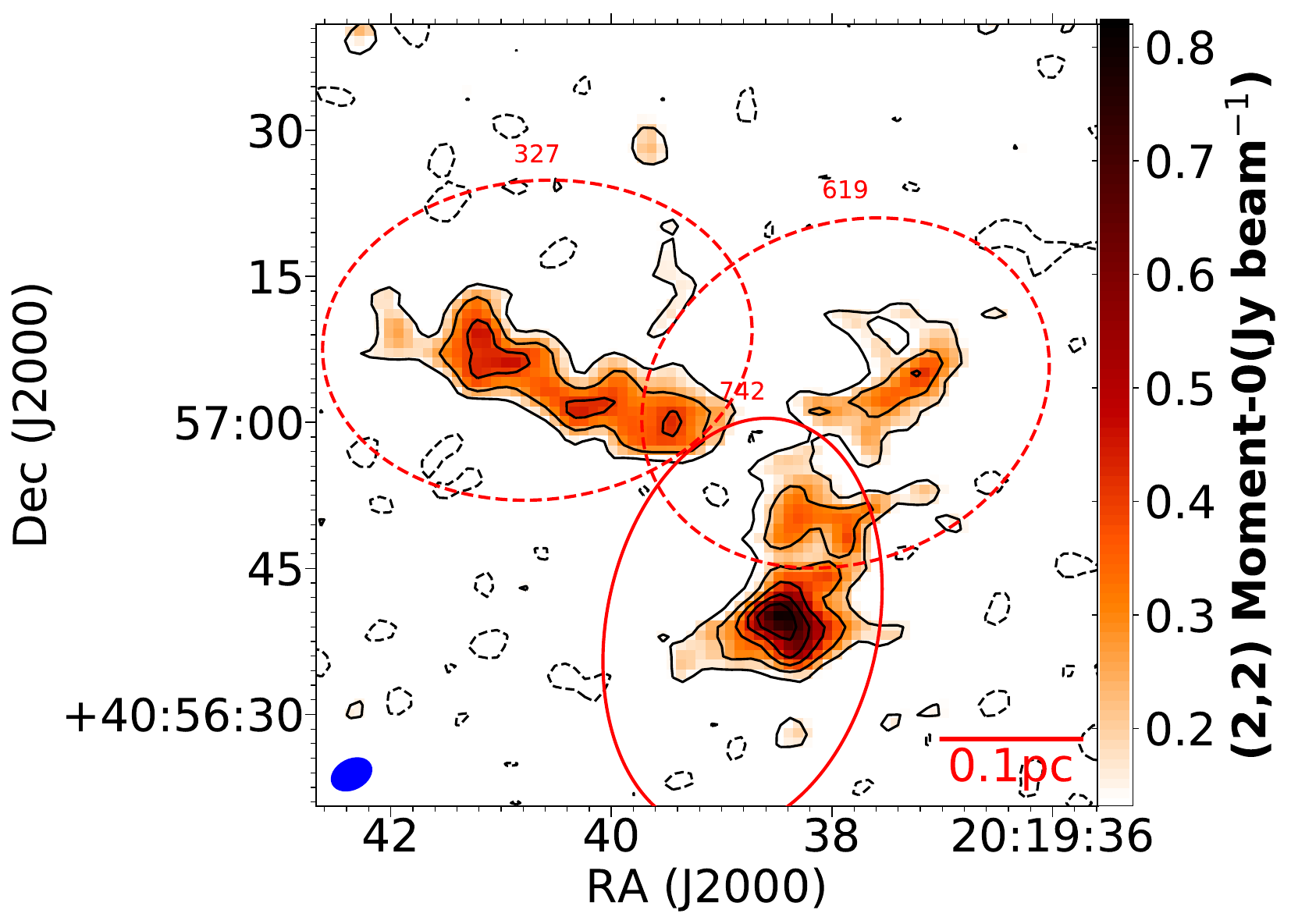} &
\includegraphics[width=.3\textwidth]{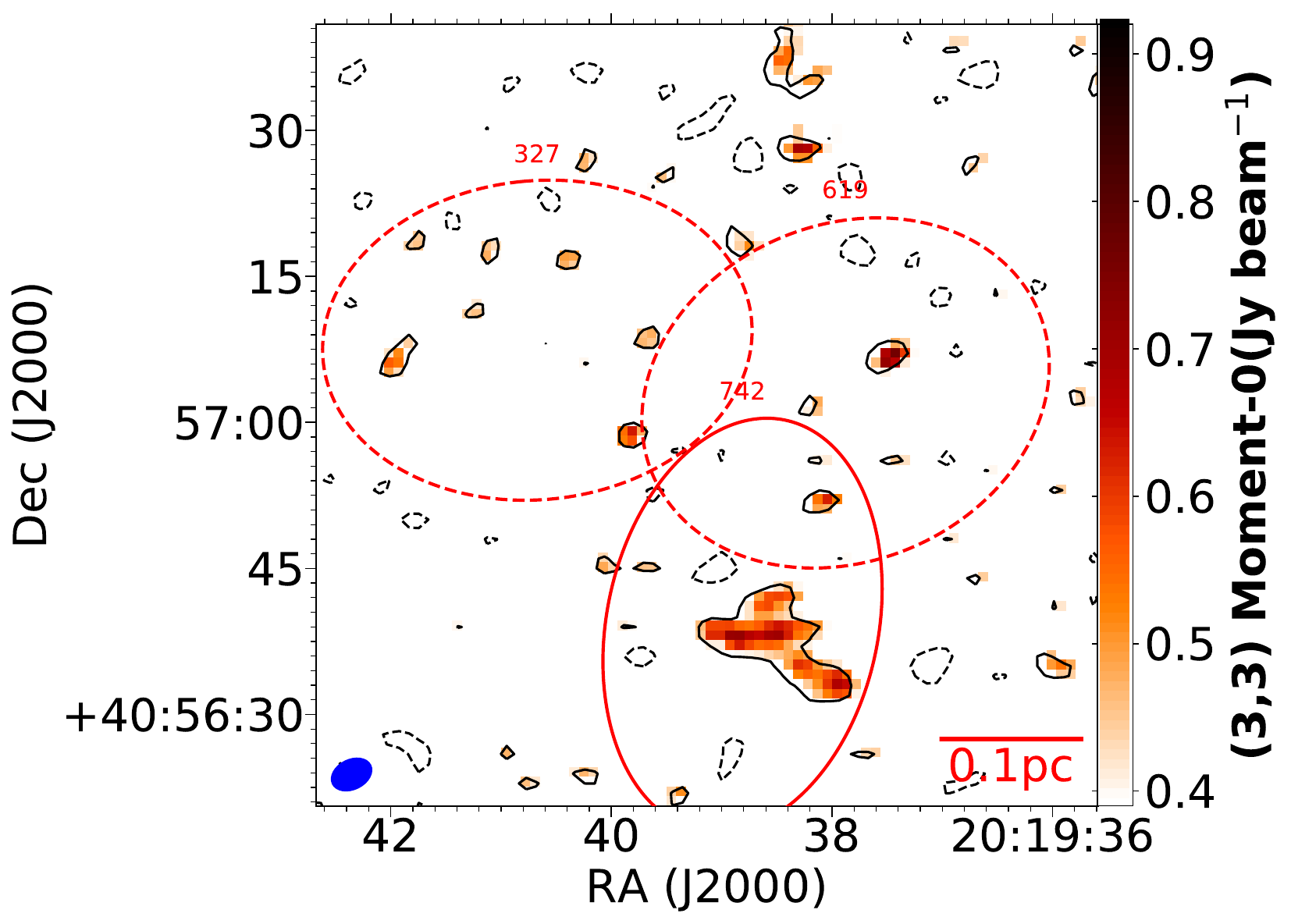} \\
 & Field 13 & \\
\includegraphics[width=.3\textwidth]{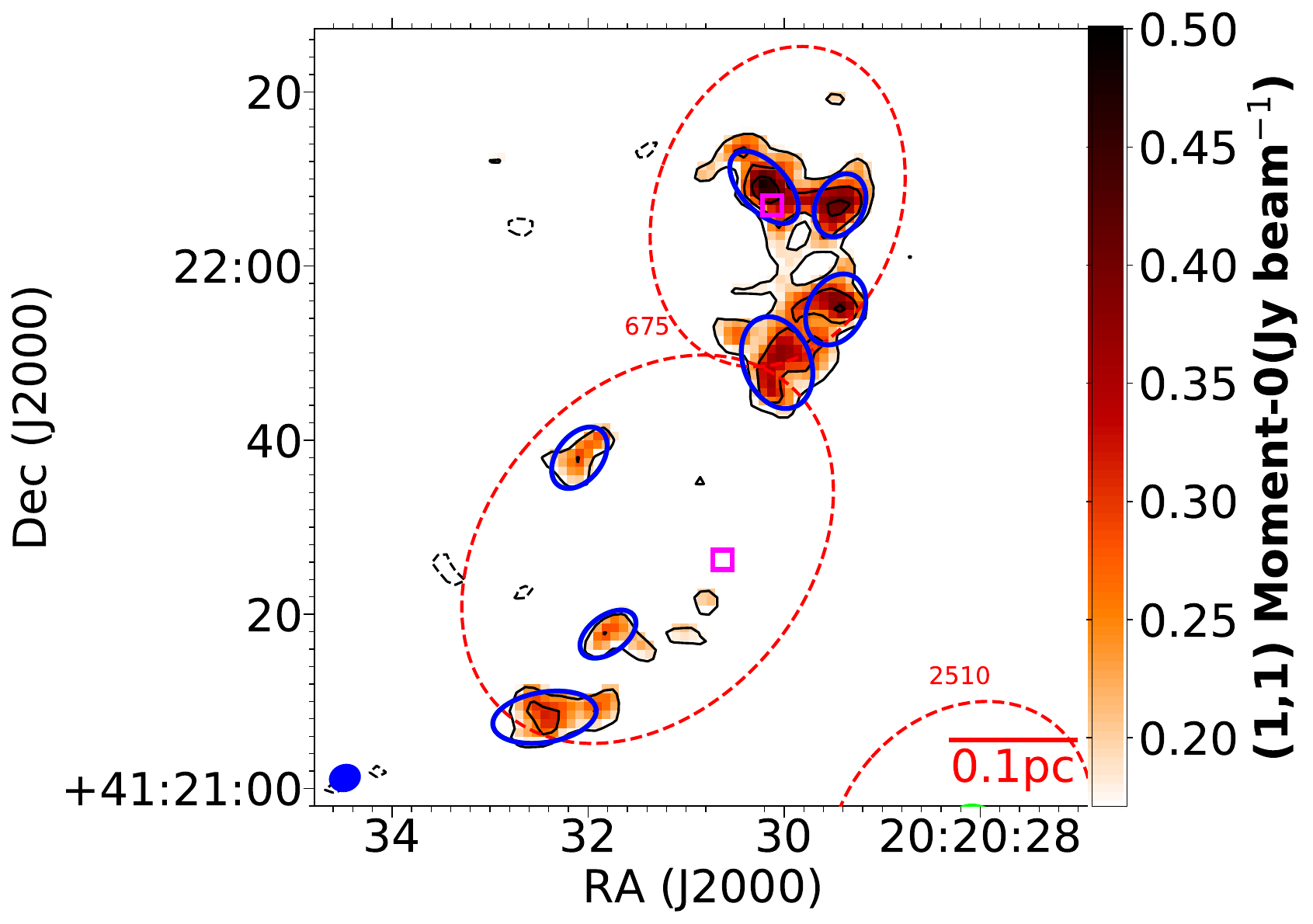} &
\includegraphics[width=.3\textwidth]{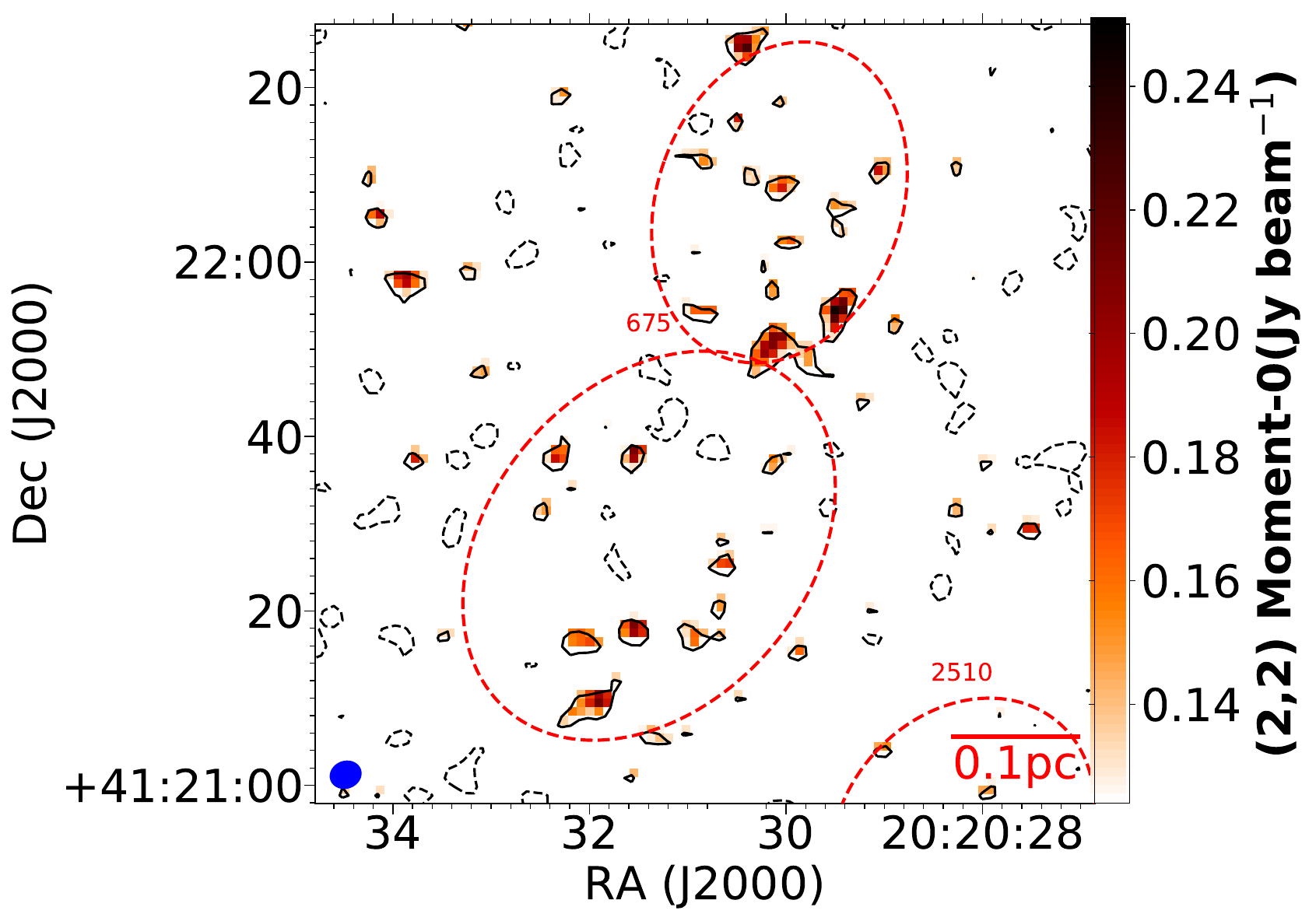} &
 \\
 & Field 14 & \\
\includegraphics[width=.3\textwidth]{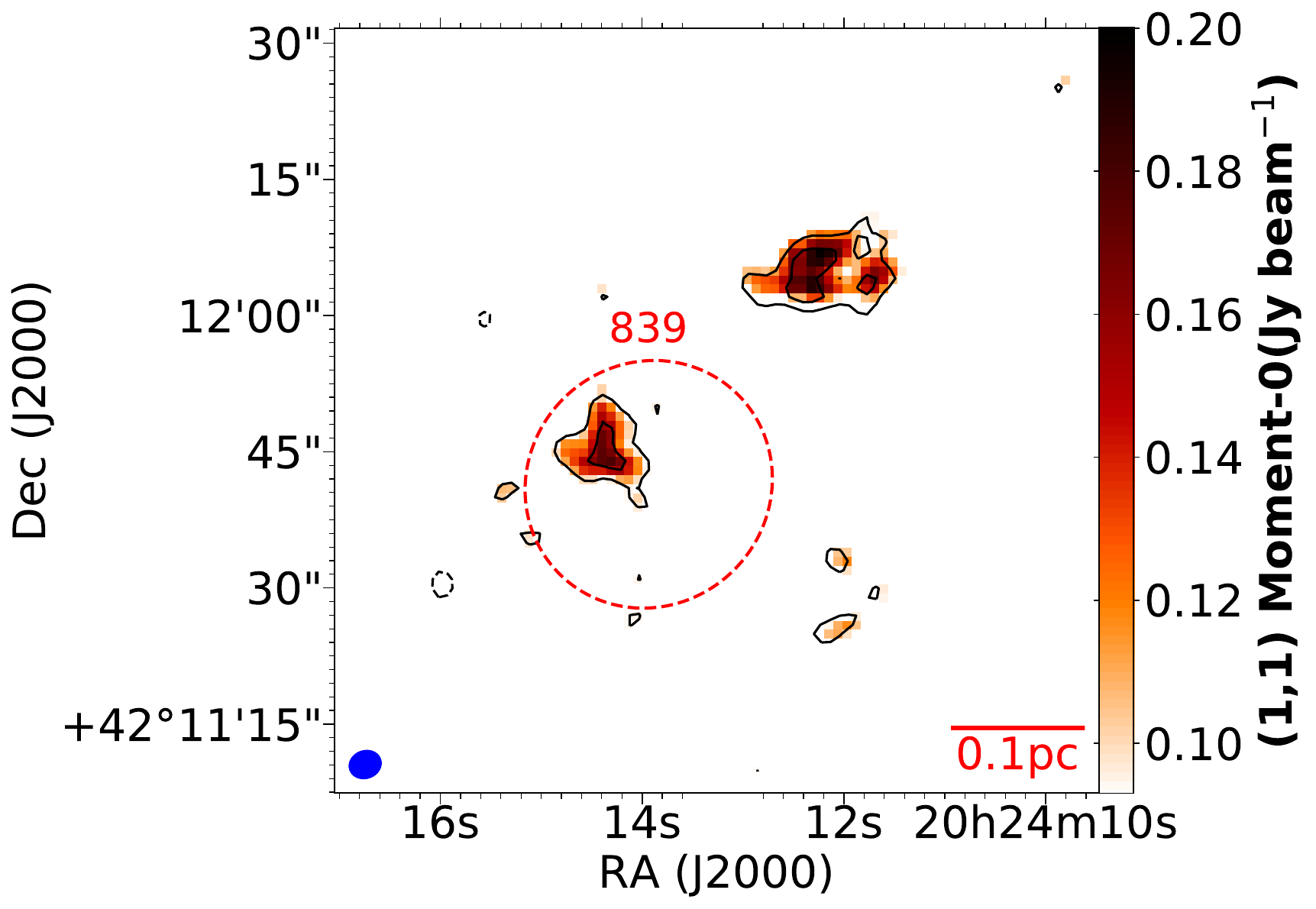} &
 &
 \\
 & Field 15 & \\
\includegraphics[width=.3\textwidth]{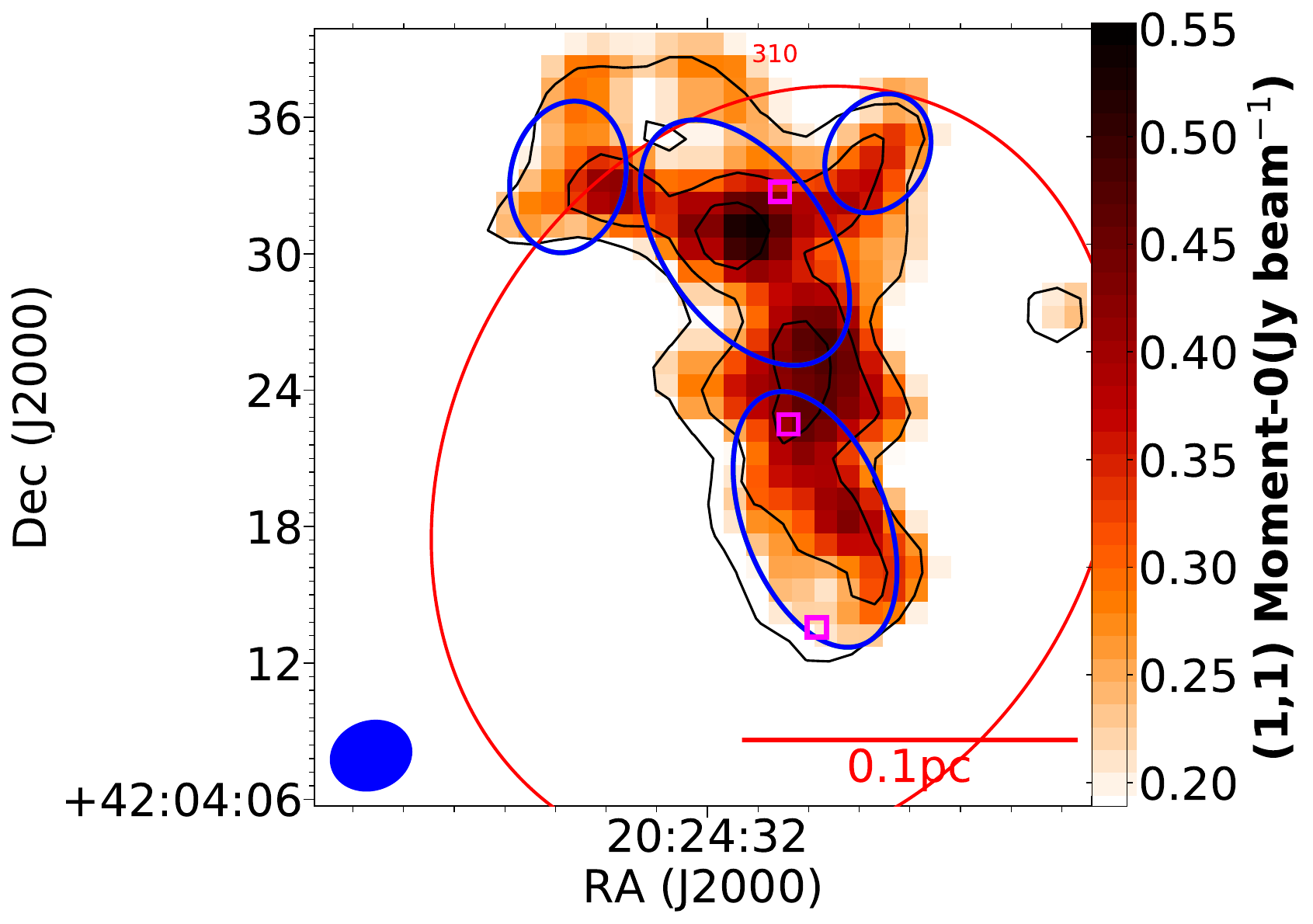} &
\includegraphics[width=.3\textwidth]{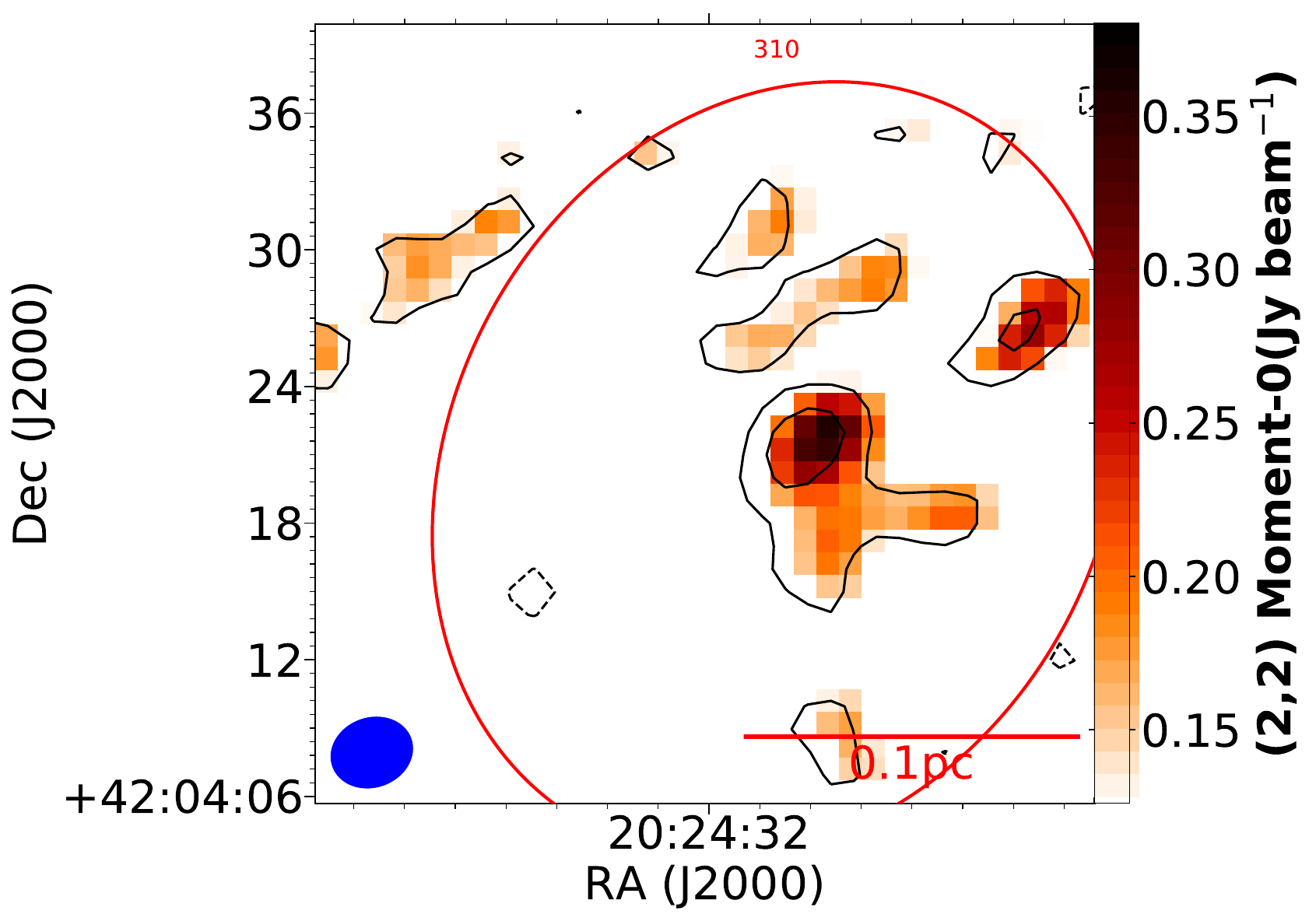} &
\includegraphics[width=.3\textwidth]{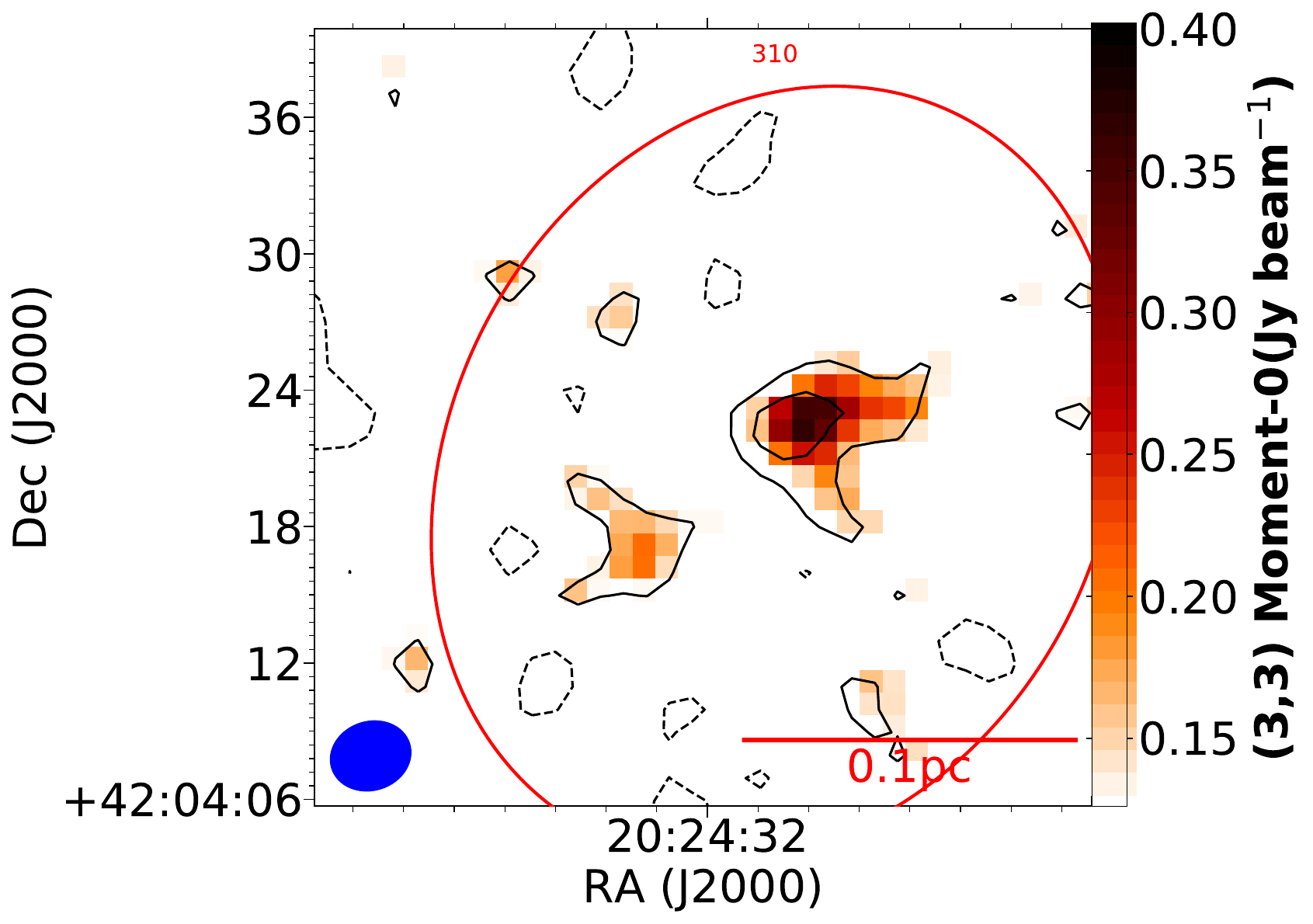} \\
 & Field 16 & \\
\includegraphics[width=.3\textwidth]{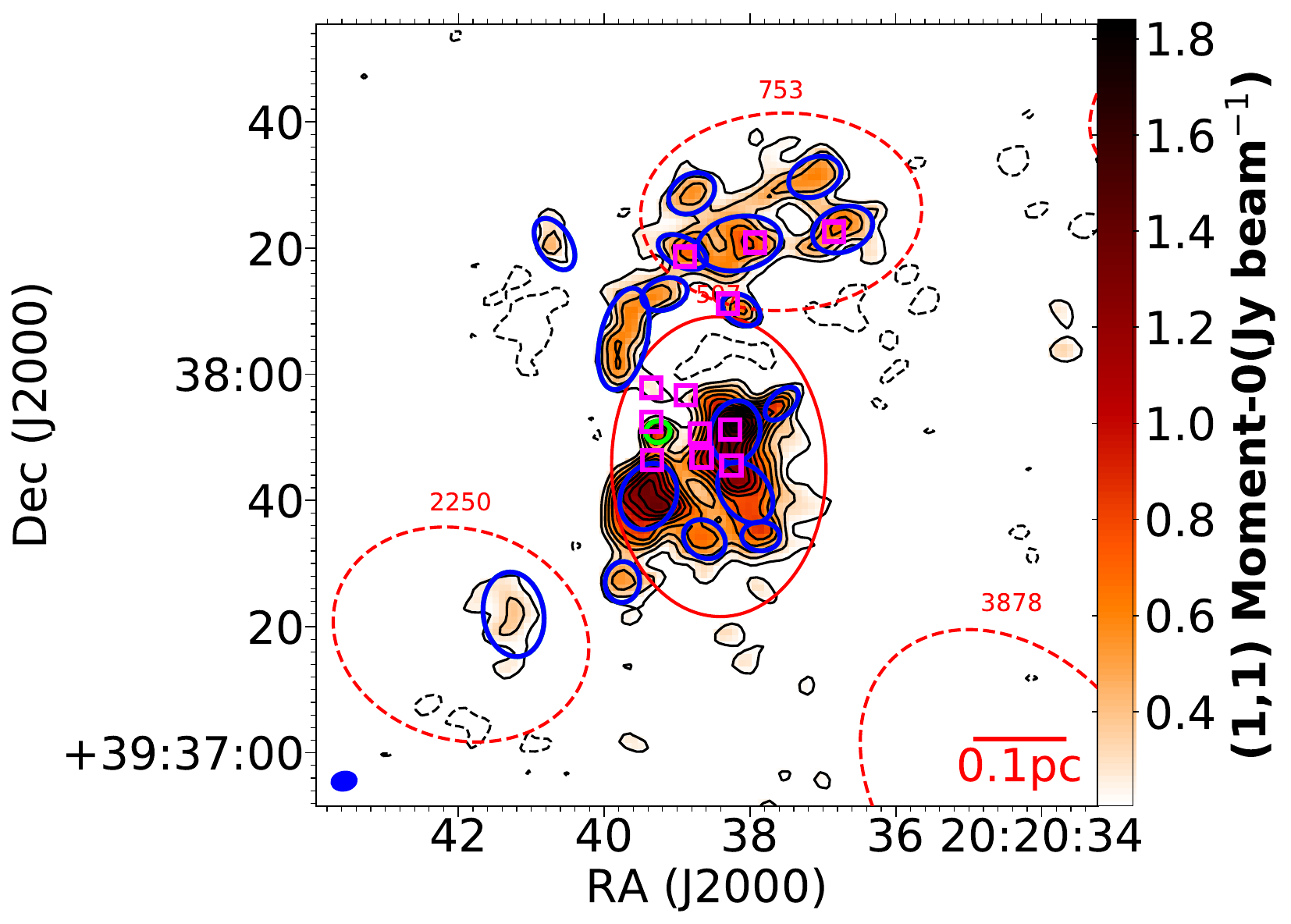} &
\includegraphics[width=.3\textwidth]{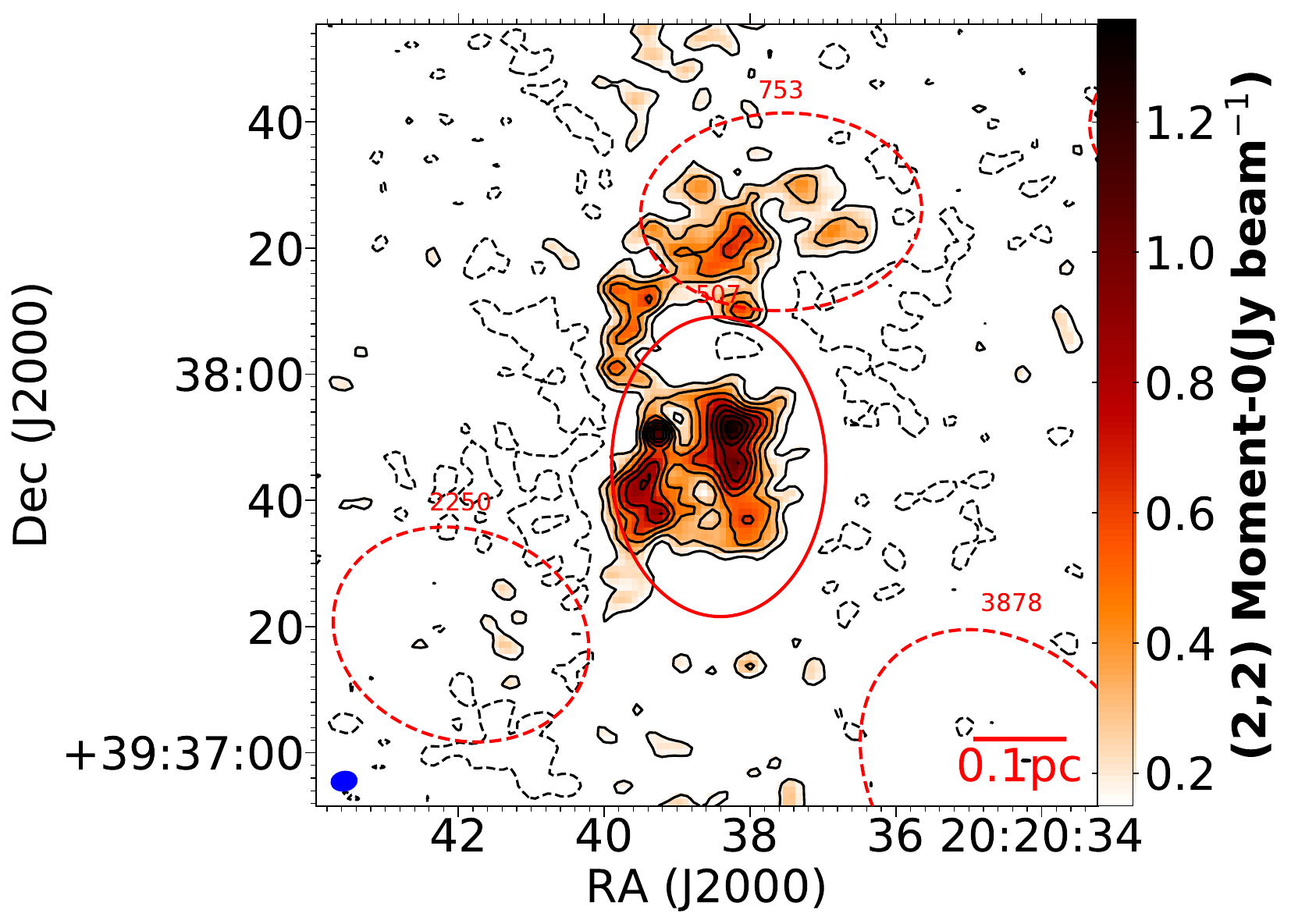} &
\includegraphics[width=.3\textwidth]{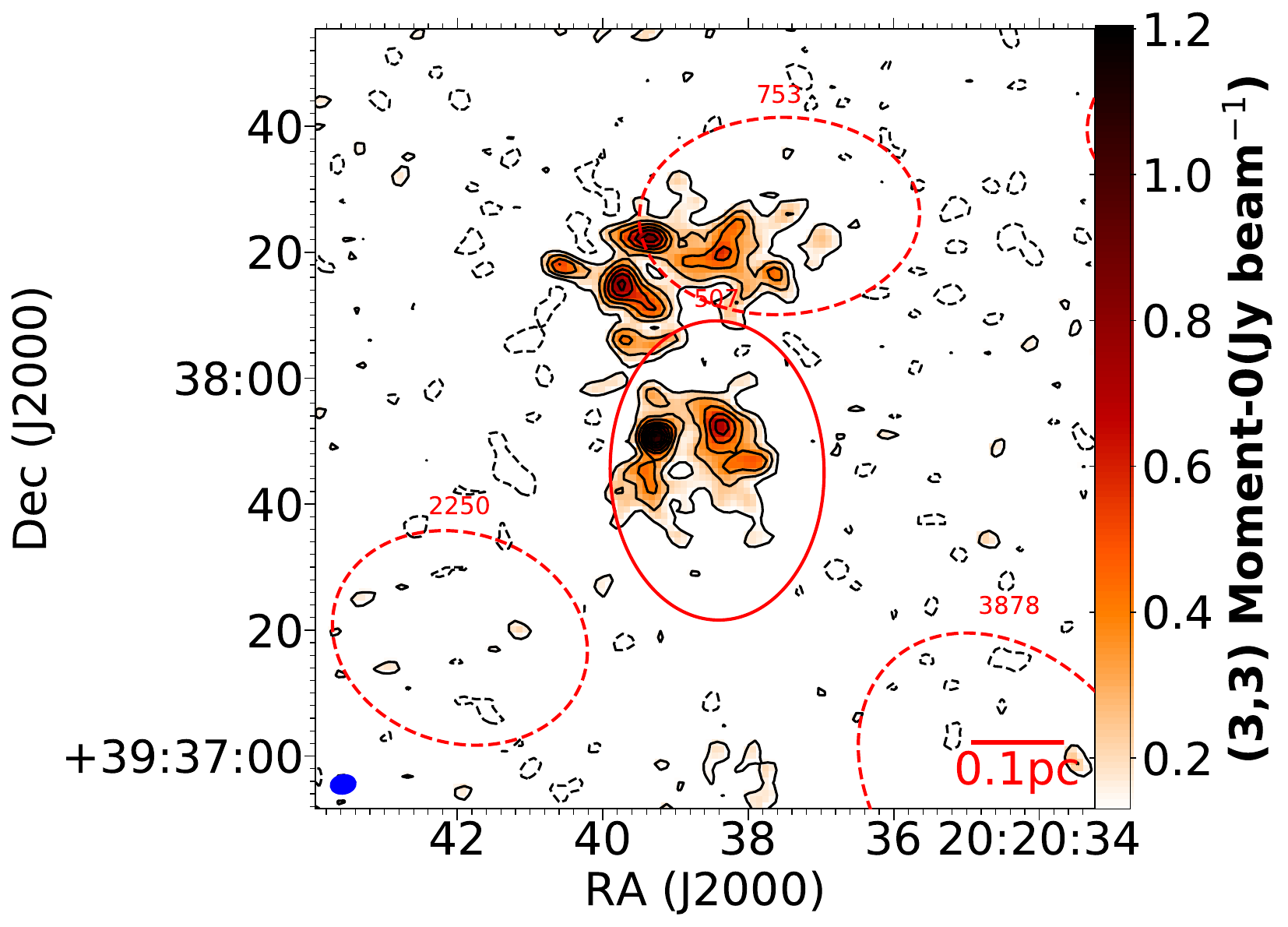} \\
 & Field 17 & \\
\includegraphics[width=.3\textwidth]{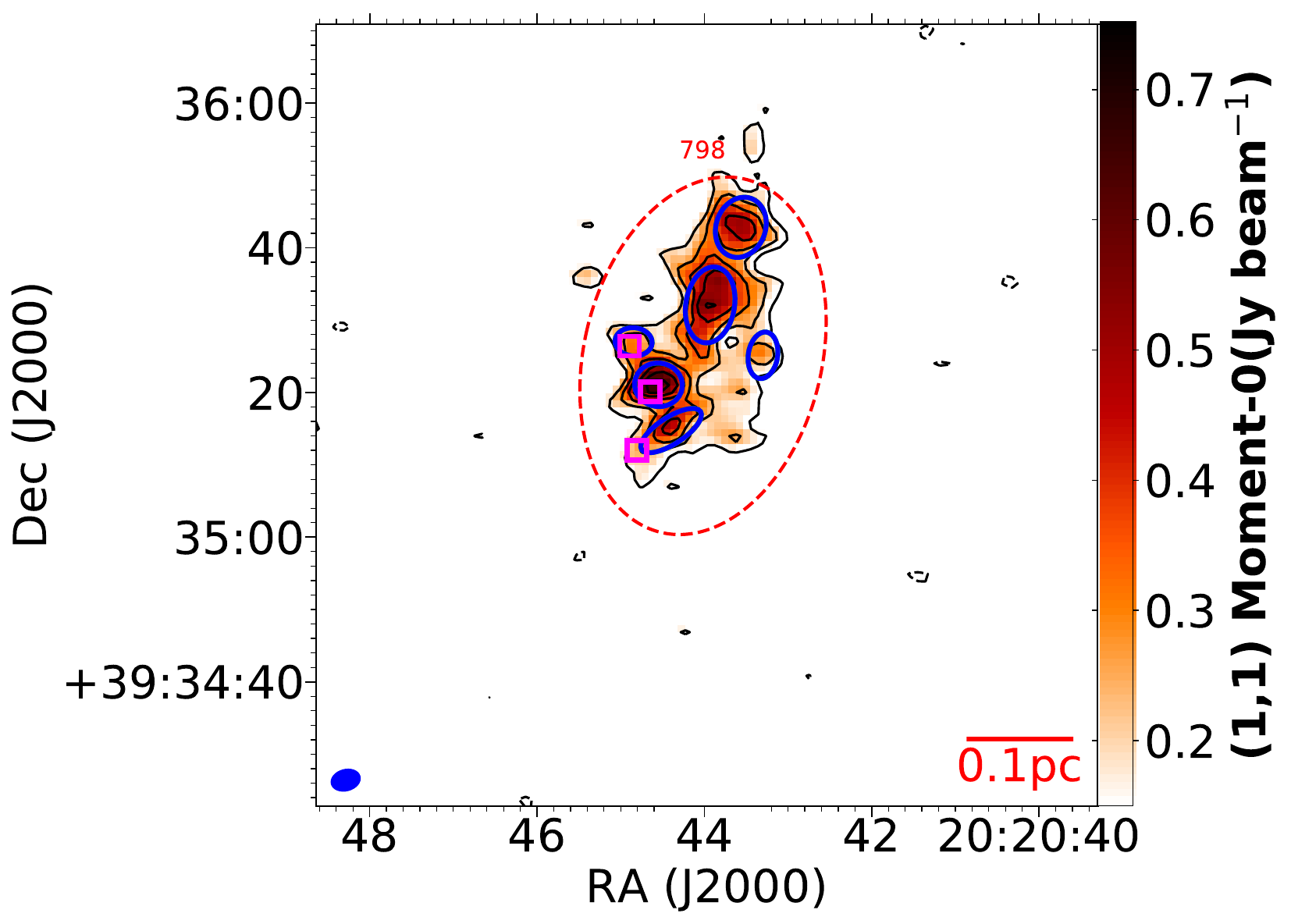} &
\includegraphics[width=.3\textwidth]{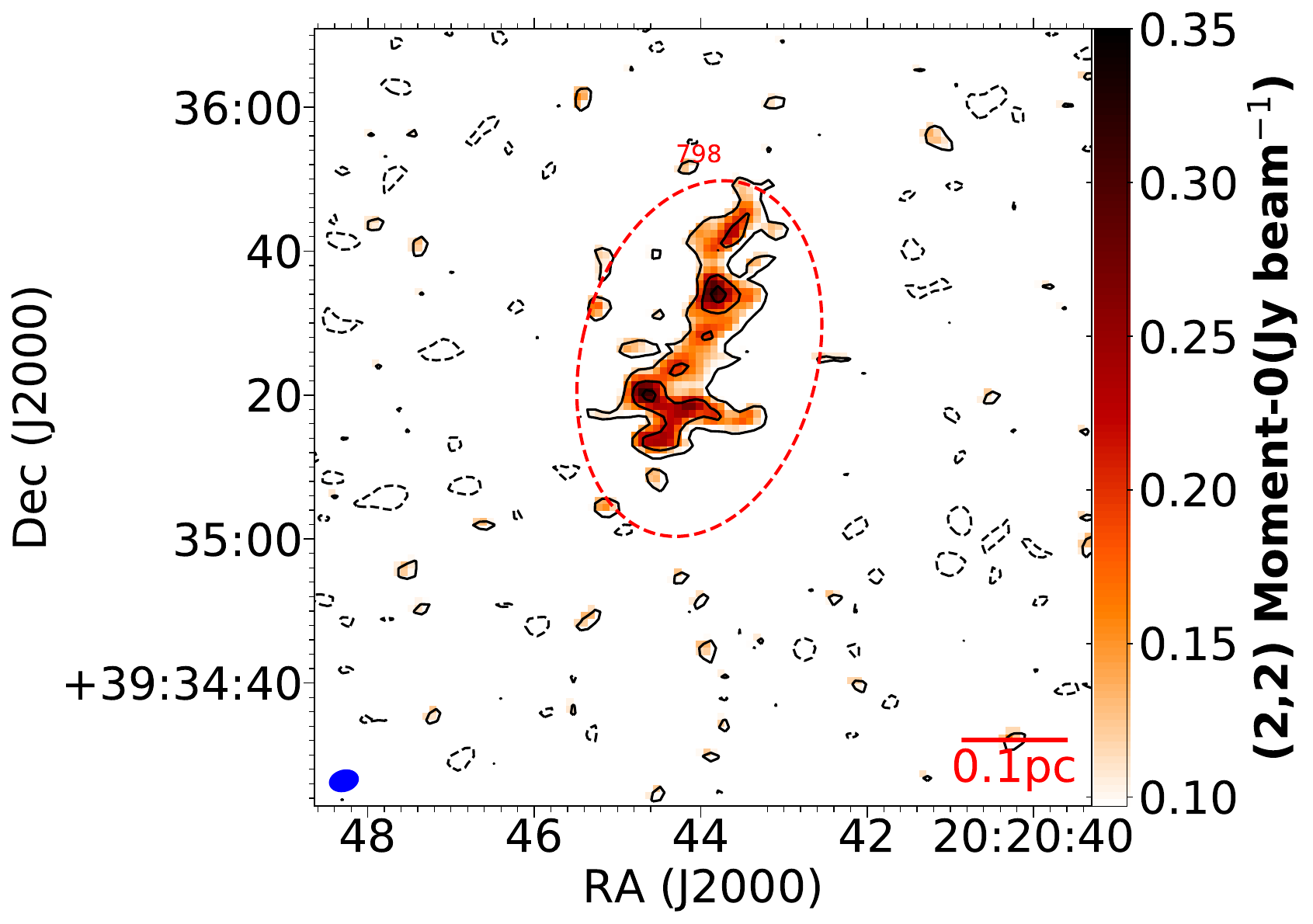} &
 \\
 & Field 18 & \\
\includegraphics[width=.3\textwidth]{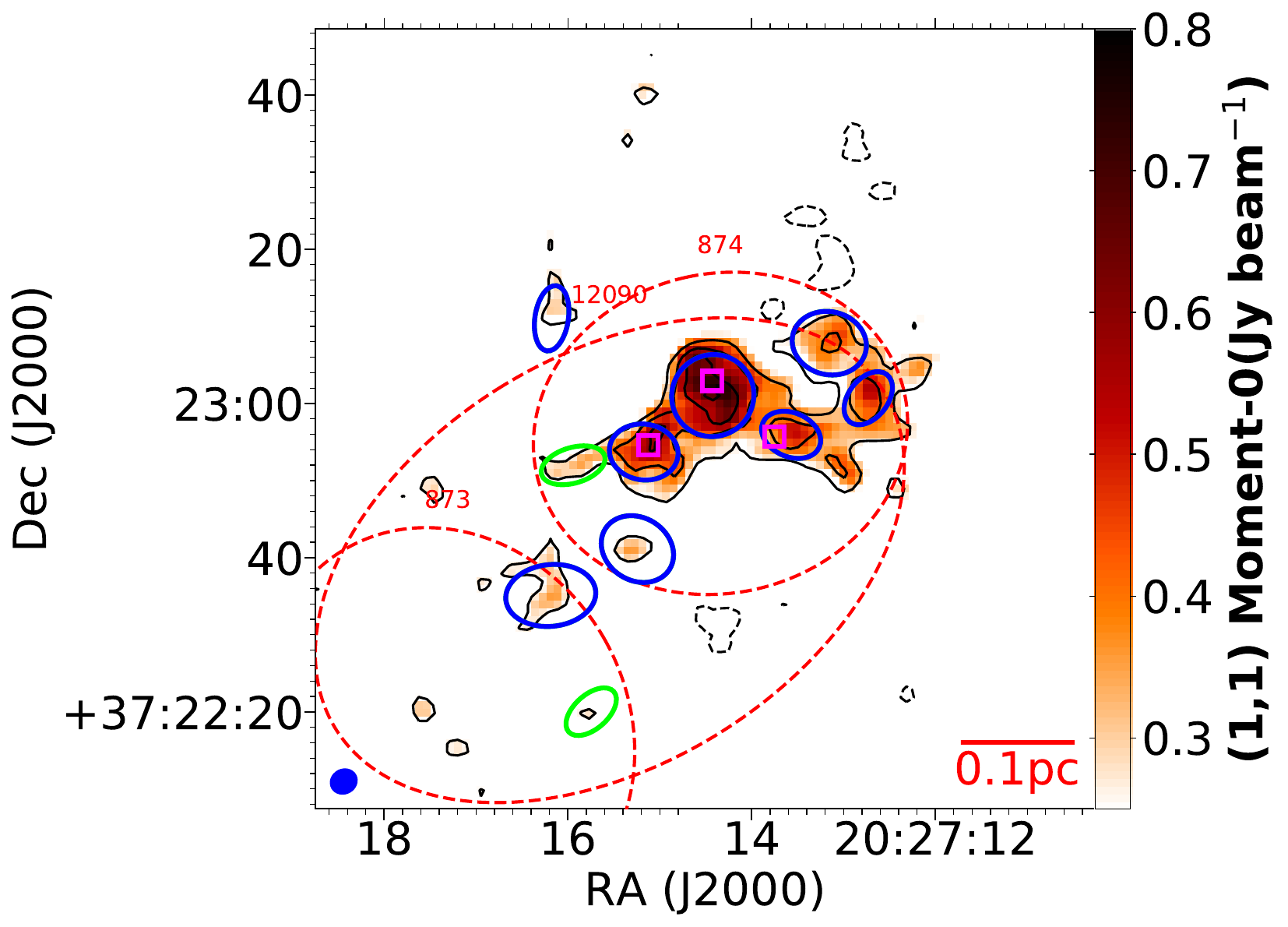} &
\includegraphics[width=.3\textwidth]{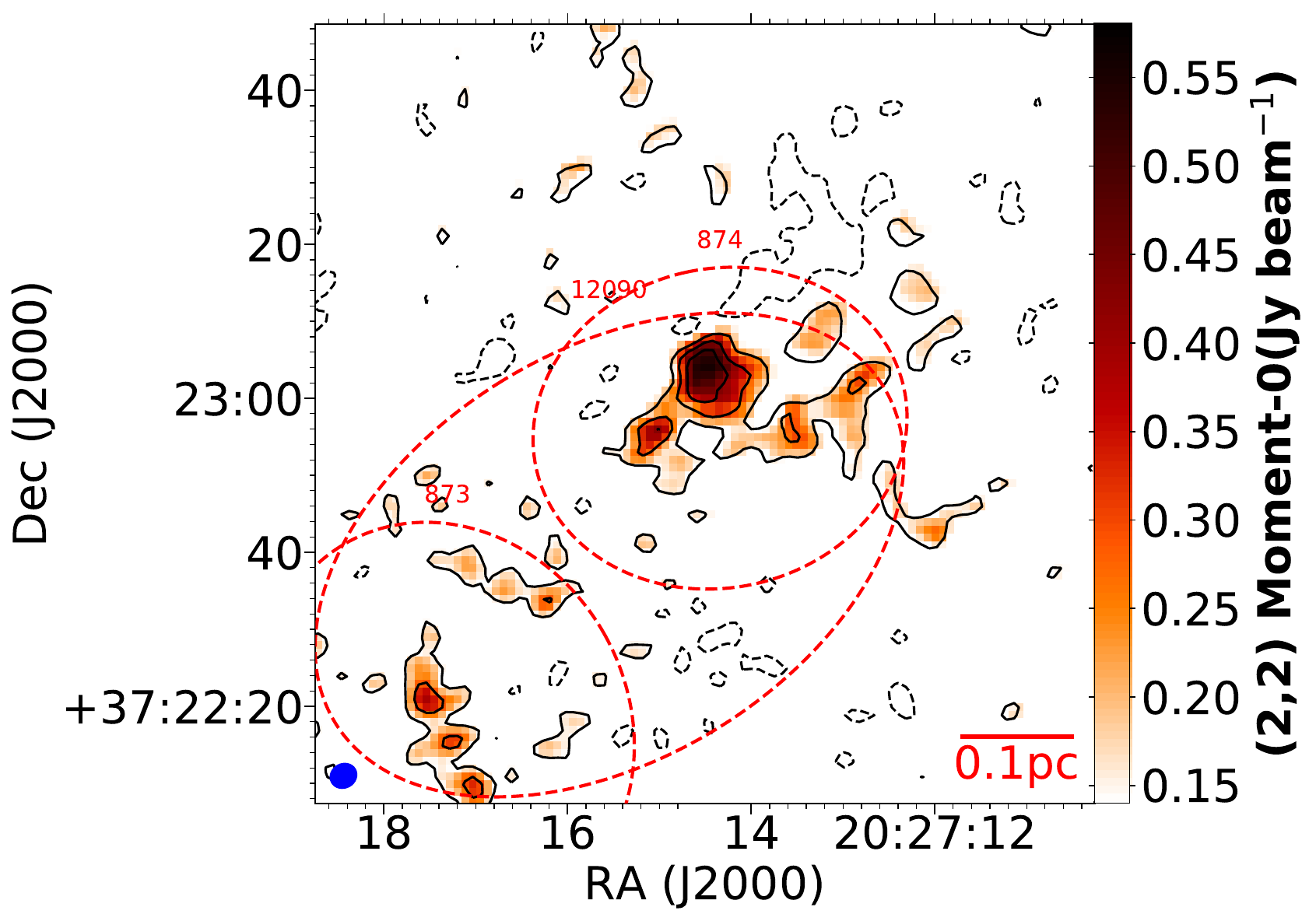} &
 \\
 & Field 19 & \\
\includegraphics[width=.3\textwidth]{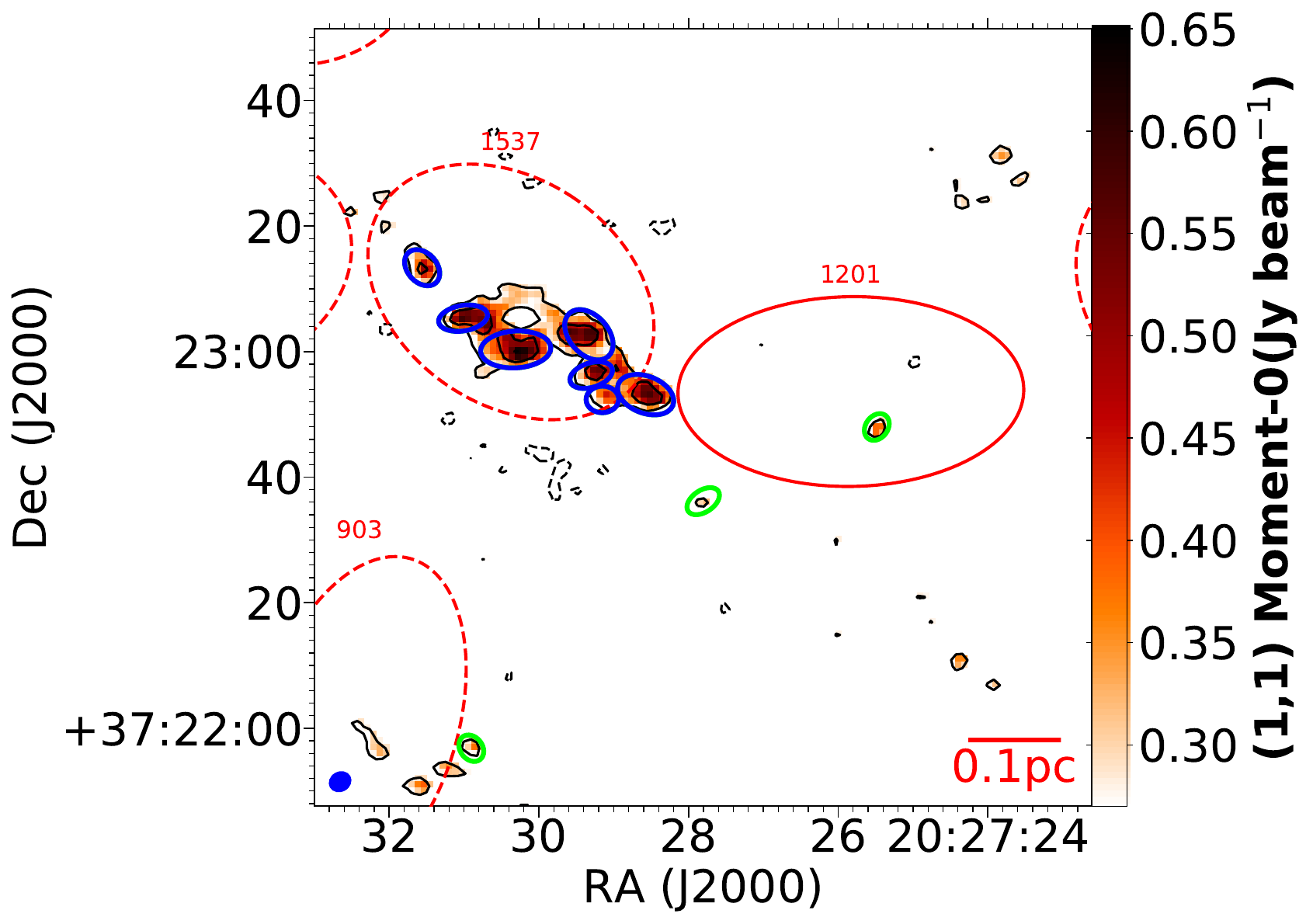} &
\includegraphics[width=.3\textwidth]{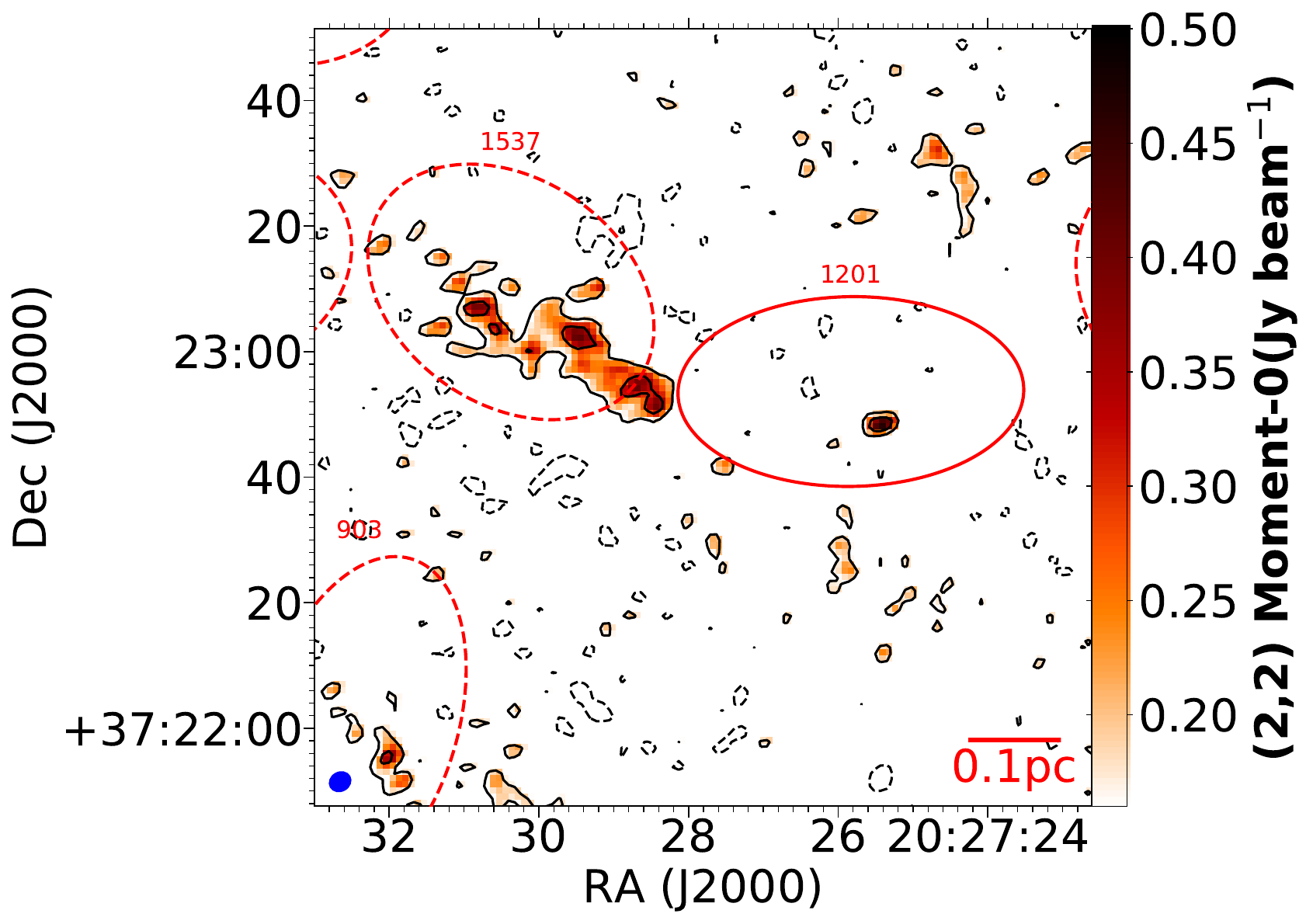} &
 \\
 & Field 20 & \\
\includegraphics[width=.3\textwidth]{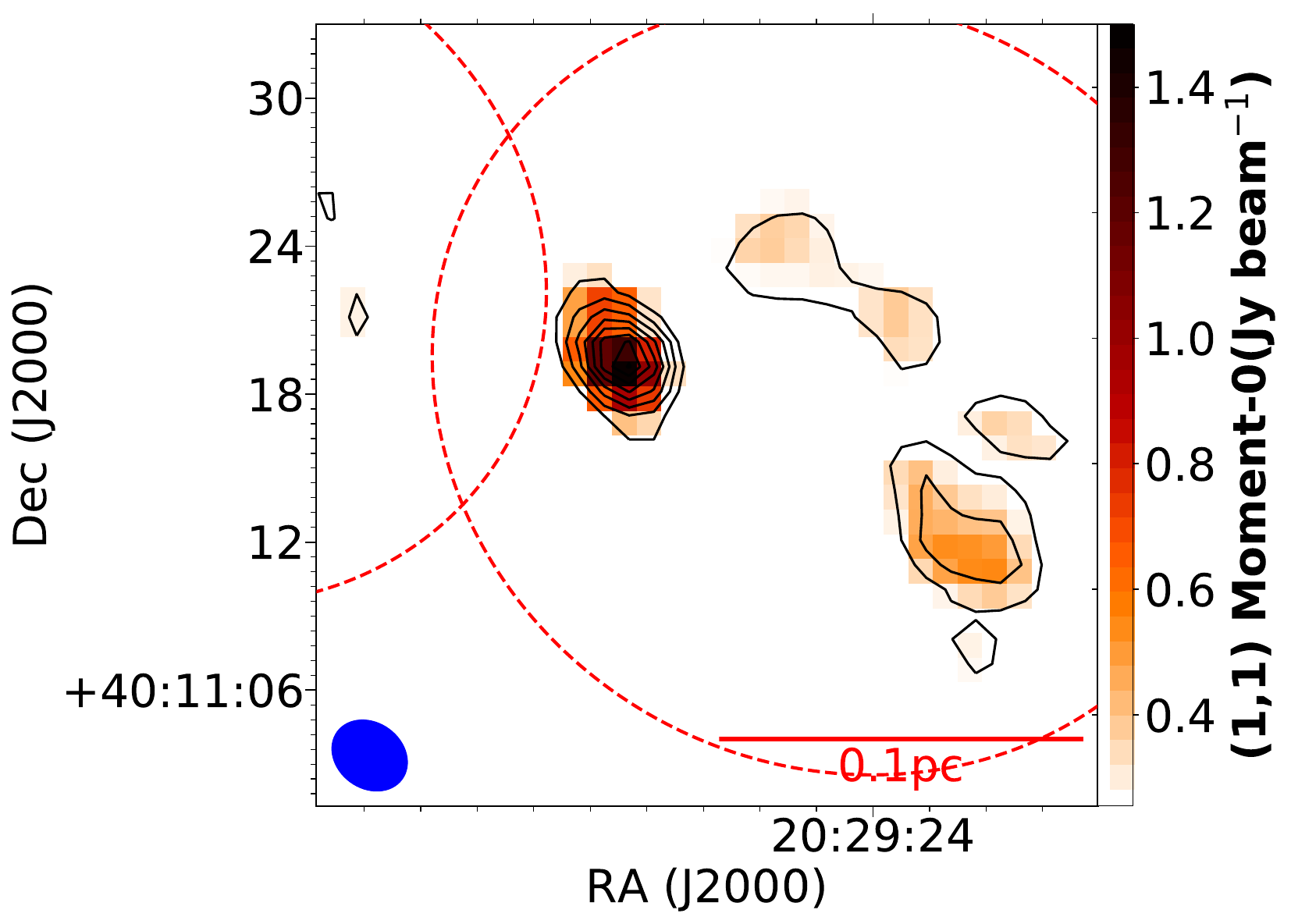} &
\includegraphics[width=.3\textwidth]{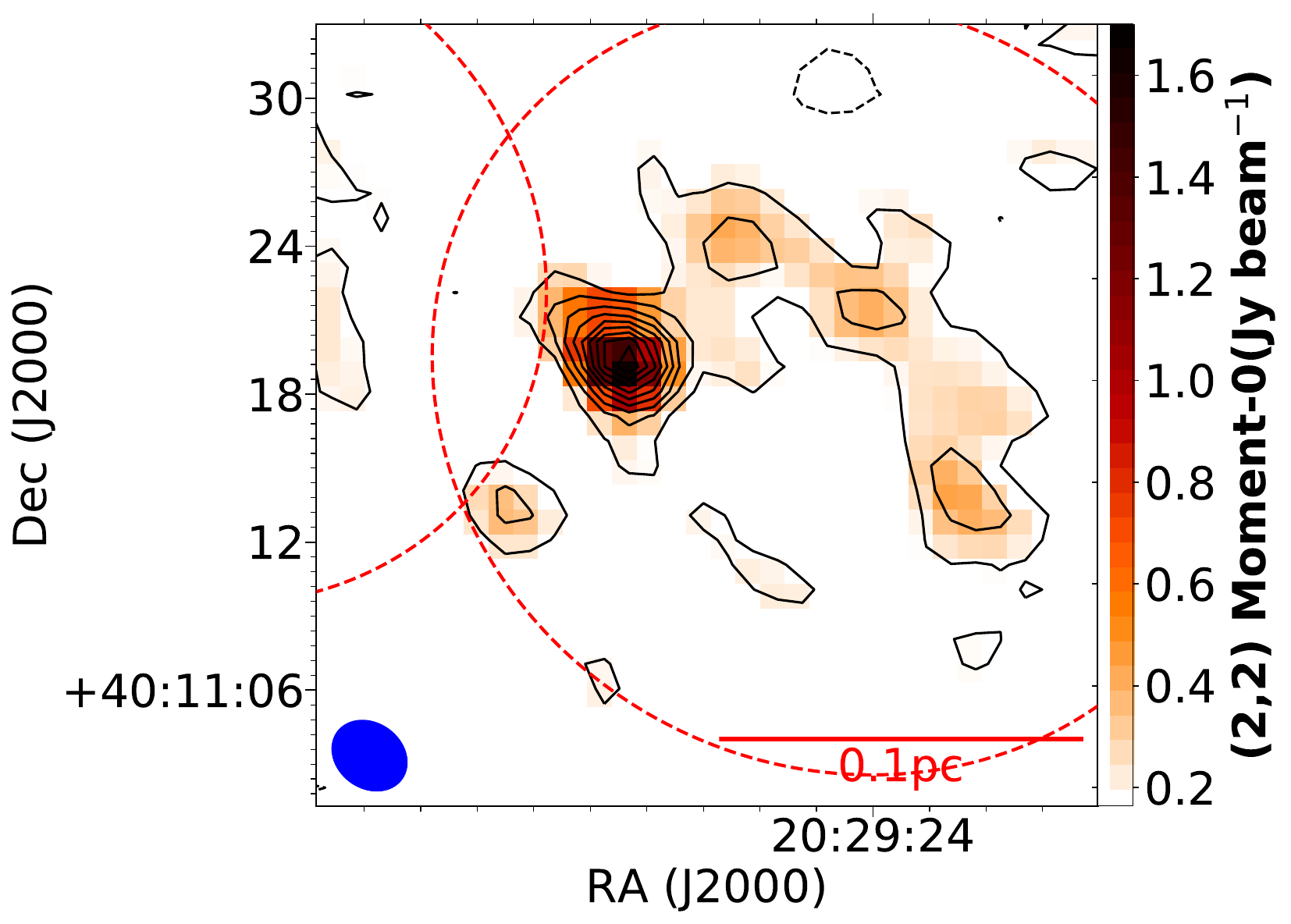} &
\includegraphics[width=.3\textwidth]{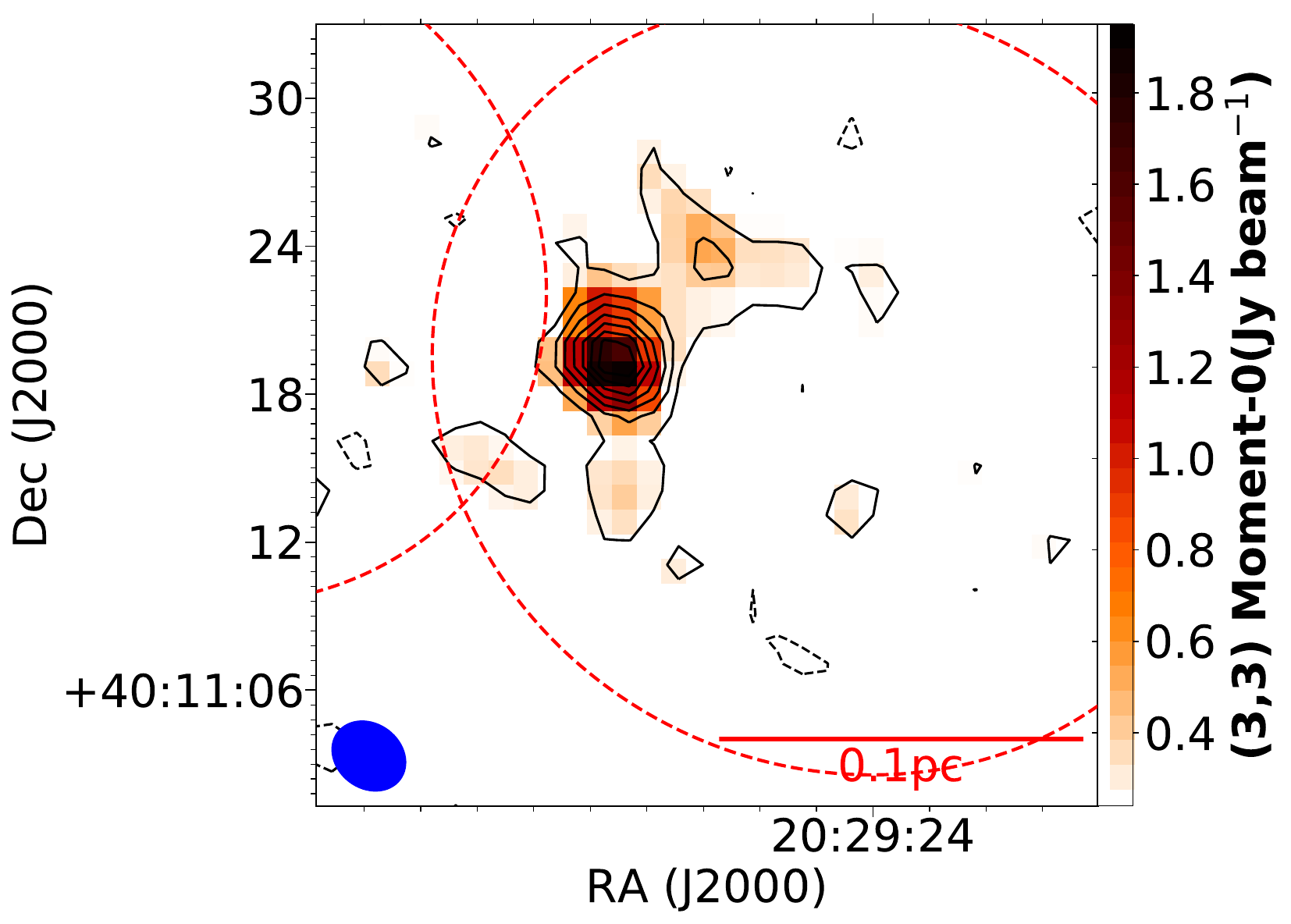} \\
\includegraphics[width=.3\textwidth]{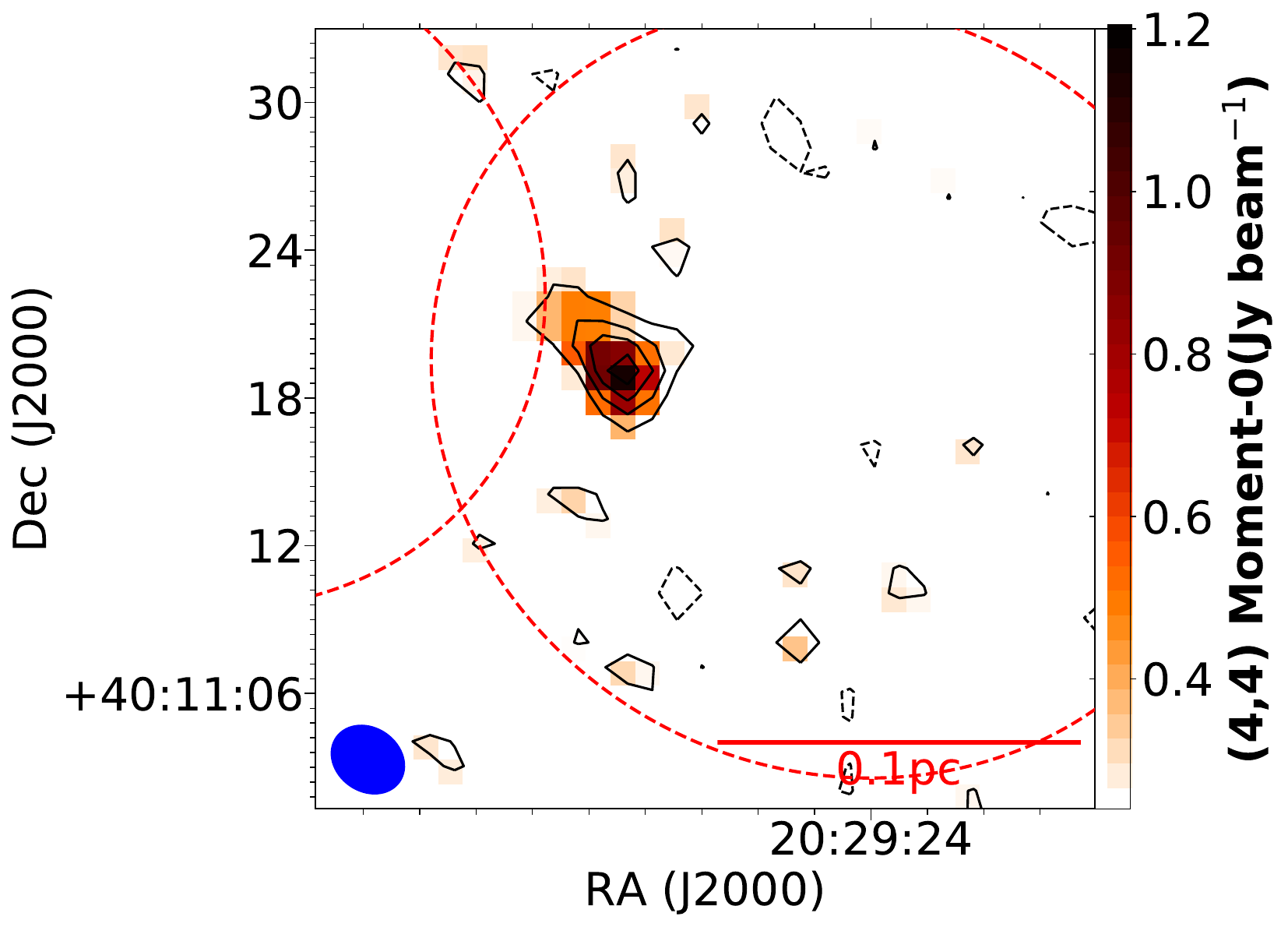} &
\includegraphics[width=.3\textwidth]{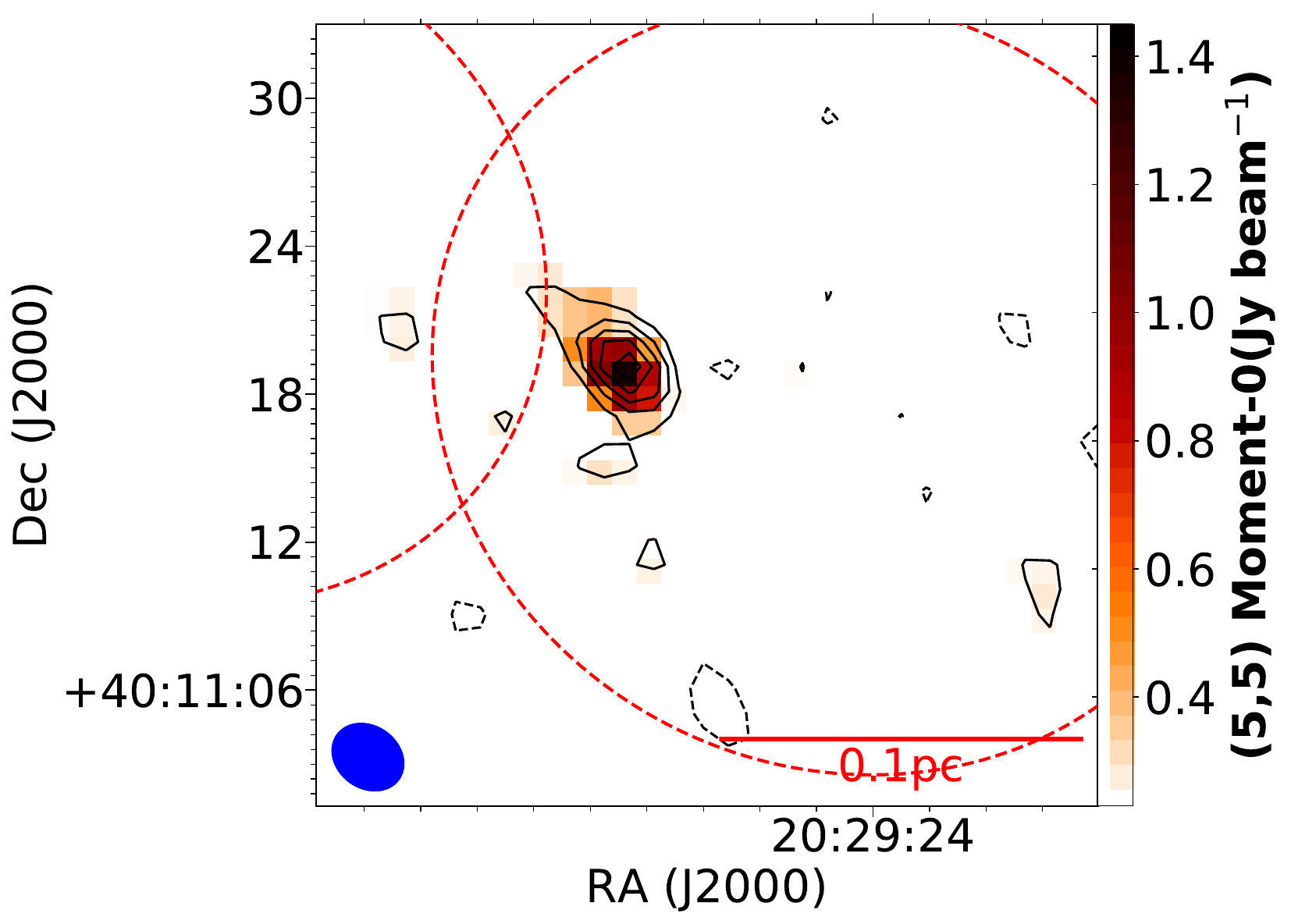} &
 \\
 & Field 21 & \\
\includegraphics[width=.3\textwidth]{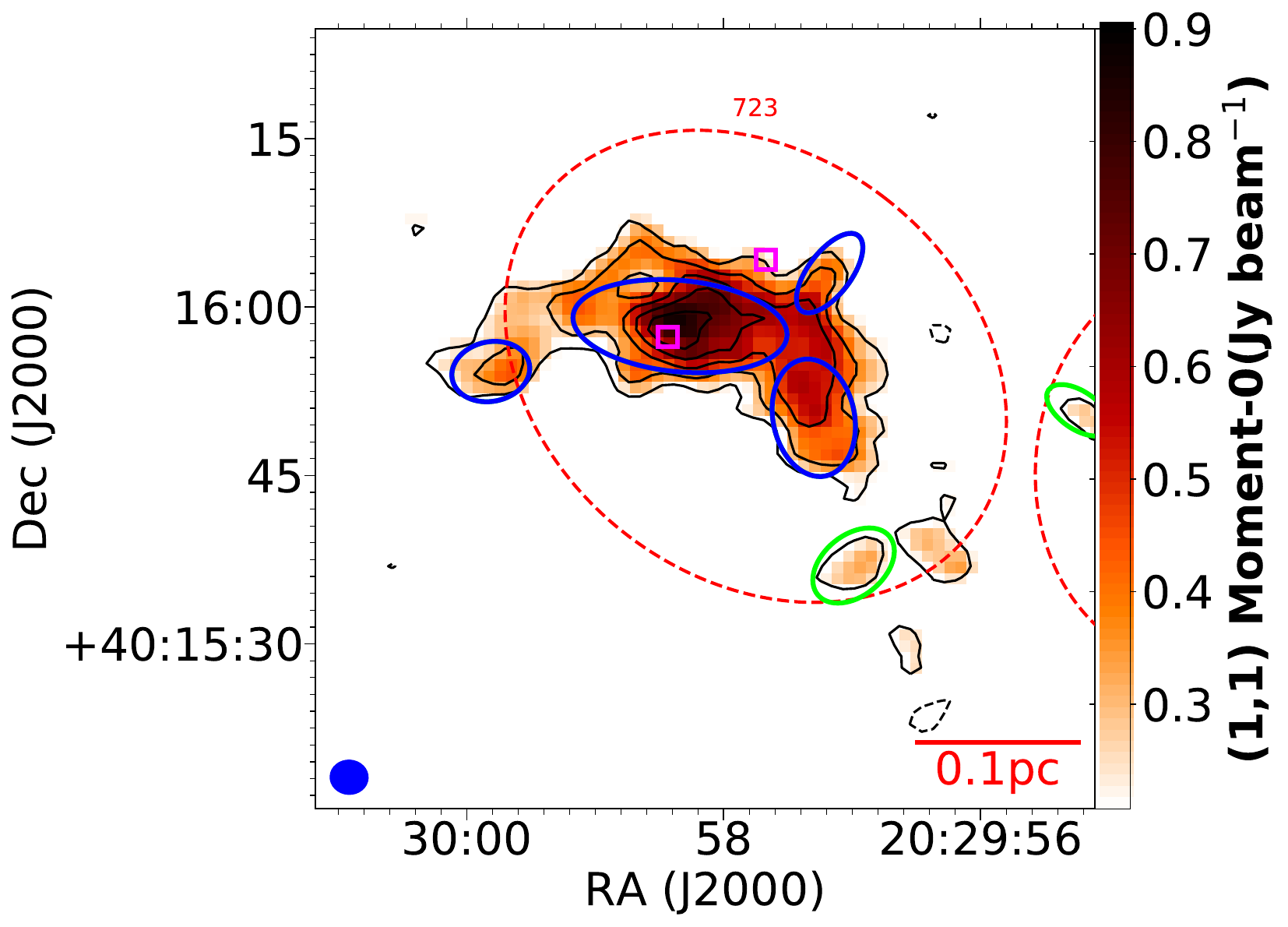} &
\includegraphics[width=.3\textwidth]{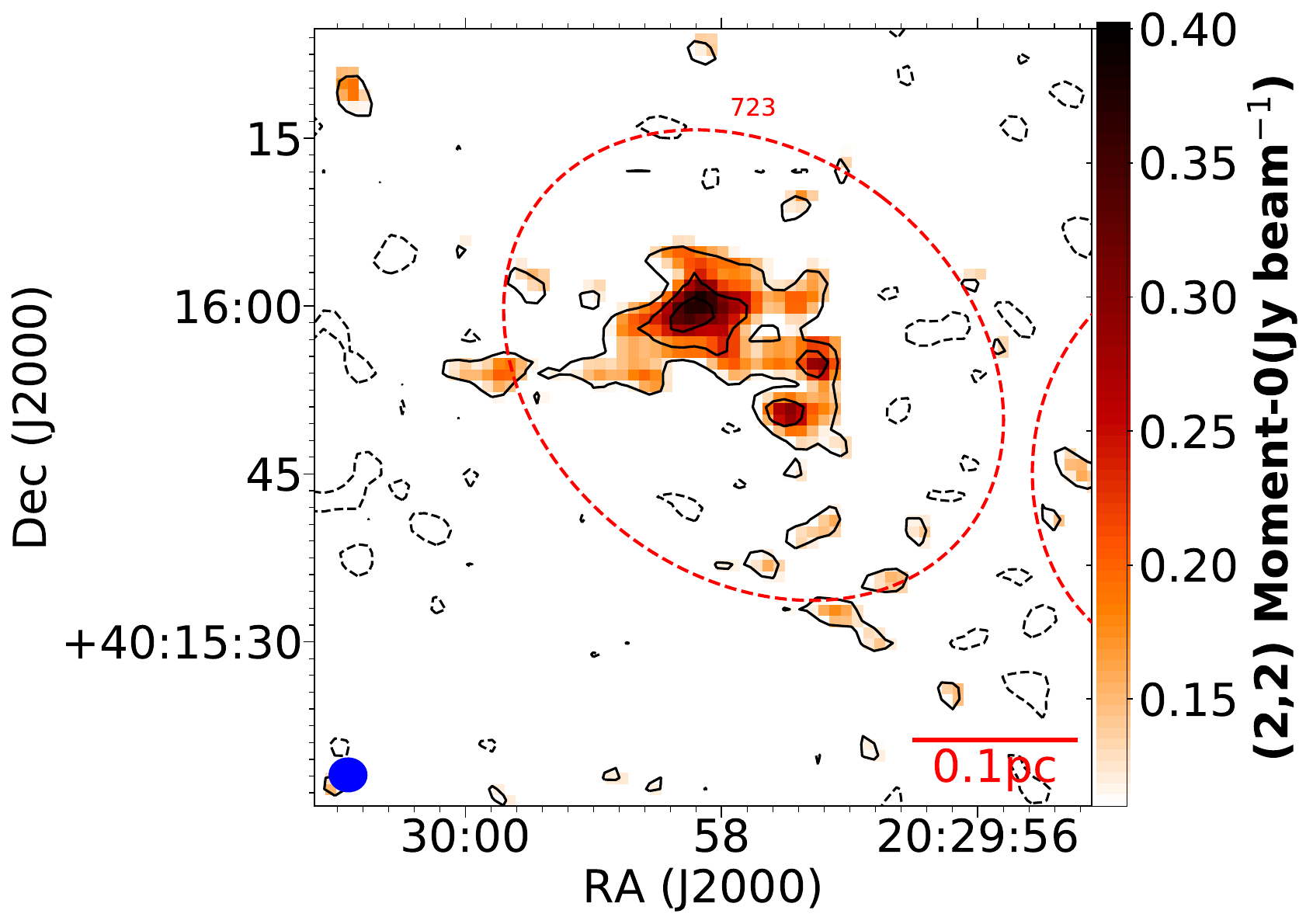} &
 \\
 & Field 22 & \\
\includegraphics[width=.3\textwidth]{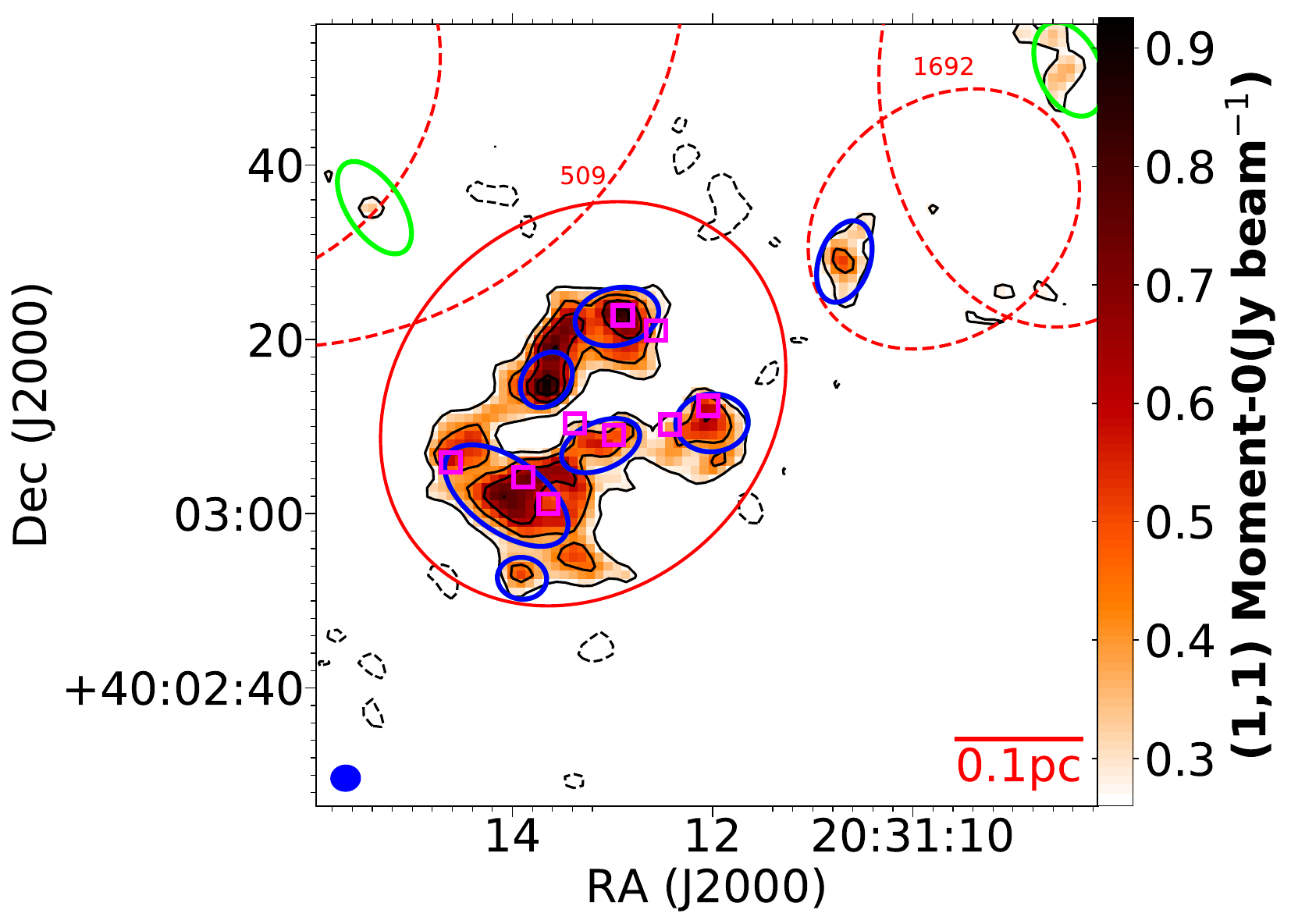} &
\includegraphics[width=.3\textwidth]{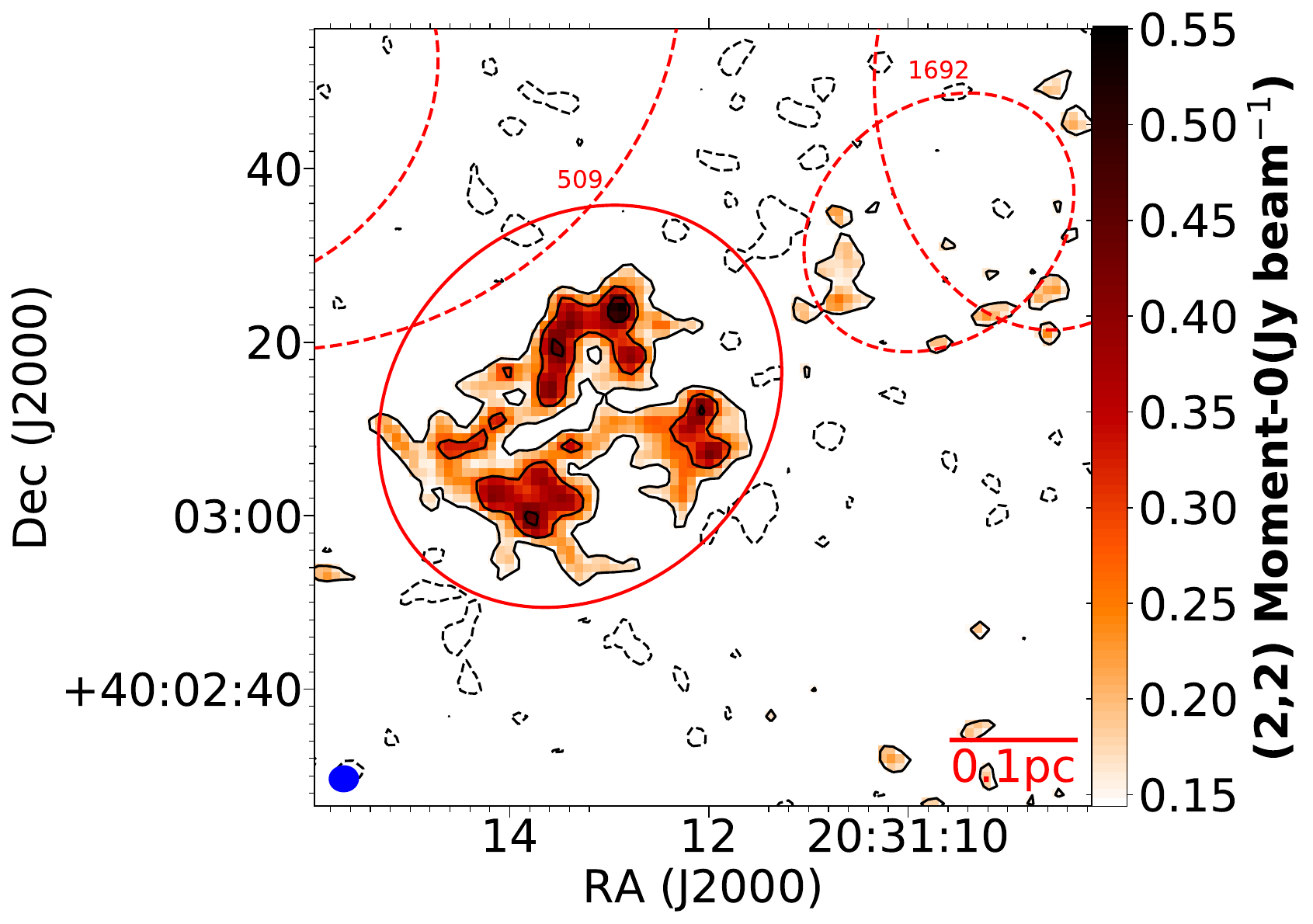} &
\includegraphics[width=.3\textwidth]{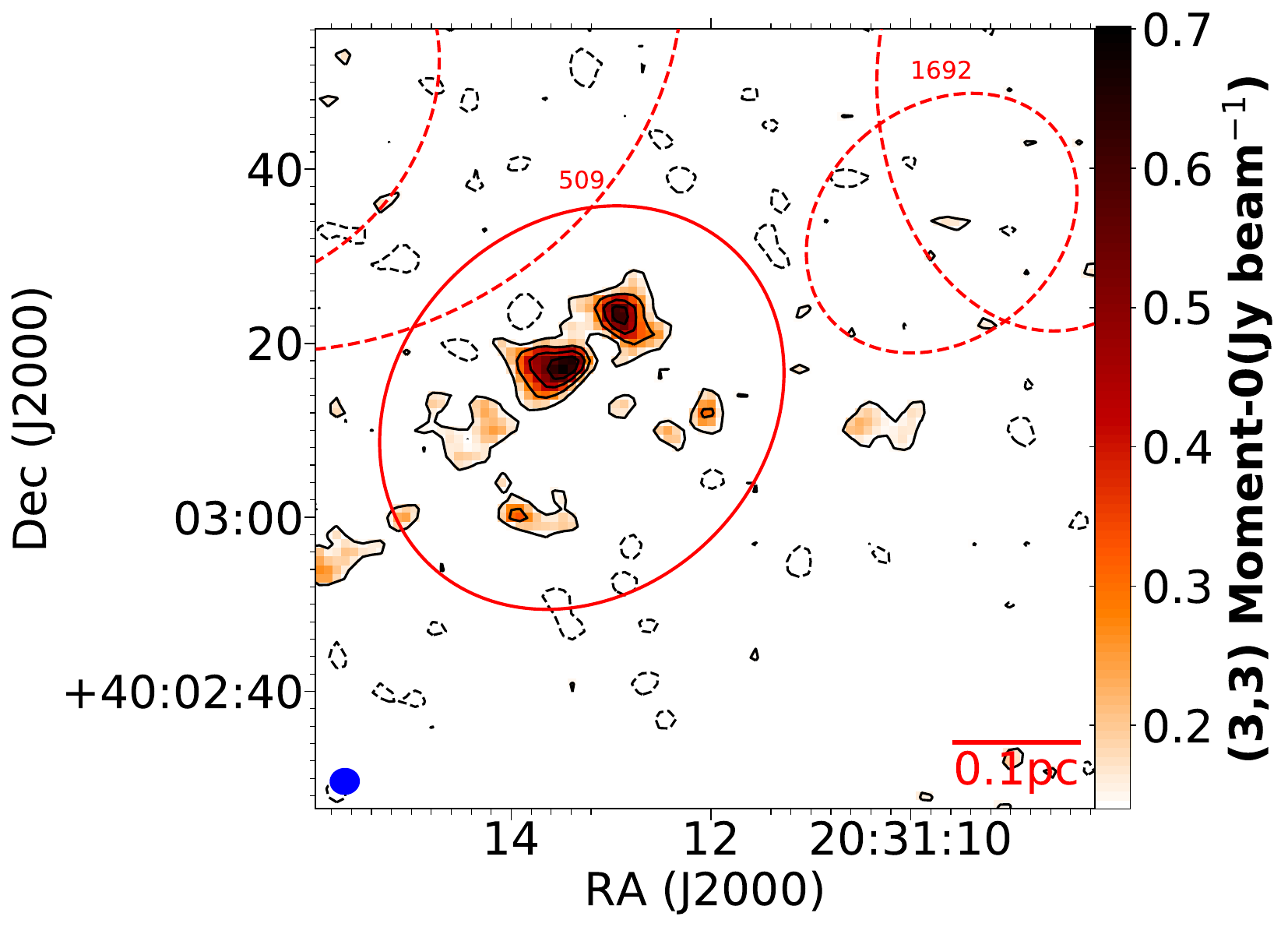} \\
 & Field 23 & \\
\includegraphics[width=.3\textwidth]{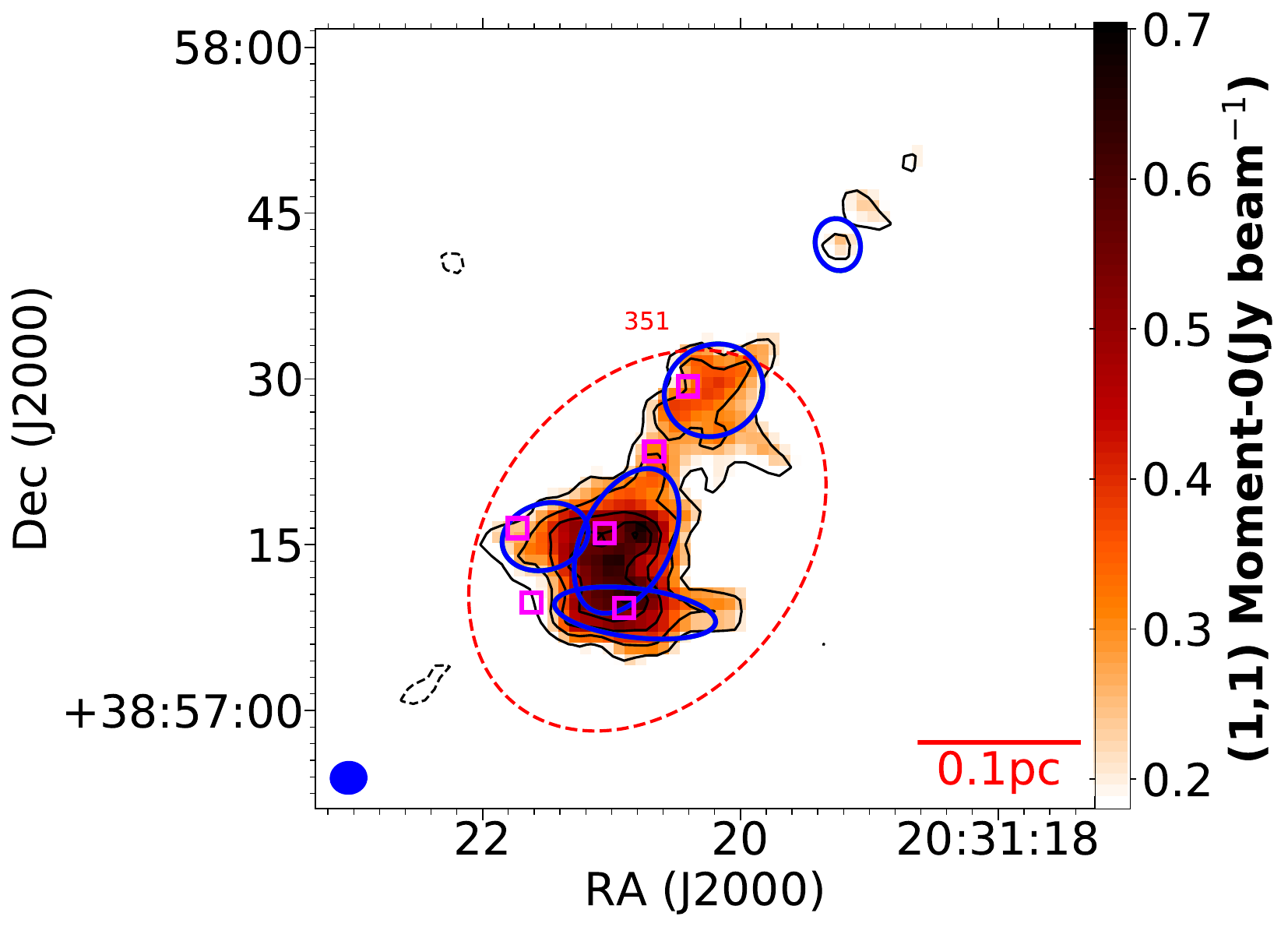} &
\includegraphics[width=.3\textwidth]{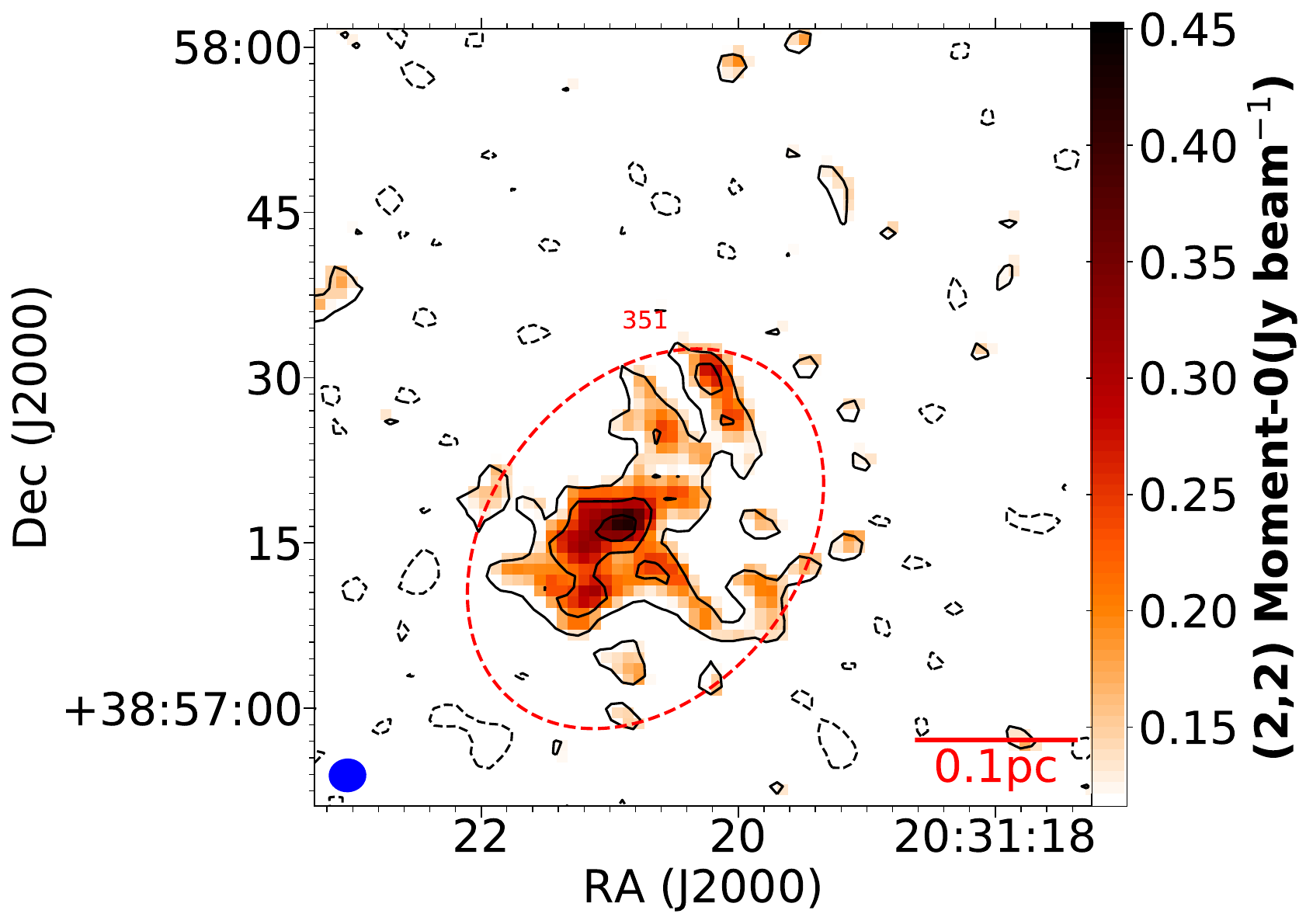} &
\includegraphics[width=.3\textwidth]{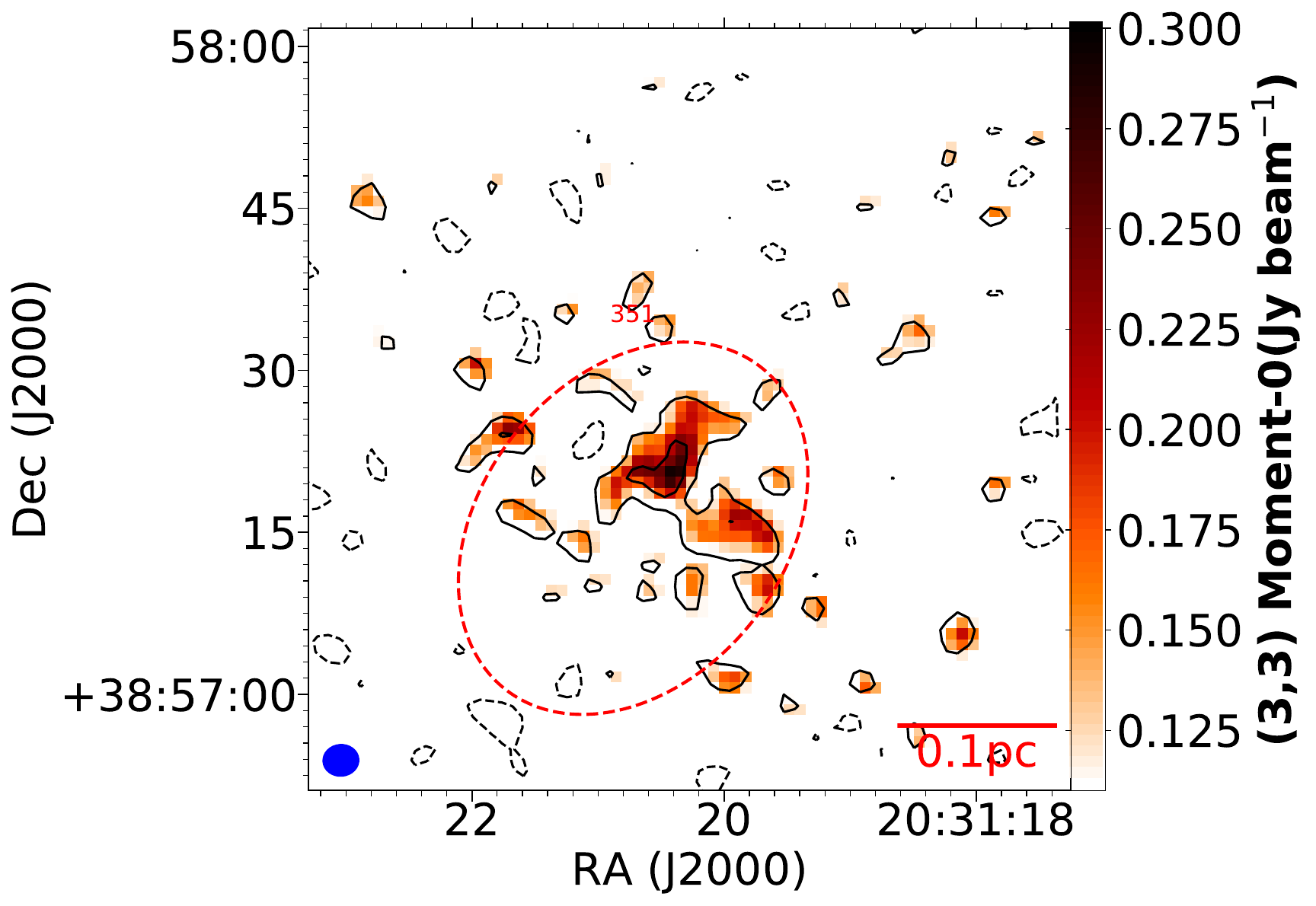} \\
 & Field 24 & \\
\includegraphics[width=.3\textwidth]{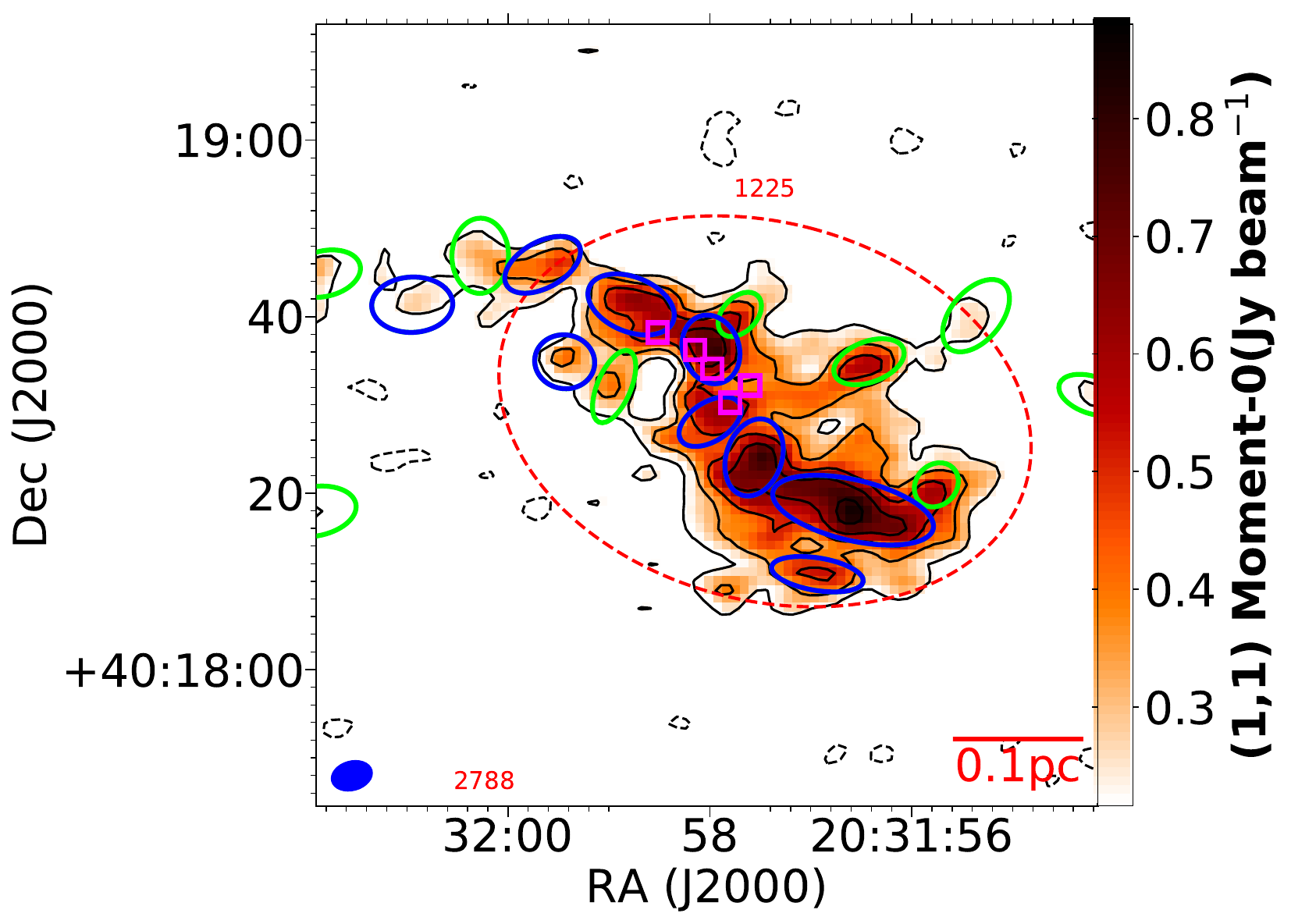} &
\includegraphics[width=.3\textwidth]{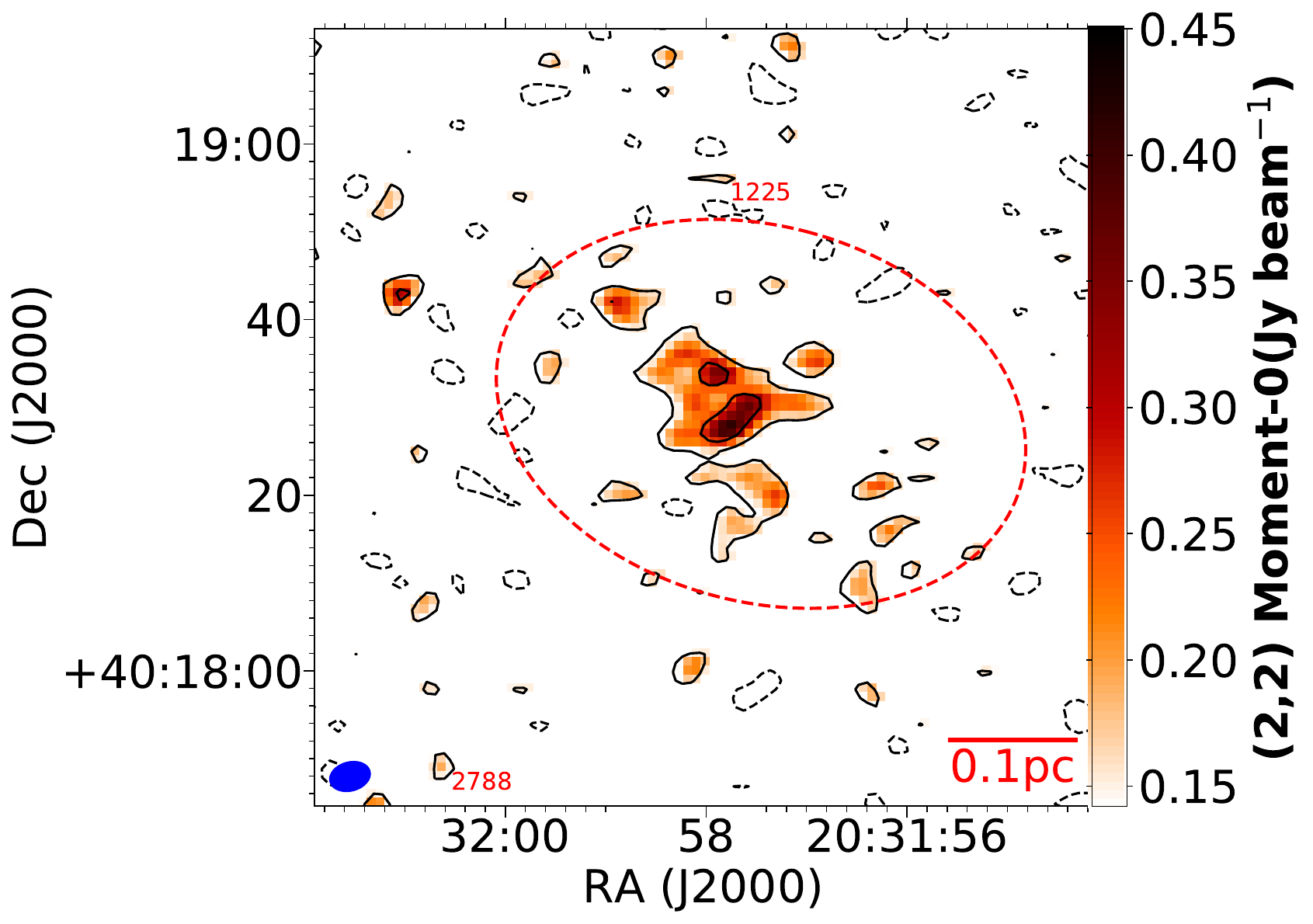} &
 \\
 & Field 25 & \\
\includegraphics[width=.3\textwidth]{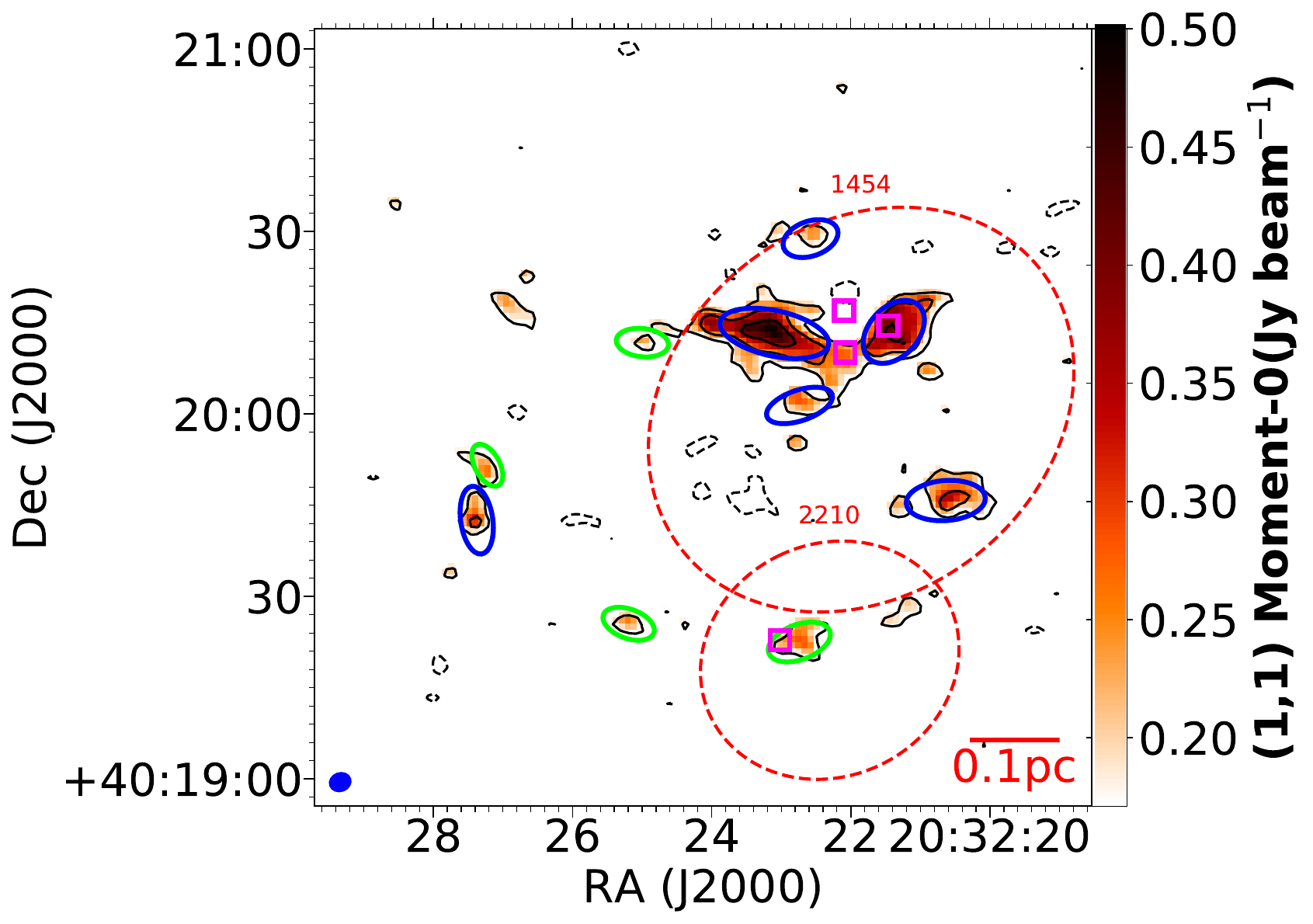} &
\includegraphics[width=.3\textwidth]{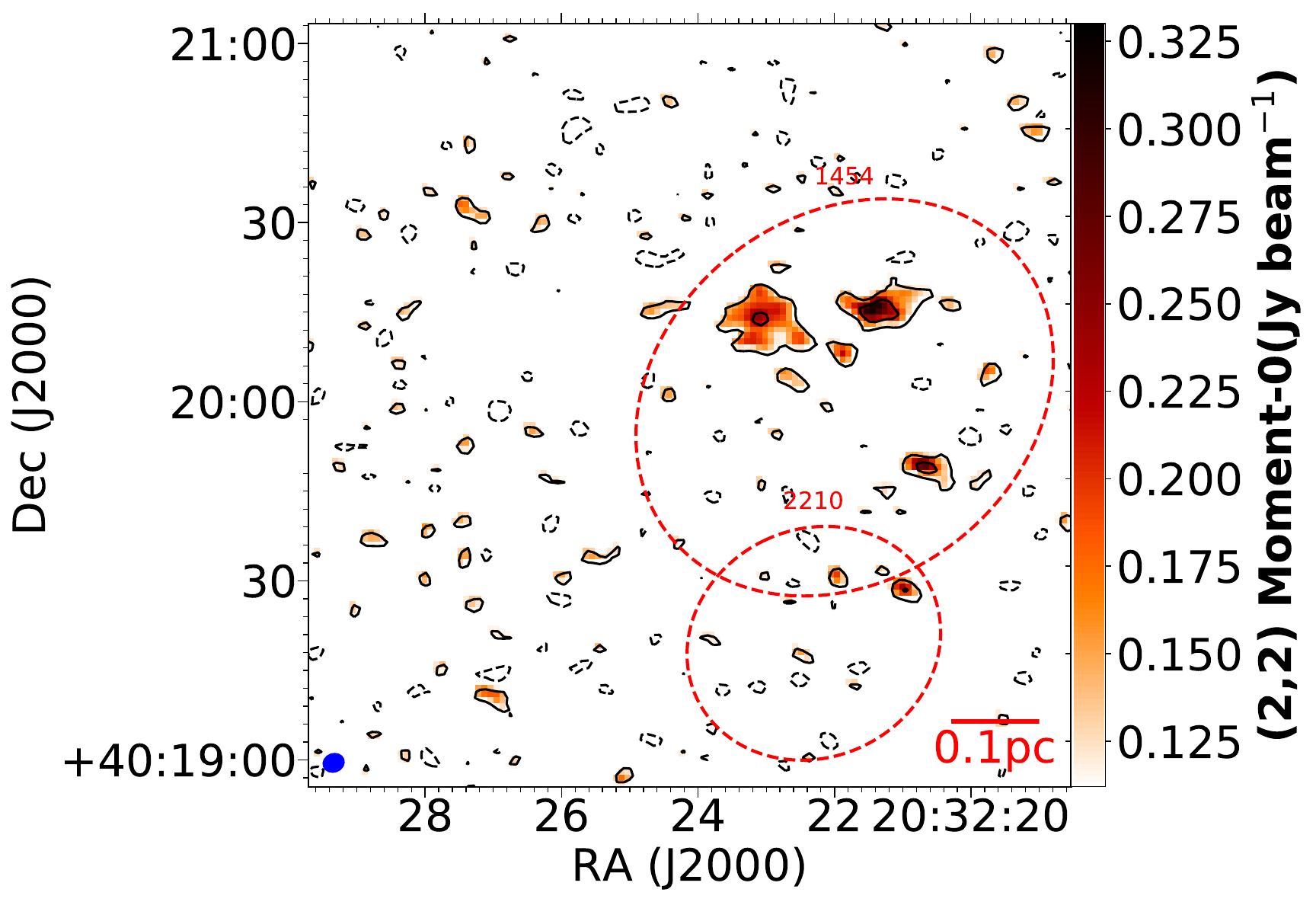} &
 \\
 & Field 26 & \\
\includegraphics[width=.3\textwidth]{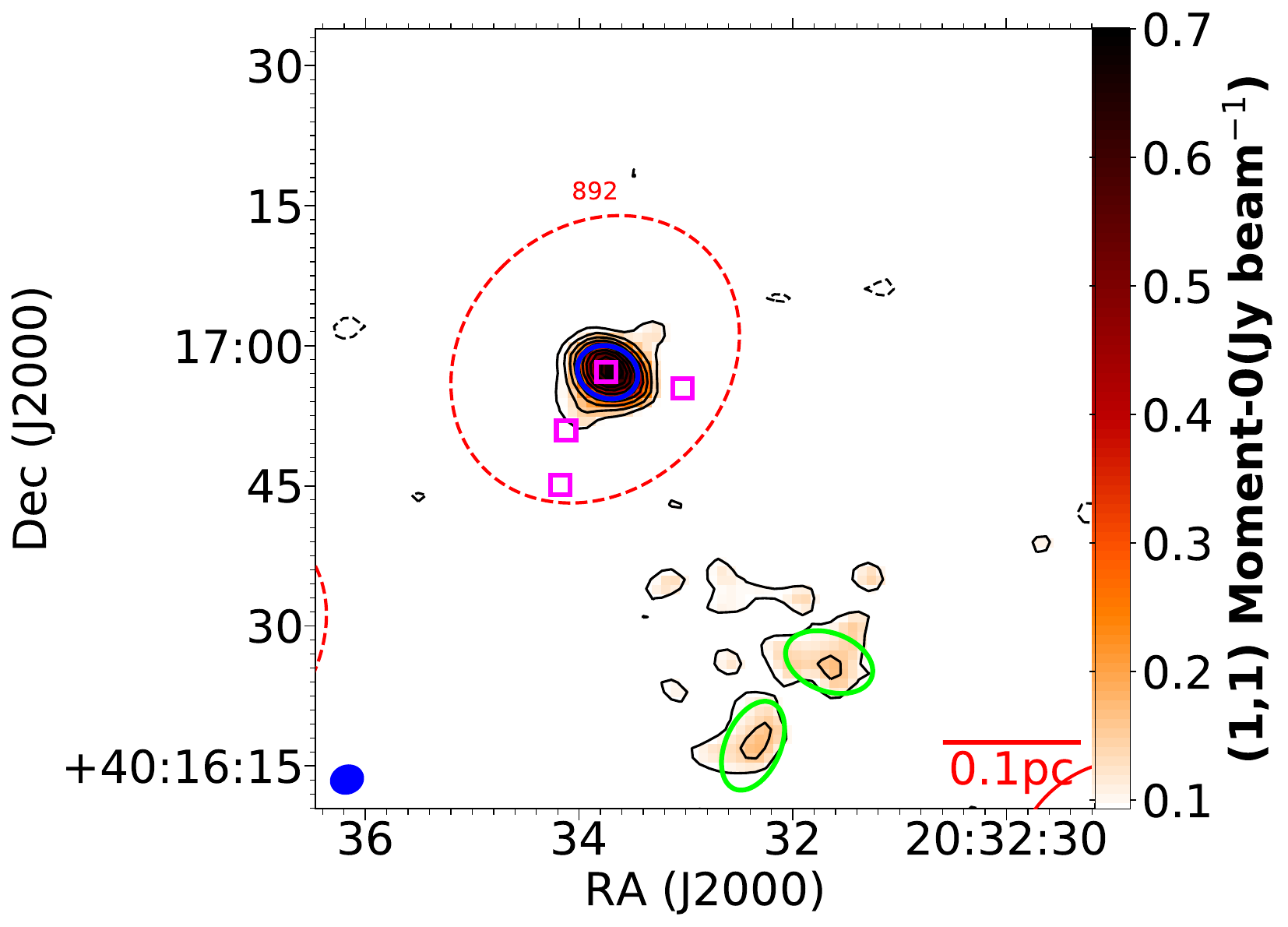} &
\includegraphics[width=.3\textwidth]{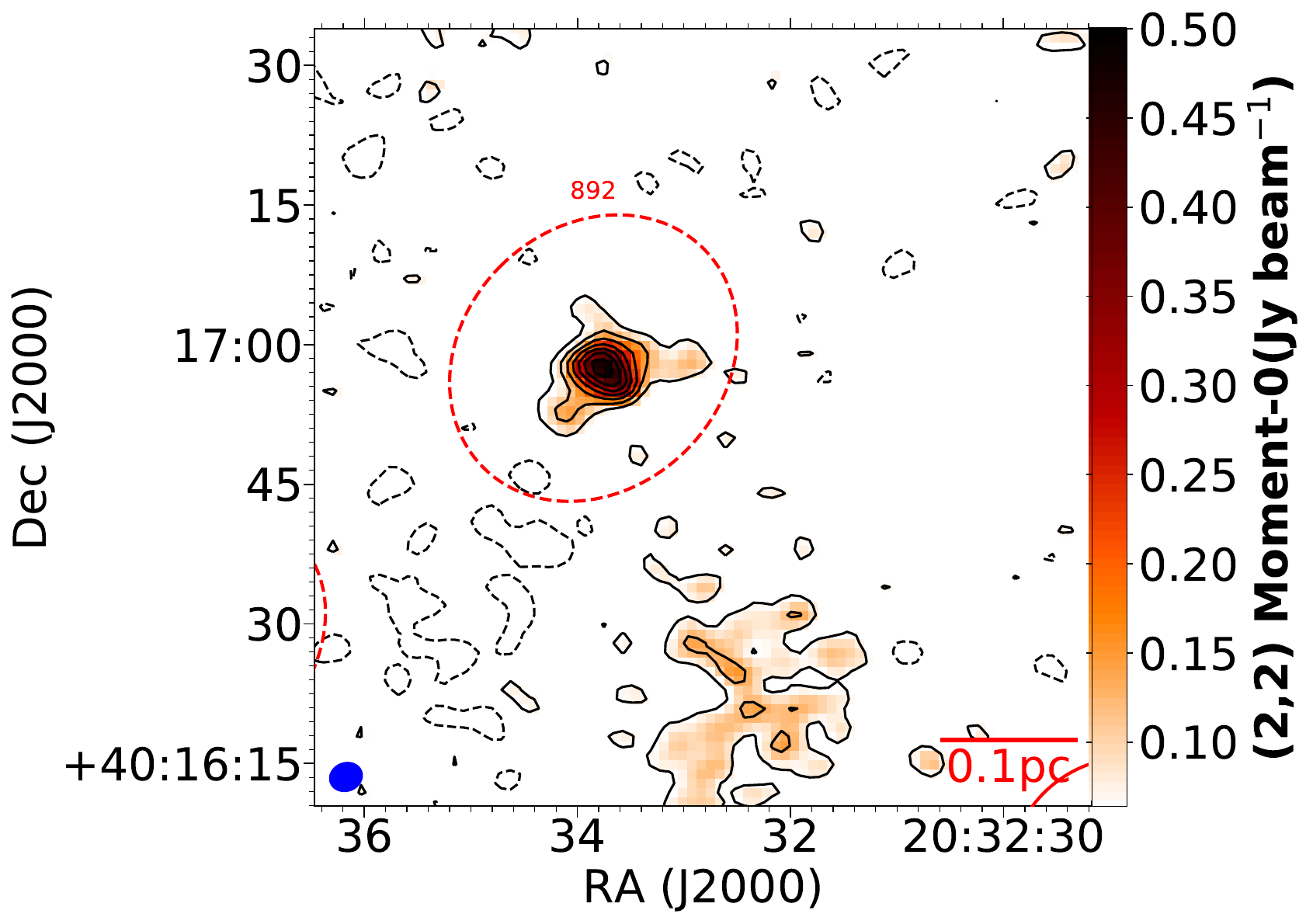} &
\includegraphics[width=.3\textwidth]{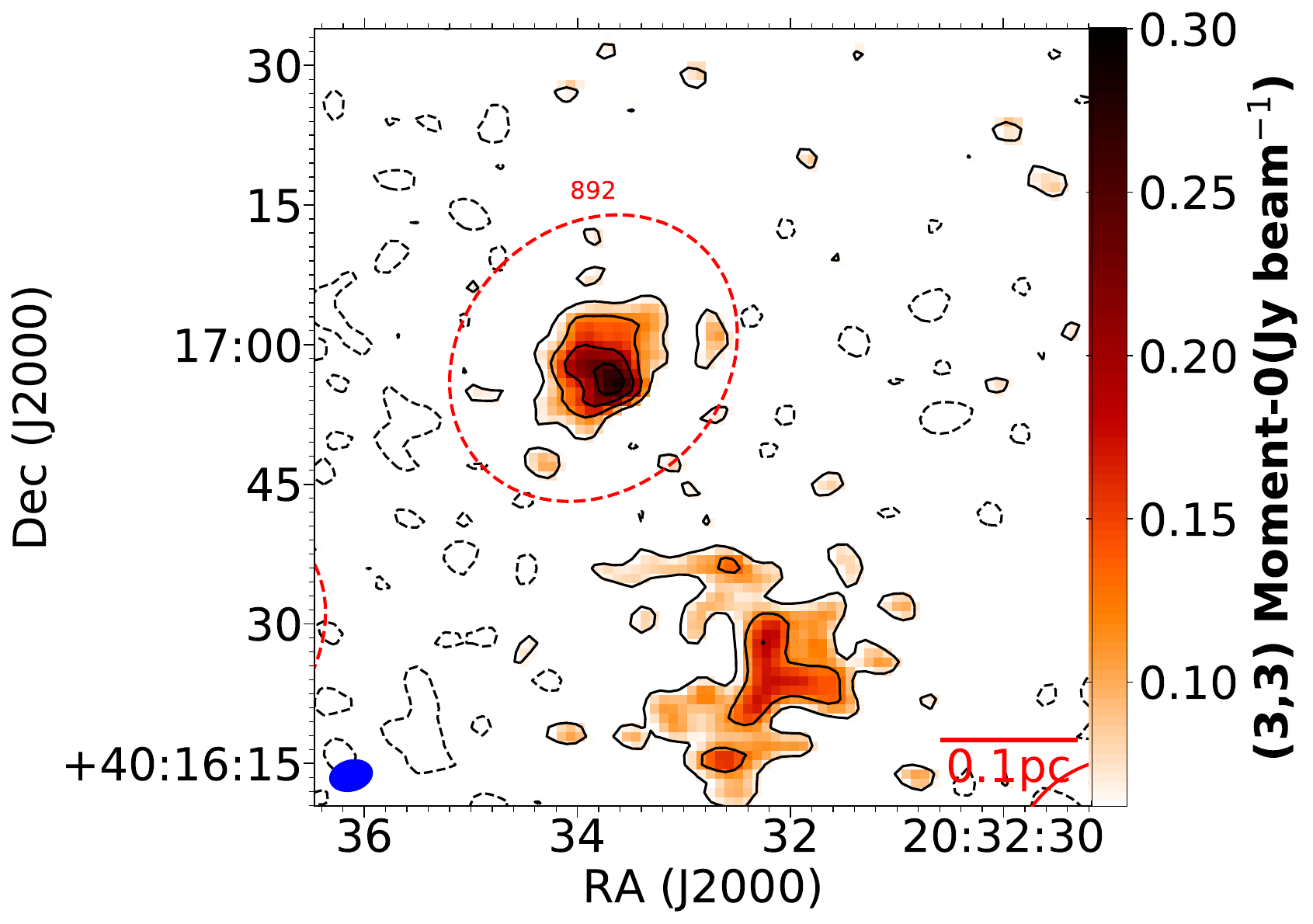} \\
 & Field 27 & \\
\includegraphics[width=.3\textwidth]{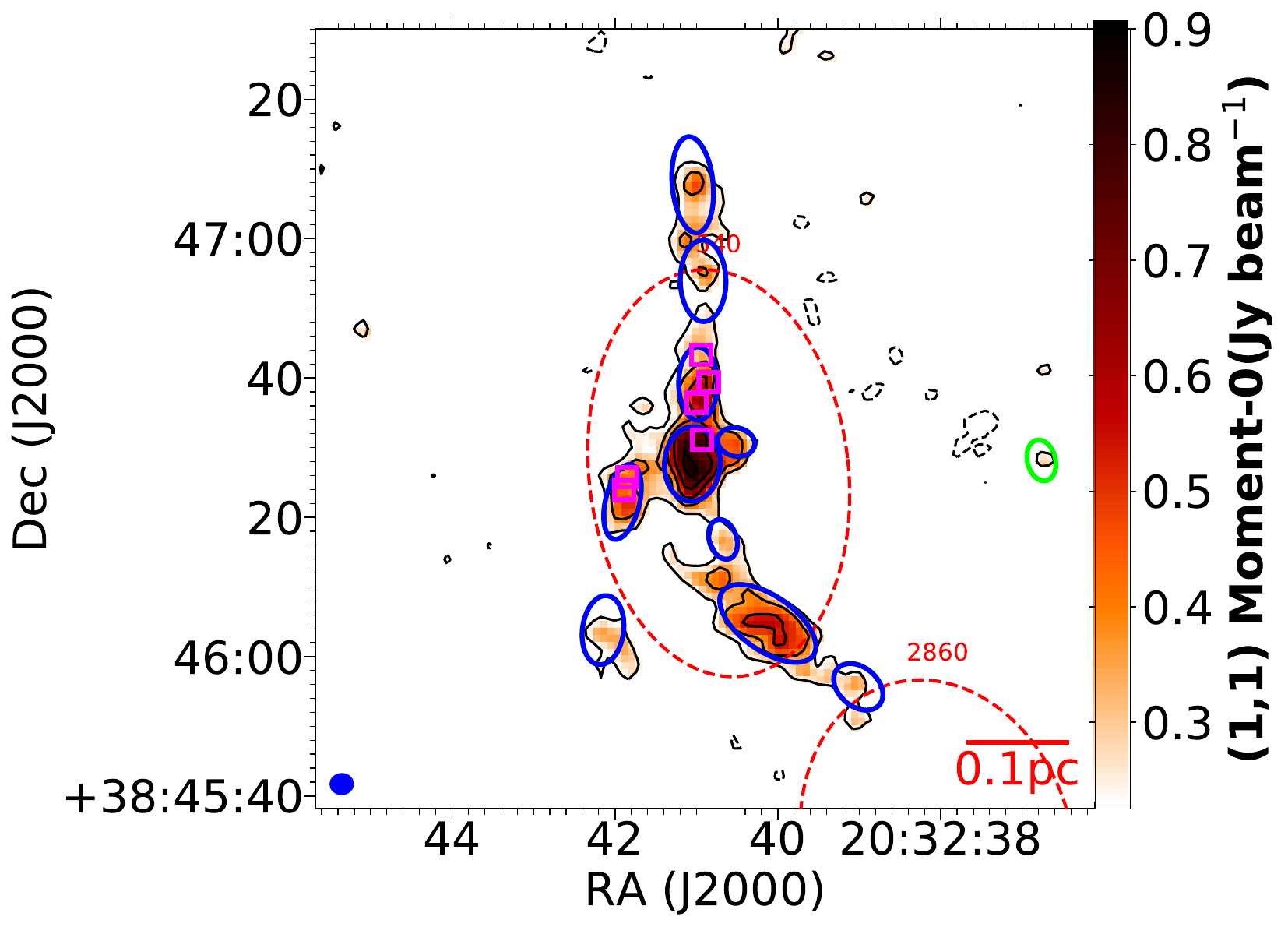} &
\includegraphics[width=.3\textwidth]{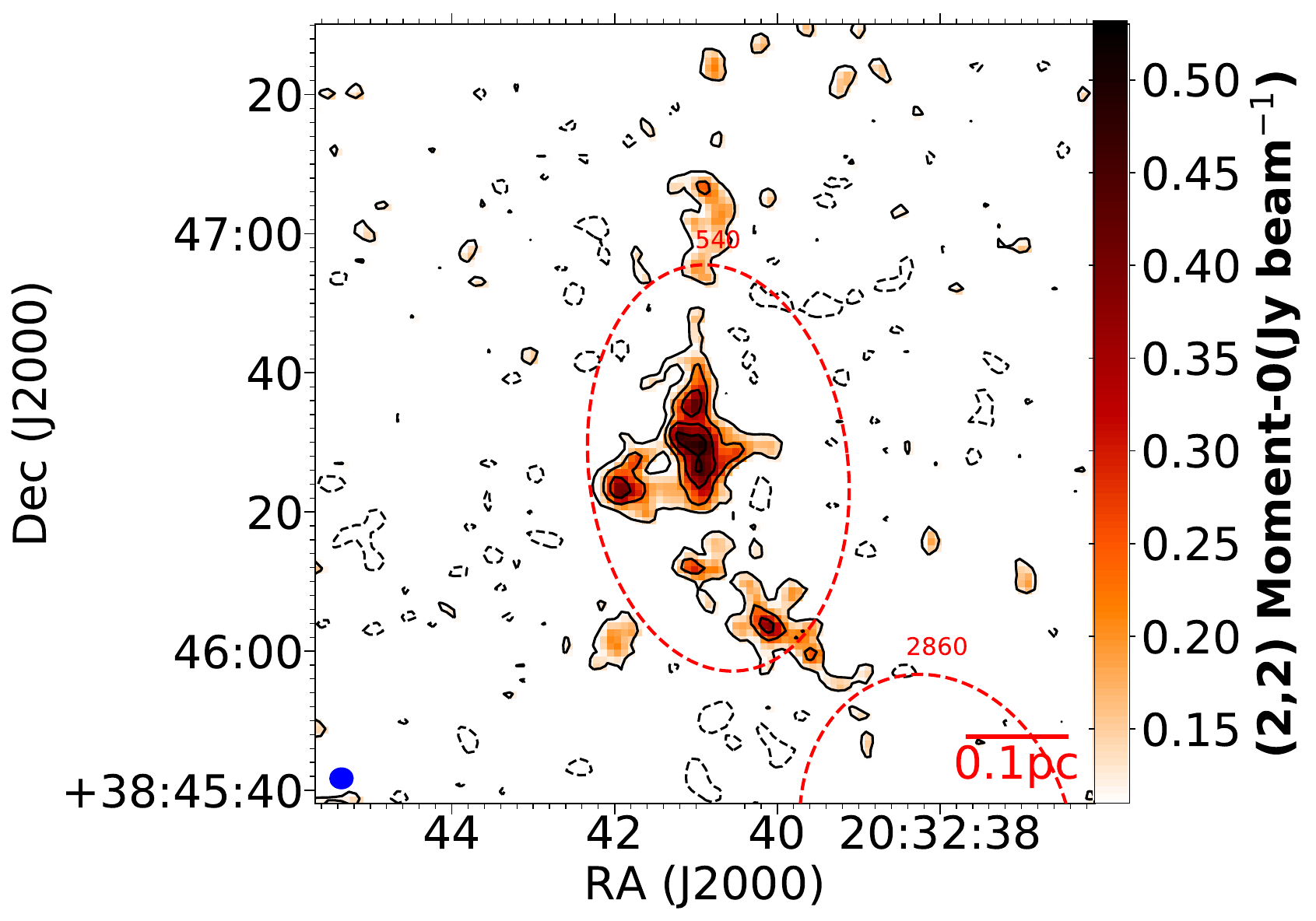} &
 \\
 & Field 28 & \\
\includegraphics[width=.3\textwidth]{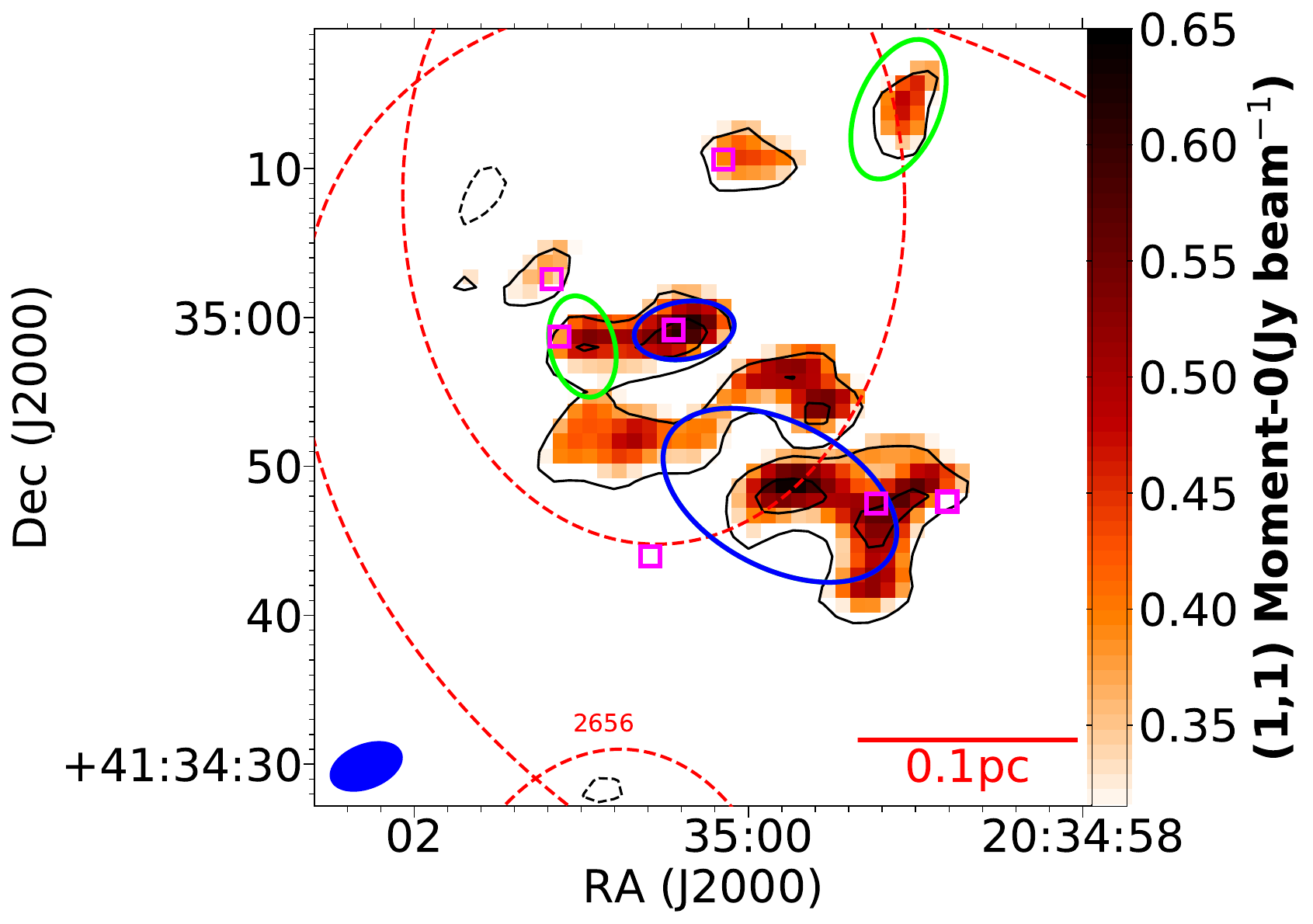} &
\includegraphics[width=.3\textwidth]{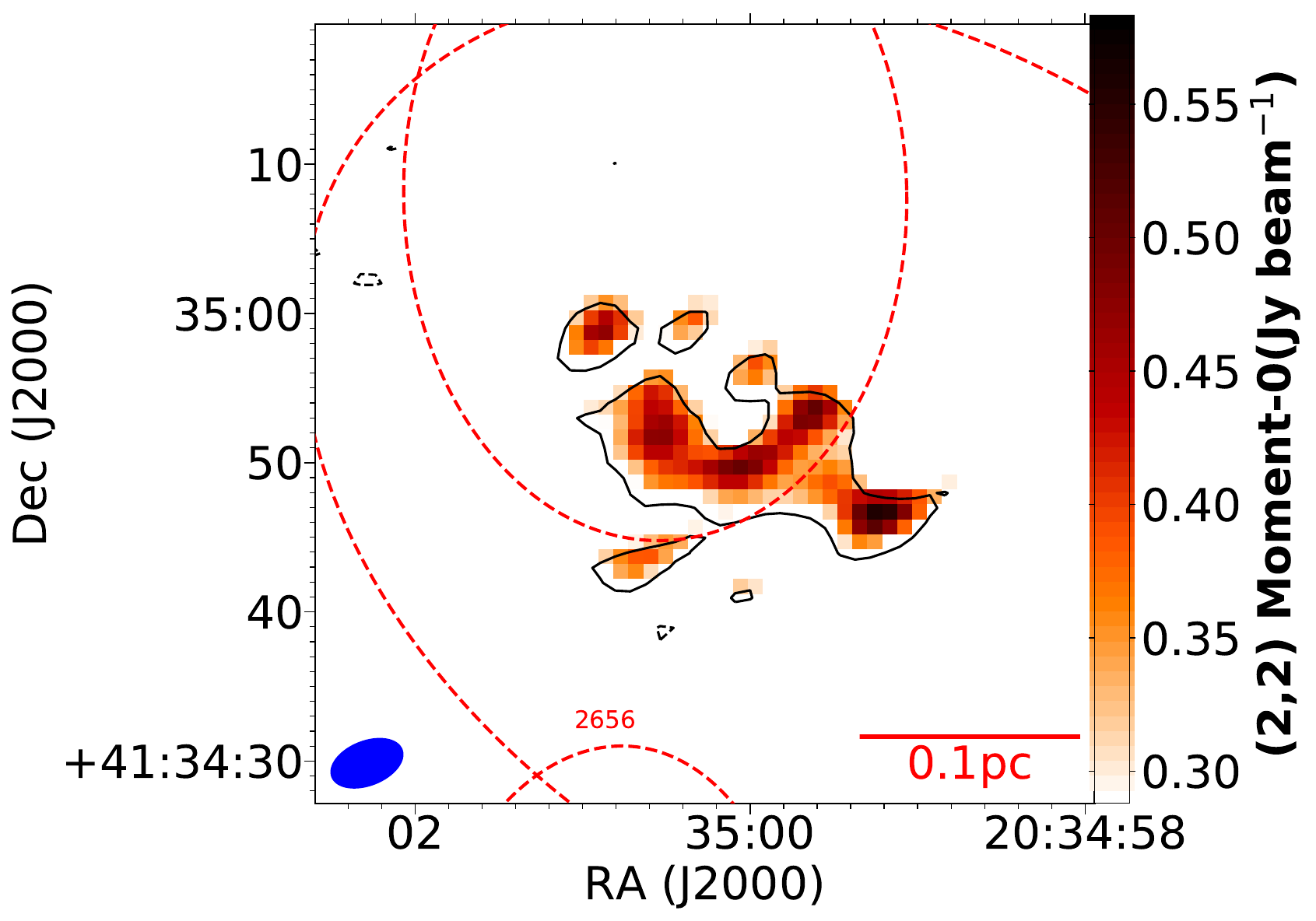} &
 \\
 & Field 29 & \\
\includegraphics[width=.3\textwidth]{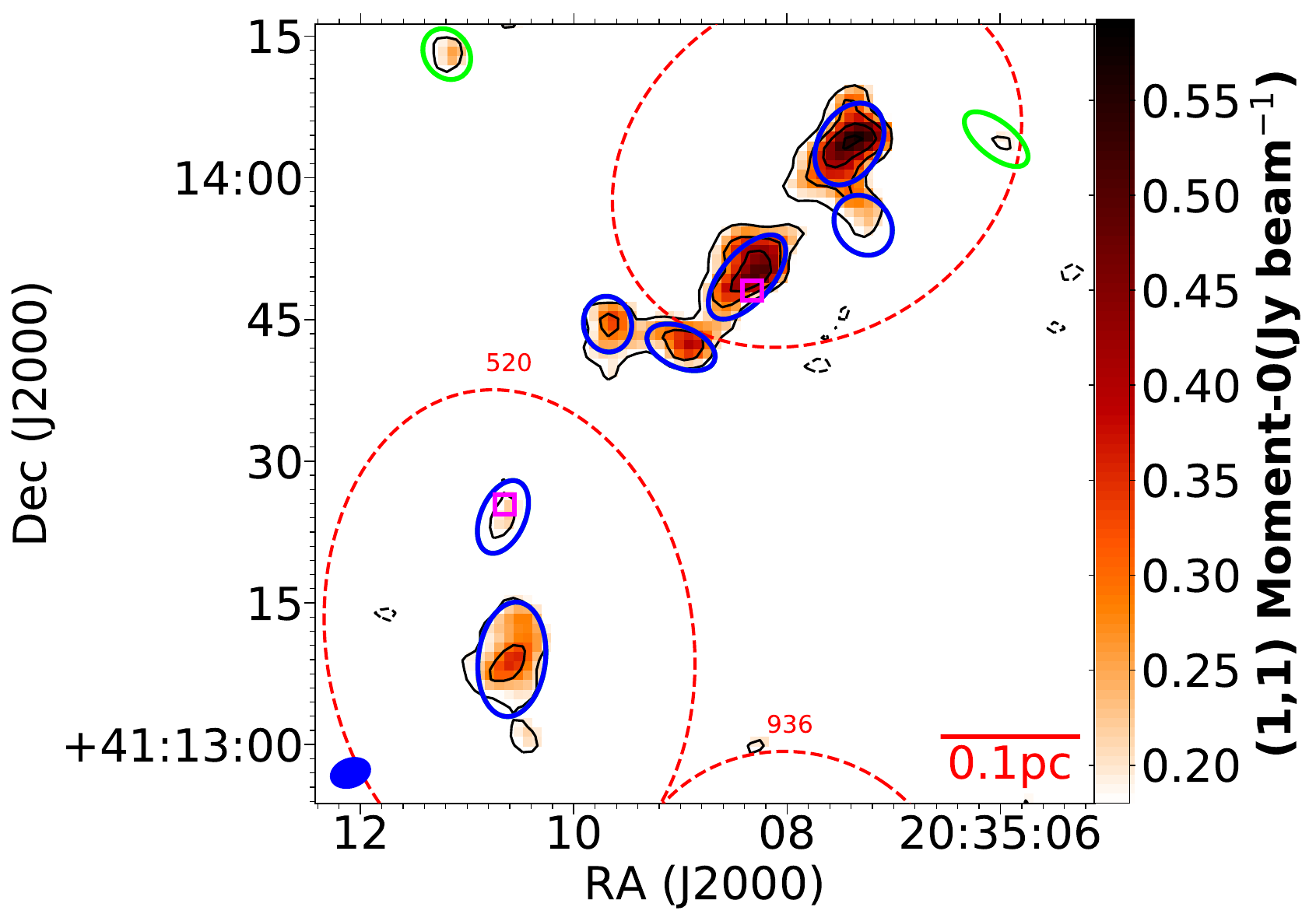} &
\includegraphics[width=.3\textwidth]{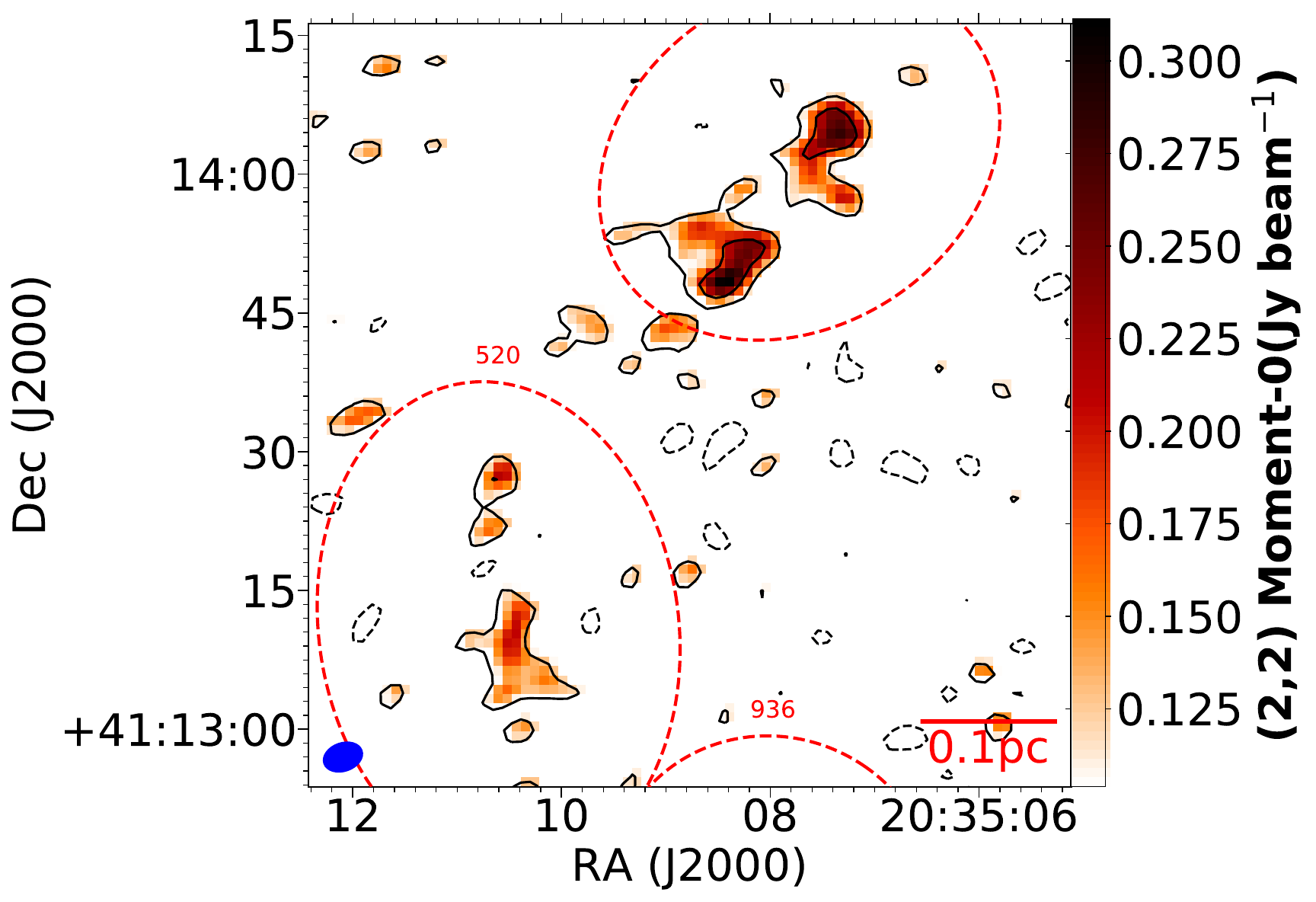} &
 \\
 & Field 31 & \\
\includegraphics[width=.3\textwidth]{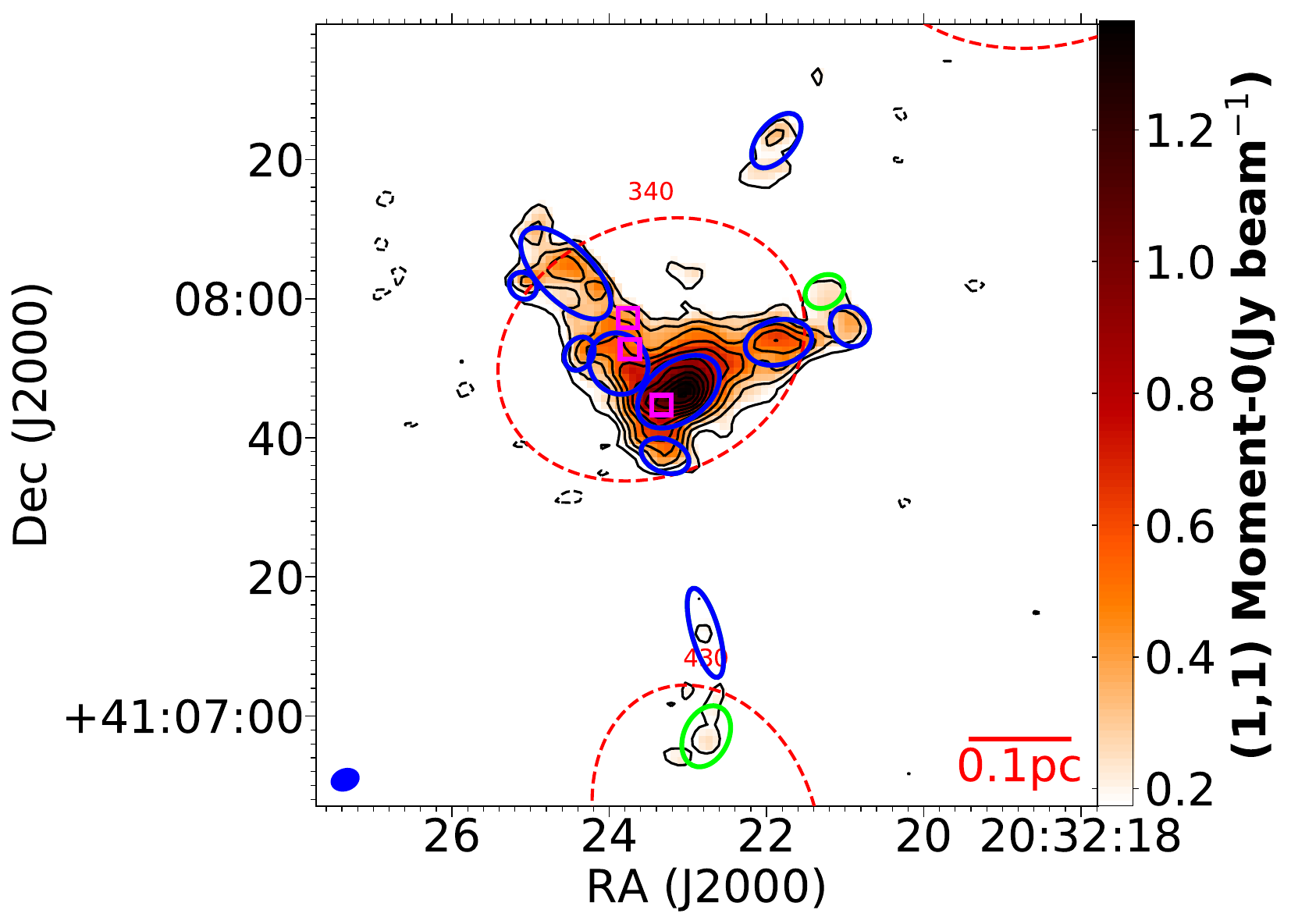} &
\includegraphics[width=.3\textwidth]{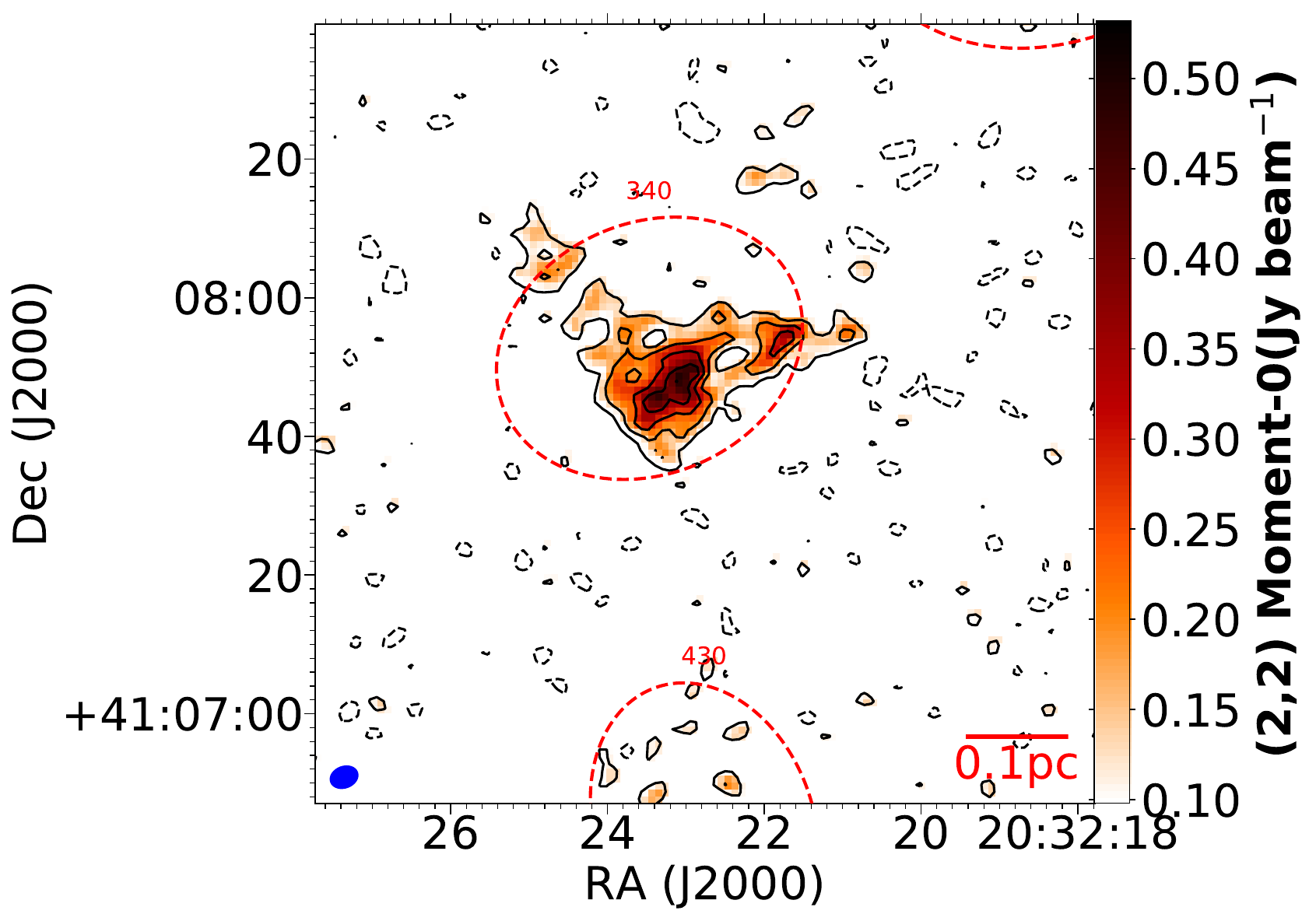} &
 \\
 & Field 32 & \\
\includegraphics[width=.3\textwidth]{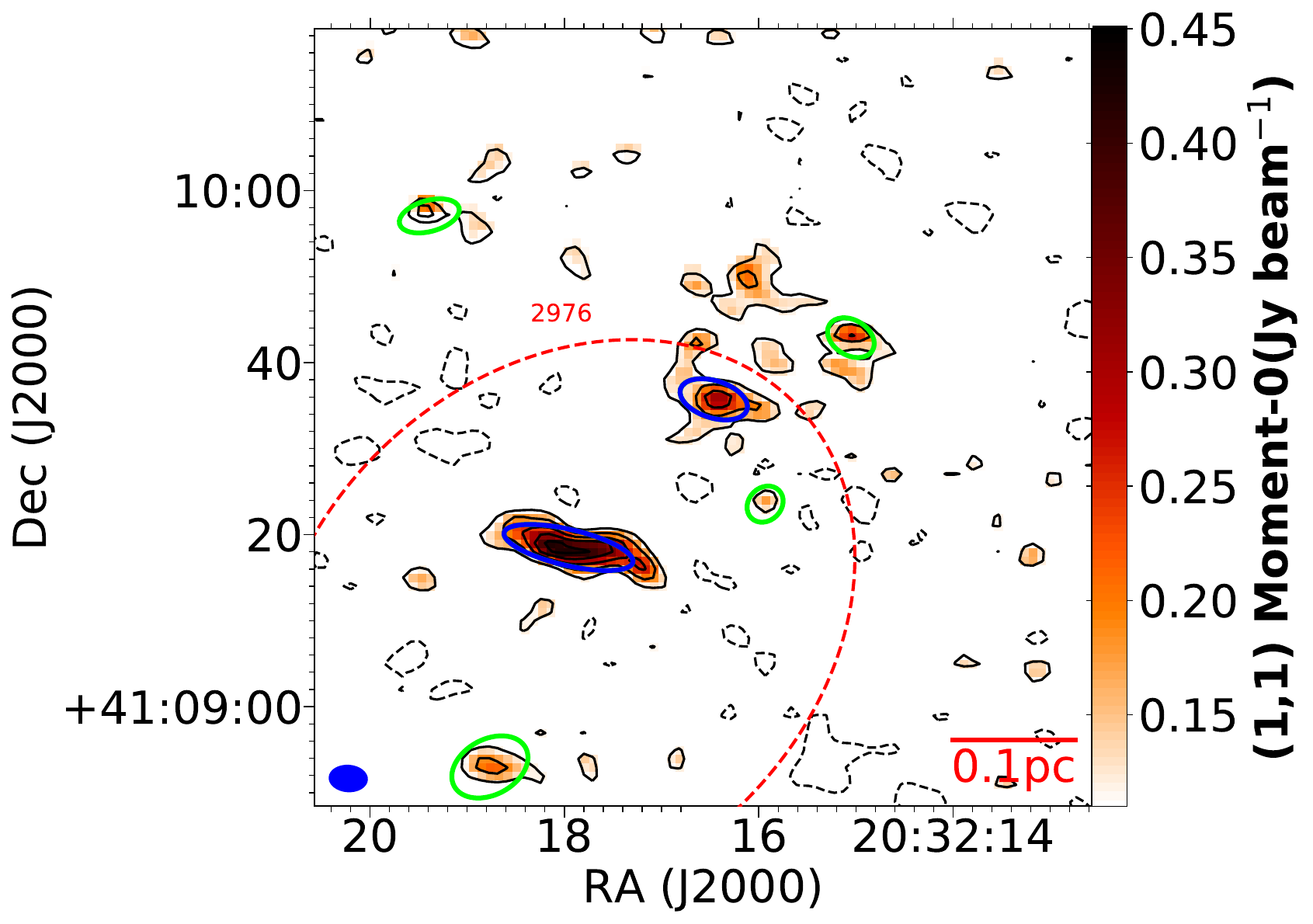} &
\includegraphics[width=.3\textwidth]{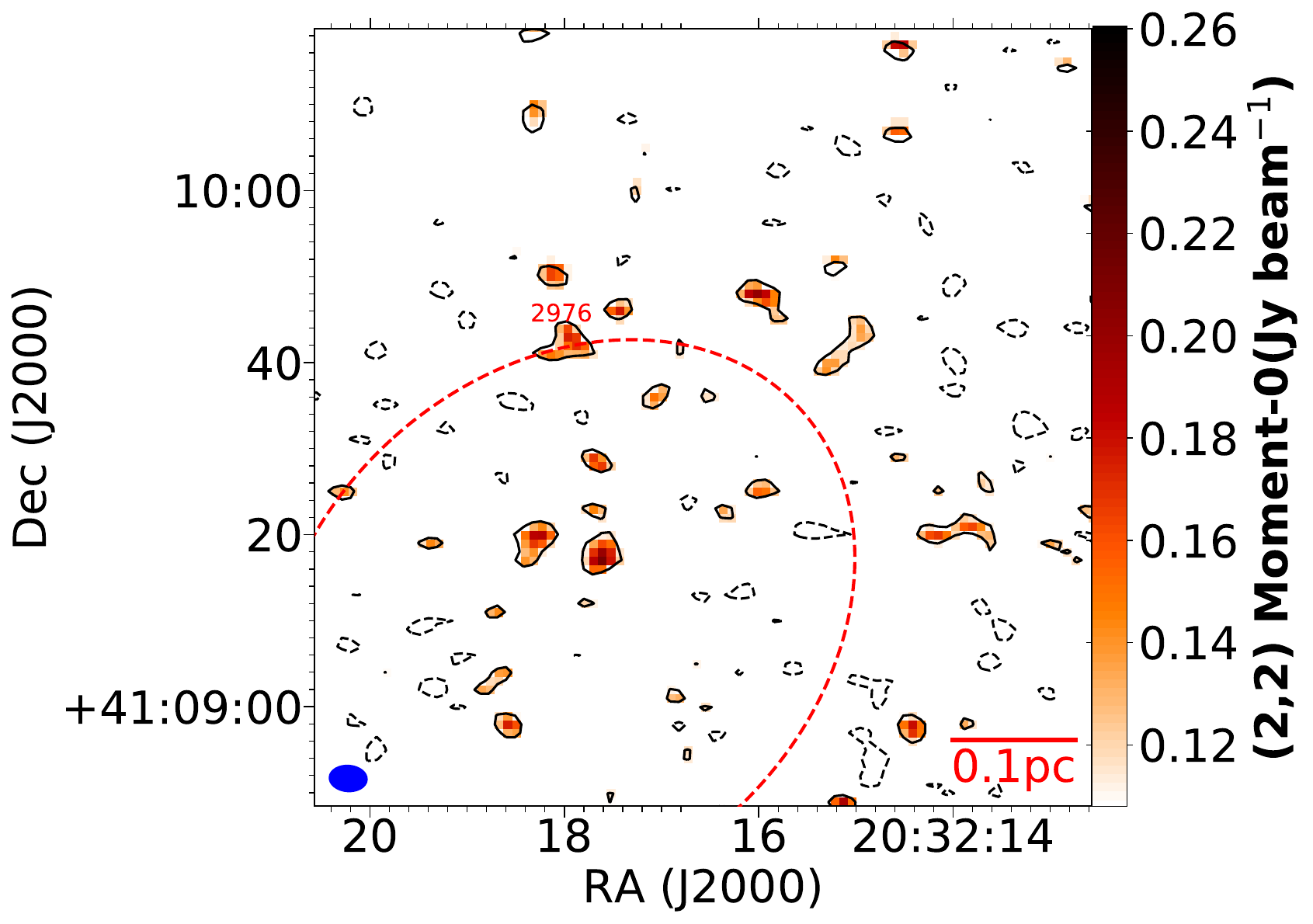} &
 \\
 & Field 33 & \\
\includegraphics[width=.3\textwidth]{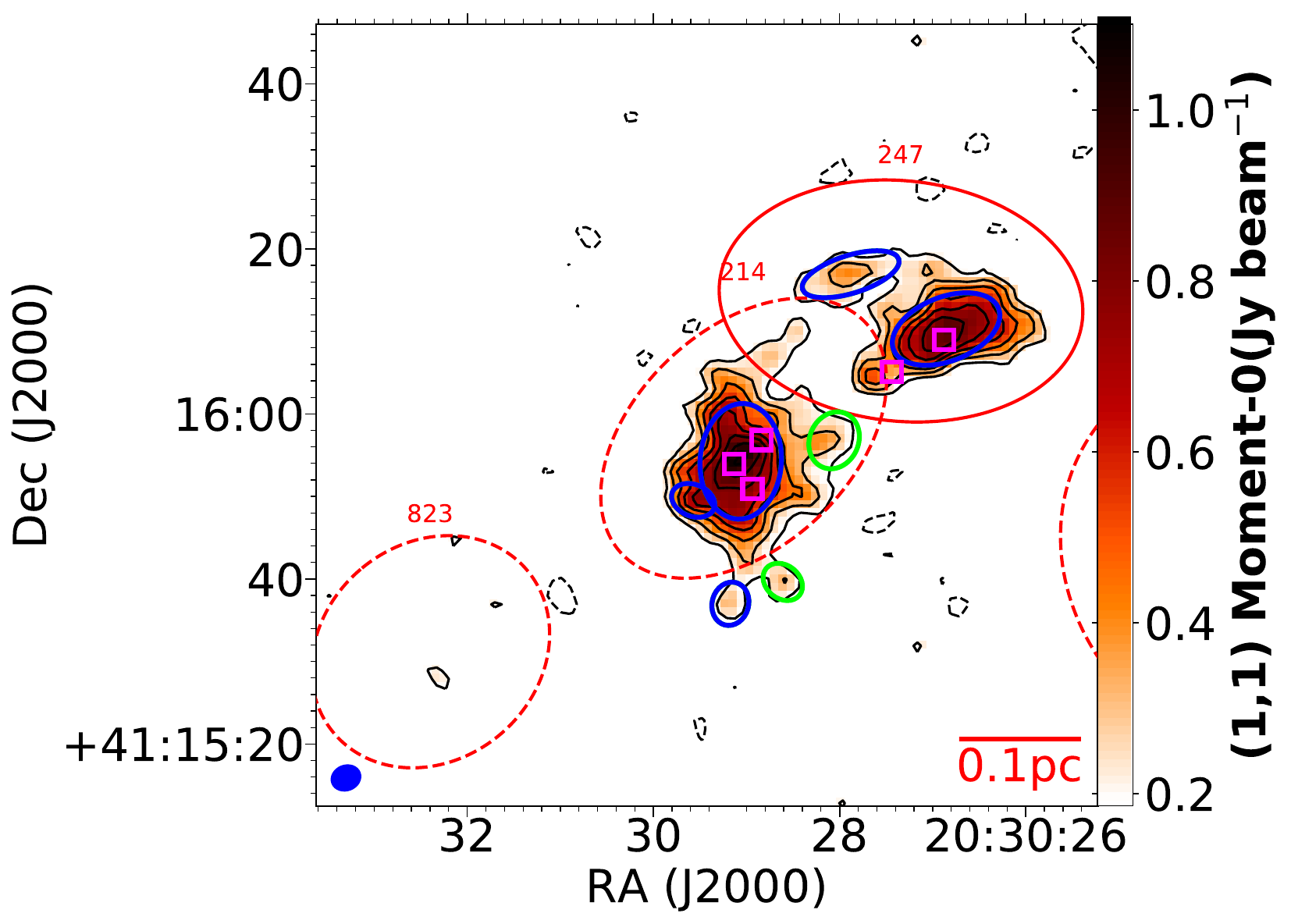} &
\includegraphics[width=.3\textwidth]{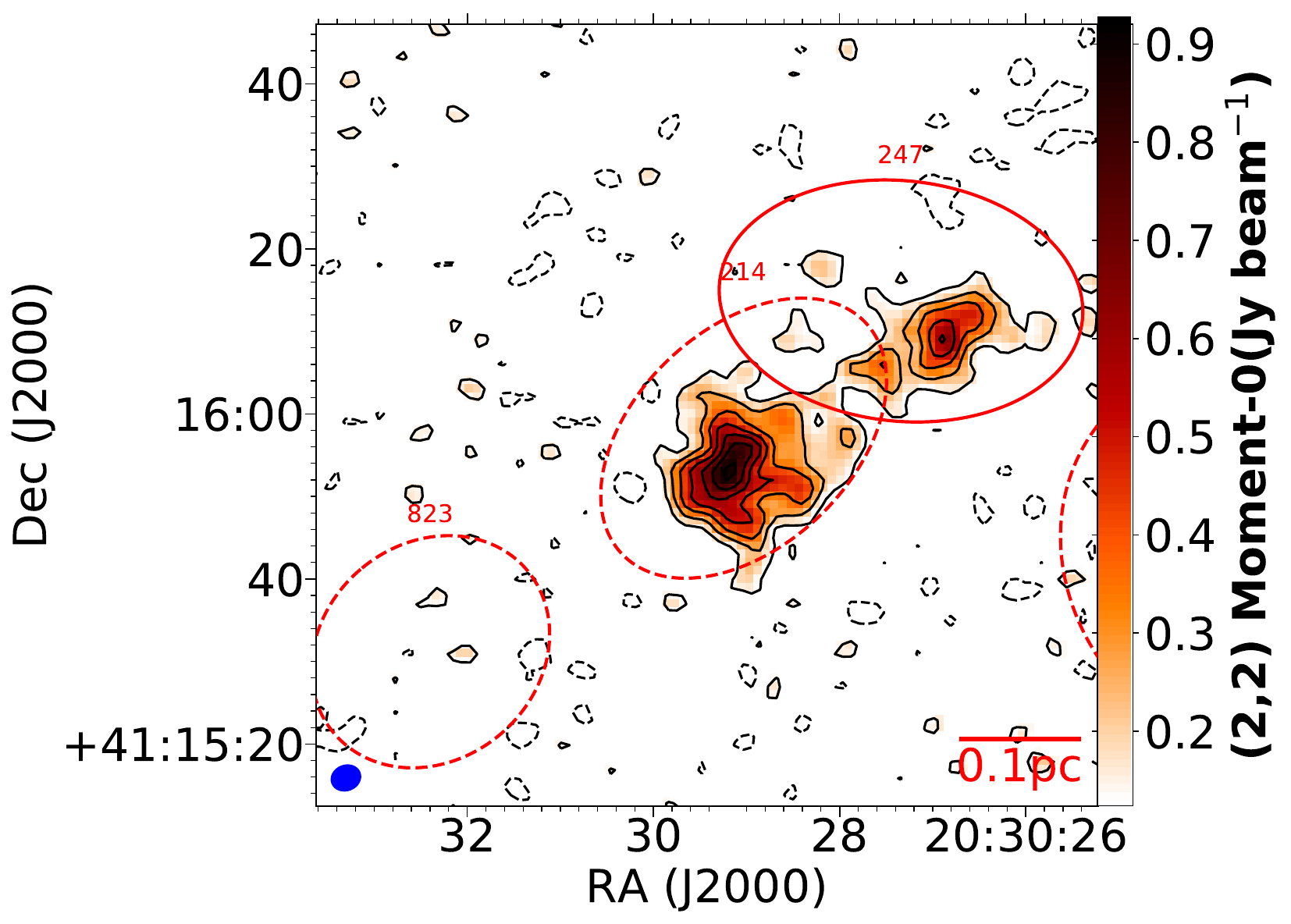} &
\includegraphics[width=.3\textwidth]{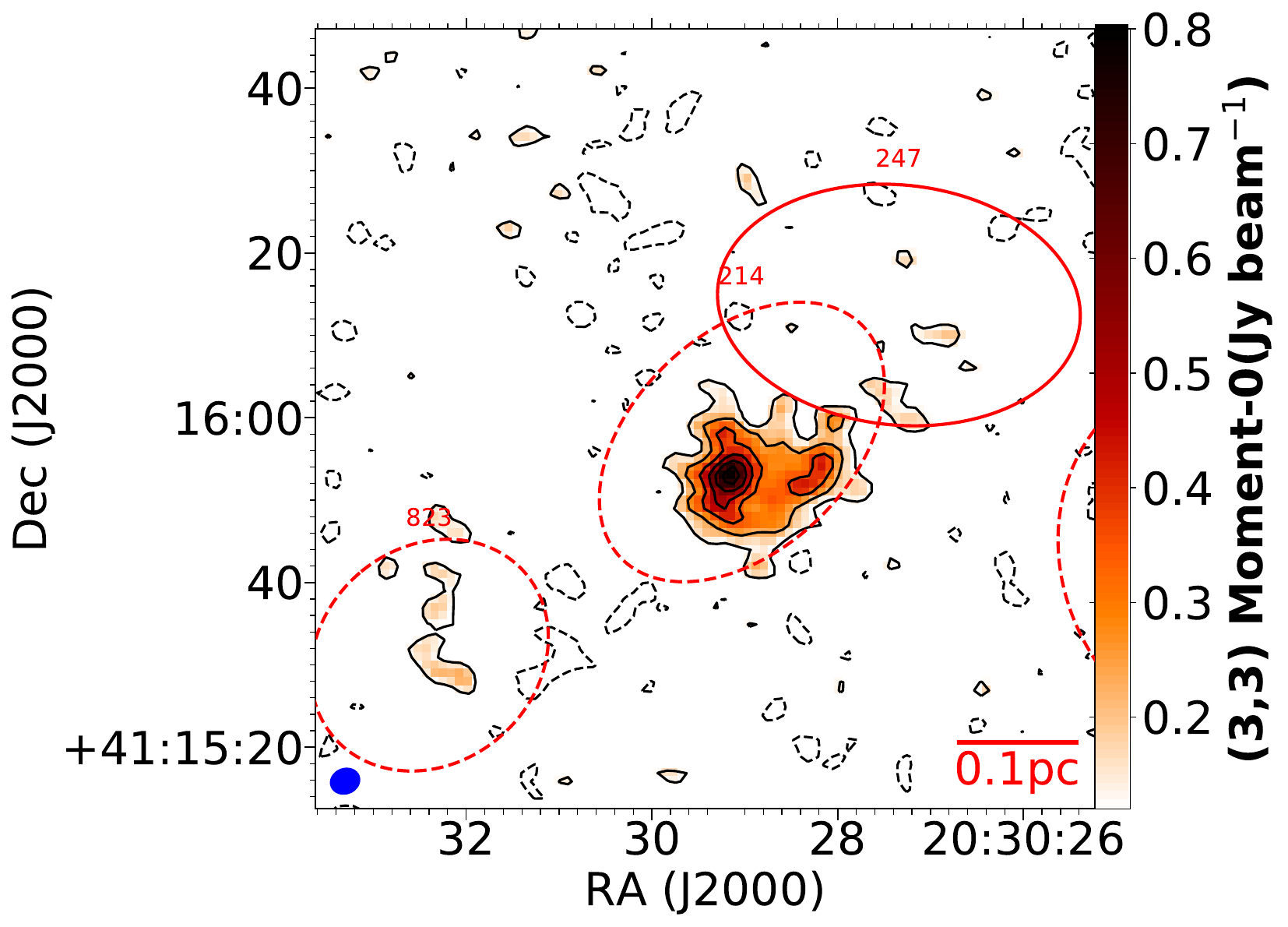} \\
 & Field 34 & \\
\includegraphics[width=.3\textwidth]{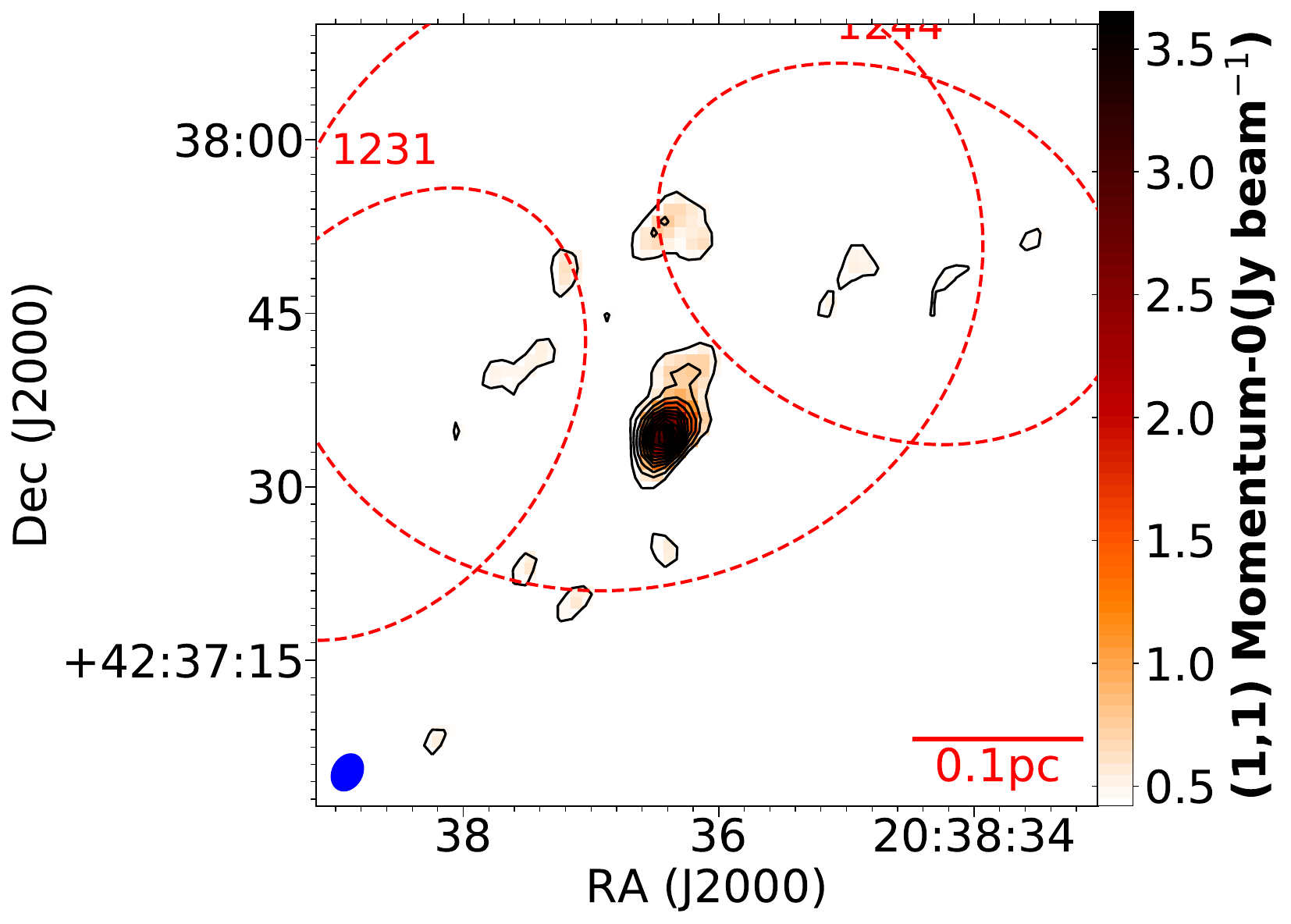} &
\includegraphics[width=.3\textwidth]{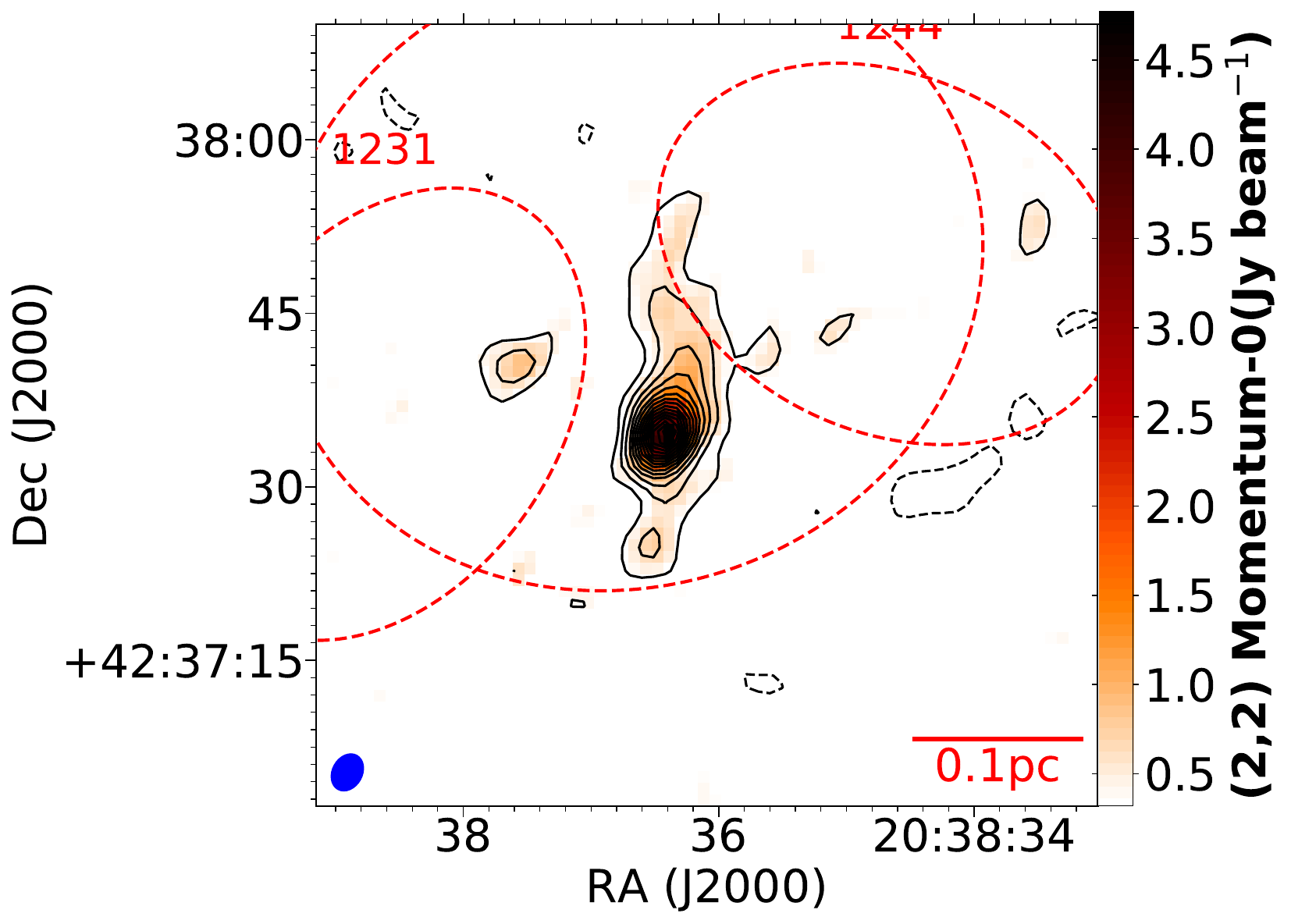} &
 \\
 & Field 37 & \\
\end{longtable*}

\section{Distributions of kinetic temperature, centroid velocity and velocity dispersion}\label{app:B}

\begin{longtable}{@{}ccc@{}}
\endhead
\caption*{\textbf{Fig. B.1.} continued.}
\endfoot
\caption*{\textbf{Fig. B.1.} Distributions of kinetic temperature (left panel), centroid velocity (middle panel), and velocity dispersion (right panel) derived from the simultaneous fitting of \nh\ \oneone\ and \twotwo\ inversion lines. The ellipses are the same as those in Fig. \ref{fig:A.1}. \label{fig:B.1}}
\endlastfoot
\includegraphics[width=.3\textwidth]{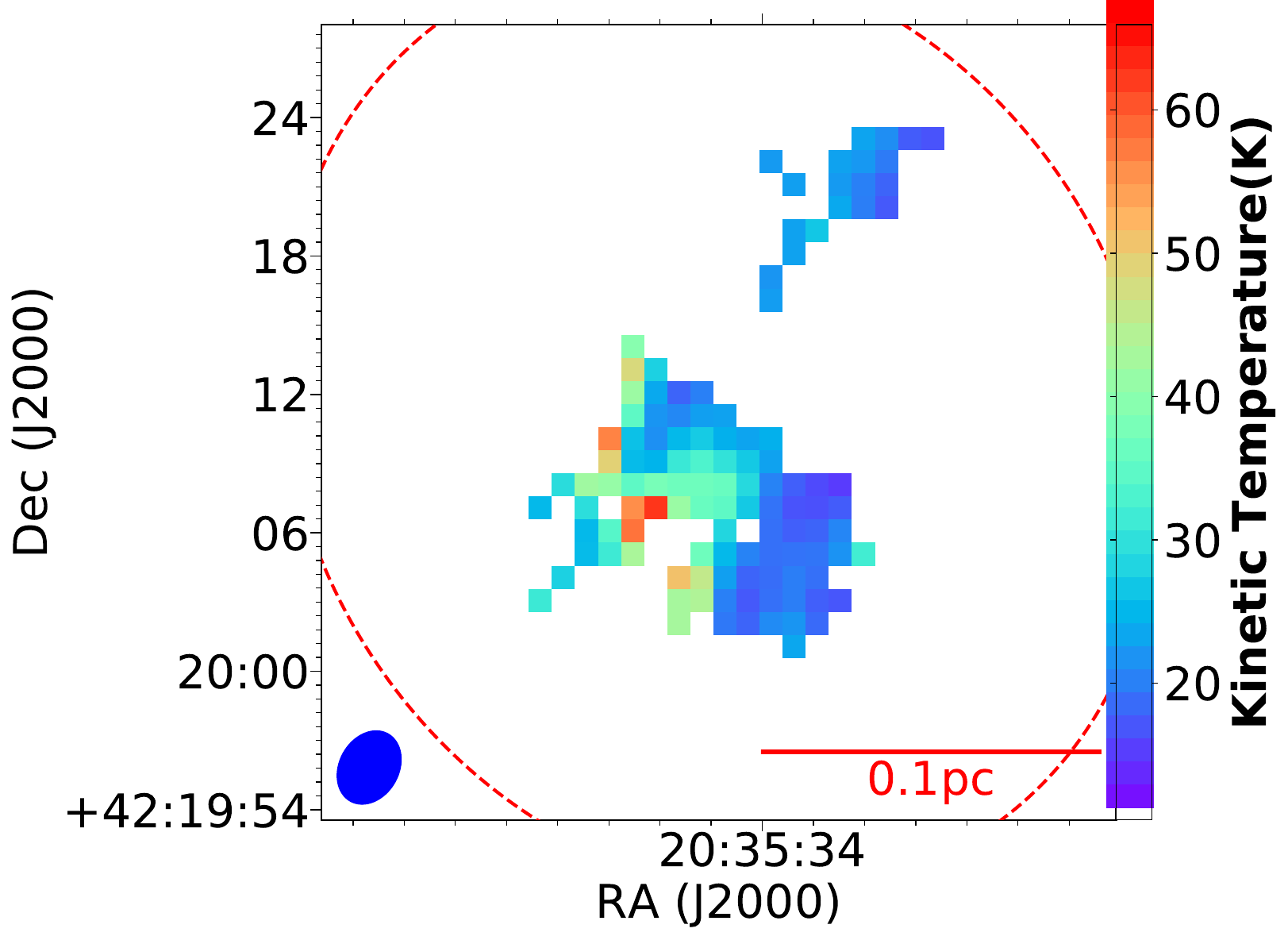} &
\includegraphics[width=.3\textwidth]{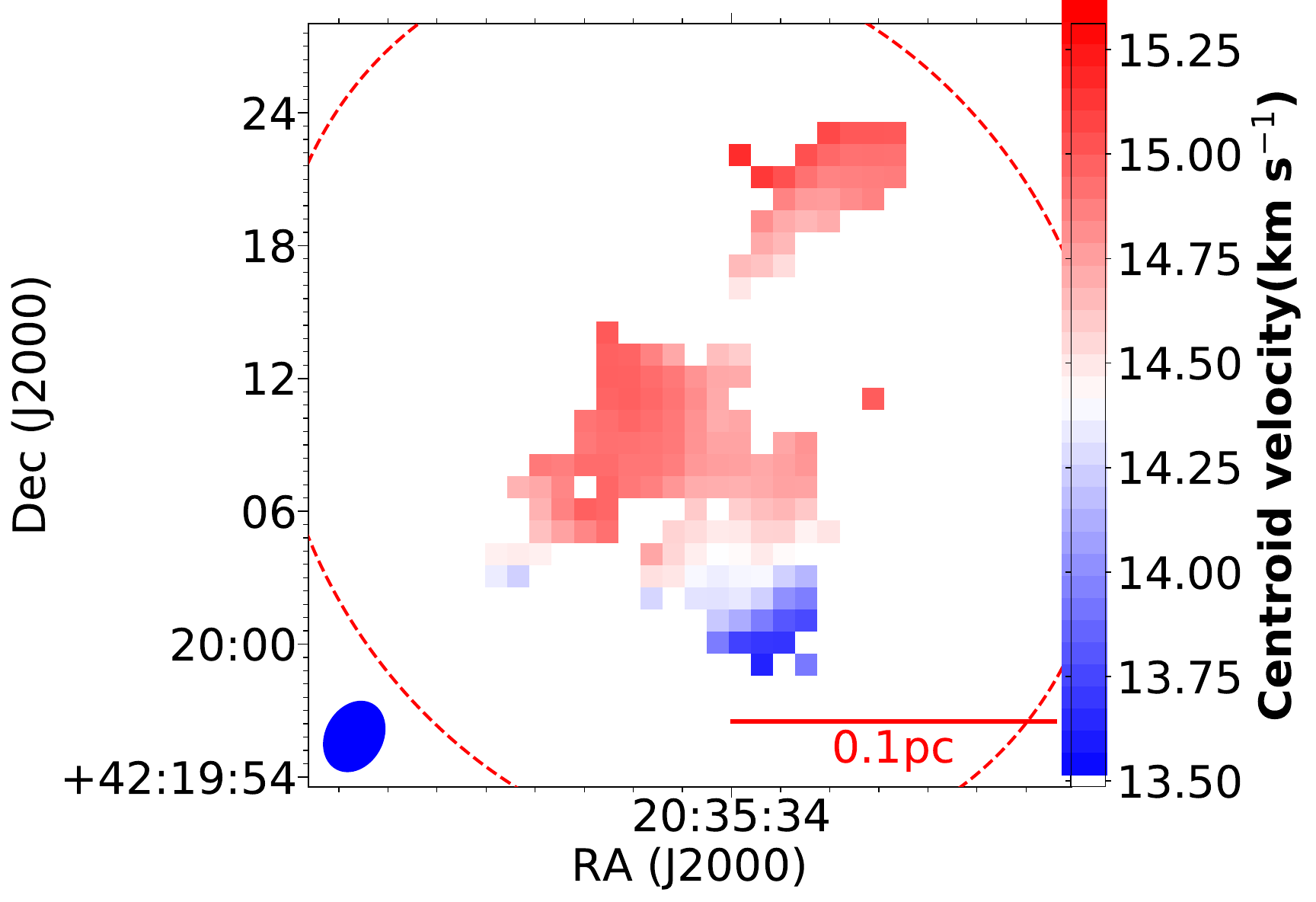} &
\includegraphics[width=.3\textwidth]{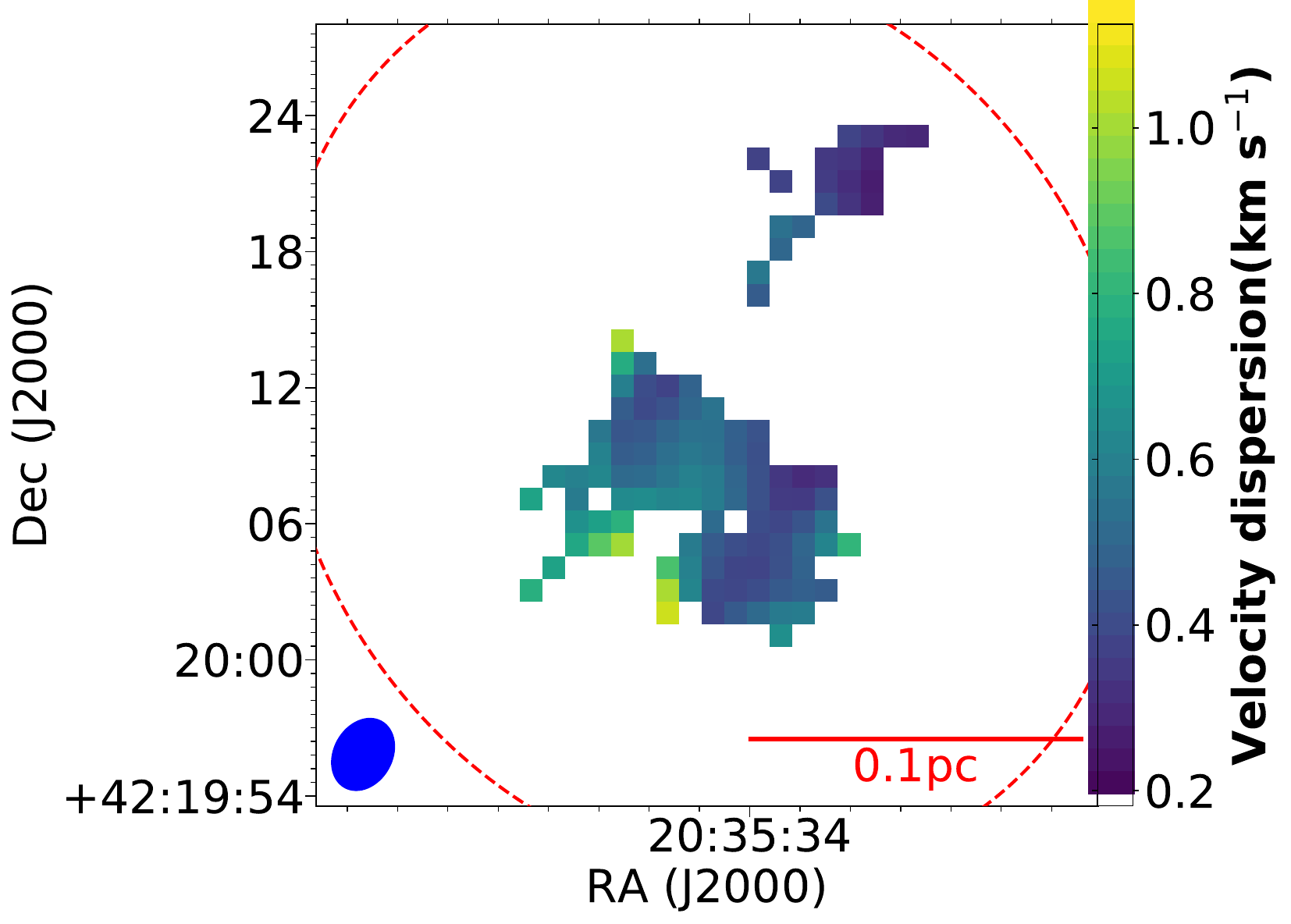} \\
 & Field 1 & \\
\includegraphics[width=.3\textwidth]{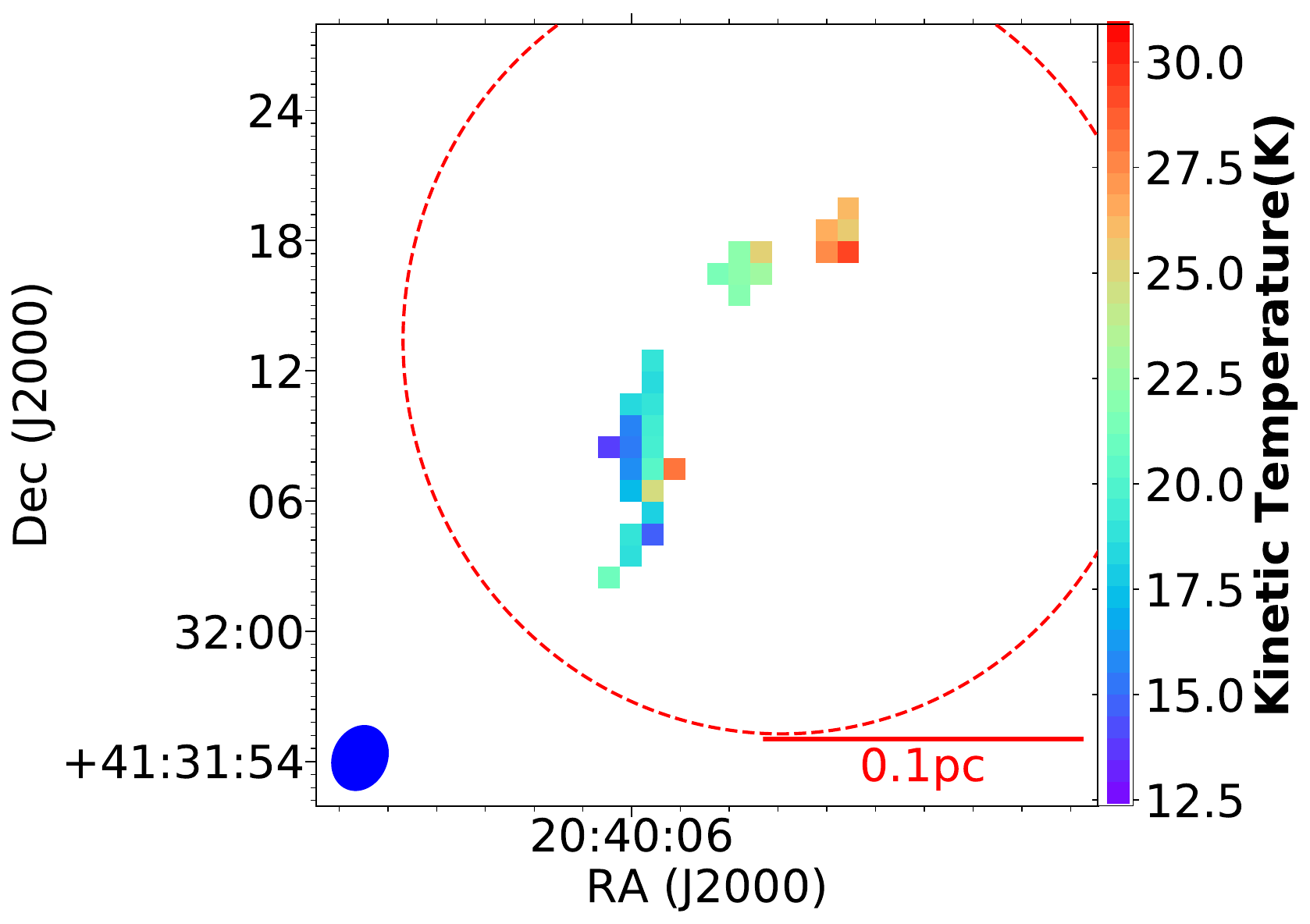} &
\includegraphics[width=.3\textwidth]{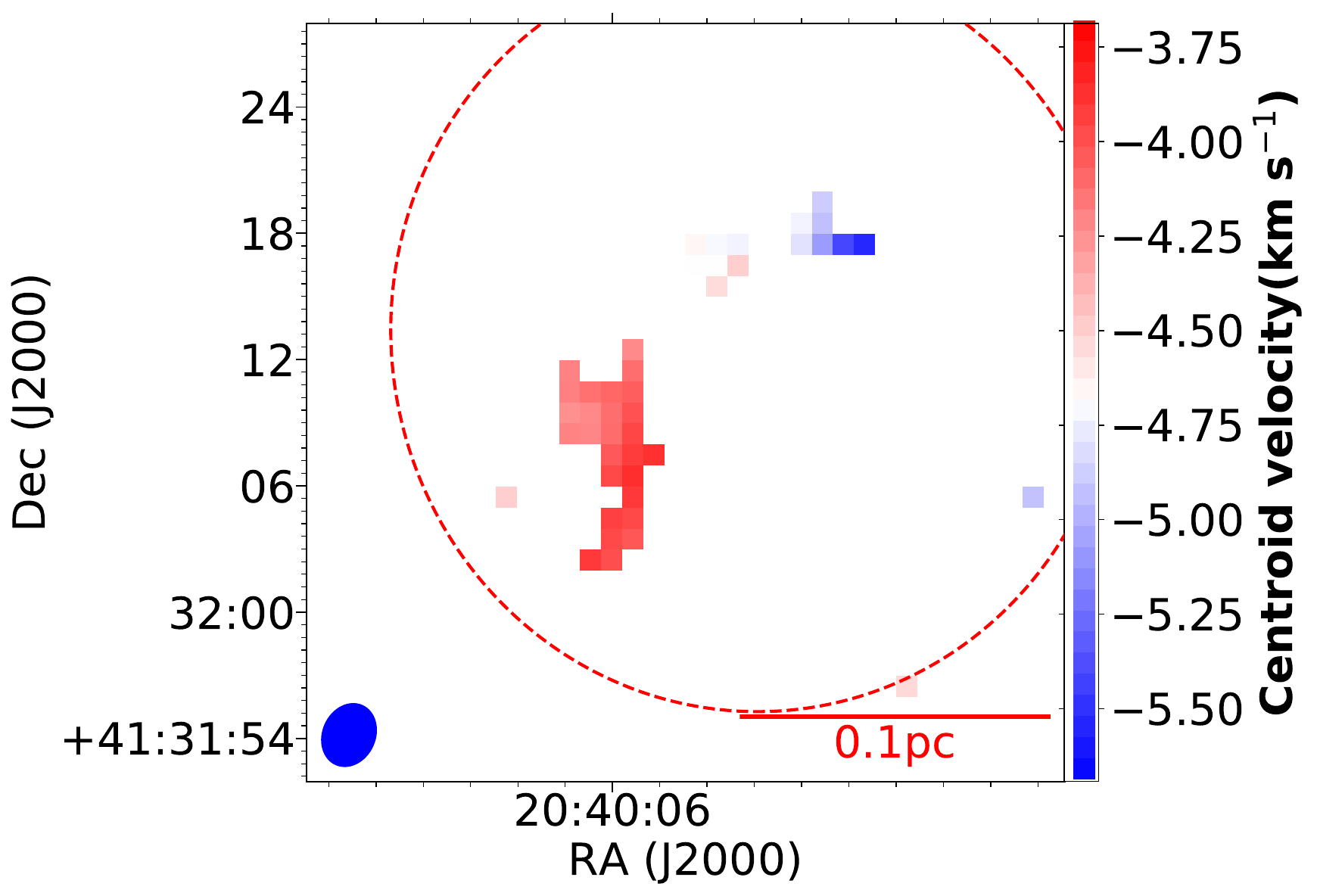} &
\includegraphics[width=.3\textwidth]{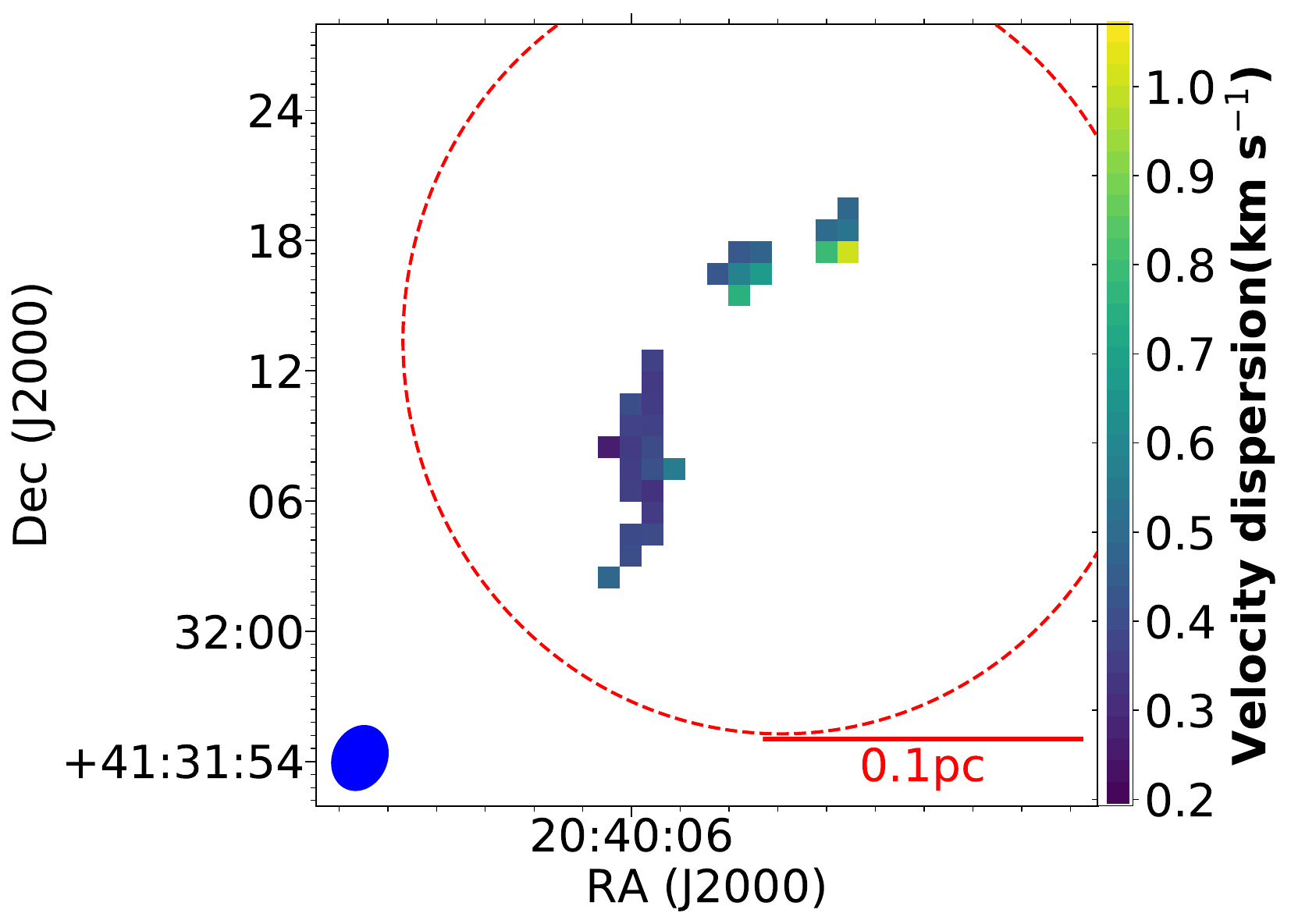} \\
 & Field 2 & \\
\includegraphics[width=.3\textwidth]{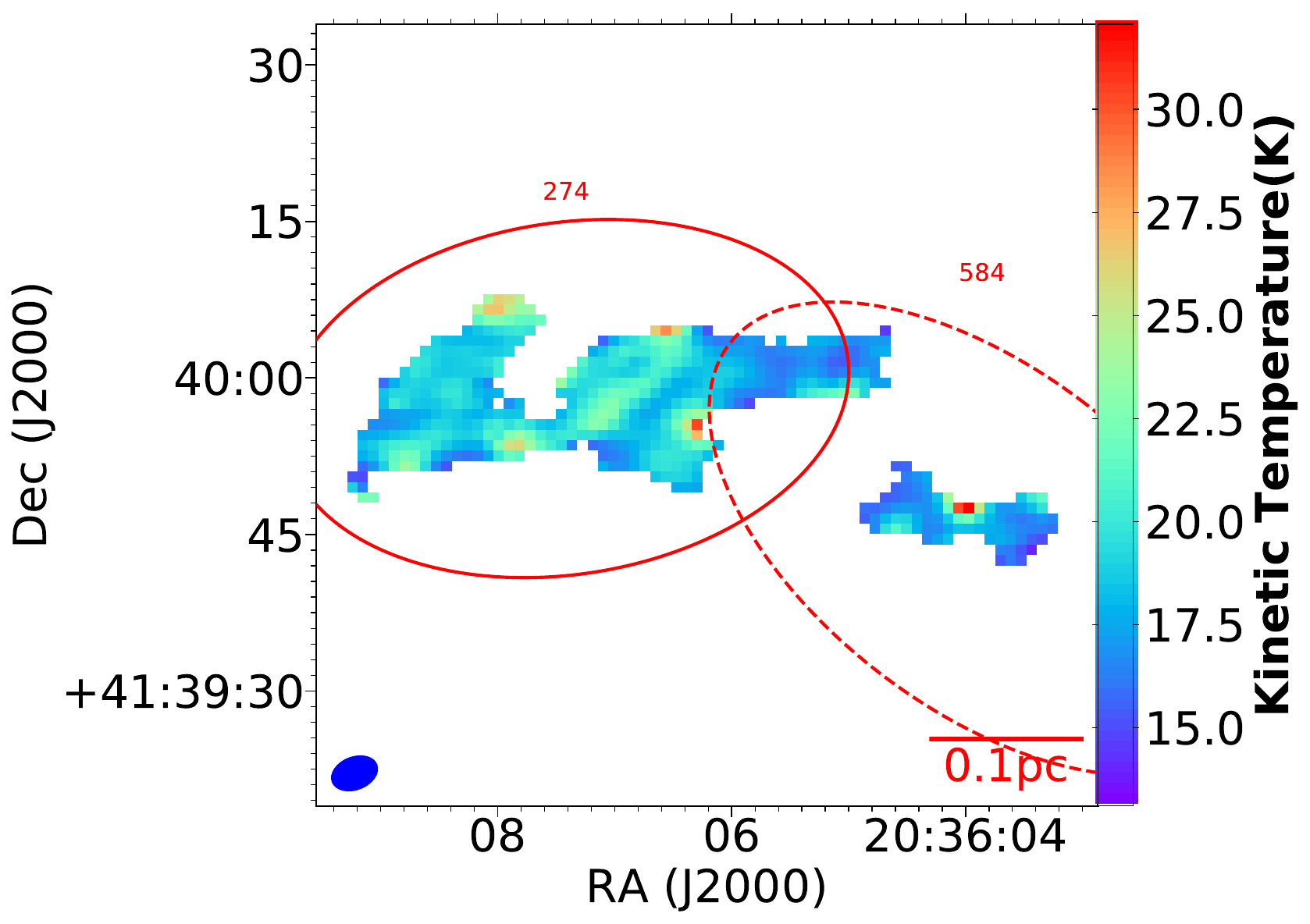} &
\includegraphics[width=.3\textwidth]{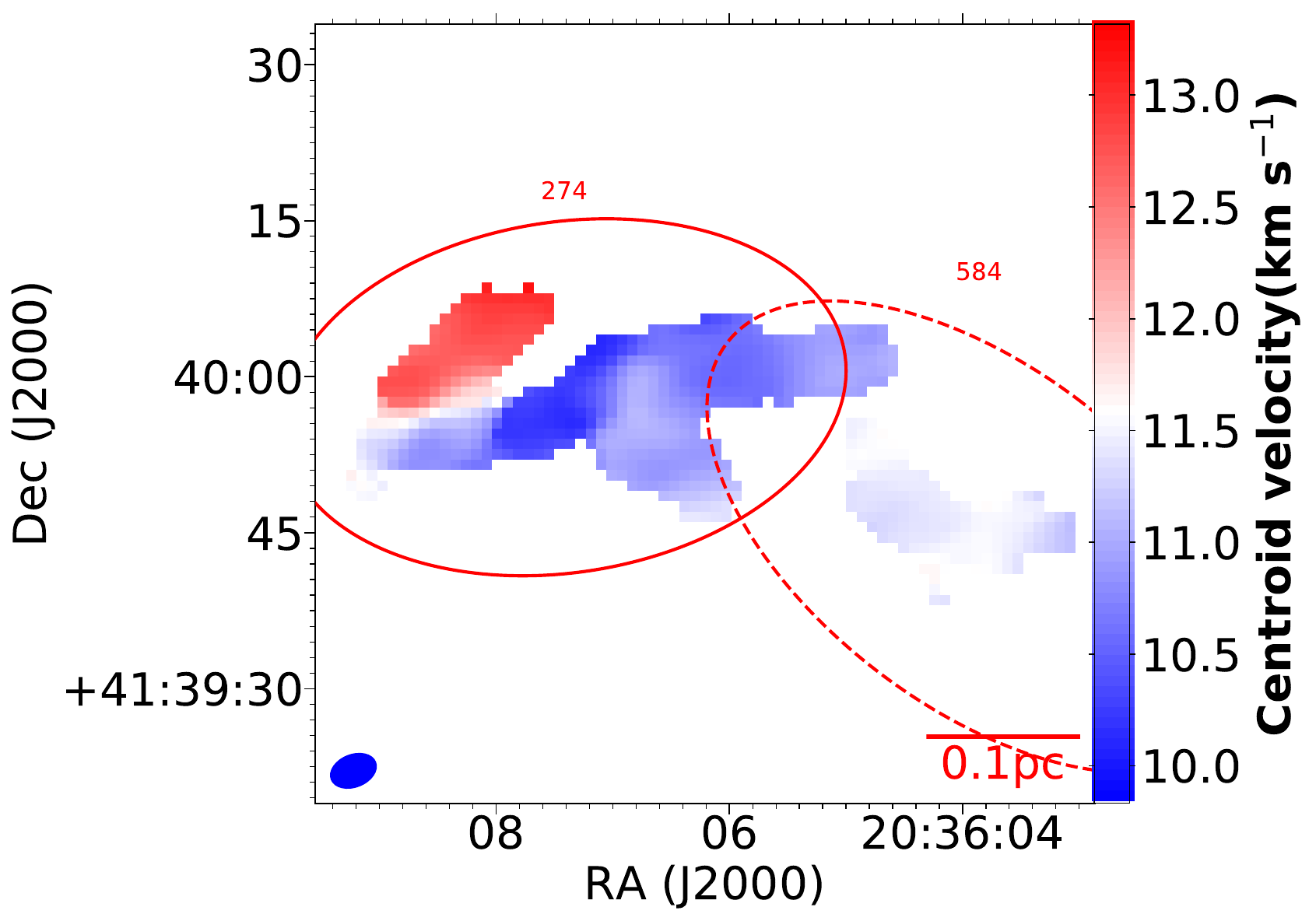} &
\includegraphics[width=.3\textwidth]{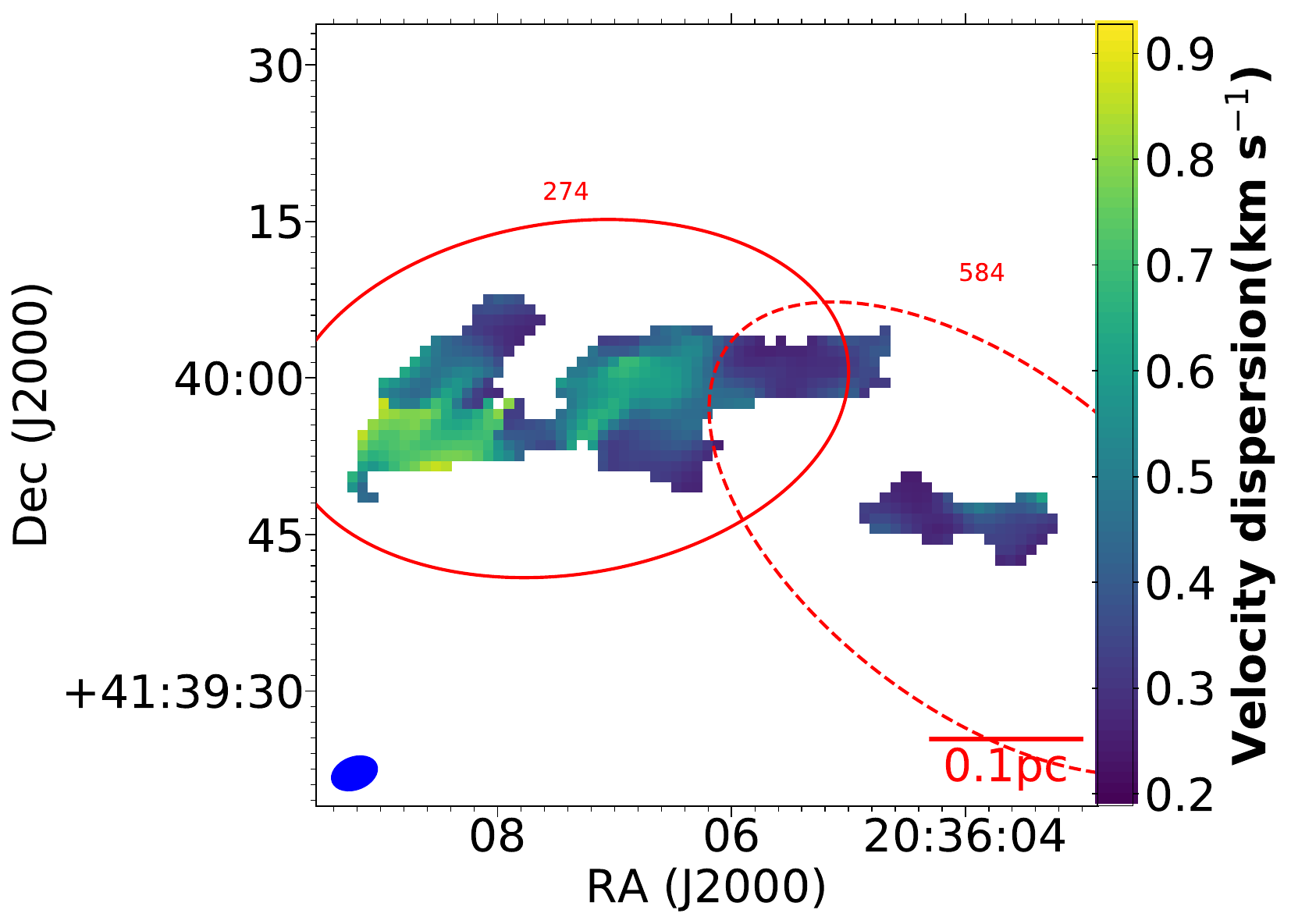} \\
 & Field 3 & \\
\includegraphics[width=.3\textwidth]{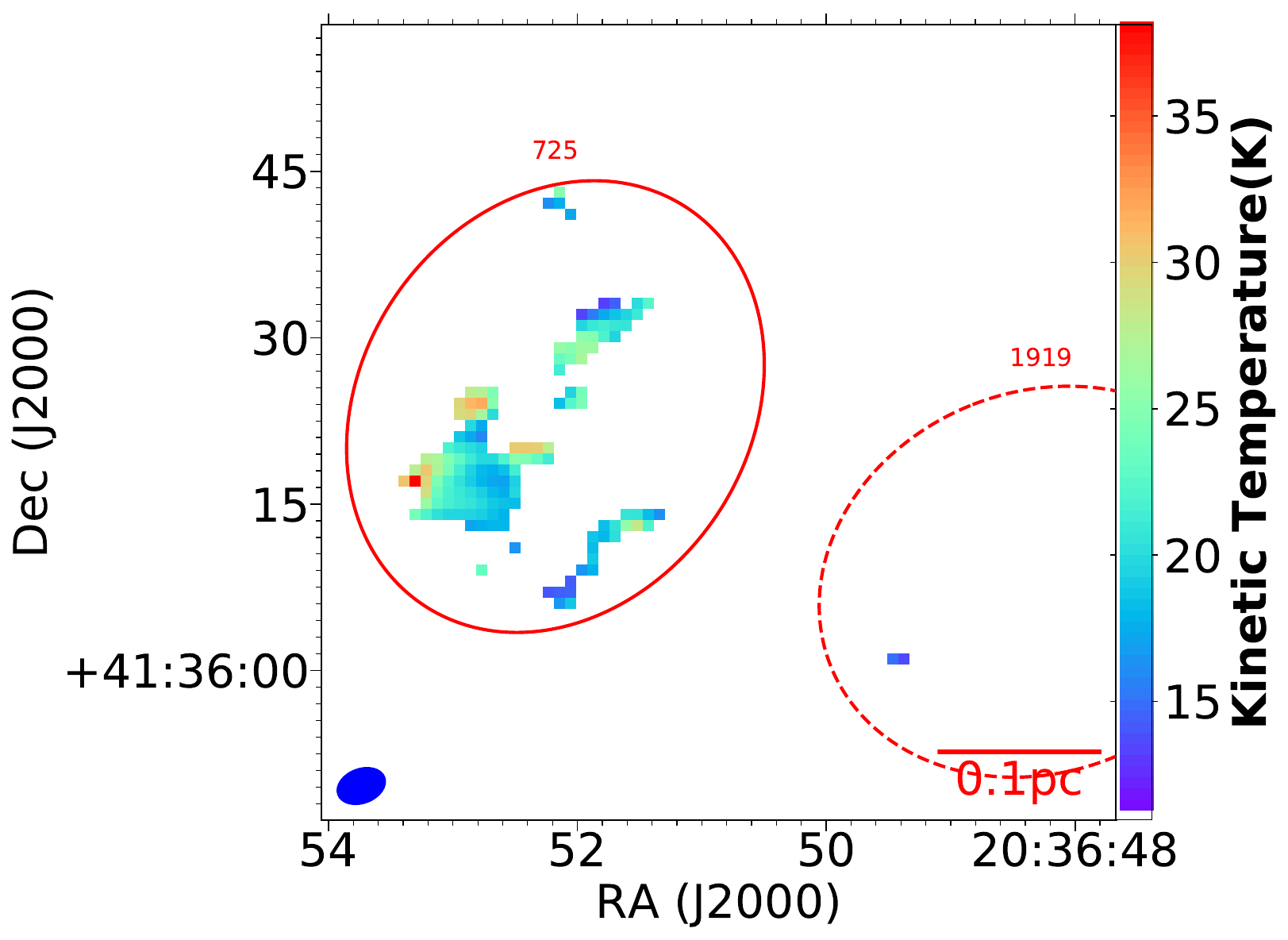} &
\includegraphics[width=.3\textwidth]{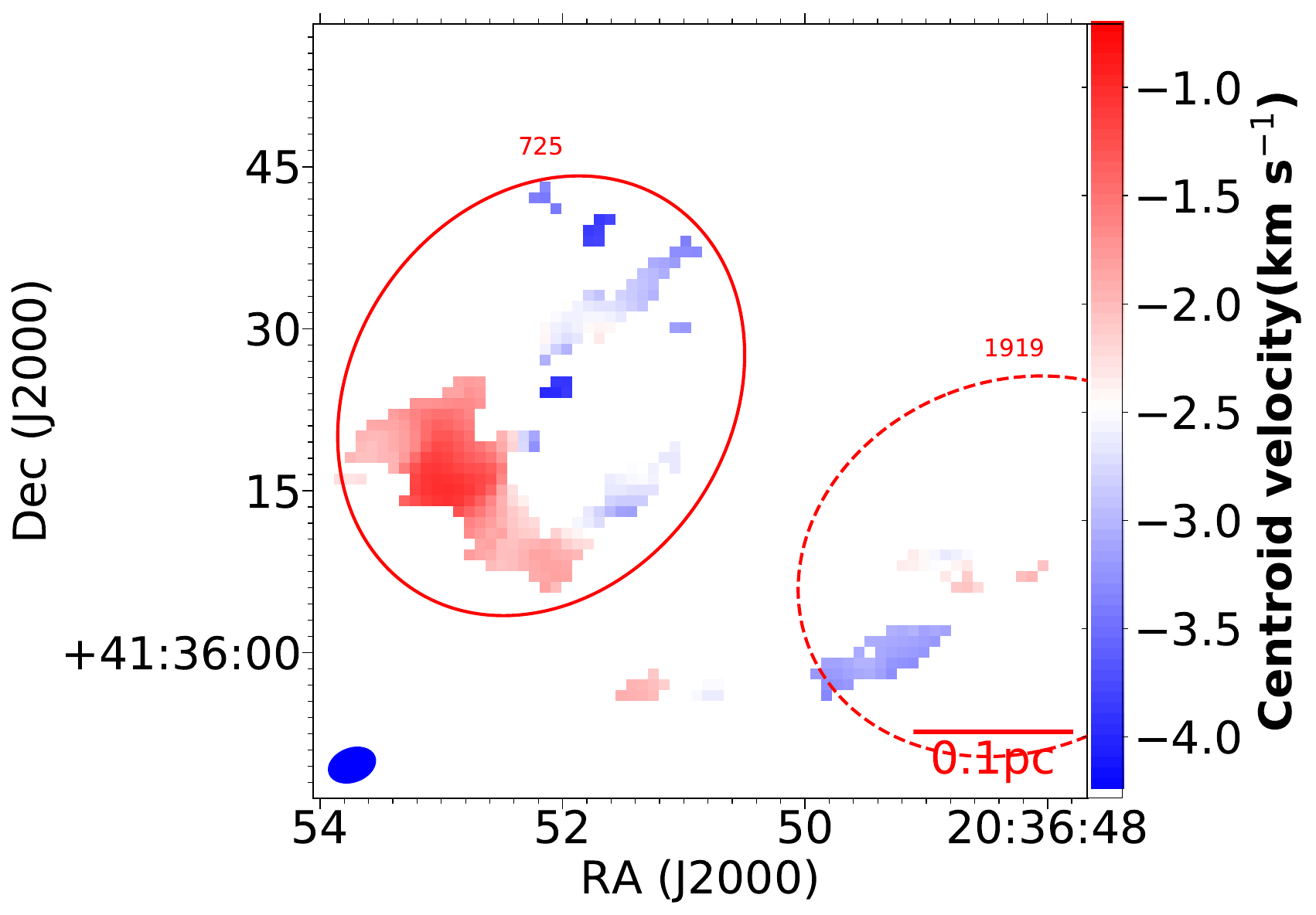} &
\includegraphics[width=.3\textwidth]{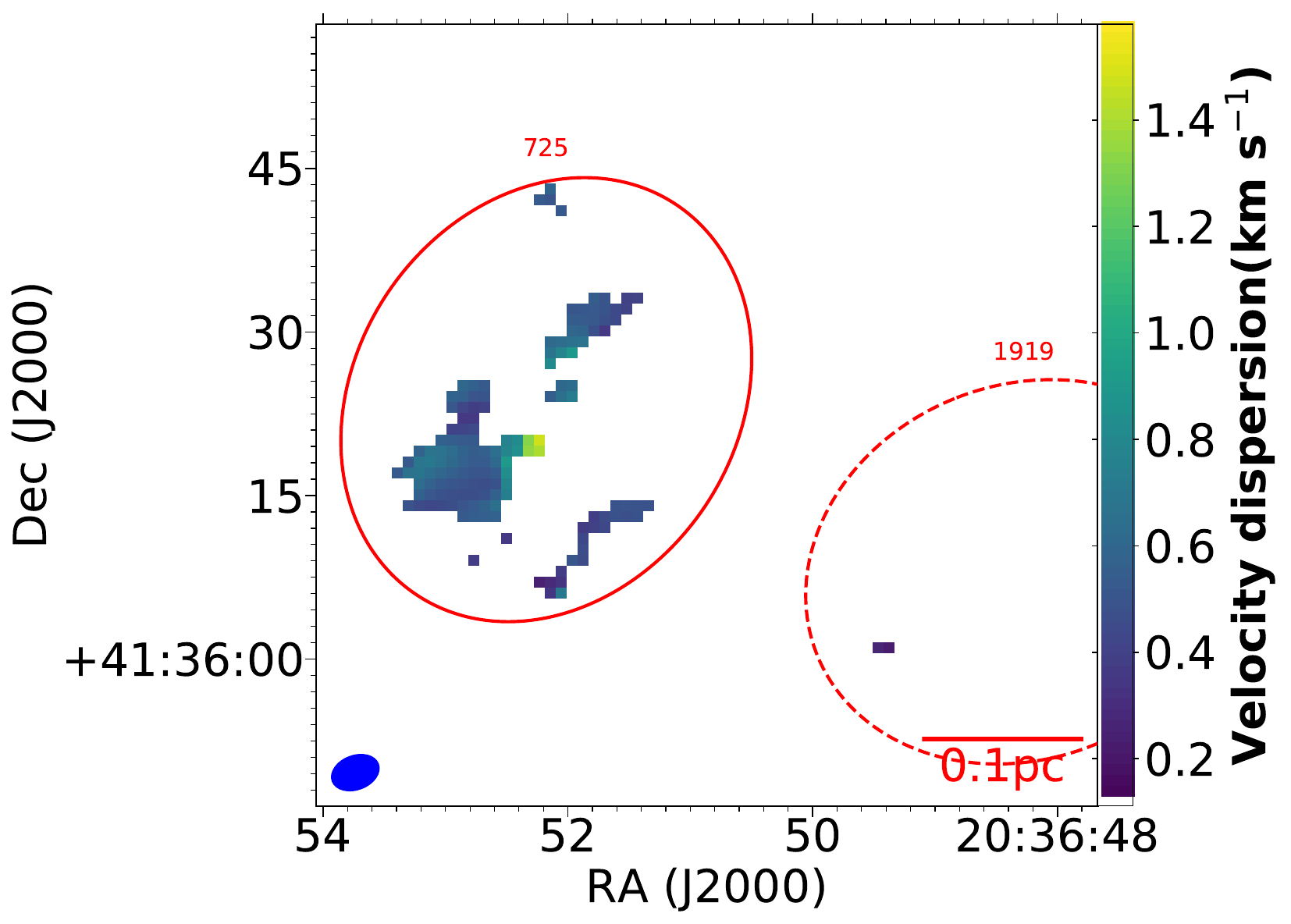} \\
 & Field 4 & \\
\includegraphics[width=.3\textwidth]{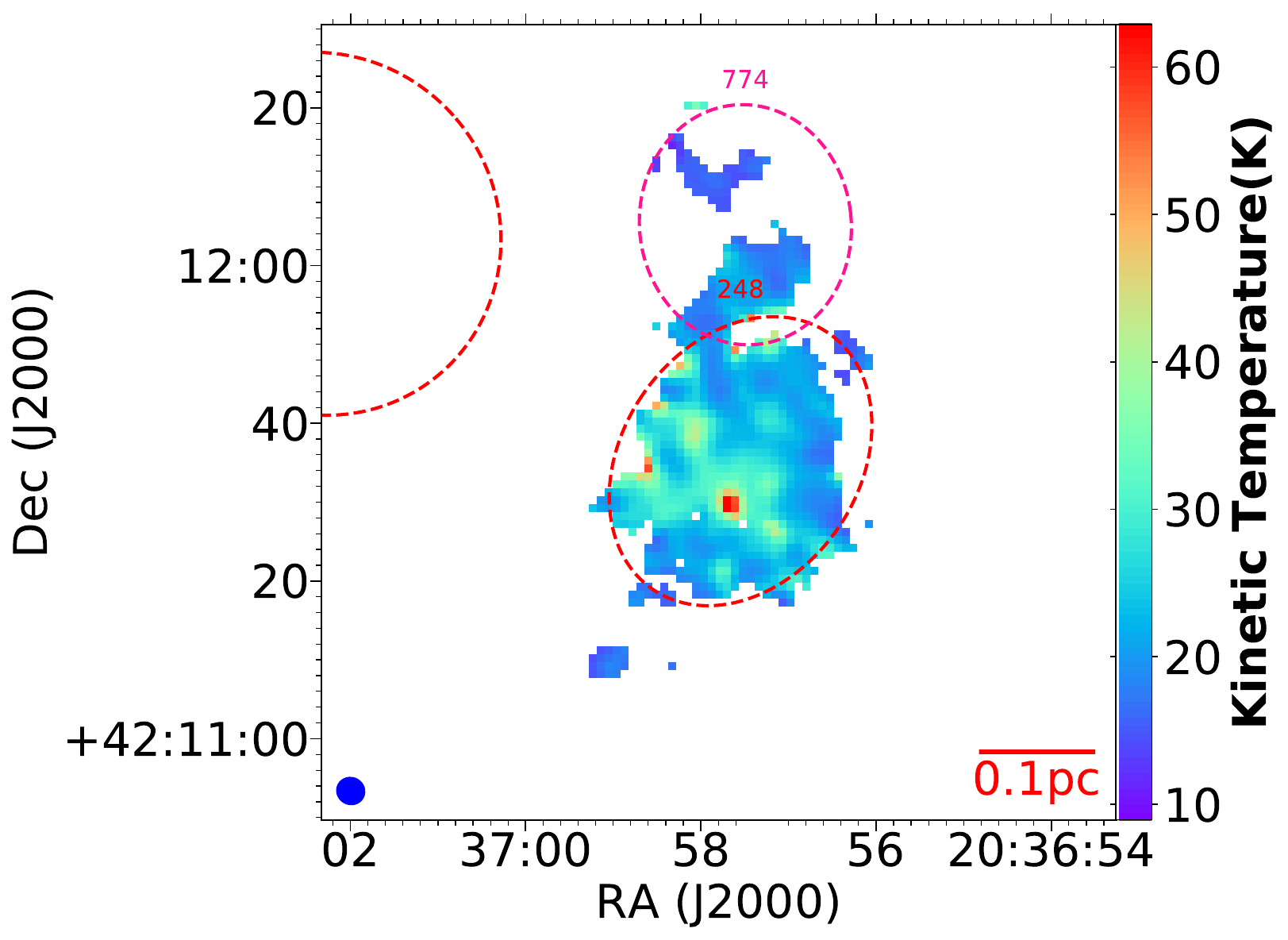} &
\includegraphics[width=.3\textwidth]{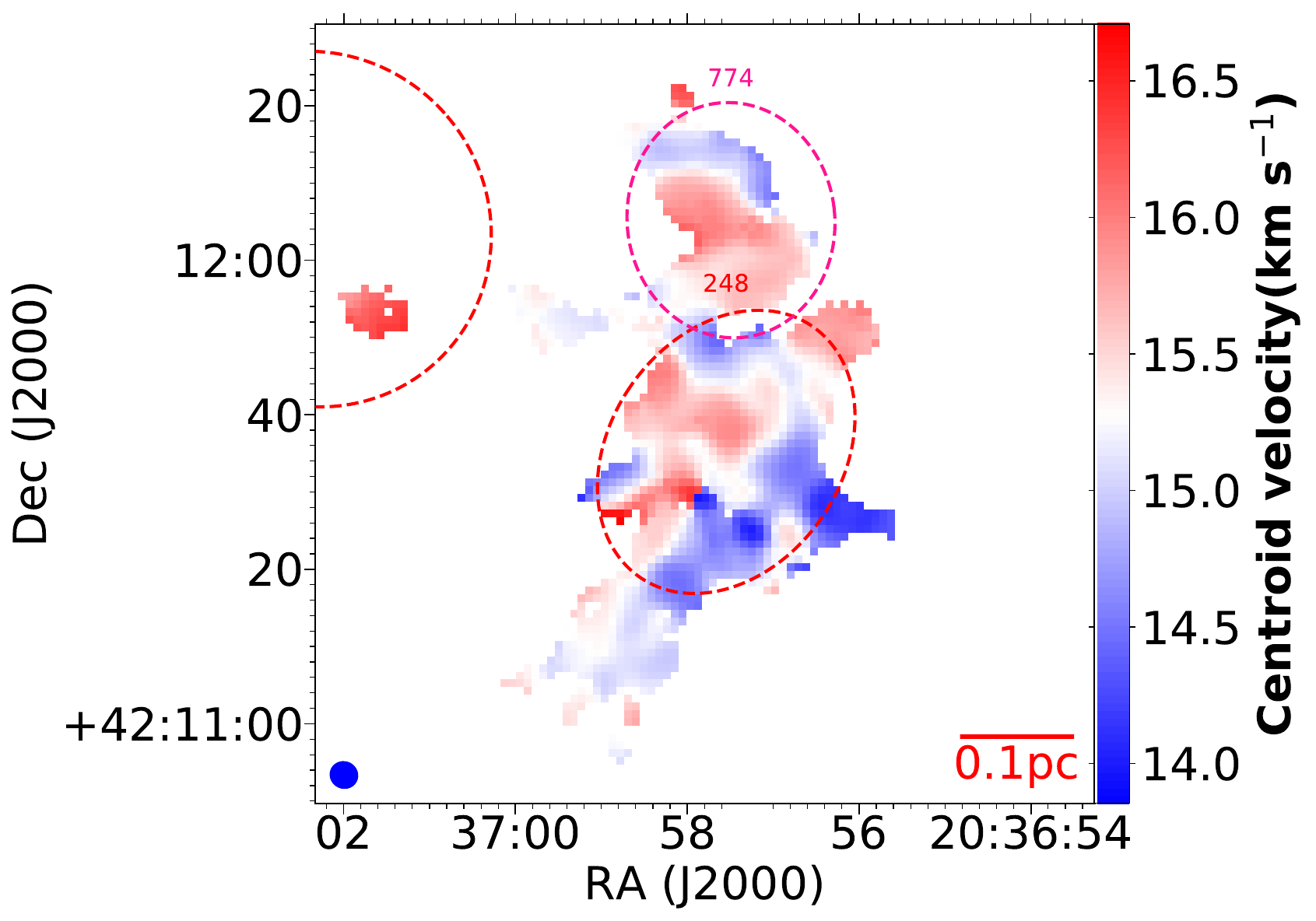} &
\includegraphics[width=.3\textwidth]{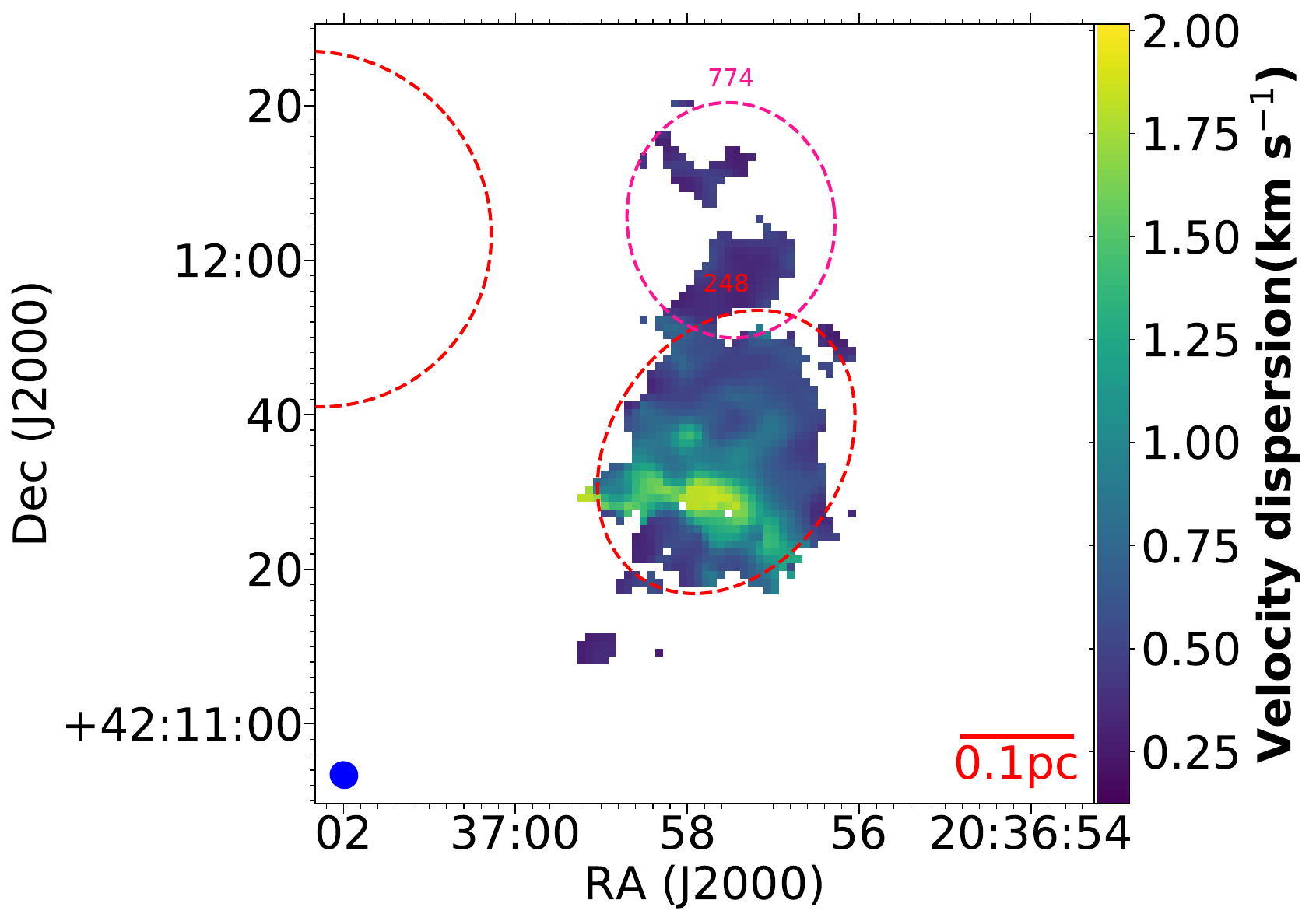} \\
 & Field 5 & \\
\includegraphics[width=.3\textwidth]{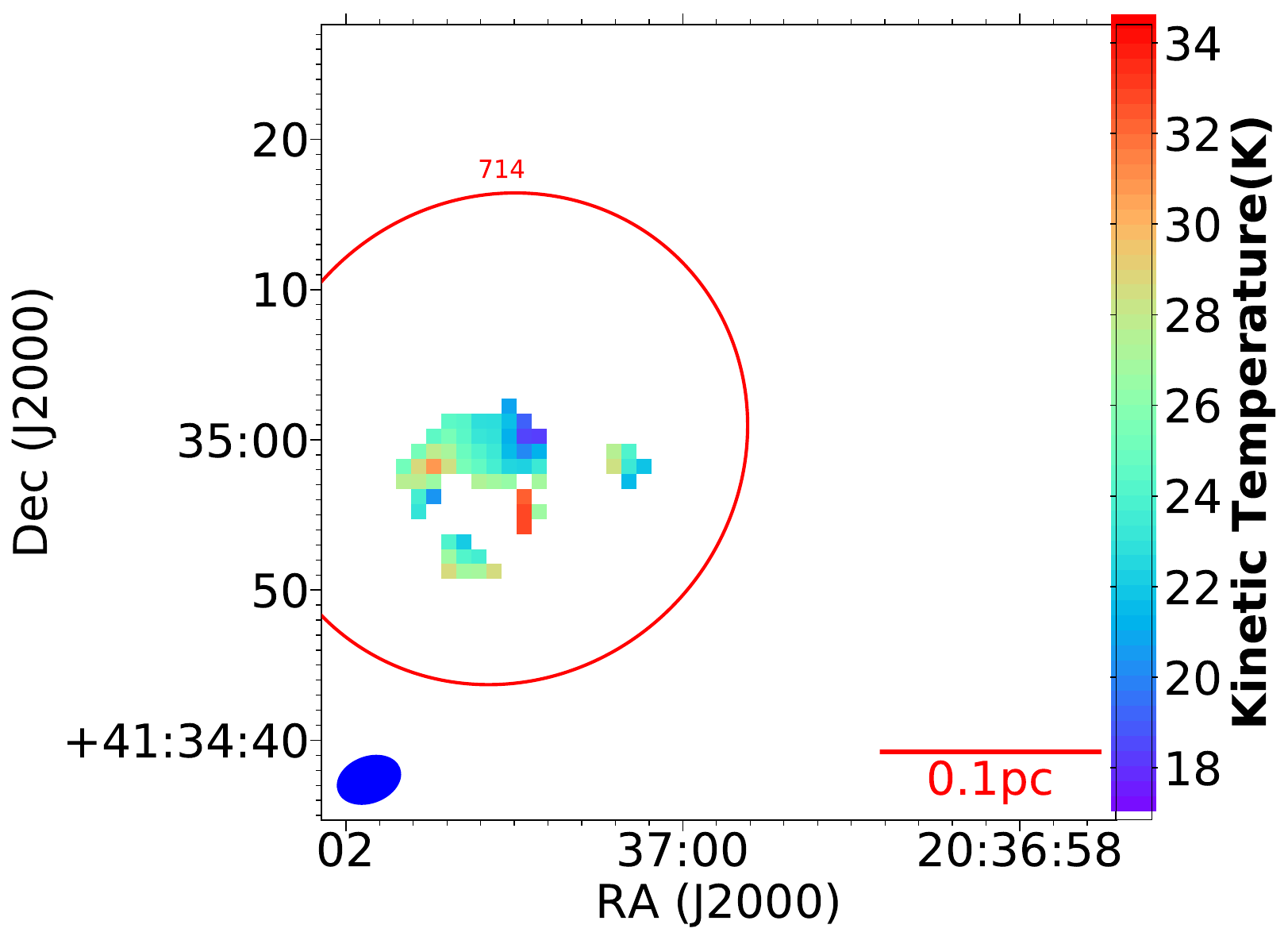} &
\includegraphics[width=.3\textwidth]{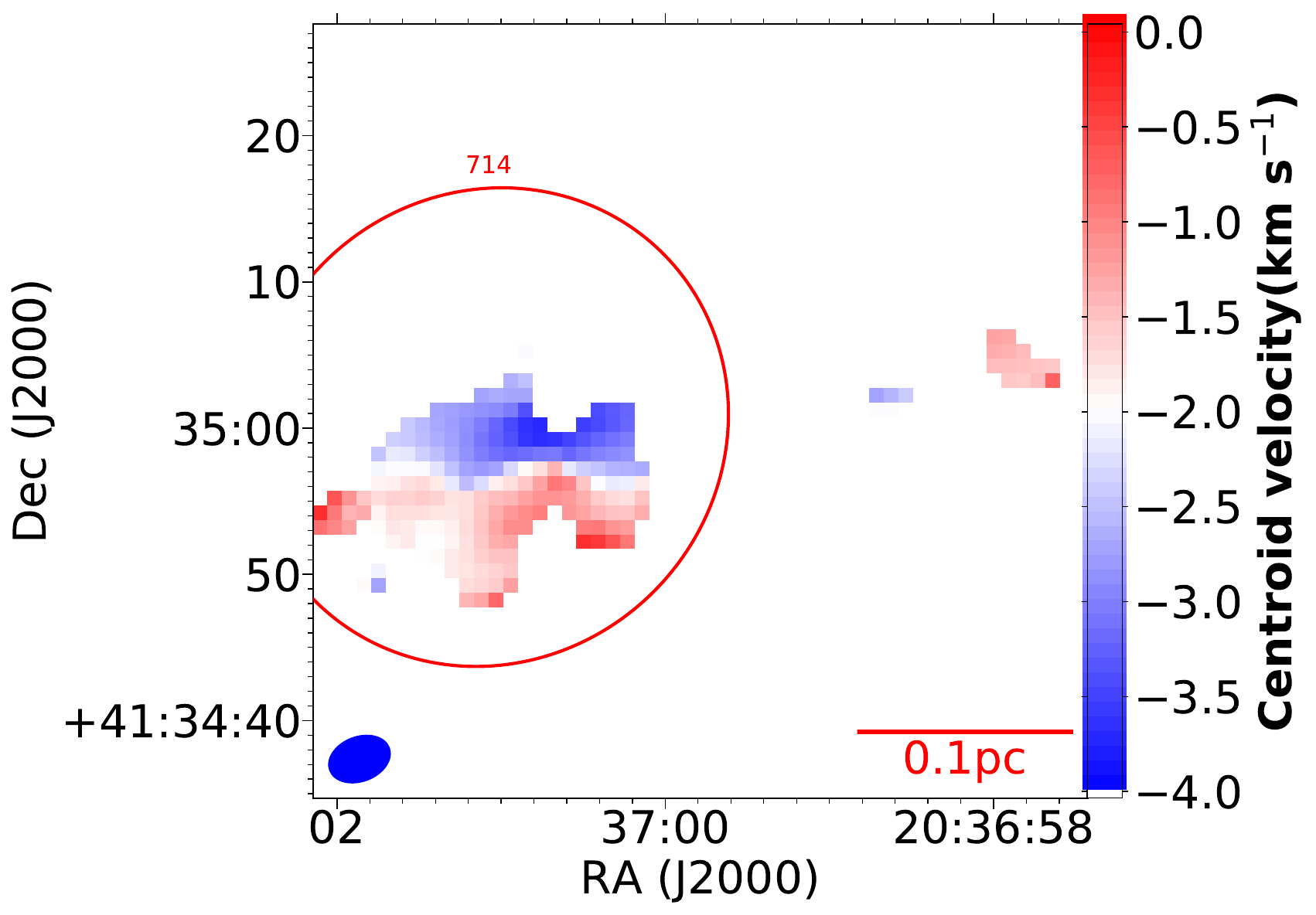} &
\includegraphics[width=.3\textwidth]{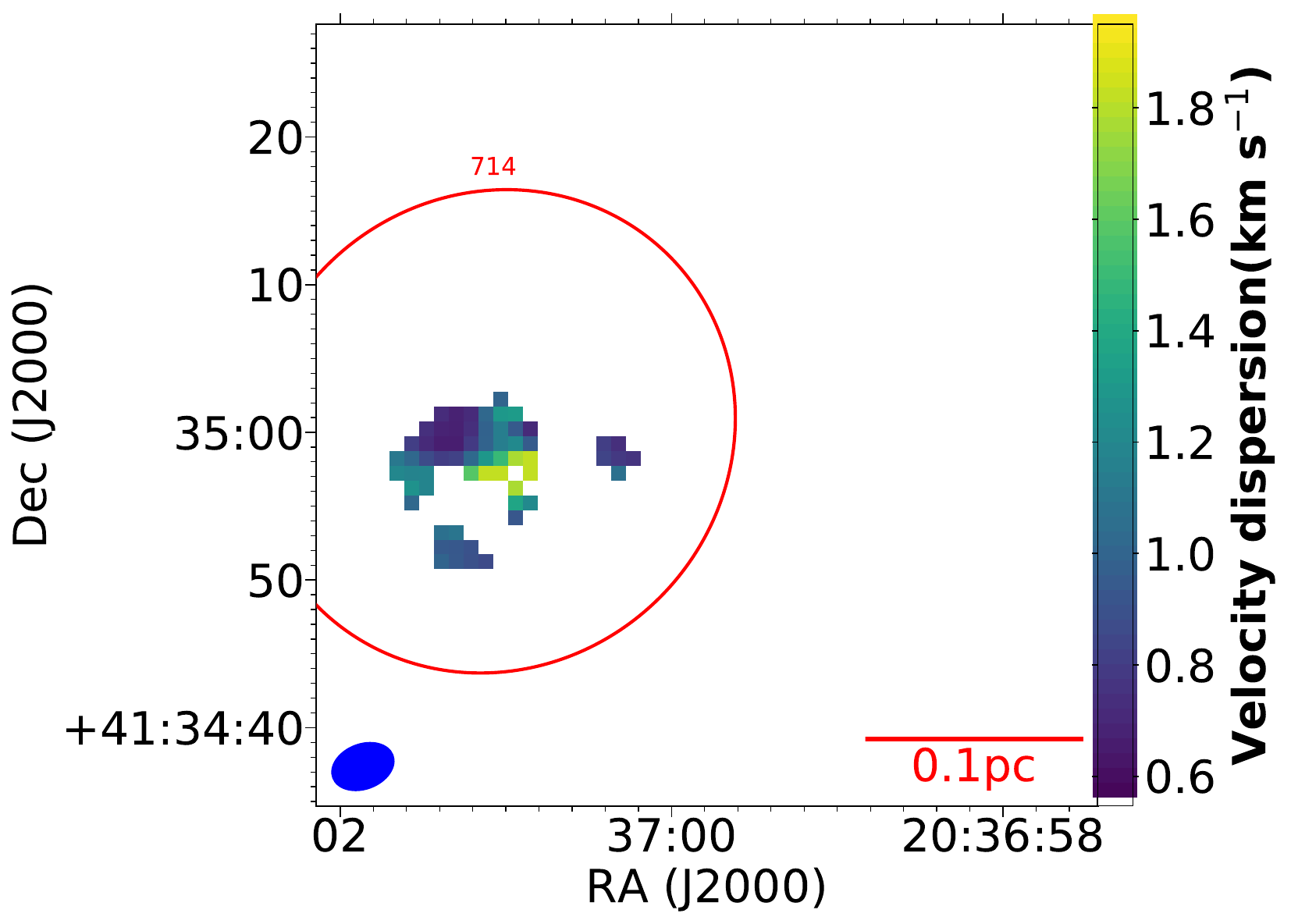} \\
 & Field 6 & \\
\includegraphics[width=.3\textwidth]{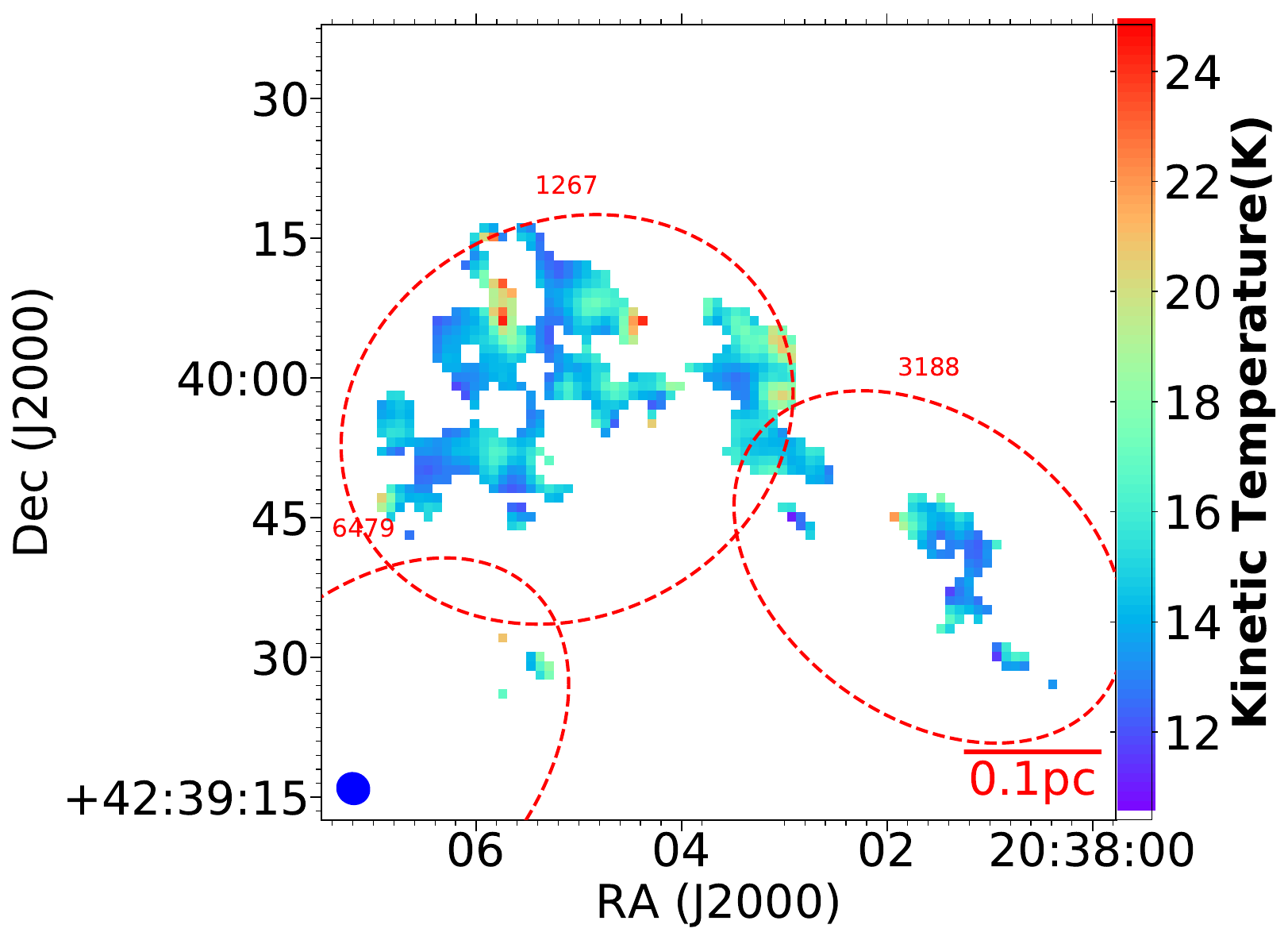} &
\includegraphics[width=.3\textwidth]{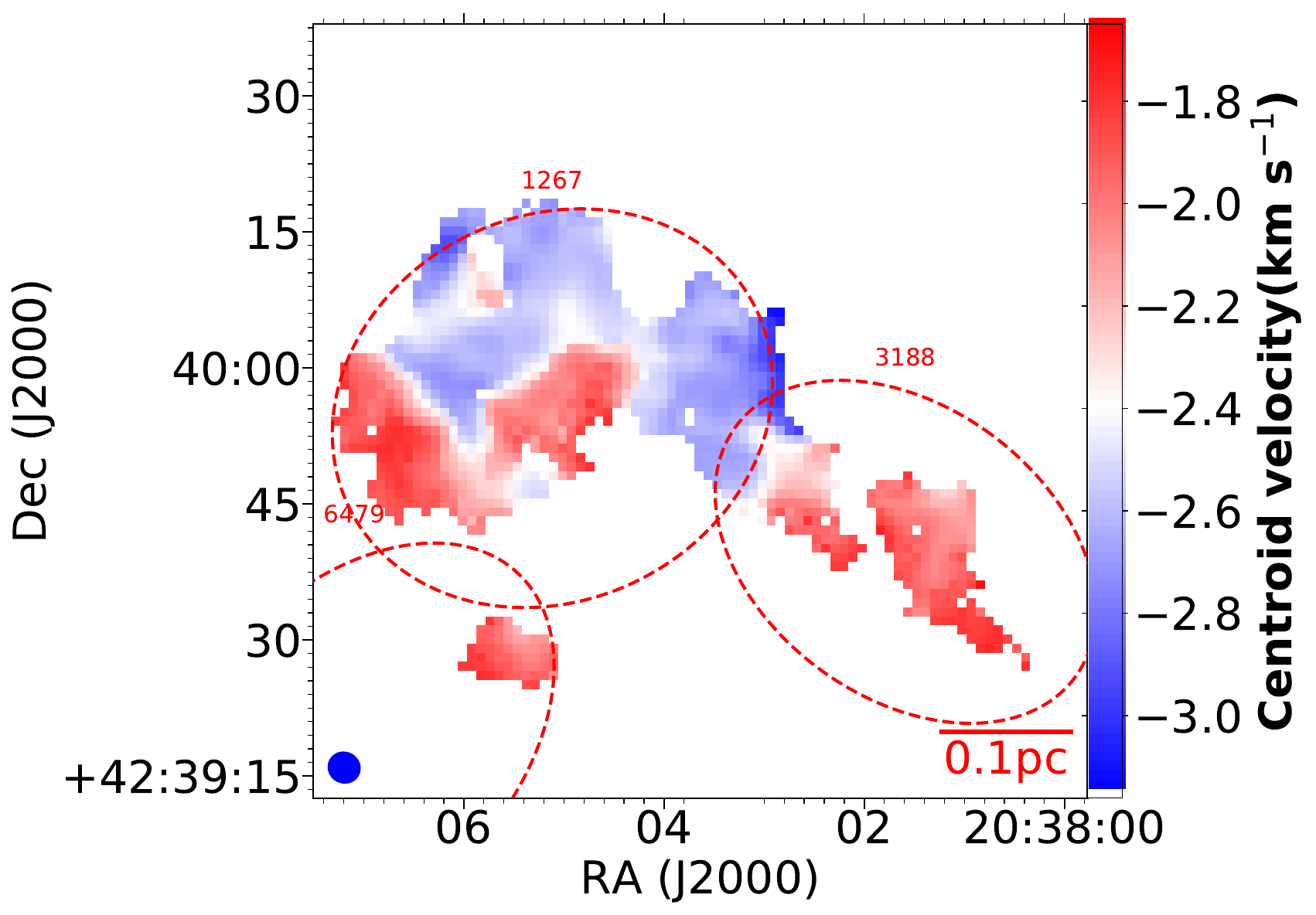} &
\includegraphics[width=.3\textwidth]{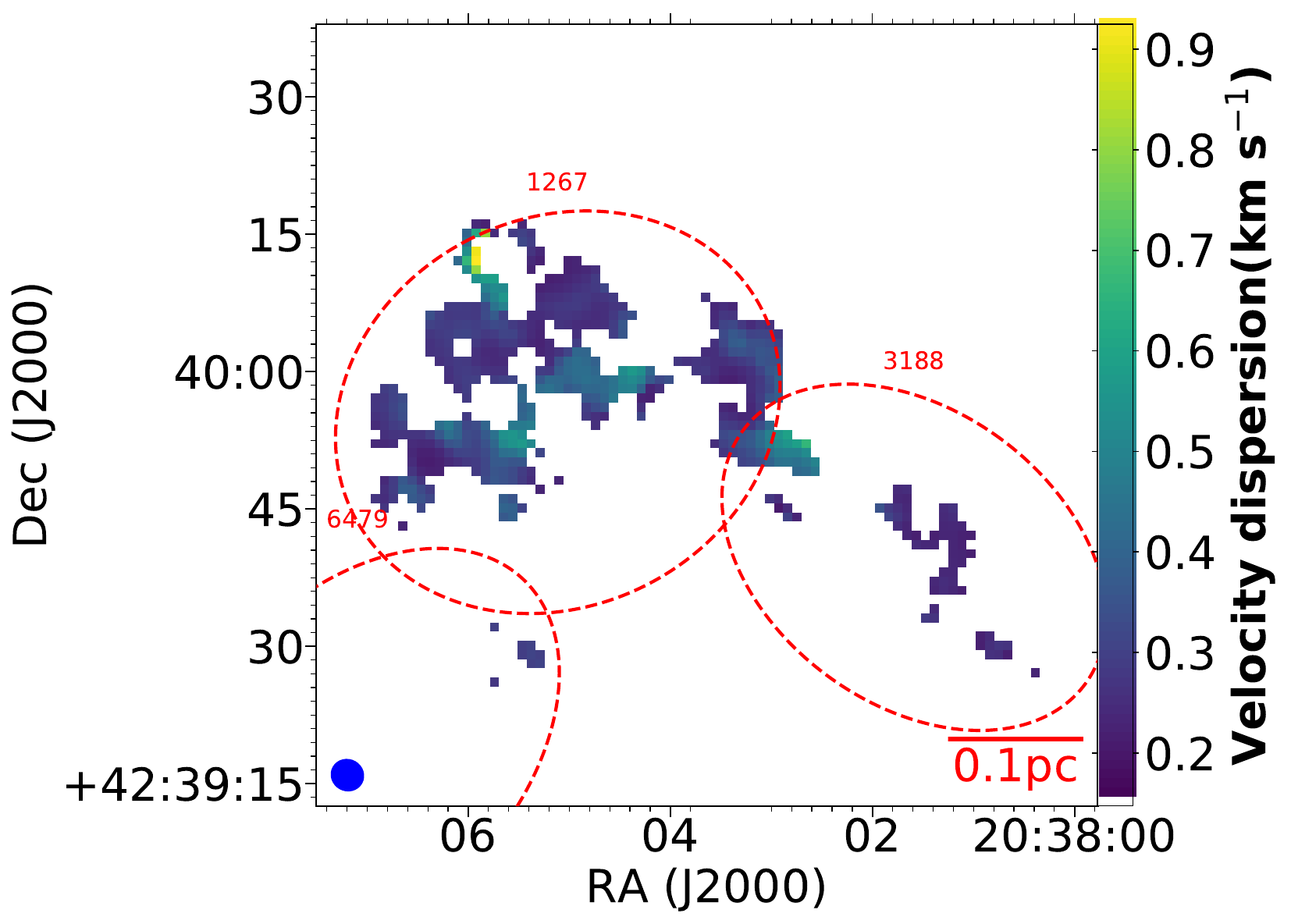} \\
 & Field 7 & \\
\includegraphics[width=.3\textwidth]{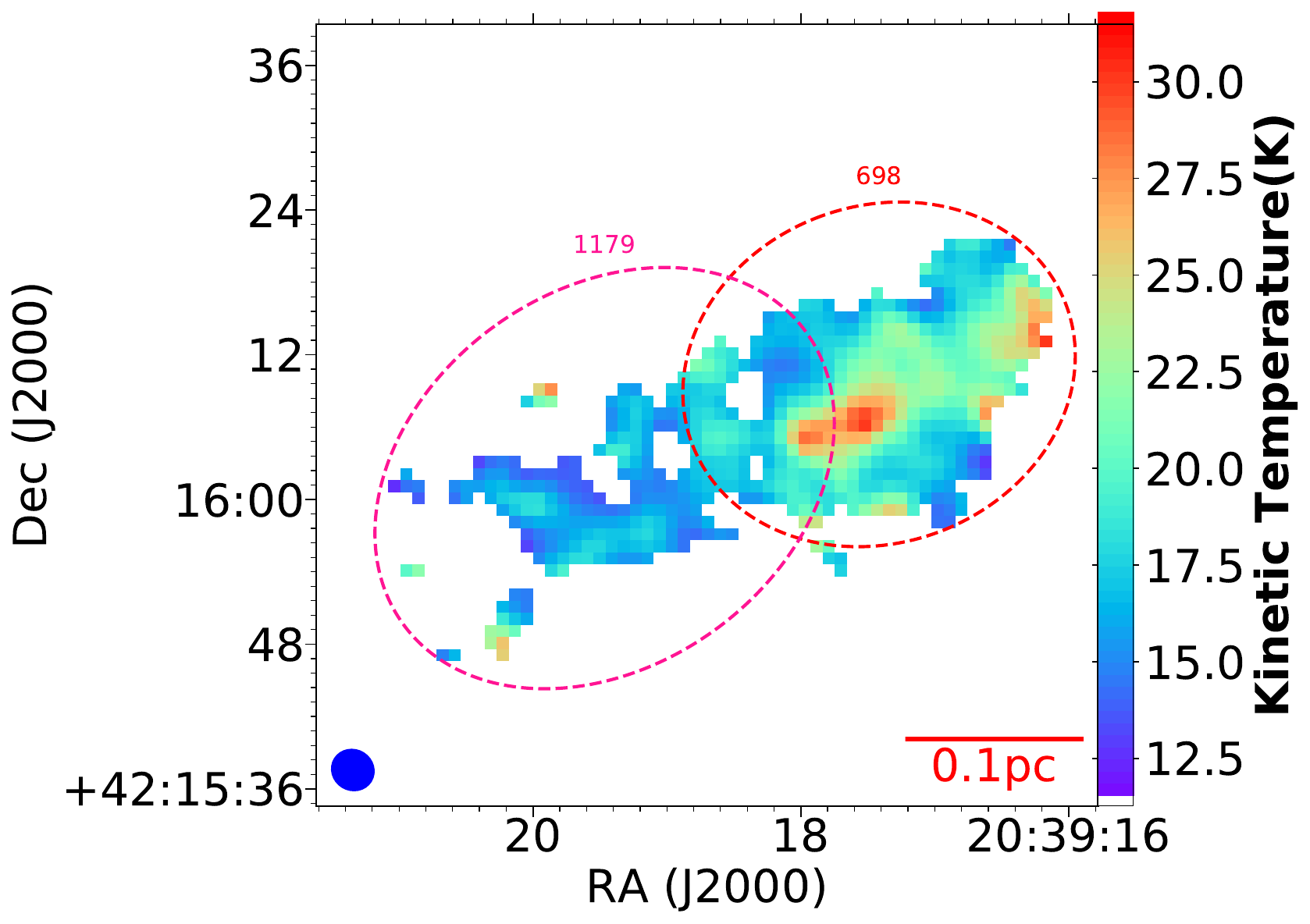} &
\includegraphics[width=.3\textwidth]{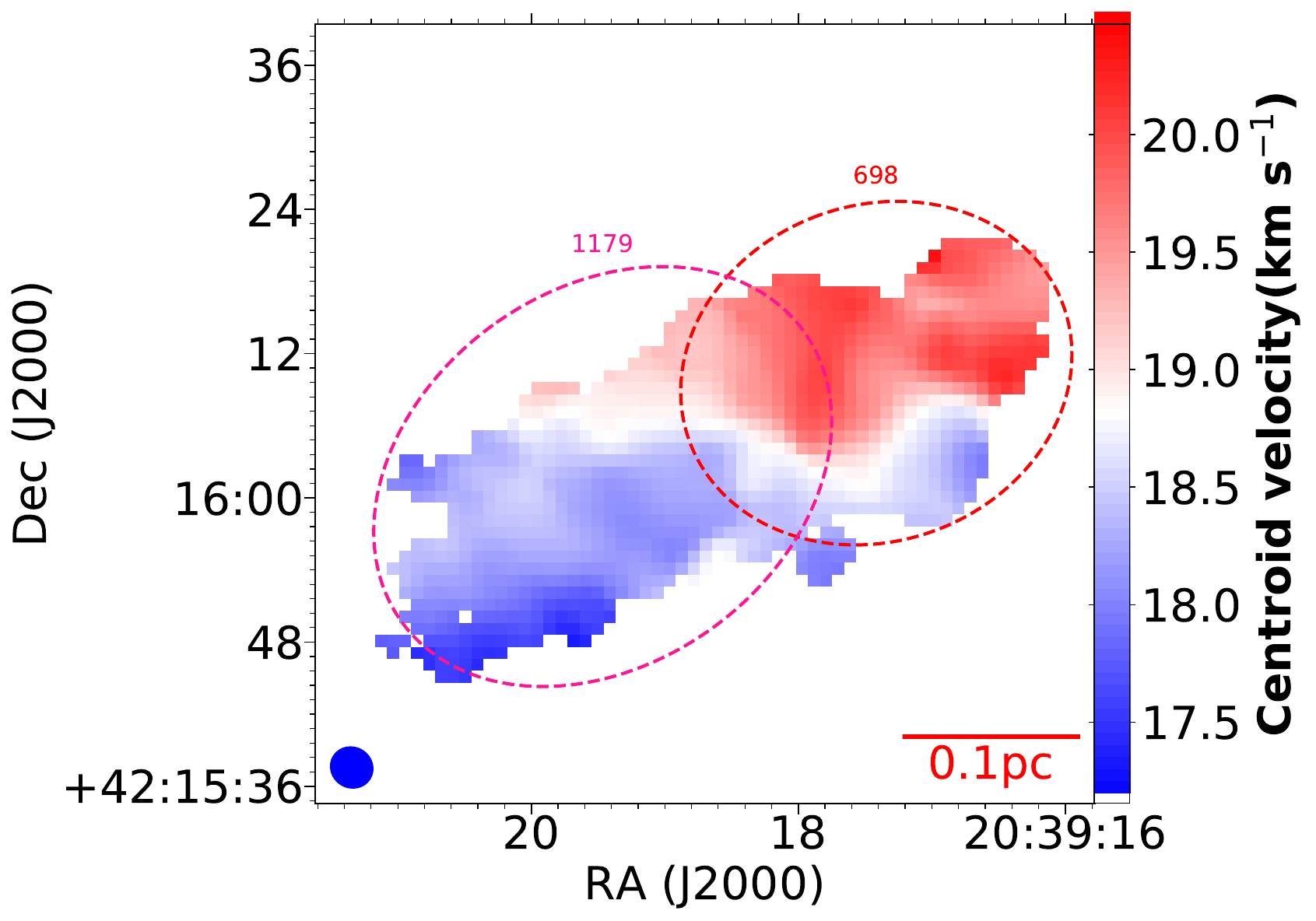} &
\includegraphics[width=.3\textwidth]{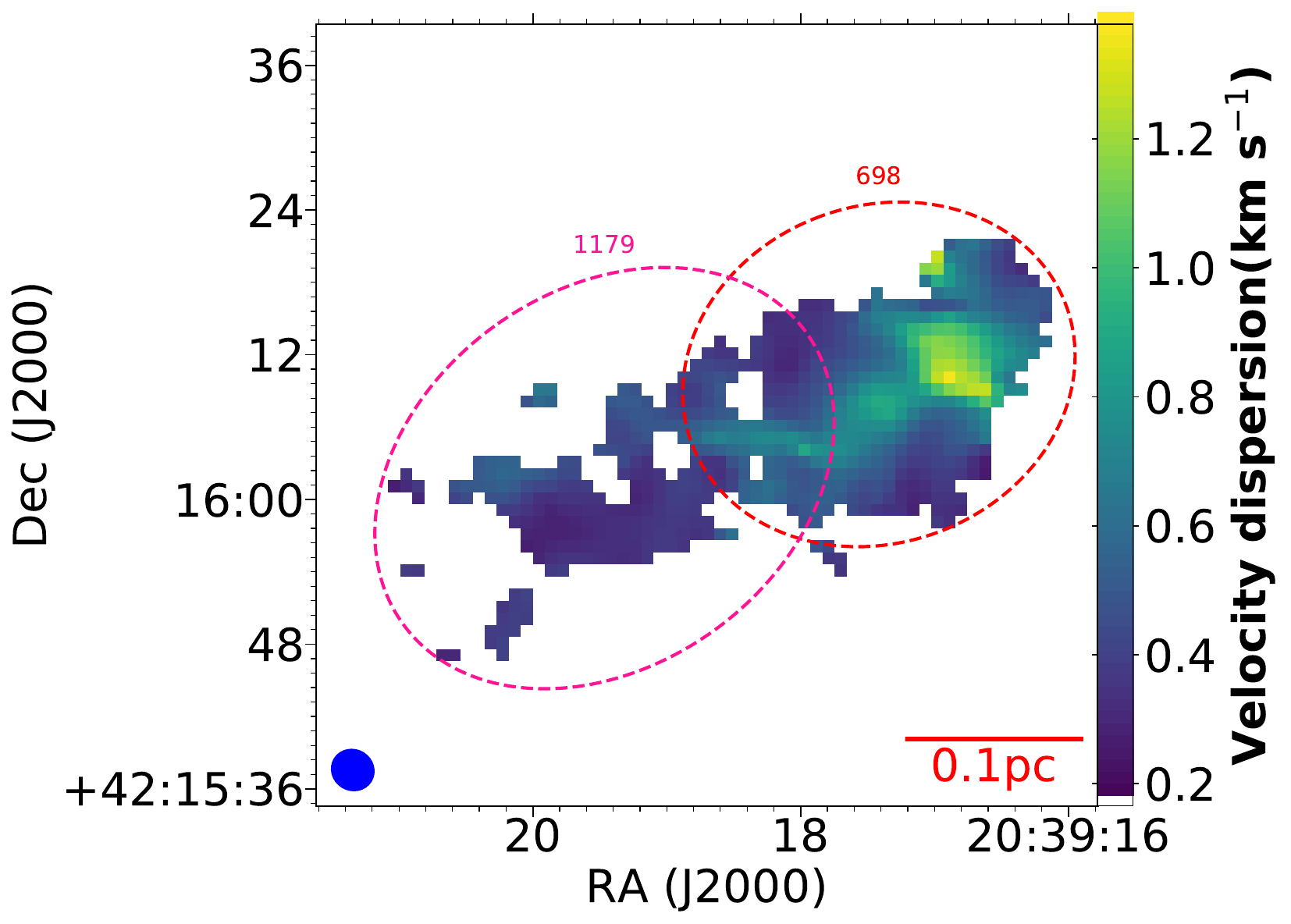} \\
 & Field 8 & \\
\includegraphics[width=.3\textwidth]{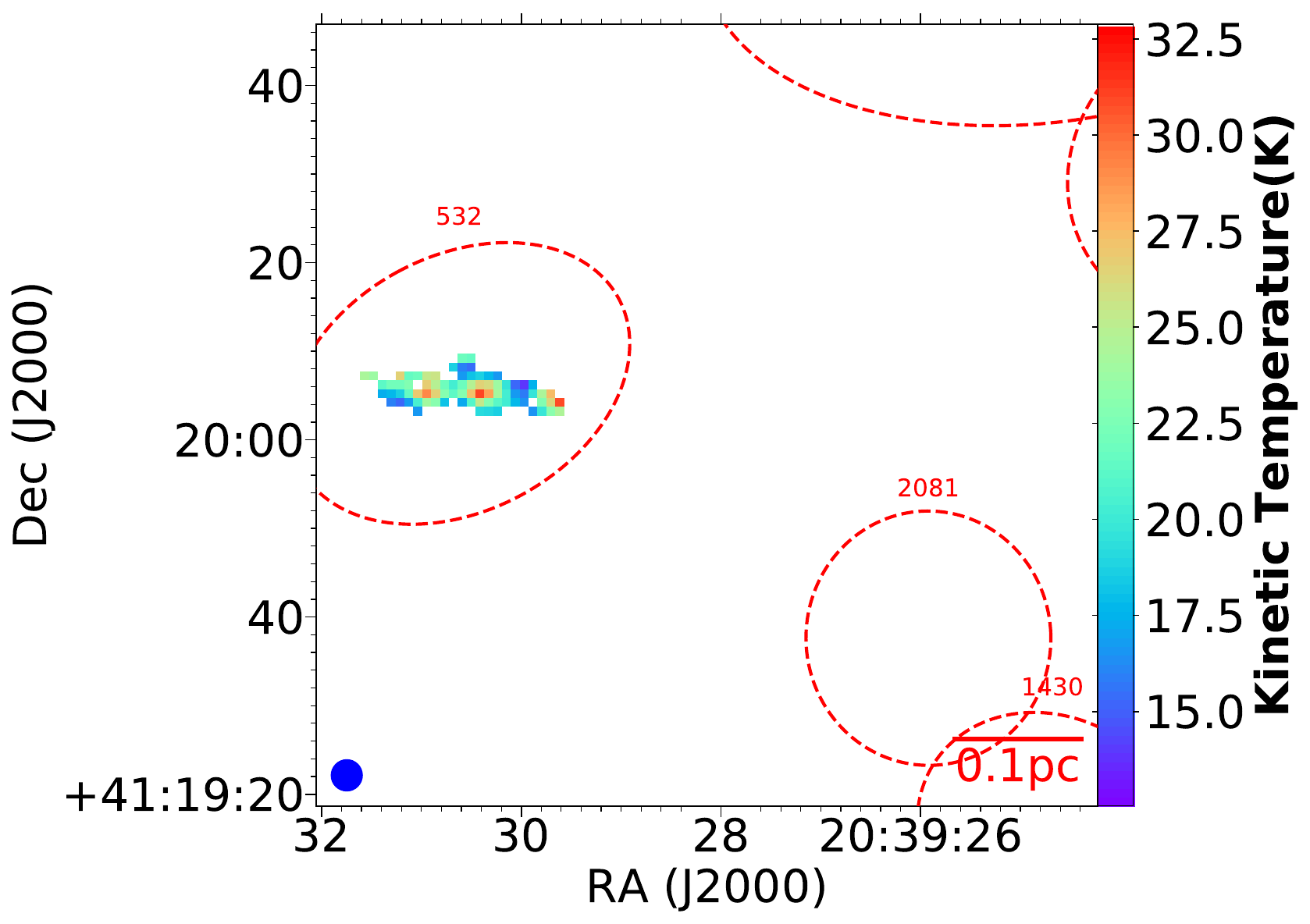} &
\includegraphics[width=.3\textwidth]{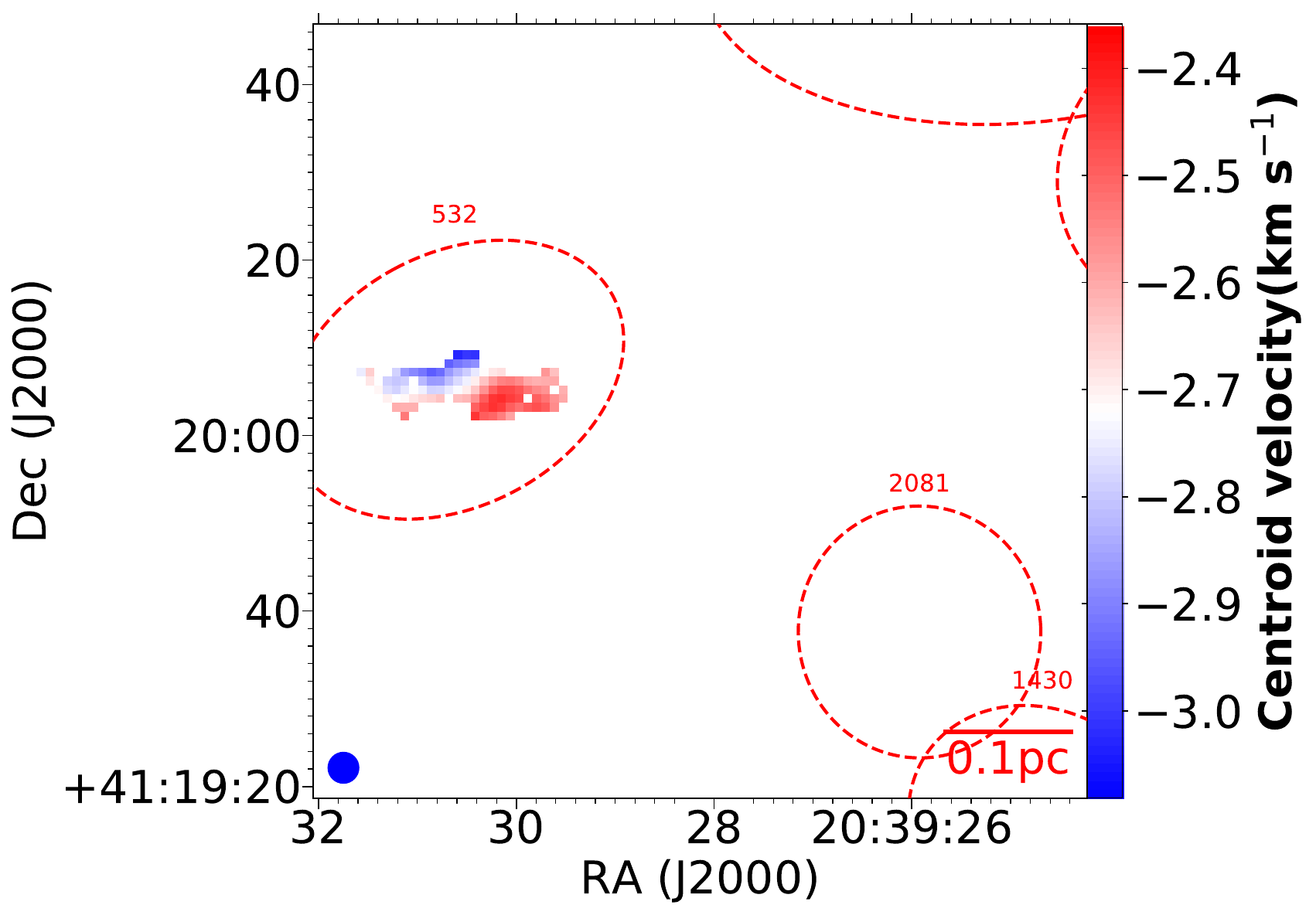} &
\includegraphics[width=.3\textwidth]{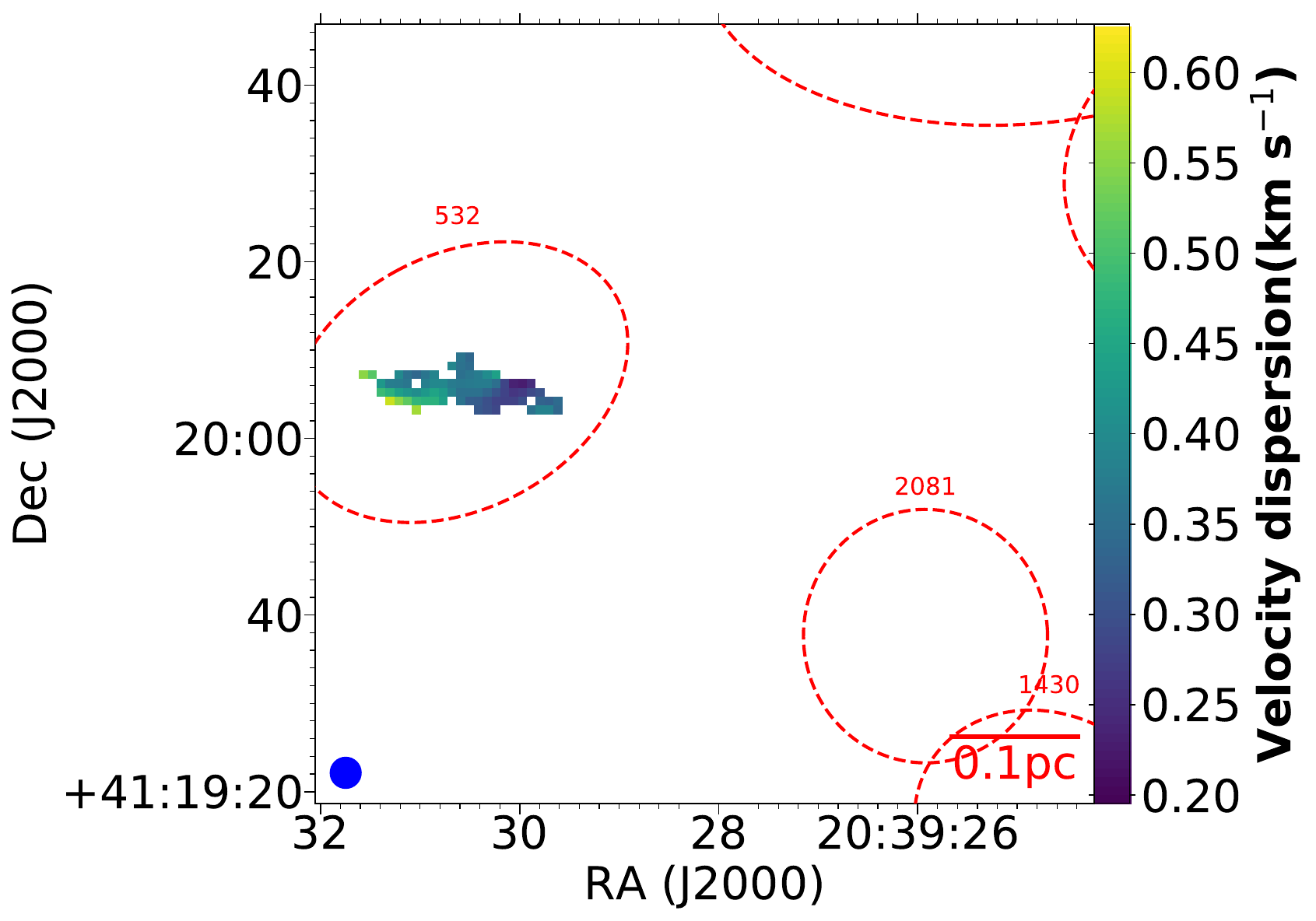} \\
 & Field 9 & \\
\includegraphics[width=.3\textwidth]{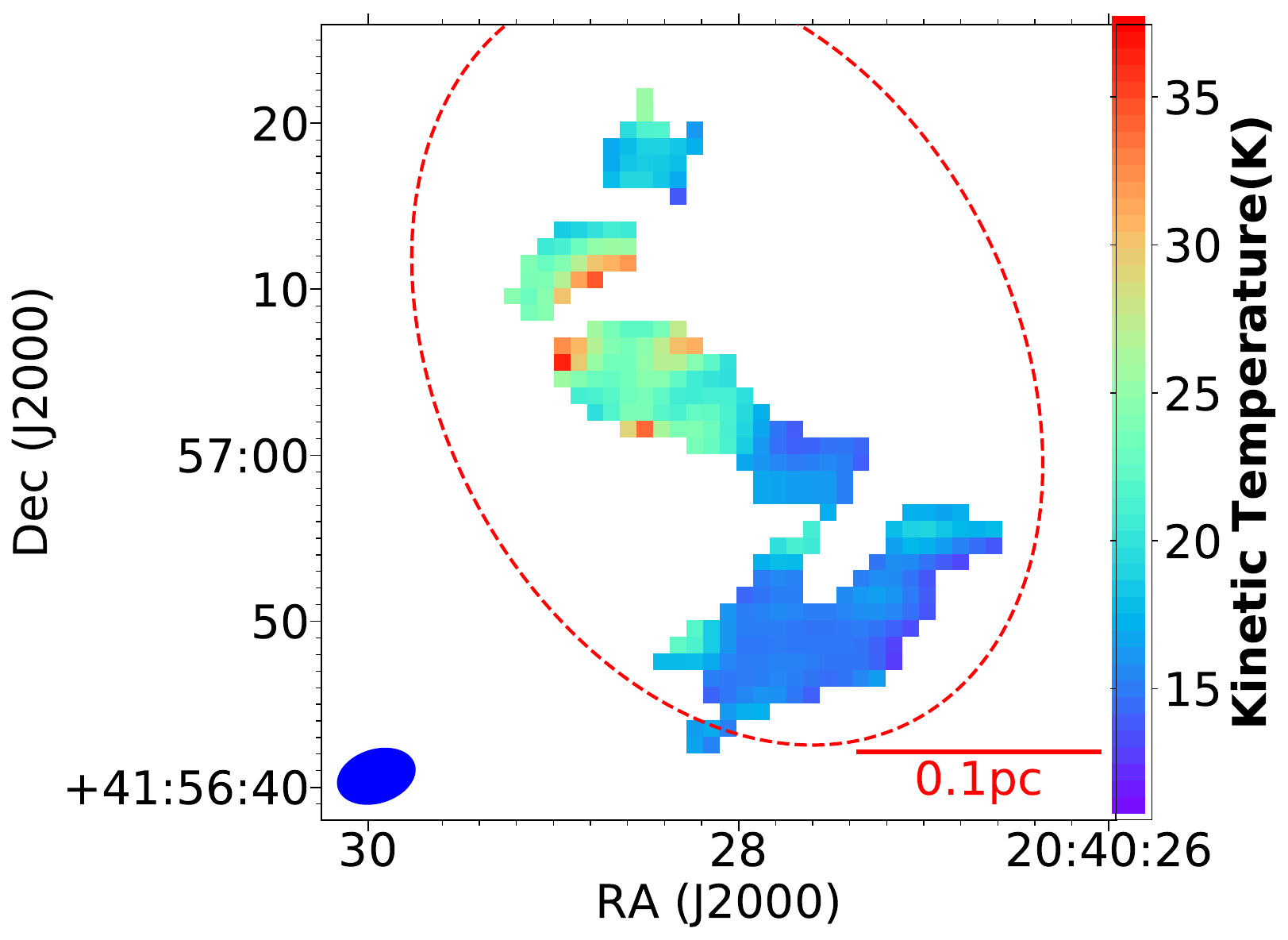} &
\includegraphics[width=.3\textwidth]{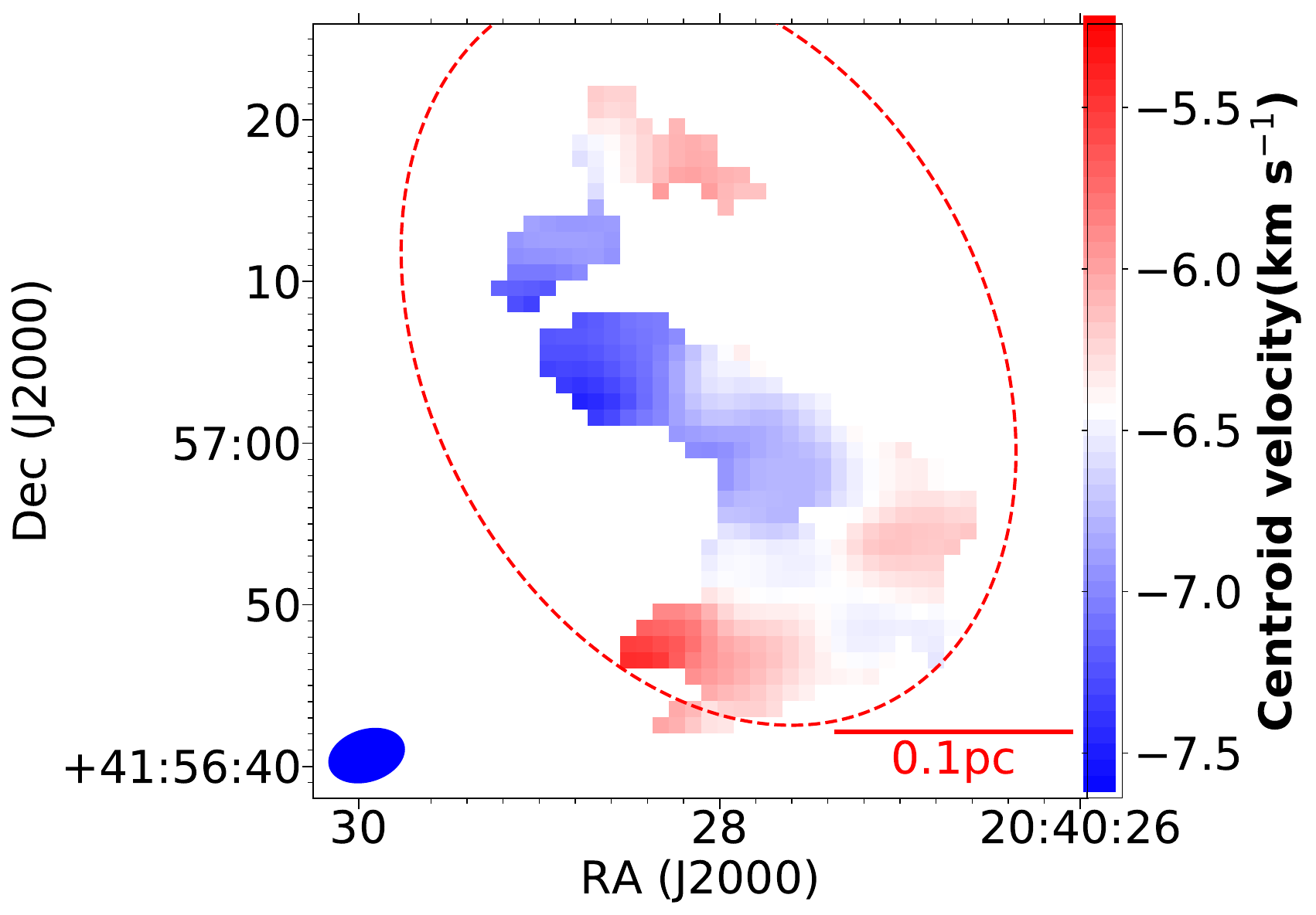} &
\includegraphics[width=.3\textwidth]{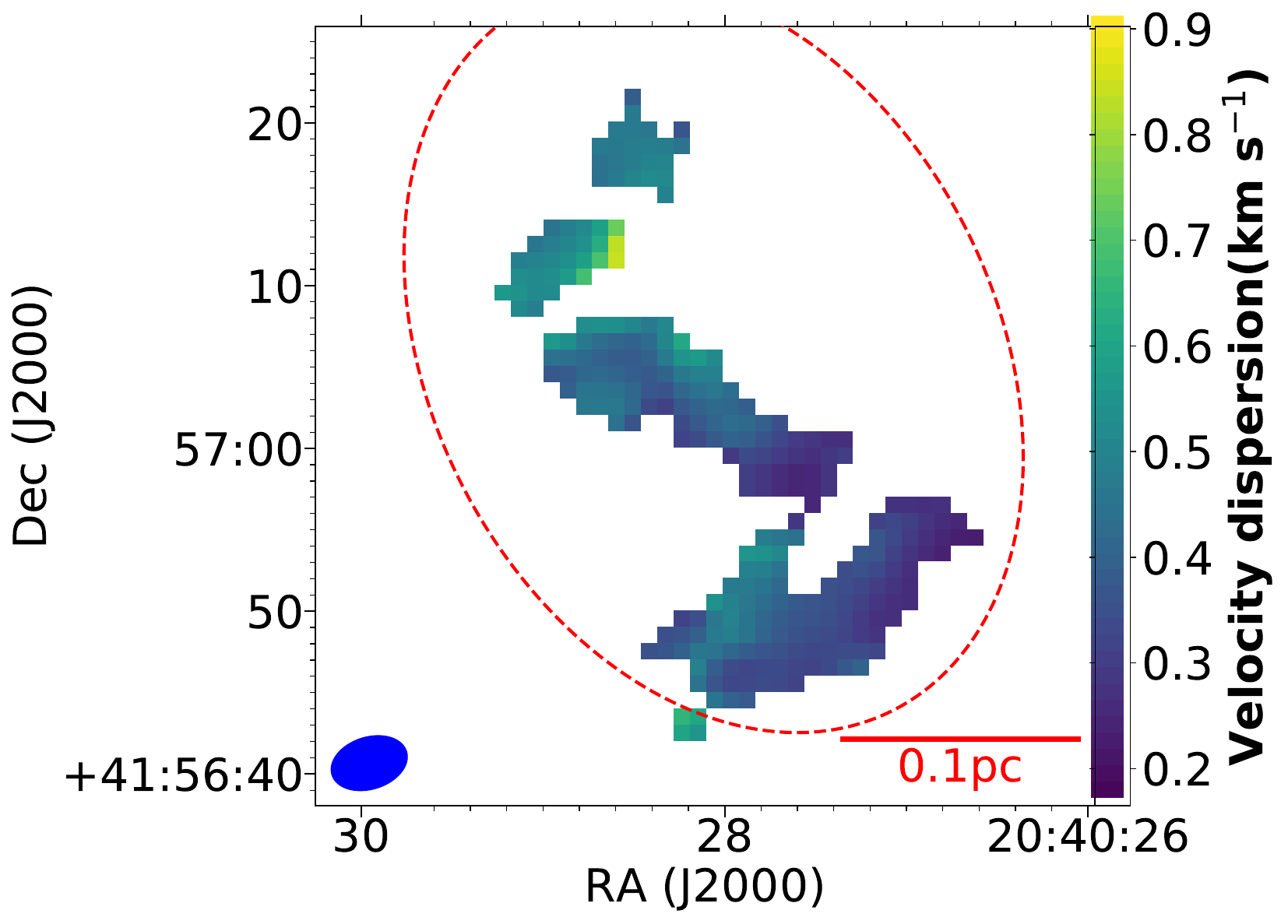} \\
 & Field 10 & \\
\includegraphics[width=.3\textwidth]{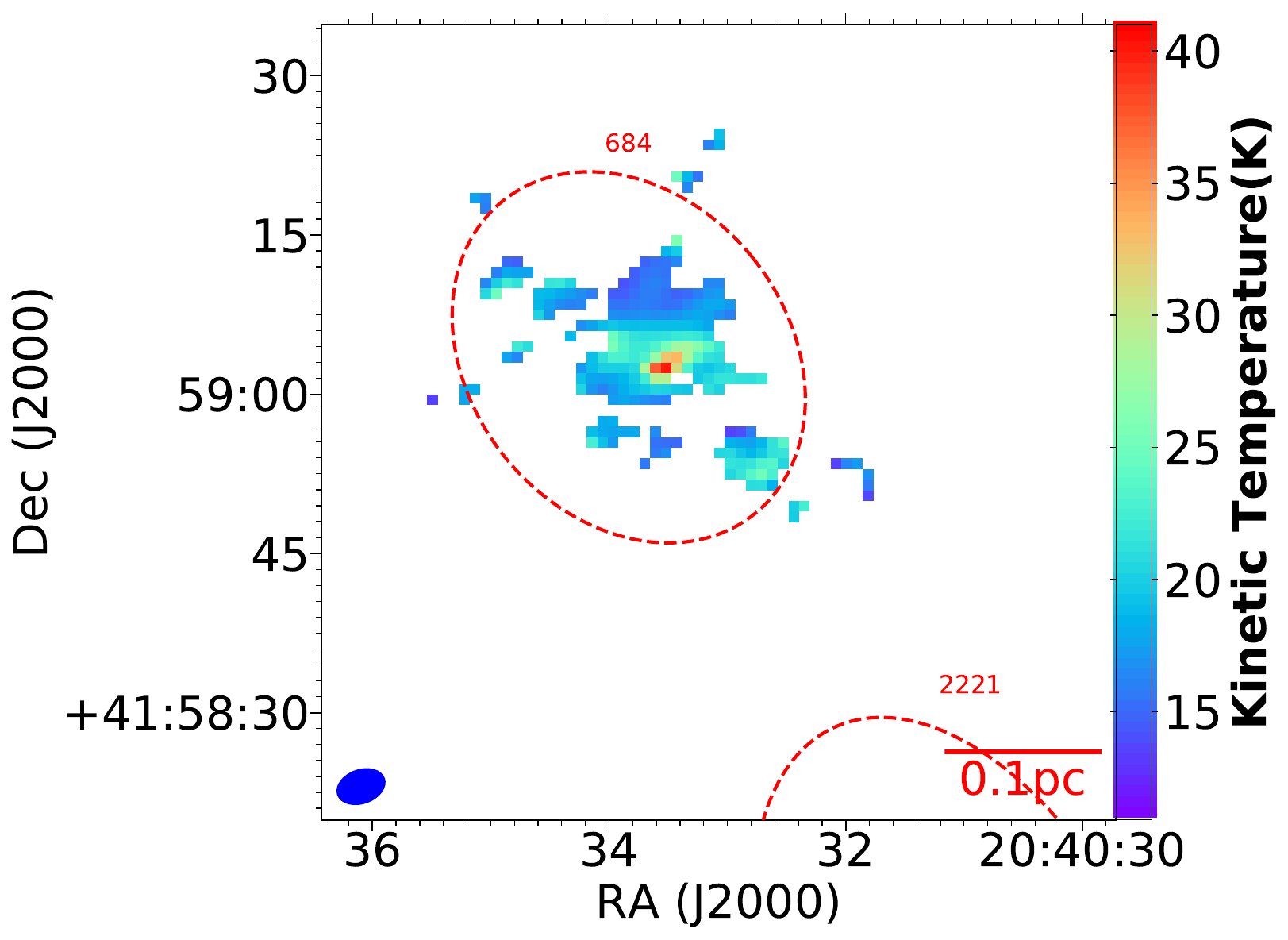} &
\includegraphics[width=.3\textwidth]{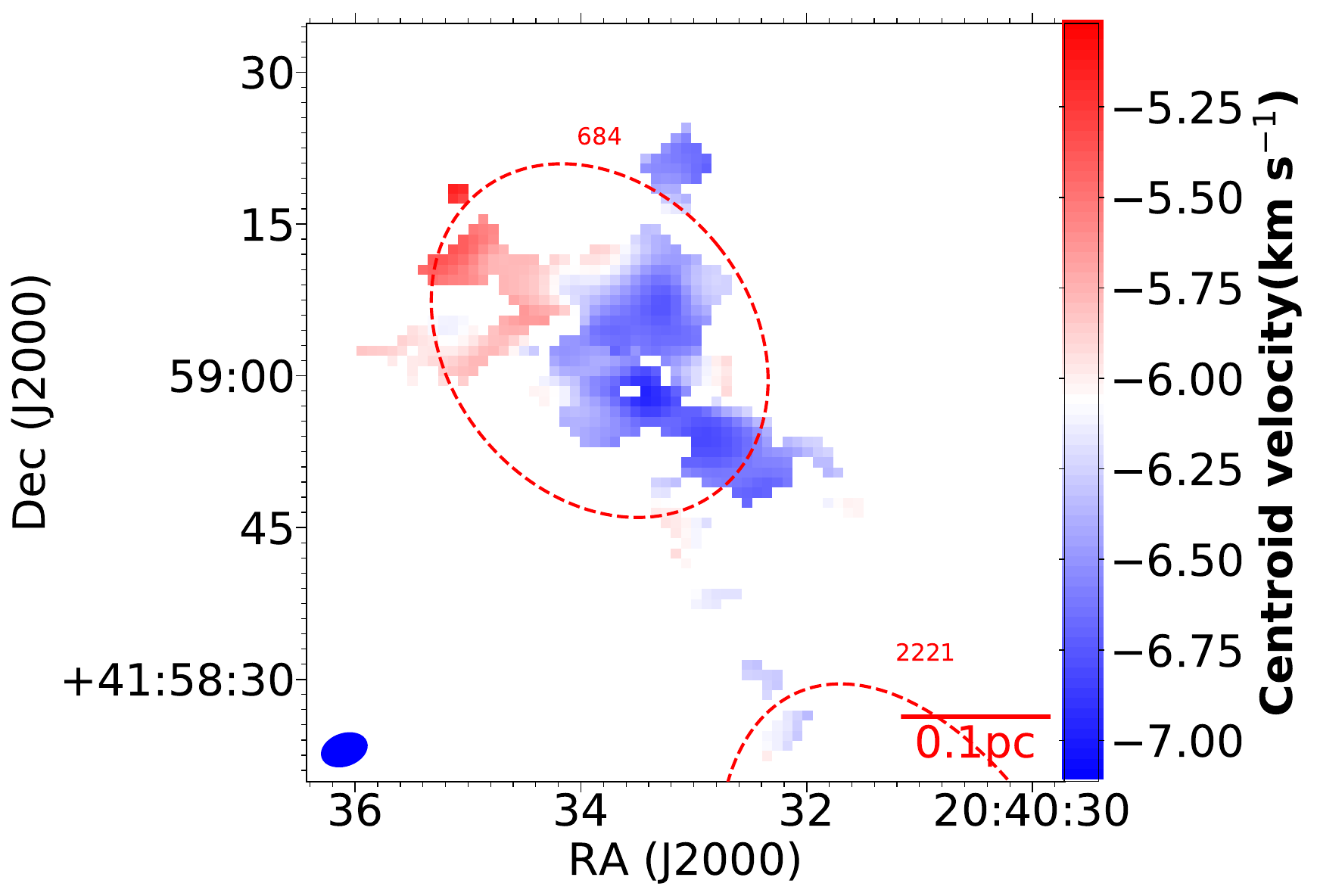} &
\includegraphics[width=.3\textwidth]{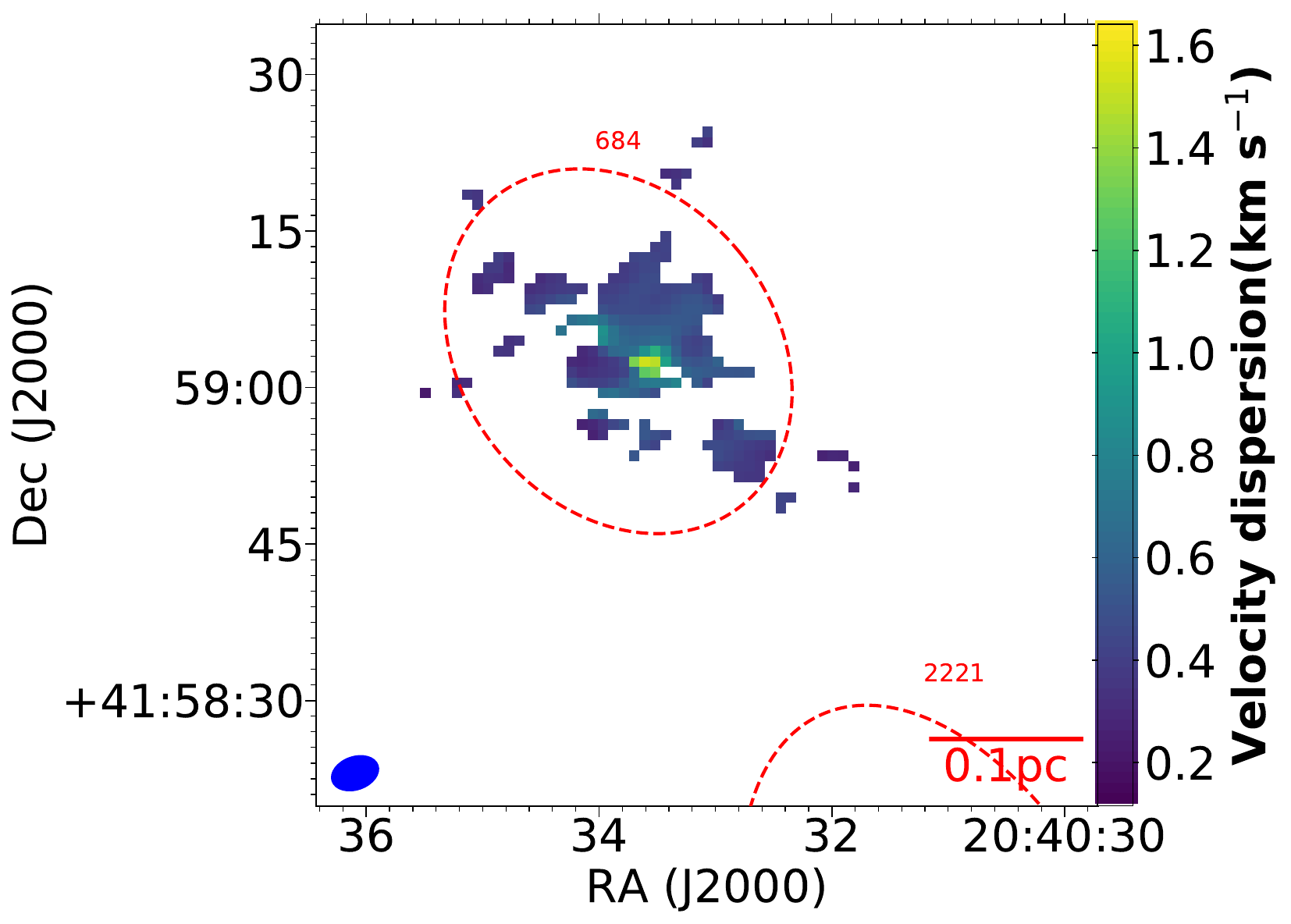} \\
 & Field 11 & \\
\includegraphics[width=.3\textwidth]{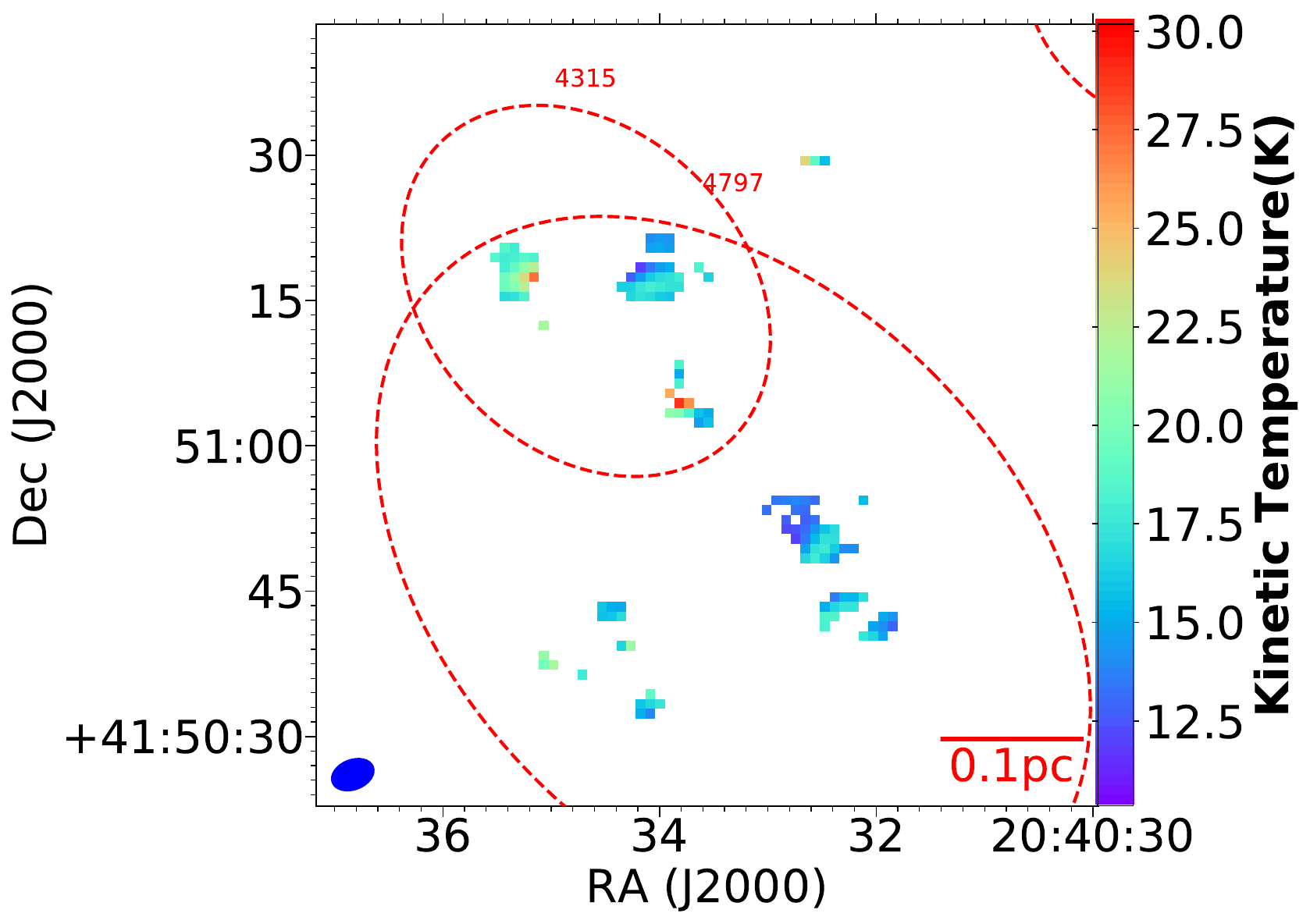} &
\includegraphics[width=.3\textwidth]{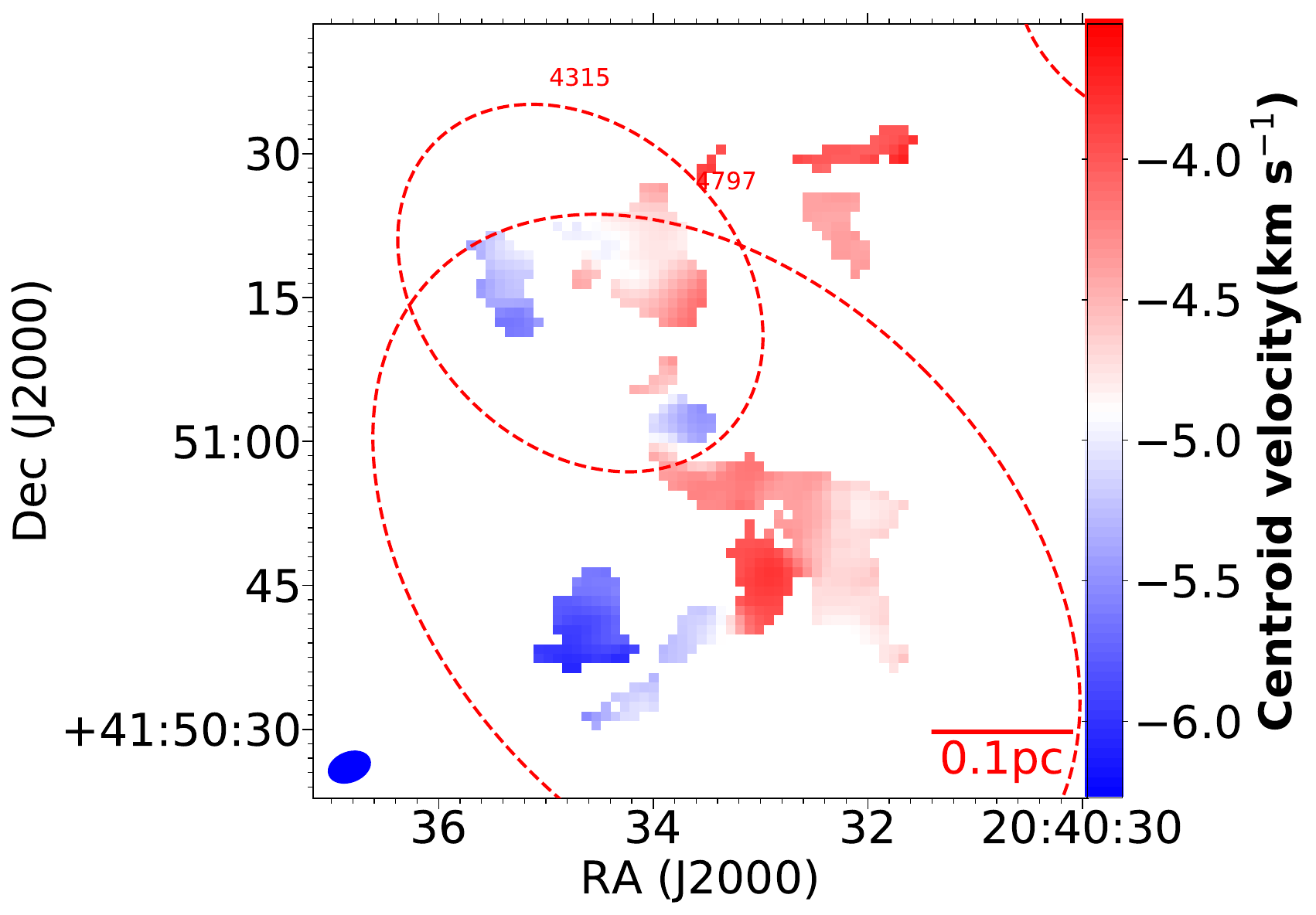} &
\includegraphics[width=.3\textwidth]{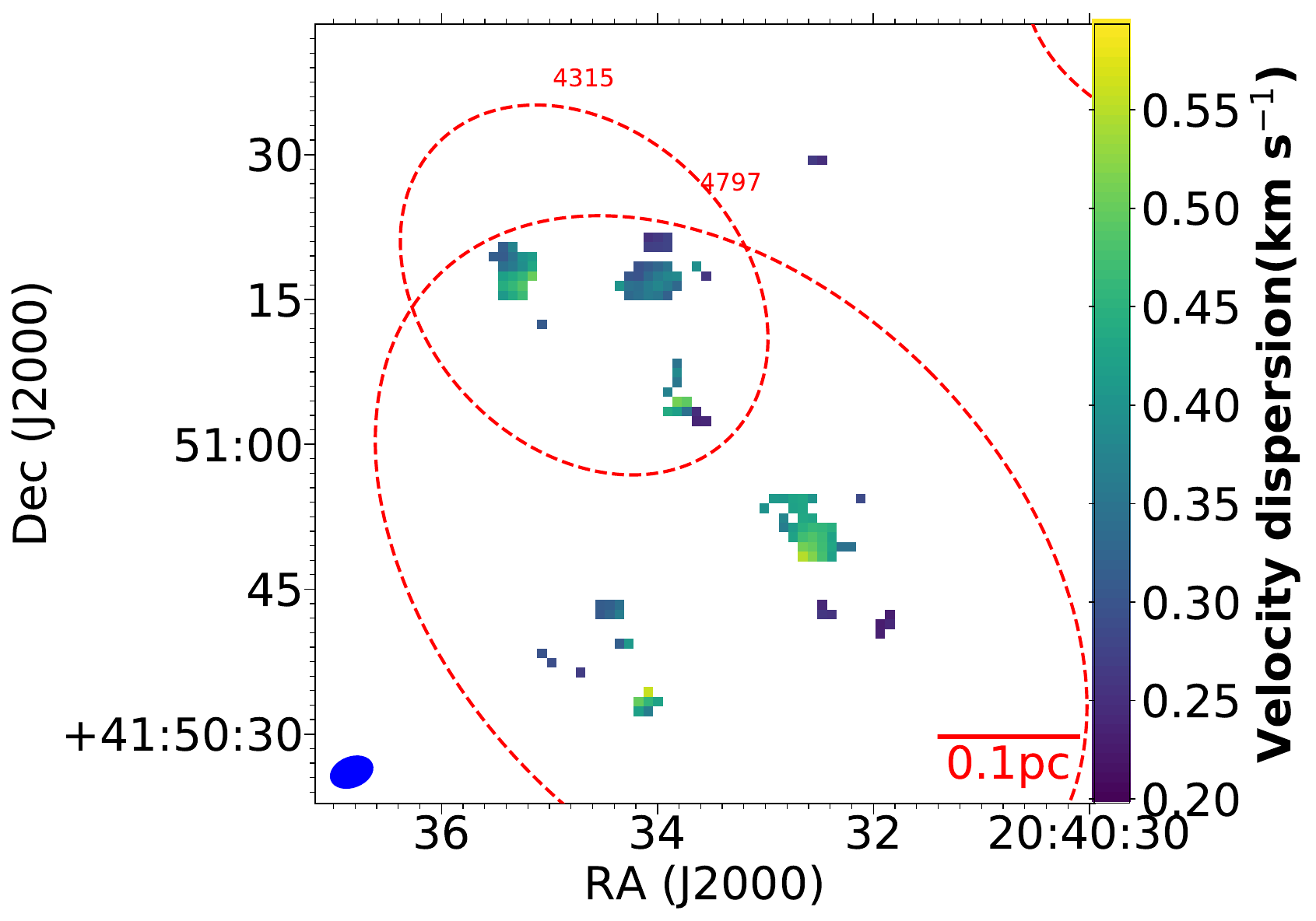} \\
 & Field 12 & \\
\includegraphics[width=.3\textwidth]{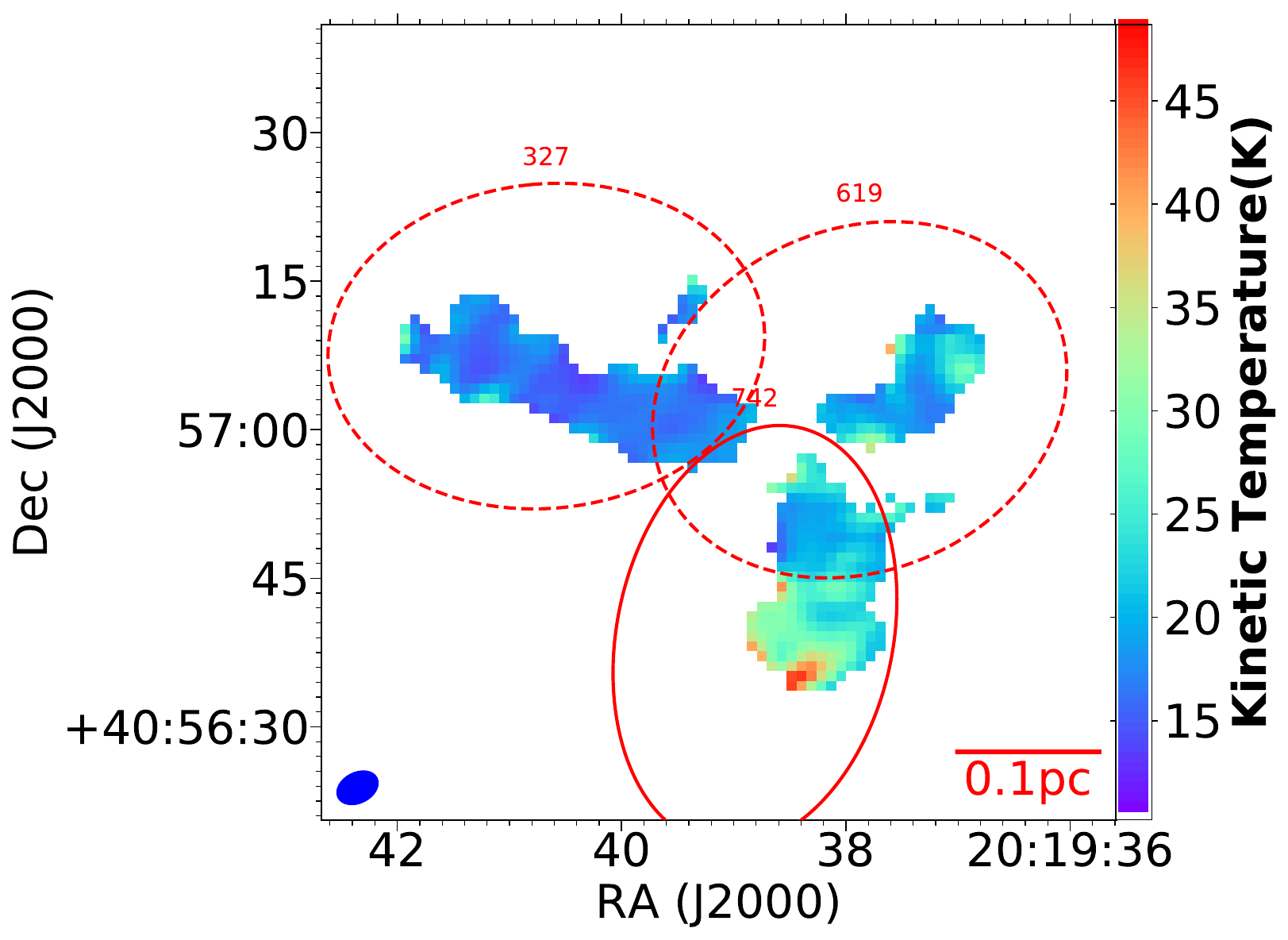} &
\includegraphics[width=.3\textwidth]{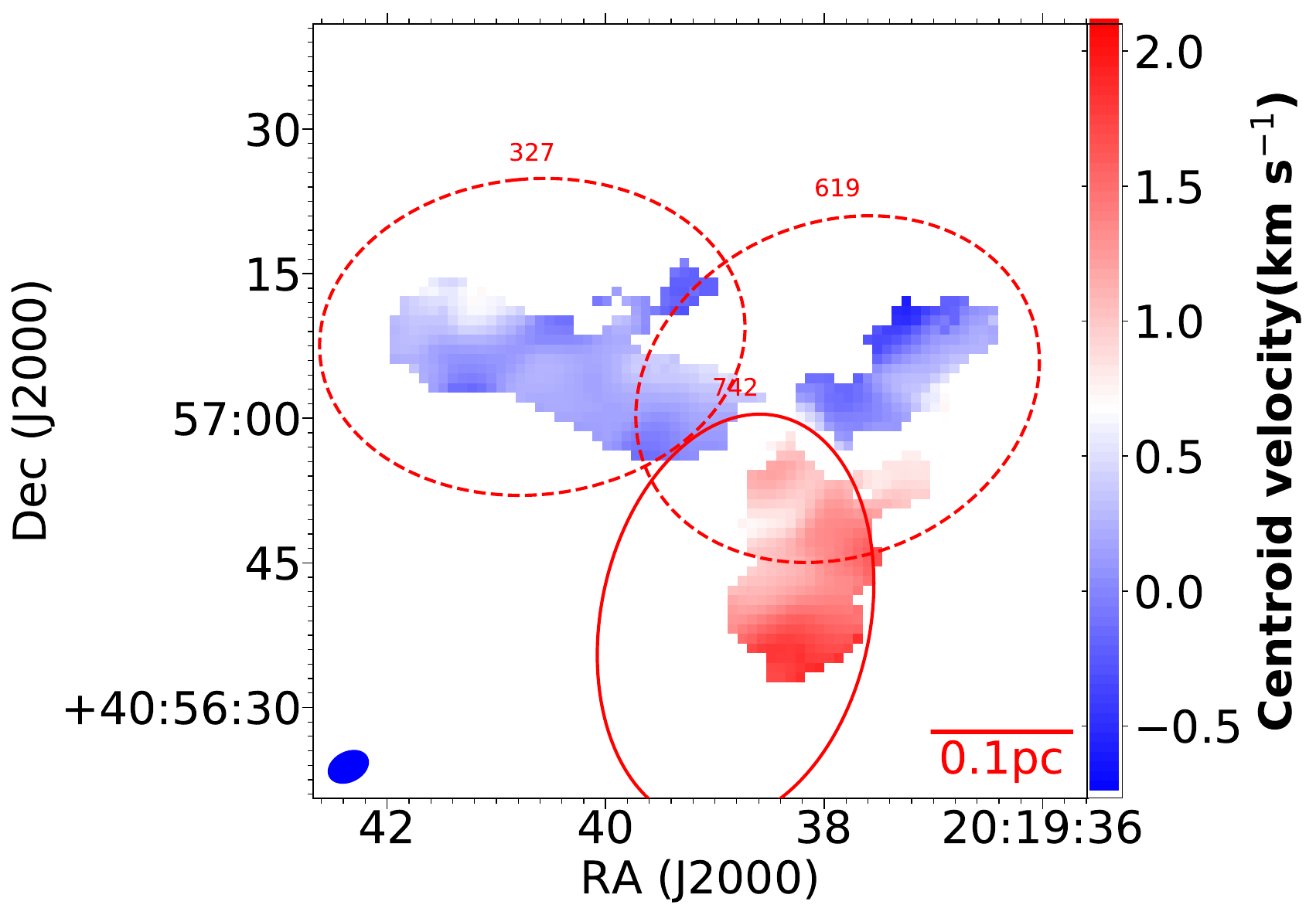} &
\includegraphics[width=.3\textwidth]{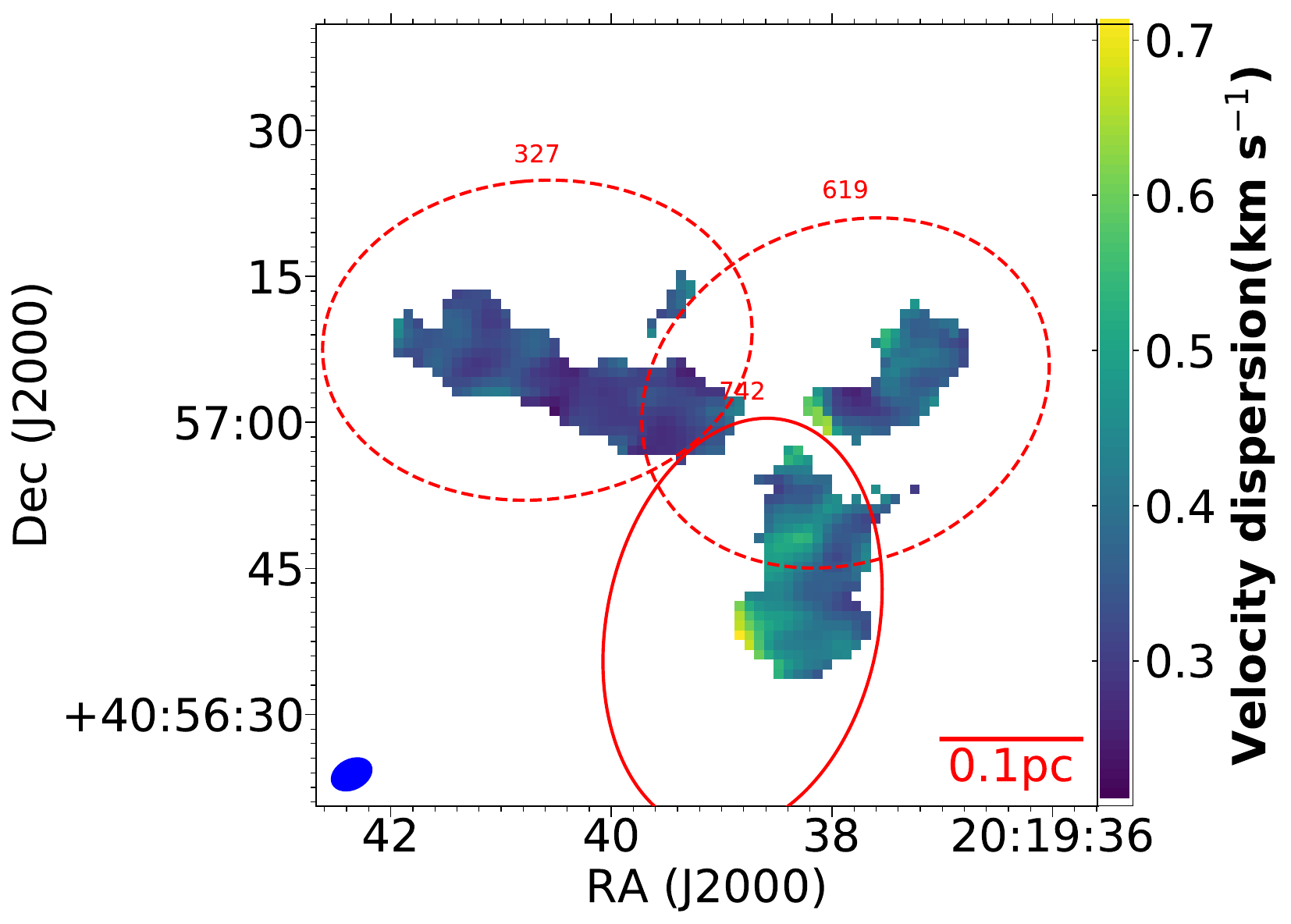} \\
 & Field 13 & \\
\includegraphics[width=.3\textwidth]{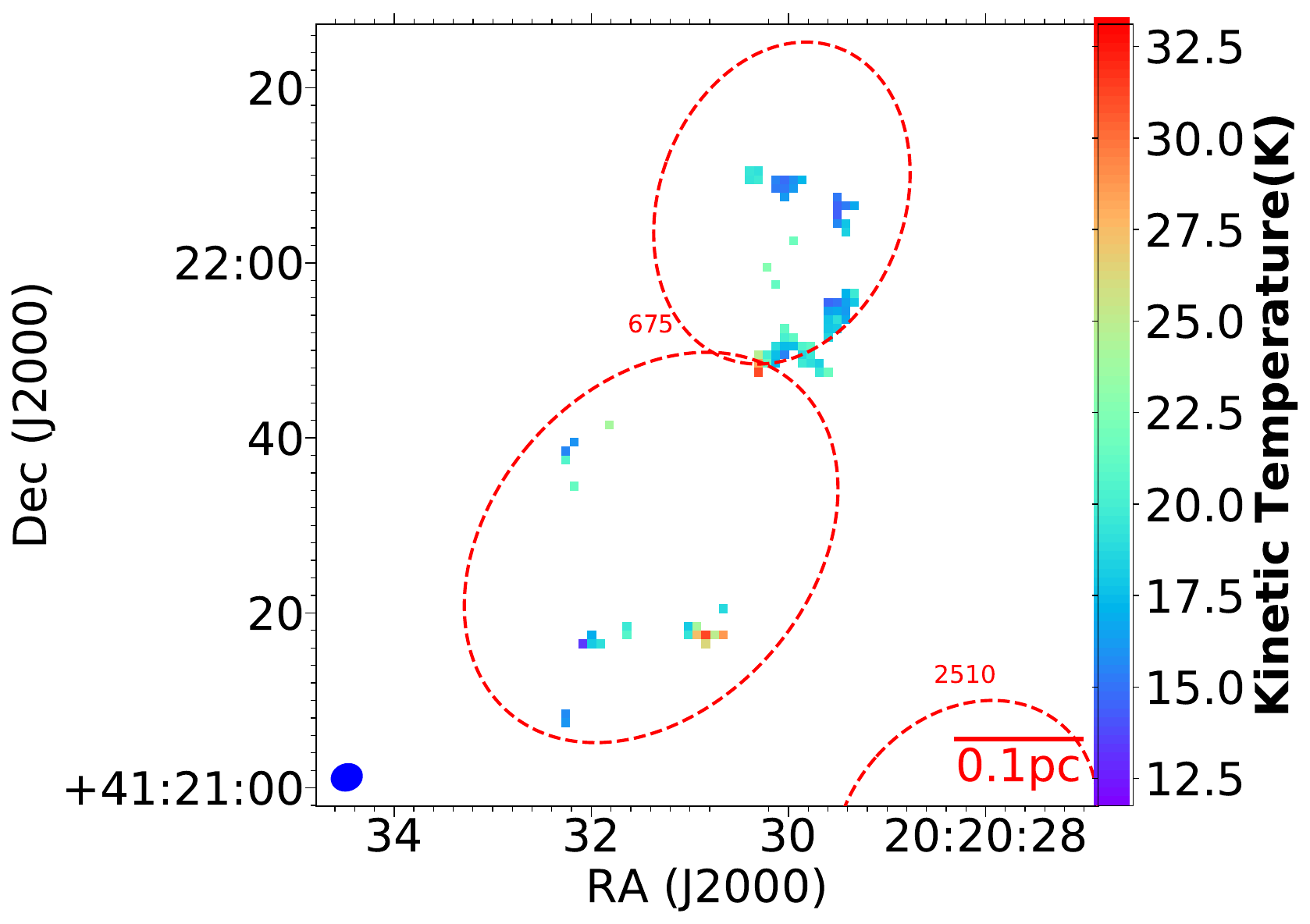} &
\includegraphics[width=.3\textwidth]{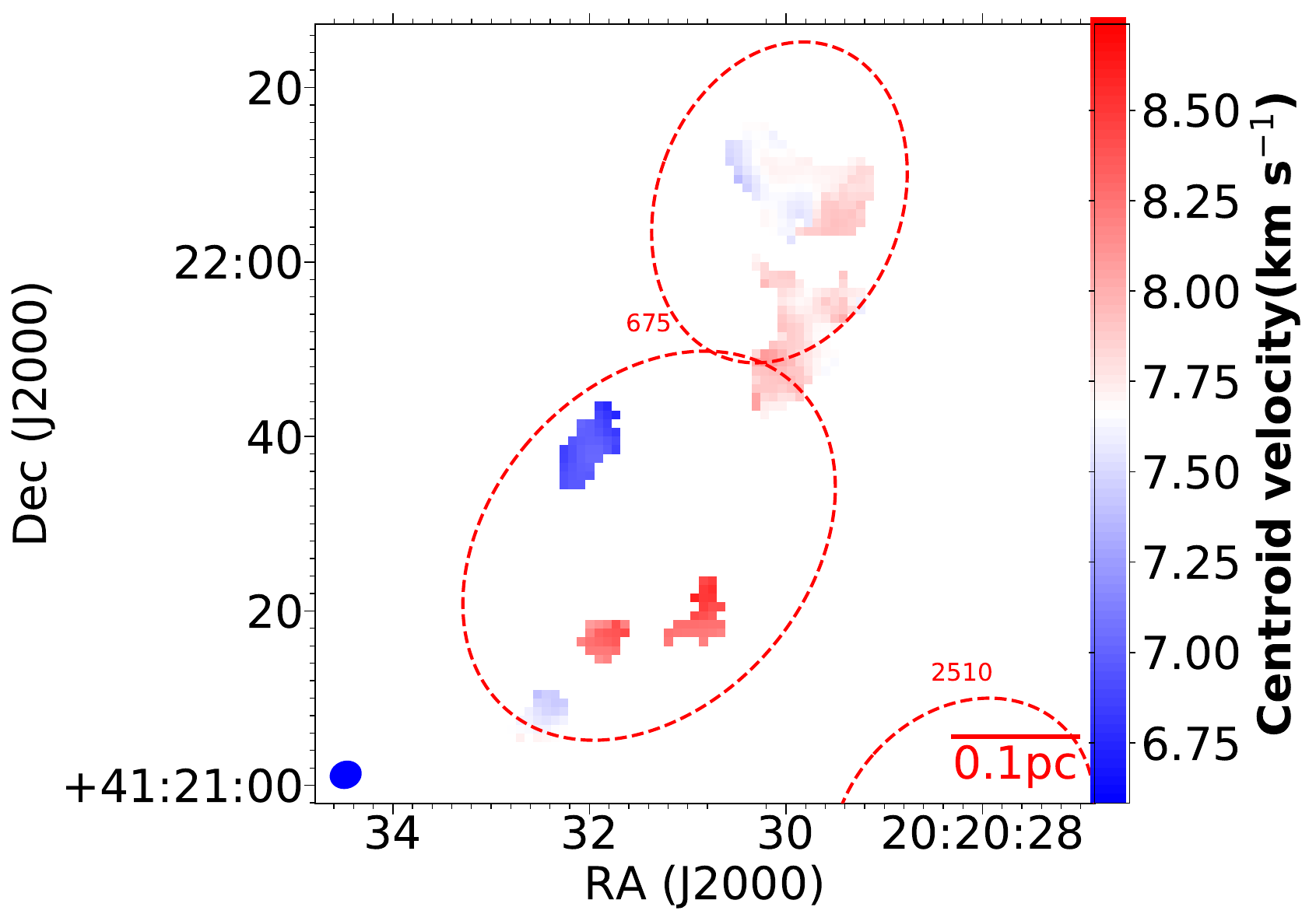} &
\includegraphics[width=.3\textwidth]{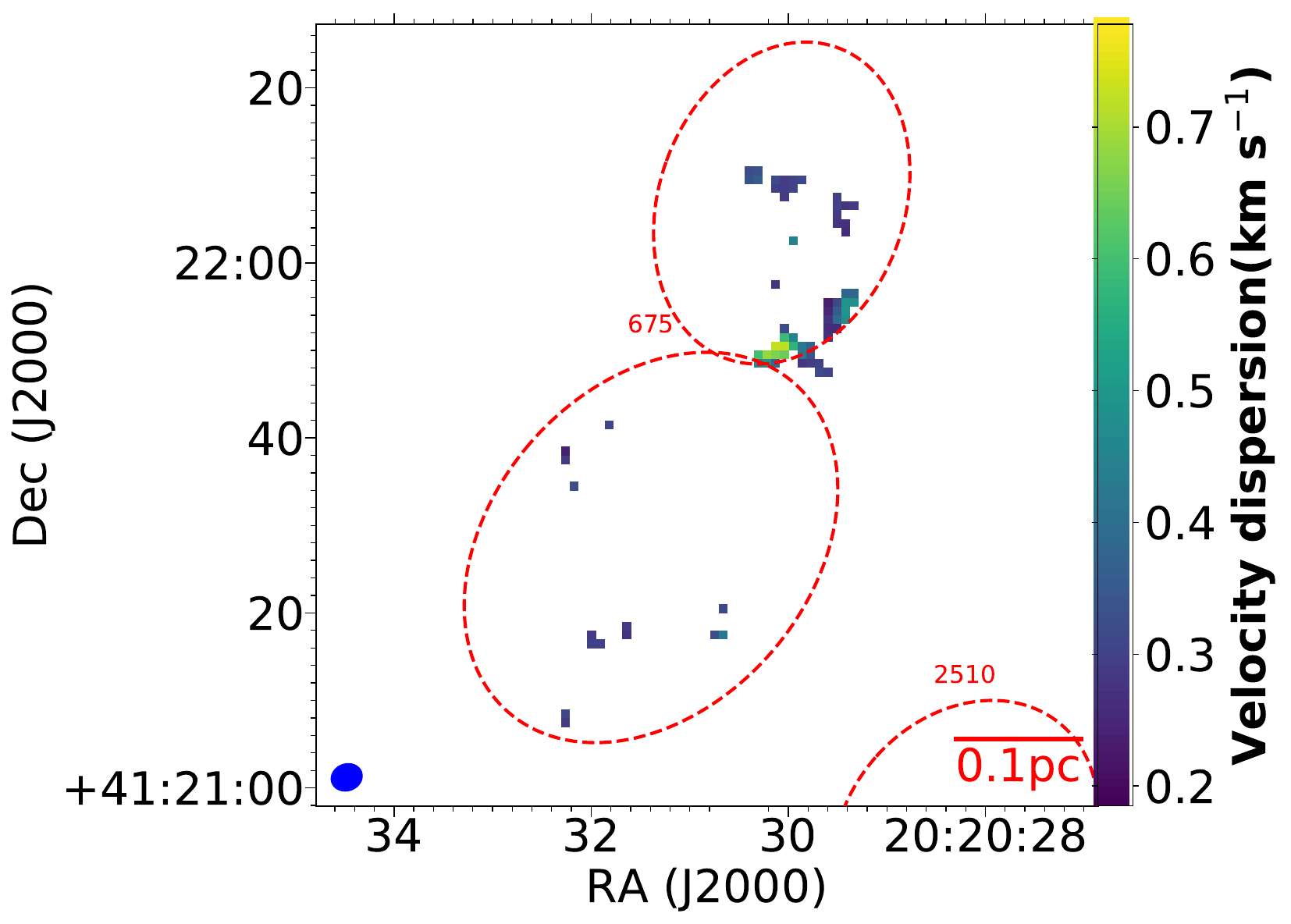} \\
 & Field 14 & \\
\includegraphics[width=.3\textwidth]{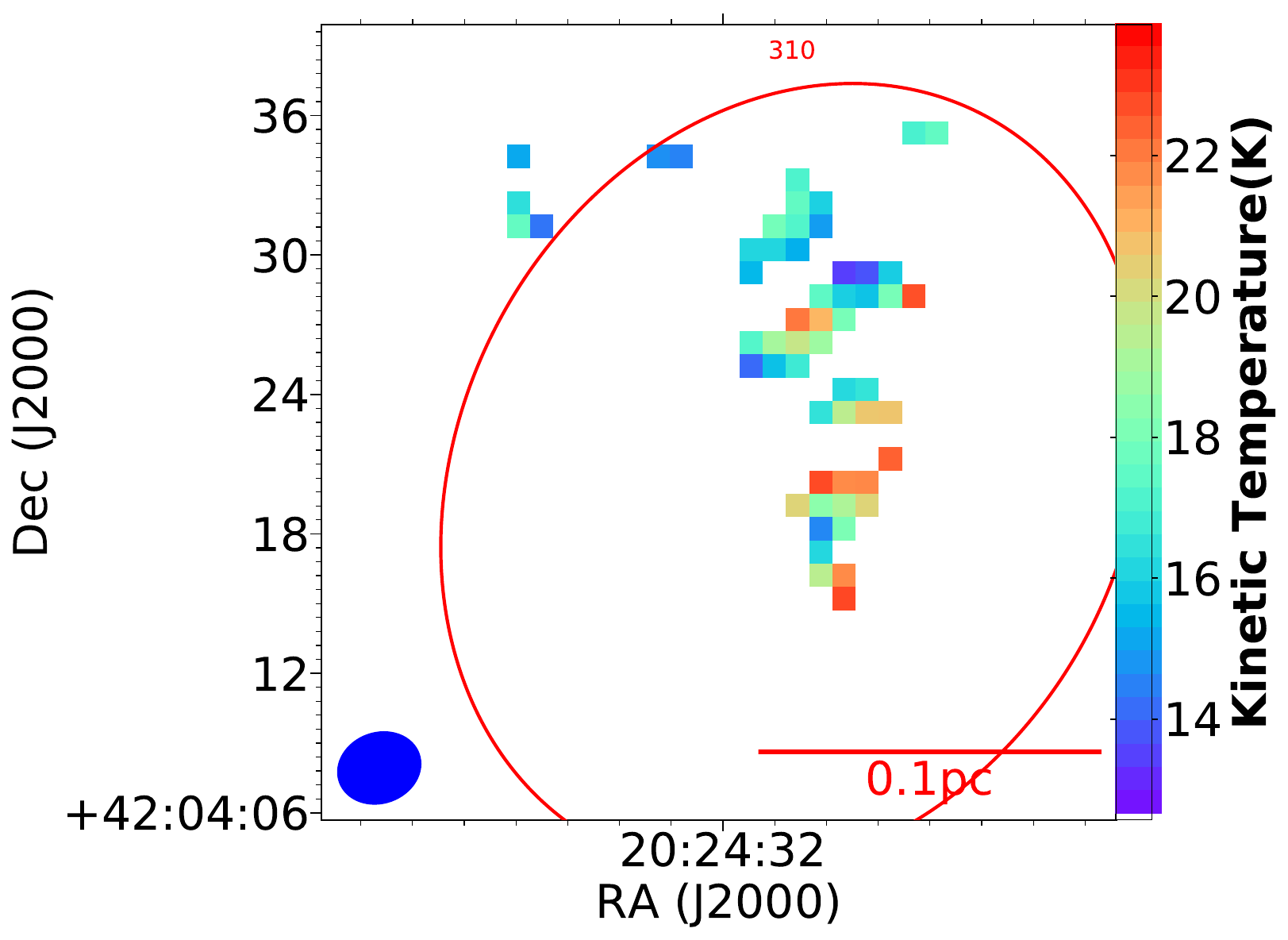} &
\includegraphics[width=.3\textwidth]{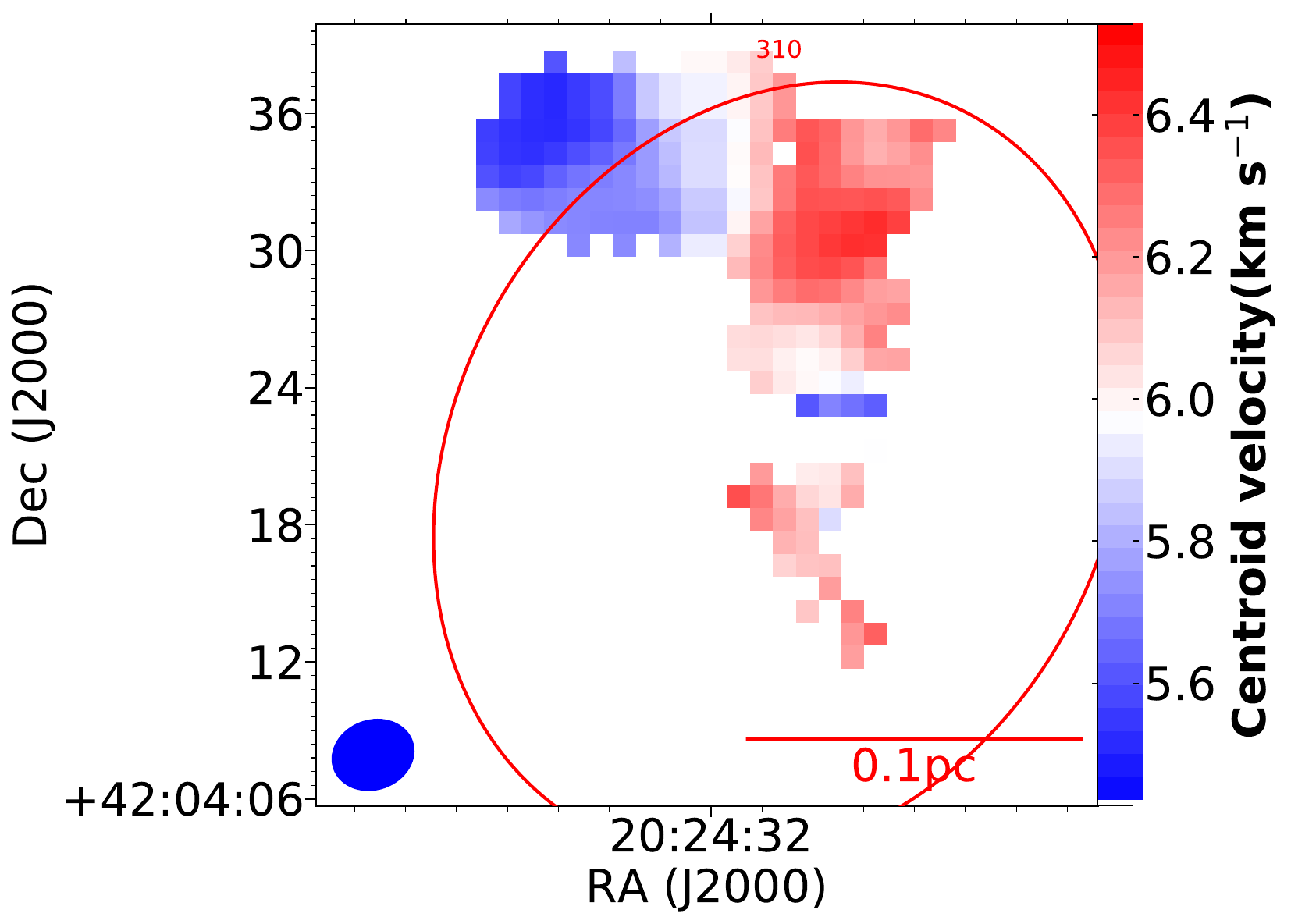} &
\includegraphics[width=.3\textwidth]{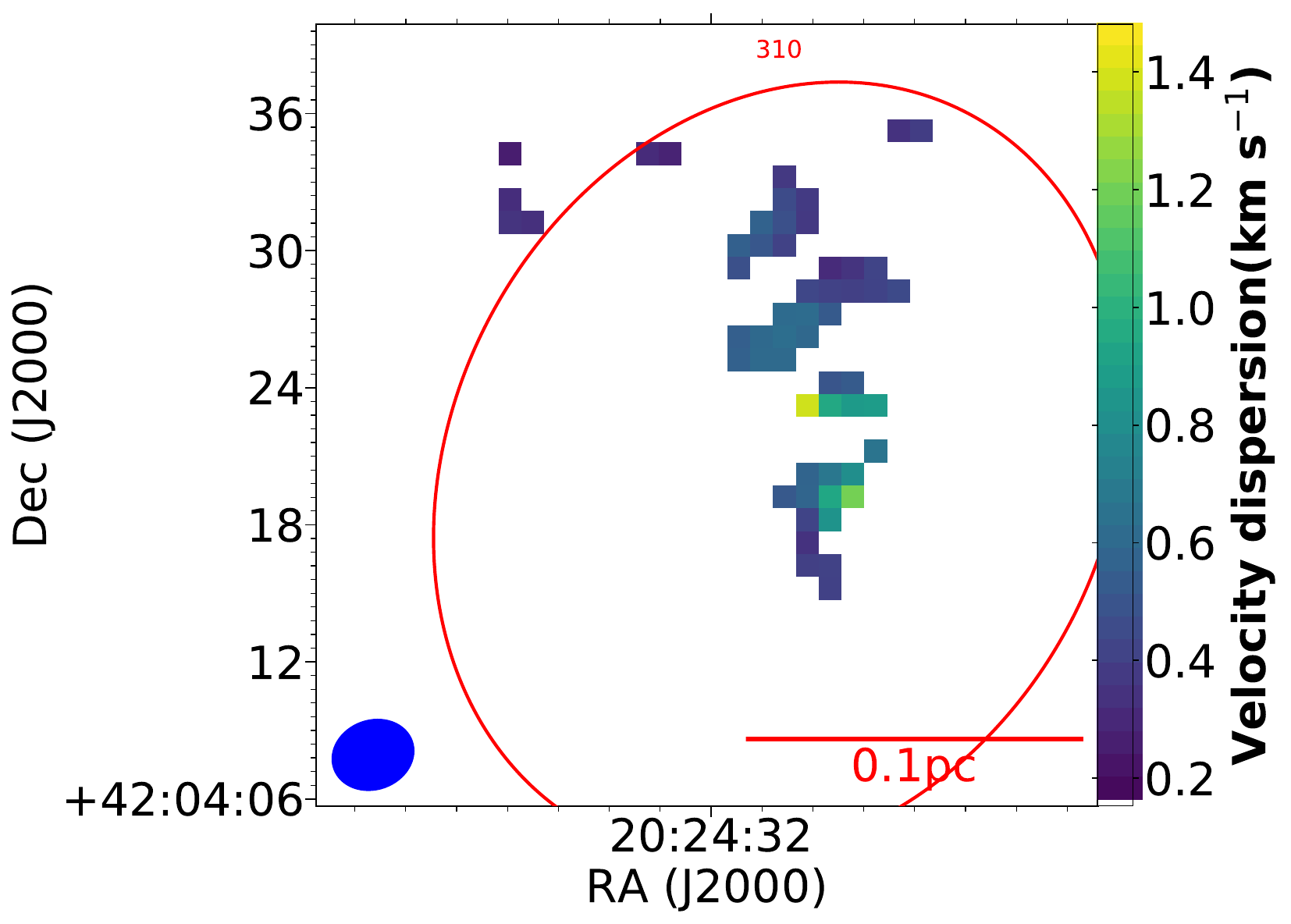} \\
 & Field 16 & \\
\includegraphics[width=.3\textwidth]{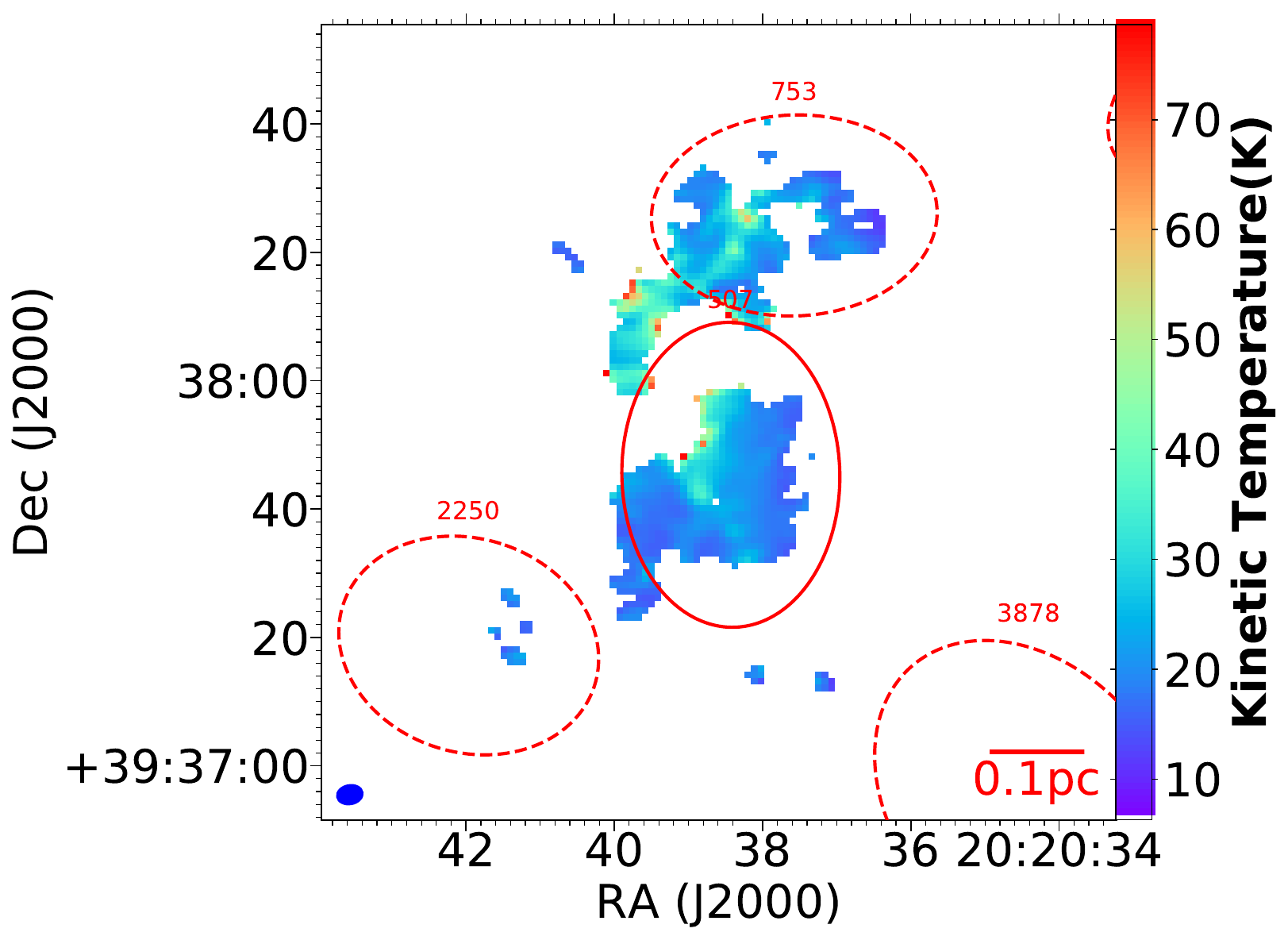} &
\includegraphics[width=.3\textwidth]{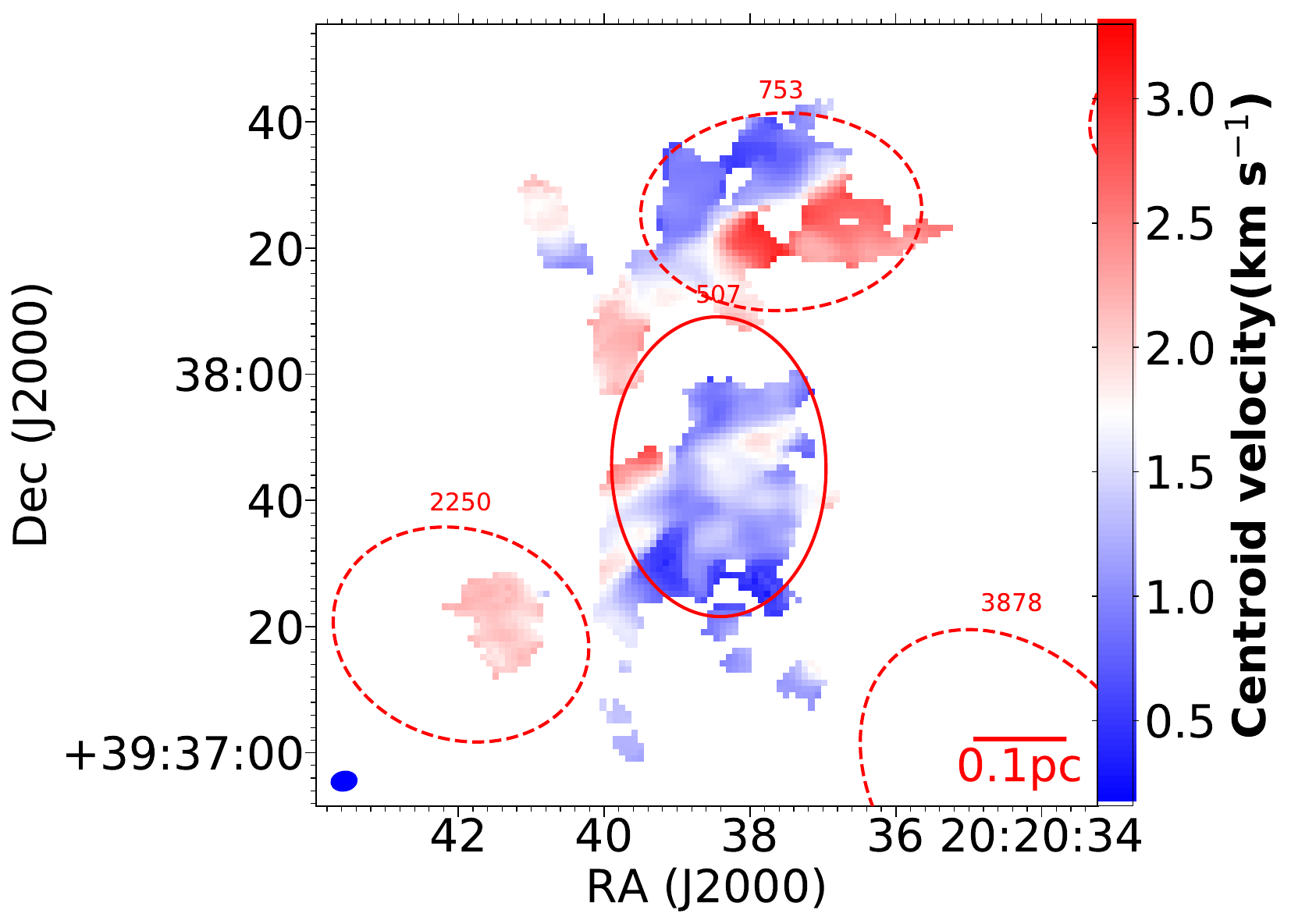} &
\includegraphics[width=.3\textwidth]{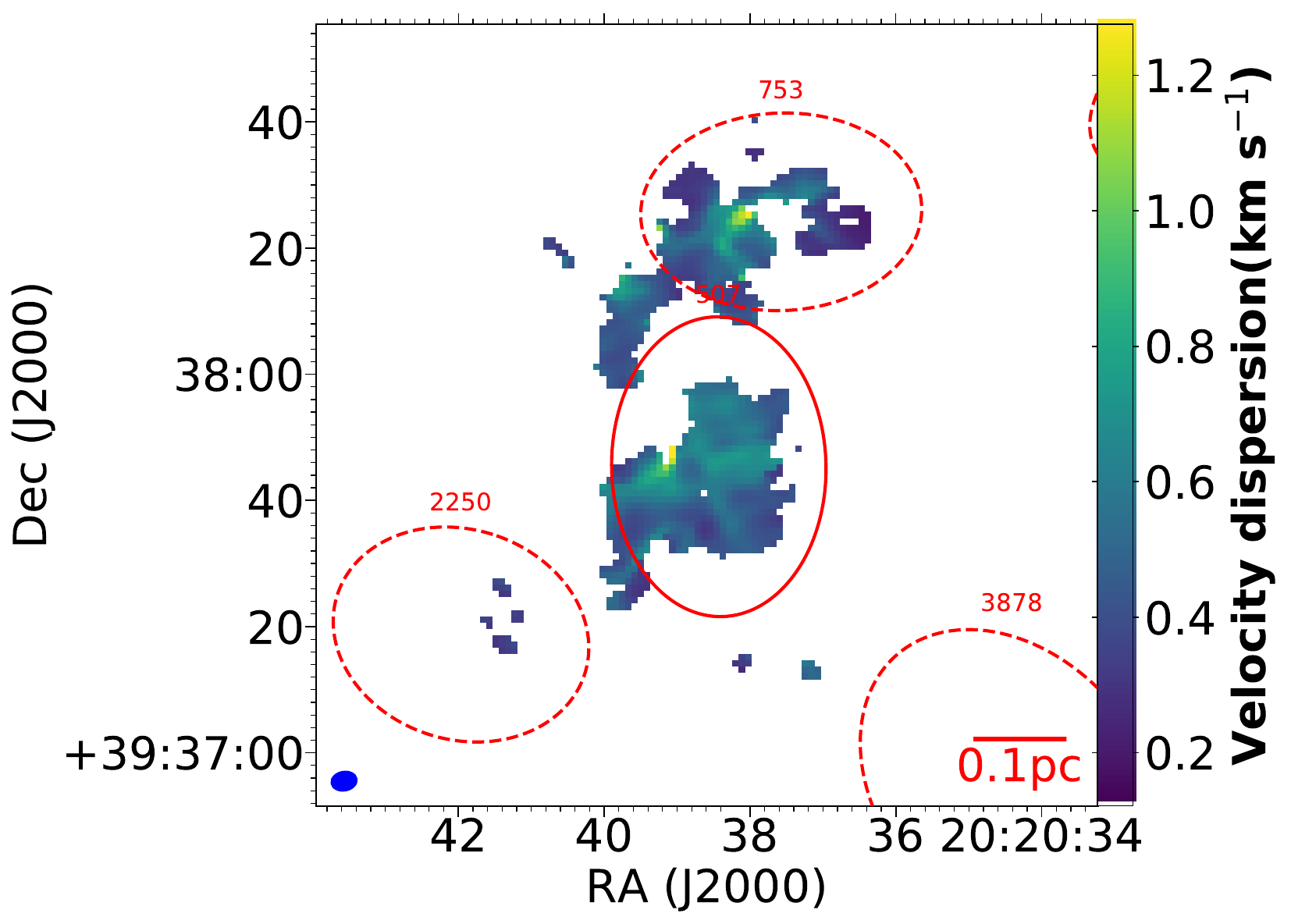} \\
 & Field 17 & \\
\includegraphics[width=.3\textwidth]{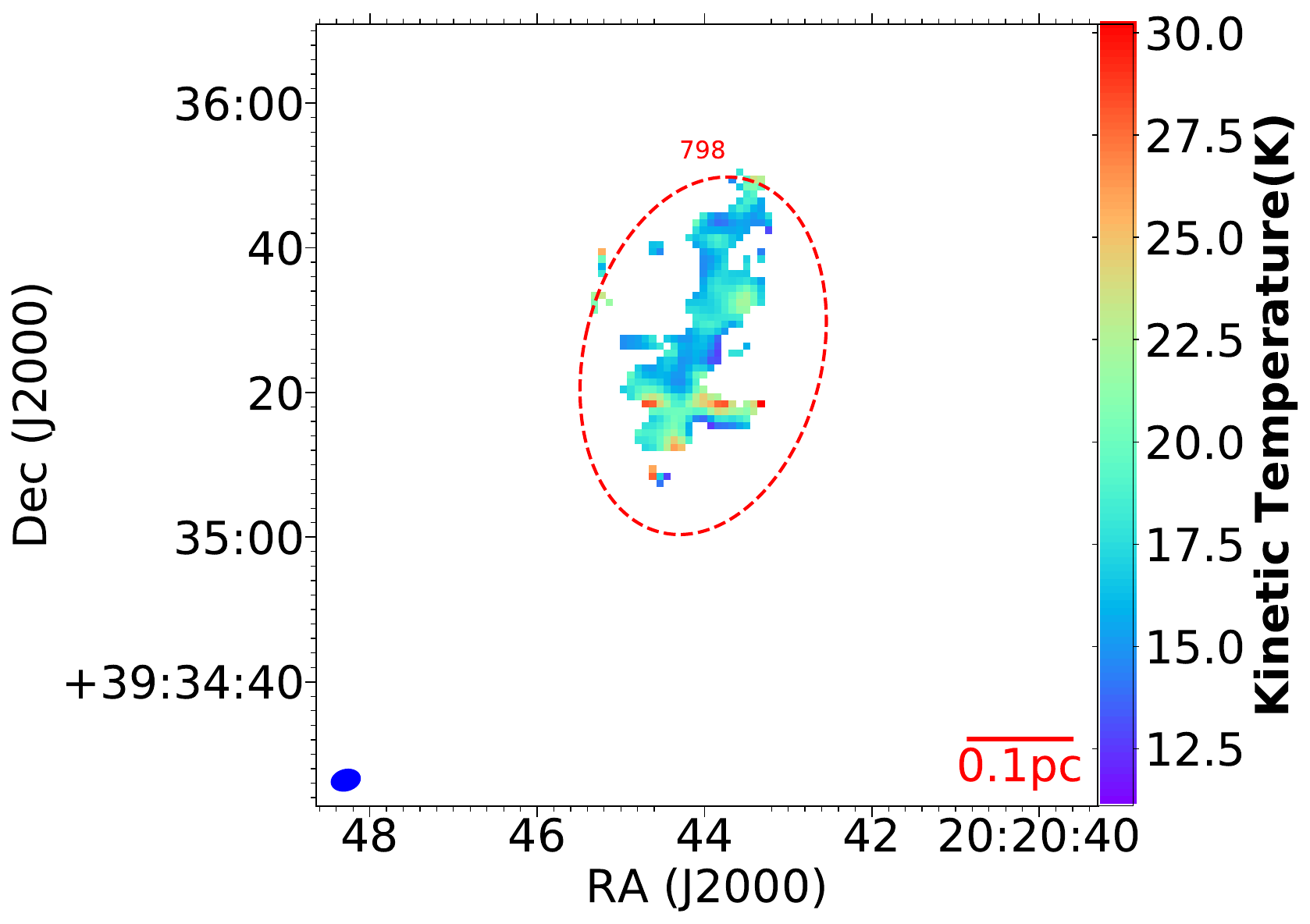} &
\includegraphics[width=.3\textwidth]{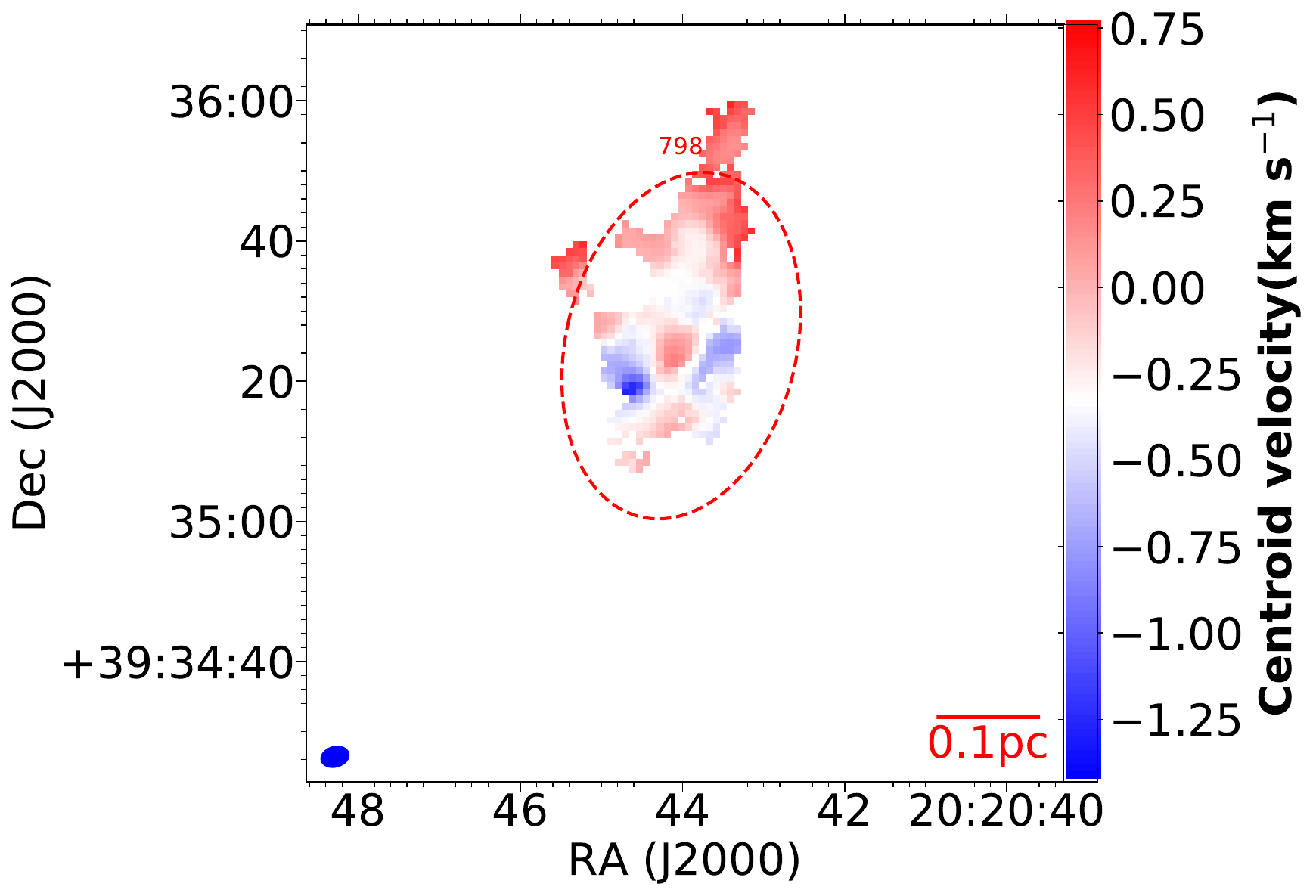} &
\includegraphics[width=.3\textwidth]{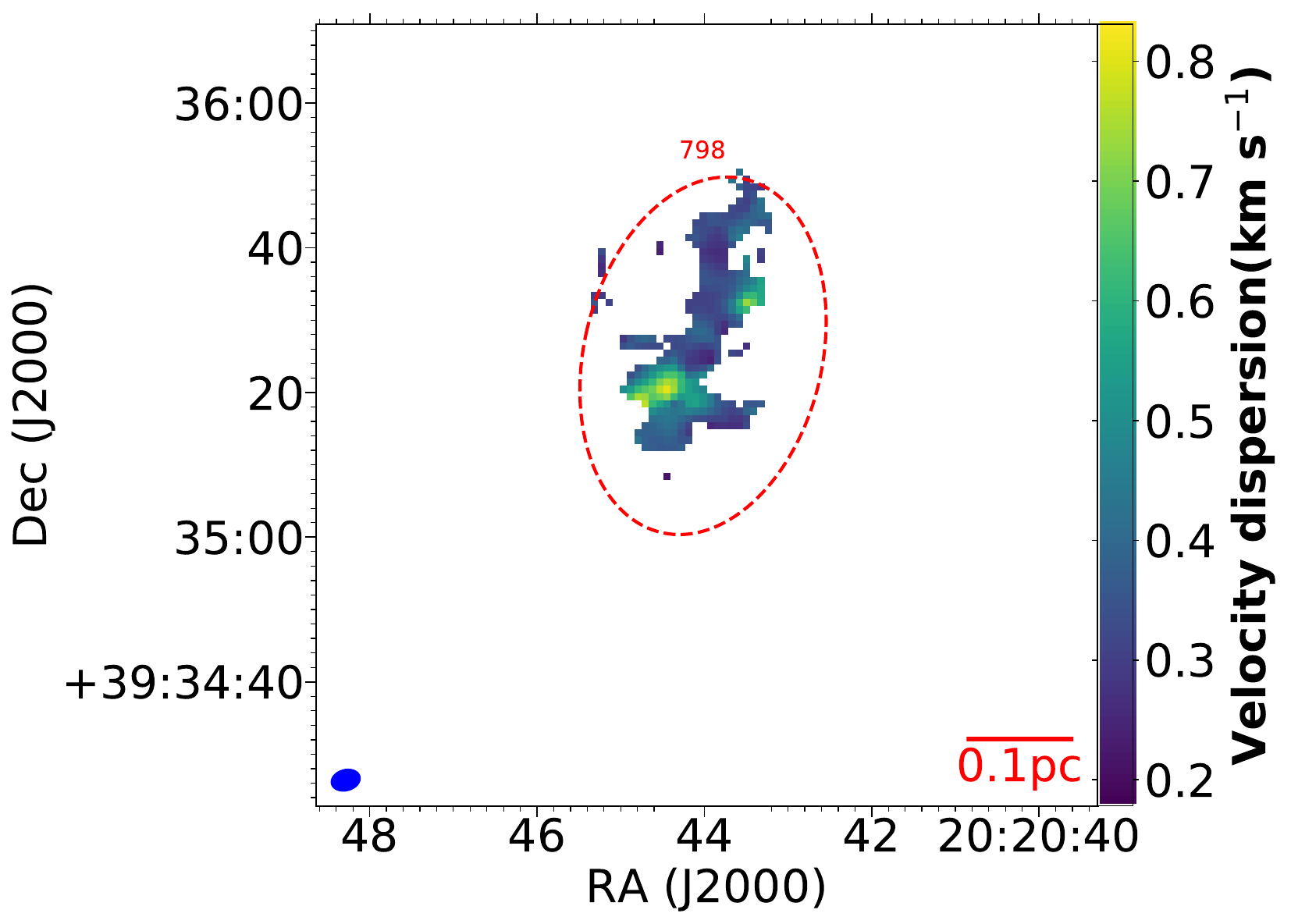} \\
 & Field 18 & \\
\includegraphics[width=.3\textwidth]{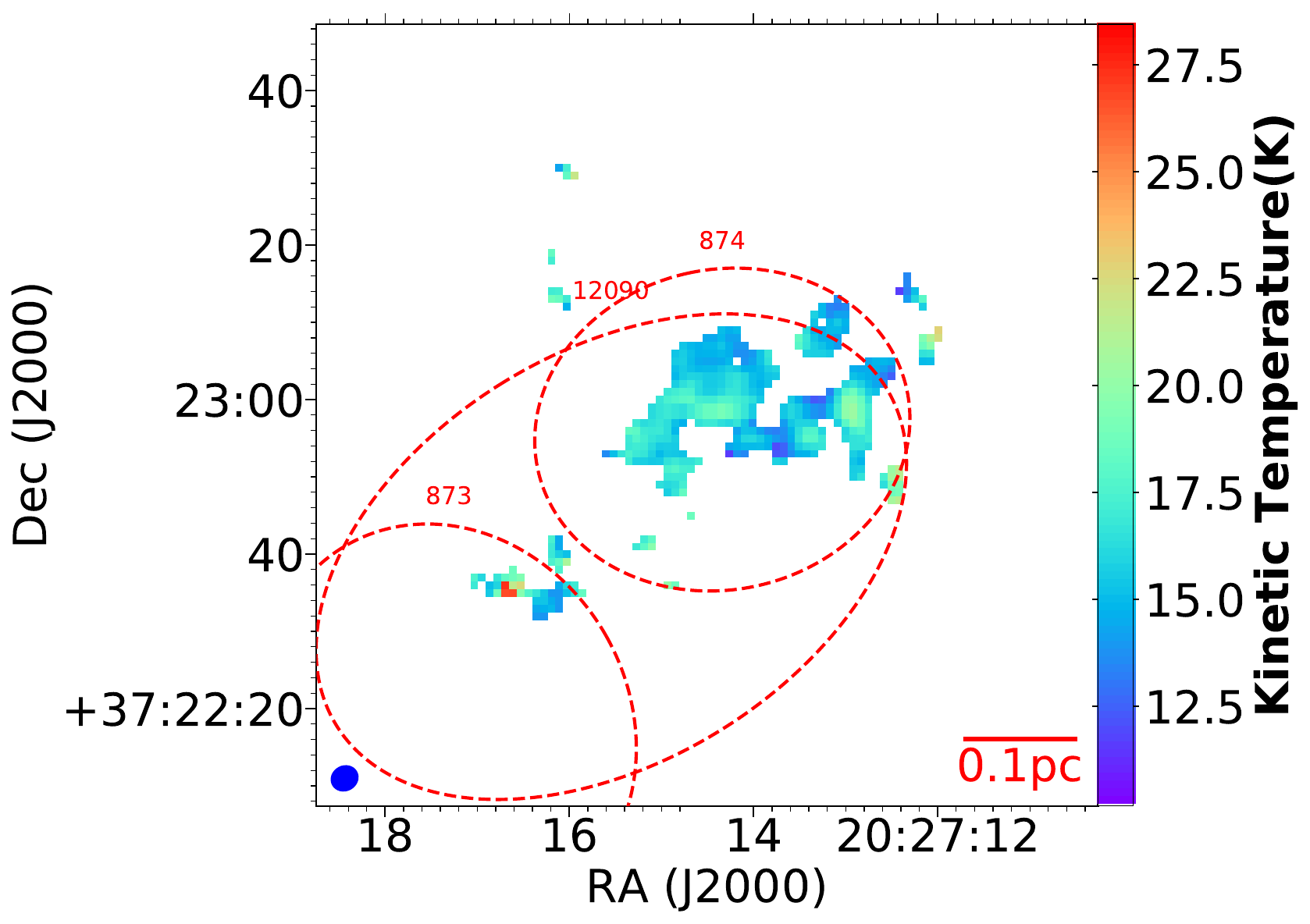} &
\includegraphics[width=.3\textwidth]{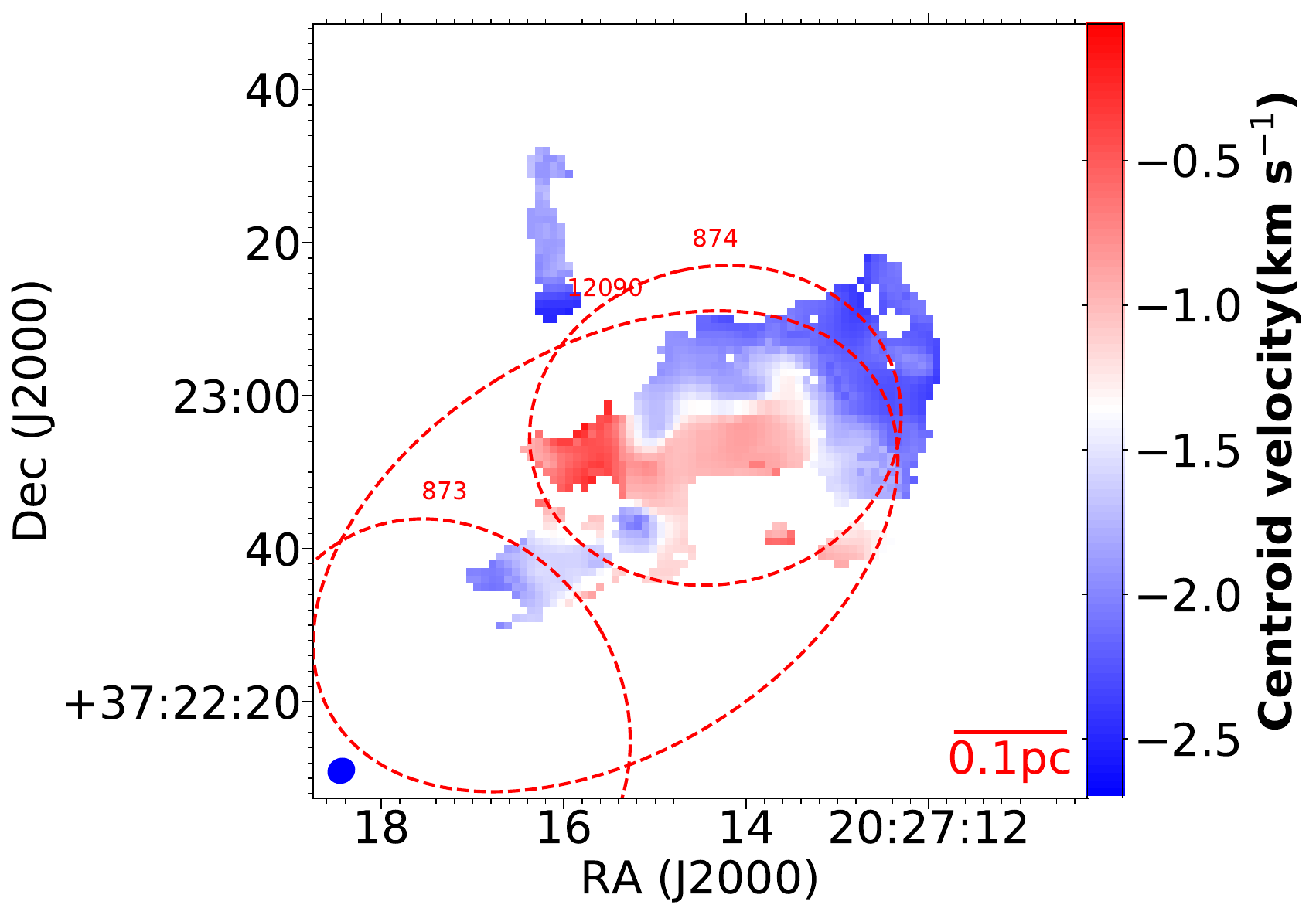} &
\includegraphics[width=.3\textwidth]{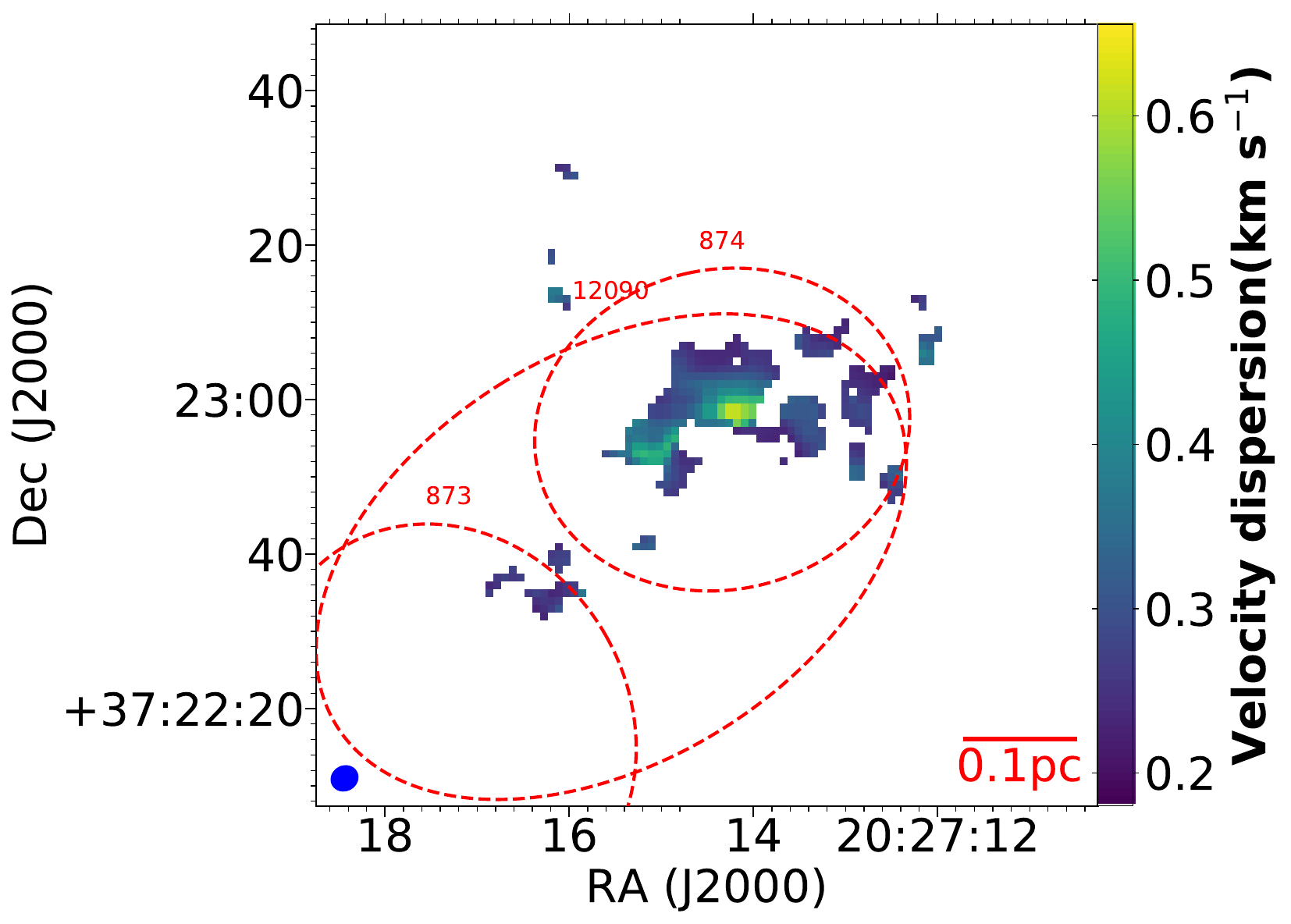} \\
 & Field 19 & \\
\includegraphics[width=.3\textwidth]{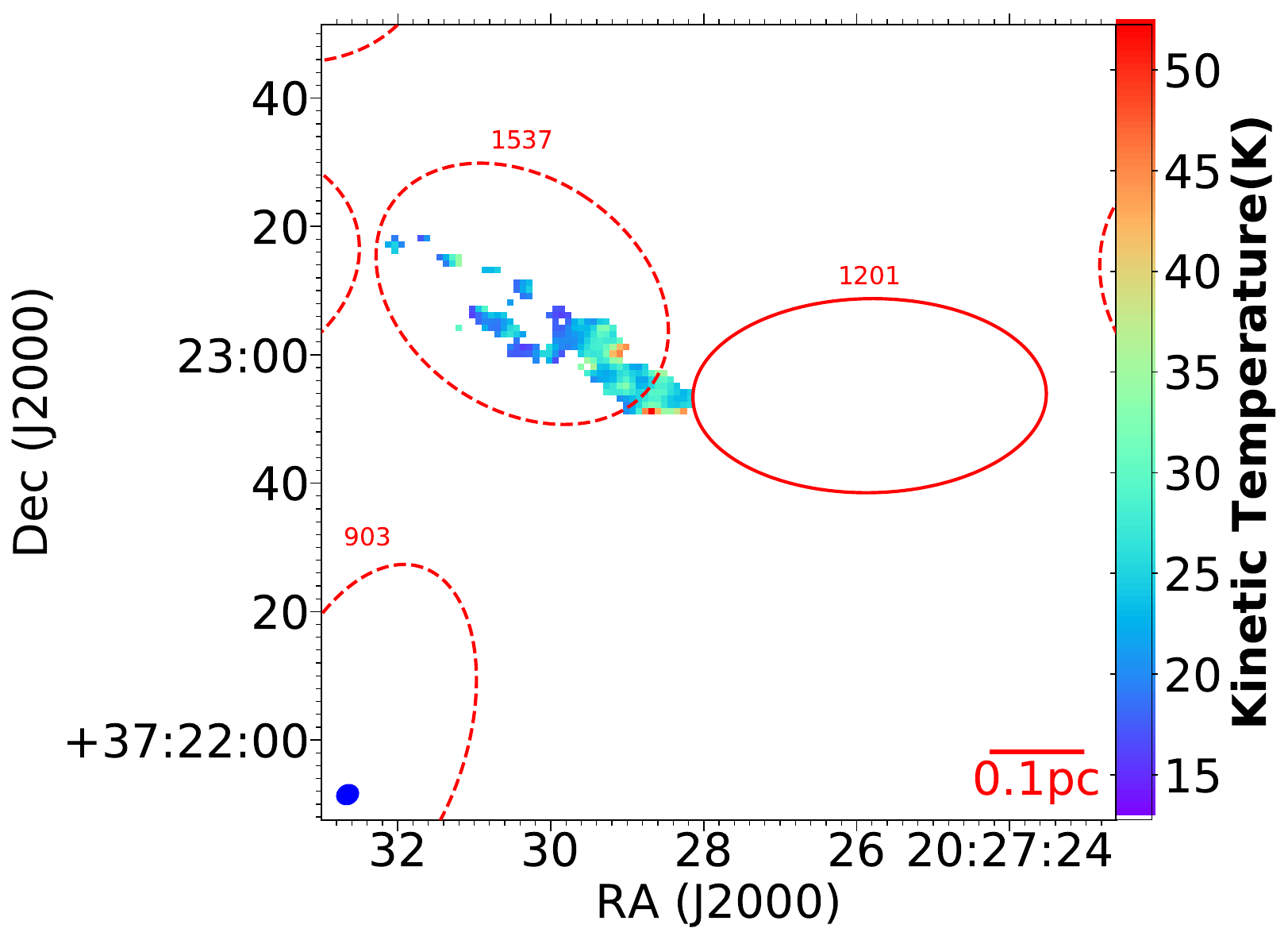} &
\includegraphics[width=.3\textwidth]{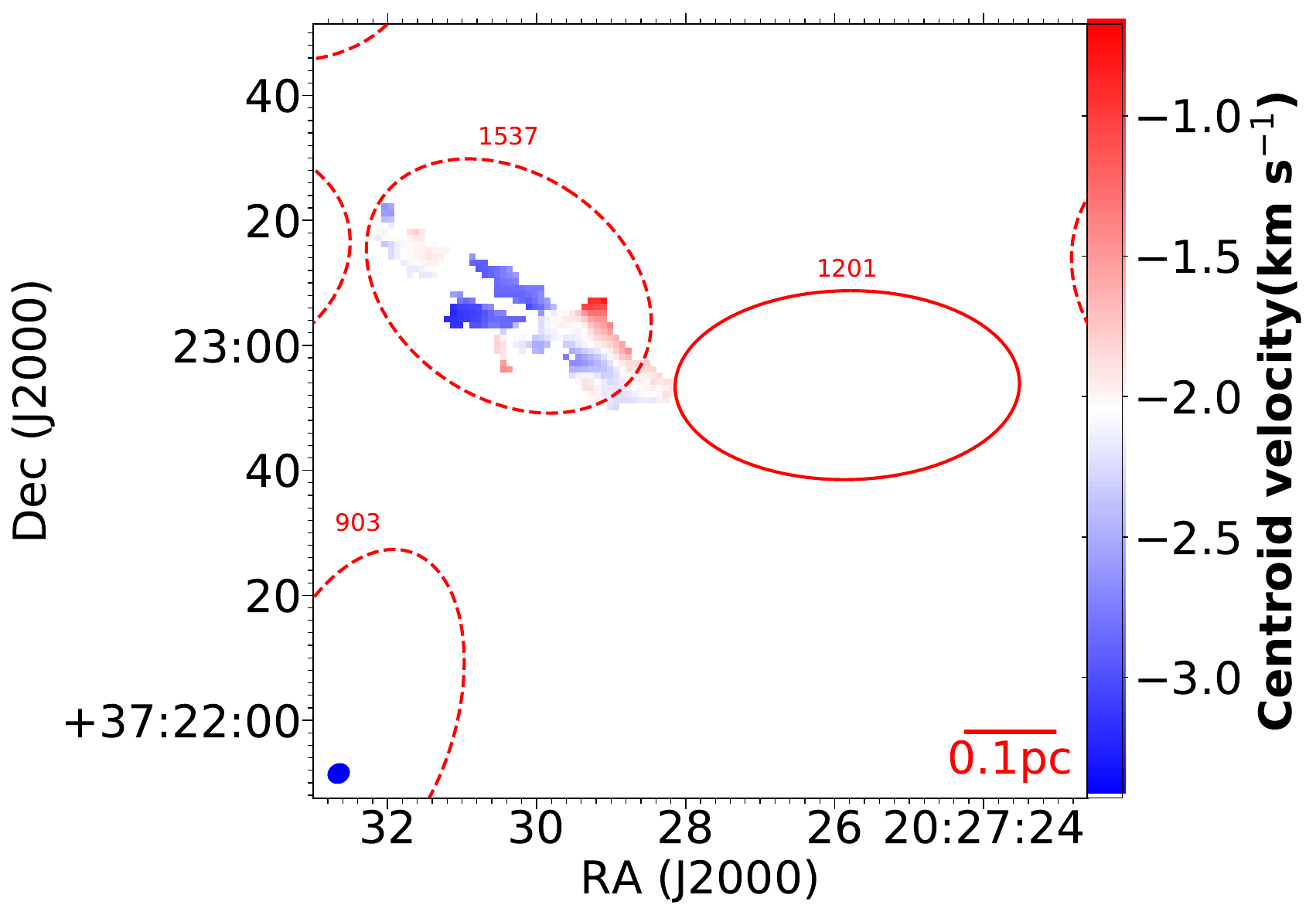} &
\includegraphics[width=.3\textwidth]{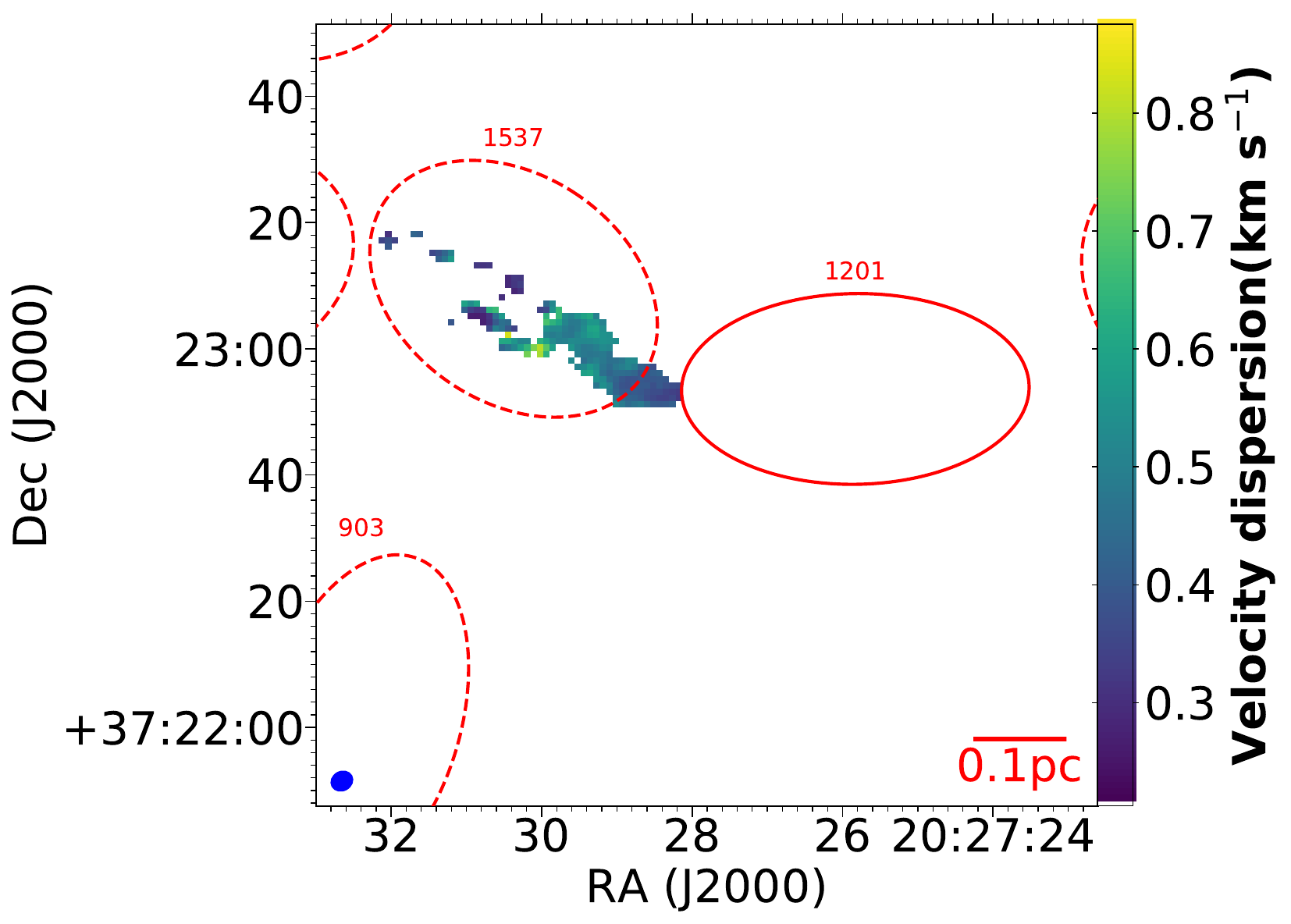} \\
 & Field 20 & \\
\includegraphics[width=.3\textwidth]{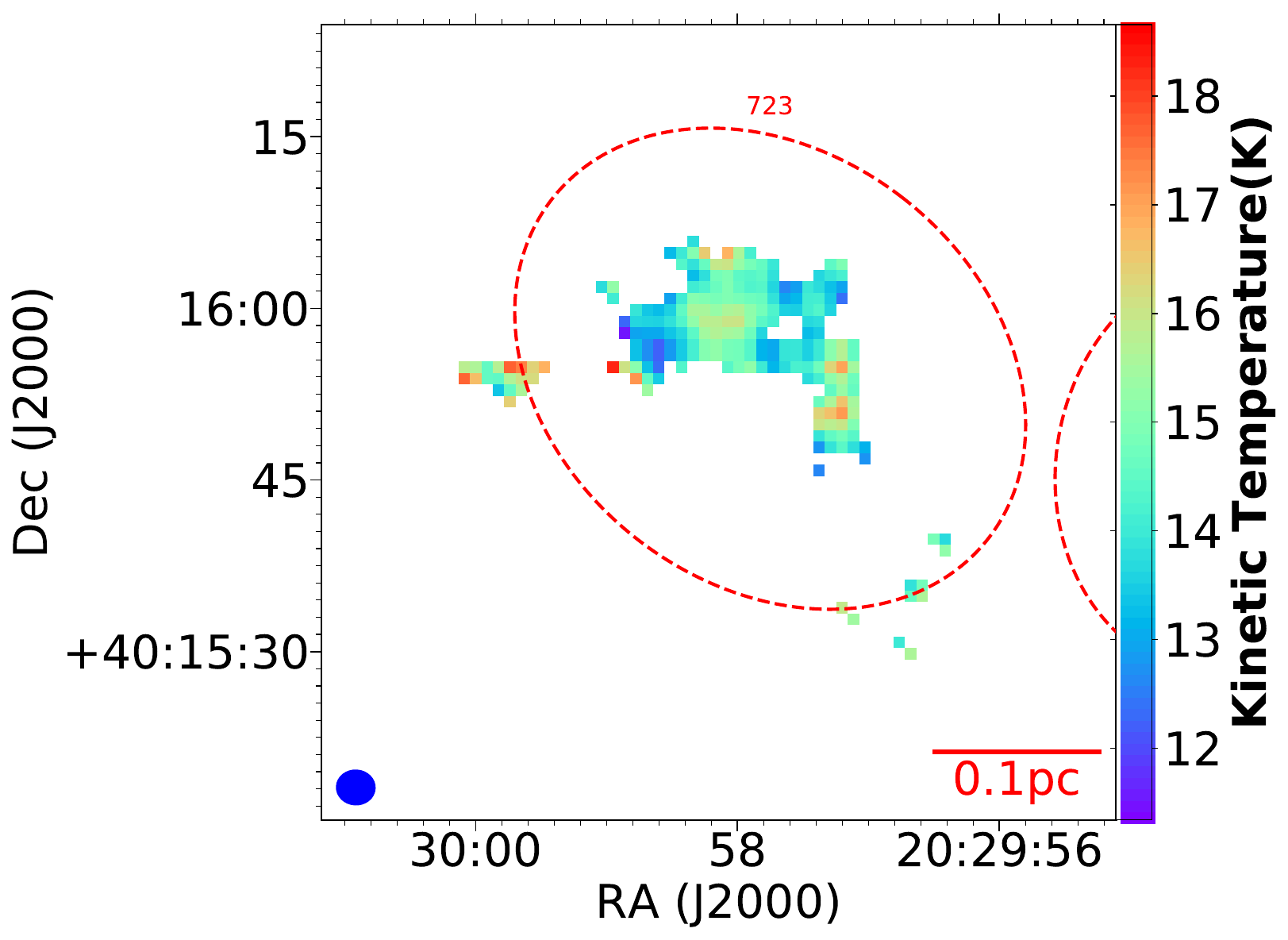} &
\includegraphics[width=.3\textwidth]{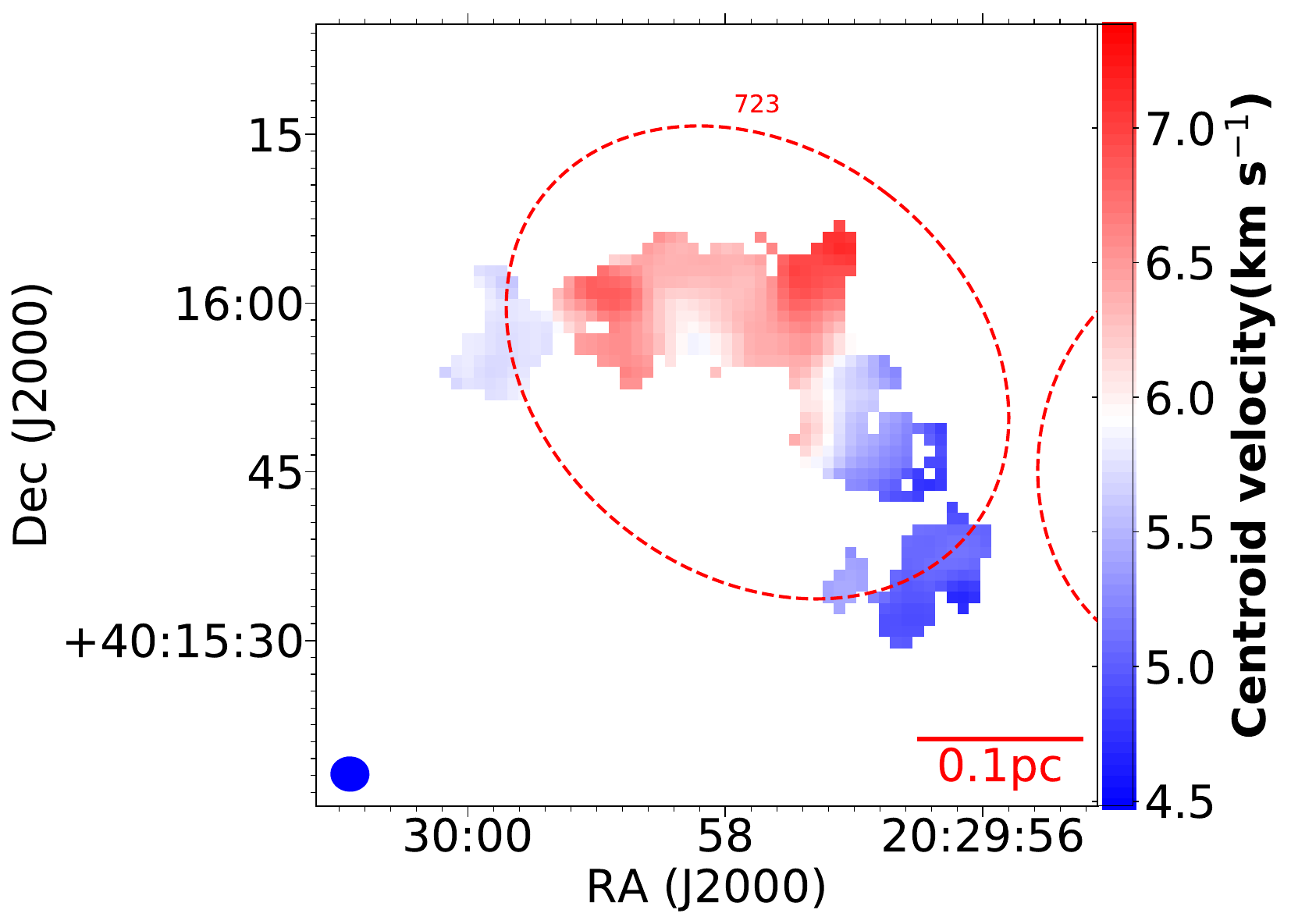} &
\includegraphics[width=.3\textwidth]{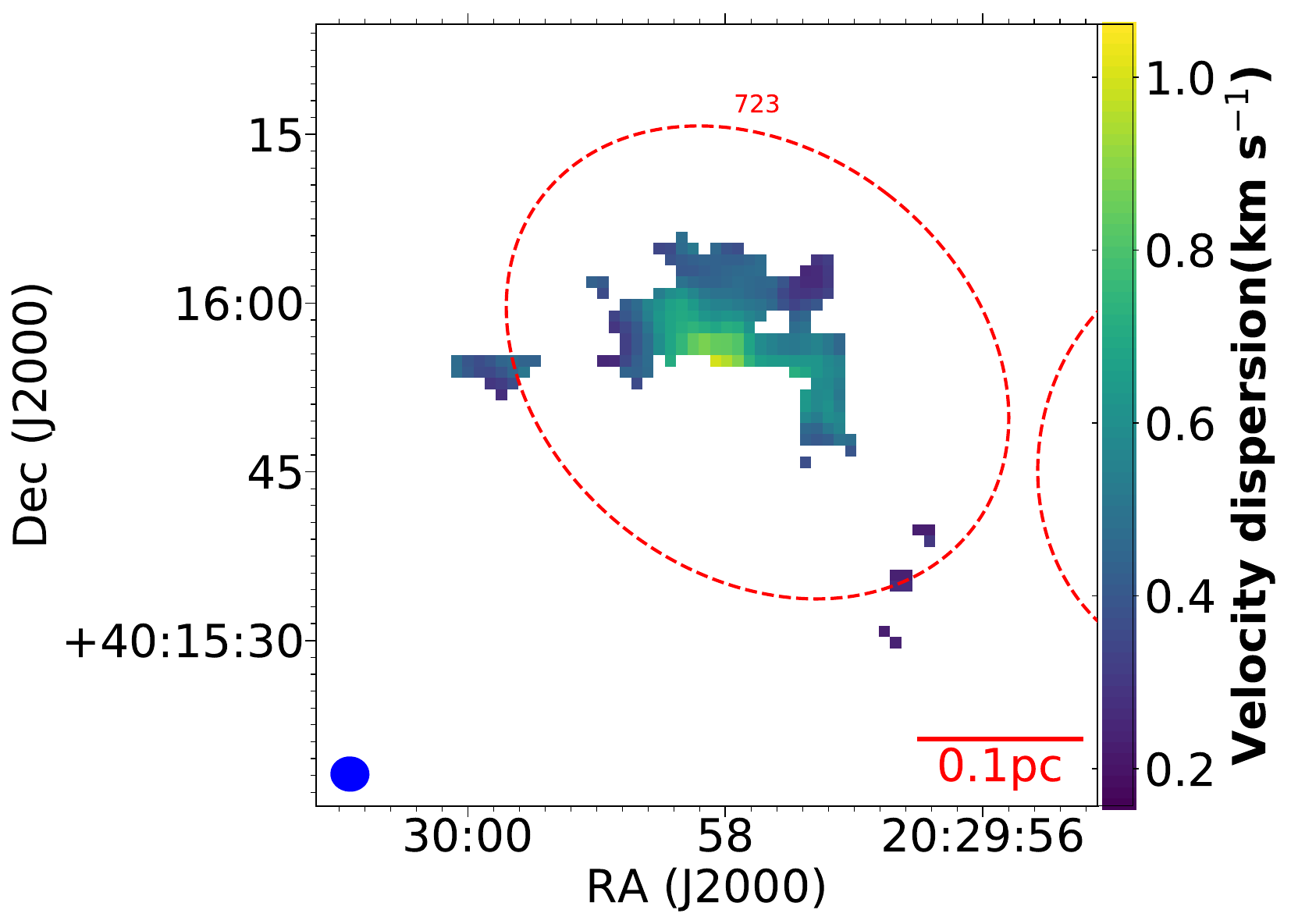} \\
 & Field 22 & \\
\includegraphics[width=.3\textwidth]{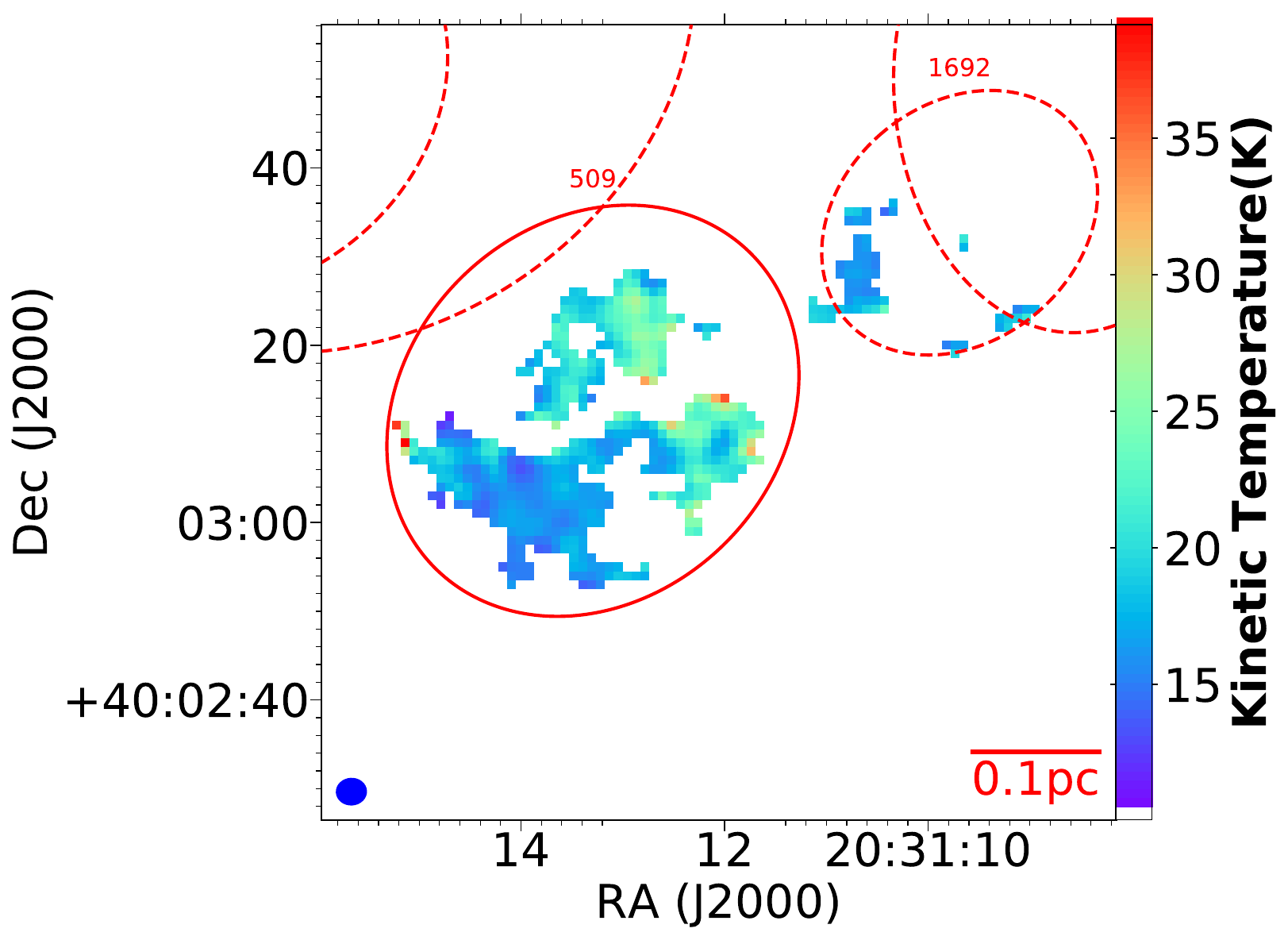} &
\includegraphics[width=.3\textwidth]{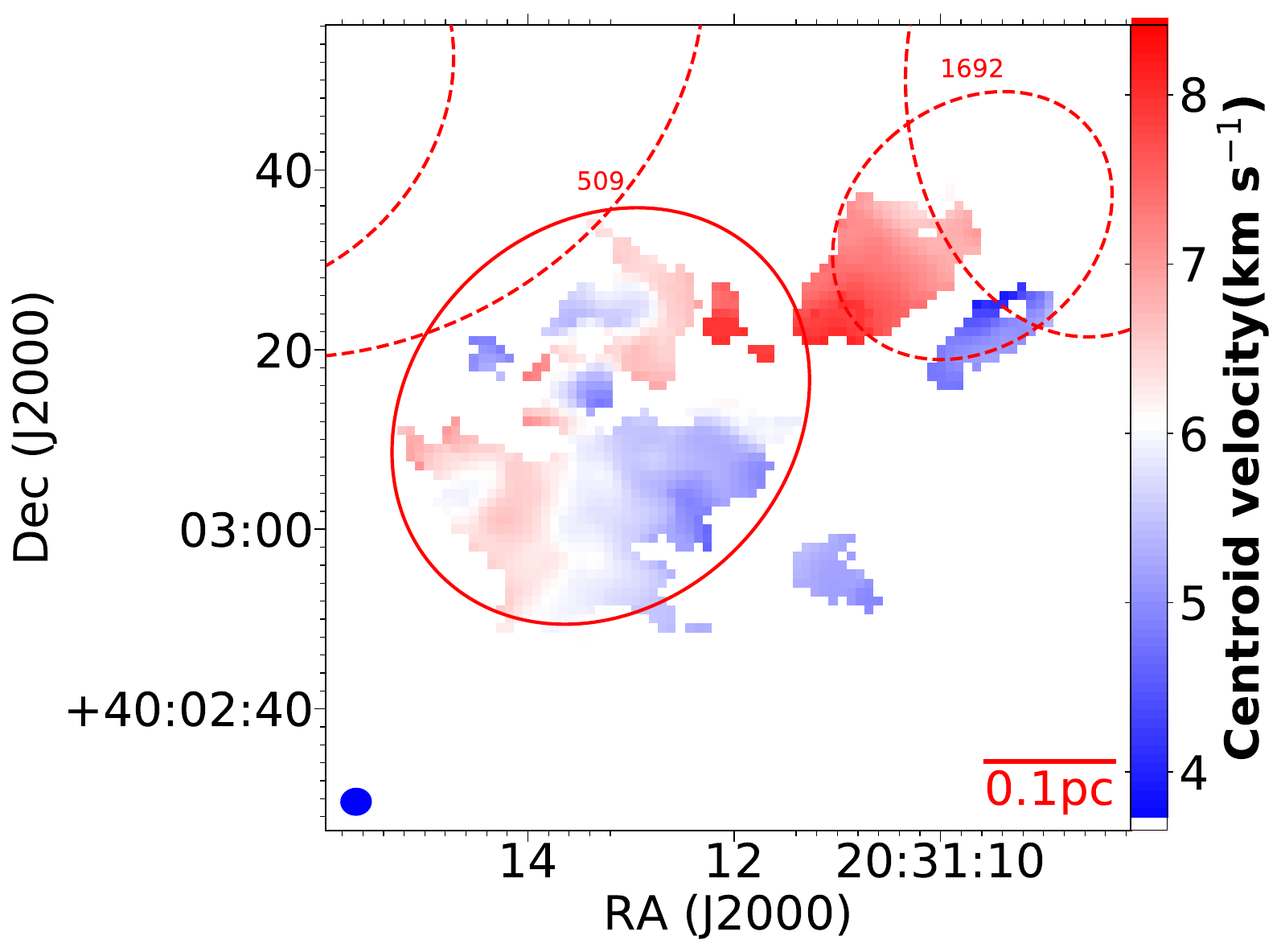} &
\includegraphics[width=.3\textwidth]{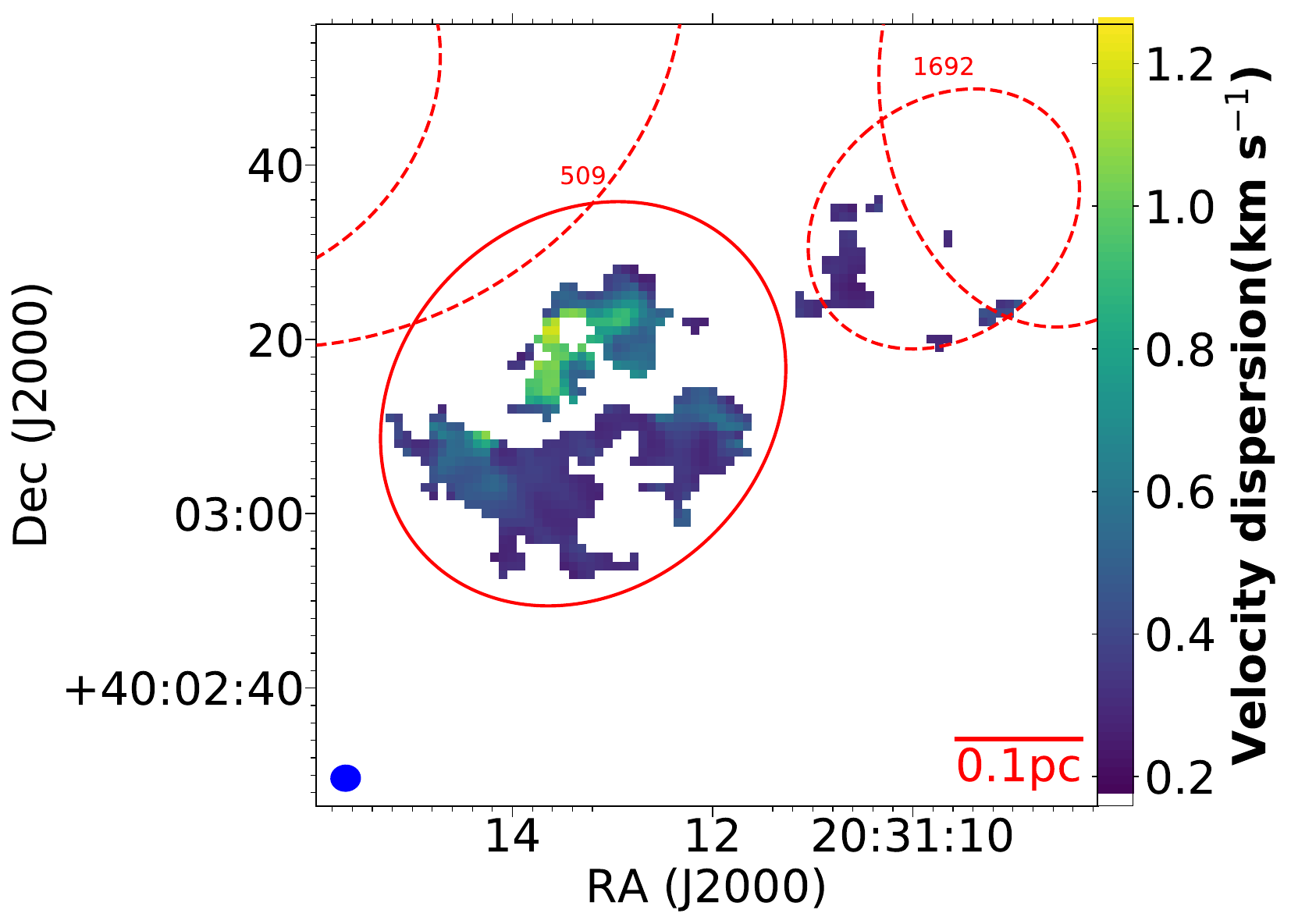} \\
 & Field 23 & \\
\includegraphics[width=.3\textwidth]{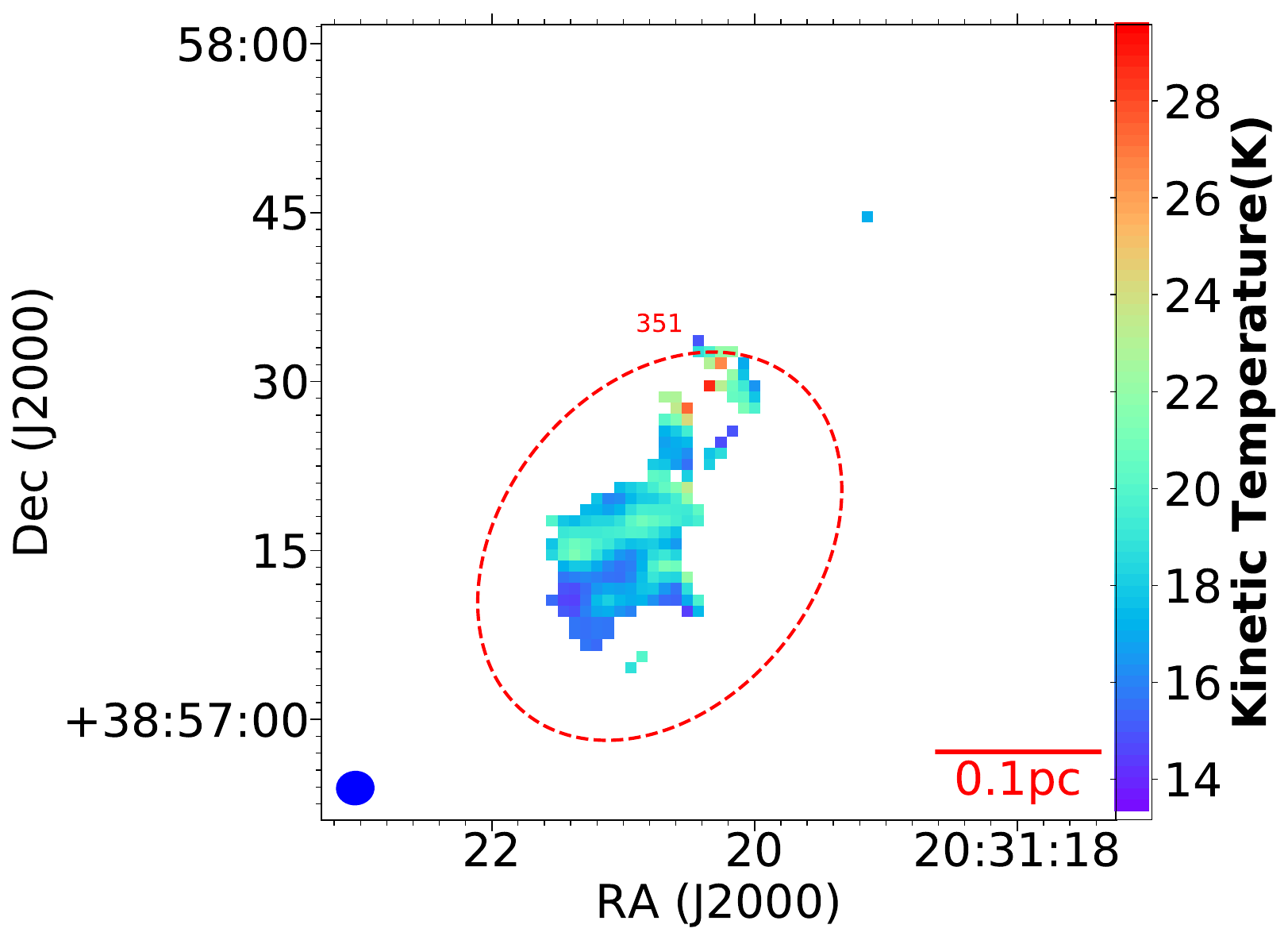} &
\includegraphics[width=.3\textwidth]{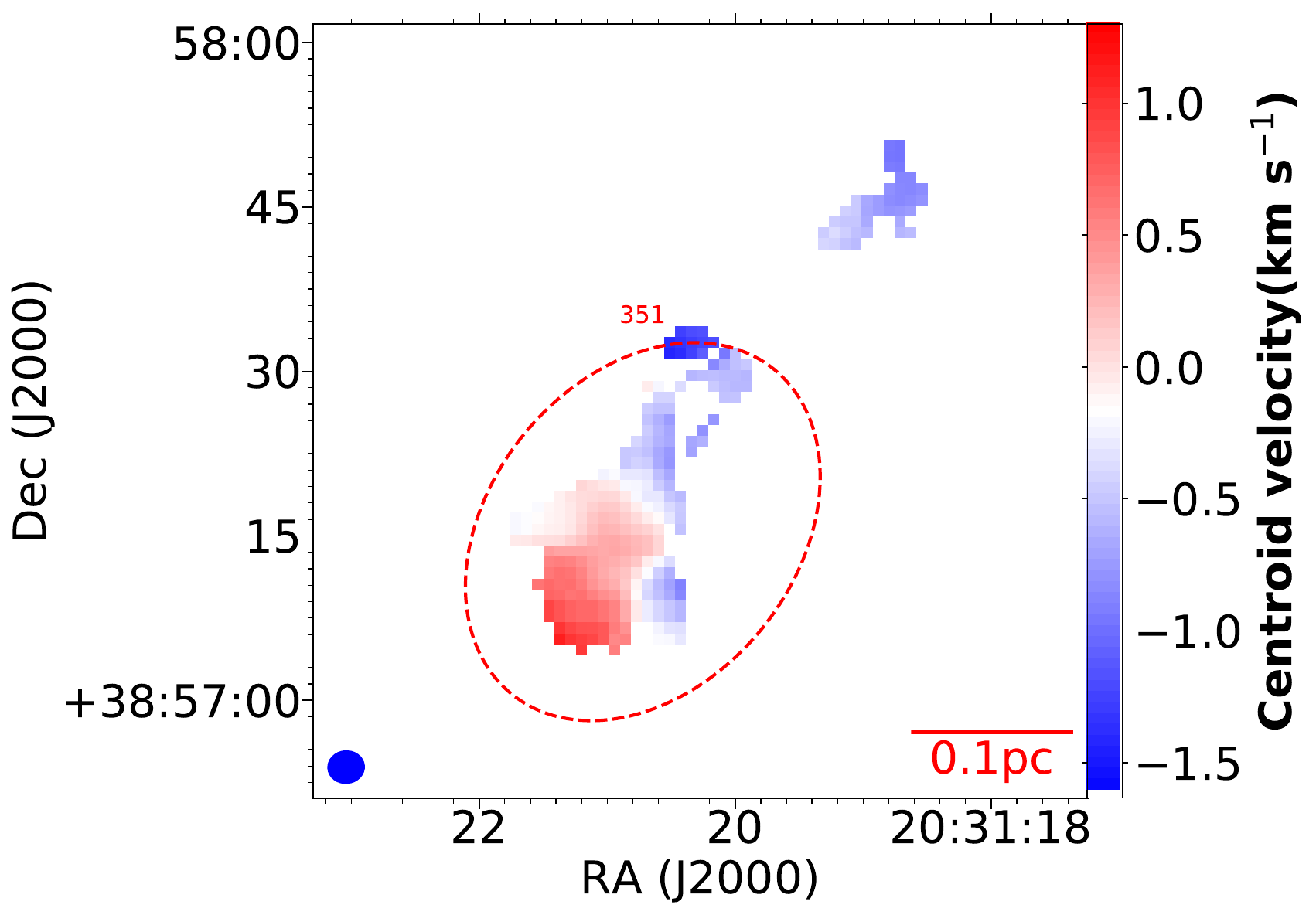} &
\includegraphics[width=.3\textwidth]{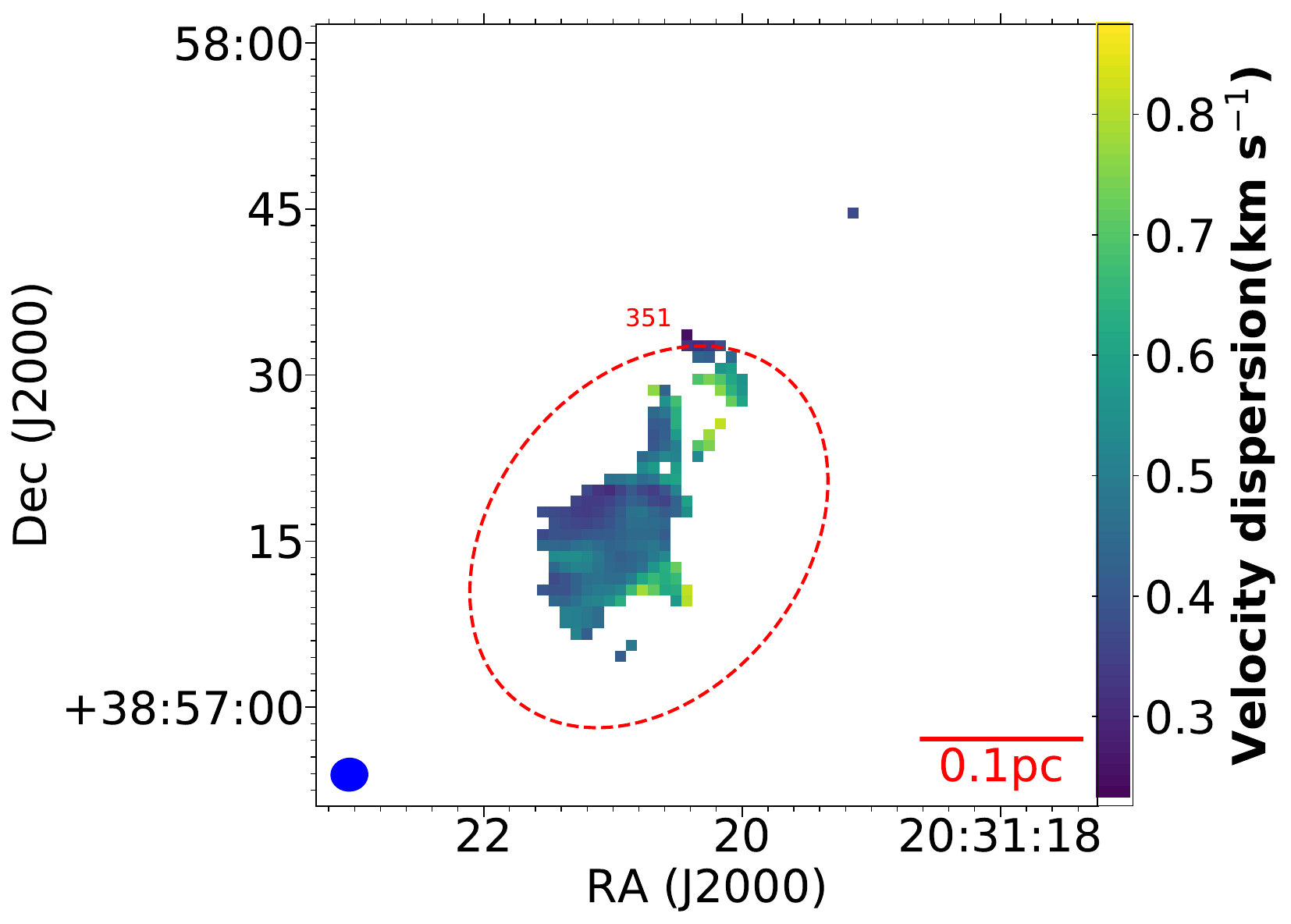} \\
 & Field 24 & \\
\includegraphics[width=.3\textwidth]{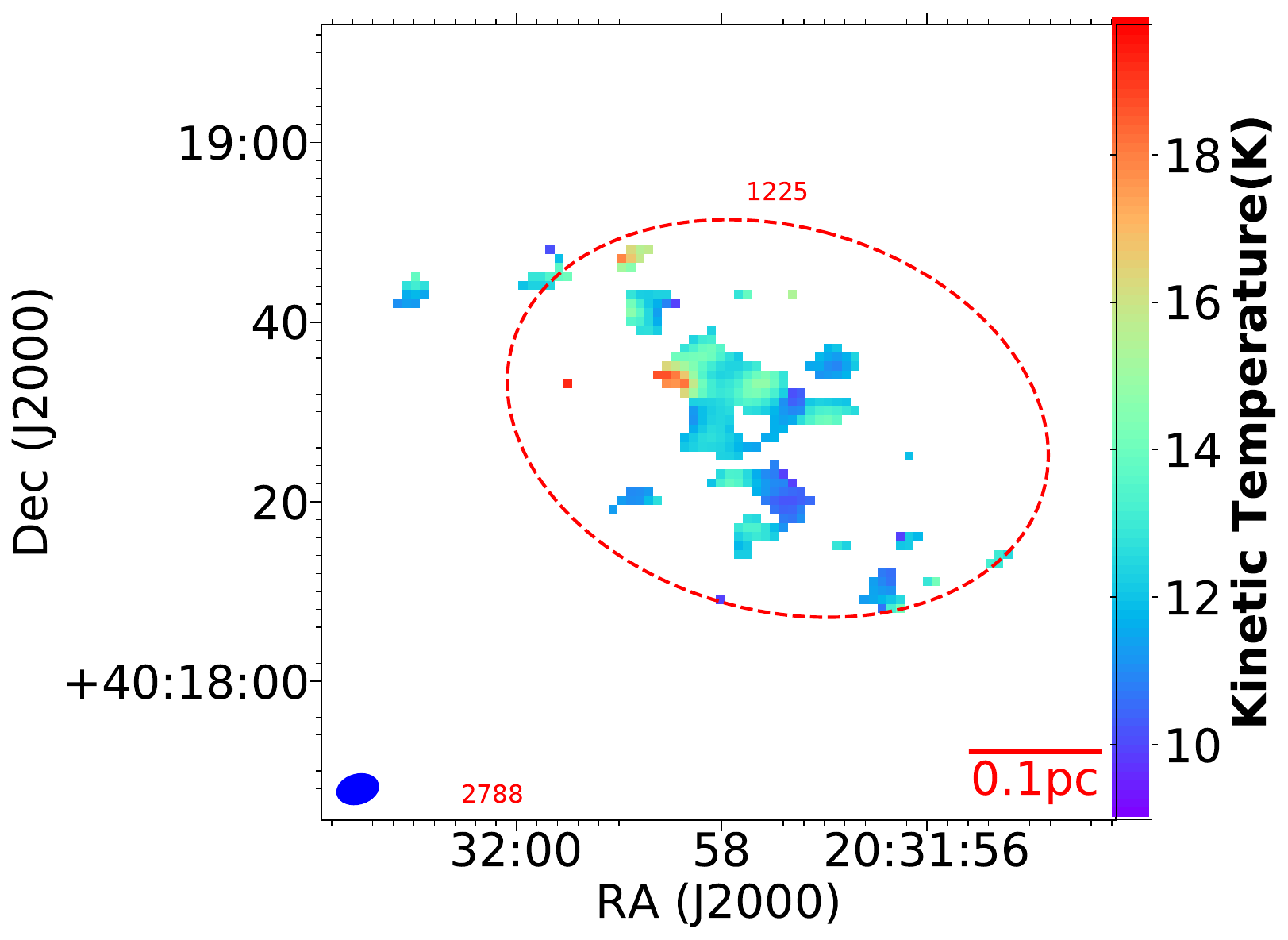} &
\includegraphics[width=.3\textwidth]{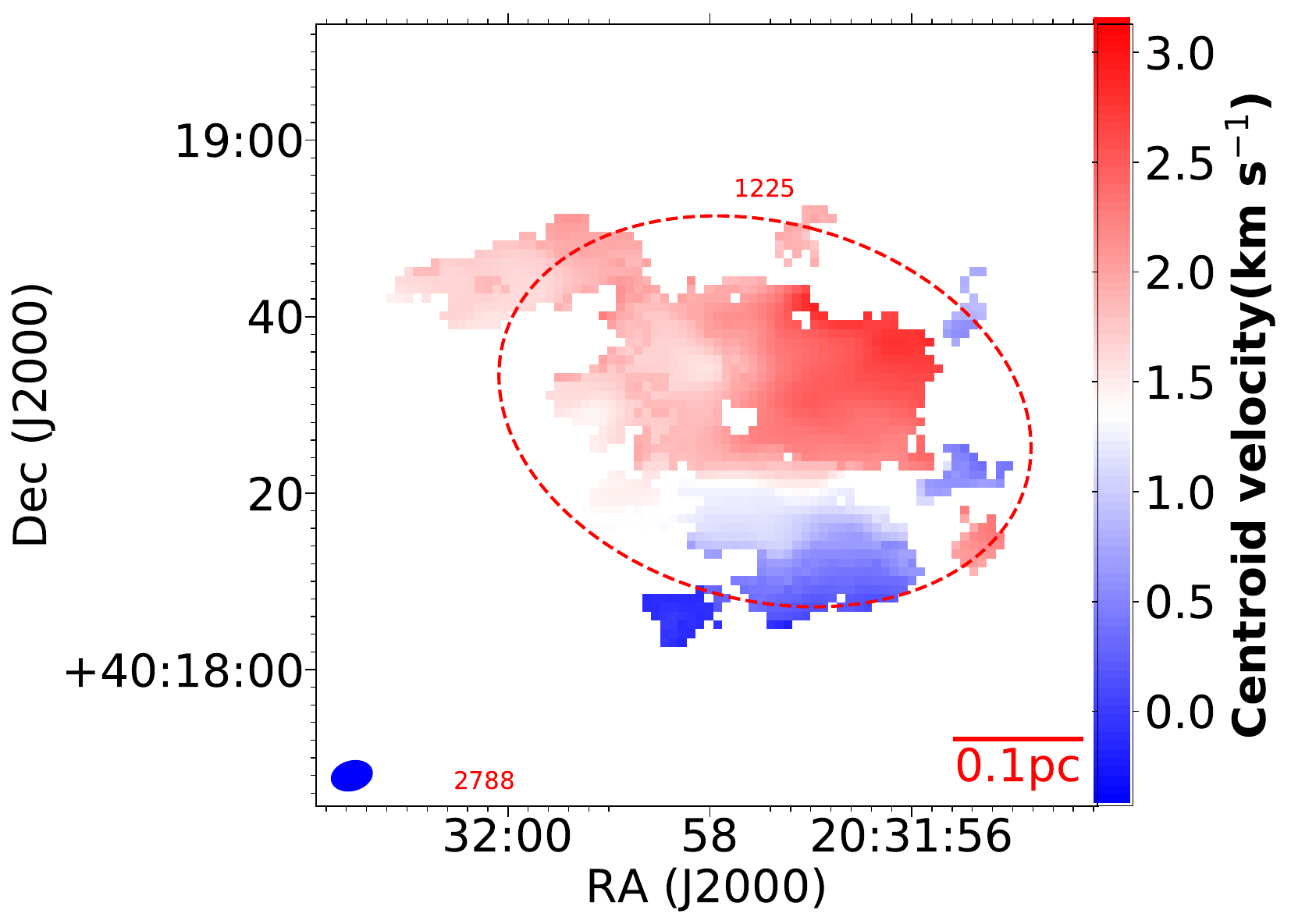} &
\includegraphics[width=.3\textwidth]{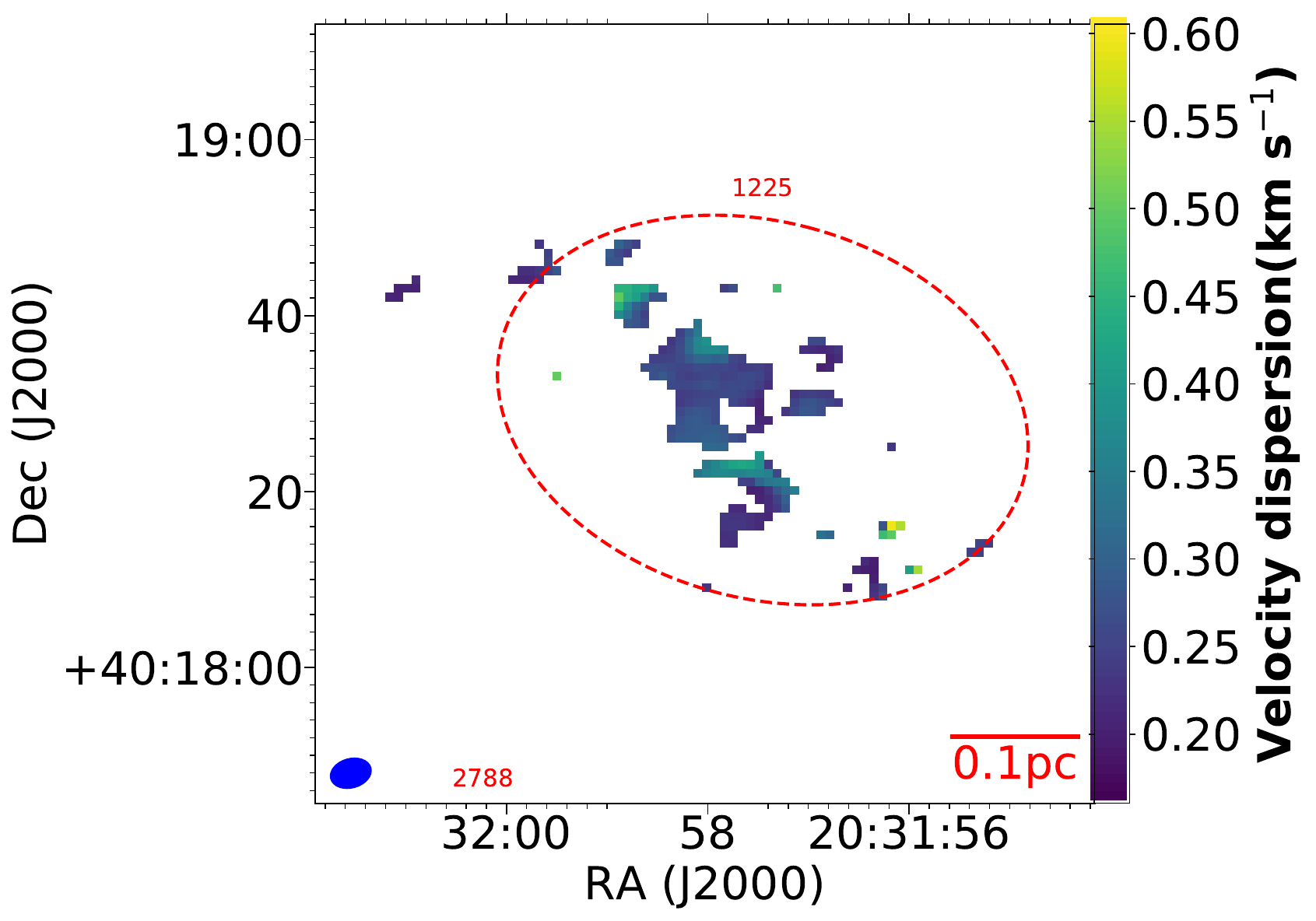} \\
 & Field 25 & \\
\includegraphics[width=.3\textwidth]{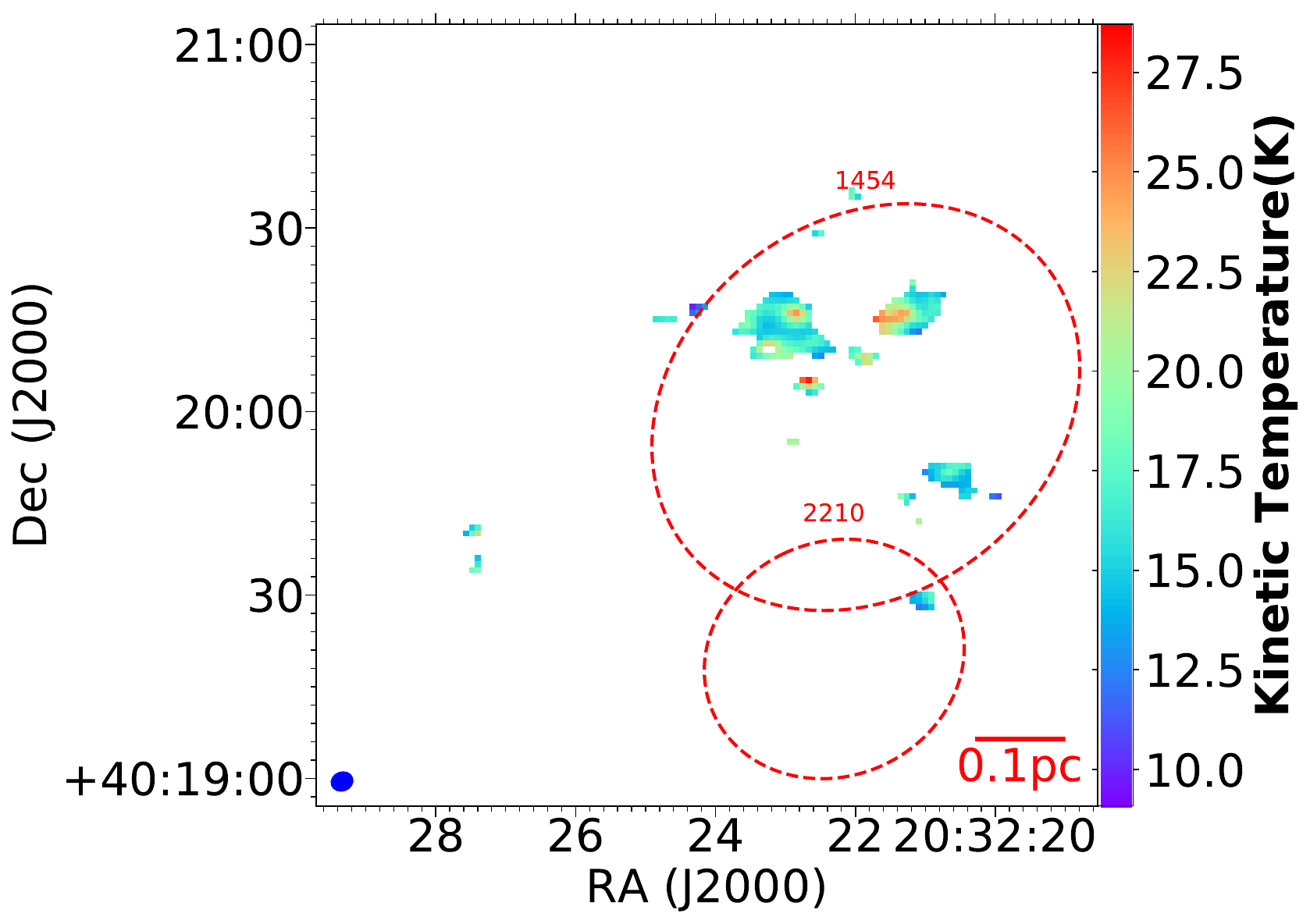} &
\includegraphics[width=.3\textwidth]{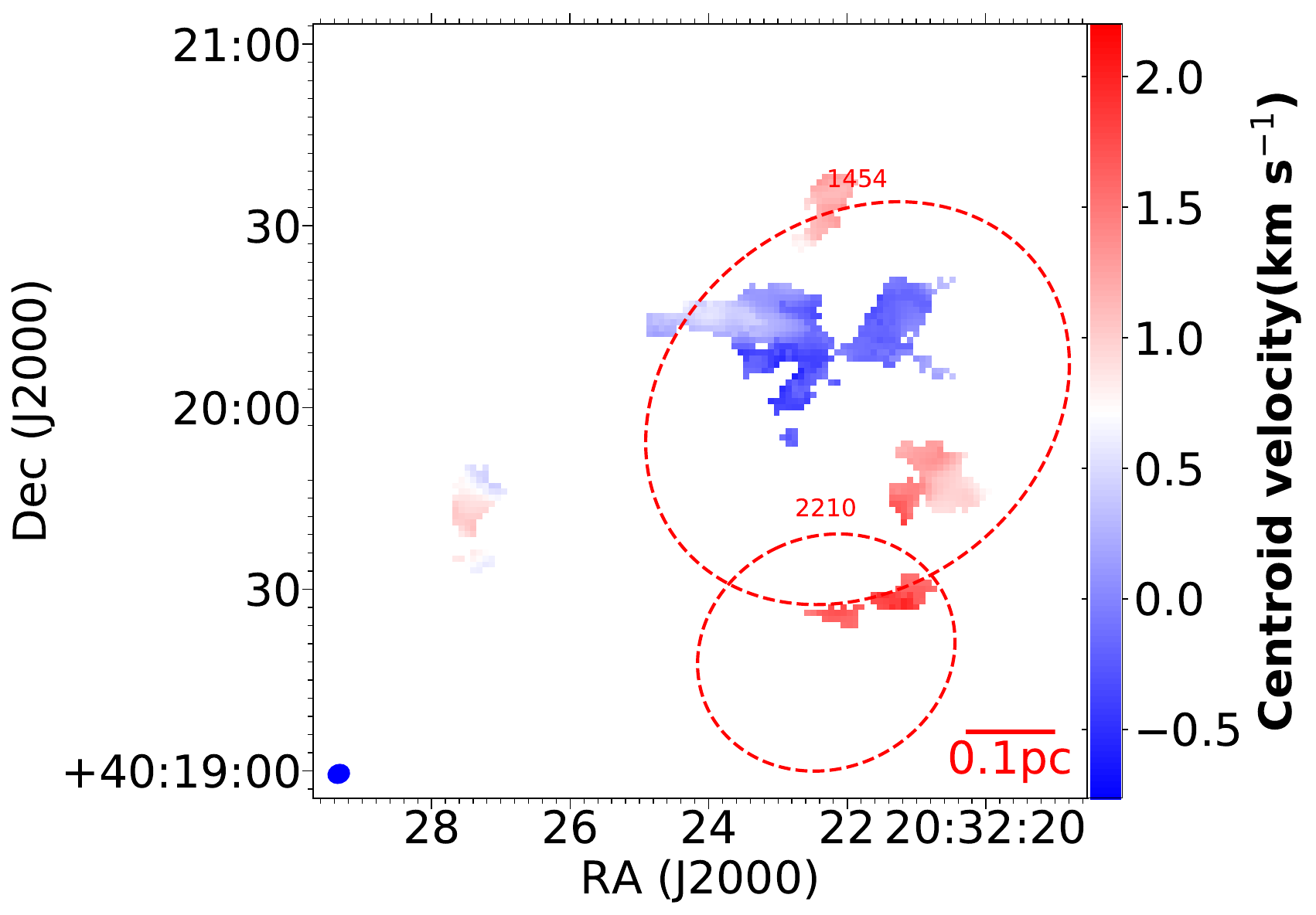} &
\includegraphics[width=.3\textwidth]{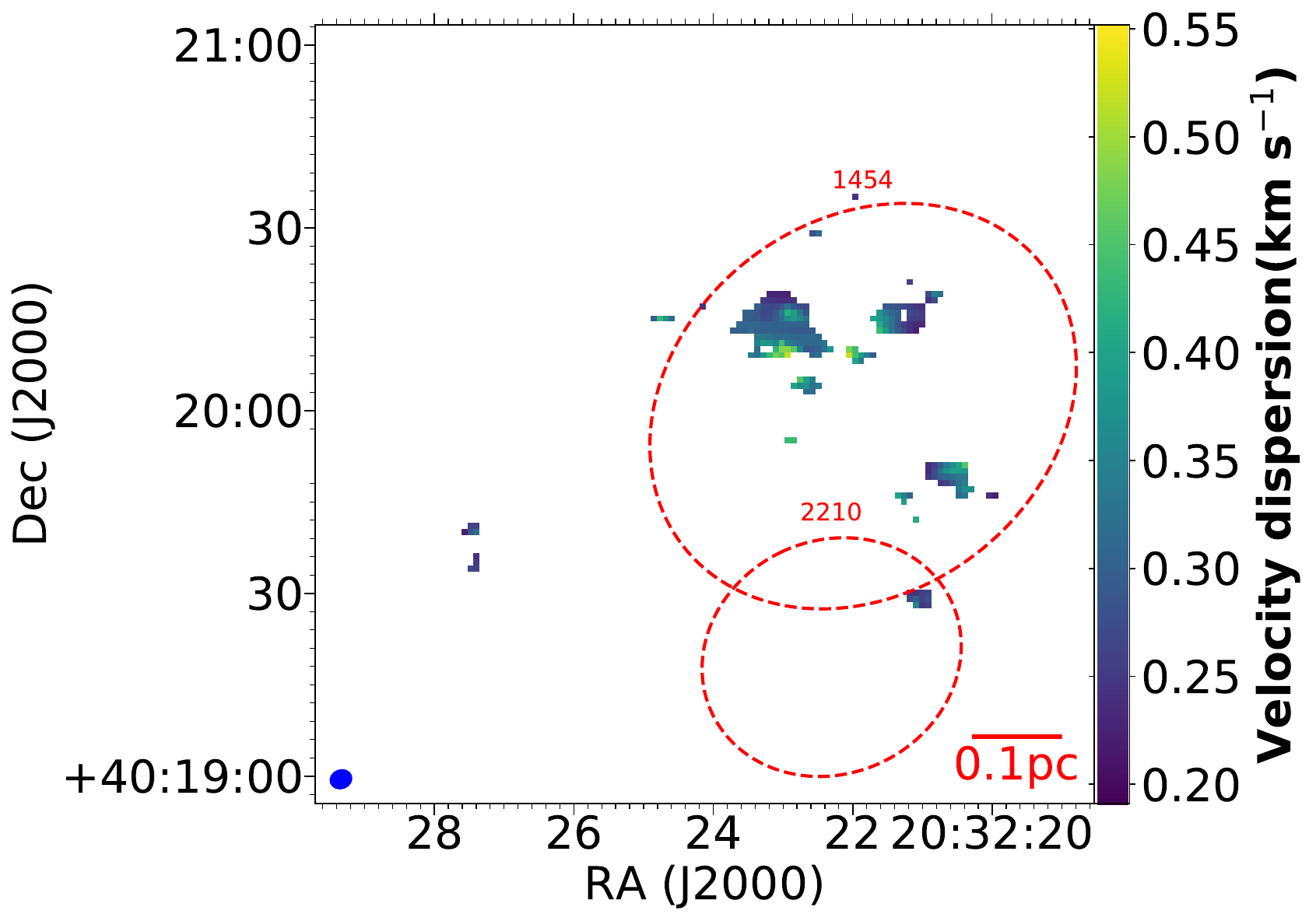} \\
 & Field 26 & \\
\includegraphics[width=.3\textwidth]{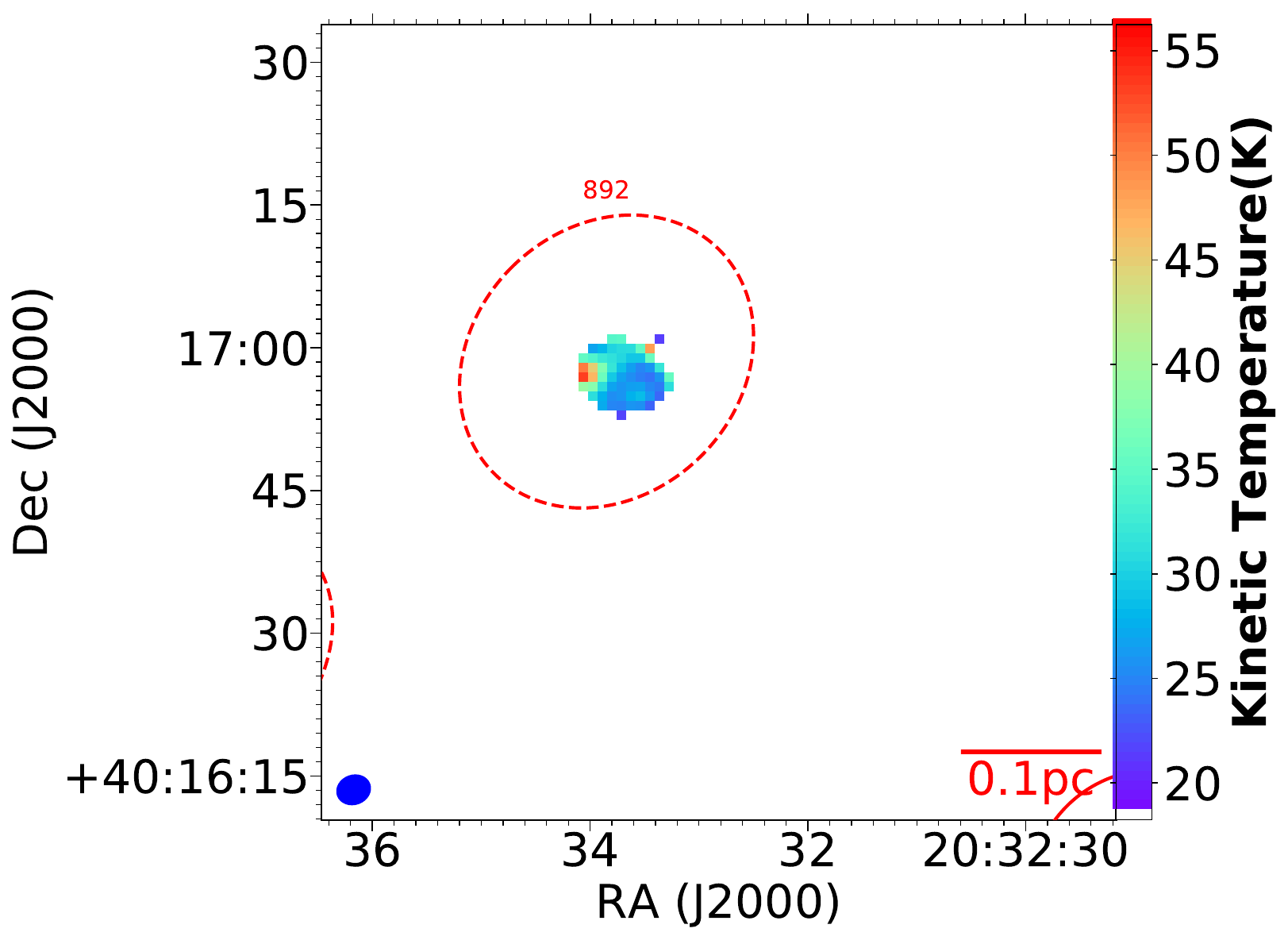} &
\includegraphics[width=.3\textwidth]{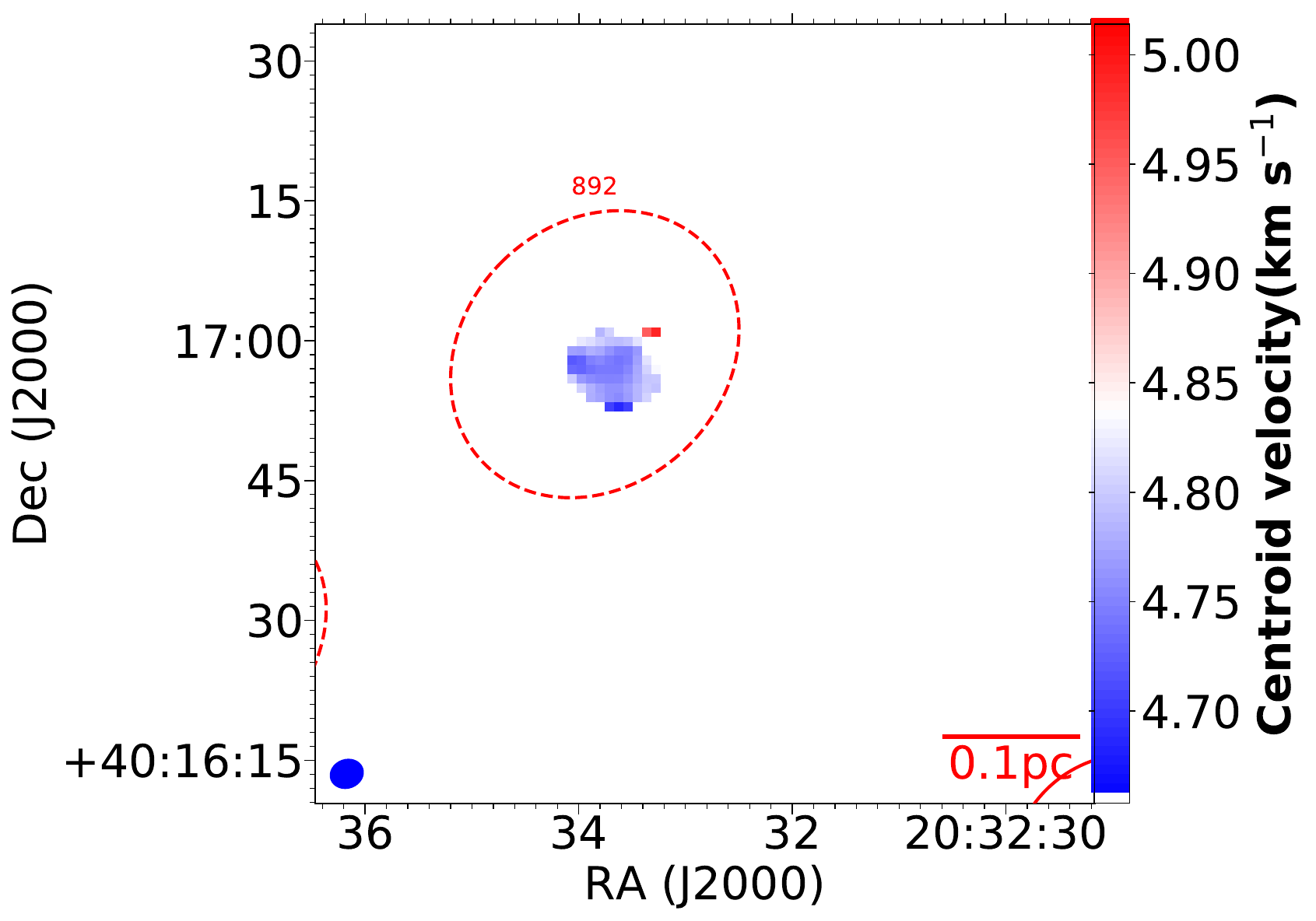} &
\includegraphics[width=.3\textwidth]{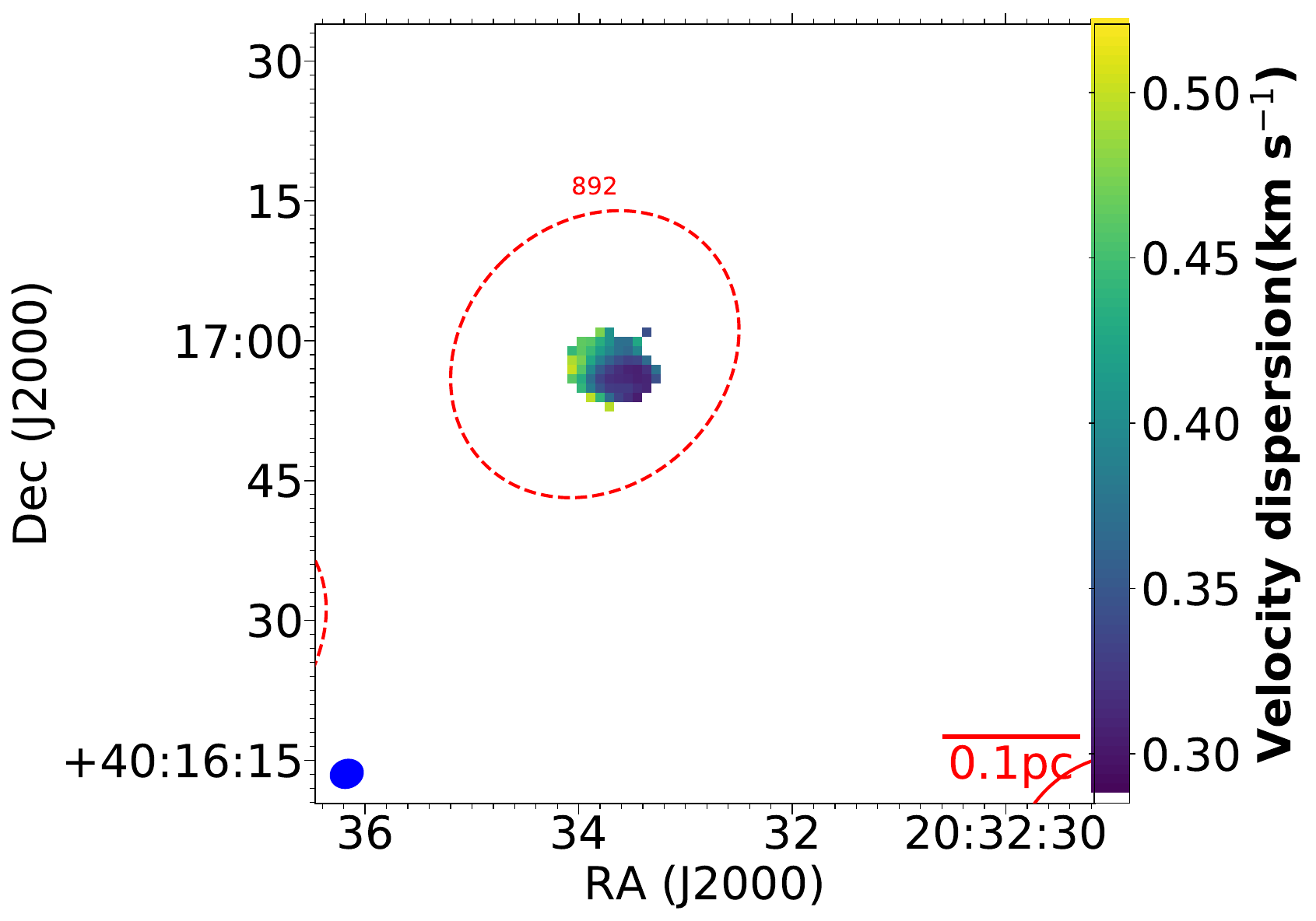} \\
 & Field 27 & \\
\includegraphics[width=.3\textwidth]{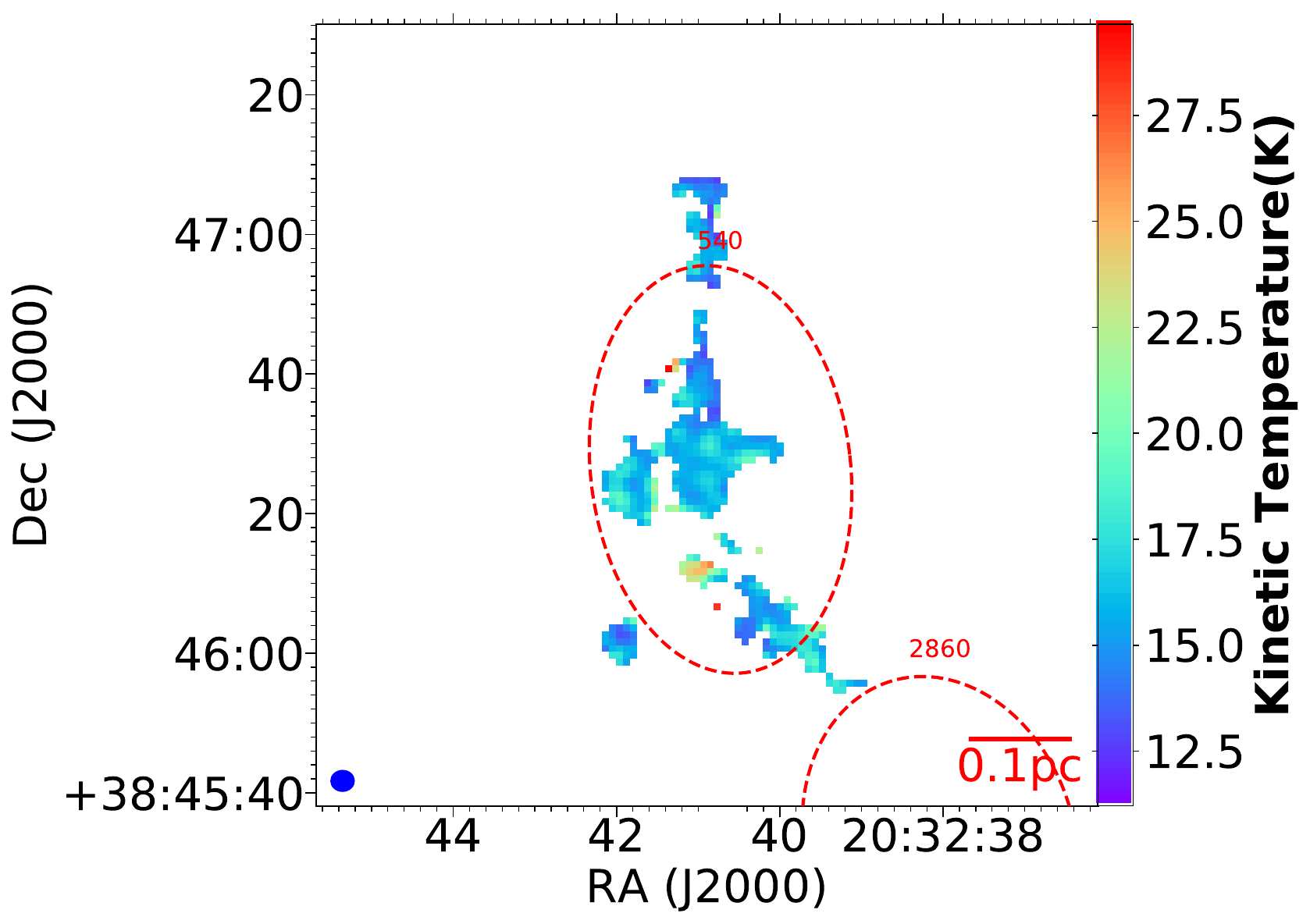} &
\includegraphics[width=.3\textwidth]{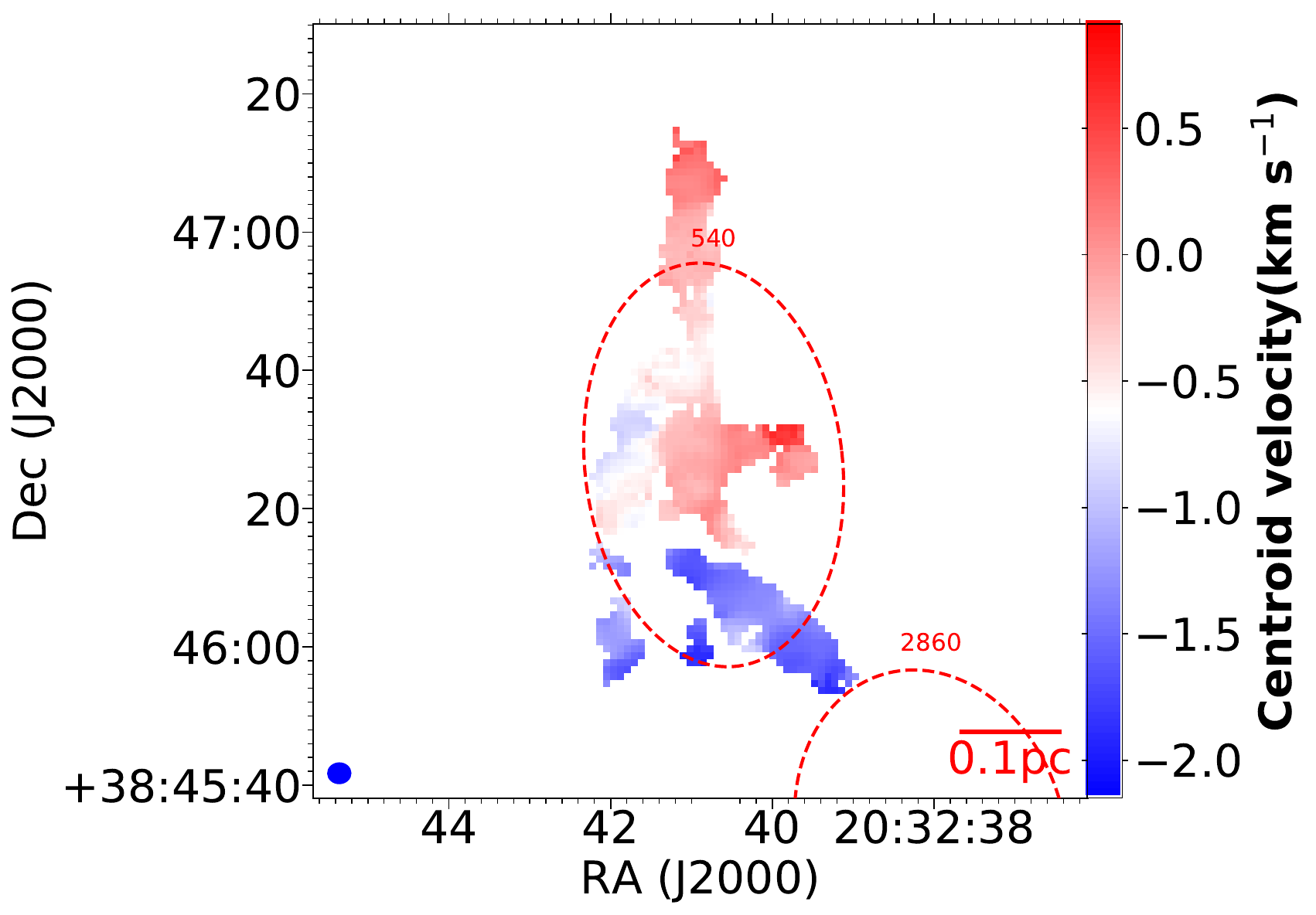} &
\includegraphics[width=.3\textwidth]{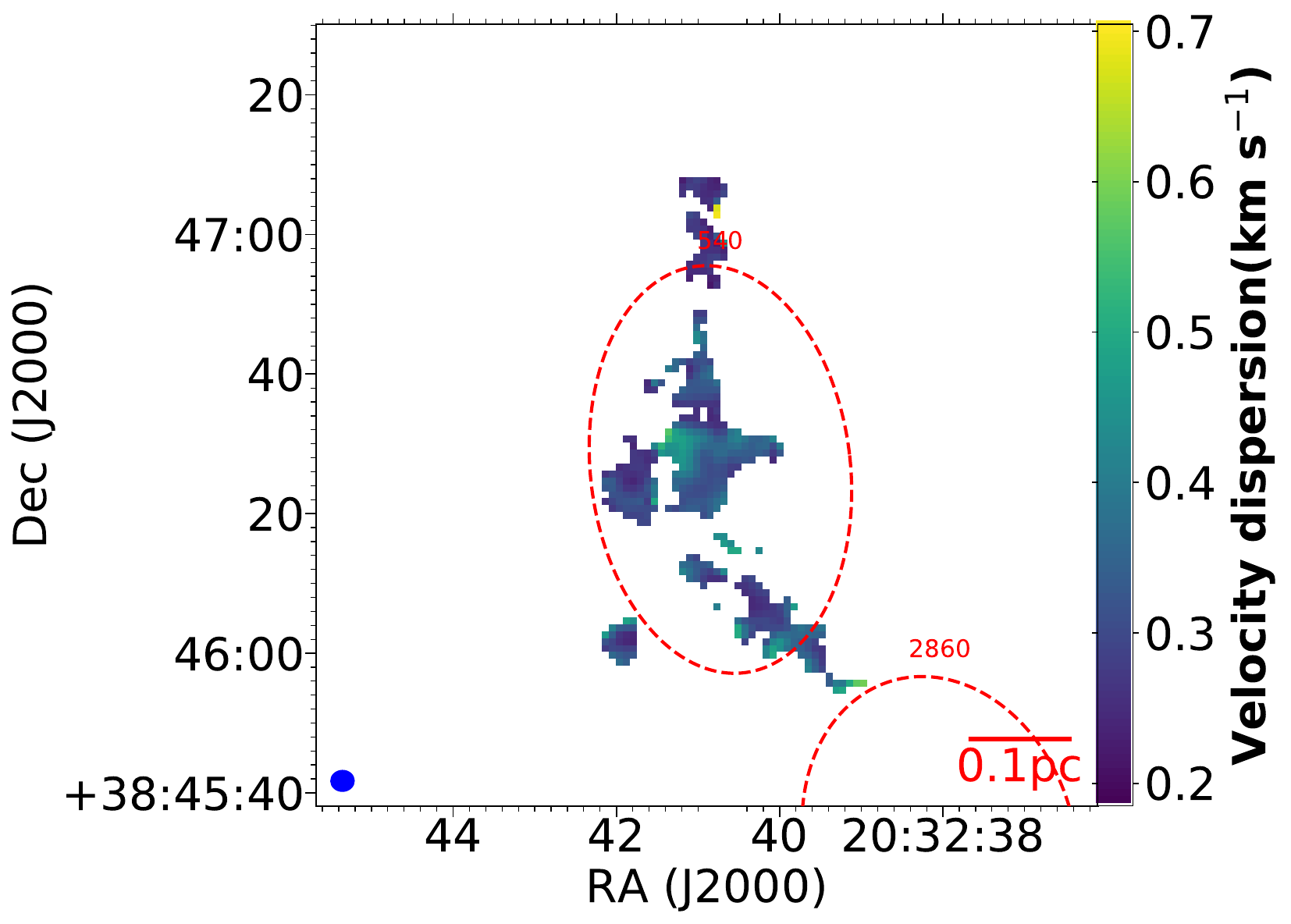} \\
 & Field 28 & \\
\includegraphics[width=.3\textwidth]{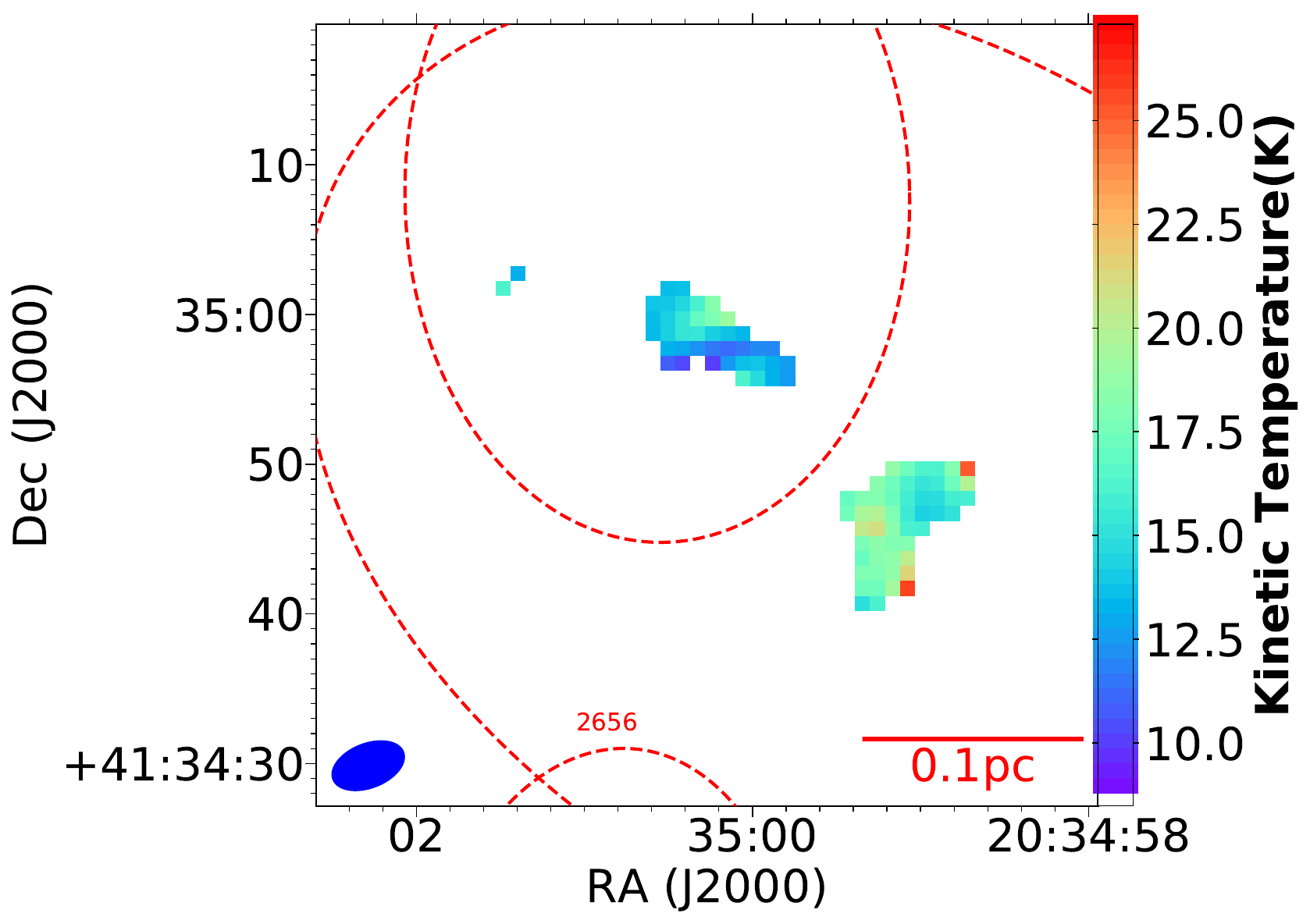} &
\includegraphics[width=.3\textwidth]{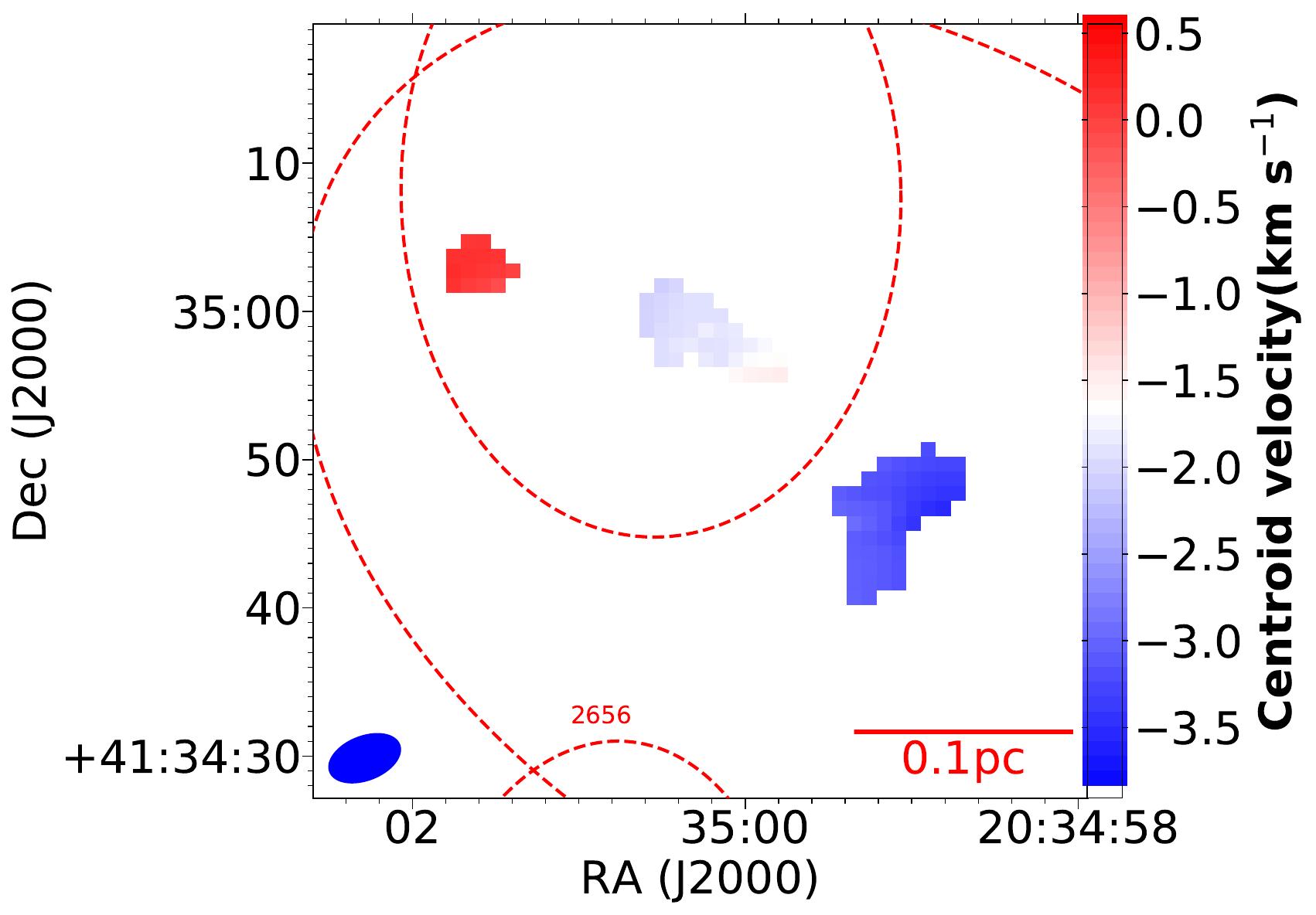} &
\includegraphics[width=.3\textwidth]{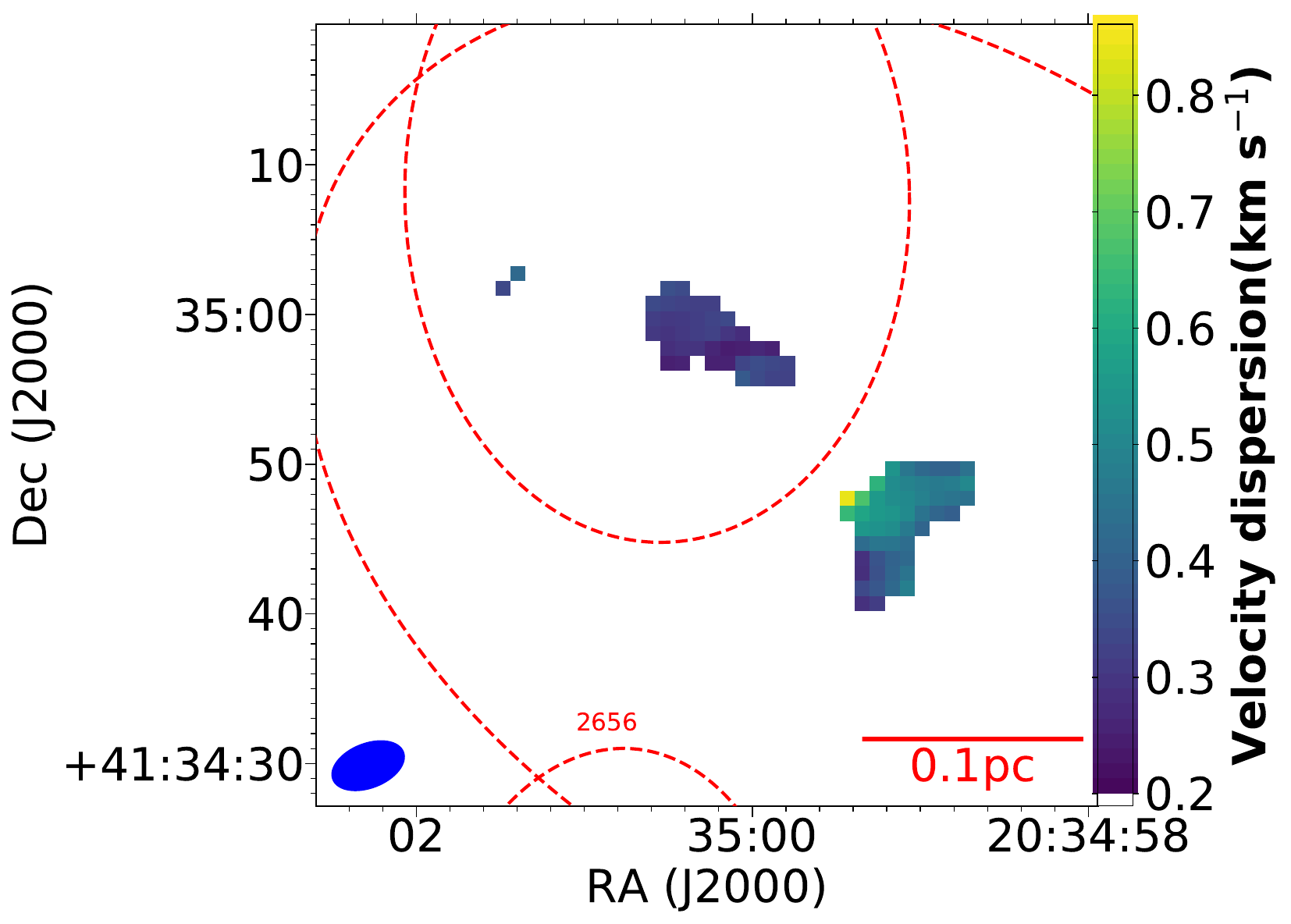} \\
 & Field 29 & \\
\includegraphics[width=.3\textwidth]{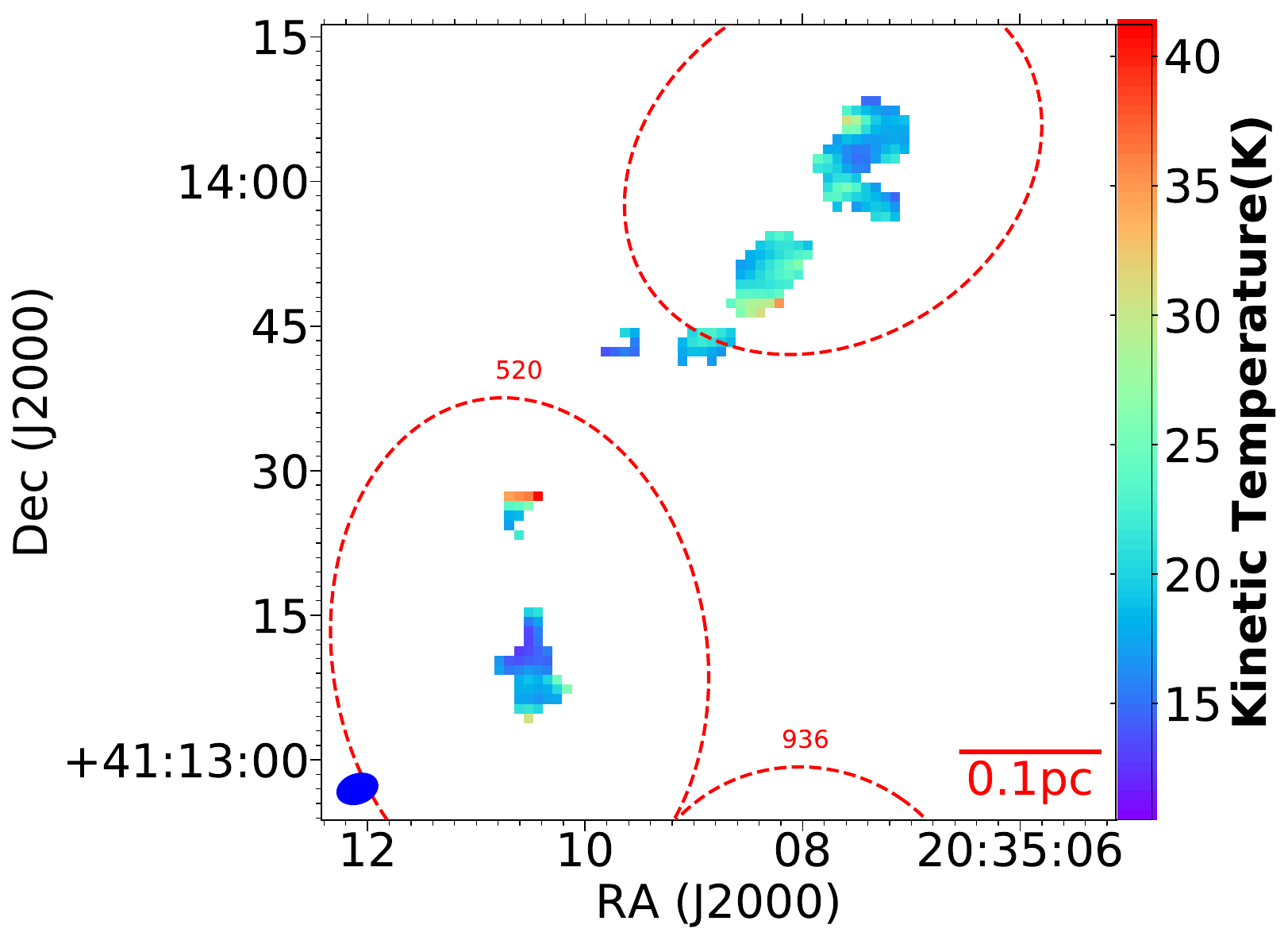} &
\includegraphics[width=.3\textwidth]{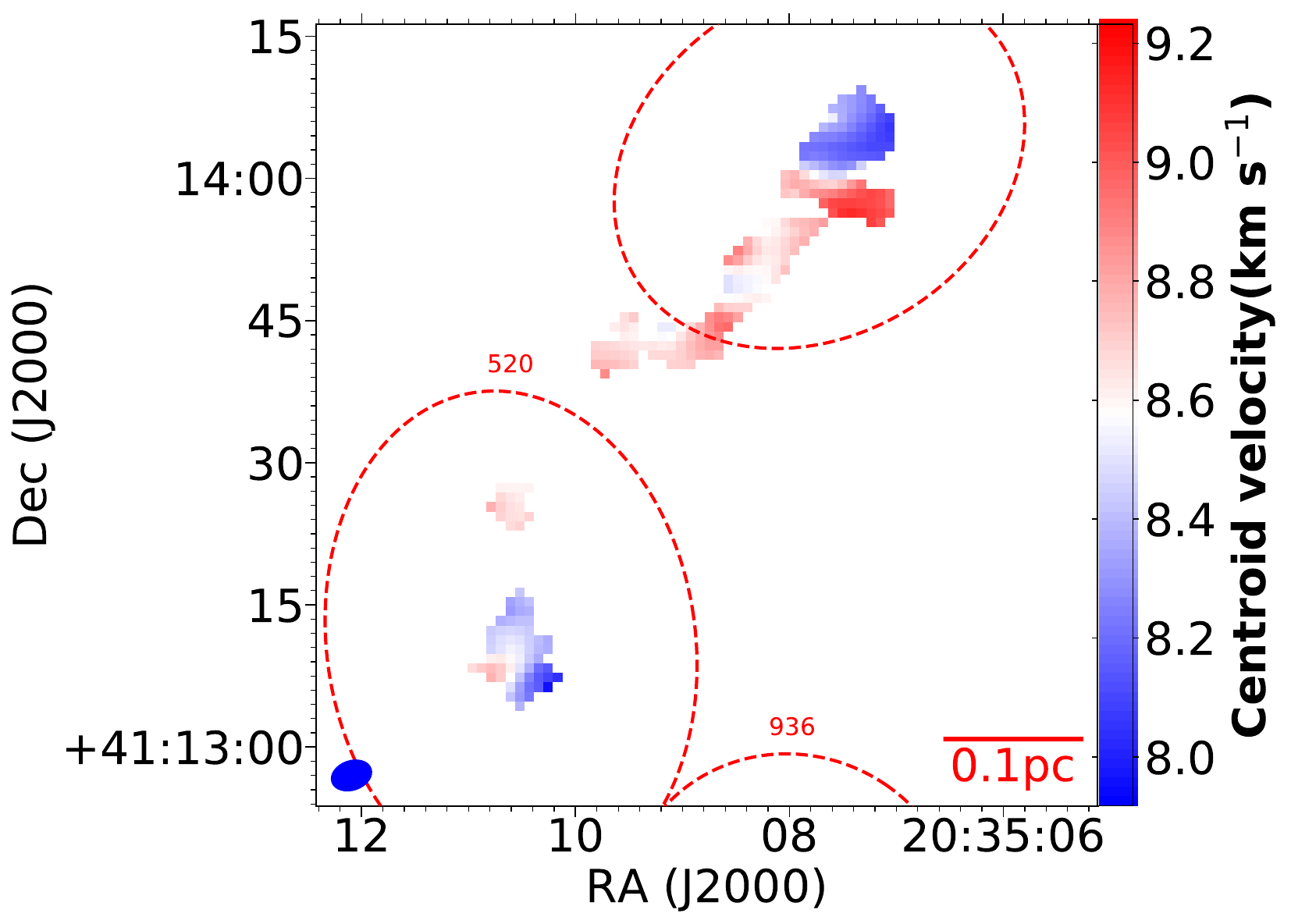} &
\includegraphics[width=.3\textwidth]{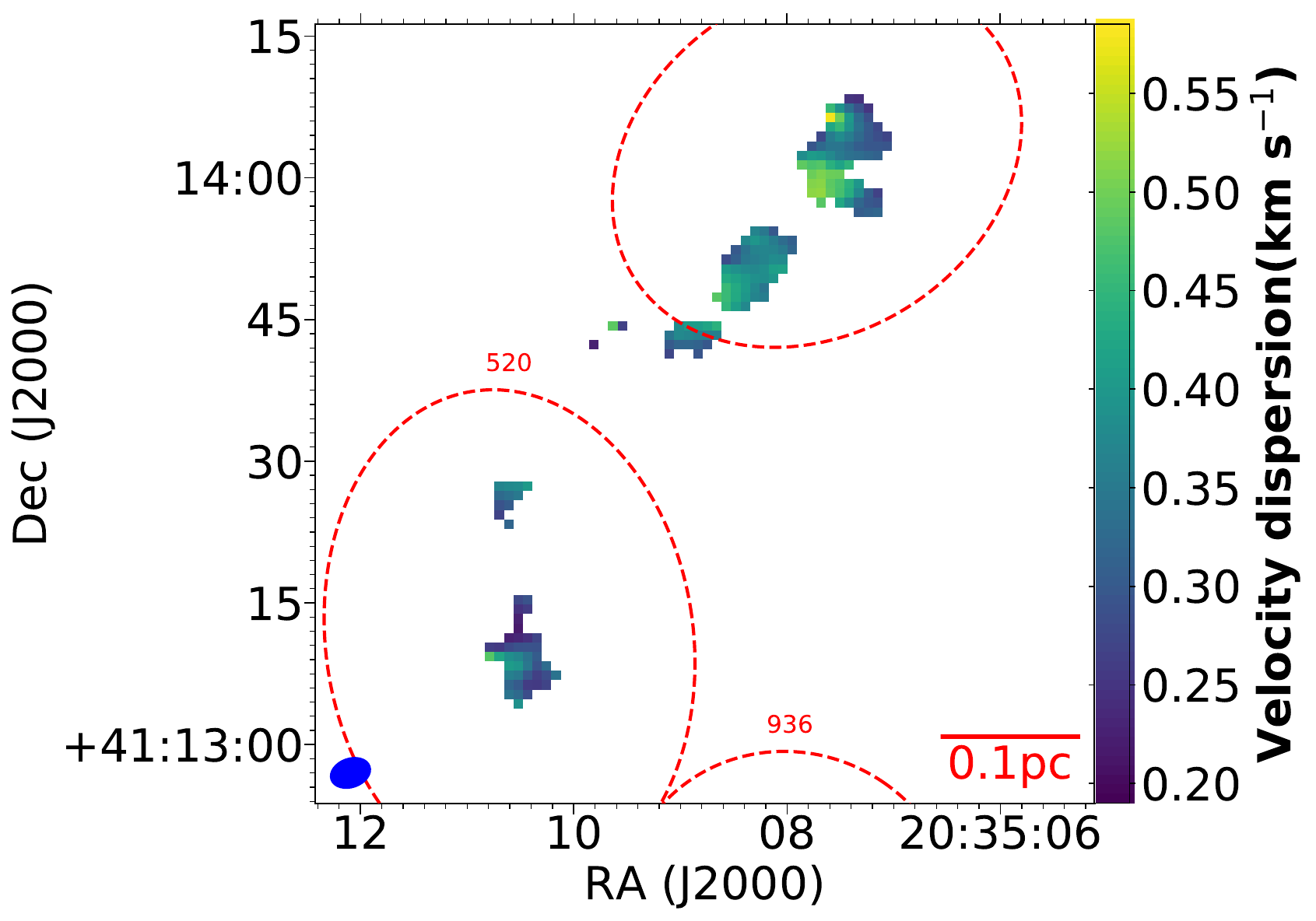} \\
 & Field 31 & \\
\includegraphics[width=.3\textwidth]{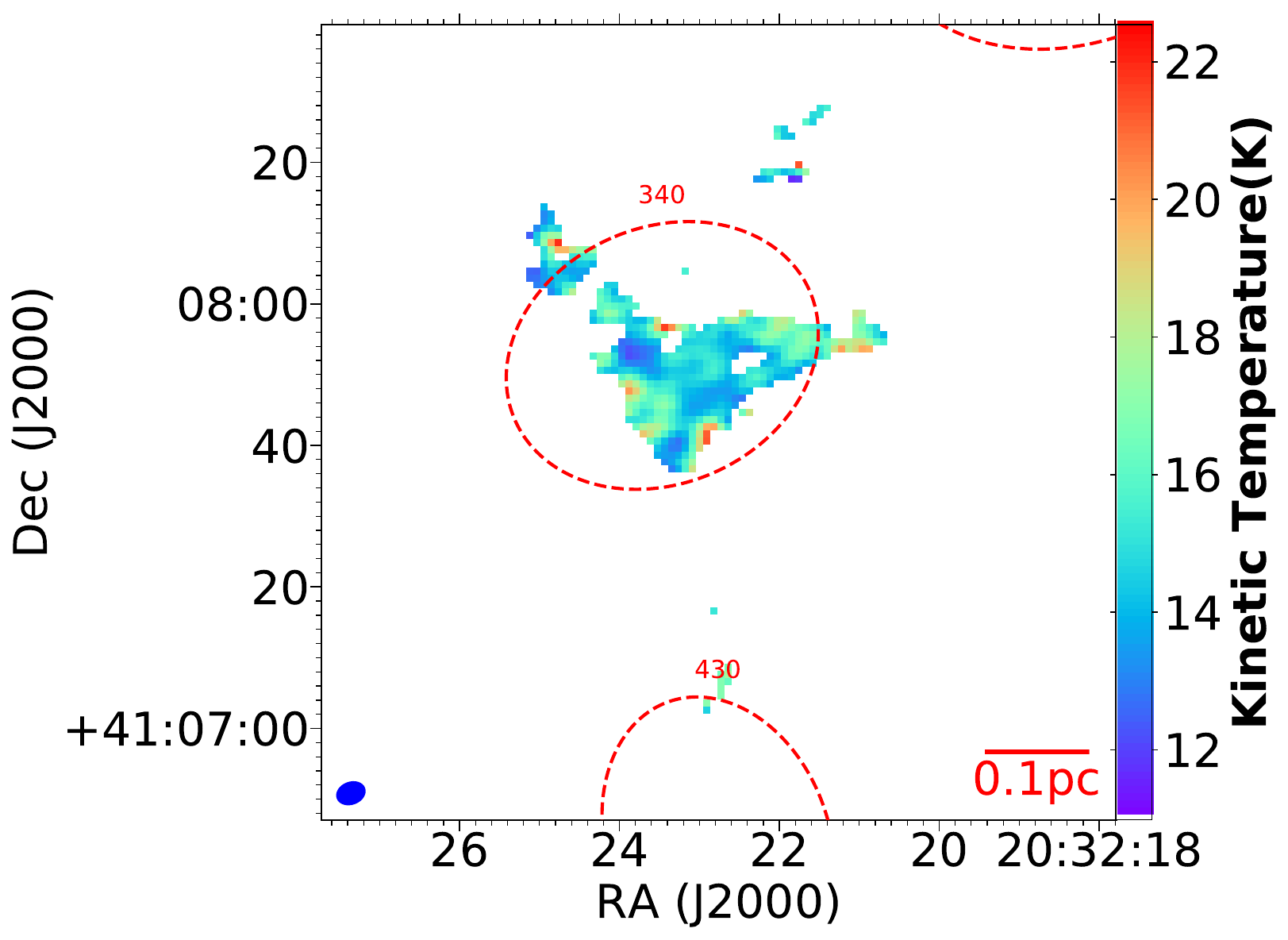} &
\includegraphics[width=.3\textwidth]{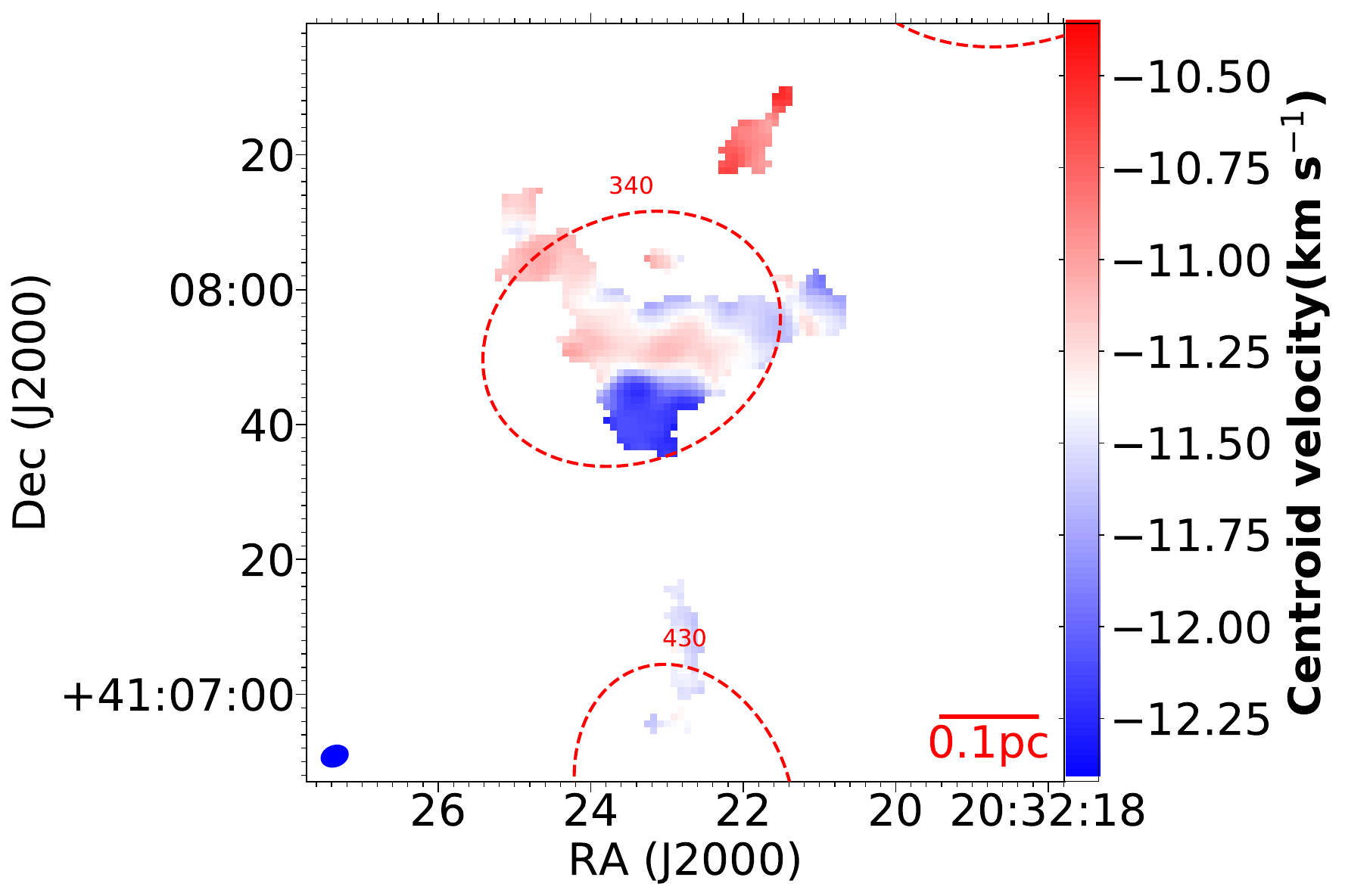} &
\includegraphics[width=.3\textwidth]{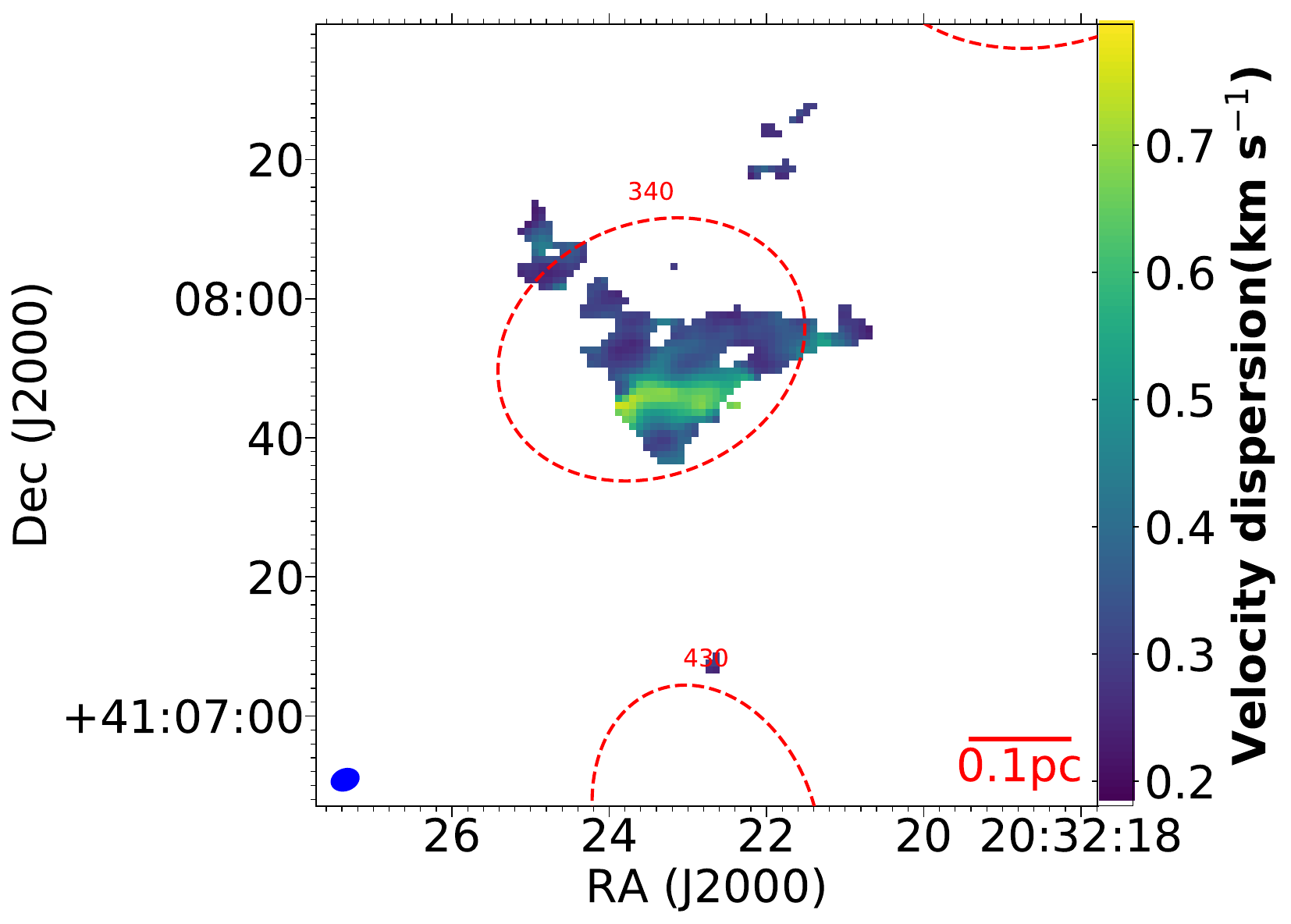} \\
 & Field 32 & \\
\includegraphics[width=.3\textwidth]{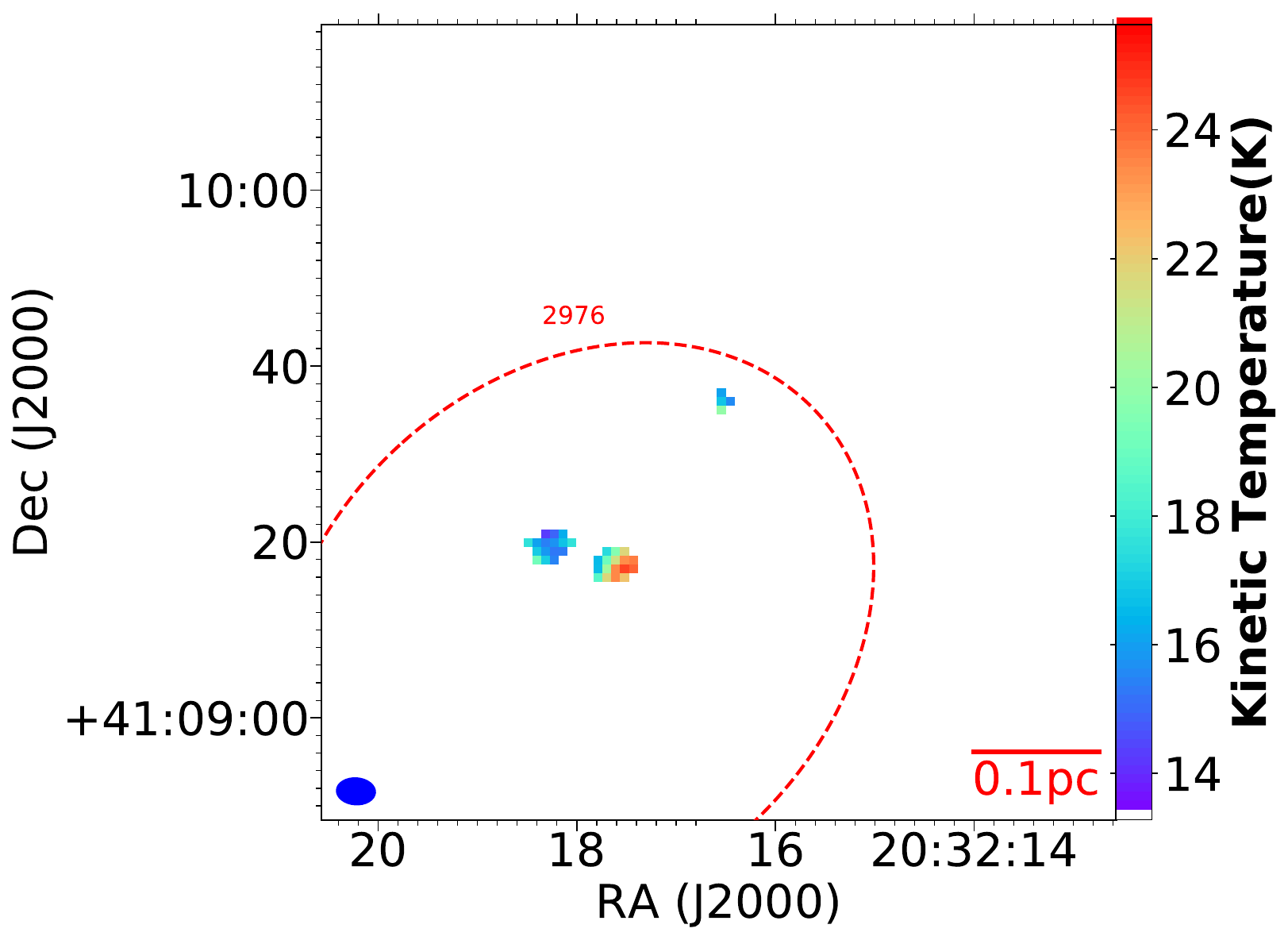} &
\includegraphics[width=.3\textwidth]{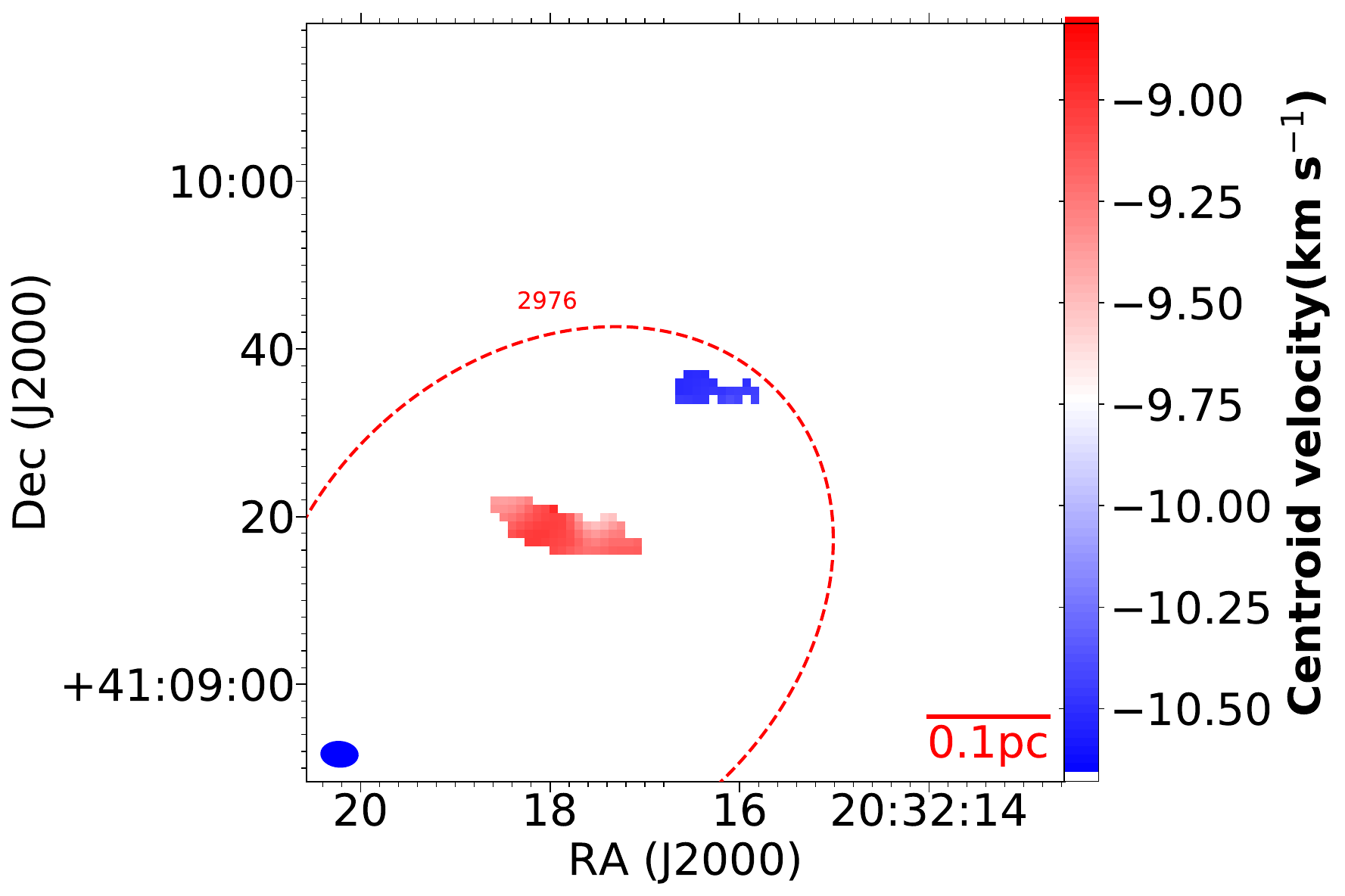} &
\includegraphics[width=.3\textwidth]{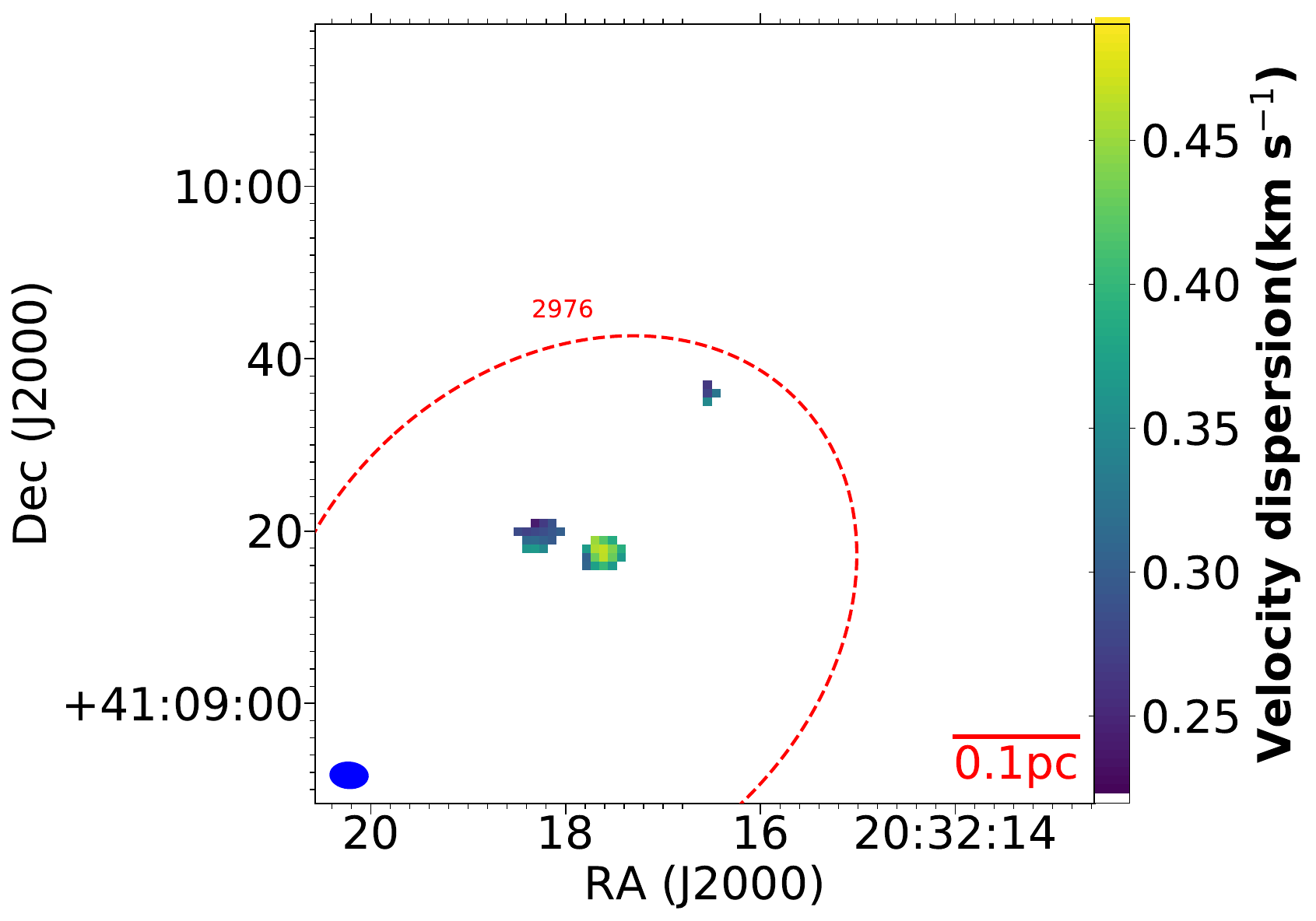} \\
 & Field 33 & \\
\includegraphics[width=.3\textwidth]{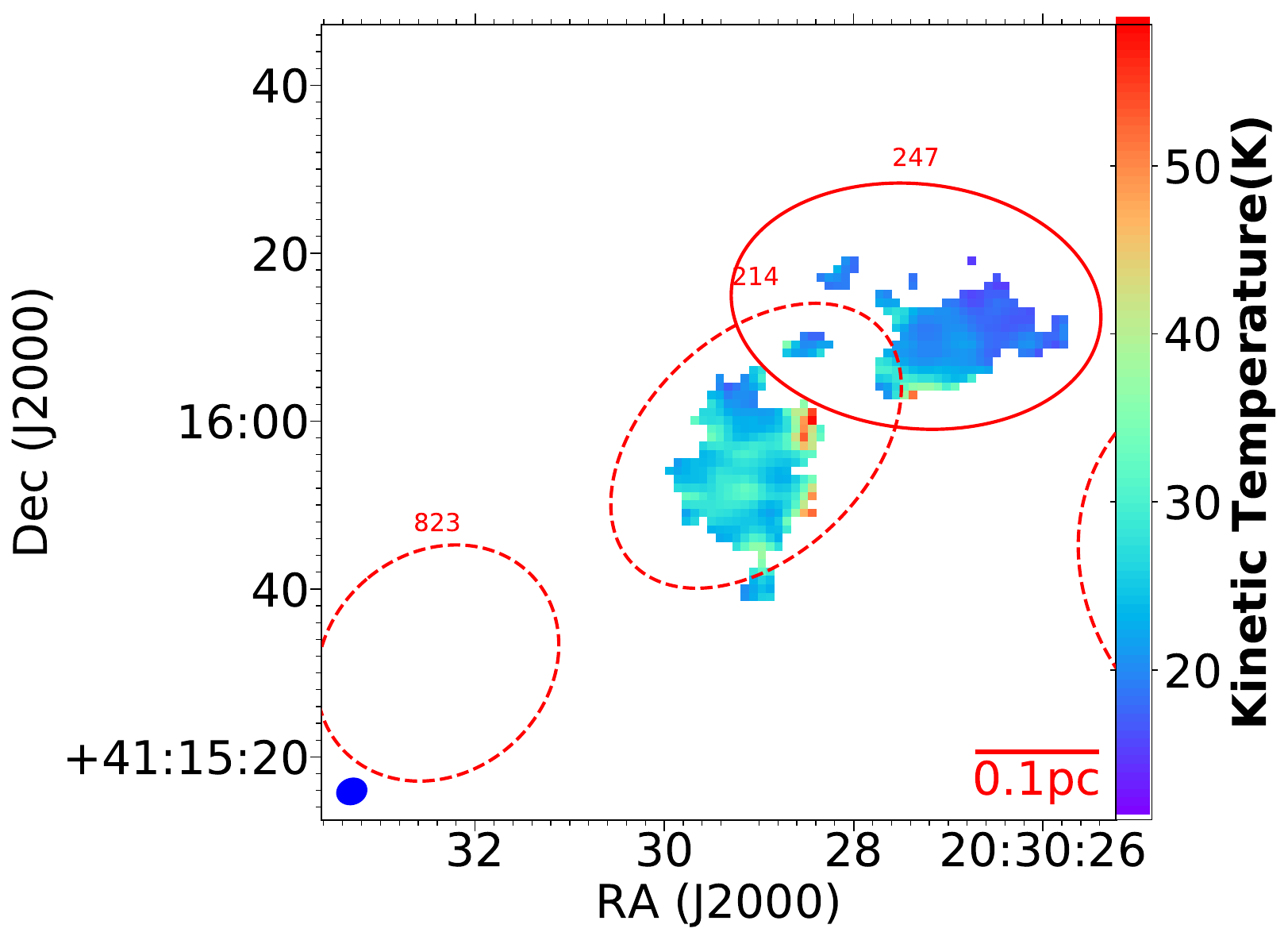} &
\includegraphics[width=.3\textwidth]{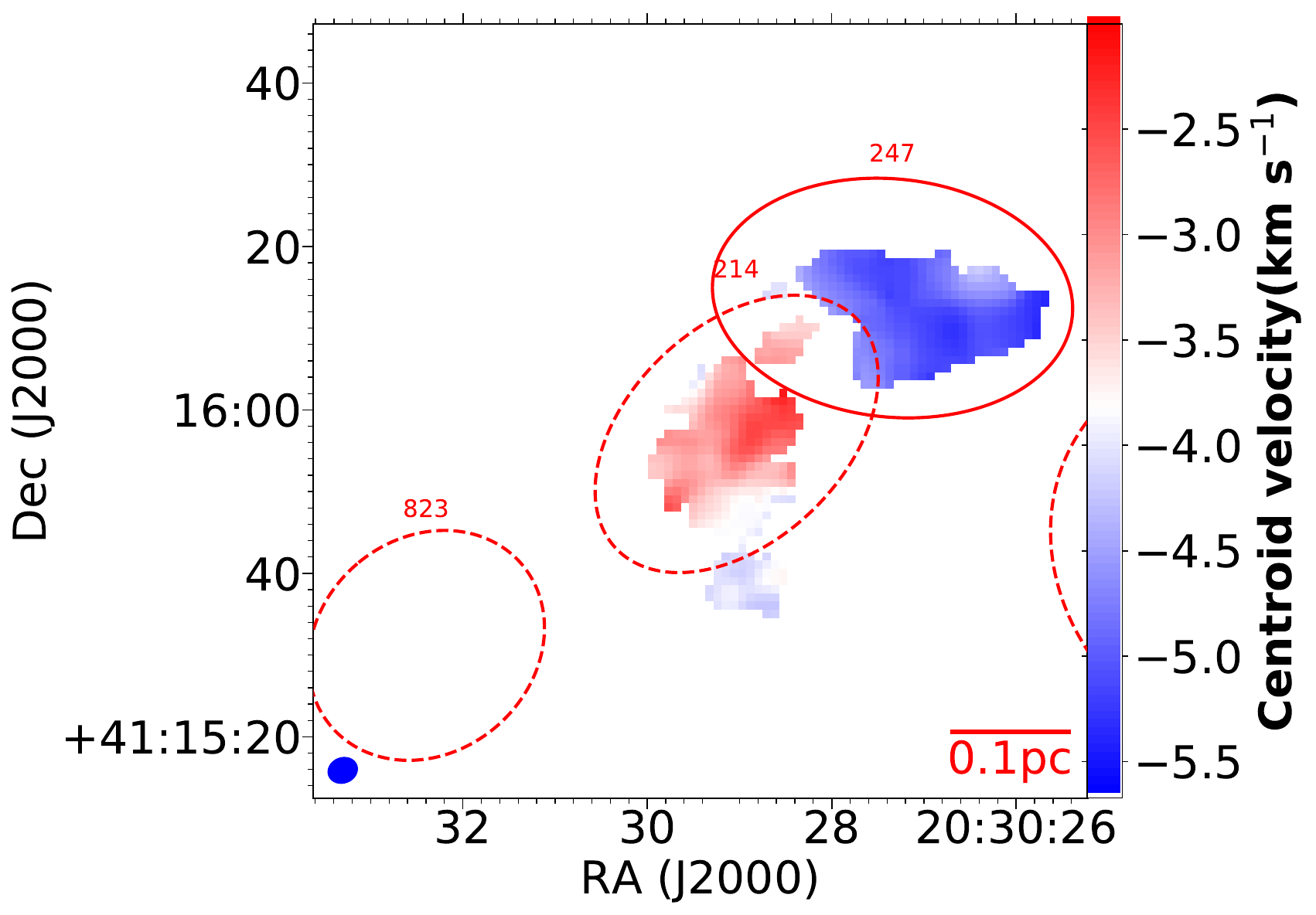} &
\includegraphics[width=.3\textwidth]{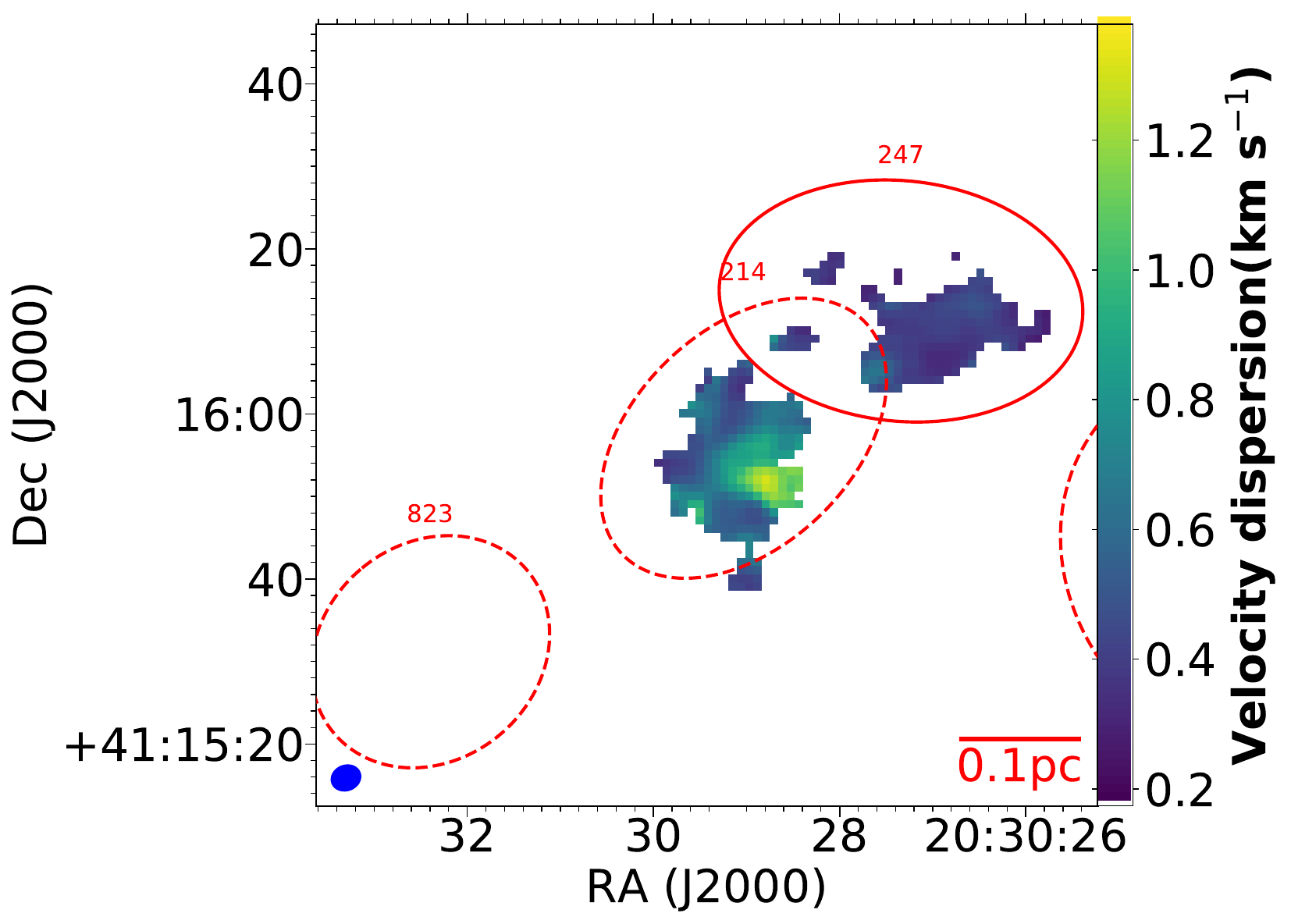} \\
 & Field 34 & \\
\includegraphics[width=.3\textwidth]{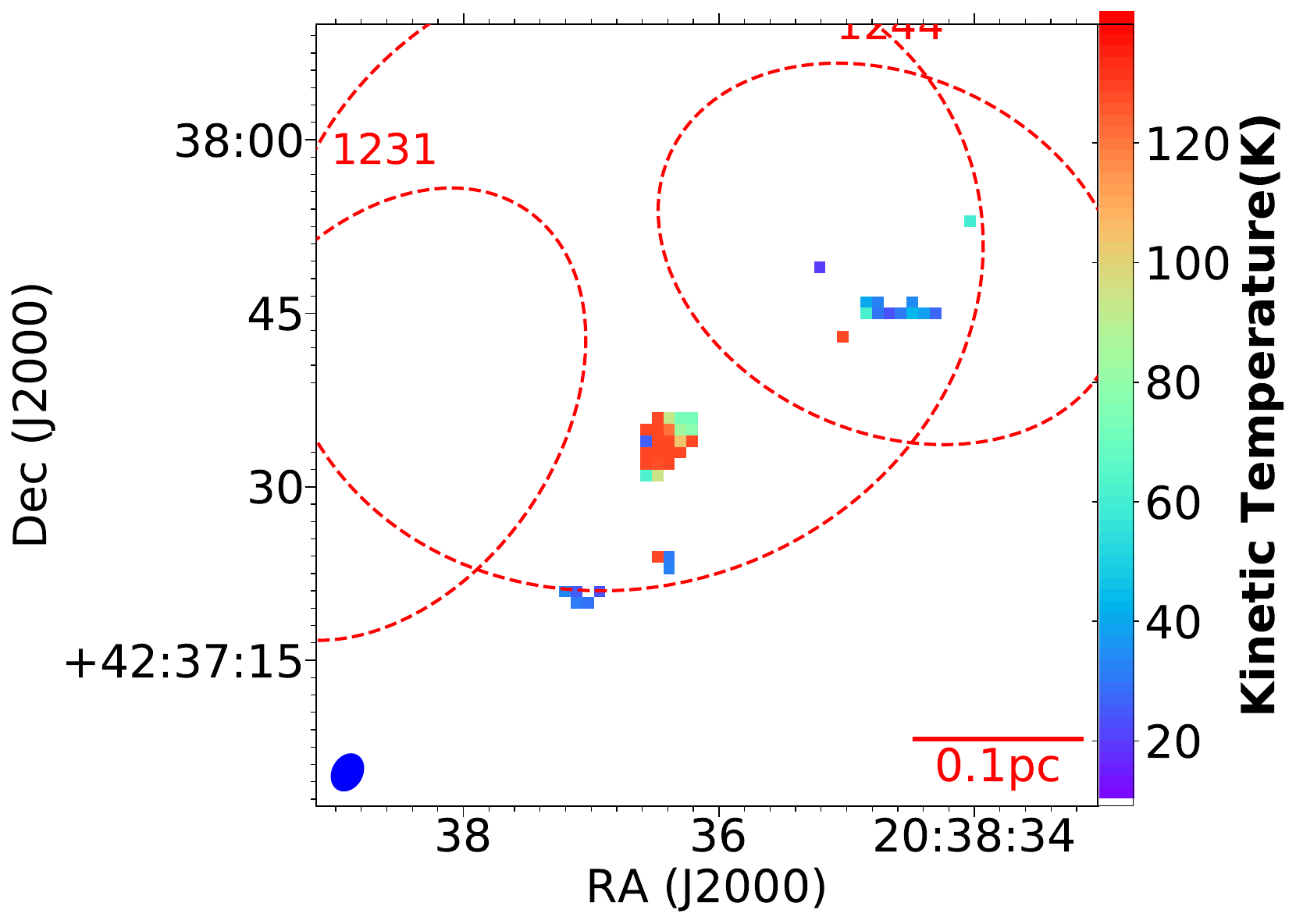} &
\includegraphics[width=.3\textwidth]{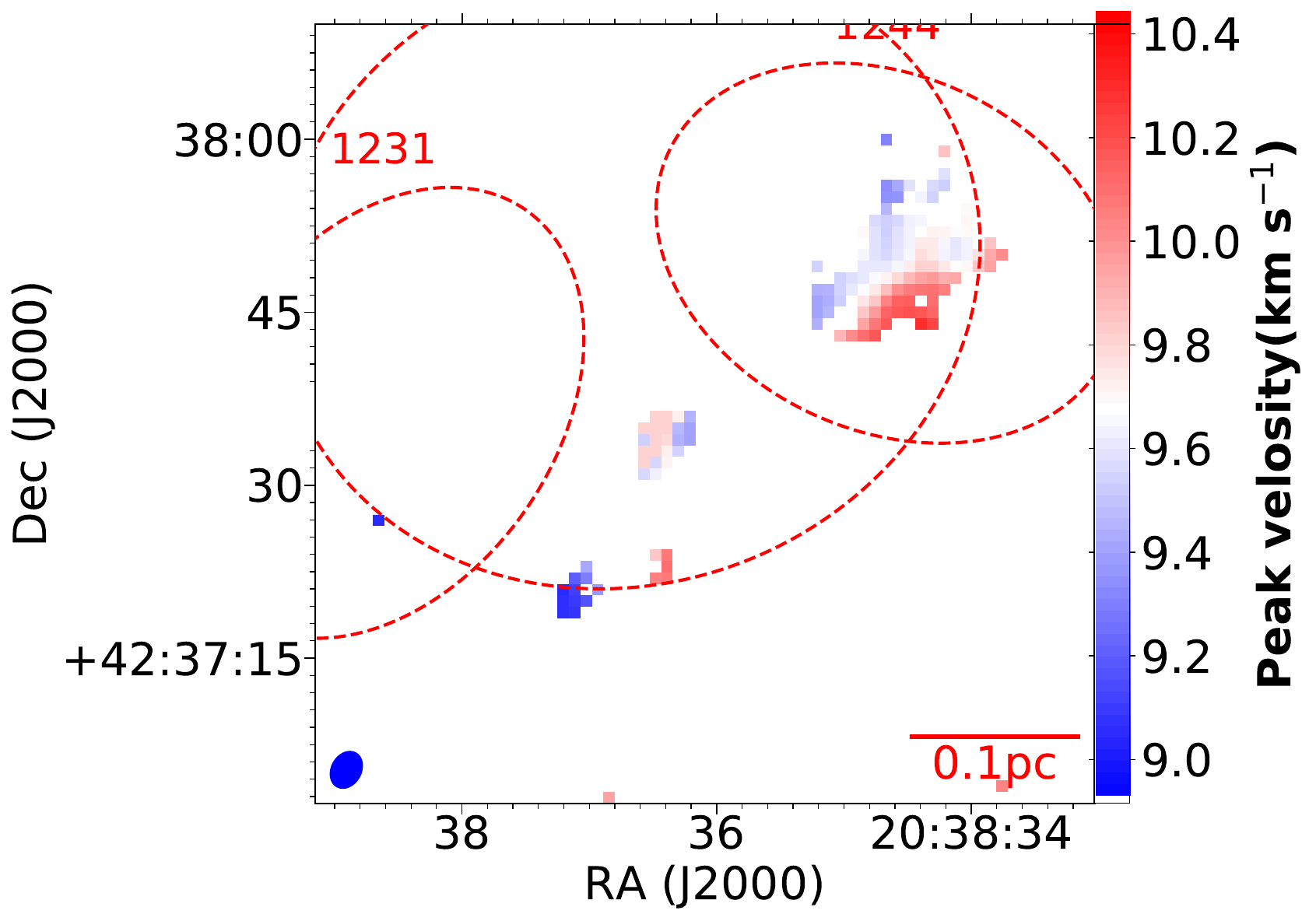} &
\includegraphics[width=.3\textwidth]{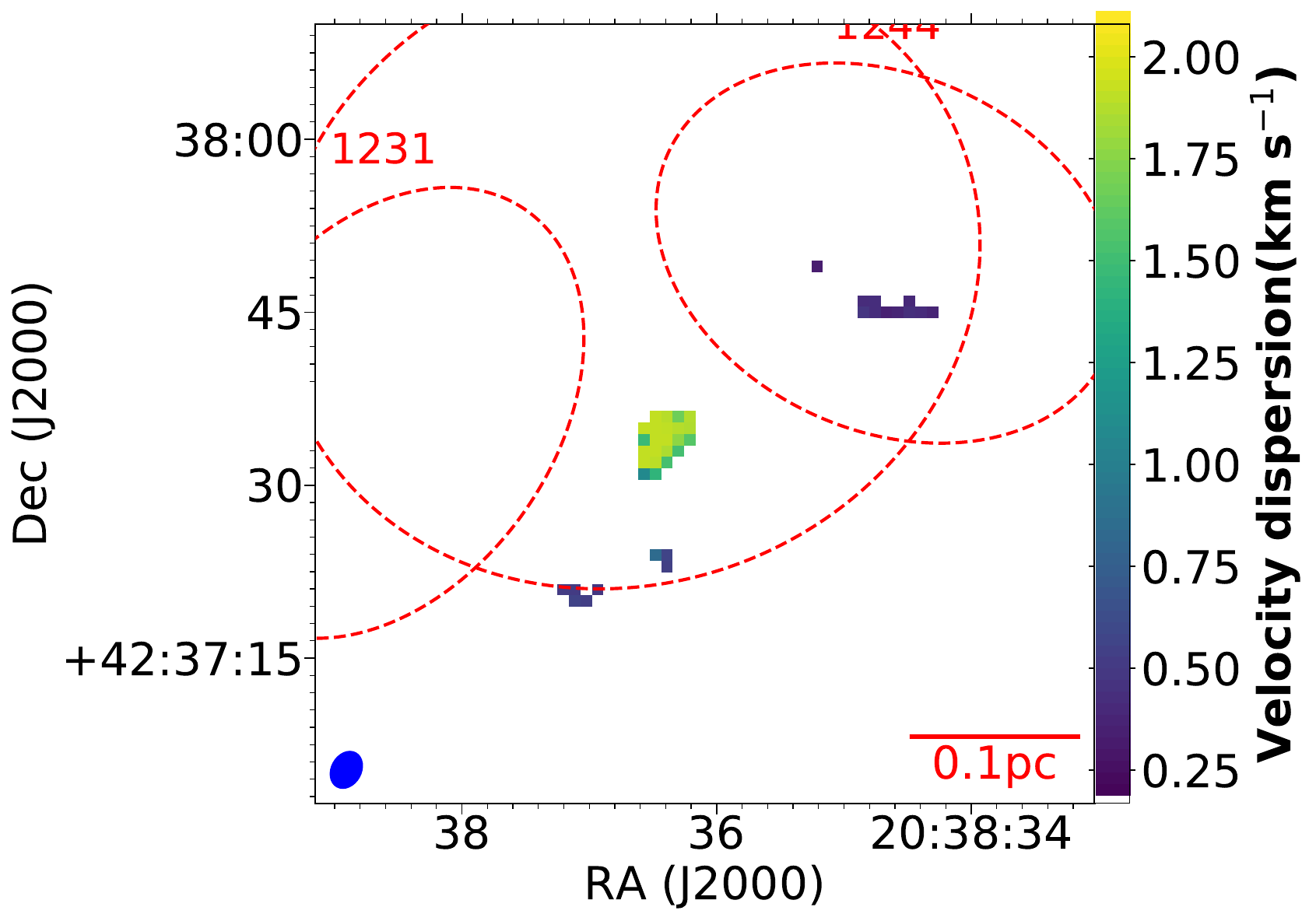} \\
 & Field 37 - Component 1 & \\
\includegraphics[width=.3\textwidth]{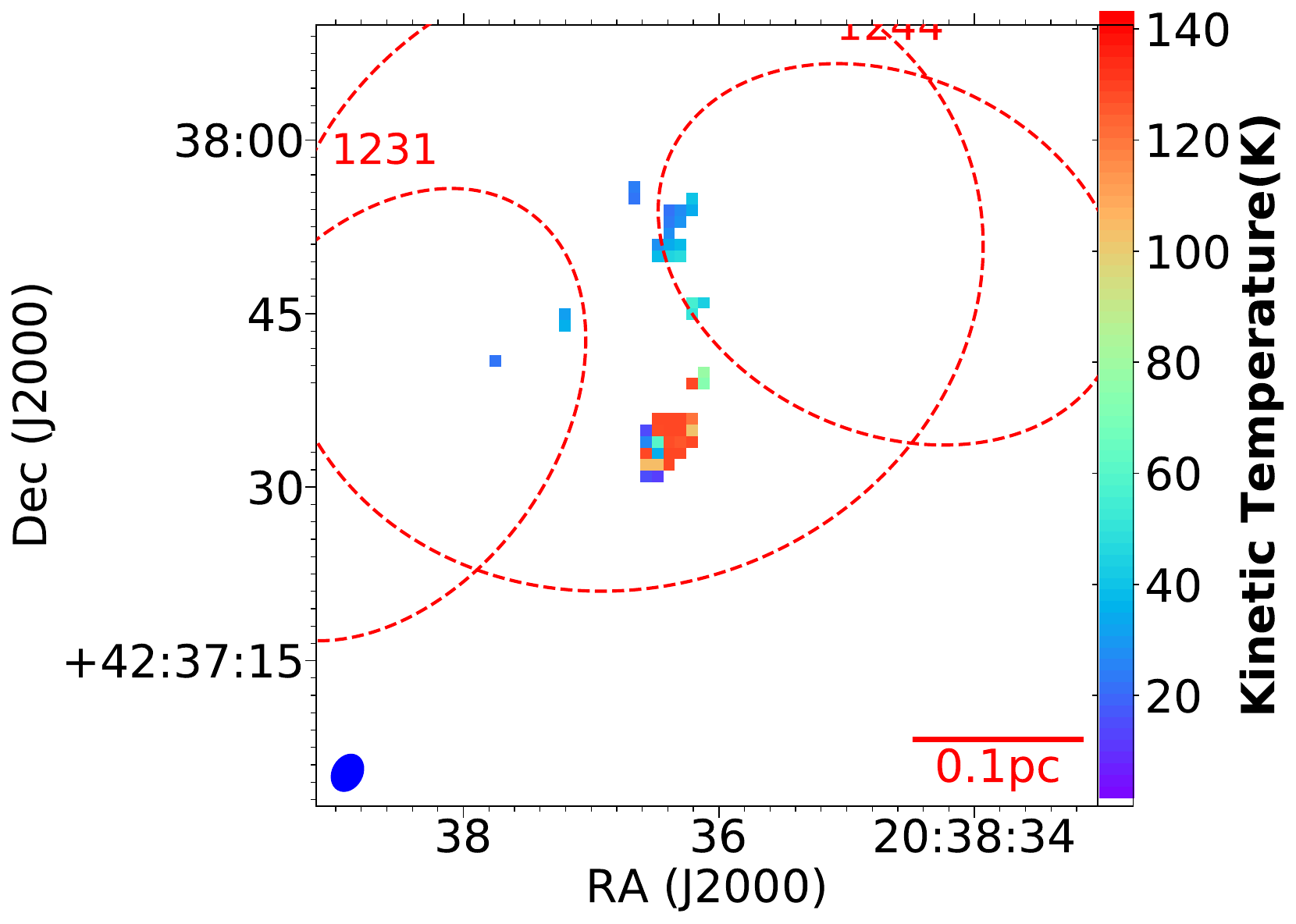} &
\includegraphics[width=.3\textwidth]{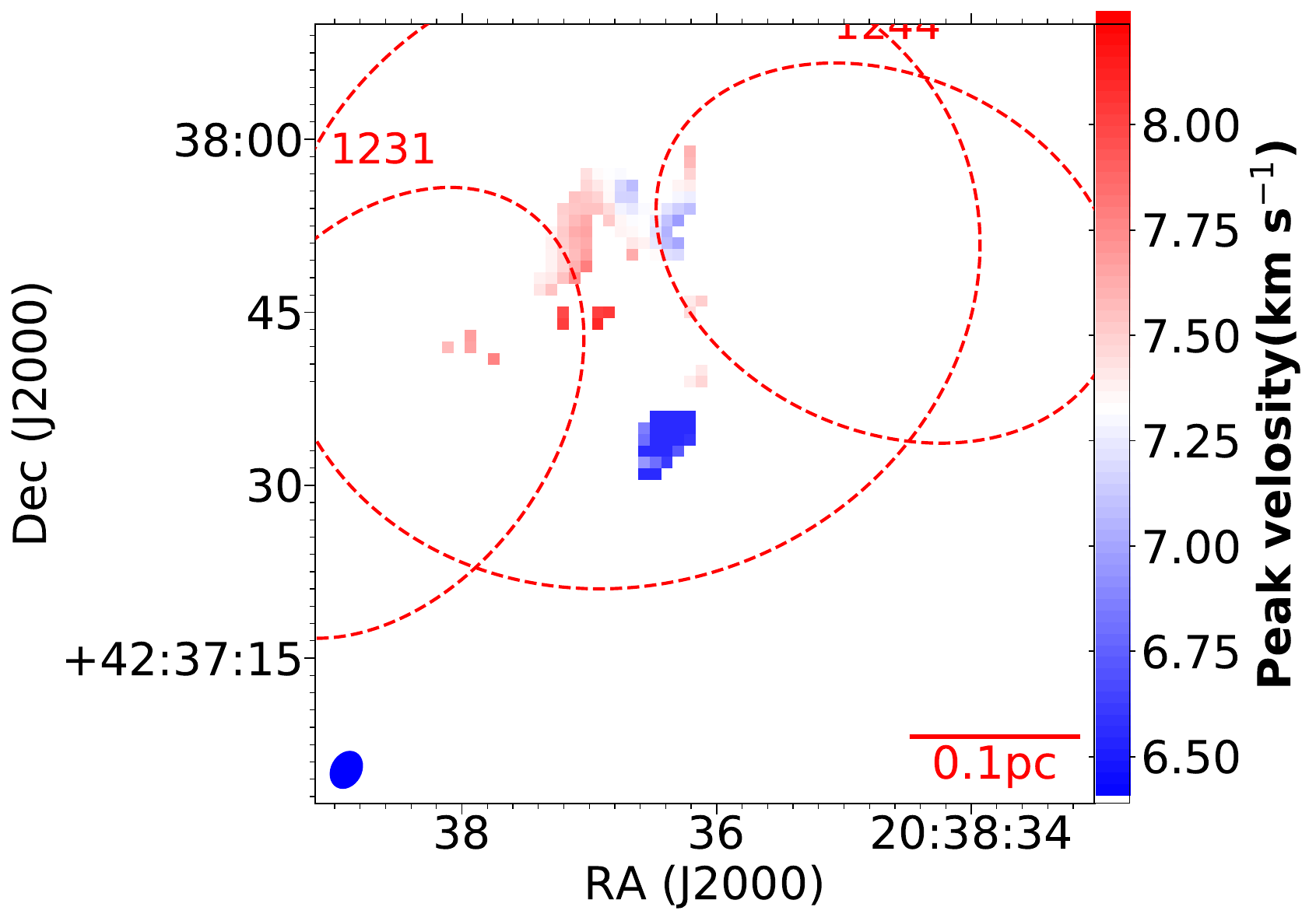} &
\includegraphics[width=.3\textwidth]{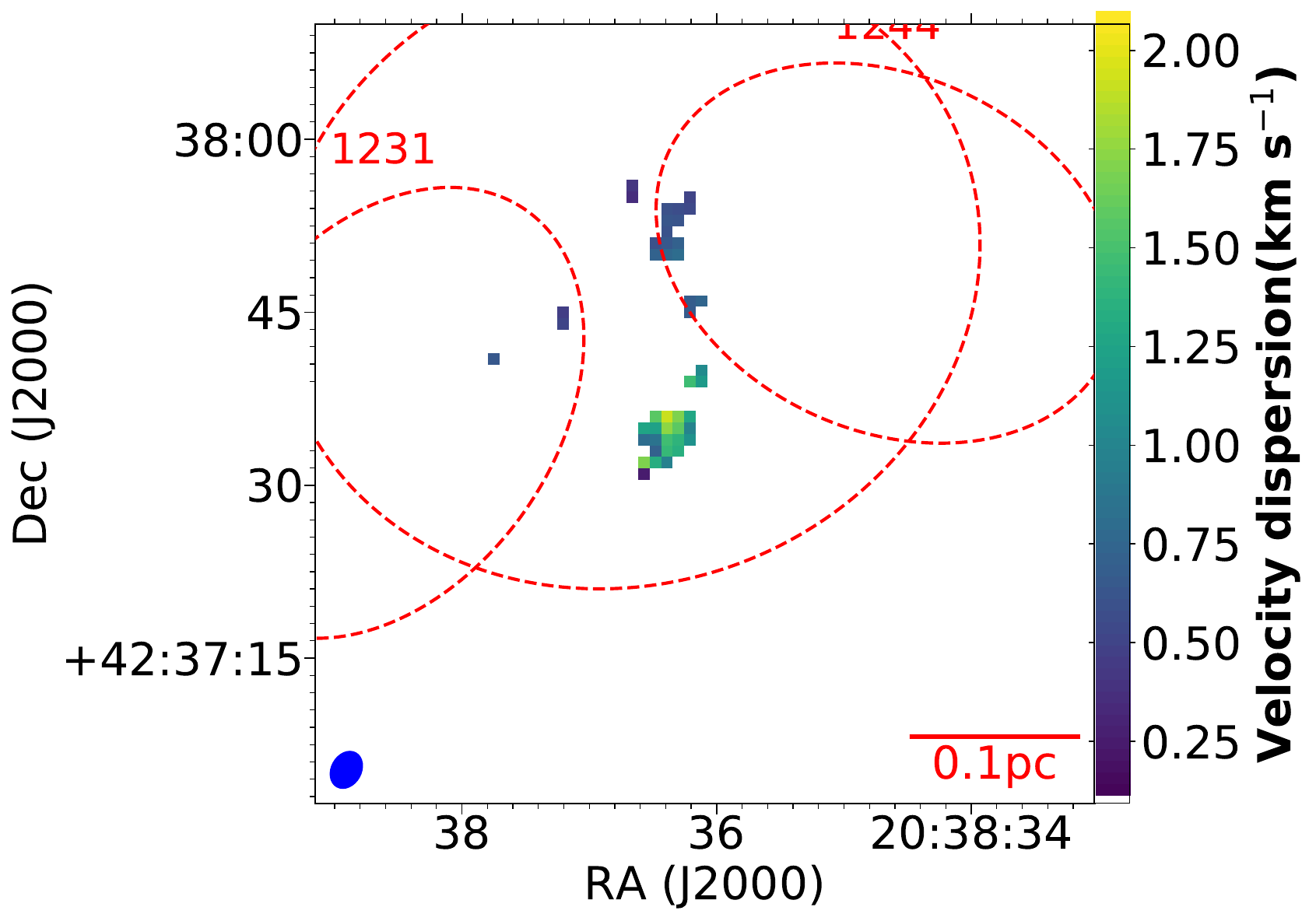} \\
 & Field 37 - Component 2 & \\
\end{longtable}

\section{Compositive images}\label{app:C}

\begin{longtable}{@{}ccc@{}}
\endhead
\caption*{\textbf{Fig. C.1.} continued.}
\endfoot
\caption*{\textbf{Fig. C.1.} Compositive images of each field. The background is the \emph{Spitzer} three-color images using 8\um\ (R), 4.5\um\ (G), and 3.6\um\ (B) bands. The column density contours and \nh\ (1,1) integrated emission contours are over-plotted with green and orange. Cyan and magenta crosses mark the positions of any \water\ masers or radio continuum sources, respectively. The ellipses are the same as those in Fig. \ref{fig:A.1}. \label{fig:C.1}}
\endlastfoot
\includegraphics[width=.3\textwidth]{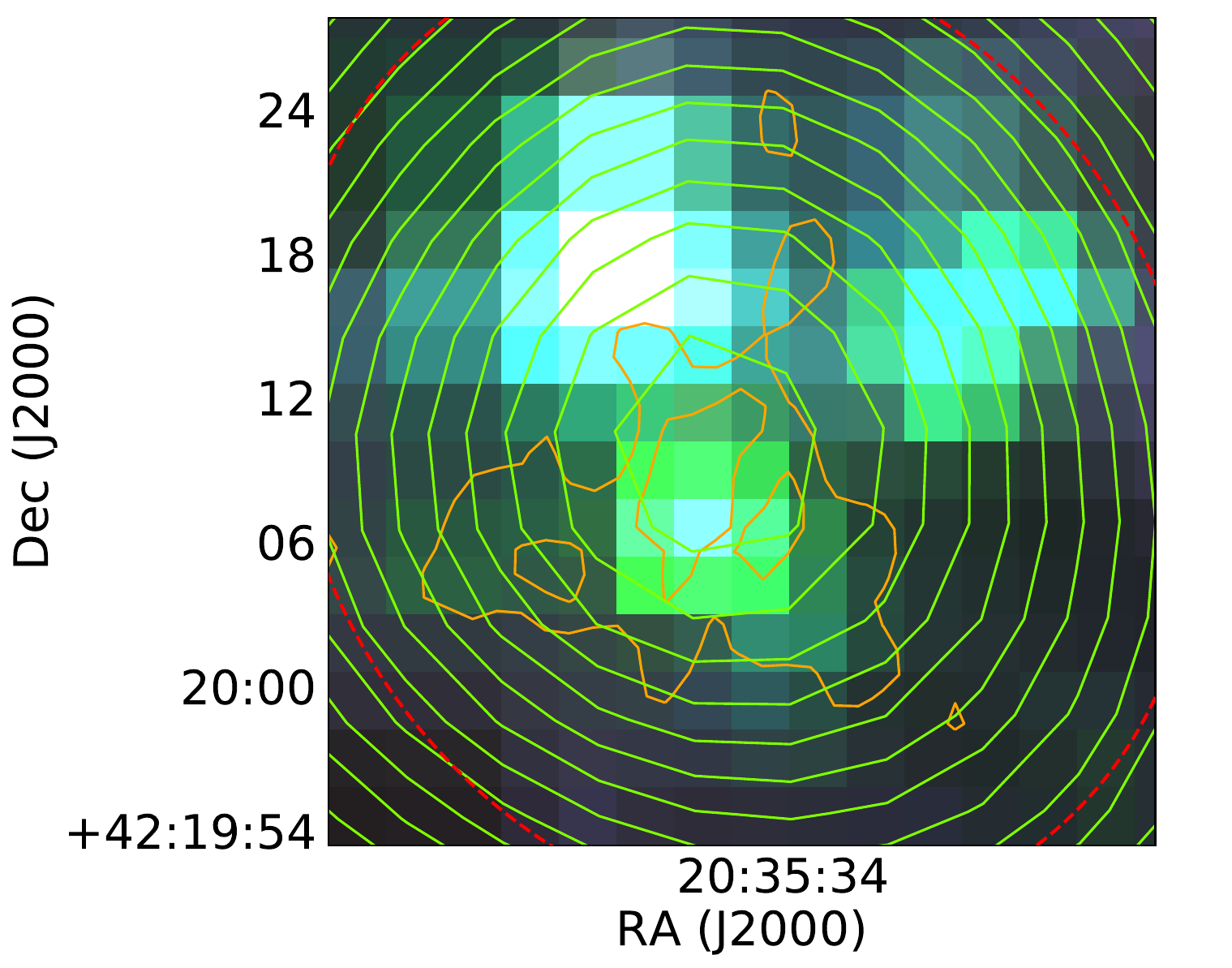} & 
\includegraphics[width=.3\textwidth]{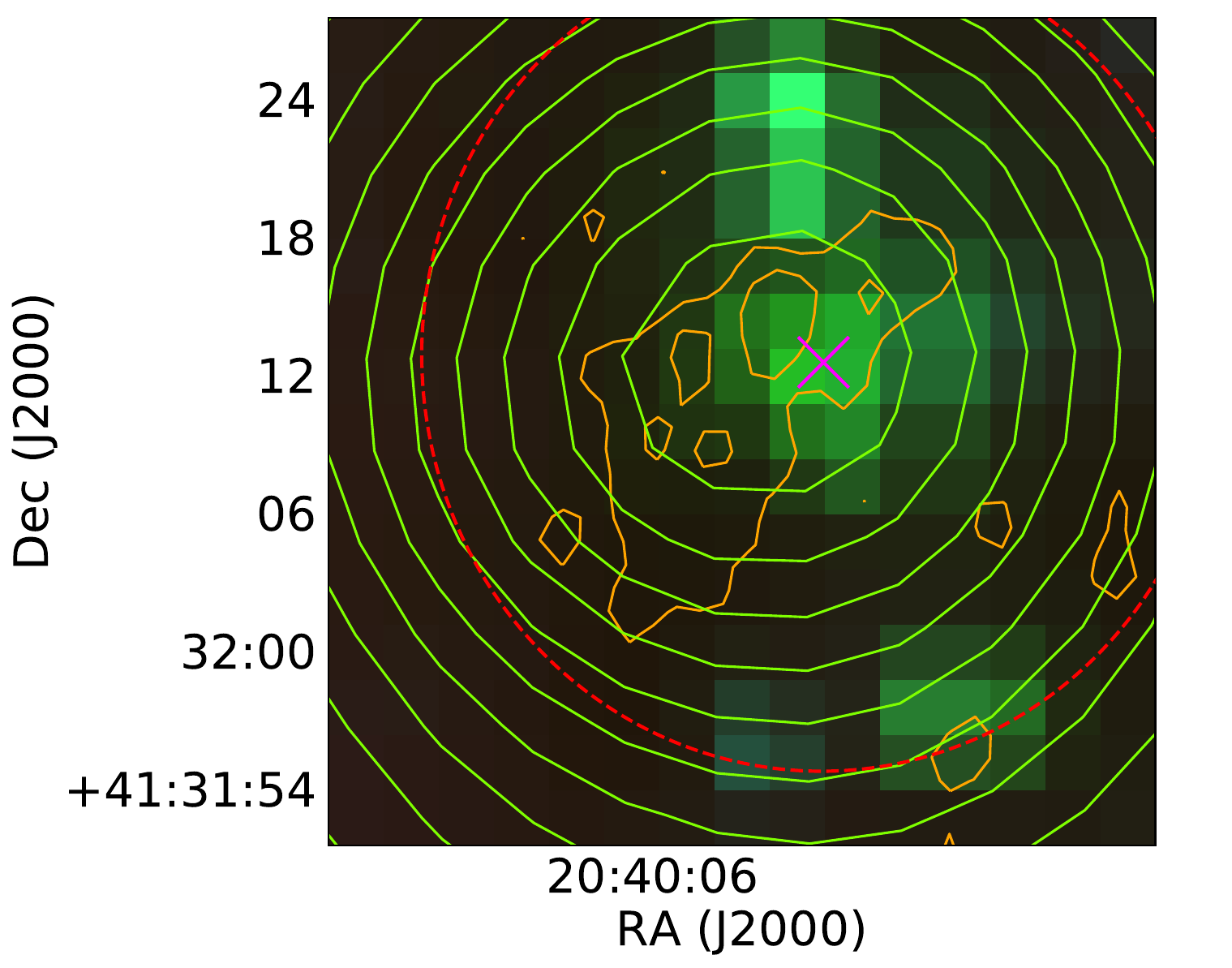} & 
\includegraphics[width=.3\textwidth]{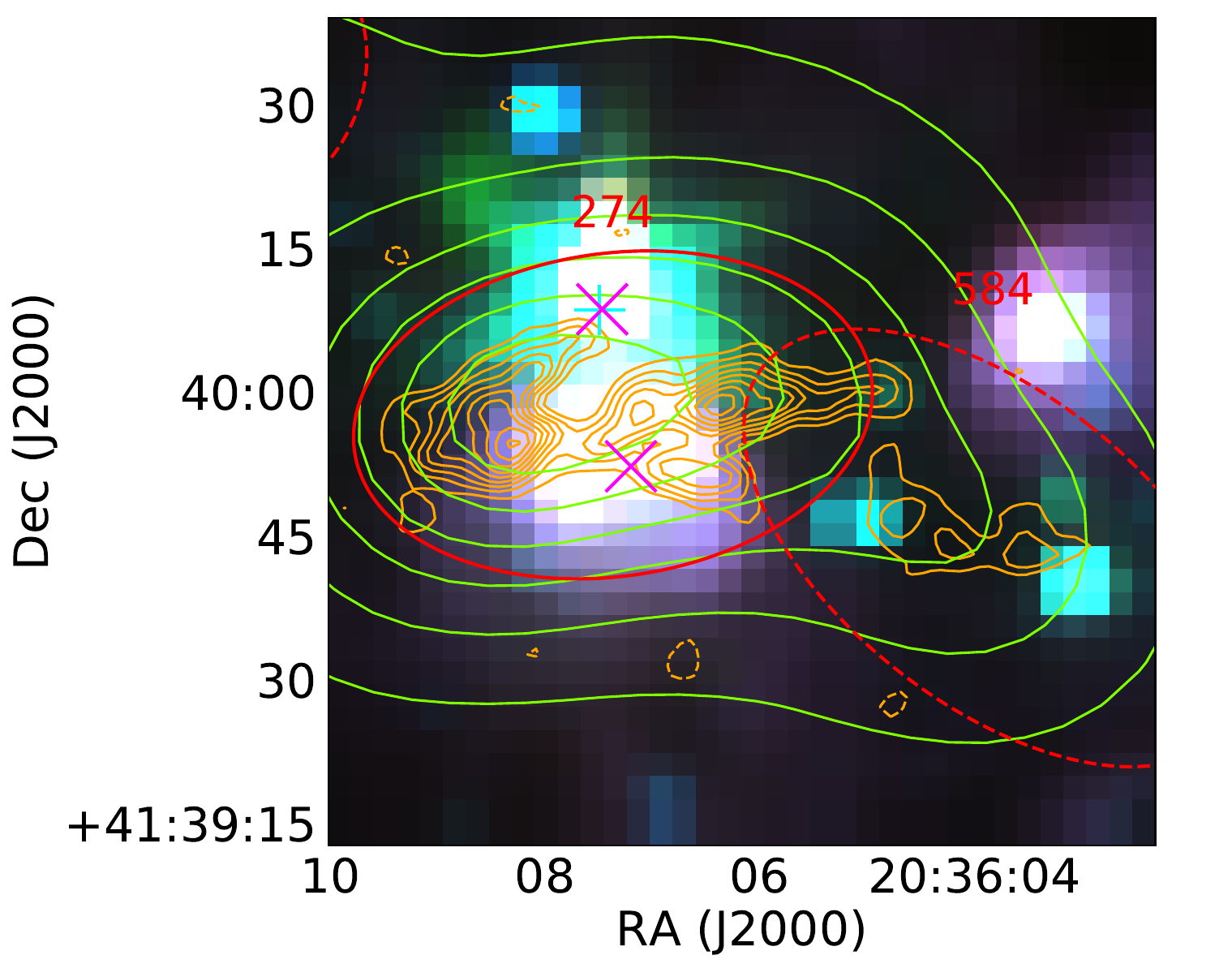} \\ 
Field 1 & Field 2 & Field 3 \\ 
\includegraphics[width=.3\textwidth]{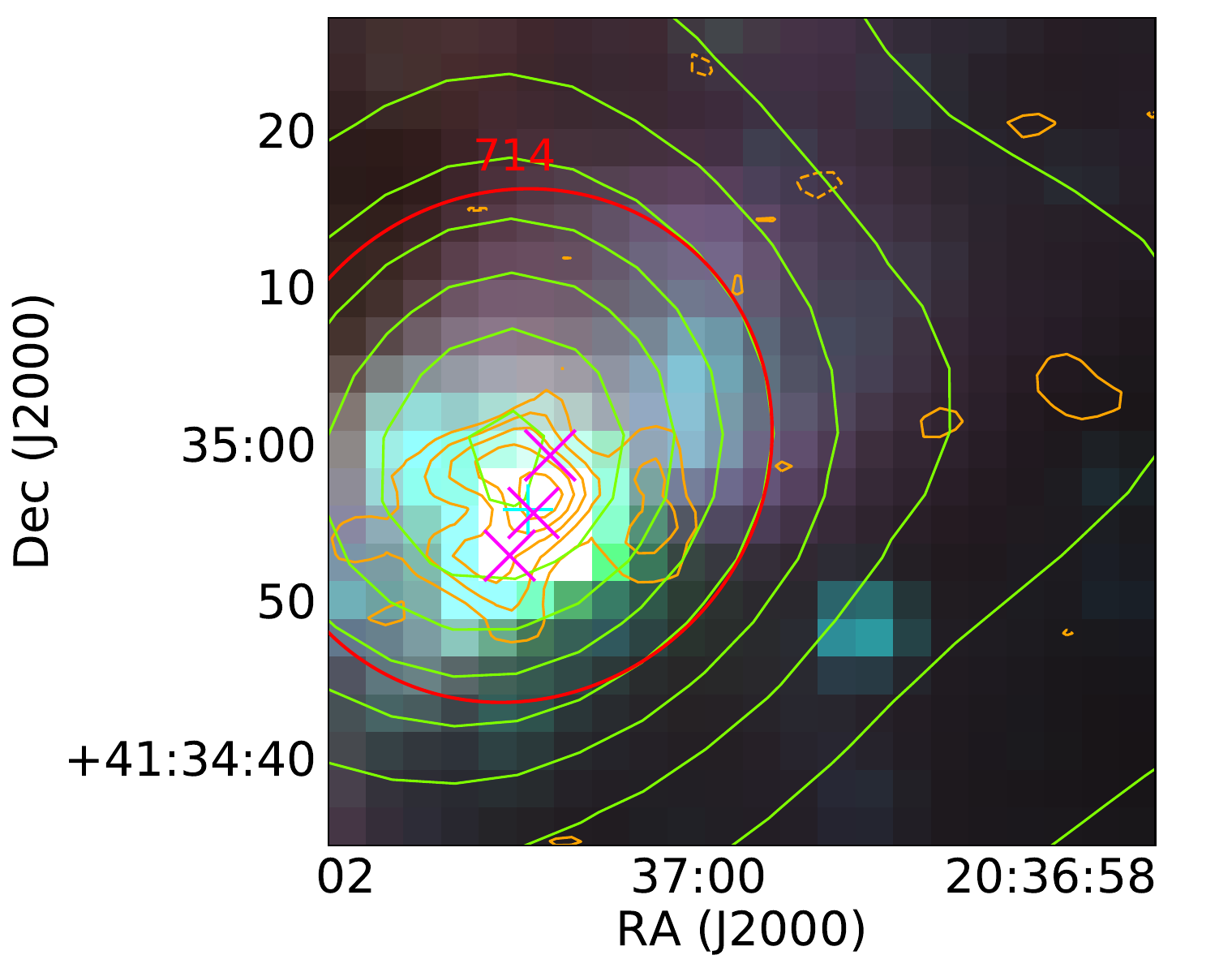} & 
\includegraphics[width=.3\textwidth]{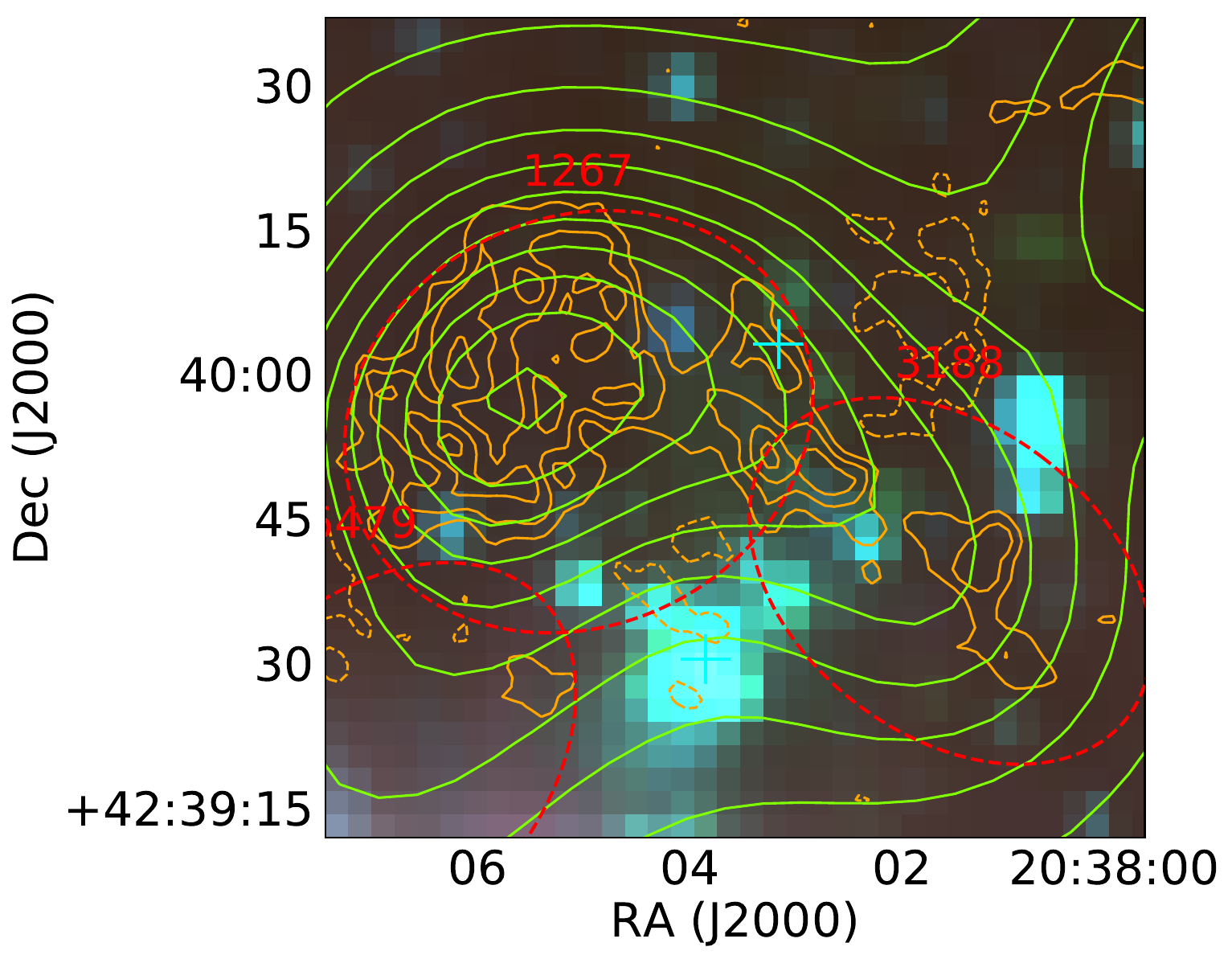} & 
\includegraphics[width=.3\textwidth]{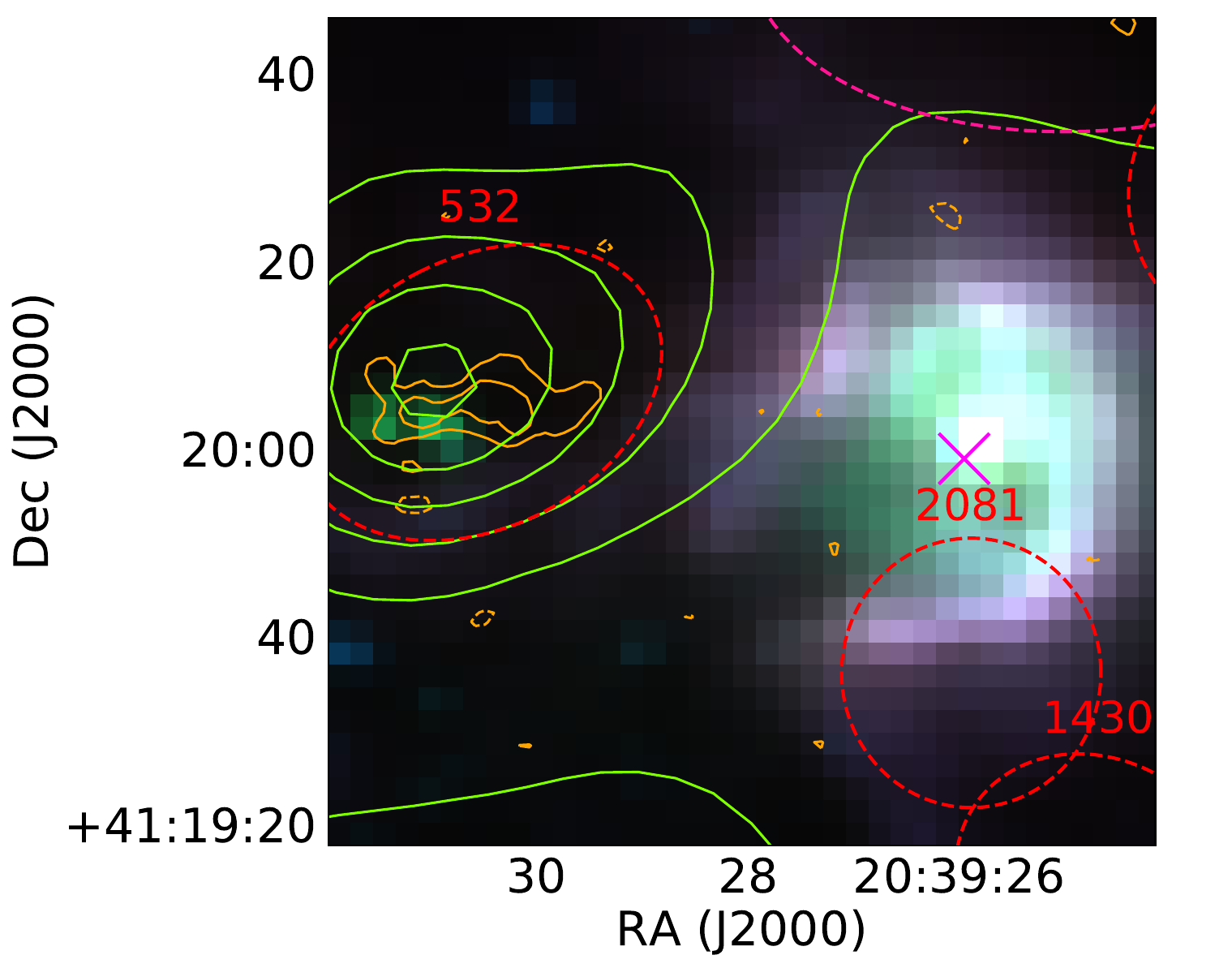} \\ 
Field 6 & Field 7 & Field 9 \\ 
\includegraphics[width=.3\textwidth]{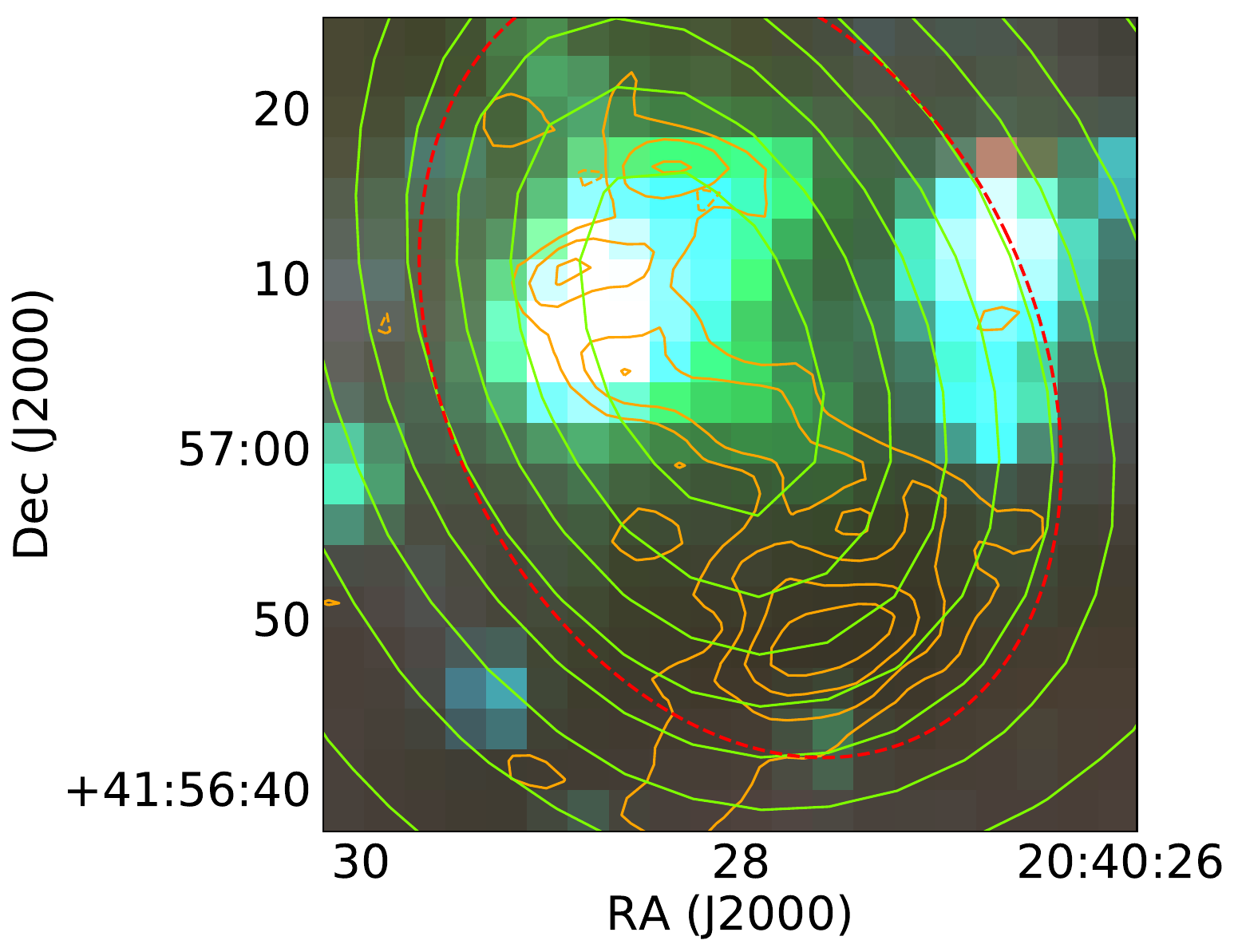} & 
\includegraphics[width=.3\textwidth]{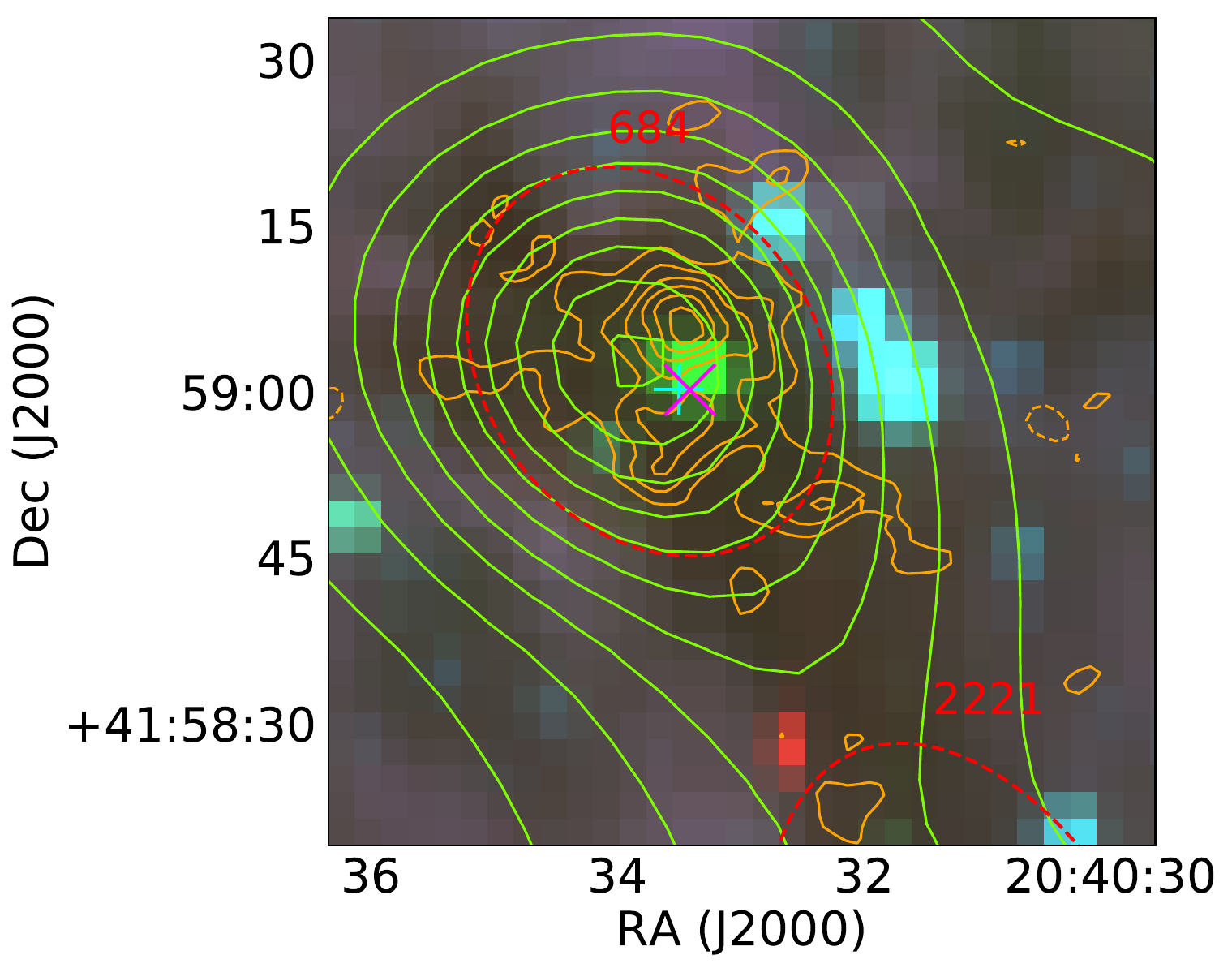} & 
\includegraphics[width=.3\textwidth]{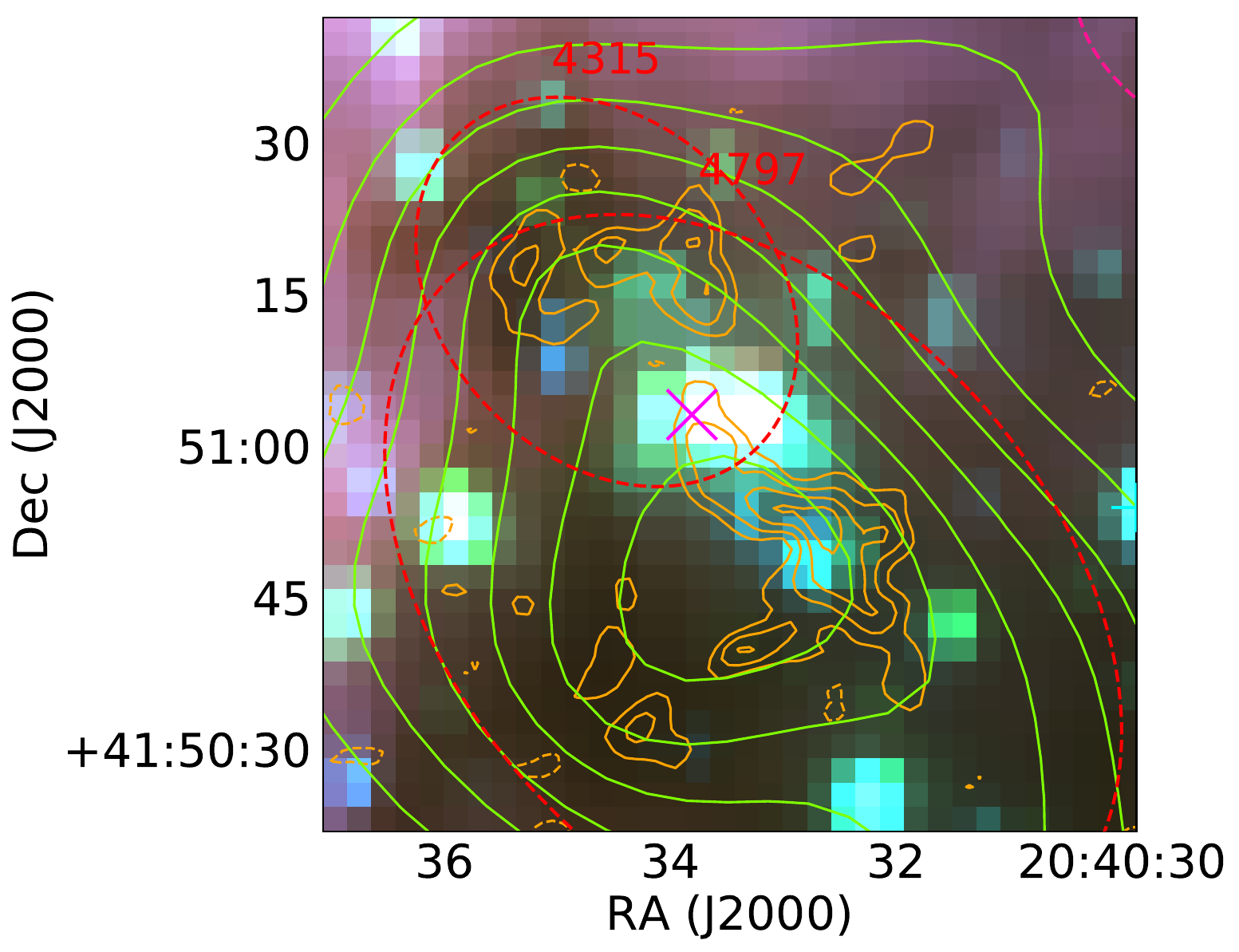} \\ 
Field 10 & Field 11 & Field 12 \\ 
\includegraphics[width=.3\textwidth]{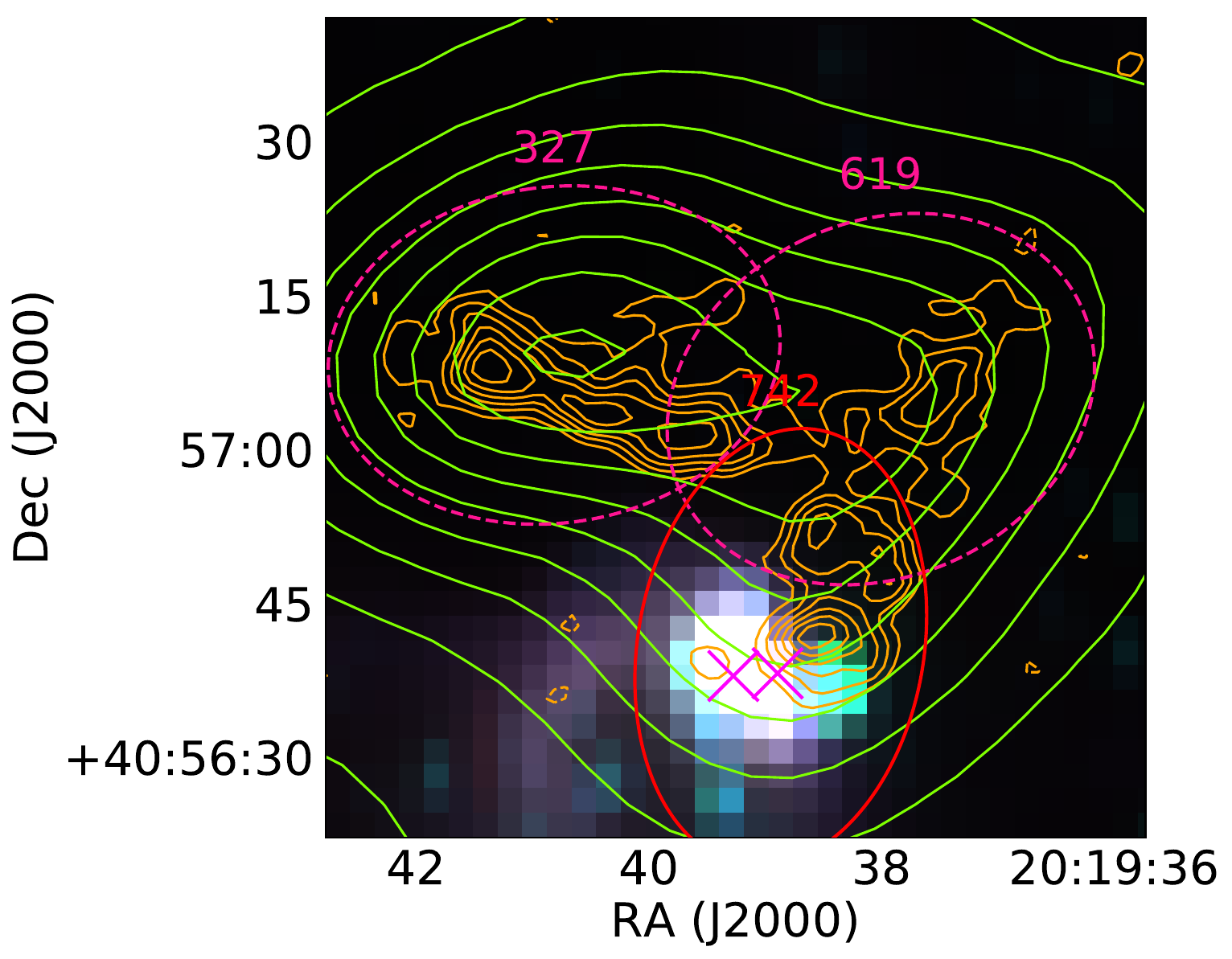} & 
\includegraphics[width=.3\textwidth]{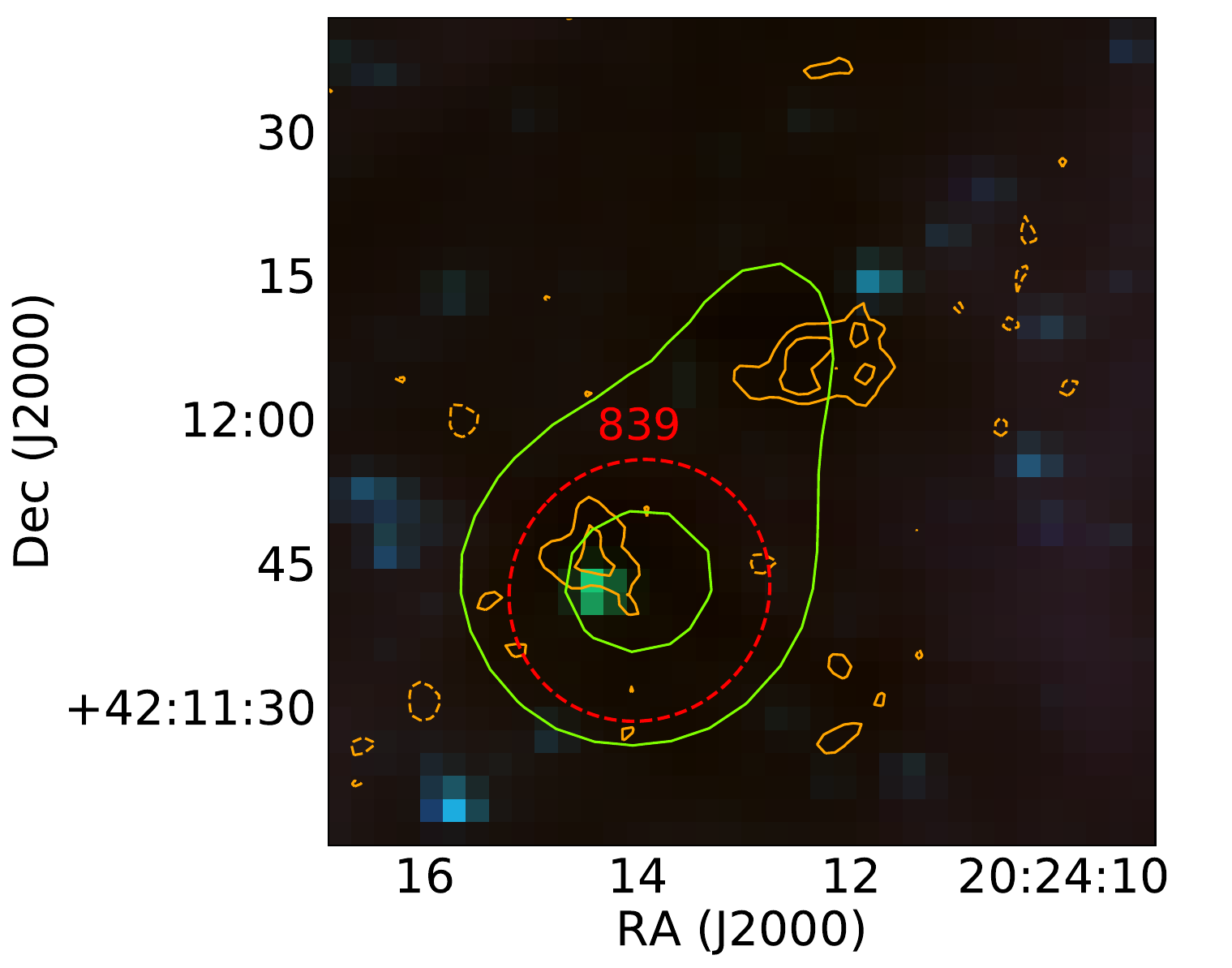} & 
\includegraphics[width=.3\textwidth]{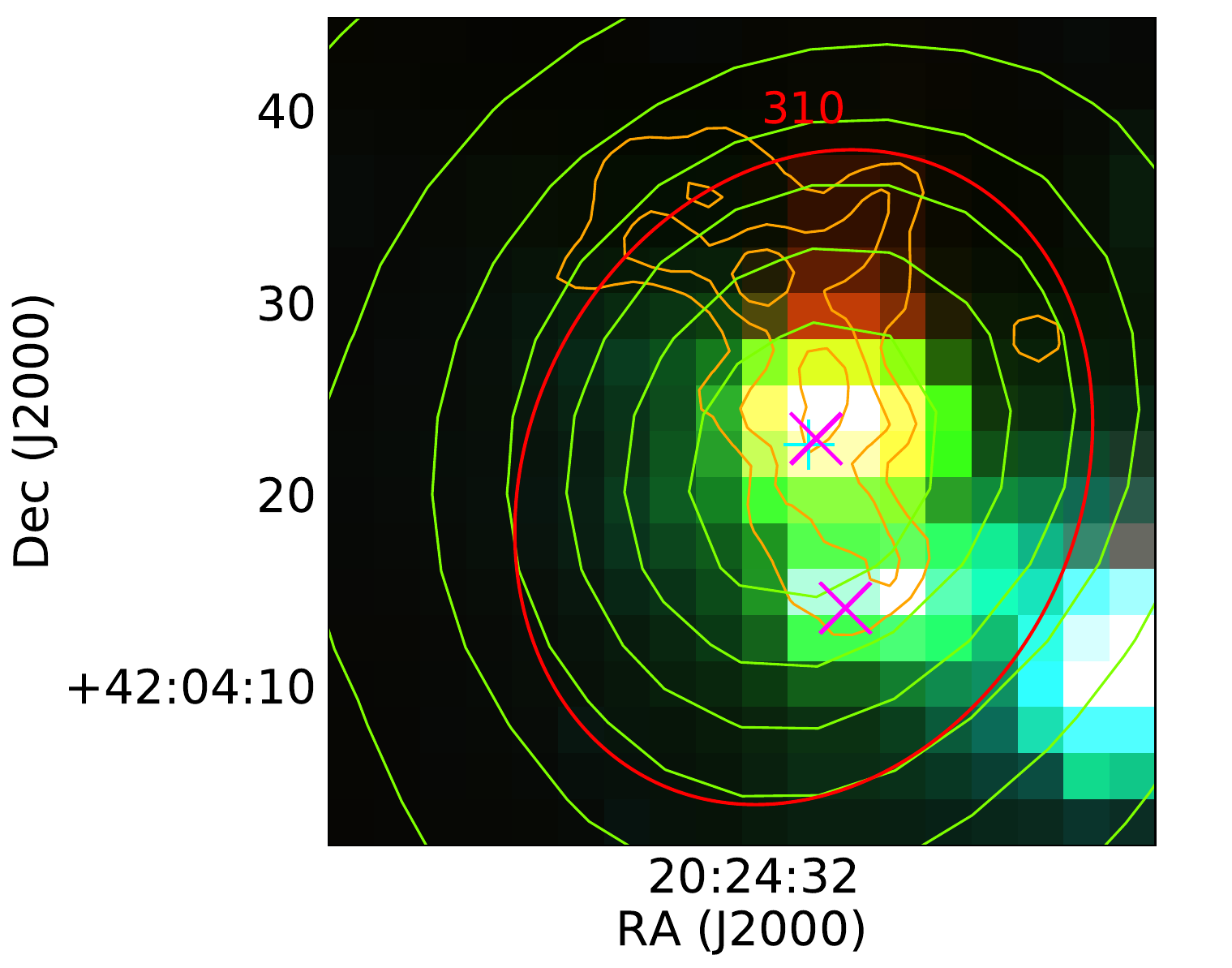} \\ 
Field 13 & Field 15 & Field 16 \\ 
\includegraphics[width=.3\textwidth]{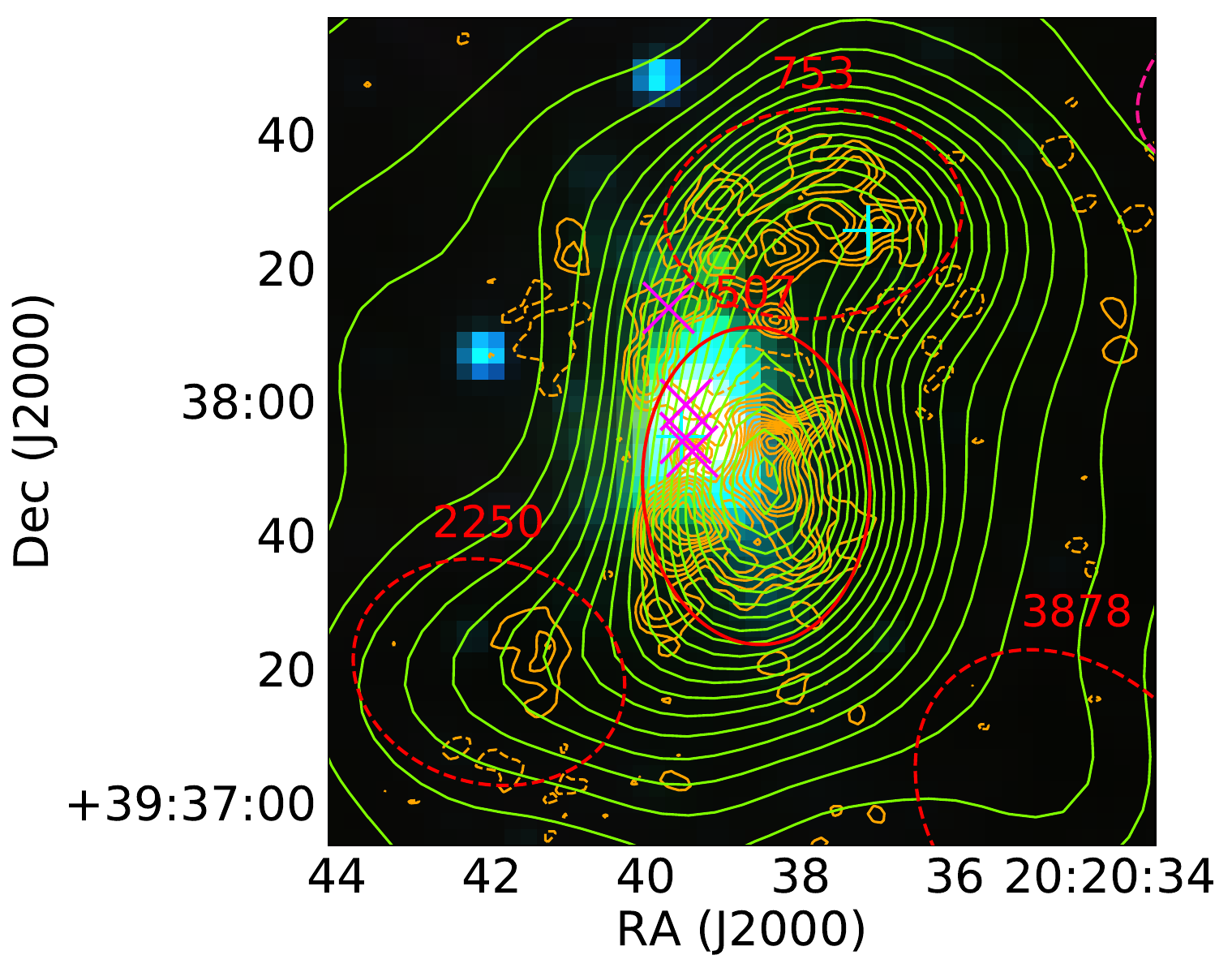} & 
\includegraphics[width=.3\textwidth]{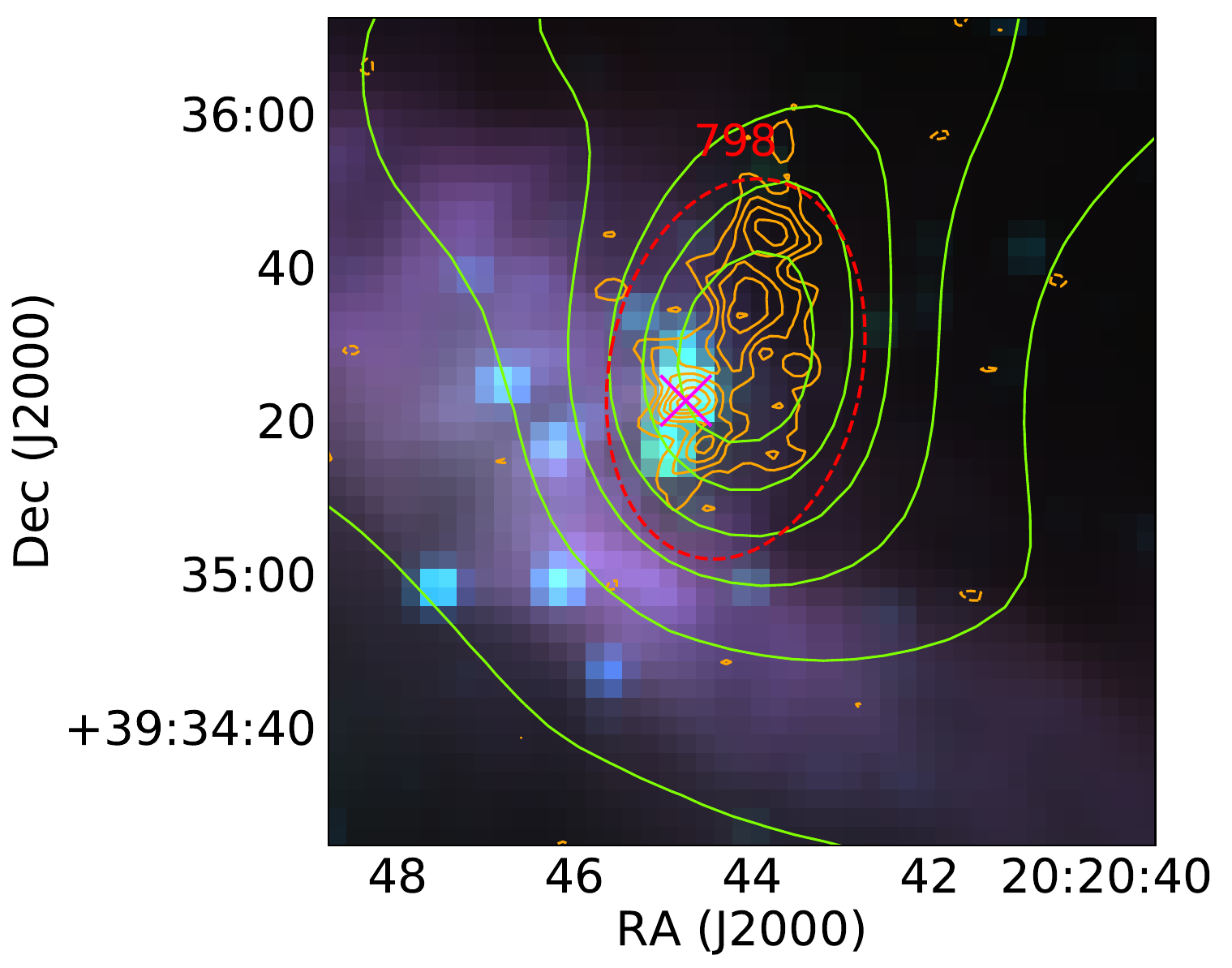} & 
\includegraphics[width=.3\textwidth]{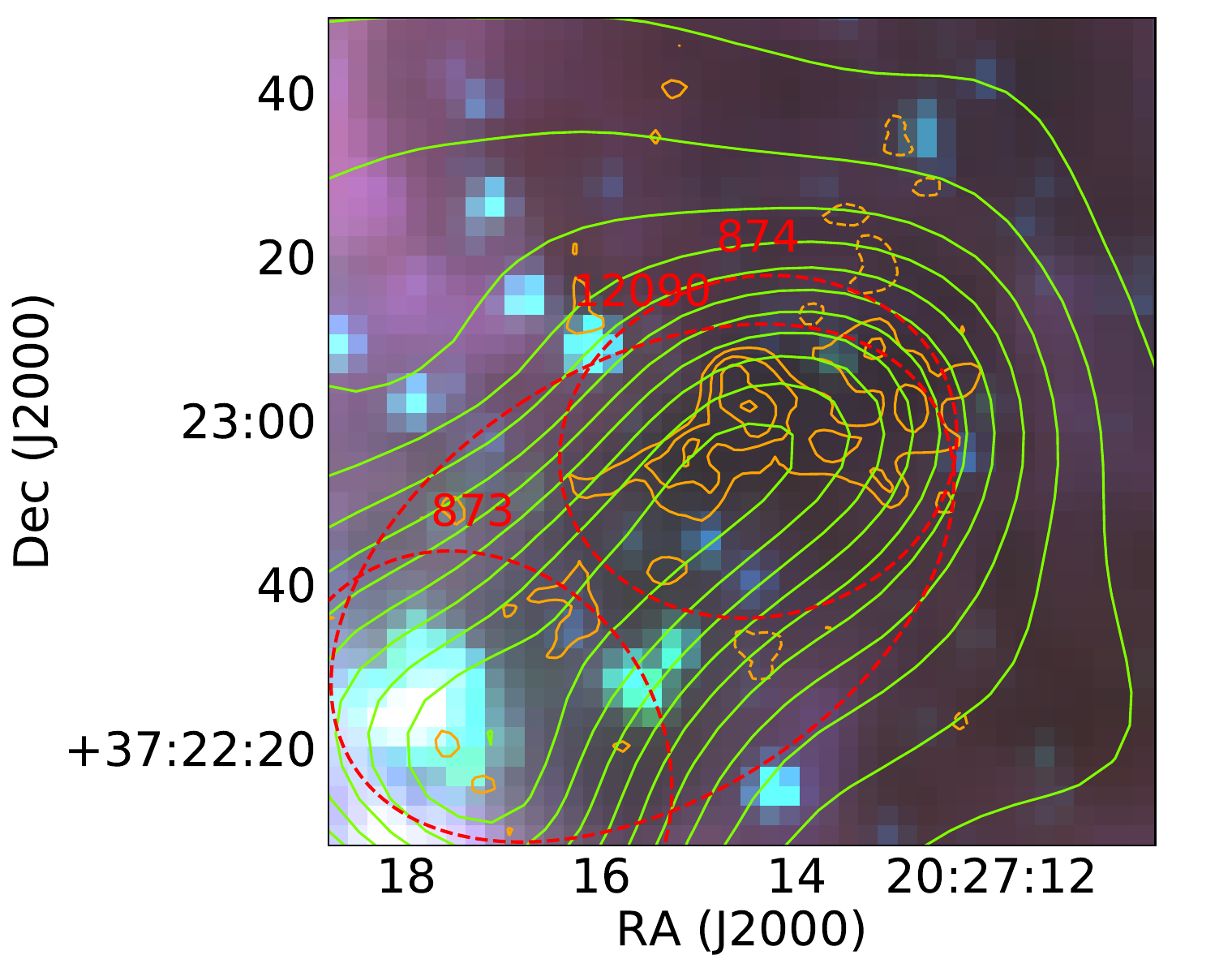} \\ 
Field 17 & Field 18 & Field 19 \\ 
\includegraphics[width=.3\textwidth]{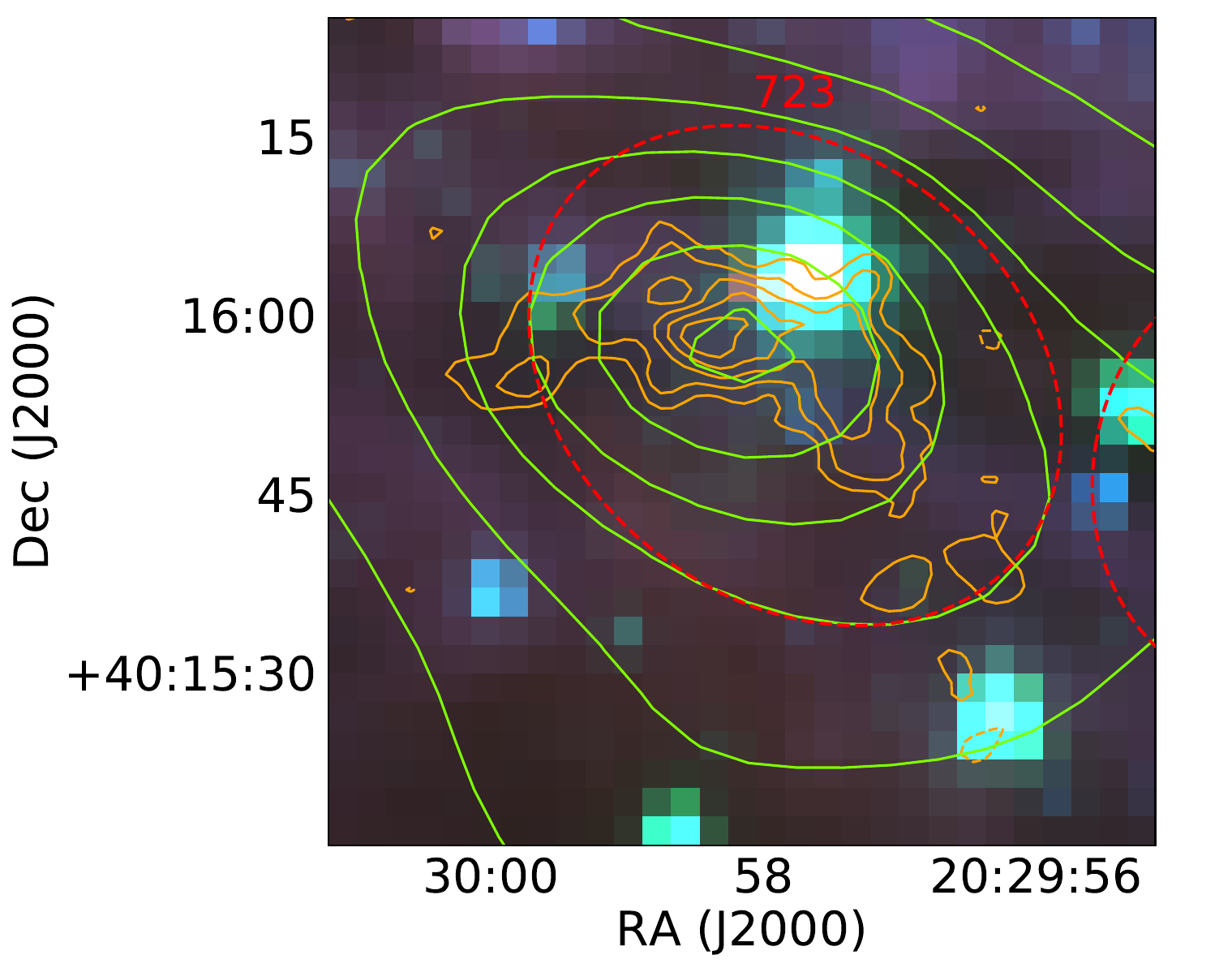} & 
\includegraphics[width=.3\textwidth]{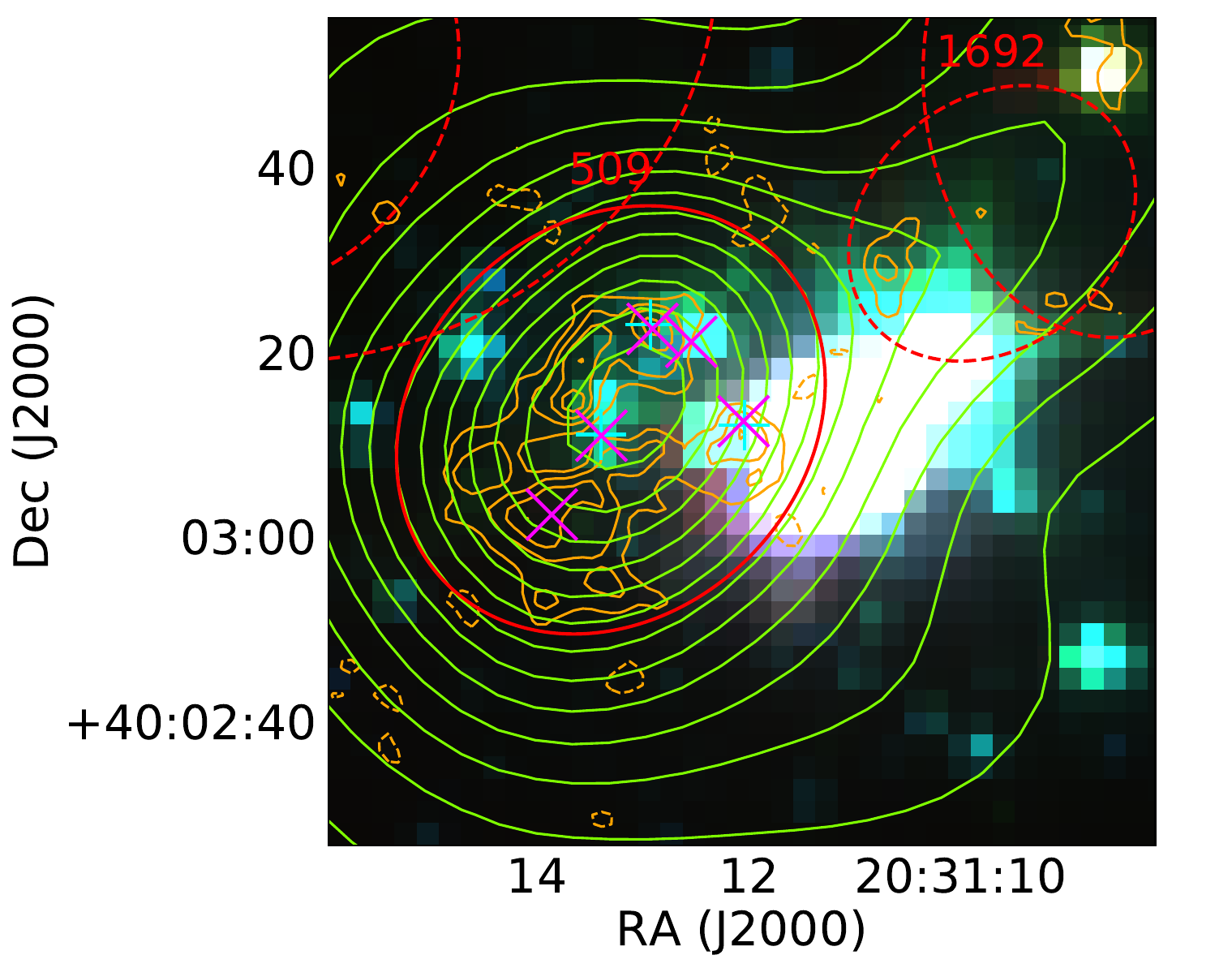} & 
\includegraphics[width=.3\textwidth]{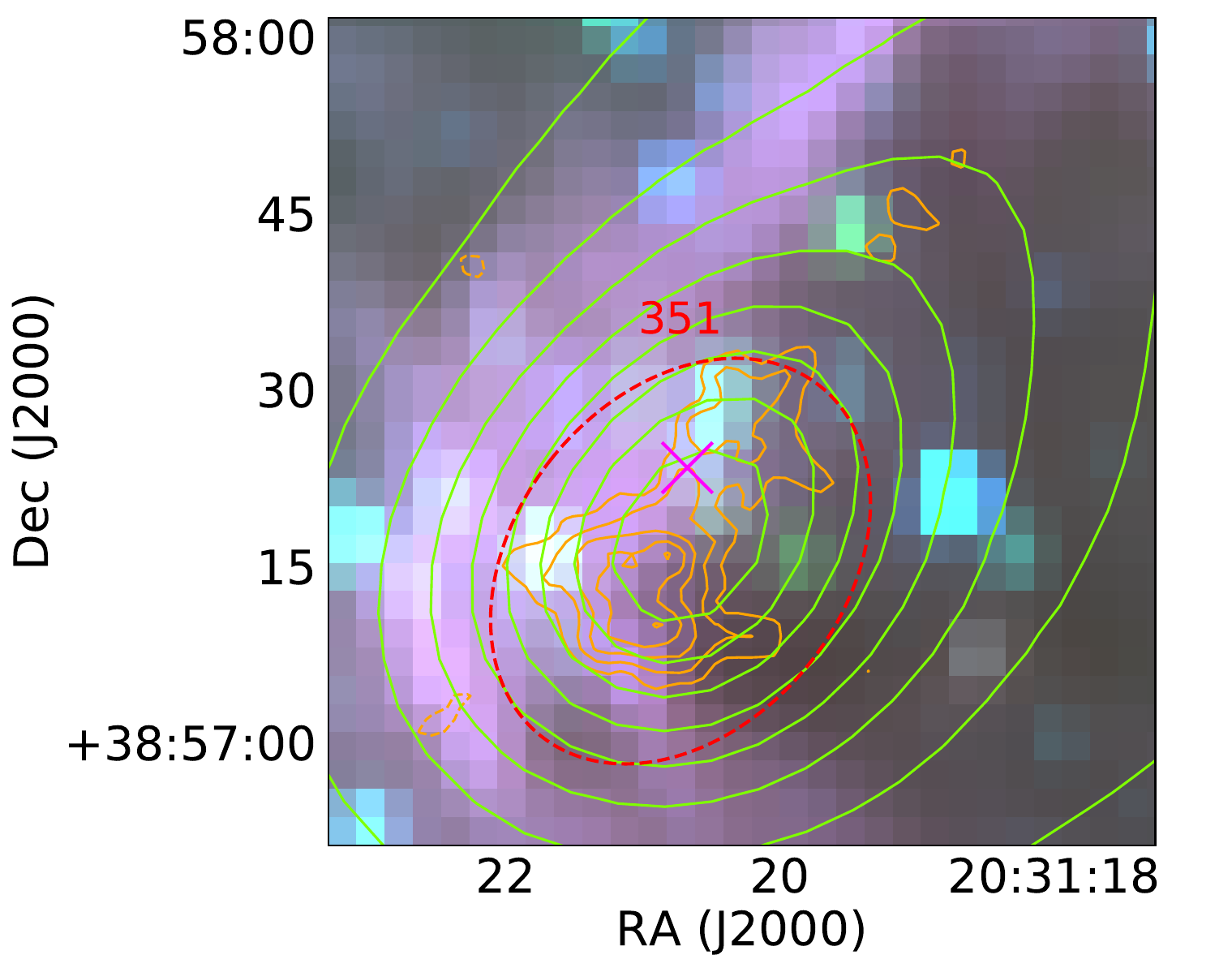} \\ 
Field 22 & Field 23 & Field 24 \\ 
\includegraphics[width=.3\textwidth]{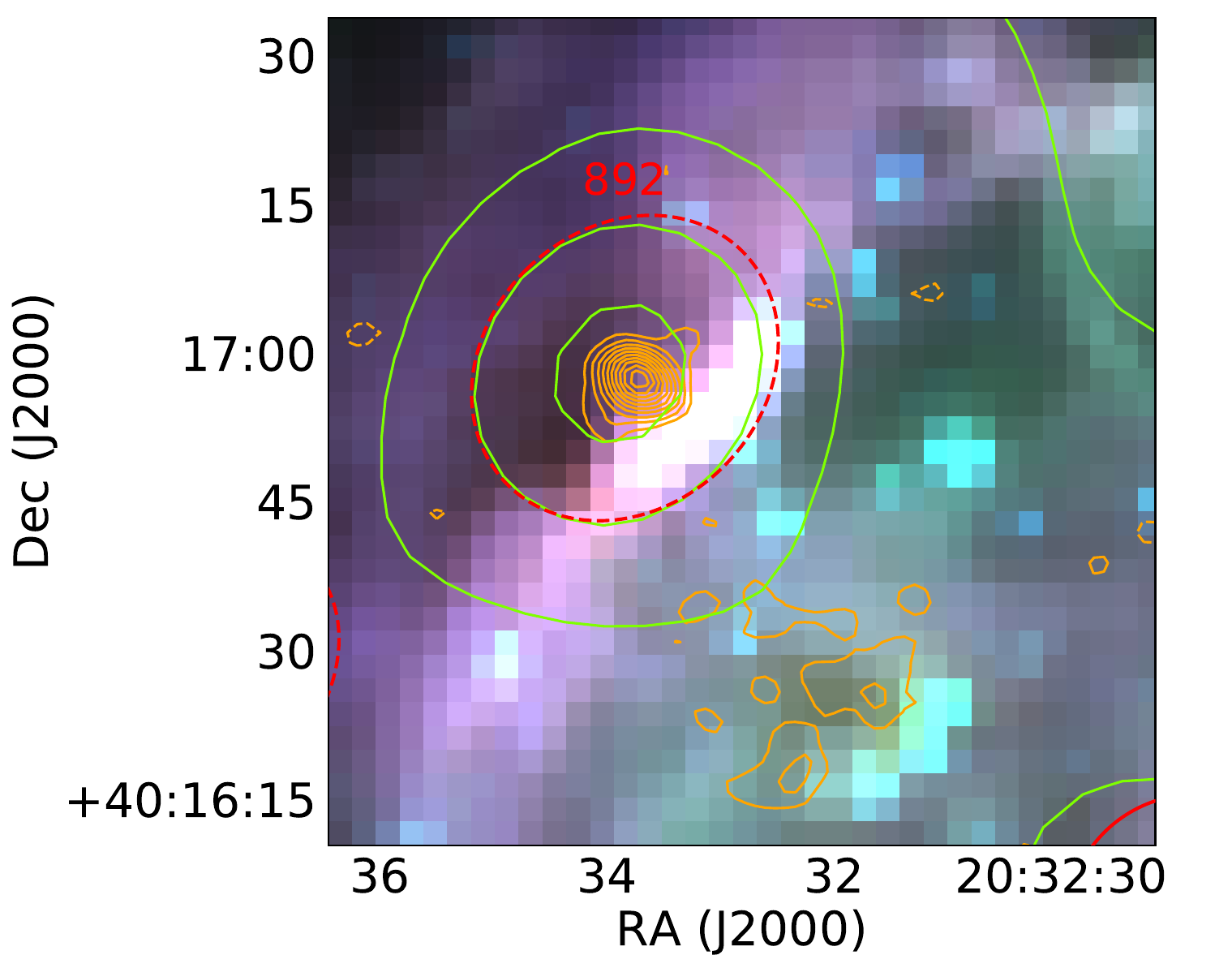} & 
\includegraphics[width=.3\textwidth]{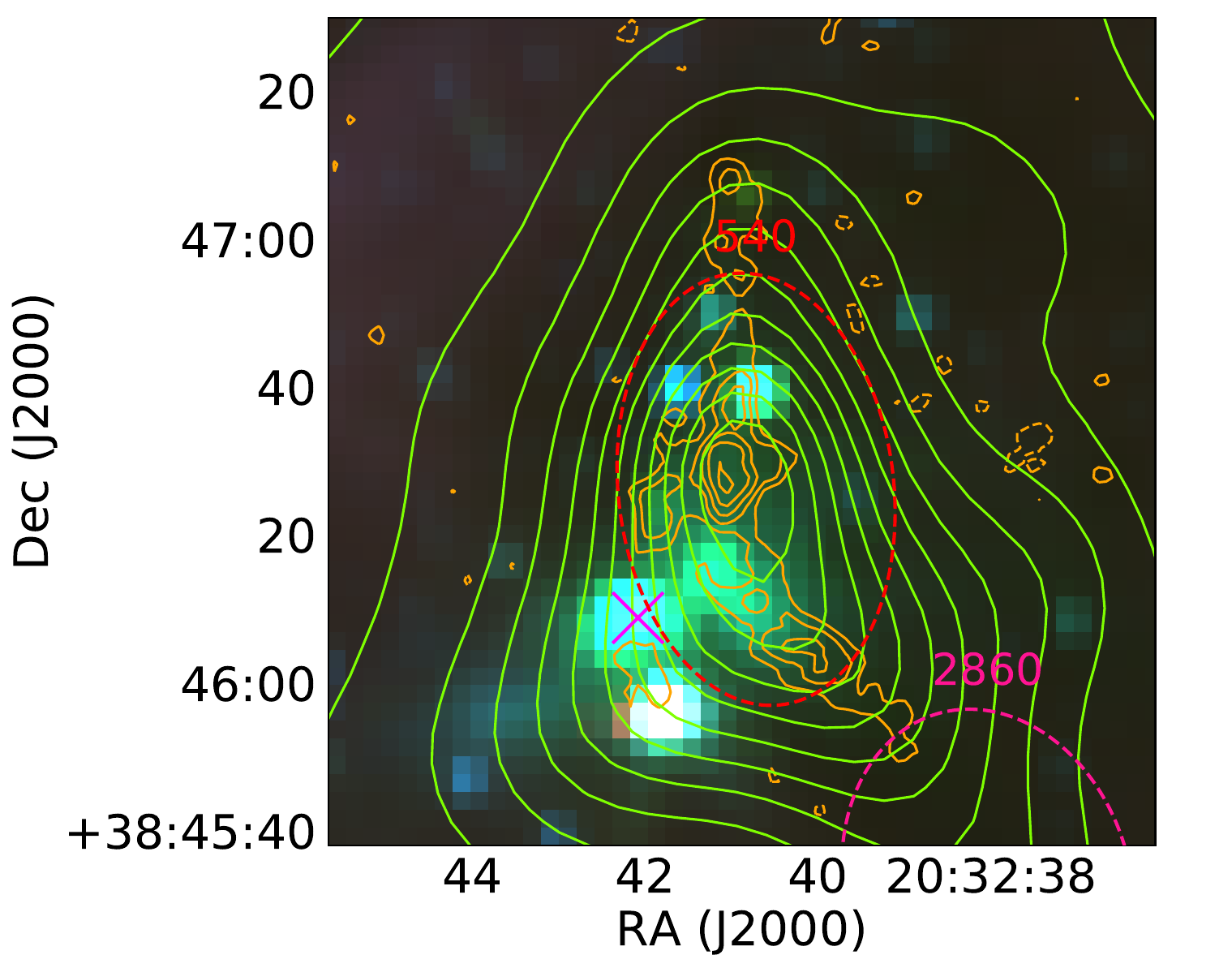} & 
\includegraphics[width=.3\textwidth]{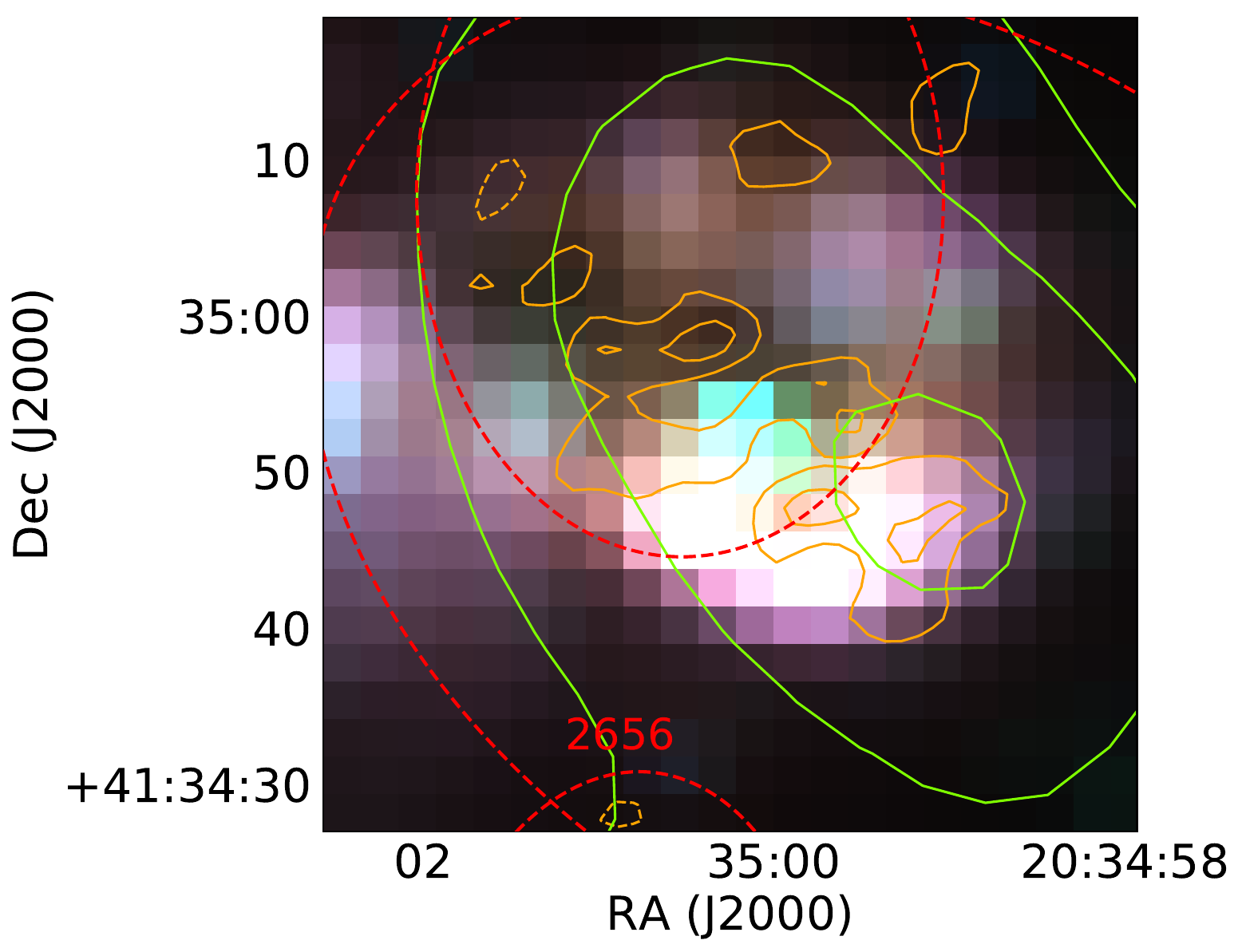} \\ 
Field 27 & Field 28 & Field 29 \\ 
\includegraphics[width=.3\textwidth]{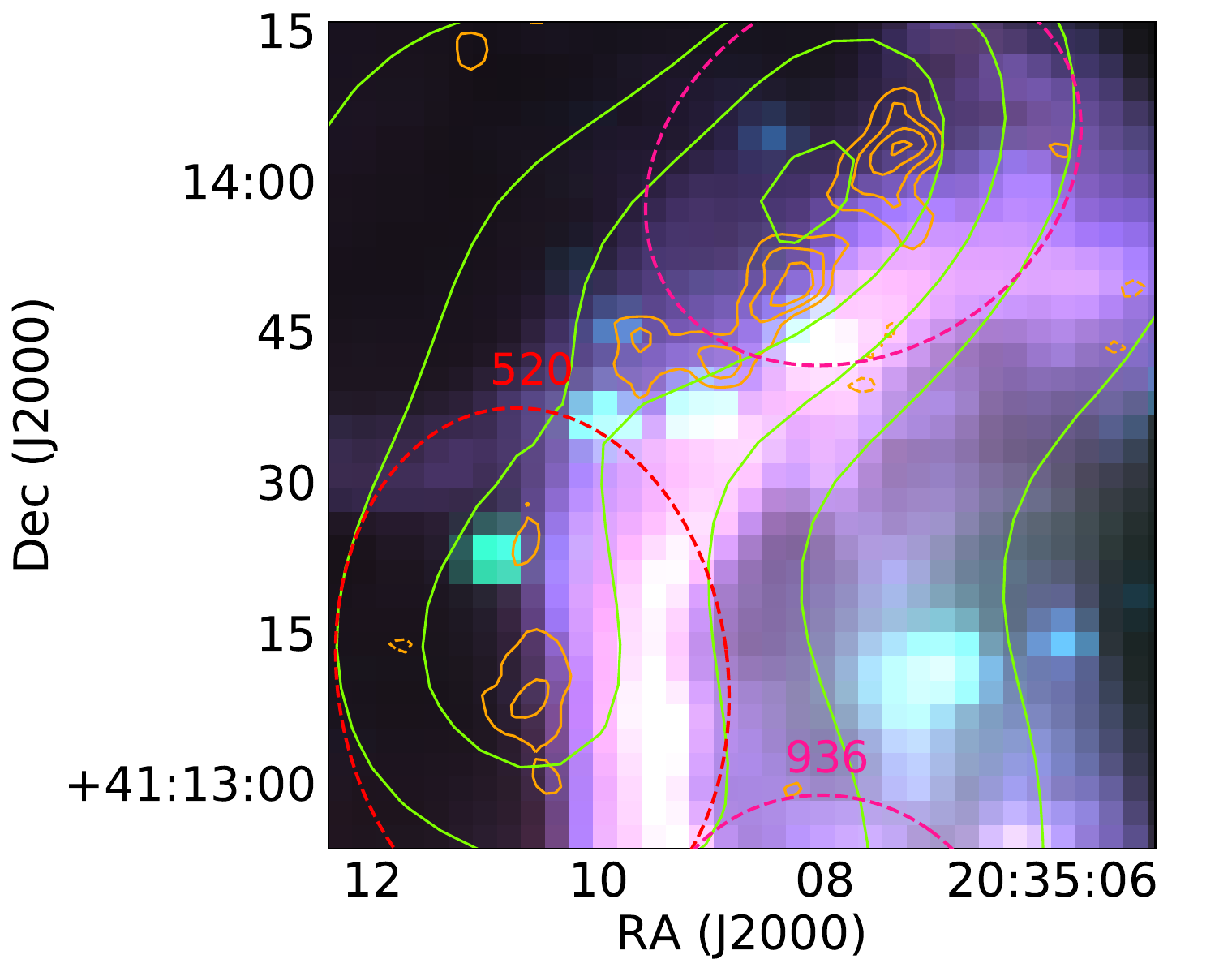} & 
\includegraphics[width=.3\textwidth]{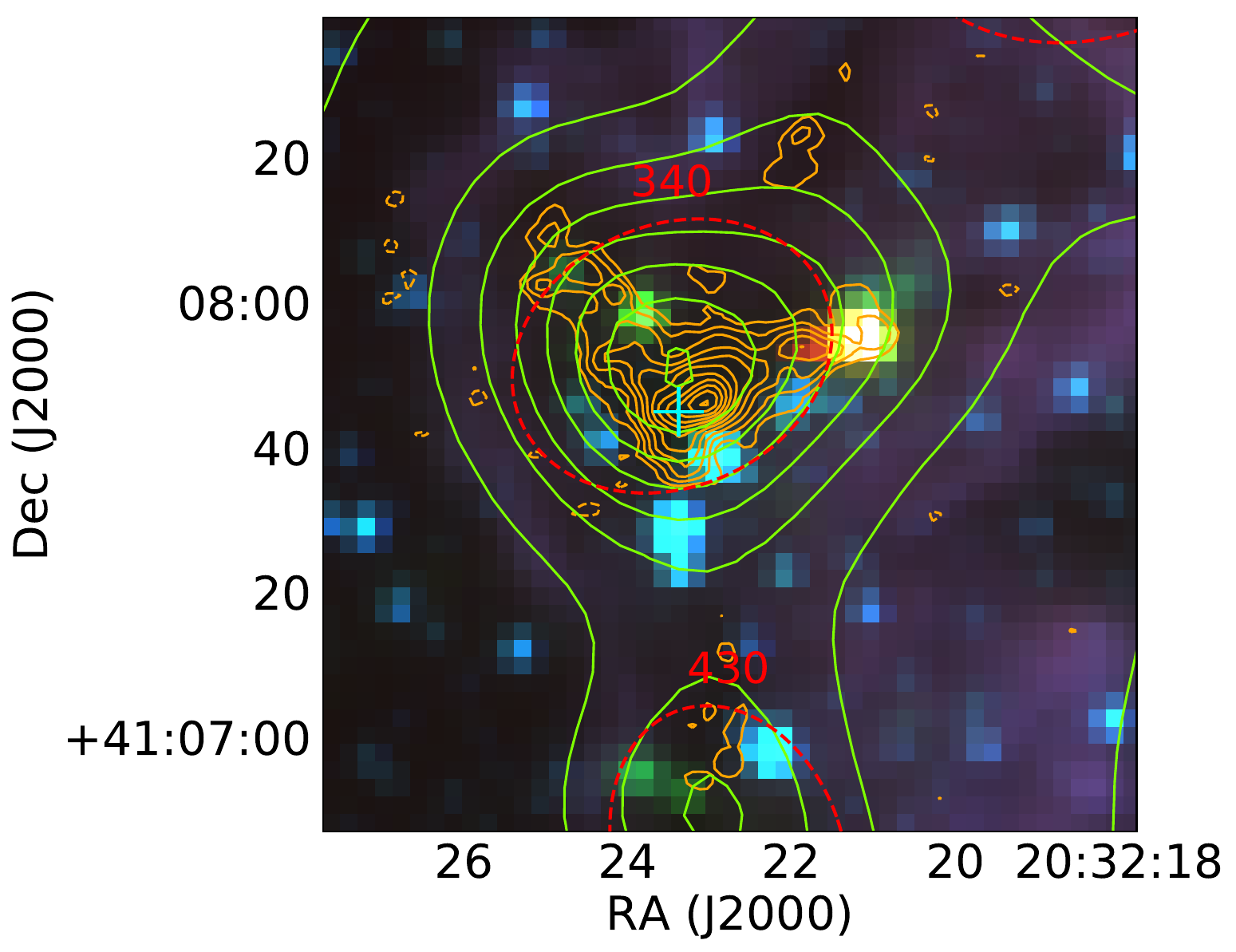} & 
\includegraphics[width=.3\textwidth]{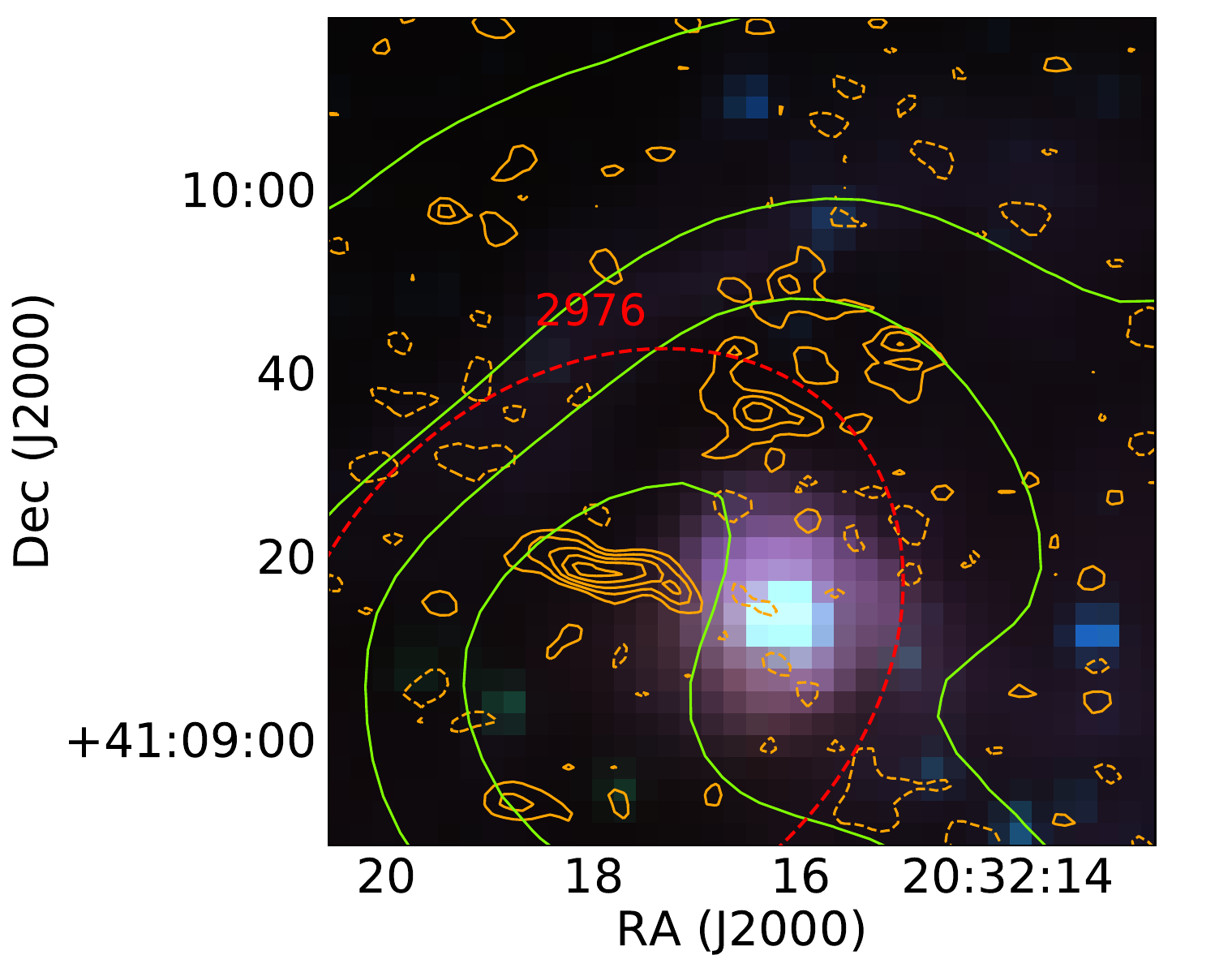} \\ 
Field 31 & Field 32 & Field 33 \\ 
\includegraphics[width=.3\textwidth]{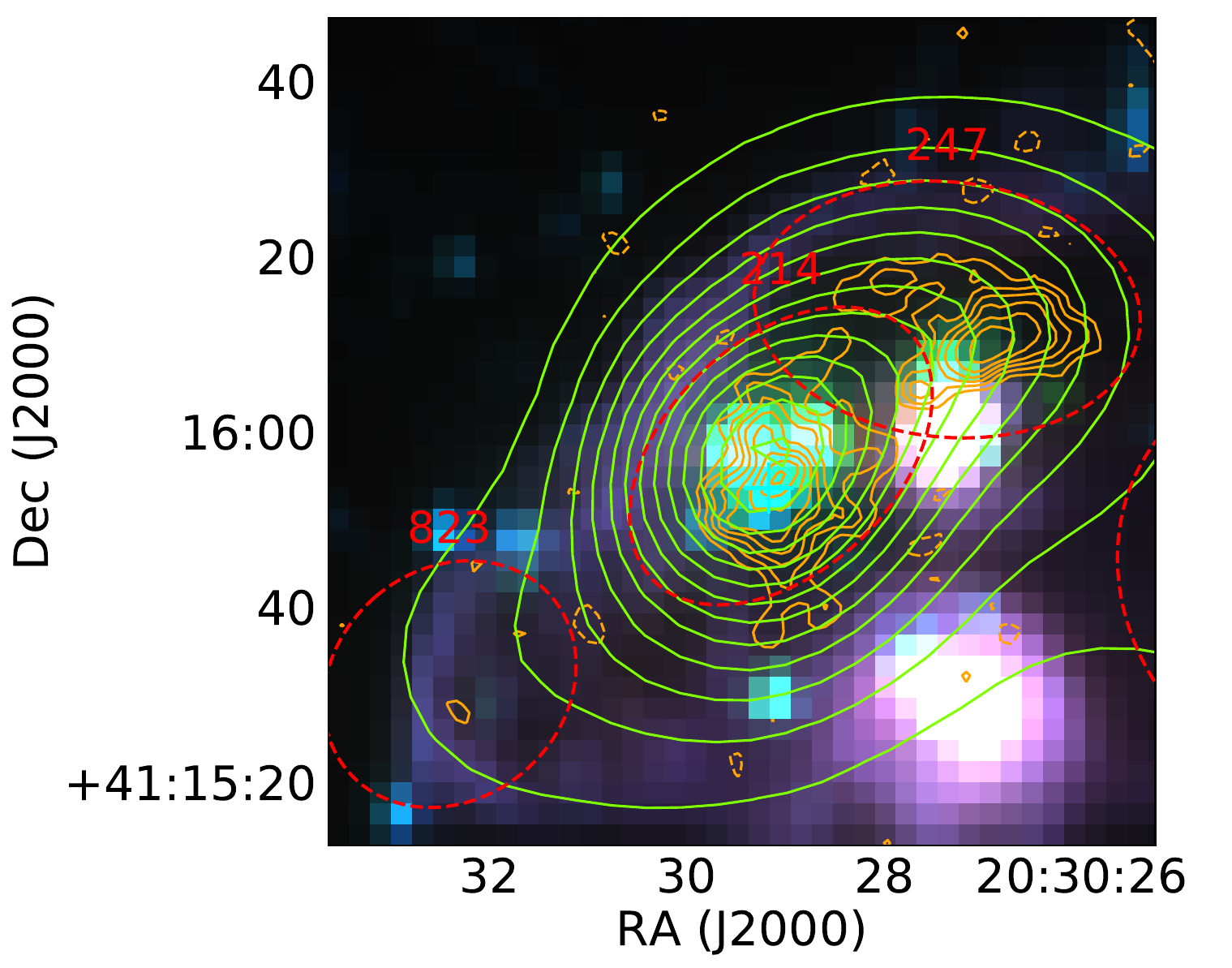} & 
\includegraphics[width=.3\textwidth]{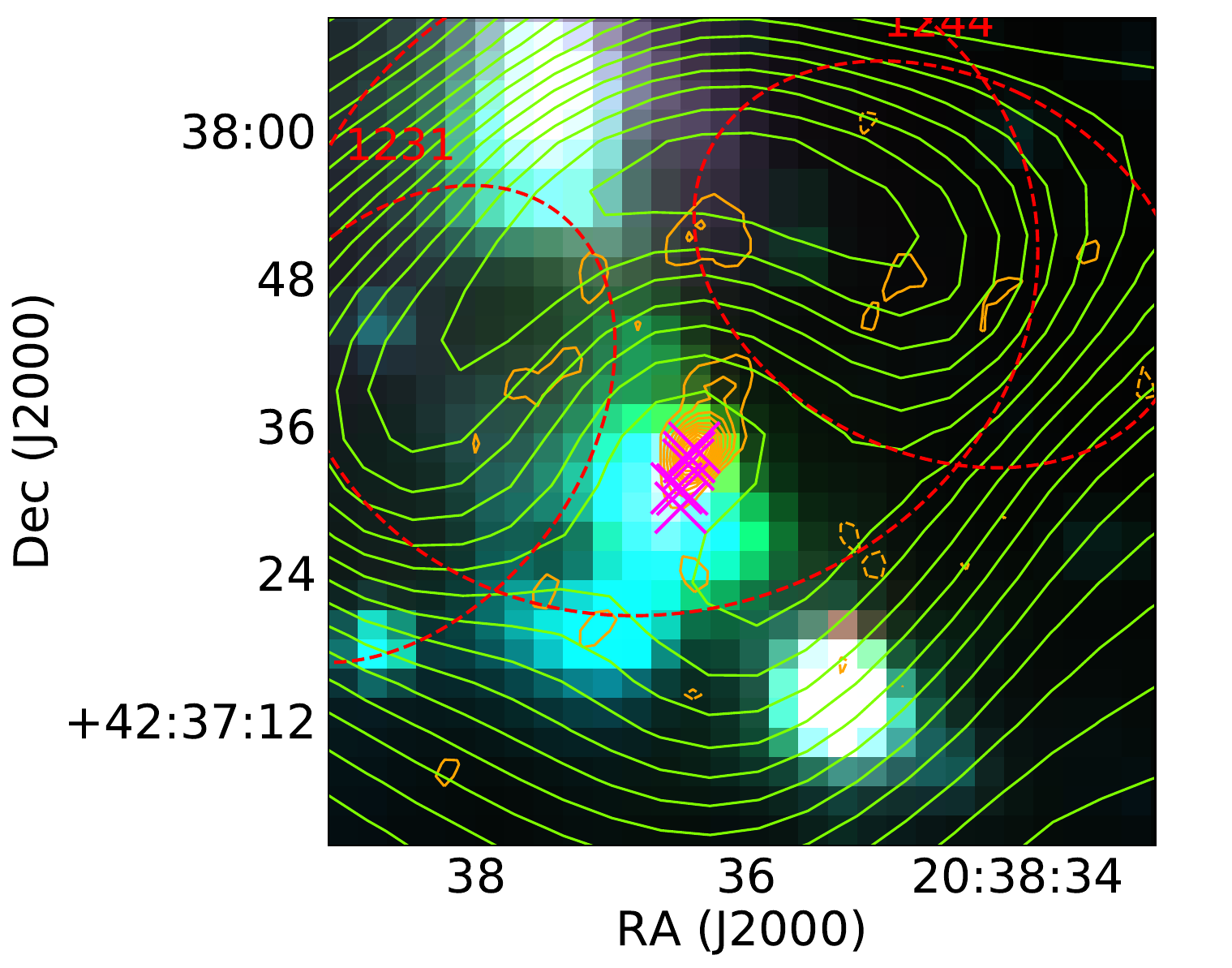} & 
 \\ 
Field 34 & Field 37 &  \\ 
\end{longtable}

\section{Maps of non-thermal velocity dispersion}\label{app:D}

\begin{longtable}{@{}ccc@{}}
\caption*{\textbf{Fig. D.1.} continued.}
\endfoot
\caption*{\textbf{Fig. D.1.} Nonthermal velocity dispersion maps of other fields. The green enclosing areas show the regions with Mach numbers larger than 1. The ellipses are the same as those in Fig. \ref{fig:A.1}. \label{fig:D.1}}
\endlastfoot
\includegraphics[width=.3\textwidth]{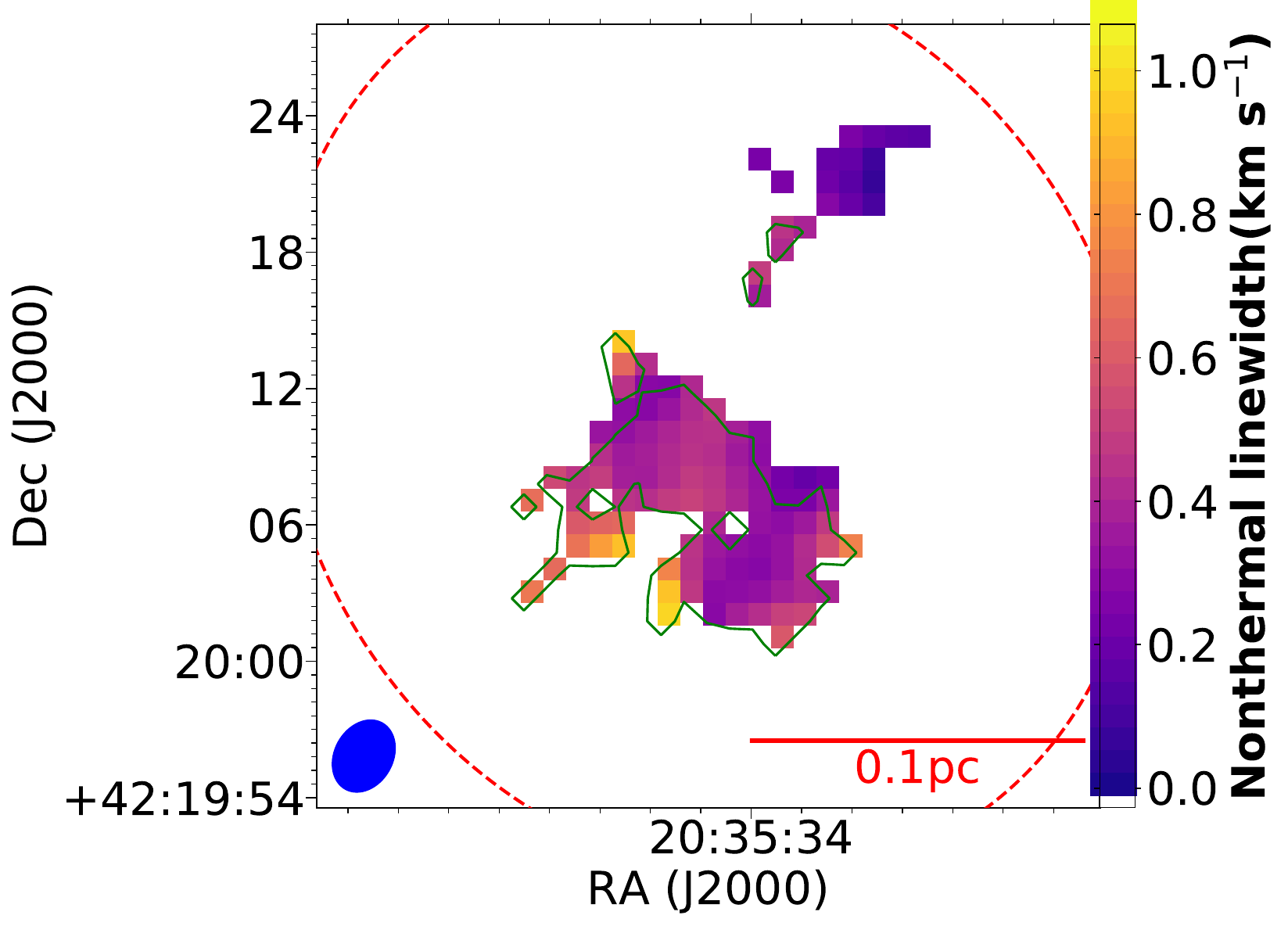} & 
\includegraphics[width=.3\textwidth]{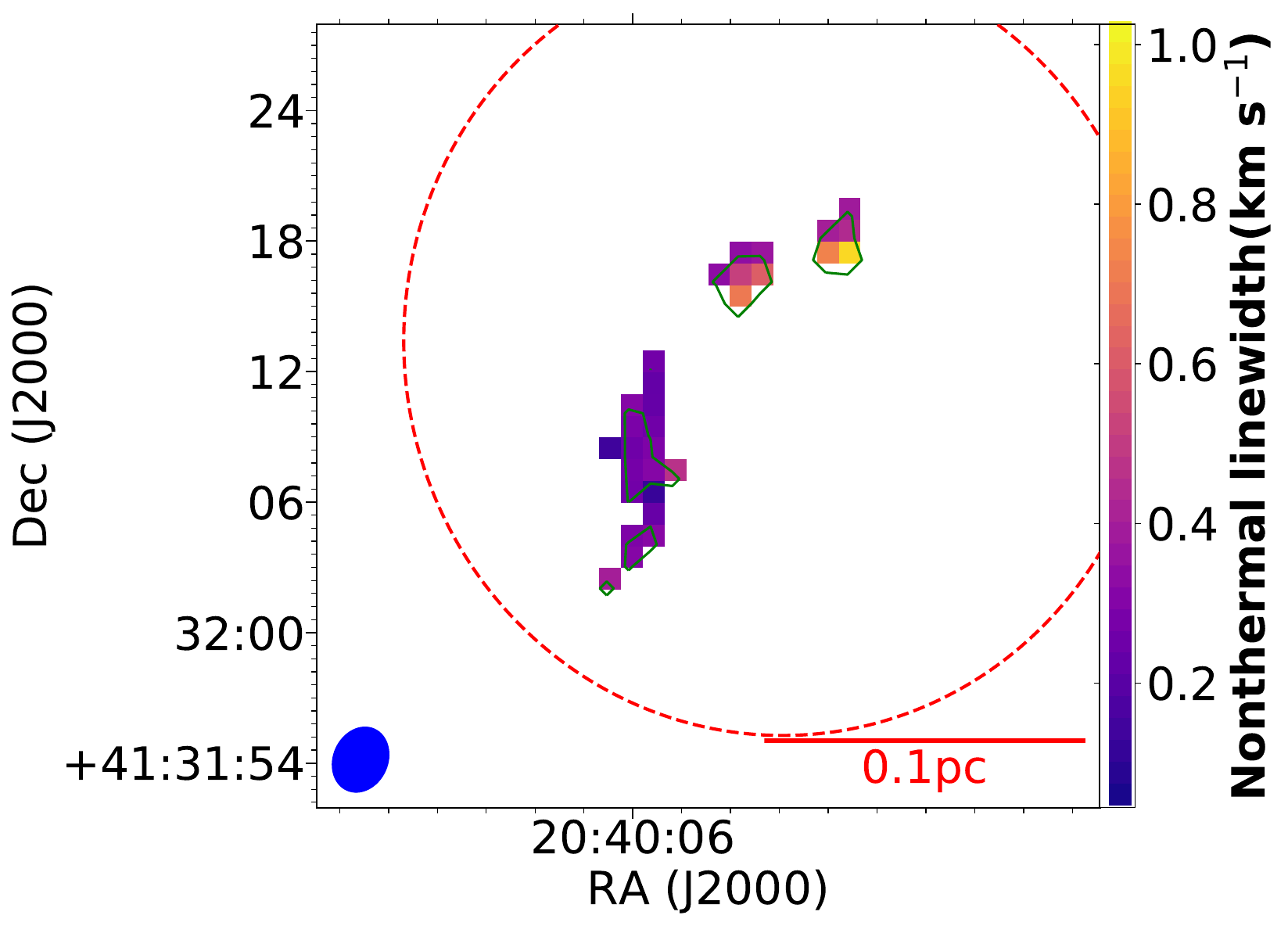} & 
\includegraphics[width=.3\textwidth]{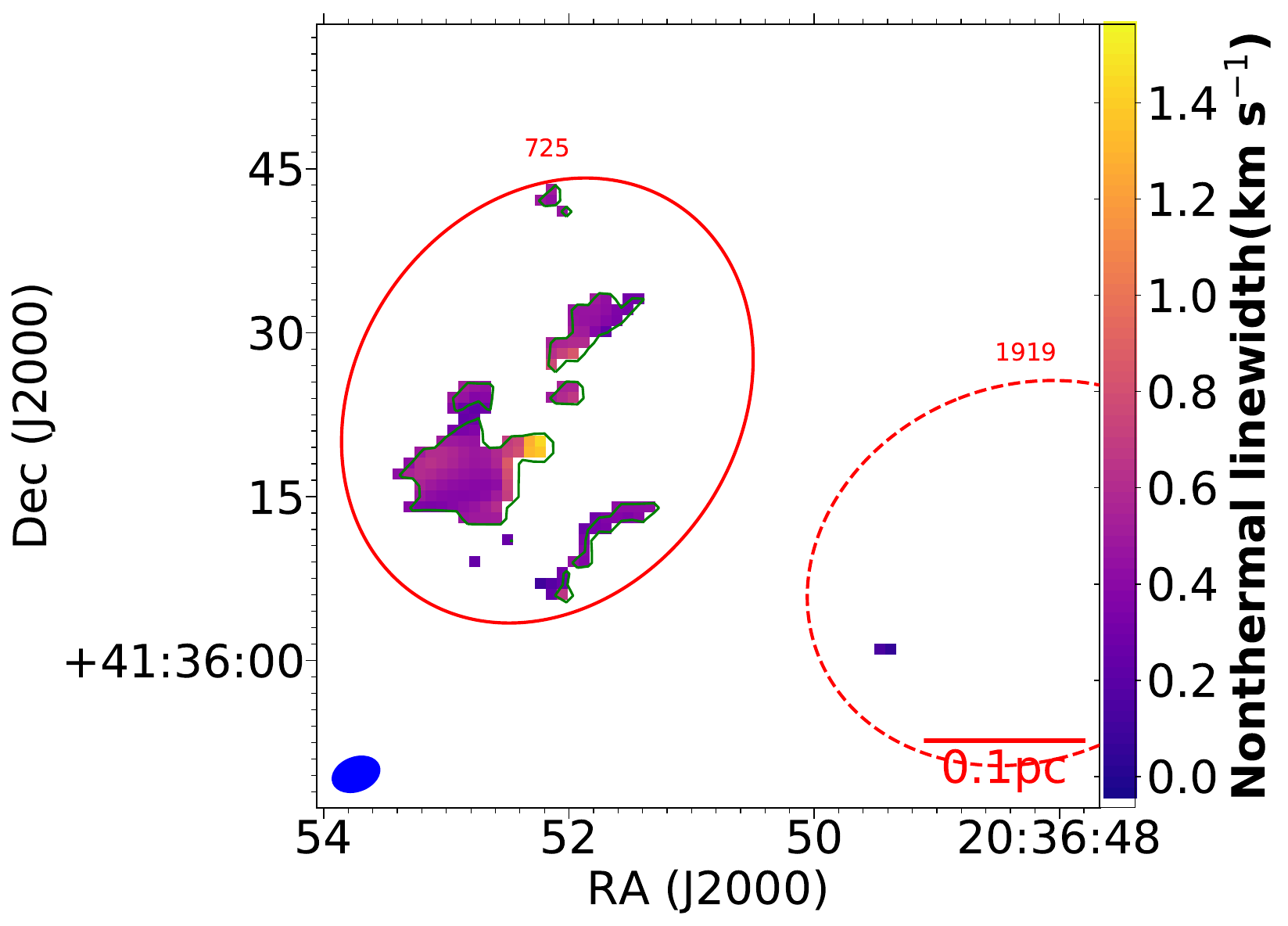} \\ 
Field 1 & Field 2 & Field 4 \\ 
\includegraphics[width=.3\textwidth]{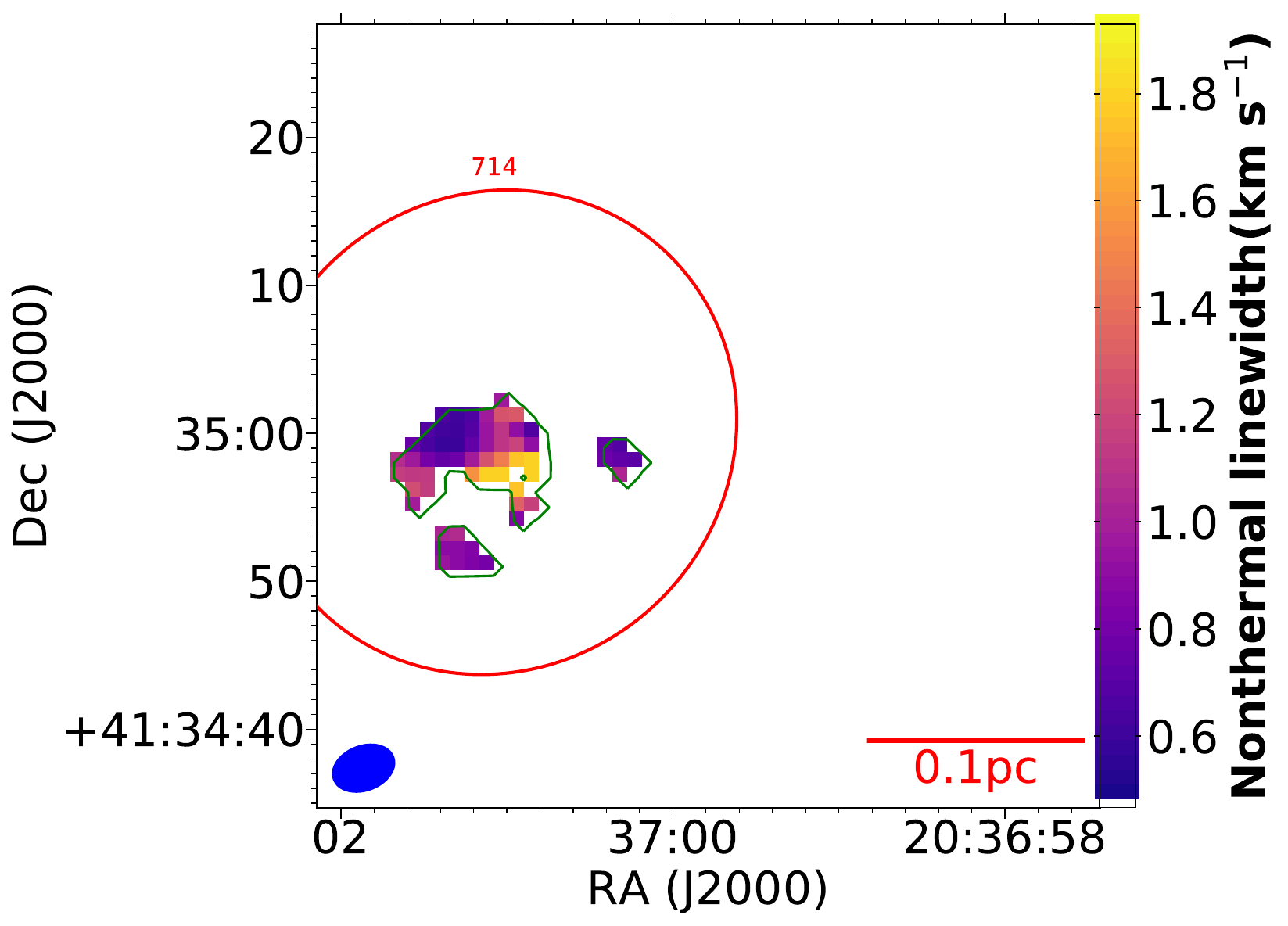} & 
\includegraphics[width=.3\textwidth]{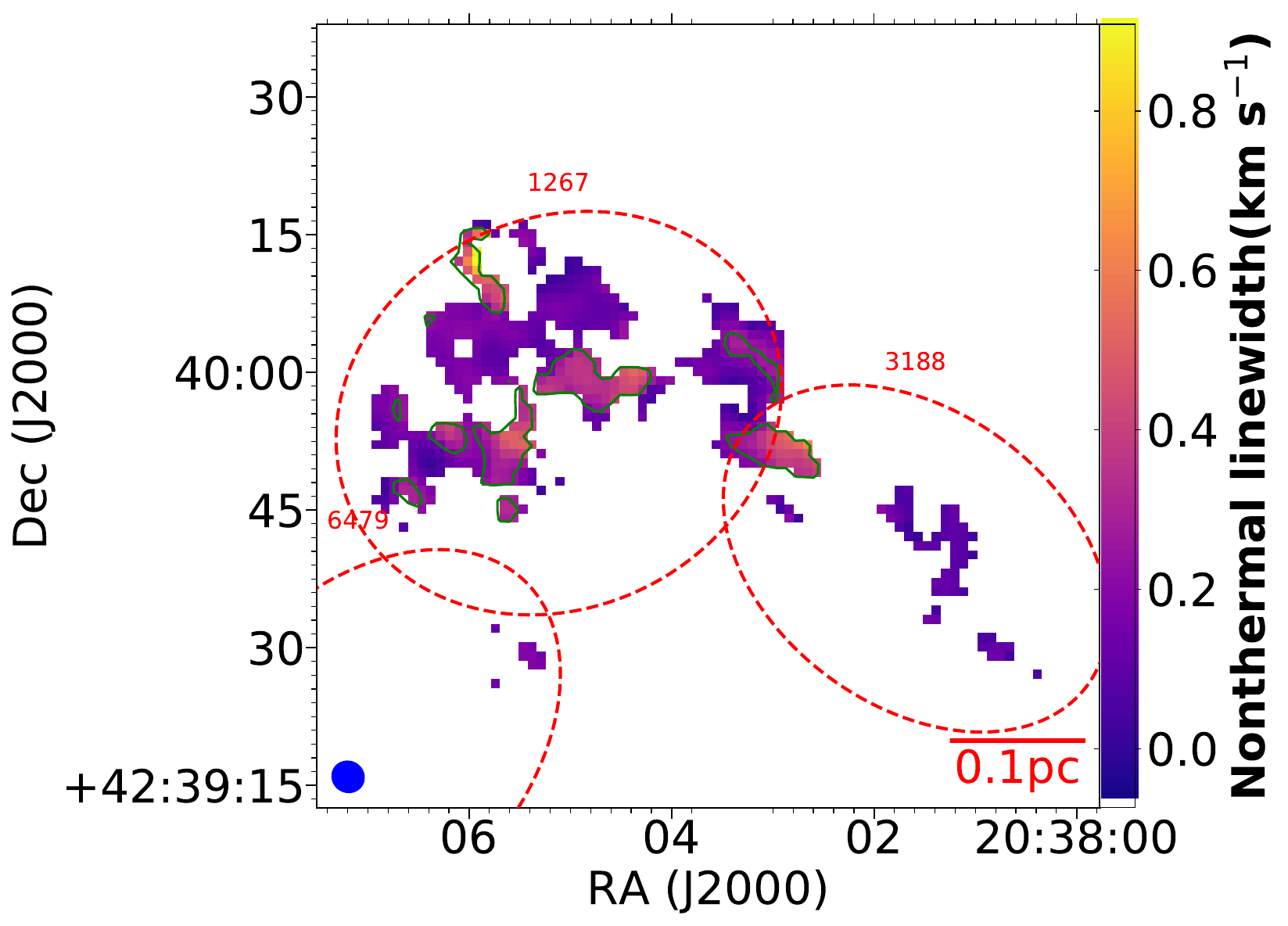} & 
\includegraphics[width=.3\textwidth]{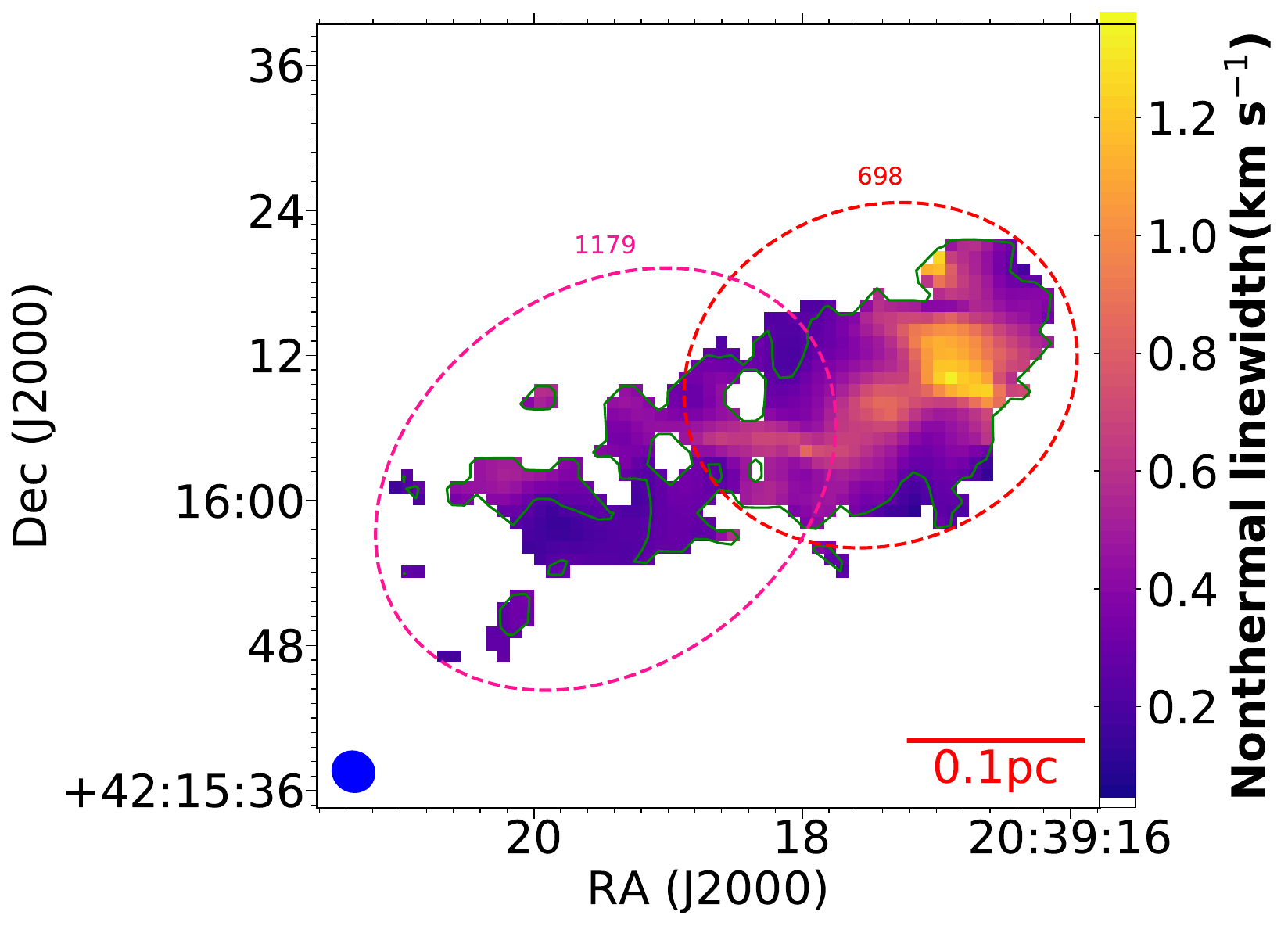} \\ 
Field 6 & Field 7 & Field 8 \\ 
\includegraphics[width=.3\textwidth]{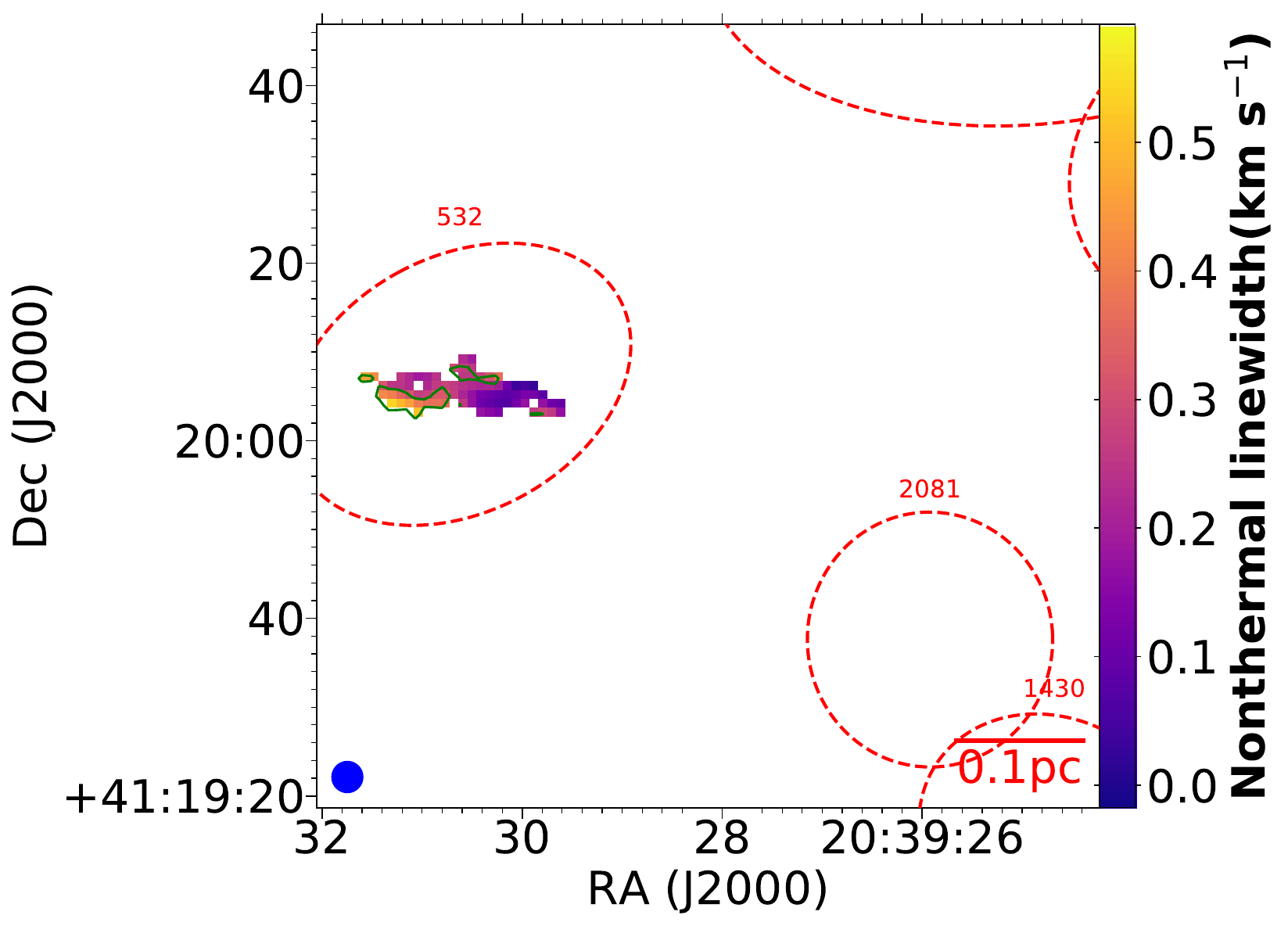} & 
\includegraphics[width=.3\textwidth]{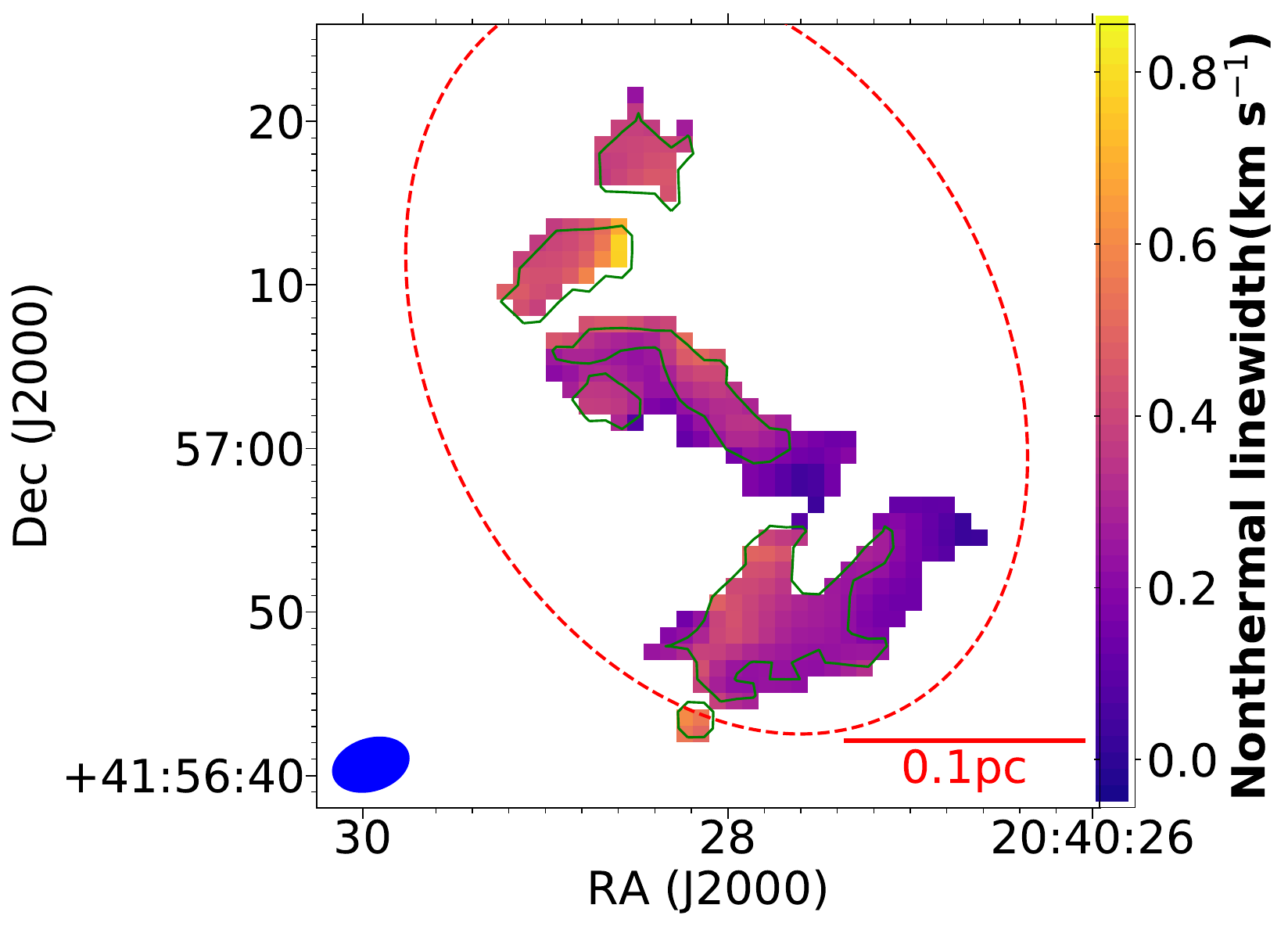} & 
\includegraphics[width=.3\textwidth]{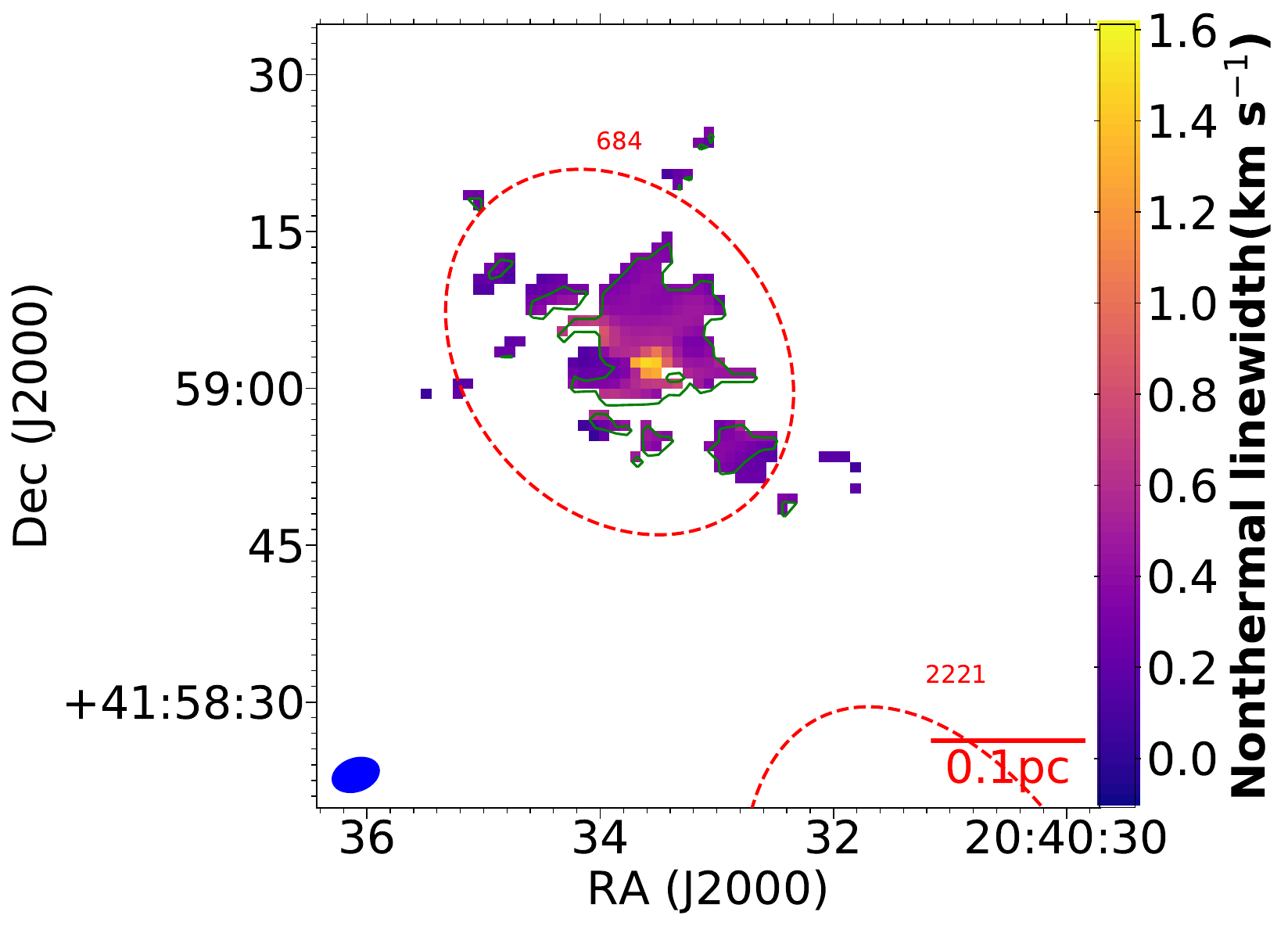} \\ 
Field 9 & Field 10 & Field 11 \\ 
\includegraphics[width=.3\textwidth]{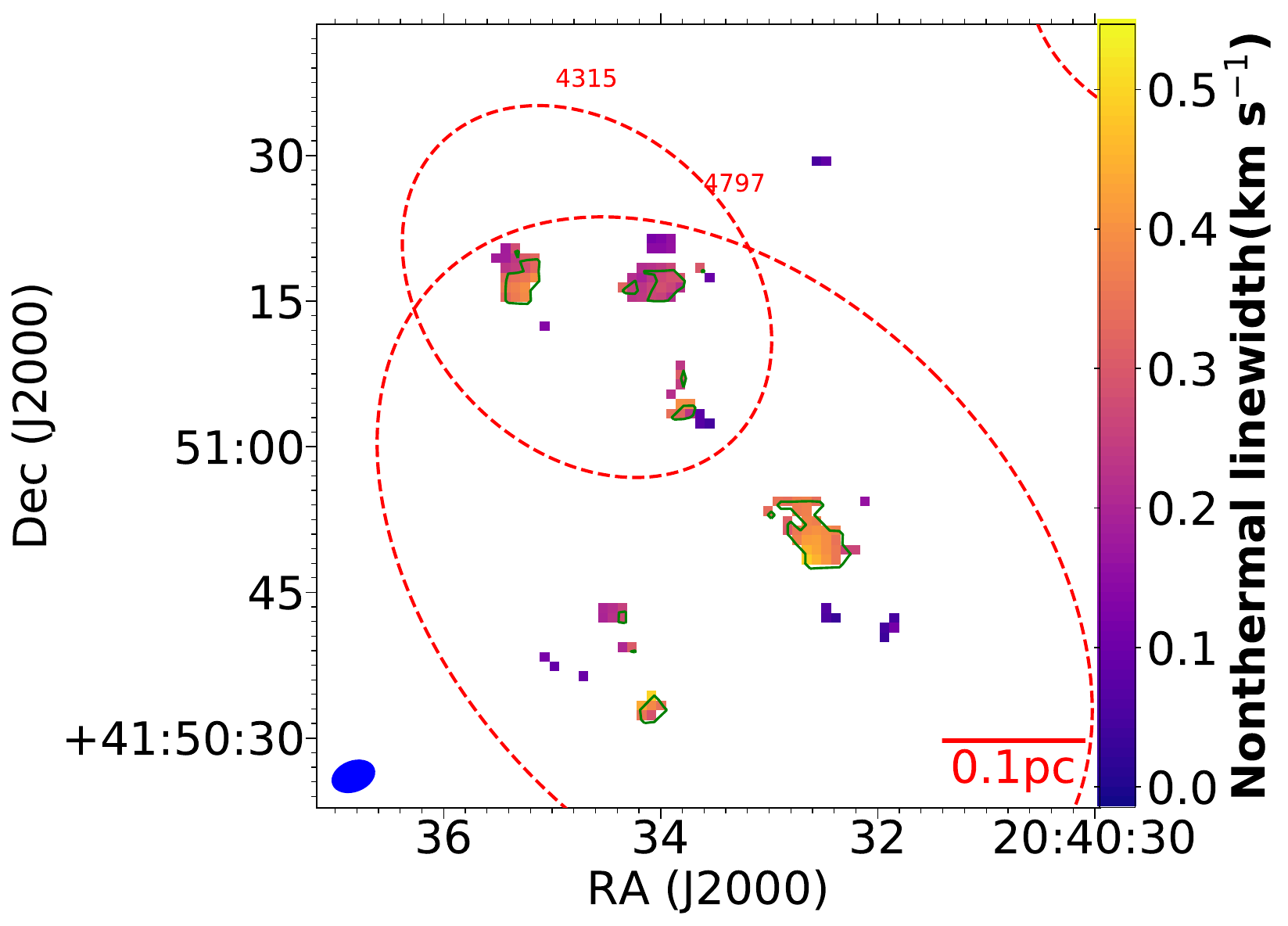} & 
\includegraphics[width=.3\textwidth]{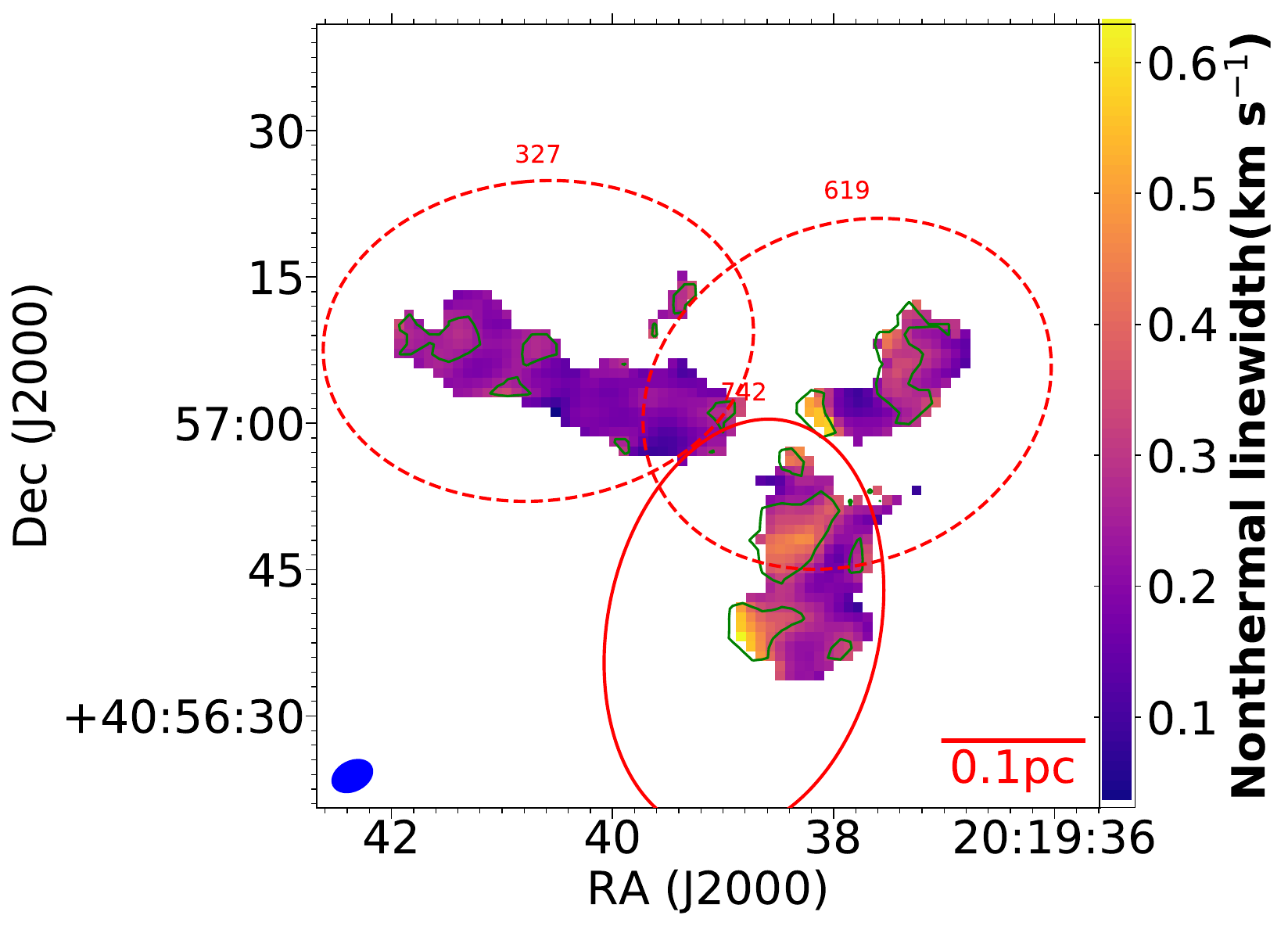} & 
\includegraphics[width=.3\textwidth]{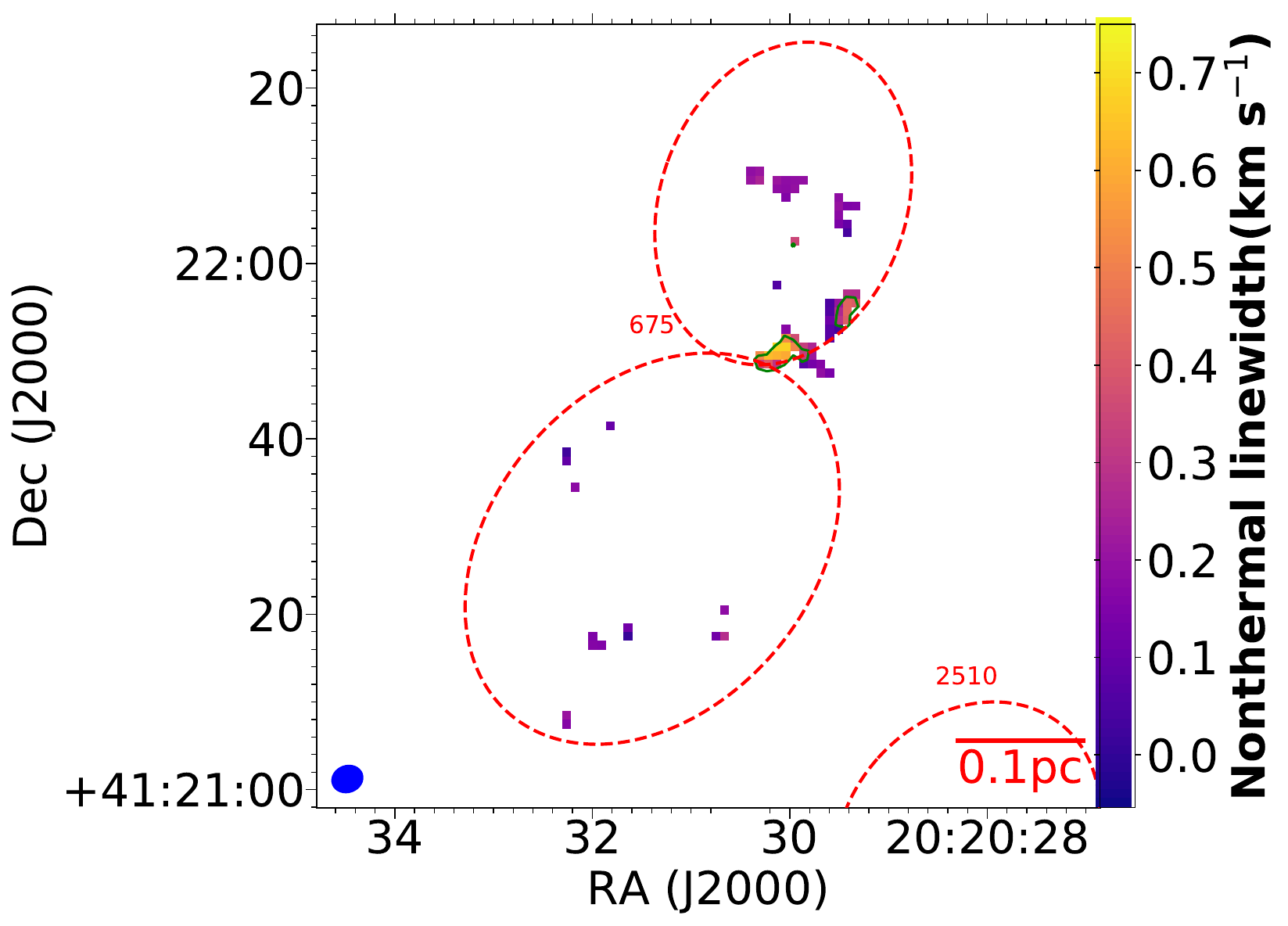} \\ 
Field 12 & Field 13 & Field 14 \\ 
\includegraphics[width=.3\textwidth]{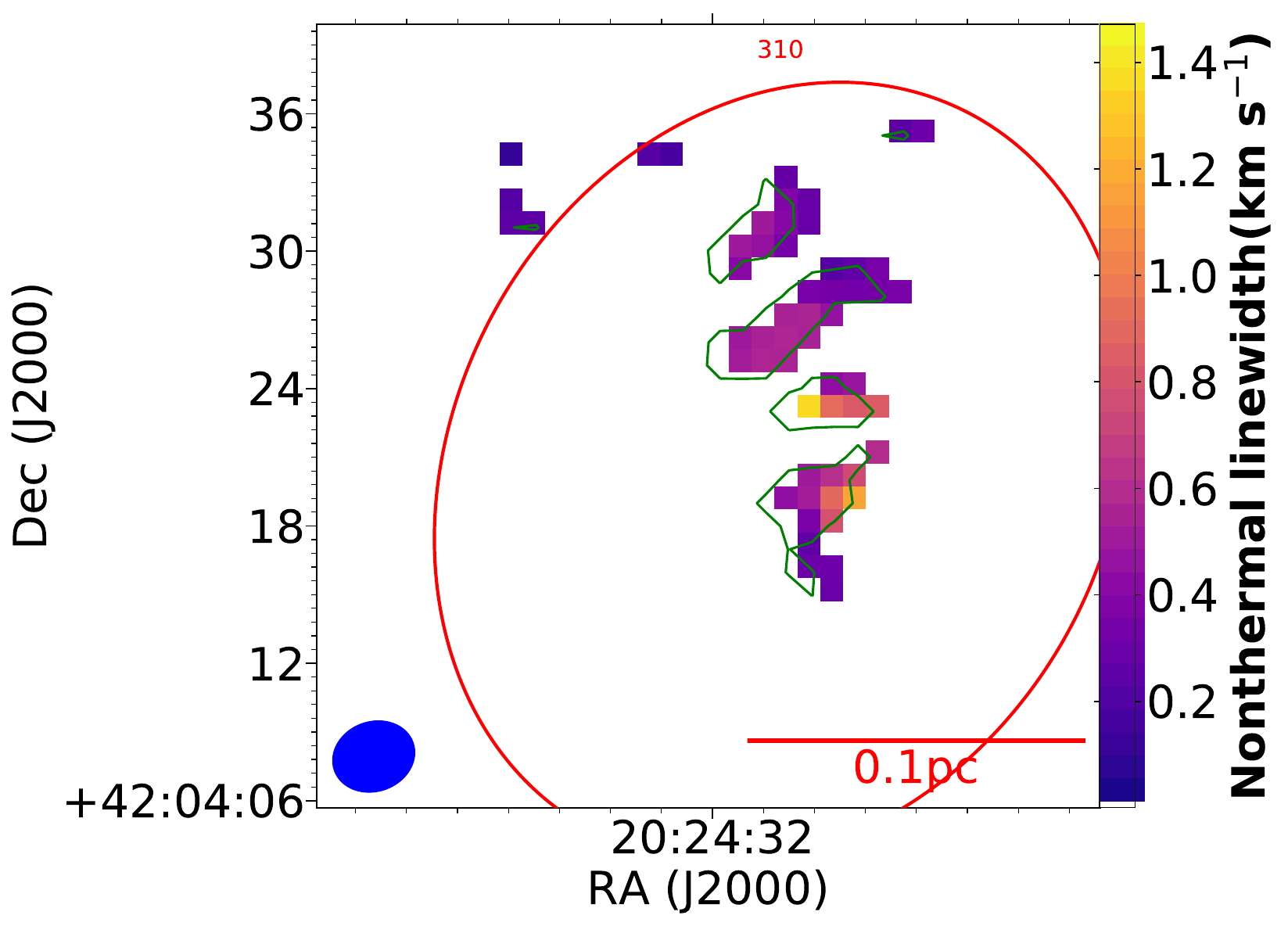} & 
\includegraphics[width=.3\textwidth]{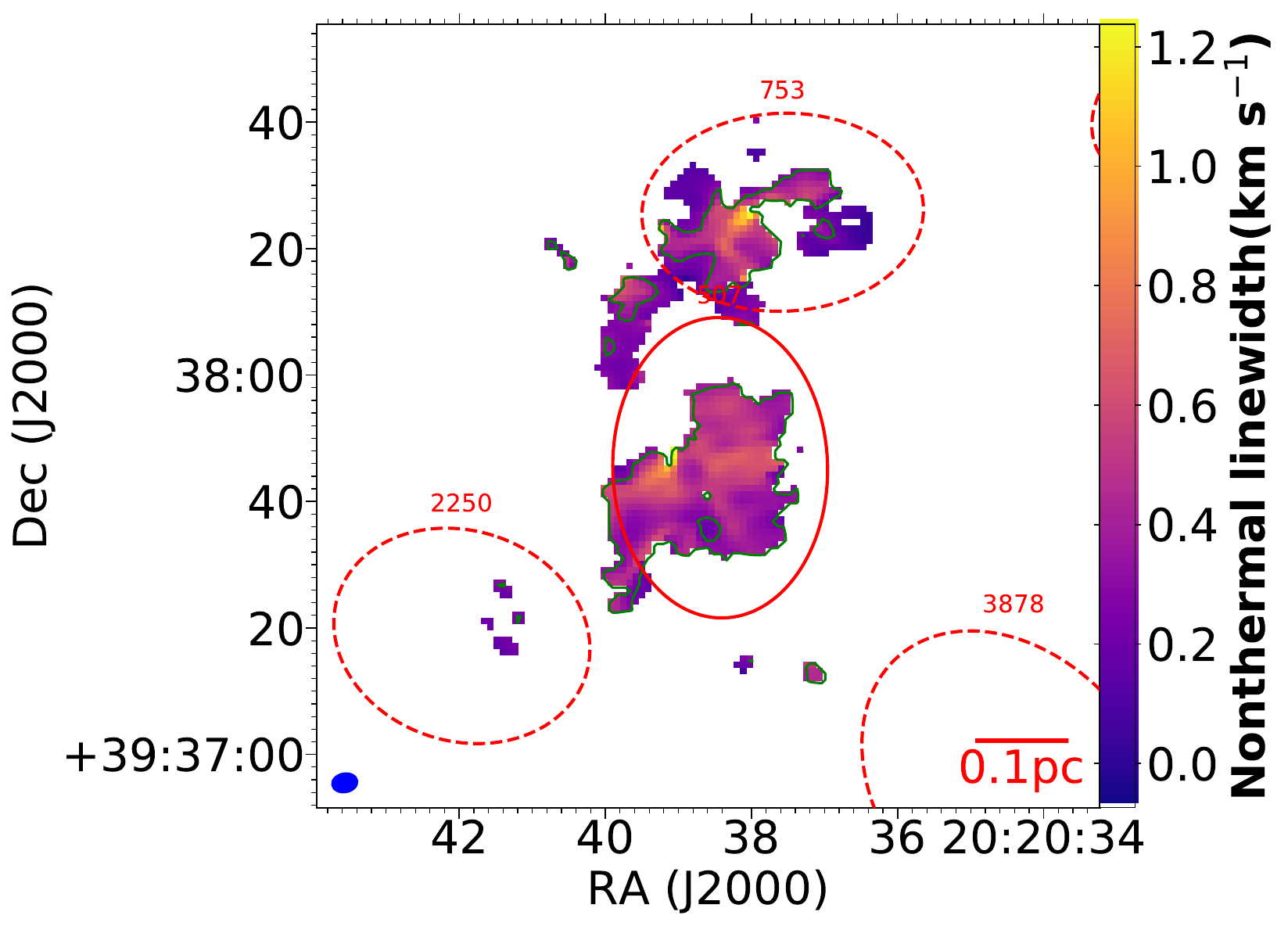} & 
\includegraphics[width=.3\textwidth]{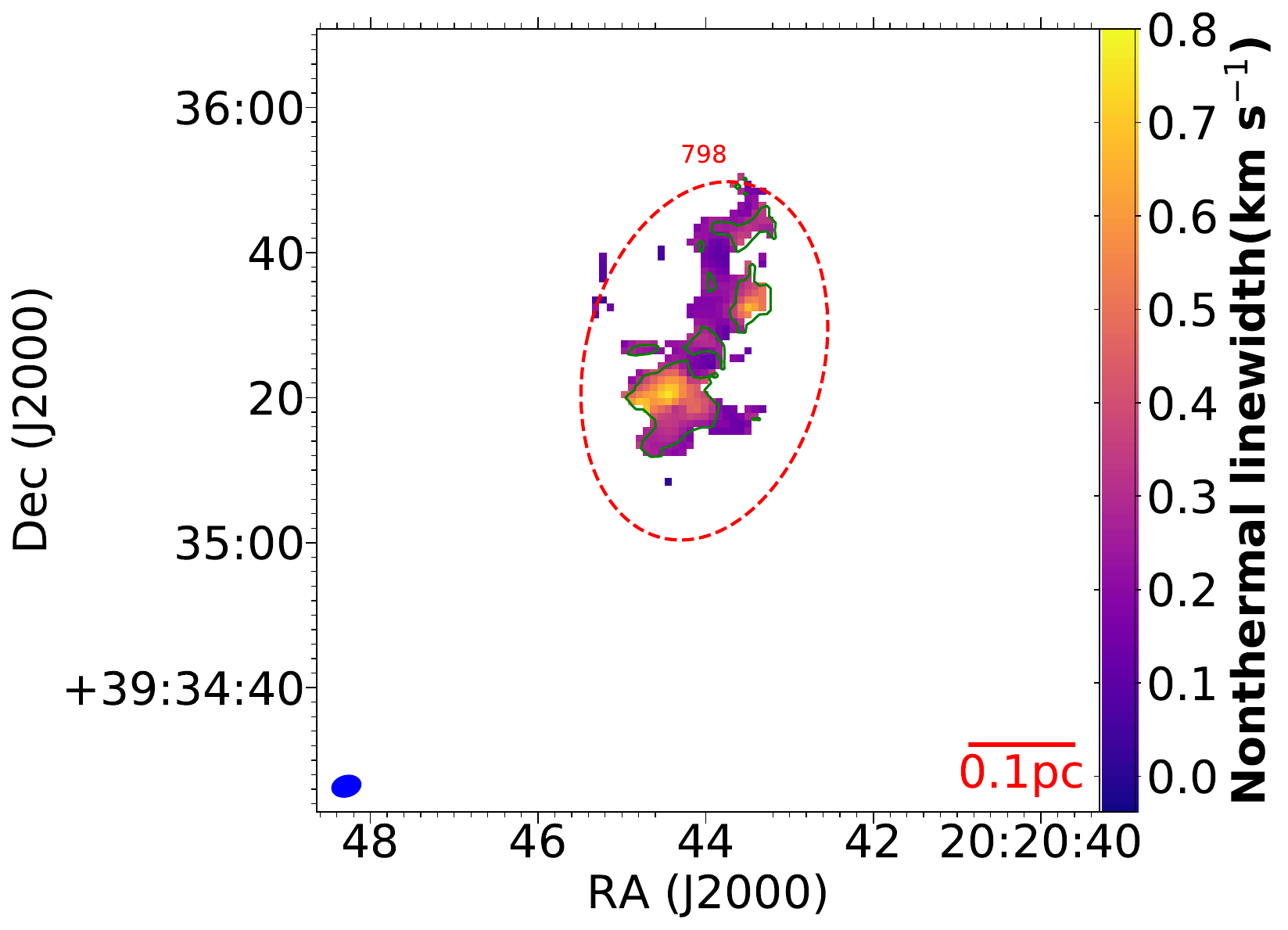} \\ 
Field 16 & Field 17 & Field 18 \\ 
\includegraphics[width=.3\textwidth]{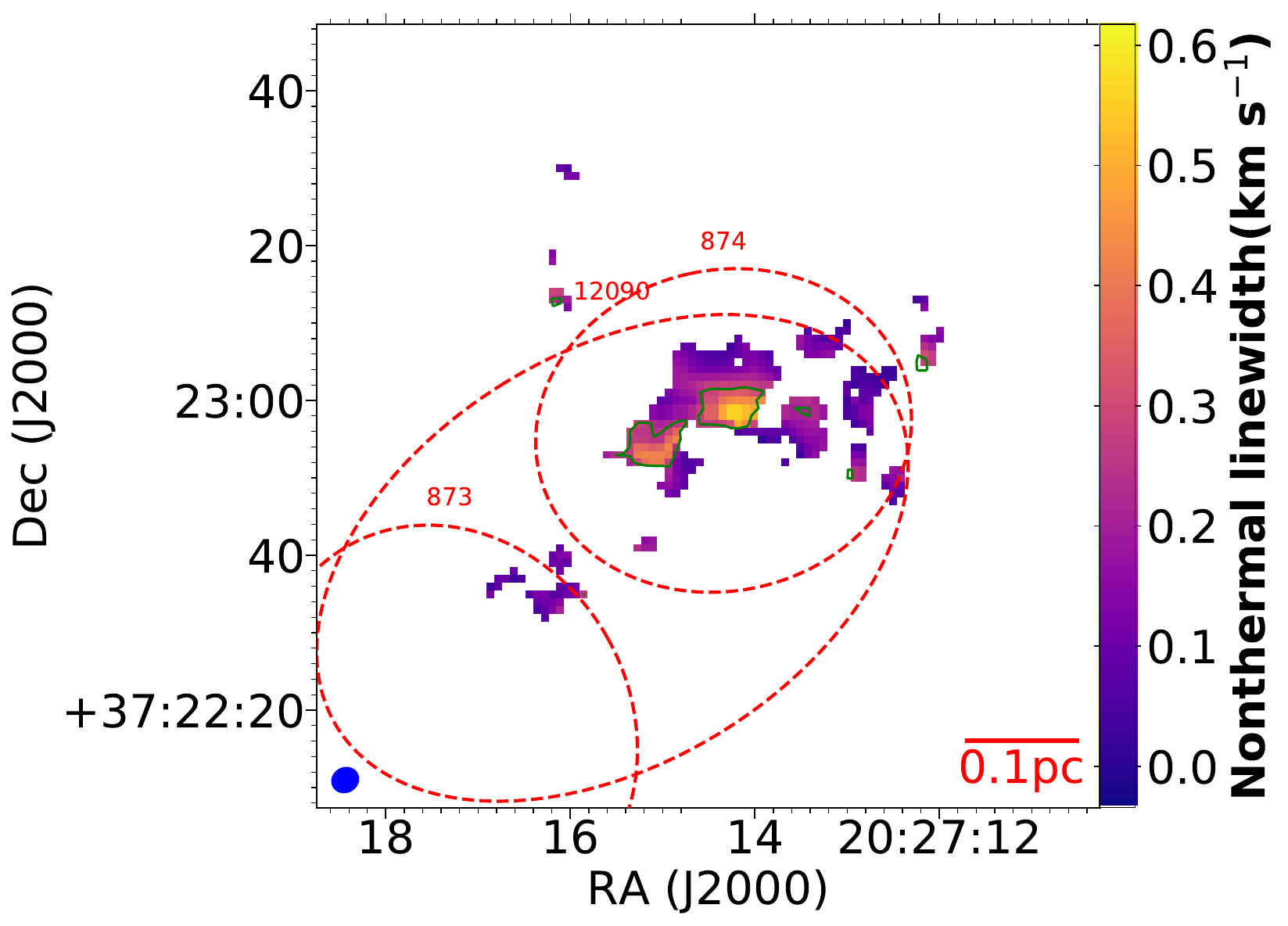} & 
\includegraphics[width=.3\textwidth]{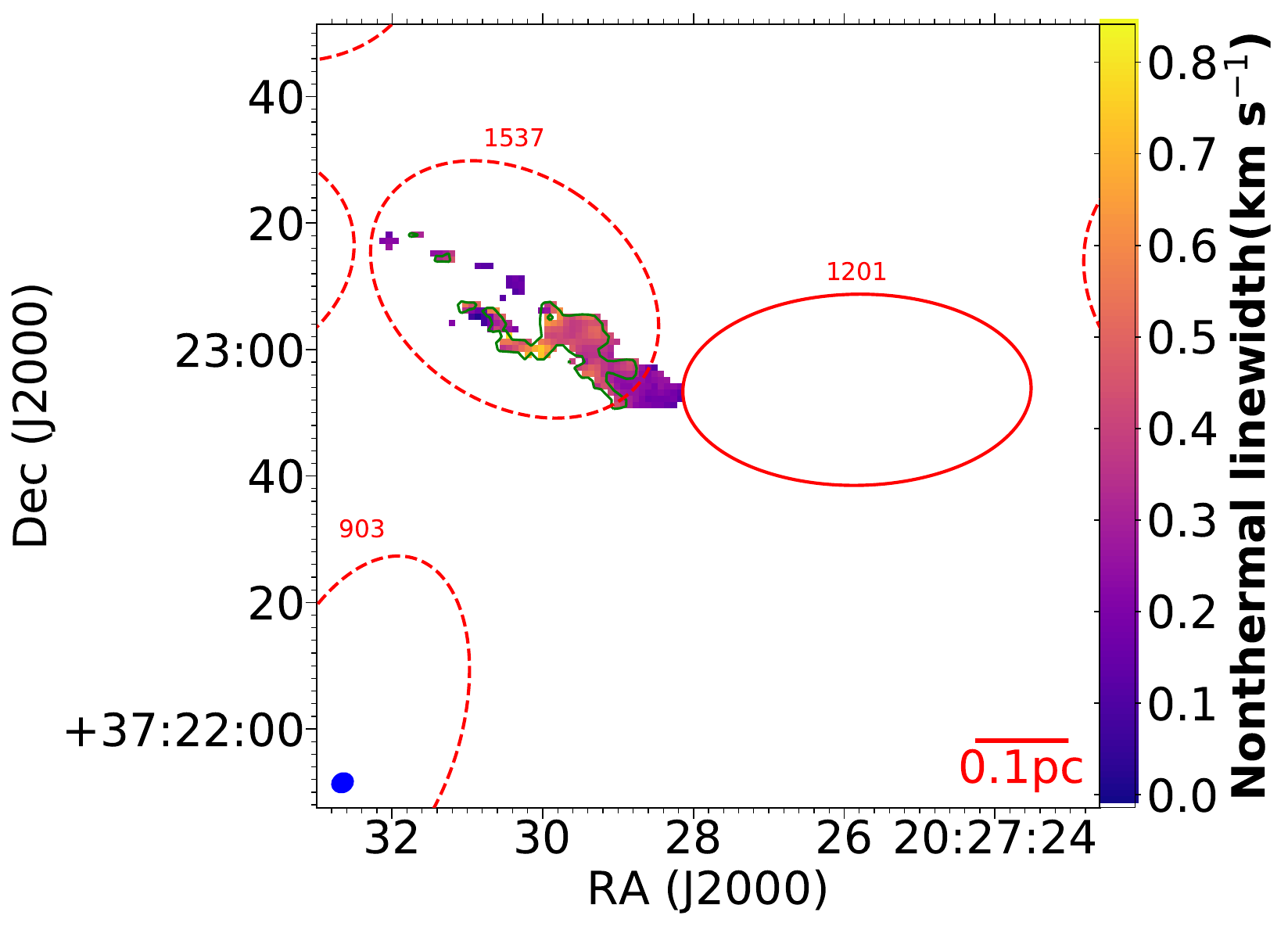} & 
\includegraphics[width=.3\textwidth]{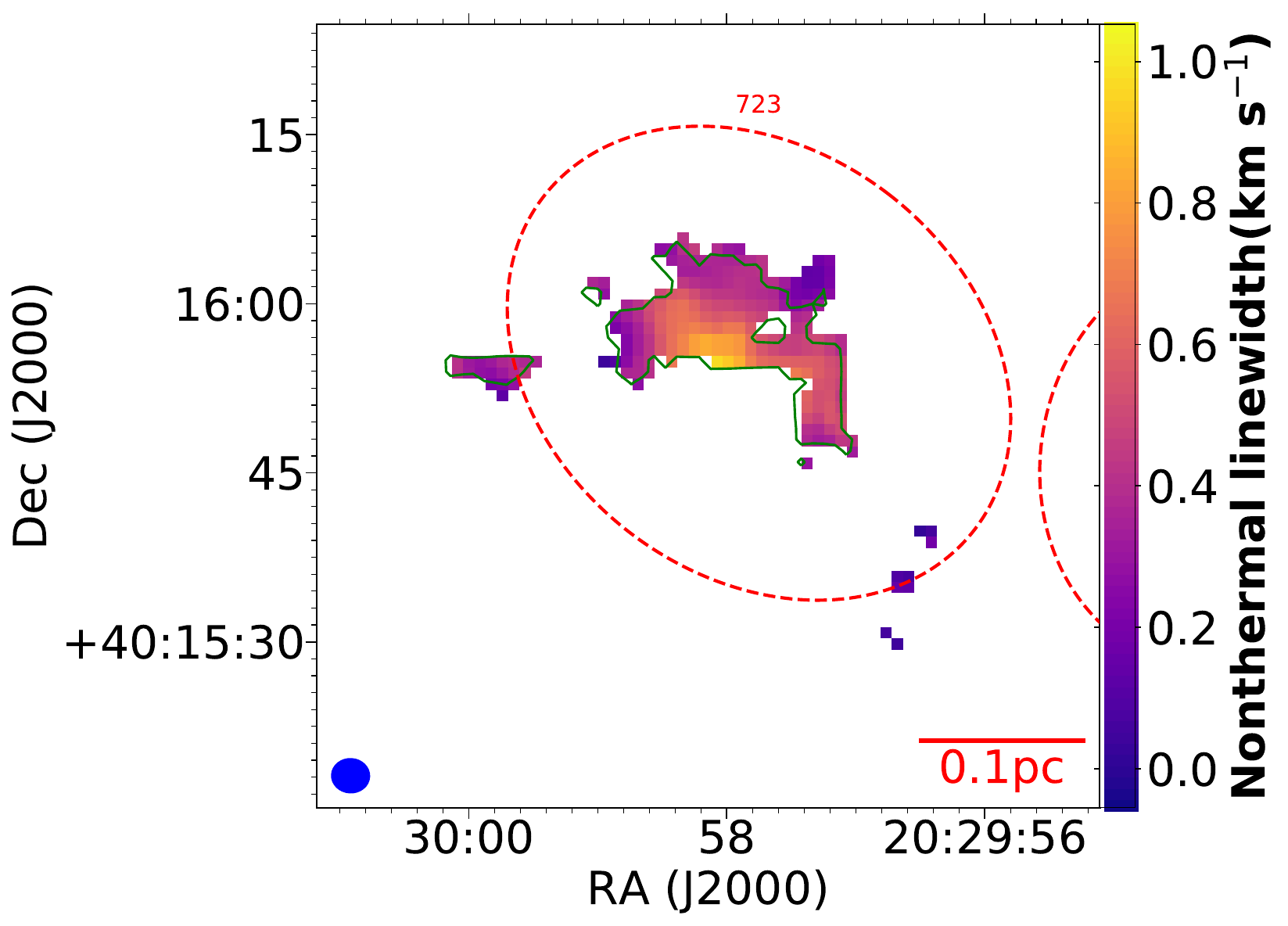} \\ 
Field 19 & Field 20 & Field 22 \\ 
\includegraphics[width=.3\textwidth]{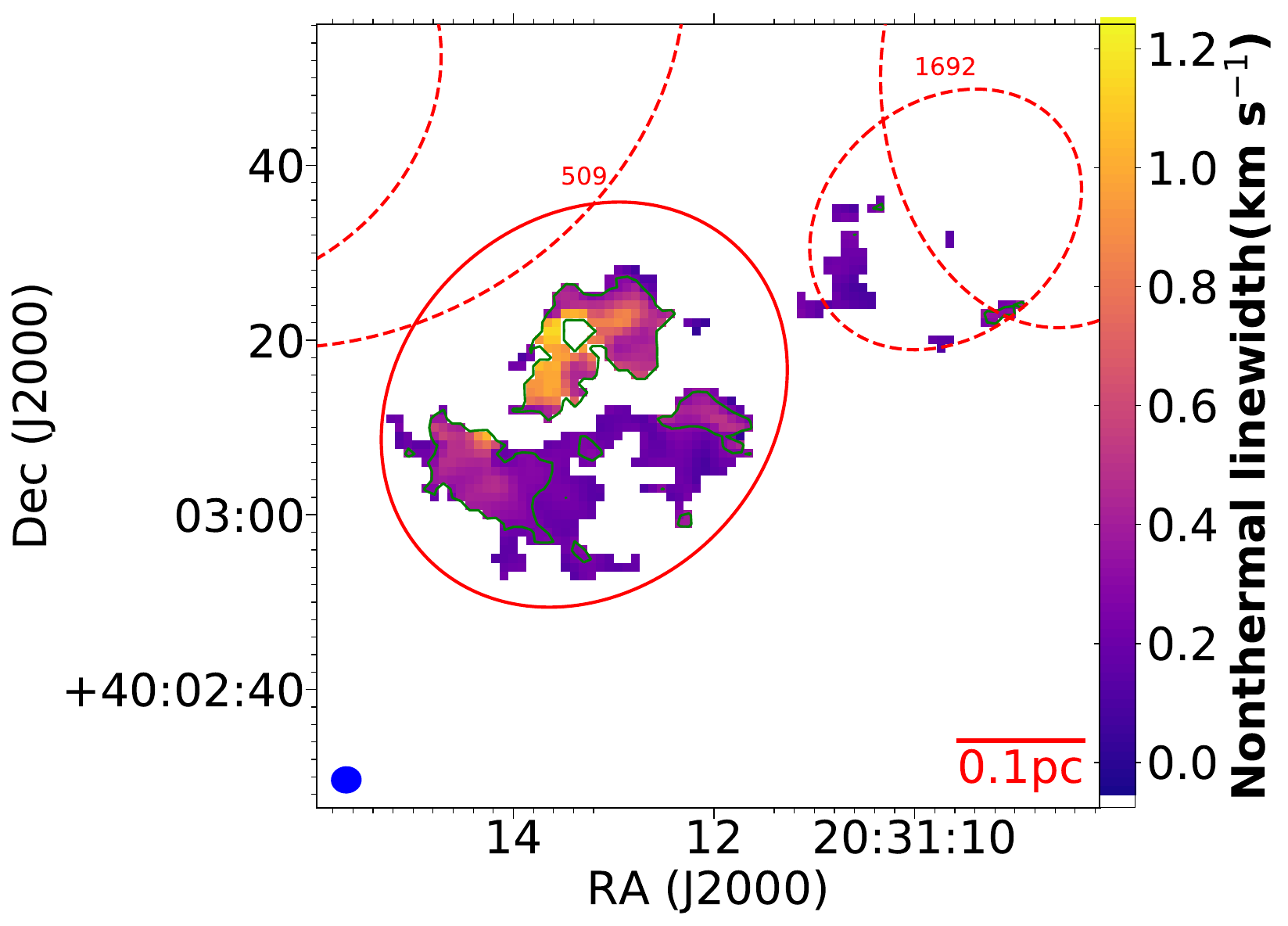} & 
\includegraphics[width=.3\textwidth]{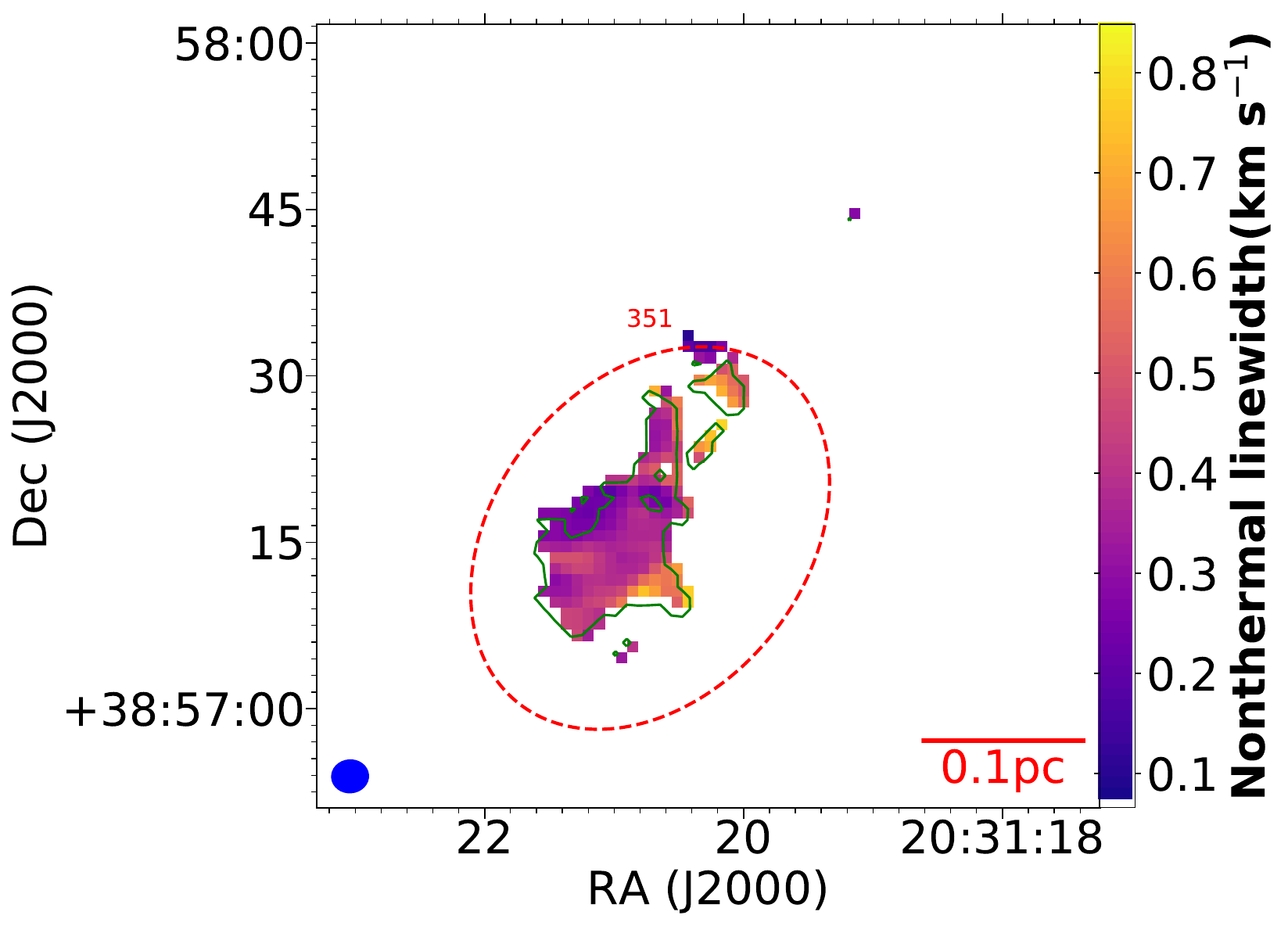} & 
\includegraphics[width=.3\textwidth]{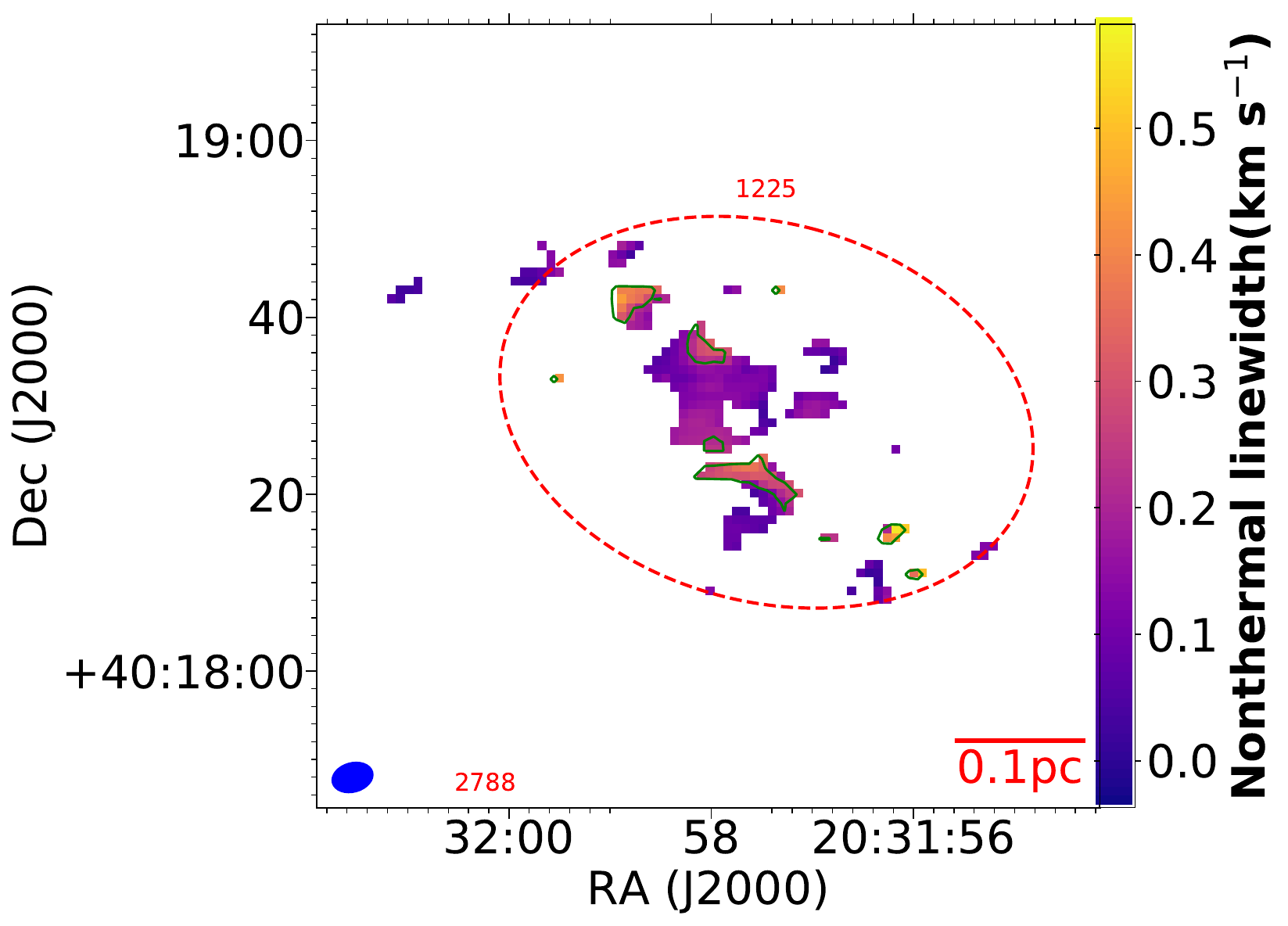} \\ 
Field 23 & Field 24 & Field 25 \\ 
\includegraphics[width=.3\textwidth]{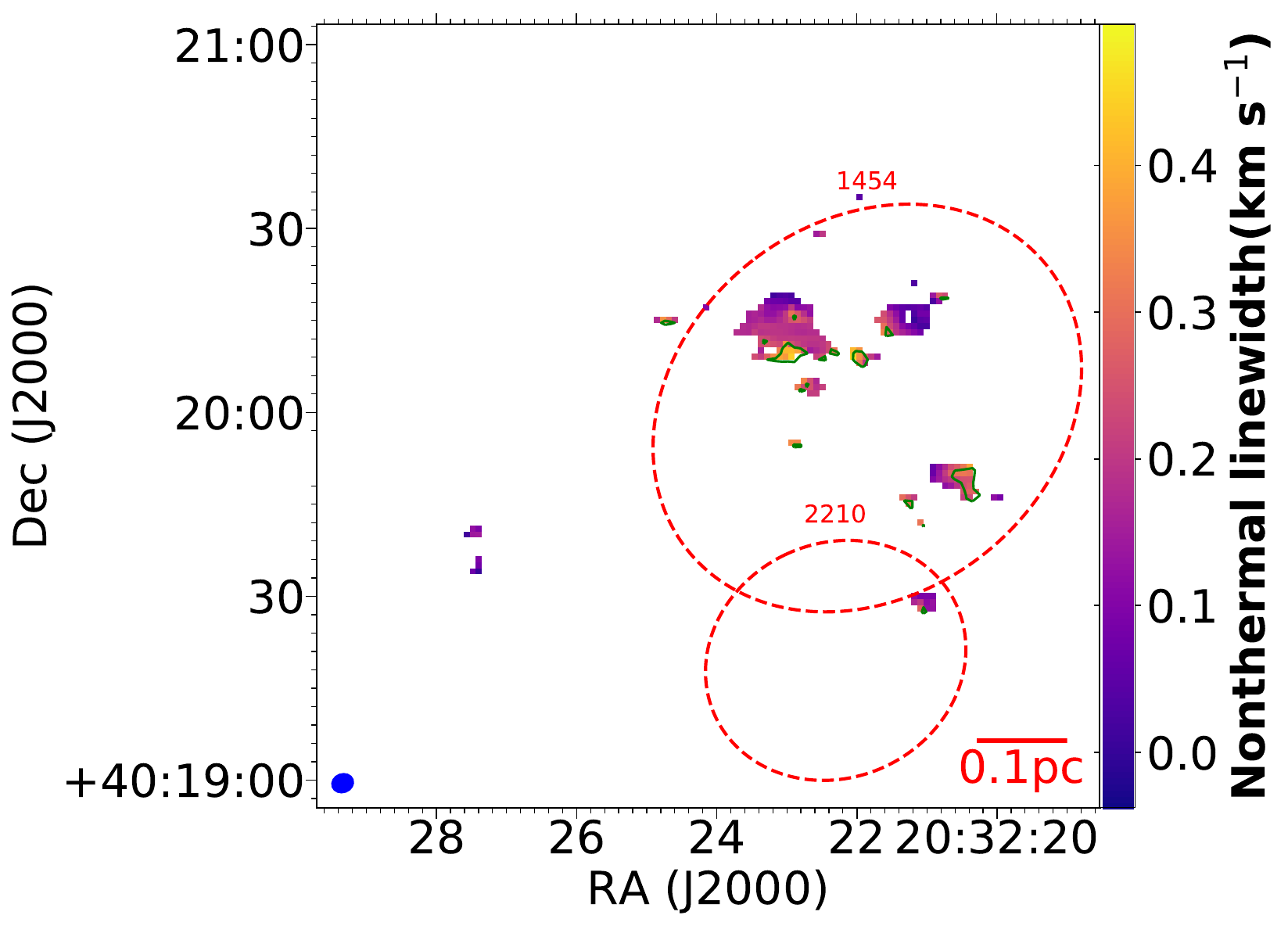} & 
\includegraphics[width=.3\textwidth]{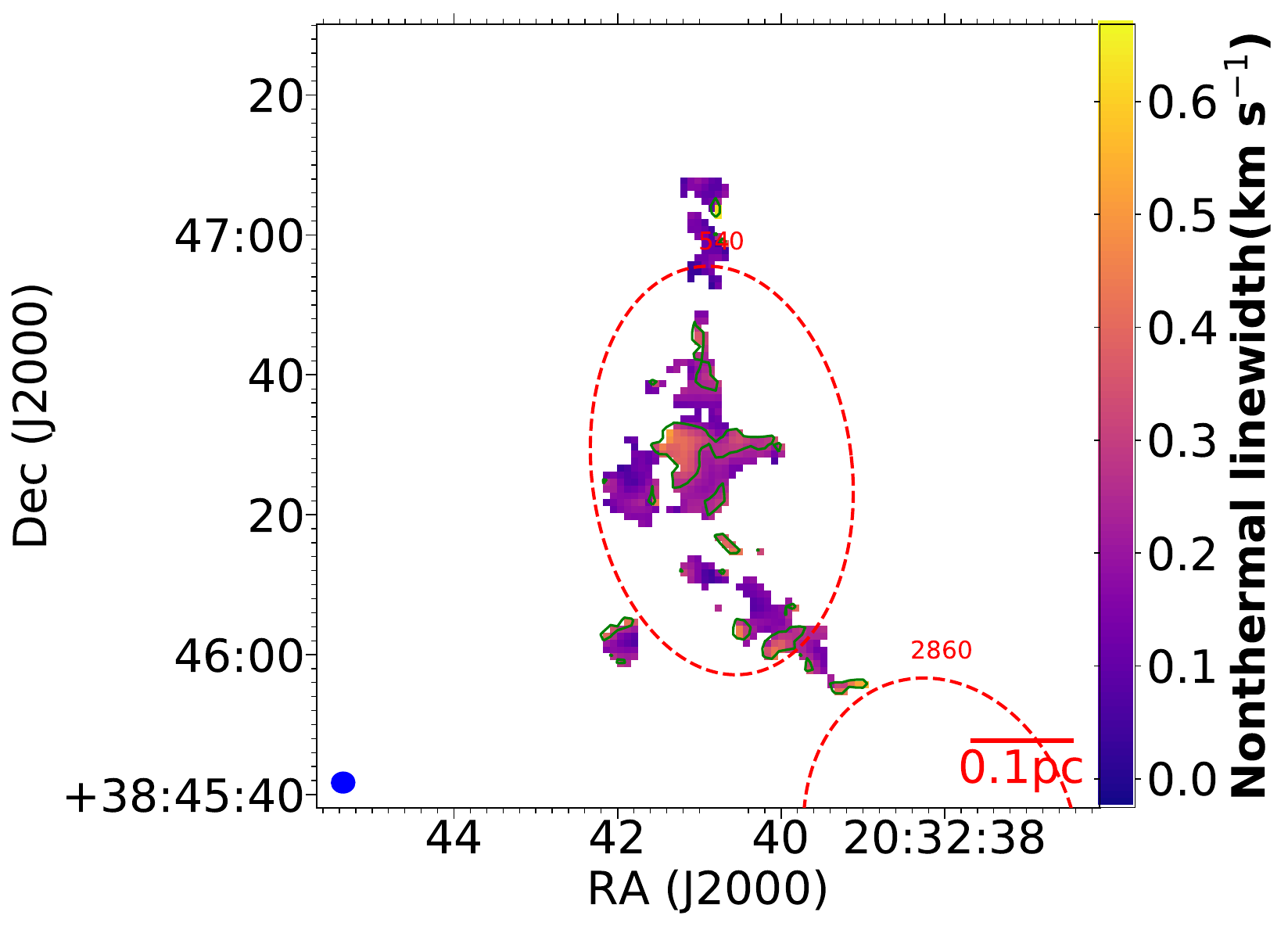} & 
\includegraphics[width=.3\textwidth]{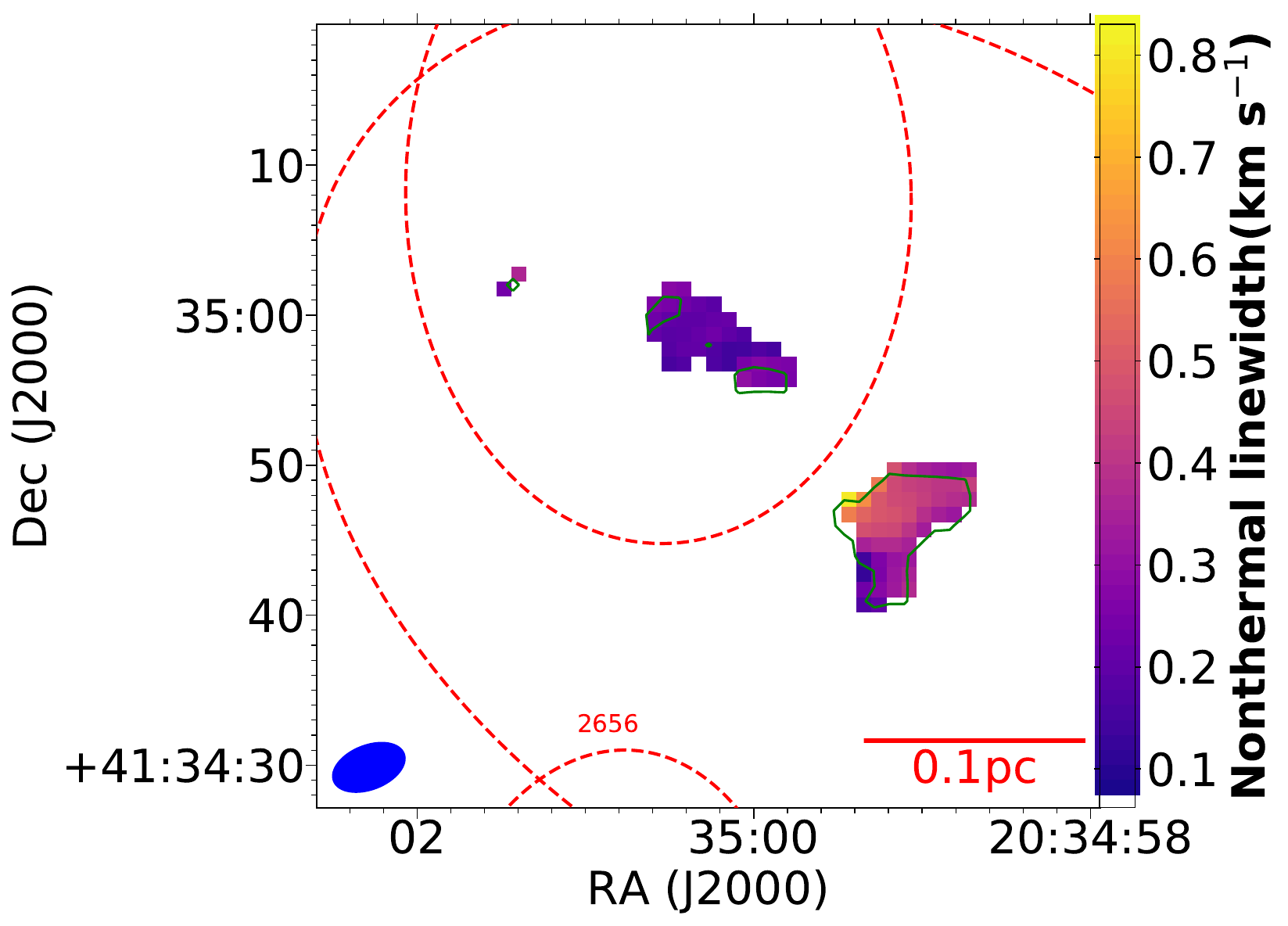} \\ 
Field 26 & Field 28 & Field 29 \\ 
\includegraphics[width=.3\textwidth]{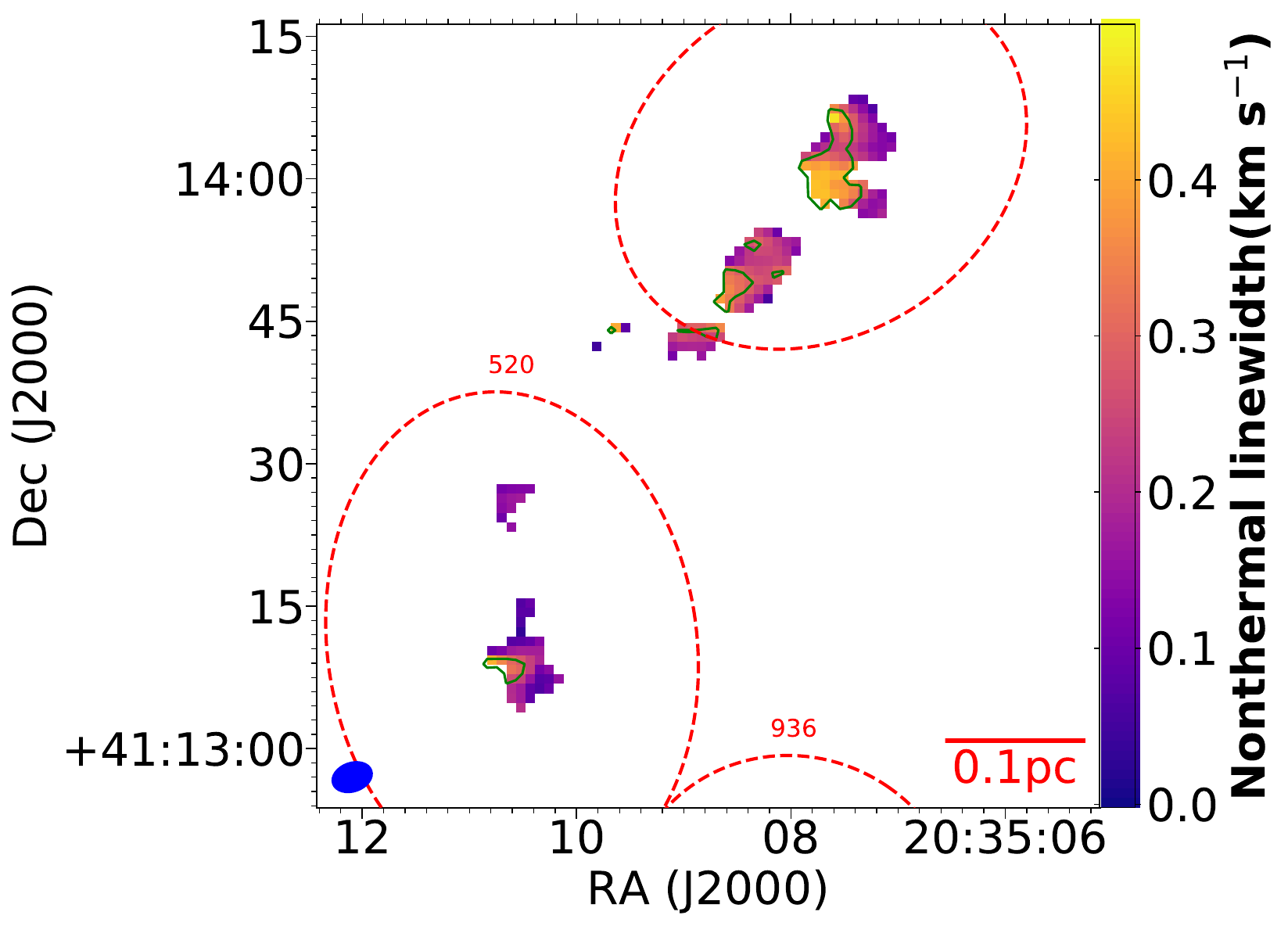} & 
\includegraphics[width=.3\textwidth]{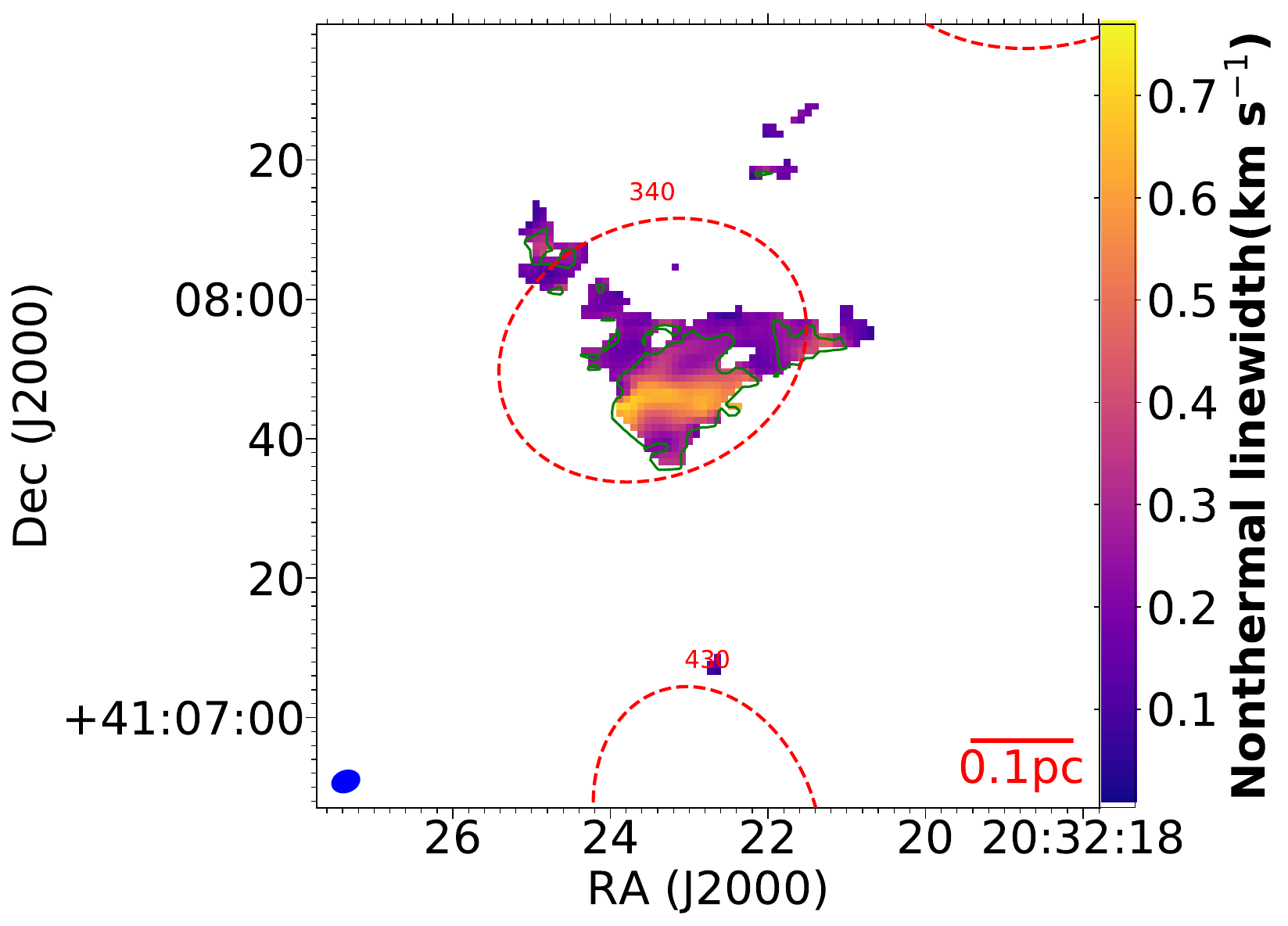} & 
\includegraphics[width=.3\textwidth]{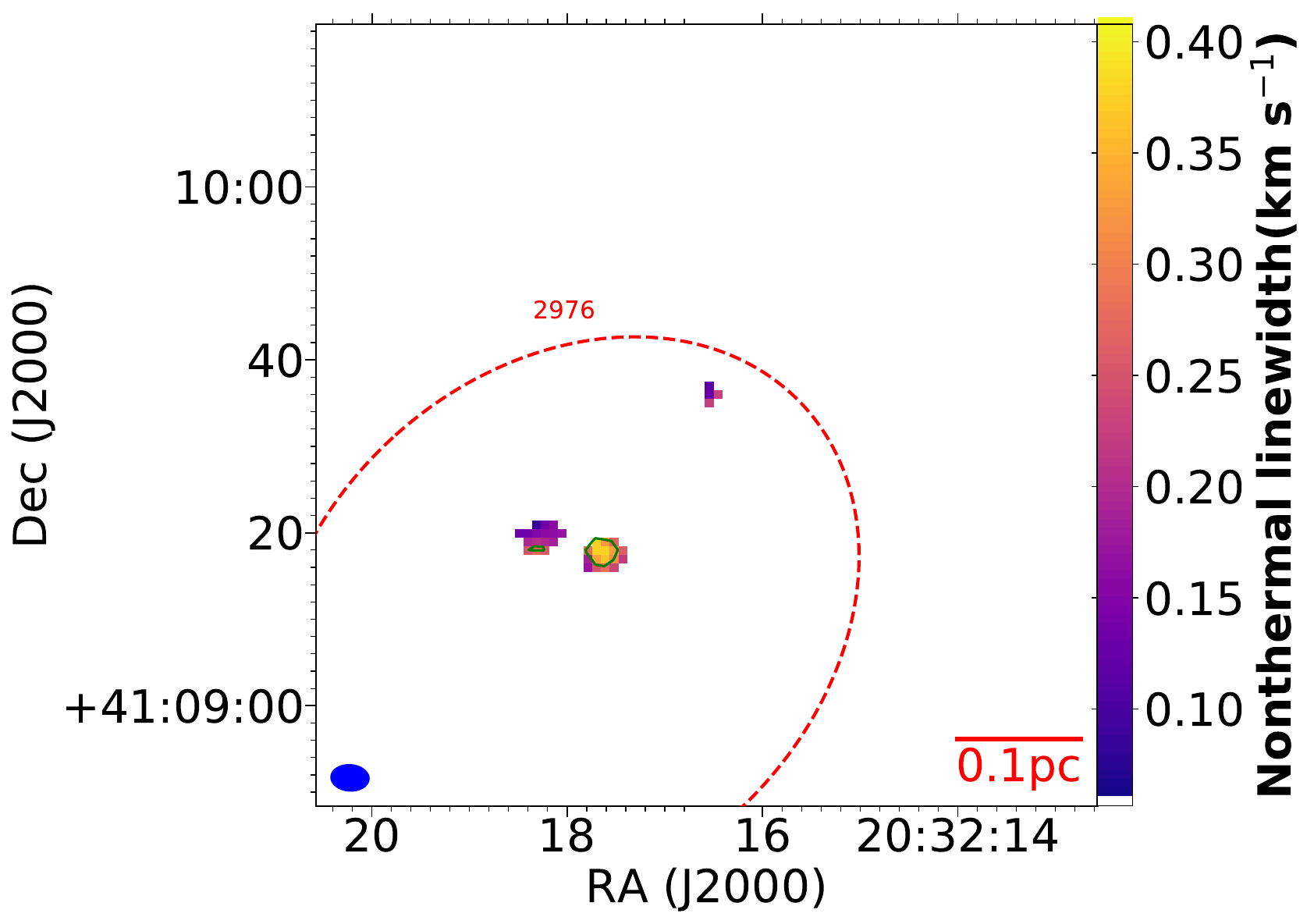} \\ 
Field 31 & Field 32 & Field 33 \\ 
\includegraphics[width=.3\textwidth]{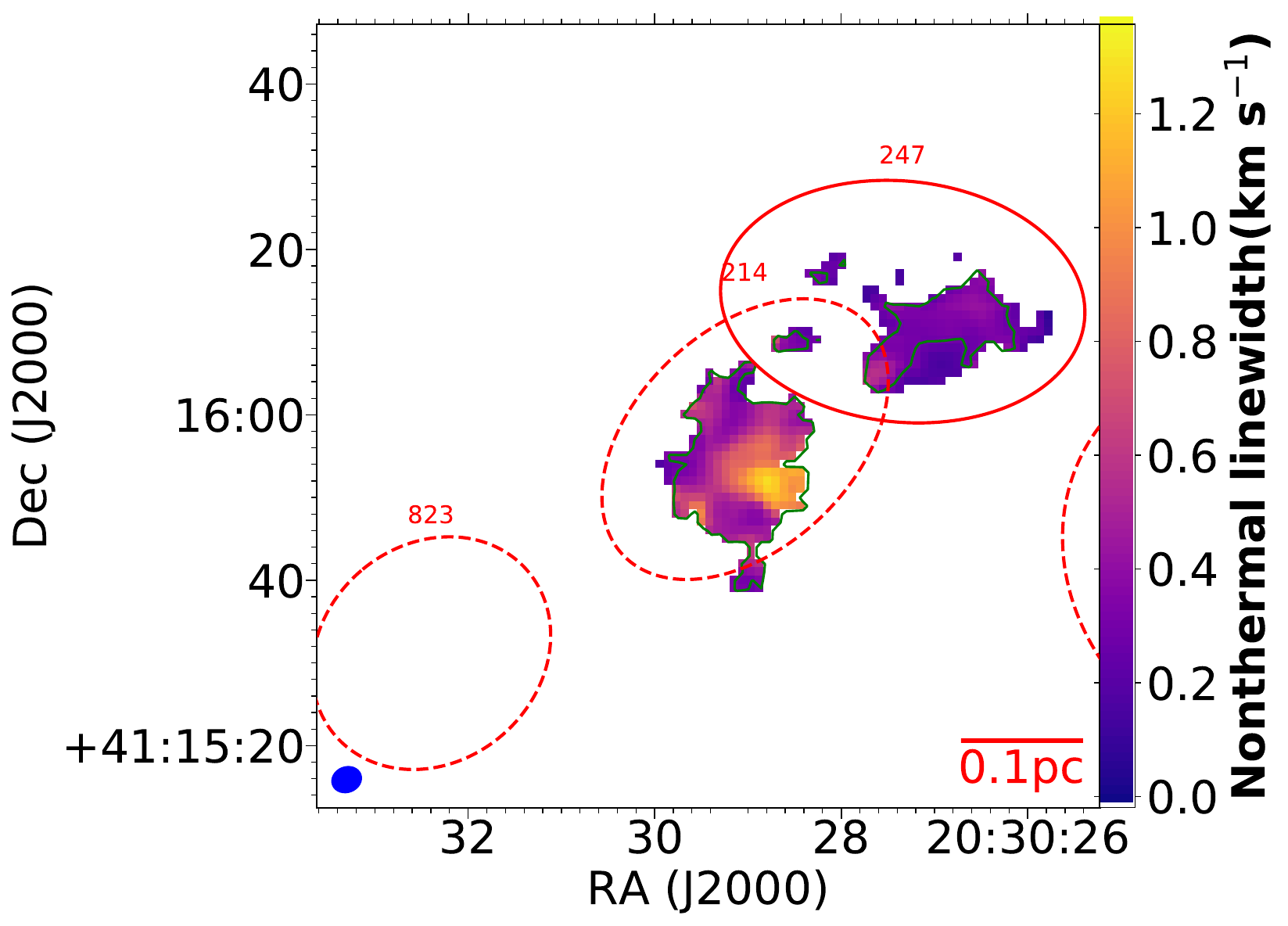} & 
\includegraphics[width=.3\textwidth]{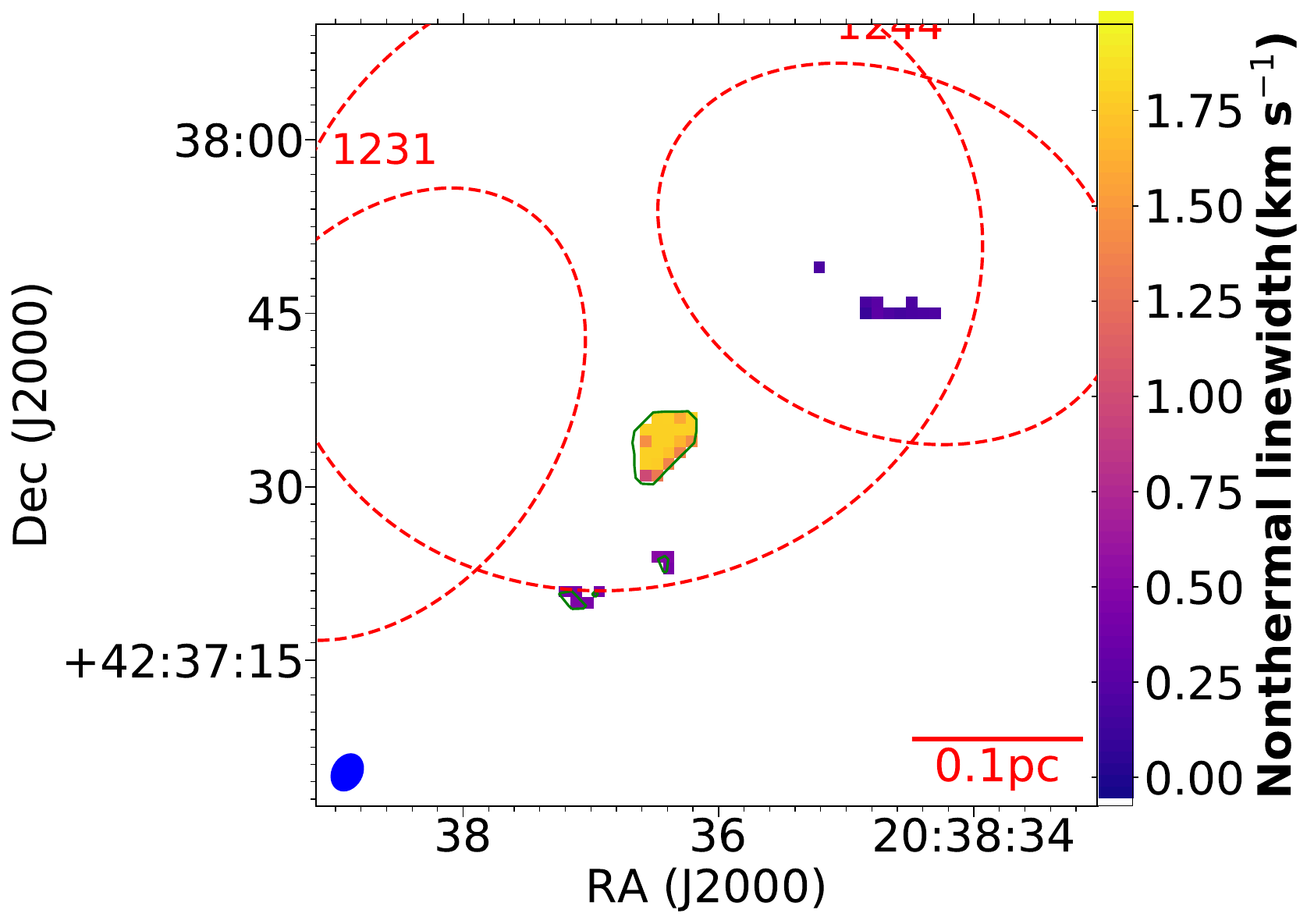} & 
\includegraphics[width=.3\textwidth]{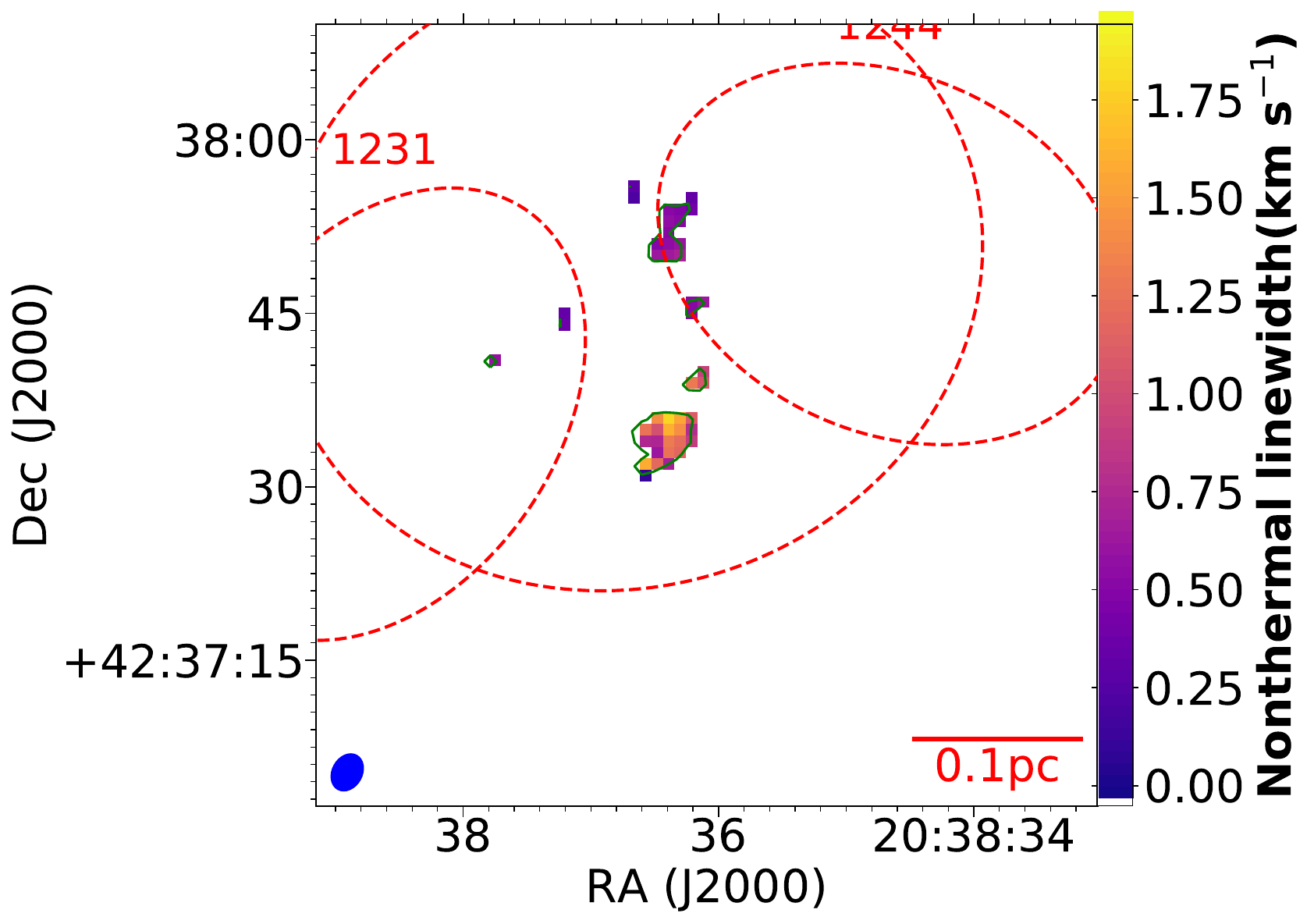} \\ 
Field 34 & Field 37 - Component 1 & Field 37 - Component 2 \\ 
\end{longtable}

\section{Derivation of different velocity dispersions} \label{app:E}

There are three velocity dispersions used in our analyses of MDCs and \nh\ fragments: the total velocity dispersion $\sigma_v$, the nonthermal velocity dispersion including the bulk motions $\sigma_{v, {\rm NT}}$, and the nonthermal velocity dispersion excluding the bulk motions $\sigma_{v, {\rm ng}}$.

First, for the total velocity dispersion, $\sigma_v$, both the velocity dispersion of a certain beam, $\sigma_{v,i}$, and the gradient of \vlsr\ need to be taken into account. The mean value of $\sigma_{v,i}$ is calculated as root mean square (RMS) weighed by the integrated flux of the \oneone\ line; that is,
\begin{equation}
\overline{\sigma_{v,i}} = \sqrt{\Sigma(w\times\sigma_{v,i}^2)/\Sigma(w)},
\end{equation}
where $w$ is the integrated flux of the \nh\ \oneone\ line from their moment-0 maps. The contribution of \vlsr\ is derived as 
\begin{equation}
\Delta v_{\rm lsr} = \sqrt{\Sigma[w\times(v_{{\rm lsr},i}-\overline{v_{{\rm lsr},i}})^2]/\Sigma(w)},
\end{equation}
where $v_{{\rm lsr},i}$ is the centroid velocity of a certain beam, and $\overline{v_{{\rm lsr},i}}=\Sigma(w\times v_{{\rm lsr},i})/\Sigma(w)$ is the weighted mean value $v_{{\rm lsr},i}$. Here, $\sigma_v$ is then derived as 
\begin{equation}
\sigma_v = \sqrt{\overline{\sigma_{v,i}}^2+\Delta v_{\rm lsr}^2}.
\end{equation}
Generally, $\sigma_v$, including all the observed motions, is used in the virial analysis.

Similarly, $\sigma_{v, {\rm NT}}$ is calculated as the weighted RMS value from the nonthermal velocity dispersion maps combined with the contribution of \vlsr, which is
\begin{equation}
\sigma_{v, {\rm NT}} = \sqrt{\Sigma(w\times\sigma_{v,NTi}^2)/\Sigma(w)+\Delta v_{\rm lsr}^2},
\end{equation}
where $\sigma_{v,NTi}$ is the nonthermal velocity dispersion of a certain beam.

In order to calculate $\sigma_{v, {\rm ng}}$, we fit the bulk motions of $v_{\rm bm}$ with a 2D linear function on the centroid velocity maps following the method provided in \citet{1993ApJ...406..528G}:
\begin{equation}
v_{\rm bm}=v_0+a\times\delta\alpha+b\times\delta\beta,
\end{equation}
where $\delta\alpha$ and $\delta\beta$ are the offsets in right ascension (RA) and declination (Dec). Then, the contribution of the bulk motion is estimated as
\begin{equation}
\Delta v_{\rm bm}=\sqrt{\Sigma[w\times(v_{\rm bm,i}-\overline{v_{\rm bm}})^2]/\Sigma(w)},
\end{equation}
where $v_{\rm bm,i}$ is the fitted velocity of a certain beam. Deducting $\Delta v_{\rm bm}$ from $\sigma_{v, {\rm NT}}$, we can obtain $\sigma_{v, {\rm ng}}$ as 
\begin{equation}
\sigma_{v, {\rm ng}} = \sqrt{\sigma_{v, {\rm NT}}^2 - \Delta v_{\rm bm}^2}.
\end{equation}
Both $\sigma_{v, {\rm NT}}$ and $\sigma_{v, {\rm ng}}$ are used in the calculation of the Mach numbers of different molecular structures.

\end{appendix}
\end{document}